\documentclass[twocolumn,twocolappendix]{aastex631}
\usepackage{amsmath}

\received{October 14, 2022}
\revised{December 6, 2022}
\accepted{January 6, 2023}

\shorttitle{}
\shortauthors{Yamada et al.}
\graphicspath{{./}{figures/}}
\begin{document}

\title{Hard X-Ray to Radio Multiwavelength SED Analysis of Local U/LIRGs in GOALS Sample with Self-consistent AGN Model Including Polar-dust Component}

\correspondingauthor{Satoshi Yamada}
\email{satoshi.yamada@riken.jp}

\author[0000-0002-9754-3081]{Satoshi Yamada}
\affiliation{RIKEN Cluster for Pioneering Research, 2-1 Hirosawa, Wako, Saitama 351-0198, Japan}
\affiliation{Department of Astronomy, Kyoto University, Kitashirakawa-Oiwake-cho, Sakyo-ku, Kyoto 606-8502, Japan}

\author[0000-0001-7821-6715]{Yoshihiro Ueda}
\affiliation{Department of Astronomy, Kyoto University, Kitashirakawa-Oiwake-cho, Sakyo-ku, Kyoto 606-8502, Japan}

\author[0000-0002-8653-020X]{Mart\'{\i}n Herrera-Endoqui}
\affiliation{Instituto de Astronom\'{\i}a sede Ensenada, Universidad Nacional Aut\'onoma de M\'exico, Km 107, Carret. Tij.-Ens., Ensenada, 22060, BC, M\'exico}

\author[0000-0002-3531-7863]{Yoshiki Toba}
\affiliation{National Astronomical Observatory of Japan, 2-21-1 Osawa, Mitaka, Tokyo 181-8588, Japan}
\affiliation{Department of Astronomy, Kyoto University, Kitashirakawa-Oiwake-cho, Sakyo-ku, Kyoto 606-8502, Japan}
\affiliation{Department of Physics, Nara Women’s University, Kitauoyanishi-machi, Nara, Nara
630-8506, Japan}
\affiliation{Academia Sinica Institute of Astronomy and Astrophysics, 11F of Astronomy-Mathematics
Building, AS/NTU, No.1, Section 4, Roosevelt Road, Taipei 10617, Taiwan}
\affiliation{Research Center for Space and Cosmic Evolution, Ehime University, 2-5 Bunkyo-cho, Matsuyama, Ehime 790-8577, Japan}

\author[0000-0002-7562-485X]{Takamitsu Miyaji}
\affiliation{Instituto de Astronom\'{\i}a sede Ensenada, Universidad Nacional Aut\'onoma de M\'exico, Km 107, Carret. Tij.-Ens., Ensenada, 22060, BC, M\'exico}

\author[0000-0002-5701-0811]{Shoji Ogawa}
\affiliation{Department of Astronomy, Kyoto University, Kitashirakawa-Oiwake-cho, Sakyo-ku, Kyoto 606-8502, Japan}

\author[0000-0001-6653-779X]{Ryosuke Uematsu}
\affiliation{Department of Astronomy, Kyoto University, Kitashirakawa-Oiwake-cho, Sakyo-ku, Kyoto 606-8502, Japan}

\author[0000-0002-0114-5581]{Atsushi Tanimoto}
\affiliation{Graduate School of Science and Engineering, Kagoshima University, Kagoshima 890-0065, Japan}

\author[0000-0001-6186-8792]{Masatoshi Imanishi}
\affiliation{National Astronomical Observatory of Japan, 2-21-1 Osawa, Mitaka, Tokyo 181-8588, Japan}
\affiliation{Department of Astronomical Science, Graduate University for Advanced Studies (SOKENDAI), 2-21-1 Osawa, Mitaka, Tokyo 181-8588, Japan}

\author[0000-0001-5231-2645]{Claudio Ricci}
\affiliation{N\'ucleo de Astronom\'{\i}a de la Facultad de Ingenier\'{\i}a, Universidad Diego Portales, Av. Ej\'ercito Libertador 441, Santiago, Chile}
\affiliation{Kavli Institute for Astronomy and Astrophysics, Peking University, Beijing 100871, People's Republic of China}




\begin{abstract}
We conduct a hard X-ray to radio multiwavelength spectral energy distribution (SED) decomposition for 57 local luminous and ultraluminous infrared galaxies (U/LIRGs) observed with Nuclear Spectroscopic Telescope Array and/or Swift/Burst Alert Telescope in GOALS \citep{Armus2009} sample.
We modify the latest SED-fitting code X-CIGALE by implementing the infrared (IR) CLUMPY model, allowing the multiwavelength study with the X-ray torus model (XCLUMPY) self-consistently.
Adopting the torus parameters obtained by the X-ray fitting \citep{Yamada2021}, we estimate the properties of host galaxies, active galactic nucleus (AGN) tori, and polar dust.
The star formation rates (SFRs) become larger with merger stage and most of them are above the main sequence. 
The SFRs are correlated with radio luminosity, indicating starburst emission is dominant in the radio band.
Although polar-dust extinction is much smaller than torus extinction, the UV-to-IR (mainly IR) polar dust luminosities are $\sim$2 times larger than the torus ones.
The polar-dust temperature decreases while the physical size, estimated by the temperature and dust sublimation radius, increases with AGN luminosity from a few tens of parsec (early mergers) to kiloparsec scales (late mergers), where the polar dust is likely the expanding (i.e., evolving) dusty outflows.
The comparison between SFRs and intrinsic AGN luminosities suggests that the starbursts occur first and AGNs arise later, and overall their growth rates follow the simultaneous coevolution relation that can establish the local galaxy--SMBH mass relation.
We confirm the coexistence of intense starbursts, AGNs, and large-scale outflows in late mergers, supporting a standard AGN feedback scenario.

\end{abstract}



%%%%%%%%%%%%%%%%%%%%%%%%%%%%%%%%
%%% Section 1 : Introduction %%%
%%%%%%%%%%%%%%%%%%%%%%%%%%%%%%%%

\section{Introduction}\label{S1-intro}
Galaxies and supermassive black holes (SMBHs) at their centers show a tight correlation between the masses of bulges ($M_{\rm bulge}$) and SMBHs ($M_{\rm BH}$), indicating that they have coevolved by regulating each other's growth (see e.g., \citealt{Marconi2003,Kormendy2013}).
Due to a huge scale gap between galaxies ($\sim$1~kpc) and SMBHs ($\ll$1~pc), the mechanism of their physical connection has been controversial \citep[e.g.,][]{Hopkins2008,Madau2014}.
For understanding the mechanism of coevolution, the merging galaxies have received attention since the merger can extract the angular momentum of the gas and trigger obscuration and rapid accretion onto SMBHs \citep[e.g.,][]{Koss2010,Koss2012,Koss2018,Kocevski2015,Lansbury2017,Ricci2017bMNRAS,Ricci2021b,Yamada2018,Yamada2021}.

According to the major merger scenario, galaxies in the most active phase of mergers become luminous infrared galaxies (LIRGs; $L_{\rm 8-1000\mu m} \geq 10^{11} L_{\odot}$) and ultraluminous infrared galaxies (ULIRGs; $L_{\rm 8-1000\mu m} \geq 10^{12} L_{\odot}$), combined as U/LIRGs \citep{Sanders1996}.
Their large bolometric luminosities are derived from the starbursts and/or active galactic nuclei (AGNs) surrounded by gas and dust, most of which are radiated in the infrared (IR) band.
After quenching the star-forming activities, U/LIRGs are thought to transit to unobscured quasars or elliptical galaxies \citep[e.g.,][]{Hopkins2008}.

The AGNs in U/LIRGs are well studied in the X-ray to radio multiwavelength bands (e.g., \citealt{U2022} for a review). 
When AGNs are heavily obscured, they radiate dimmed UV-to-near-IR emission compared to the AGN luminosity identified in the mid-IR and X-ray bands \citep[see e.g.,][]{Hickox2018}.
The mid-IR studies reveal that the AGNs in U/LIRGs are deeply buried by a large amount of gas and dust, where even the direction of the lowest dust column-density can be opaque to the ionizing UV photons  \citep[e.g.,][]{Imanishi2006,Imanishi2008,Lee2012,Yamada2019}.
Hard X-ray observations, which are less affected by the contamination of starburst emission, can identify the AGNs with large hydrogen column densities ($N_{\rm H}$).
For local U/LIRGs observed with Nuclear Spectroscopic Telescope Array (NuSTAR; \citealt{Harrison2013}), \citet{Ricci2017bMNRAS,Ricci2021b} analyze the broadband X-ray spectra and find the large fraction of Compton-thick (CT; $N_{\rm H} > 10^{24}$~cm$^{-2}$) AGNs in late mergers.
By conducting the X-ray spectroscopy with the X-ray clumpy torus model (XCLUMPY; \citealt{Tanimoto2019}), \citet{Yamada2021} present the individual torus covering fractions, supporting the buried AGN structure in the late mergers.

\citet{Yamada2021} report that the AGNs in the final phase of mergers show high Eddington ratios ($\lambda_{\rm Edd}$) and the signatures of multiphase outflows at subparsec to kiloparsec scales: that is, ultrafast outflows (UFOs), ionized outflows, and molecular outflows.
The UFOs are extremely fast ($\sim$0.1--0.3$c$) and highly ionized winds at $\sim$0.01~pc (e.g., \citealt{Tombesi2015,Mizumoto2019,Smith2019}).
The ionized outflows are the fast ($\sim$1000~km~s$^{-1}$) winds with a size of kiloparsecs detected by optical and near-IR spectroscopy (e.g., \citealt{Rich2015,Cortijo-Ferrero2017b,Kakkad2018,Boettcher2020,Fluetsch2021}).
Molecular outflows are cold gas winds on $\sim$400~pc with a velocity of $\sim$500~km~s$^{-1}$ discovered at far-IR and submillimeter wavelengths (e.g., \citealt{Spoon2013,Veilleux2013,Cicone2014,Gonzalez-Alfonso2017,Laha2018}).
However, the properties of outflows in many U/LIRGs are still unclear because these methods using the blue-shifted emission/absorption lines have difficulty detecting weak outflows.
Therefore, the new methods to systematically reveal the outflow properties are necessary to present the schematic picture of the merger-driven coevolution.

The observations of outflowing dust \citep[e.g.,][]{Honig2019,Venanzi2020} may be an ideal means for the systematic studies on the outflows in U/LIRGs.
Recent mid-IR observations with high spatial resolutions detect the extended dust emission along the polar direction, called polar dust (e.g., \citealt{Asmus2016,Stalevski2017}).
An issue at this point is that, due to the poor understanding of the polar dust structure, it is not even clear whether the polar dust is (1) the galactic ISM or dust in a narrow line region (NLR) only being illuminated by the AGN or (2) the ``outflowing'' dusty winds launched from the inner edge of the torus.
\citet{Asmus2019} finds a positive correlation between their physical sizes and the Eddington ratios, supporting that the polar dust structure may be related to the AGN activities.
The analytical model of polar dust emission has been implemented into the latest multiwavelength spectral energy distribution (SED) models (e.g., X-CIGALE; \citealt{Boquien2019,Yang2020}).
These models help us to estimate the properties of polar dust (e.g., \citealt{Toba2021b,Buat2021}), even though the degeneracy of torus and polar dust emission will provide uncertainties.

The combination of X-ray and IR observations will become the best way to constrain the properties of polar dust.
The broadband X-ray study with XCLUMPY for several tens of Swift/Burst Alert Telescope (BAT)-selected AGNs by \citet{Ogawa2021} successfully constrains the torus covering fractions ($C_{\rm T}$).
These estimates are well consistent with the typical $\lambda_{\rm Edd}$--$C_{\rm T}$ relation (\citealt{Ricci2017bMNRAS}), while are much smaller than those estimated with the IR CLUMPY model (\citealt{Nenkova2008a,Nenkova2008b}).
They indicate that the difference can be explained by the presence of the polar dust, which should have small effects on the obscuration in the X-ray band but re-radiate the large IR luminosities.
Thus, the updated CLUMPY model consisting of the clumpy torus and polar dust components will provide rich information on polar dust by applying the torus parameters obtained by the X-ray fitting with XCLUMPY \citep{Yamada2021}.

In this study, we generate the updated CLUMPY model that can reproduce the IR emission from the clumpy torus and polar dust, and perform the multiwavelength (hard X-ray to radio) SED decomposition for the sample of 57 local U/LIRGs in \citet{Yamada2021}.
We first investigate the features of multiwavelength radiation, which are helpful to understand the environments from the nucleus to galactic scales.
This will be incidentally useful for future works to explore buried AGNs in U/LIRGs in the distant universe.
Taking into account these results, we examine the polar dust structures and their relation to the AGN activities, and finally discuss the coevolution process of the galaxies, SMBHs, and outflows in the merger phase.

This paper is organized as follows.
In Section~\ref{S2-sample} and Section~\ref{S3-data}, we show the sample and their photometric data of multiwavelength observations, respectively.
Section~\ref{S4-model} describes the implementation of the updated CLUMPY model we create and the best-fit parameters of the multiwavelength SED decomposition.
Section~\ref{S5-results} illustrates the features of multiwavelength emission to understand the characteristics of the host galaxies and AGNs in U/LIRGs.
In Section~\ref{S6-discussion1}, we discuss the structure of the polar dust in U/LIRGs based on the results of the multiwavelength SED analysis.
Section~\ref{S7-discussion2} provides the discussion on the coevolution process of galaxies, SMBHs, and outflows in U/LIRGs.
The main results are summarized in Section~\ref{S8-summary}.
In this paper, we adopt the cosmology of a flat universe with $H_{\rm 0} = 70$~km~s$^{-1}$~Mpc$^{-1}$, $\Omega_{\rm M} = 0.3$, and $\Omega_{\rm \Lambda} = 0.7$. 
Throughout this paper, the uncertainties are at the 1$\sigma$ level unless otherwise stated, and the initial mass function (IMF) of \citet{Chabrier2003} is assumed.\footnote{When the values of the star formation rates (SFR) and stellar masses ($M_*$) are referred from previous works, we correct these values assuming \citet{Salpeter1955} by decreasing 0.15 and 0.24 dex, or values assuming \citet{Kroupa2001} by subtracting 0.02 and 0.03 dex, respectively \citep[e.g.,][]{Santini2014,Speagle2014}.}
\\

%%%%%%%%%%%%%%%%%%%%%%%%%%
%%% Section 2 : Sample %%%
%%%%%%%%%%%%%%%%%%%%%%%%%%

\section{Sample} \label{S2-sample}
The sample selection of 57 local U/LIRGs observed with the hard X-ray observations is described as follows.
We focus on the Great Observatories All-sky LIRG Survey (GOALS; \citealt{Armus2009}), which consists of 180 LIRGs and 22 ULIRGs in the local universe at redshifts $z < 0.088$.
They are contained in the IRAS Revised Bright Galaxy Sample (RBGS; \citealt{Sanders2003}), a complete sample of 629 extragalactic objects having 60~$\mu$m fluxes above 5.24~Jy at Galactic latitudes $|b| > 5$\degr.
These U/LIRGs have been observed in the multiwavelength bands, by the IR telescopes of Spitzer (e.g., \citealt{Imanishi2007,Diaz-Santos2010,Petric2011,Inami2013,Stierwalt2013}), AKARI (e.g., \citealt{Imanishi2008,Lee2012,Inami2018}),
Herschel (e.g., \citealt{Diaz-Santos2013,Diaz-Santos2014,Chu2017,Lu2017}), and X-ray telescopes of Chandra (e.g., \citealt{Iwasawa2011,Torres-Alba2018}), and NuSTAR (\citealt{Teng2015,Ricci2017bMNRAS,Ricci2021b,Privon2020,Yamada2021}).
Here, we select the same targets as in \citet{Yamada2021}, the 57 local U/LIRGs containing 84 individual galaxies observed with the hard X-ray telescopes NuSTAR and/or Swift/BAT.
According to the detailed X-ray spectral analysis combining the available soft X-ray observations, our targets are comprised of 40 AGNs (two unobscured AGNs, 21 obscured AGNs, 16 CT AGNs, one jet-dominated AGN), and 44 starburst-dominant or hard X-ray--undetected sources.

Here, the source coordinates, redshift, merger stages, and the projected separation between the two nuclei are taken from Table~1 of \citet{Yamada2021}.
Based on high-spatial-resolution images (e.g., \citealt{Stierwalt2013}), the merger stages are classified into five stages:
stage A (galaxy pairs before a first encounter), stage B (post-first encounter with symmetric galaxy disks but showing signature of tidal tails), stage C (showing some signatures of mergers, such as tidal tails, amorphous disks), stage D (two nuclei within a common envelope), and stage N (no signatures of mergers or massive neighbors).
In the same manner as \citet{Yamada2021}, we hereafter call stage A--B galaxies as early mergers, stage C--D as late mergers, and stage N as nonmergers.

\clearpage
\startlongtable
\begin{deluxetable*}{lllcccccccc}
\label{T1_targets}
\tablecaption{Basic Information of our Sample}
%\tablewidth{700pt}
\tabletypesize{\footnotesize}
\tablehead{
\colhead{ID} &
\colhead{IRAS Name} &
\colhead{Object Name} &
\colhead{$z$} &
\colhead{$M$} &
\colhead{$D_{12}$} &
\colhead{$D_{12}$} &
\colhead{AGN} &
\colhead{Type} &
\colhead{log($L_{\rm IR}$)} &
\colhead{IRS}\\ 
& & & & & (arcsec) & (kpc) & & & ($L_{\odot}$) &
}
\decimalcolnumbers
\startdata
ID01 & F00085$-$1223 & NGC~34 & 0.0196 & D & S & S & Y & Obs & 11.49 & Y\\
ID02 & F00163$-$1039$^a$ & MCG$-$02-01-052/MCG$-$02-01-051 & 0.0272 & B & $\cdots$ & $\cdots$ & $\cdots$ & $\cdots$ & $\cdots$ & $\cdots$ \\
ID03 & F00163$-$1039 N & MCG$-$02-01-052 & 0.0273 & B & 56.1 & 30.7 & n & n & [11.48] & n\\
ID04 & F00163$-$1039 S & MCG$-$02-01-051 & 0.0271 & B & 56.1 & 30.7 & n & n & 11.48 & Y\\
ID05 & F00344$-$3349 & ESO~350$-$38 & 0.0206 & C & 5.2 & 2.2 & Y? & n & 11.28 & Y\\
ID06 & F00402$-$2349$^a$ & NGC~232/NGC~235 & 0.0224 & B & $\cdots$ & $\cdots$ & $\cdots$ & $\cdots$ & $\cdots$ & $\cdots$ \\
ID07 & F00402$-$2349 W & NGC~232 & 0.0226 & B & 120.8 & 54.7 & n & n & [11.44]: & Y\\
ID08 & F00402$-$2349 E & NGC~235 & 0.0222 & B & 120.8 & 54.7 & Y & Obs & [11.44]: & Y\\
ID09 & F00506+7248 & MCG+12-02-001 & 0.0157 & C & 0.9 & 0.3 & Y & CT & 11.50 & Y\\
ID10 & F01053$-$1746$^b$ & IC~1623A/IC~1623B & 0.0202 & C & 15.7 & 6.4 & n & n/n & 11.71 & n/Y\\
ID11 & F02071$-$1023$^a$ & NGC~833/NGC~835 & 0.0132 & A & $\cdots$ & $\cdots$ & $\cdots$ & $\cdots$ & $\cdots$ & $\cdots$ \\
ID12 & F02071$-$1023 W2 & NGC~833 & 0.0129 & A & 56.4 & 15.3 & Y & Obs & [11.05]: & Y\\
ID13 & F02071$-$1023 W & NGC~835 & 0.0136 & A & 56.4 & 15.3 & Y & Obs & [11.05]: & Y\\
ID14 & F02071$-$1023 E & NGC~838 & 0.0128 & A & 148.8 & 39.1 & n & n & [11.05]: & Y\\
ID15 & F02071$-$1023 S & NGC~839 & 0.0129 & A & 148.8 & 39.1 & n & n & [11.05]: & Y\\
ID16 & F02401$-$0013 & NGC~1068 & 0.0038$^\dagger$ & N & n & n & Y & CT & 11.40 & Y\\
ID17 & F03117+4151$^a$ & UGC~2608/UGC~2612 & 0.0276 & N & $\cdots$ & $\cdots$ & $\cdots$ & $\cdots$ & $\cdots$ & $\cdots$ \\
ID18 & F03117+4151 N & UGC~2608 & 0.0233 & N & n & n & Y & CT & 11.41 & Y\\
ID19 & F03117+4151 S & UGC~2612 & 0.0318 & N & n & n & n & n & [11.41] & Y\\
ID20 & F03164+4119 & NGC~1275 & 0.0176 & N & n & n & Y & Jet & 11.26 & Y\\
ID21 & F03316$-$3618 & NGC~1365 & 0.0055$^\dagger$ & N & n & n & Y & Obs & 11.00 & Y\\
ID22 & F04454$-$4838 & ESO~203$-$1 & 0.0529 & B & 7.5 & 7.7 & n & n & 11.86 & Y\\
ID23 & F05054+1718$^a$ & CGCG~468-002W/CGCG~468-002E & 0.0171 & B & $\cdots$ & $\cdots$ & $\cdots$ & $\cdots$ & $\cdots$ & $\cdots$ \\
ID24 & F05054+1718 W & CGCG~468-002W & 0.0175 & B & 29.5 & 10.3 & Y & Obs & [11.22]: & Y\\
ID25 & F05054+1718 E & CGCG~468-002E & 0.0168 & B & 29.5 & 10.3 & n & n & [11.22]: & Y\\
ID26 & F05189$-$2524 & IRAS~F05189$-$2524 & 0.0426 & D & S & S & Y & Obs & 12.16 & Y\\
ID27 & F06076$-$2139$^b$ & IRAS~F06076$-$2139/ & 0.0374 & C & 8.3 & 6.2 & Y & Obs/n & 11.65 & Y/n\\
 & & 2MASS~06094601$-$2140312 \\
ID28 & F08354+2555 & NGC~2623 & 0.0185 & D & S & S & Y & Obs & 11.60 & Y\\
ID29 & F08520$-$6850$^b$ & ESO~060$-$IG016~West/East & 0.0451 & B & 15.4 & 13.6 & Y & n/Obs & 11.82 & n/Y\\
ID30 & F08572+3915 & IRAS~F08572+3915 & 0.0580 & D & 4.4 & 5.6 & Y & Obs & 12.16 & Y\\
ID31 & F09320+6134 & UGC~5101 & 0.0394 & D & S & S & Y & Obs & 12.01 & Y\\
ID32 & F09333+4841$^a$ & MCG+08-18-012/MCG+08-18-013 & 0.0255 & A & $\cdots$ & $\cdots$ & $\cdots$ & $\cdots$ & $\cdots$ & $\cdots$ \\
ID33 & F09333+4841 W & MCG+08-18-012 & 0.0252 & A & 65.3 & 33.6 & n & n & [11.34] & n\\
ID34 & F09333+4841 E & MCG+08-18-013 & 0.0259 & A & 65.3 & 33.6 & n & n & 11.34 & Y\\
ID35 & F10015$-$0614$^a$ & MCG$-$01-26-013/NGC~3110 & 0.0165 & A & $\cdots$ & $\cdots$ & $\cdots$ & $\cdots$ & $\cdots$ & $\cdots$ \\
ID36 & F10015$-$0614 S & MCG$-$01-26-013 & 0.0161 & A & 108.8 & 36.5 & n & n & [11.37] & n\\
ID37 & F10015$-$0614 N & NGC~3110 & 0.0169 & A & 108.8 & 36.5 & n & n & 11.37 & Y\\
ID38 & F10038$-$3338 & ESO~374$-$IG032 & 0.0340 & D & S & S & Y? & n & 11.78 & Y\\
ID39 & F10257$-$4339 & NGC~3256 & 0.0094 & D & 5.1 & 1.0 & n & n & 11.64 & Y\\
ID40 & F10565+2448 & IRAS~F10565+2448 & 0.0431 & D & 7.4 & 6.7 & n & n & 12.08 & Y\\
\tablebreak
ID41 & F11257+5850$^b$ & NGC~3690~West/East & 0.0103 & C & 22.0 & 4.6 & Y & CT/n & 11.93 & Y/Y\\
ID42 & F12043$-$3140$^b$ & ESO~440$-$58/MCG$-$05-29-017 & 0.0230 & B & 11.8 & 5.5 & n & n/n & 11.43 & Y/Y\\
ID43 & F12112+0305 & IRAS~F12112+0305 & 0.0733 & D & 3.0 & 4.1 & n & n & 12.36 & Y\\
ID44 & F12243$-$0036$^a$ & NGC~4418/MCG+00-32-013 & 0.00735 & A & $\cdots$ & $\cdots$ & $\cdots$ & $\cdots$ & $\cdots$ & $\cdots$ \\
ID45 & F12243$-$0036 NW & NGC~4418 & 0.0073$^\dagger$ & A & 179.9 & 29.4 & Y? & n & 11.19 & Y\\
ID46 & F12243$-$0036 SE & MCG+00-32-013 & 0.0074$^\dagger$ & A & 179.9 & 29.4 & n & n & [11.19] & n\\
ID47 & F12540+5708 & Mrk~231 & 0.0422 & D & S & S & Y & Obs & 12.57 & Y\\
ID48 & F12590+2934$^b$ & NGC~4922S/NGC~4922N & 0.0238 & C & 22.1 & 10.6 & Y & n/Obs & 11.38 & n/Y\\
ID49 & F13126+2453 & IC~860 & 0.0112 & N & n & n & n & n & 11.14 & Y\\
ID50 & 13120$-$5453 & IRAS~13120$-$5453 & 0.0308 & D & S & S & Y & CT & 12.32 & Y\\
ID51 & F13188+0036 & NGC~5104 & 0.0186 & N & n & n & n & n & 11.27 & Y\\
ID52 & F13197$-$1627 & MCG$-$03-34-064 & 0.0165 & N & n & n & Y & Obs & 11.28 & Y\\
ID53 & F13229$-$2934 & NGC~5135 & 0.0137 & N & n & n & Y & CT & 11.30 & Y\\
ID54 & F13362+4831$^b$ & Mrk~266B/Mrk~266A & 0.0278 & B & 10.1 & 5.6 & Y & CT/Obs & 11.56 & Y/Y\\
ID55 & F13428+5608 & Mrk~273 & 0.0378 & D & 0.9 & 0.7 & Y & Obs & 12.21 & Y\\
ID56 & F14348$-$1447 & IRAS~F14348$-$1447 & 0.0827 & D & 3.4 & 5.2 & Y & CT & 12.39 & Y\\
ID57 & F14378$-$3651 & IRAS~F14378$-$3651 & 0.0681 & D & S & S & n & Y & 12.23 & Y\\
ID58 & F14544$-$4255$^b$ & IC~4518A/IC~4518B & 0.0159 & B & 35.5 & 11.5 & Y & Obs/n & 11.23 & Y/n\\
ID59 & F15250+3608 & IRAS~F15250+3608 & 0.0552 & D & 0.7 & 0.8 & n & n & 12.08 & Y\\
ID60 & F15327+2340$^b$ & Arp~220W/Arp~220E & 0.0181 & D & 1.0 & 0.4 & Y & CT/n & 12.28 & Y(u)\\
ID61 & F16504+0228$^b$ & NGC~6240S/NGC~6240N & 0.0245 & D & 1.7 & 0.8 & Y & CT & 11.93 & Y(u)\\
ID62 & F16504+0228$^a$ & NGC~6285/NGC~6286 & 0.0186 & B & $\cdots$ & $\cdots$ & $\cdots$ & $\cdots$ & $\cdots$ & $\cdots$ \\
ID63 & F16577+5900 N & NGC~6285 & 0.0190 & B & 91.0 & 34.5 & n & n & [11.37] & Y\\
ID64 & F16577+5900 S & NGC~6286 & 0.0183 & B & 91.0 & 34.5 & Y & CT & 11.37 & Y\\
ID65 & F17138$-$1017 & IRAS~F17138$-$1017 & 0.0173 & D & S & S & Y & Obs & 11.49 & Y\\
ID66 & F18293$-$3413 & IRAS~F18293$-$3413 & 0.0182 & N & S & S & n & n & 11.88 & Y\\
ID67 & F19297$-$0406 & IRAS~F19297$-$0406 & 0.0857 & D & S & S & n & n & 12.45 & Y\\
ID68 & F20221$-$2458$^b$ & NGC~6907/NGC~6908 & 0.0104 & B & 43.9 & 9.4 & n & n/n & 11.11 & Y/n\\
ID69 & 20264+2533$^a$ & NGC~6921/MCG+04-48-002 & 0.0142 & A & $\cdots$ & $\cdots$ & $\cdots$ & $\cdots$ & $\cdots$ & $\cdots$ \\
ID70 & 20264+2533 W & NGC~6921 & 0.0145 & A & 91.4 & 26.5 & Y & CT & [11.11] & n\\
ID71 & 20264+2533 E & MCG+04-48-002 & 0.0139 & A & 91.4 & 26.5 & Y & Obs & 11.11 & Y\\
ID72 & F20550+1655$^b$ & II~Zw~096/IRAS~F20550+1655~SE & 0.0353 & C & 11.6 & 8.1 & n & n/n & 11.94 & Y/Y\\
ID73 & F20551$-$4250 & ESO~286$-$19 & 0.0430 & D & S & S & n & n & 12.06 & Y\\
ID74 & F21453$-$3511 & NGC~7130 & 0.0162 & N & n & n & Y & CT & 11.42 & Y\\
ID75 & F23007+0836$^a$ & NGC~7469/IC~5283 & 0.0162 & A & $\cdots$ & $\cdots$ & $\cdots$ & $\cdots$ & $\cdots$ & $\cdots$ \\
ID76 & F23007+0836 S & NGC~7469 & 0.0163 & A & 79.7 & 26.2 & Y & Unobs & 11.65 & Y\\
ID77 & F23007+0836 N & IC~5283 & 0.0160 & A & 79.7 & 26.2 & n & n & [11.65] & n\\
ID78 & F23128$-$5919 & ESO~148$-$2 & 0.0446 & C & 4.5 & 3.9 & Y & CT & 12.06 & Y\\
ID79 & F23157+0618 & NGC~7591 & 0.0165 & N & n & n & n & n & 11.12 & Y\\
ID80 & F23254+0830$^a$ & NGC~7674/MCG+01-59-081 & 0.0292 & A & $\cdots$ & $\cdots$ & $\cdots$ & $\cdots$ & $\cdots$ & $\cdots$ \\
\tablebreak
ID81 & F23254+0830 W & NGC~7674 & 0.0289 & A & 33.3 & 19.5 & Y & Obs & 11.56 & Y\\
ID82 & F23254+0830 E & MCG+01-59-081 & 0.0295 & A & 33.3 & 19.5 & n & n & [11.56] & n\\
ID83 & 23262+0314$^a$ & NGC~7679/NGC~7682 & 0.0171 & A & $\cdots$ & $\cdots$ & $\cdots$ & $\cdots$ & $\cdots$ & $\cdots$ \\
ID84 & 23262+0314 W & NGC~7679 & 0.0171 & A & 269.8 & 93.8 & Y & Unobs & 11.11 & Y\\
ID85 & 23262+0314 E & NGC~7682 & 0.0171 & A & 269.8 & 93.8 & Y & Obs & [11.11] & n\\
\enddata
\tablecomments{Columns:
(1) Target ID;
(2) IRAS name. ``a'' marks 13 total systems of resolved pairs to our targets.
``b'' marks 12 U/LIRGs (24 individual galaxies) that are too close to be separated by the Hershel PACS 70 $\mu$m images or UV-to-near-IR images; 
(3) object name;
(4) redshift from NASA/IPAC Extragalactic Database (NED);
(5) merger stage based on the high-spatial-resolution images \citep[e.g.,][]{Stierwalt2013};
(6--7) projected separation between the two nuclei in arcseconds and kiloparsecs. S and n mean that a single nucleus is observed in merging and nonmerging U/LIRGs, respectively;
(8) Y and n mark the presence of a hard X-ray--detected AGN or not, respectively.
Y? marks AGN candidates among the hard X-ray--undetected sources identified by the multiwavelength SED analysis (see Section~\ref{subsub4-2-4-BIC});
(9) X-ray classification (Unobs = unobscured AGN, Obs = obscured AGN, CT = CT AGN, Jet = jet-dominant AGN, and n = hard X-ray–undetected sources);
(10) logarithmic total IR luminosity in units of $L_{\odot}$ \citep{Armus2009}. 
Values in brackets should be upper limits due to contamination from nearby much brighter (or equally bright with the suffix ``:'') IR sources;
(11) Y and n mark detection and nondetection with Spitzer/IRS in the 10--20 $\mu$m (SH) or 19--38 $\mu$m (LH) band, respectively \citep{Alonso-Herrero2012,Mazzarella2012,Inami2013}. 
The (u) means that the two nuclei are not divided.
All information in Column (2--11) are referred from \citet{Yamada2021}.
}
\tablenotetext{\dagger}{Throughout the paper, redshift-independent measurements of the luminosity distance are utilized for the multiwavelength results (e.g., luminosities) of the closest objects at $z < 0.01$, NGC 1068 (14.4~Mpc; \citealt{Tully1988}; \citealt{Bauer2015}), NGC 1365 (17.3~Mpc; \citealt{Venturi2018}), and NGC~4418/MCG$+$00-32-013 (34~Mpc; e.g., \citealt{Ohyama2019}).}
\tablenotetext{\ }{(This table is available in its entirety in machine-readable form.)\\}
\end{deluxetable*}

Some of the pair galaxies are too close to be separated by the other wavelength bands.
Particularly, the spatial distributions of far-IR emissions within the GOALS sources are poorly determined because of the limitations in the angular resolution of pre-Herschel data (IRAS, ISO, and AKARI).
\citet{Chu2017} provided the total system fluxes and component fluxes (where possible) in all Herschel bands for the GOALS sample.
The beam profiles have a mean FWHM value of 5\farcs6 for the Photodetector Array Camera and Spectrometer (PACS; \citealt{Pilbratt2010}) 70~$\mu$m band, the shortest wavelength band of Herschel data.
After the cross-matching with multiwavelength catalogs (Section~\ref{S3-data}), we treat each interacting pair as a total system for 11 U/LIRGs (22 individual galaxies) that are too close (with a projected separation of $\lesssim$20\arcsec) to be resolved by the Hershel PACS images, whose flux densities of individual galaxies are summed in all wavelength bands.
Since we do not obtain any fluxes for IC~4518B but for a total system in the UV-to-near-IR bands, the other wavelength photometries are also combined as total fluxes of IC~4518A and IC~4518B (35\farcs5).
Here, the 12 pairs are unresolved among 84 galaxies, and thus our sample has 72 individual sources in 57 local U/LIRGs.

Finally, to investigate the difference between the results of multiwavelength SED analysis by using the combined photometry of a total system and separated photometries of the individual galaxies, we duplicately add the 13 total systems of resolved pairs (with a projected separation of $\gtrsim$30\arcsec) to our targets (see Section~\ref{sub4-4-results}).
Therefore, our sample host 85 sources consisting of 72 individual sources and 13 systems of resolved pairs.
\\

%%%%%%%%%%%%%%%%%%%%%%%%%%%%%%%%%%%%%%%%%%%%%%%
%%% Section 3 : Multiwavelength Observation %%%
%%%%%%%%%%%%%%%%%%%%%%%%%%%%%%%%%%%%%%%%%%%%%%%

\section{Multiwavelength Observations} \label{S3-data}
For the 57 U/LIRGs (containing 84 individual sources in \citealt{Yamada2021}), we compiled the multiwavelength data from hard X-ray to radio bands.
As noted in Section~\ref{S2-sample}, the system fluxes of the close interacting pairs are calculated by adding the component fluxes.
We note that neither the upper limits nor low-significant ($<$5$\sigma$) values were utilized unless otherwise stated.
While considering the spatial resolutions of individual instruments, we took care to utilize the multiwavelength photometries whose apertures sufficiently cover the fluxes derived from the extended emission of each target.
In the following sections, we describe the multiwavelength data for each telescope.
The characteristics of the multiwavelength surveys are summarized in Table~\ref{T2_catalogs}.

% Table 2
\begin{deluxetable*}{llccccc}
\label{T2_catalogs}
\tablecaption{Characteristics of the Multiwavelength Catalogs Utilized in This Work }
\tabletypesize{\scriptsize}
\tablehead{
\colhead{Class} &
\colhead{Instrument} &
\colhead{Band} &
\colhead{Wavelength} &
\colhead{$\Delta$R (FWHM)} &
\colhead{Area Coverage} &
\colhead{Ref.}
}
\decimalcolnumbers
\startdata
Hard X-ray & Swift/BAT  & 14--195~keV & 0.0064--0.0886 nm & 1170\arcsec & All sky       & 1,2,3 \\
           & NuSTAR     &   3--79~keV &   0.016--0.413 nm & 18\arcsec & Targeted obs. & 3,4 \\
       & Suzaku/HXD-PIN &  10--70~keV &   0.018--0.124 nm & $\cdots$  & Targeted obs. & 3,5 \\
\hline
Soft X-ray & XMM-Newton & 0.1--15~keV & 0.083--12.40 nm & $\sim$6\arcsec   & Targeted obs. & 3,6 \\
           & Chandra    & 0.2--10~keV & 0.12--6.20 nm & $\lesssim$0\farcs5 & Targeted obs. & 3,7 \\
           & Suzaku/XIS & 0.2--12~keV & 0.10--6.20 nm & 96\arcsec--120\arcsec & Targeted obs. & 3,5 \\
           & Swift/XRT  & 0.3--10~keV & 0.12--4.13 nm & 8\farcs8           & Targeted obs. & 1,3 \\
\hline
UV         & GALEX      & FUV & 152.8 (134.4--178.6) nm & 4\farcs2  & 24,790 deg$^2$ & 8,9 \\
           & GALEX      & NUV & 231.0 (177.1--283.1) nm & 5\farcs2  & 24,790 deg$^2$ & 8,9 \\
\hline
Optical    & Pan-STARRS & $g$ & 481.1 (394.3--559.3) nm & 1\farcs47 & $\delta \geq -30$\degr\ (3$\pi$) & 10,11 \\
           & Pan-STARRS & $r$ & 615.6 (538.6--703.6) nm & 1\farcs31 & $\delta \geq -30$\degr\ (3$\pi$) & 10,11 \\
           & Pan-STARRS & $i$ & 750.4 (677.8--830.4) nm & 1\farcs19 & $\delta \geq -30$\degr\ (3$\pi$) & 10,11 \\
           & Pan-STARRS & $z$ & 866.9 (802.8--934.6) nm & 1\farcs14 & $\delta \geq -30$\degr\ (3$\pi$) & 10,11 \\
           & Pan-STARRS & $y$ &961.3 (910.1--1083.9) nm & 1\farcs09 & $\delta \geq -30$\degr\ (3$\pi$) & 10,11 \\
           & SkyMapper  & $u$ & 350.0 (306.7--386.7) nm & 3\farcs1 & $\delta \lesssim 0$\degr\ (2$\pi$) & 10,12,13 \\
           & SkyMapper  & $v$ & 387.9 (355.0--421.7) nm & 2\farcs9 & $\delta \lesssim 0$\degr\ (2$\pi$) & 10,12,13 \\
           & SkyMapper  & $g$ & 501.6 (410.3--657.0) nm & 2\farcs6 & $\delta \lesssim 0$\degr\ (2$\pi$) & 10,12,13 \\
           & SkyMapper  & $r$ & 607.7 (492.5--723.2) nm & 2\farcs4 & $\delta \lesssim 0$\degr\ (2$\pi$) & 10,12,13 \\
           & SkyMapper  & $i$ & 773.3 (692.9--864.7) nm & 2\farcs3 & $\delta \lesssim 0$\degr\ (2$\pi$) & 10,12,13 \\
           & SkyMapper  & $z$ &912.0 (815.9--1067.9) nm & 2\farcs3 & $\delta \lesssim 0$\degr\ (2$\pi$) & 10,12,13 \\
           & SDSS       & $u$ & 355.1 nm     & 1\farcs53 & 14,555 deg$^2$ & 14,15\\
           & SDSS       & $g$ & 468.6 nm     & 1\farcs44 & 14,555 deg$^2$ & 14,15\\
           & SDSS       & $r$ & 616.5 nm     & 1\farcs32 & 14,555 deg$^2$ & 14,15\\
           & SDSS       & $i$ & 748.1 nm     & 1\farcs26 & 14,555 deg$^2$ & 14,15\\
           & SDSS       & $z$ & 893.1 nm     & 1\farcs29 & 14,555 deg$^2$ & 14,15\\
\hline
Near-IR    & 2MASS      & $J$ & 1.235 (1.081--1.407) $\mu$m & $\sim$2\farcs5 & All sky & 10,16,17 \\
           & 2MASS      & $H$ & 1.662 (1.479--1.823) $\mu$m & $\sim$2\farcs5 & All sky & 10,16,17 \\
        & 2MASS & $K_{\rm s}$ & 2.159 (1.954--2.355) $\mu$m & $\sim$2\farcs5 & All sky & 10,16,17 \\
\hline
Mid-IR     & WISE      & W1   & 3.353 (2.754--3.872) $\mu$m  &  6\farcs1 & All sky & 10,18 \\
           & WISE      & W2   & 4.603 (3.963--5.341) $\mu$m  &  6\farcs4 & All sky & 10,18 \\
           & WISE      & W3   & 11.56 (7.443--17.261) $\mu$m &  6\farcs5 & All sky & 10,18 \\
           & WISE      & W4   & 22.09 (19.52--27.91) $\mu$m  & 12\farcs0 & All sky & 10,18 \\
           & AKARI     & S9W  & 8.228 (5.846--12.188) $\mu$m &  5\farcs5 & All sky & 10,19,20 \\
           & AKARI     & L18W & 17.61 (13.61--28.67) $\mu$m  &  5\farcs7 & All sky & 10,19,20 \\
\hline
Far-IR & Herschel & PACS blue &  70~$\mu$m (60--85) $\mu$m   &  5\farcs6 & Targeted obs.$^\dagger$ & 21,22 \\
       & Herschel & PACS green& 100~$\mu$m (85--130) $\mu$m  &  6\farcs8 & Targeted obs.$^\dagger$ & 21,22 \\
       & Herschel & PACS red  & 160~$\mu$m (130--210) $\mu$m & 11\farcs3 & Targeted obs.$^\dagger$ & 21,22 \\
       & Herschel & PSW       & 250~$\mu$m                   & 18\farcs1 & Targeted obs.$^\dagger$ & 21,23 \\
       & Herschel & PMW       & 350~$\mu$m                   & 25\farcs2 & Targeted obs.$^\dagger$ & 21,23 \\
       & Herschel & PLW       & 500~$\mu$m                   & 36\farcs6 & Targeted obs.$^\dagger$ & 21,23 \\
\hline
Radio  & VLA (VLASS) &      3.0 GHz &  99.9 mm & 2.5\arcsec & $\delta \geq -40$\degr\ (3.3$\pi$) & 24 \\
       & VLA (NVSS)  &      1.4 GHz & 214.1 mm &  45\arcsec & $\delta \geq -40$\degr\ (3.3$\pi$) & 25 \\
       & VLA (FIRST) &      1.4 GHz & 214.1 mm &   5\arcsec & 10,575~deg$^2$                     & 26 \\
       & VLA (WENSS) &  325.125 MHz & 922.1 mm &  54\arcsec & $\delta \geq +30$\degr\ ($\pi$)    & 27 \\
       & VLA (VLSSr) &     73.8 MHz &  4.06 m  &  75\arcsec & $\delta \geq -40$\degr\ (3.3$\pi$) & 28 \\
       & MOST (SUMSS)&      843 MHz & 355.6 mm &  45\arcsec & $\delta \leq -30$\degr\ ($\pi$)    & 29,30 \\
       & MWA (GLEAM) & 170--231 MHz & 1.50 (1.30--1.76) m & 120\arcsec & $\delta \leq +30$\degr\ (3$\pi$) & 31,32 \\
       & MWA (GLEAM) & 147--154 MHz & 1.99 (1.95--2.04) m & 120\arcsec & $\delta \leq +30$\degr\ (3$\pi$) & 31,32 \\
       & MWA (GLEAM) &   72--80 MHz & 3.94 (3.75--4.16) m & 120\arcsec & $\delta \leq +30$\degr\ (3$\pi$) & 31,32 \\
& GMRT (TGSS)& 147.5 (140--156) MHz & 2.03 (1.92--2.14) m &  25\arcsec & $\delta \geq -53$\degr\ (3.6$\pi$) & 33  \\
\enddata
\tablecomments{Comments:
(1) wavelength class; 
(2--3) instrument and its wavelength band;
(4) wavelength range for the X-ray bands, effective wavelength for the UV-to-far-IR bands, and the typical wavelength for the radio bands.
Bandwidth is denoted in parentheses;
(5) angular resolution;
(6) area coverage of each survey;
(7) References of column (3)--(6).
\\
\textbf{References}: 
(1) \citet{Gehrels2004}; (2) \citet{Oh2018}; (3) \citet{Yamada2021} and the references therein;
(4) \citet{Harrison2013}; (5) \citet{Mitsuda2007}; (6) \citet{Jansen2001}; 
(7) \citet{Garmire2003}; (8) \citet{Bianchi2017}; (9) \citet{Morrissey2007};
(10) Spanish Virtual Observatory (SVO) Filter Profile Servise (\url{http://svo2.cab.inta-csic.es/theory/fps/};
(11) \citet{Chambers2016};
(12) \citet{Wolf2018}; (13) \citet{Onken2019}; (14) \citet{Albareti2017};
(15) \citet{Ahumada2020} and the information (scope and image quality) on the DR16 website (\url{https://www.sdss.org/dr16/});
(16) \citet{Cohen2003}; (17) \citet{Skrutskie2006}; (18) \citet{Wright2010};
(19) \citet{Murakami2007}; 
(20) \citet{Ishihara2010};
(21) \citet{Chu2017};
(22) \citet{Poglitsch2010};
(23) \citet{Griffin2010};
(24) \citet{Gordon2021};
(25) \citet{Condon1998};
(26) \citet{Helfand2015};
(27) \citet{Rengelink1997};
(28) \citet{Lane2014};
(29) \citet{Mauch2003};
(30) \citet{Mauch2008};
(31) \citet{Hurley-Walker2017};
(32) \citet{Hurley-Walker2019};
(33) \citet{Intema2017}.\\
\tablenotetext{\dagger}{All of U/LIRGs in GOALS sample are mapped by the Herschel with both the PACS and SPIRE \citep[see][]{Chu2017}.}
}
\end{deluxetable*}

\subsection{X-ray Spectra} \label{sub3-1-Xray}
Among 57 U/LIRGs, \citet{Yamada2021} carried out the broadband X-ray spectral analysis for 49 U/LIRG systems by using all of the available NuSTAR, Chandra \citep{Garmire2003}, XMM-Newton \citep{Jansen2001}, and Suzaku \citep{Mitsuda2007} data observed by 2020 April.
They also performed the analysis of Swift/X-Ray Telescope (XRT; \citealt{Burrows2005}) data when no other soft X-ray data were obtained, and utilized the Swift/BAT spectra in the 105-month catalog \citep{Oh2018} if detected.
Considering that X-ray spectral works support the clumpy nature of AGN tori \citep[see e.g.,][]{LiuYuan2014,Furui2016,Tanimoto2018,Tanimoto2019,Buchner2019}, these best-fitting models were provided with a Monte Carlo-based model from a clumpy torus (XCLUMPY; \citealt{Tanimoto2019}), which enables us to constrain the torus covering fractions for individual AGNs \citep[e.g.,][]{Ogawa2019,Ogawa2021,Tanimoto2020,Tanimoto2022,Yamada2020,Yamada2021,Uematsu2021,Inaba2022}.
The broadband X-ray spectral analysis for three other sources UGC 2608, NGC 5135, and NGC 7469 was also conducted with XCLUMPY (\citealt{Yamada2020,Ogawa2021}).
Therefore, we compiled these X-ray results of 52 U/LIRG systems with XCLUMPY to compare the results from the multiwavelength SED decomposition with an updated CLUMPY model (Section~\ref{sub4-1-CLUMPY}).

For the other five U/LIRGs, \citet{Yamada2021} only analyzed the NuSTAR data because their spectra show complex features.
Instead of the XCLUMPY model, we analyzed their NuSTAR spectra ($\sim$3--70~keV) by adopting the best-fitting models in previous works for three nonmerging LIRGs NGC 1068 (M2d in \citealt{Bauer2015}), NGC~1275 (Model2 in \citealt{Rani2018}), NGC~1365 (final model for Observation 4 in \citealt{Rivers2015}), 
and two dual-AGN systems Mrk~266B/Mrk~266A (SW-R model in \citealt{Iwasawa2020})\footnote{The torus reflection component reproduced by the SW-R model \citep{Ikeda2009} is defined only in the 1--100~keV band. 
Assuming the power-law model with the same slope as in the 80--100~keV band, we approximately expanded the X-ray model to the 100--200~keV spectra in Figures~\ref{SED01-F} and \ref{SED15-F}.}
and NGC~6240S/NGC~6240N (sum of two AGN models with MYtorus in \citealt{Puccetti2016}), respectively.

The best-fitting spectral models were applied in the multiwavelength SEDs.
By following the same manner as \citet{Yamada2021}, we calculated the 5$\sigma$ upper limits in the 8--24~keV band from the NuSTAR counts for the hard X-ray--undetected sources, by using the HEASARC tool WebPIMMS v4.11a assuming a power law of $\Gamma = 1.8$ \citep[e.g.,][]{Ueda2014,Ricci2017dApJS} with Galactic absorption, whose hydrogen column density was fixed at the value of \citet{Willingale2013}. 
The 2--7~keV upper limits were provided from the XMM-Newton/MOS (MOS1 and MOS2) for IC~5283 and MCG+01-59-081, since their NuSTAR 8--24~keV fluxes were smaller than those of the interacting companion that causes the high contamination.

\subsection{UV Photometry} \label{sub3-2-UV}
The far-UV (FUV; $\lambda_{\rm eff} \sim 152.8$~nm) and near-UV (NUV; $\lambda_{\rm eff} \sim 231.0$~nm) photometries were compiled from the latest version of the Galaxy Evolution Explore (GALEX; \citealt{Martin2005,Bianchi2011}) satellite catalog (GR6plus7 data release; \citealt{Bianchi2017}).\footnote{The GALEX observations utilized in this paper can be accessed via \dataset[10.17909/t9-pyxy-kg53]{https://doi.org/10.17909/t9-pyxy-kg53}.}
The GALEX observation area covers almost all of the sky ($\gtrsim90$\%).
Cross-matching the optical positions of the individual nuclei and the UV peak positions, we extracted the GALEX photometries (FUV\_mag and NUV\_mag).
To ensure reliable data, we finally selected the data with Fexf = 0 and Nexf = 0, where the Fexf and Nexf represent the extended flags for FUV and NUV bands, respectively.

\subsection{Optical Photometry} \label{sub3-3-optical}
The optical photometries were taken from the large catalogs with three ground-base telescopes, Pan-STARRS1 (PS1) DR2 (e.g., \citealt{Chambers2016,Flewelling2020,Magnier2020a,Magnier2020b,Magnier2020c,Waters2020}), SkyMapper Southern Survey DR1 and DR2 \citep{Wolf2018,Onken2019}, and SDSS DR16 \citep{Ahumada2020}.
The current coverage areas are $\sim$3/4 sky (north of $\delta$ = $-$30\degr) for Pan-STARRS, $\sim$1/2 sky (south of $\delta$ = 0\degr) for SkyMapper, and $\sim$1/3 sky (a large part of the northern area) for SDSS.
We preferentially used the Pan-STARRS mean object magnitudes in $g$, $r$, $i$, $z$, and $y$ bands.
Since most of our targets are extended sources, we calculated the flux densities from their Kron magnitudes \citep{Kron1980}, which were corrected for the missing fluxes (about 10\%)\footnote{\url{https://outerspace.stsci.edu/display/PANSTARRS/PS1+Kron+photometry+of+extended+sources}} by multiplying 100/90 (or $-$0.115 mag).
When the Pan-STARRS data were not available, we applied the SkyMapper data in $u$, $v$, $g$, $r$, $i$, and $z$ bands.
Since the Petrosian magnitudes for extended sources were given in the SkyMapper catalog, we adopted them and corrected for the missing fluxes by $-$0.130 mag \citep{Graham2005} assuming the typical Sersic index to be 3 (e.g., \citealt{Haan2011a}).
We finally chose the SDSS photometry (cmodelMag) in $u$, $g$, $r$, $i$, and $z$ bands for the other sources, and for NGC 4418/MCG+00-32-013 and Mrk~266B/Mrk~266A because the Pan-STARRS photometries of these two systems were $>$0.4~mag fainter than those of SDSS values.

We extracted the optical photometry from these catalogs in each of three bins, $u$--$v$, $g$--$z$, and $y$ bands.
The averaged differences of the optical magnitudes, $\overline{\Delta{\rm Mag}}$ (SkyMapper $-$ Pan-STARRS) = 0.07, 0.11, $-$0.12, $-$0.16, and $\overline{\Delta{\rm Mag}}$ (SDSS $-$ Pan-STARRS) = 0.14, 0.11, 0.02, $-$0.19 for $g$, $r$, $i$, and $z$ bands, respectively. Here, we systematically added the 10\% uncertainties ($\sim$0.115 mag) for the whole optical flux densities as a systematic error derived from the different measurements.
We excluded five pieces of the optical photometry for NGC~232 (SkyMapper $v$-band), ESO 374$-$IG032 (SkyMapper $v$-band), IRAS F12112+0305 (Pan-STARRS $y$-band), and NGC~6907/NGC~6908 (SkyMapper $u$--$v$ bands) because their flux densities were much smaller than those in other optical bands by a factor of 3--10.

\subsection{Near-IR Photometry} \label{sub3-4-NIR}
In the near-IR bands, we utilized the data of the Two Micron All Sky Survey (2MASS; \citealt{Skrutskie2006}).\footnote{The Digital Object Identifier (DOI) of the 2MASS catalog is  \dataset[10.26131/IRSA2]{https://catcopy.ipac.caltech.edu/dois/doi.php?id=10.26131/IRSA2}.}
The effective wavelengths of 2MASS filters are 1.235~$\mu$m, 1.662~$\mu$m, and 2.159~$\mu$m for $J$, $H$, and $K_{\rm s}$ bands, respectively \citep{Turner2011}.
The data of our targets were taken from the extended catalog. Since the values were not presented only for IRAS~F17138--1017, we took its values from the point source catalog.
Finally, we extracted the data with cc\_flg = ``000'' to avoid the effects of artifacts.

\subsection{Mid-IR Photometry} \label{sub3-5-MIR}
The mid-IR data were compiled from the Wide-field IR Survey Explorer (WISE; \citealt{Wright2010}) and AKARI \citep{Murakami2007}.\footnote{The DOIs of the WISE and AKARI catalogs are \dataset[10.26131/IRSA1]{https://catcopy.ipac.caltech.edu/dois/doi.php?id=10.26131/IRSA1} and \dataset[10.26131/IRSA181]{https://catcopy.ipac.caltech.edu/dois/doi.php?id=10.26131/IRSA181}, respectively.}
We employed the photometry in W1 (3.4~$\mu$m), W2 (4.6~$\mu$m), W3 (12~$\mu$m), and W4 (22~$\mu$m) bands from the ALLWISE catalog \citep{Cutri2014}, or ALLSKY catalog \citep{Cutri2012} if not obtained.
We primarily utilized the ``gmag'' (elliptical aperture magnitude), or secondarily ``mpro'' (instrumental profile-fit photometry magnitude), with w1-4sat=0 and w1-4cc\_map=0.
The AKARI fluxes in S9W (9~$\mu$m) and L18W (18~$\mu$m) bands were obtained from AKARI/IRC mid-IR all-sky Survey \citep{Ishihara2010}, where we selected the data with q\_S09=0 \& X09=0 for S9W, and q\_S18=0 \& X18=0 for L18W, respectively.

\subsection{Far-IR Photometry} \label{sub3-6-FIR}
We applied the far-IR fluxes presented by \citet{Chu2017}, using the Herschel Space Observatory \citep{Pilbratt2010} Photodetector Array Camera and Spectrometer (PACS; \citealt{Poglitsch2010}) and the Spectral and Photometric Imaging Receiver (SPIRE; \citealt{Griffin2010}).
For all the PACS bands (70~$\mu$m, 100~$\mu$m, 160~$\mu$m) and SPIRE bands (250~$\mu$m, 350~$\mu$m, 500~$\mu$m), they computed the total system fluxes and component fluxes of individual galaxies (where possible) for the entire GOALS sample.
The aperture radius was set by the band with the largest beam size for each instrument: usually the 160~$\mu$m band for PACS, and the 500~$\mu$m band for SPIRE.

\subsection{Radio Photometry} \label{sub3-7-Radio}
Radio surveys with large sky coverages have been performed over a wide frequency range \citep[see e.g.,][]{Shimwell2019,Stein2021}.
In our study, we utilized eight kinds of wide-area radio catalogs, which have been conducted with four main telescopes: Karl G. Jansky Very Large Array (VLA), 
Molonglo Observatory Synthesis Telescope (MOST; \citealt{Mills1981,Robertson1991}), 
Murchison Widefield Array (MWA; \citealt{Lonsdale2009,Tingay2013}), and
Giant Metrewave Radio Telescope (GMRT; \citealt{Swarup1991}).
The catalogs are described below:
\\

The catalogs of the VLA observations:
\begin{enumerate}
    \item[(1)] The Very Large Array Sky Survey (VLASS; \citealt{Gordon2021}), observing the entire sky north of $\delta$ = $-$40\degr\ (82\% sky region) at 3~GHz.
    
    \item[(2)] The NRAO VLA Sky Survey (NVSS; \citealt{Condon1998}) covering the sky north of $\delta$ = $-$40\degr\ at 1.4~GHz.
    
    \item[(3)] The Faint Images of the Radio Sky at Twenty-centimeters (FIRST; \citealt{Helfand2015}), observing in 10,575~deg$^2$ of sky coverage (8,444~deg$^2$ in the north Galactic cap and 2,131~deg$^2$ in the southern cap) at 1.4~GHz.\\

    \item[(4)] The Westerbork Northern Sky Survey (WENSS; \citealt{Rengelink1997}), a low-frequency radio survey that will cover the whole sky north of $\delta$ = $+$30\degr\ at 325.125~MHz.

    \item[(5)] The VLA Low-frequency Sky Survey Redux (VLSSr; \citealt{Cohen2007,Lane2014}), covering the sky north of $\delta$ = $-$40\degr\, at 73.8~MHz.
    
\end{enumerate}

The catalogs of the other radio observations:
\begin{enumerate}

    \item[(6)] The Sydney University Molonglo Sky Survey (SUMSS) catalog, Version 2.1r (updated in 2012; \citealt{Mauch2003,Mauch2008}) with MOST, covering the sky south of $\delta$ = $-$30\degr\ 
    at 843~MHz.
    The data of IRAS F13120--5453 are not listed in the original catalog but are obtained from \citet{Allison2014}.
    
    \item[(7)] The GaLactic and Extragalactic All-sky MWA (GLEAM; \citealt{Hurley-Walker2017}) survey, covering 24,831~deg$^2$ over sky south of $\delta$ = $+$30\degr\ and Galactic latitudes outside 10\degr\ of the Galactic plane, excluding some areas such as the Magellanic Clouds, across 72–231 MHz.
    \citet{Hurley-Walker2019} also presented the GLEAM catalog of the Galactic plane (2,860 deg$^2$); in which only the system of MCG+04-48-002 and NGC~6921 was contained, but their fluxes were not utilized because both galaxies are not divided by the MWA.
    To get similar data from VLA catalogs for northern-sky targets, we picked out the fluxes in the 170--231~MHz, 147--154~MHz, and 72--80~MHz bands.

    \item[(8)] The TIFR GMRT Sky Survey (TGSS; \citealt{Intema2017}), observing 36,900 deg$^2$ of the sky north of $\delta$ = $-$53\degr\ ($\sim$90\% sky), at 147.5~MHz (with a bandwidth of 140--156 MHz).\\

\end{enumerate}

We basically utilized the integrated fluxes referred from these radio catalogs.
Although only the peaked fluxes were presented in the VLSSr catalog, they should be the same as integrated fluxes for unresolved point sources because of the large beam size (75\arcsec; see Table~\ref{T2_catalogs}).
To avoid flux underestimates due to the resolved-out extended emission, we basically employed the radio catalogs excluding the two catalogs observed with the high-spatial-resolution instruments: VLASS (2\farcs5) and FIRST (5\arcsec).\footnote{Since the uncertainties of FIRST flux densities were not provided, we assume them as 10\% uncertainties.}
For the faint sources that any data are not obtained, we adopted the 5$\sigma$ upper limits from the TGSS catalog ($<$24.5~mJy at 150~MHz). Additionally, we used the fluxes of these faint sources from the VLASS (3.0~GHz) and FIRST (1.4~GHz) catalogs if detected at the $\gtrsim$5$\sigma$ level\footnote{We utilized the VLASS fluxes for the low-significance radio sources, NGC~833 (3.3$\sigma$) and MCG+01-59-081 (4.4$\sigma$).}, or the 5$\sigma$ upper limits ($<$0.44~and $<$1.00~mJy, respectively).

\subsection{Correction for Galactic Extinction} \label{sub3-8-GalExt}

The magnitudes in the FUV, NUV, $u$--$y$, $J$, $H$, $K_{\rm s}$, W1 and W2 bands were corrected for the Galactic extinction by following \citet{Schlegel1998}, hereafter SFD.
We adopted the correction factors, $R_{\lambda}$ (= $A_{\lambda}$/$E$($B-V$)$_{\rm SFD}$, where $A_{\lambda}$ is the extinction at the $\lambda$ band in units of magnitude, calculated by $A_{\lambda} = 1.086\tau_{\lambda}$) from \citet{Bianchi2017} for GALEX, \citet{Schlafly2011} for Pan-STARRS, \citet{Wolf2018} for SkyMapper, and \citet{Yuan2013} for the other telescopes:

\begin{enumerate}
    \item[\labelitemi] GALEX: 
    
    [$R_{\rm FUV}$, $R_{\rm NUV}$] = [8.06, 7.95]

    \item[\labelitemi] Pan-STARRS: 
    
    [$R_{g}$, $R_{r}$, $R_{i}$, $R_{z}$, $R_{y}$] 
    
    = [3.172, 2.271, 1.682, 1.322, 1.087]
    
    \item[\labelitemi] SkyMapper: 
    
    [$R_{u}$, $R_{v}$, $R_{g}$, $R_{r}$, $R_{i}$, $R_{z}$] 
    
    = [4.294, 4.026, 2.986, 2.288, 1.588, 1.206]

    \item[\labelitemi] SDSS: 
    
    [$R_{u}$, $R_{g}$, $R_{r}$, $R_{i}$, $R_{z}$] 
    
    = [4.39, 3.30, 2.31, 1.71, 1.29]

    \item[\labelitemi] 2MASS: 
    
    [$R_{\rm J}$, $R_{\rm H}$, $R_{\rm Ks}$] = [0.72, 0.46, 0.306]

    \item[\labelitemi] WISE: 
    
    [$R_{\rm W1}$, $R_{\rm W2}$] = [0.18, 0.16]

\end{enumerate}

The magnitudes of $E$($B-V$) reddening and the flux densities of the UV-to-radio photometries corrected for the Galactic extinction in our sample are summarized in Table~\ref{TE1_photo1} ($\lambda <$ 1~$\mu$m), Table~\ref{TE2_photo2} ($\lambda$ = 1--200~$\mu$m), and Table~\ref{TE3_photo3} ($\lambda >$ 200~$\mu$m).
\\

% Figure 1
\begin{figure*}
    \centering
   \includegraphics[keepaspectratio, scale=0.7]{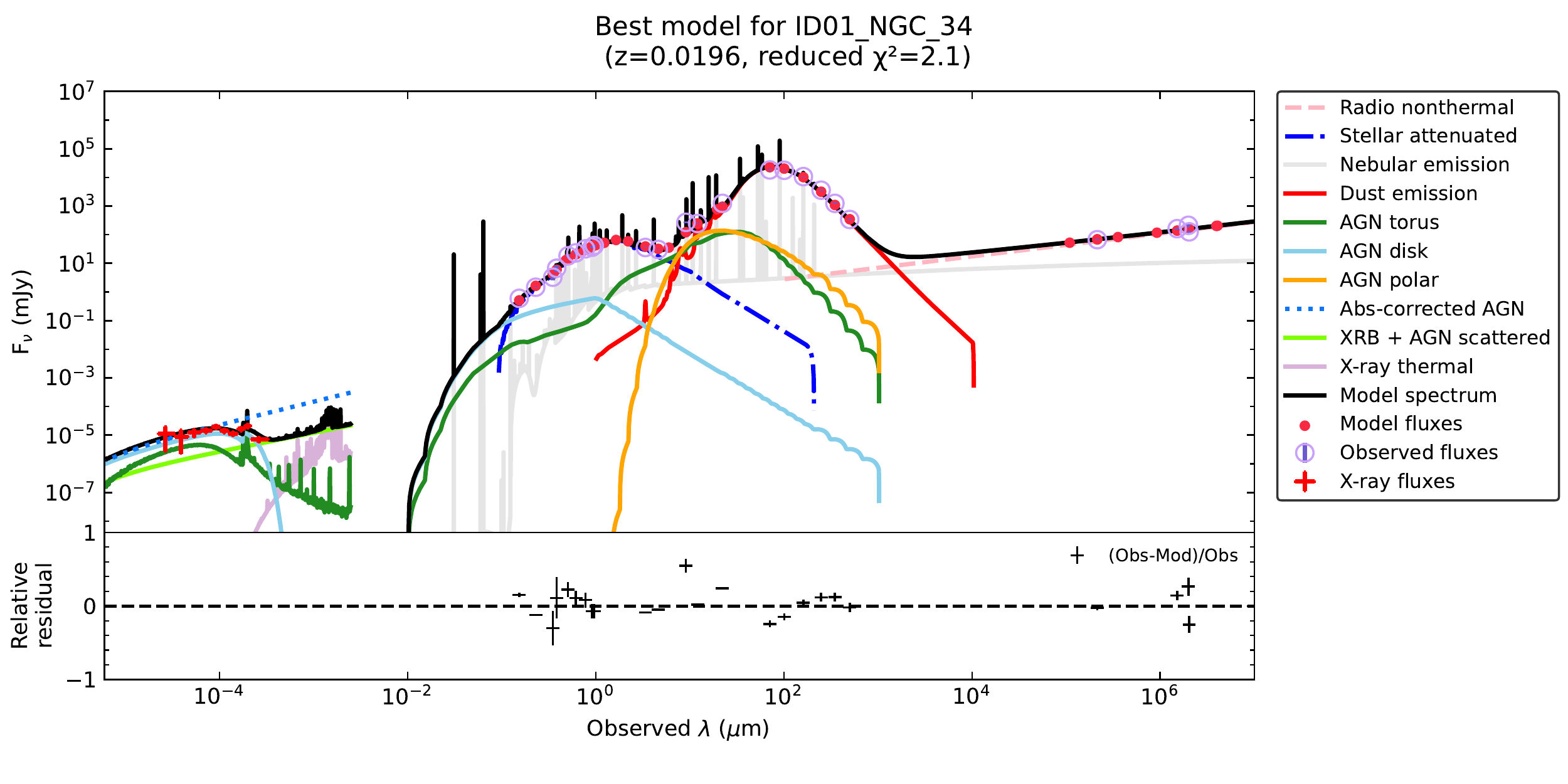}
   \includegraphics[keepaspectratio, scale=0.7]{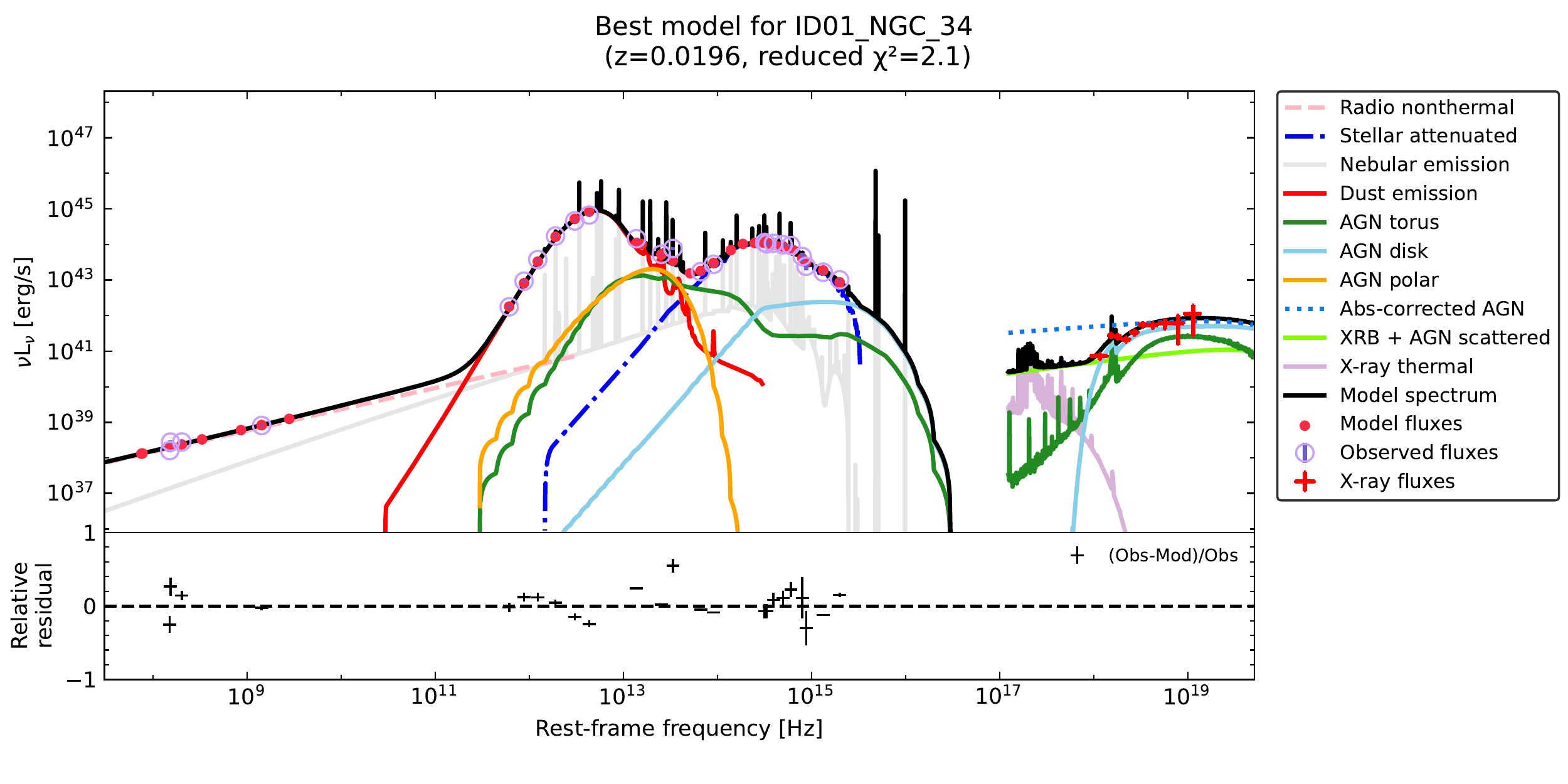}
        \caption{An example of the hard-X-ray-to-radio SED and best-fitting models for a stage-D merger NGC 34 (ID01).
        Upper panel: observed wavelength vs. flux density.
        Lower panel: rest-frame frequency vs. luminosity.
        The bottom panels show the residuals in the UV-to-radio bands.
        The individual curves show the SED models of radio non-thermal emission (pink dashed), attenuated stellar emission (blue dash-dotted), nebular emission (gray), dust emission (red), AGN torus (green), escaped AGN disk (cyan), polar dust (orange), absorption-corrected AGN X-ray emission (blue dotted), the X-ray emission from AGN scatters and X-ray binaries (light green), and X-ray thermal emission (light purple).
        Purple and red circles represent the observed and model flux densities, respectively. 
        Red crosses in the X-ray band denote the NuSTAR spectra analyzed by \citet{Yamada2021}.
        The results for all targets are given in Figures~\ref{SED01-F}--\ref{SED18-F}.
        \\}
\label{F1-SED00-F}
\end{figure*}

%%%%%%%%%%%%%%%%%%%%%%%%%%%%%%%%%%%%%%%%%%%%%%%
%%% Section 4 : Multiwavelength SED Fitting %%%
%%%%%%%%%%%%%%%%%%%%%%%%%%%%%%%%%%%%%%%%%%%%%%%

\section{Multiwavelength SED Model and Best-fit Parameters} \label{S4-model}
To best constrain the properties of polar dust, we update the latest multiwavelength SED-fitting code X-CIGALE \citep{Yang2020} by implementing the IR CLUMPY model (Section~\ref{sub4-1-CLUMPY}), whose geometry is the same as in XCLUMPY.
The UV-to-IR SEDs are analyzed with the modified code for all targets (Section~\ref{sub4-2-SED-model}).
For the radio data, we perform the fitting of radio photometry after the UV-to-IR SED decomposition (Section~\ref{sub4-3-radio-fit}). 
We finally combine the results of X-ray spectral analysis \citep{Yamada2021}, UV-to-IR SED decomposition, and radio fitting.
The validity of the best-fit parameters are evaluated in Section~\ref{sub4-4-results}.

The tables of best-fitting parameters and supplemental measurements of each target are listed in Appendix~\ref{AppendixA-best-values}.
The results of a mock analysis (i.e., the test for the reliability of the estimates, given the uncertainty of the photometry) are shown in Appendix~\ref{AppendixB-mock}.
The comparison of our results with the previous works is discussed in Appendix~\ref{AppendixC-previous-works}, and the comparison of the results with a different AGN model (SKIRTOR; \citealt{Stalevski2016}) is in Appendix~\ref{AppendixD-SKIRTOR}.
The hard X-ray to radio multiwavelength SEDs and their best-fitting models are presented in Figure~\ref{F1-SED00-F} and Figures~\ref{SED01-F}--\ref{SED18-F} (see Appendix~\ref{AppendixE-SEDs}).

\subsection{New Implementation of CLUMPY for X-CIGALE} \label{sub4-1-CLUMPY}

\subsubsection{AGN Model in Multiwavelength SED} \label{subsub4-1-1-AGN}
The combination of X-ray and IR data enables us to constrain the distribution of the gas and dust in the AGNs \citep[e.g.,][]{Hickox2018}.
For making the analysis consistent with the XCLUMPY X-ray spectra, \citet{Miyaji2019} have modified the CIGALE package, a fitting code to model the UV-to-radio SEDs of galaxies \citep{Burgarella2005,Noll2009,Boquien2019}, to include an implementation of CLUMPY \citep{Nenkova2008a,Nenkova2008b}.
They conduct the X-ray spectral analysis with XCLUMPY and the UV-to-IR SED decomposition with CLUMPY as a module, and successfully estimate the torus parameters of an AKARI-selected CT AGN, such as X-ray absorbing column, torus optical depth, torus angular width, and inclination angle.
Recent works \citep{Tanimoto2020,Ogawa2021} show that these clumpy torus models mostly present the larger covering fraction in the IR band than those in X-rays, consistent with the presence of the polar dust (see e.g., \citealt{Asmus2016,Asmus2019,Lopez-Gonzaga2016a,Lopez-Gonzaga2016b,Honig2017,Honig2019,Lyu2018,Buat2021,Toba2021b,Andonie2022}).

\citet{Yang2020} develop a new X-ray module for CIGALE, allowing it to fit SEDs, named X-CIGALE (but their X-ray model does not consider the absorption from the target, our Galaxy, and/or the intergalactic medium).\footnote{\citet{Yang2022} improve the X-CIGALE code mainly related to AGN intrinsic X-ray anisotropy, X-ray binary emission, AGN accretion-disk SED shape, and AGN radio emission (new version as CIGALE v2022.1).}
Torus models are mainly classified by their dust distribution such as smooth, clumpy, and two-phase (clumpy and smooth) models.
The current version of X-CIGALE (CIGALE v2020.0) includes two AGN models of a classical smooth torus model by \citet{Fritz2006} and a modern two-phase model (SKIRTOR) by \citet{Stalevski2012,Stalevski2016}.
They also implement the model of polar-dust extinction (e.g., \citealt{Lyu2018,Honig2019}).
For maintaining energy conservation in the scheme of X-CIGALE, the polar dust model considers the radiative energy absorbed by their dust.
The model of polar dust assumes the isotropic dust re-radiation and the ``gray body'' model (e.g., \citealt{Casey2012}).

Since a clumpy distribution of the dust is necessary to prevent the destruction of grains \citep[e.g.,][]{Krolik1988}, many works adopt the SKIRTOR model in X-CIGALE \citep[e.g.,][]{Yang2020,Yang2022,Buat2021}.
However, conducting the SED fit by a self-consistent AGN model with X-CIGALE and XCLUMPY is difficult because the torus component of SKIRTOR has different geometry and parameter definitions from those of XCLUMPY. 
Thus we use the CLUMPY module by combining the SKIRTOR's polar dust component.

% Figure 2
\begin{figure*}
    \centering
   \includegraphics[keepaspectratio, scale=0.64]{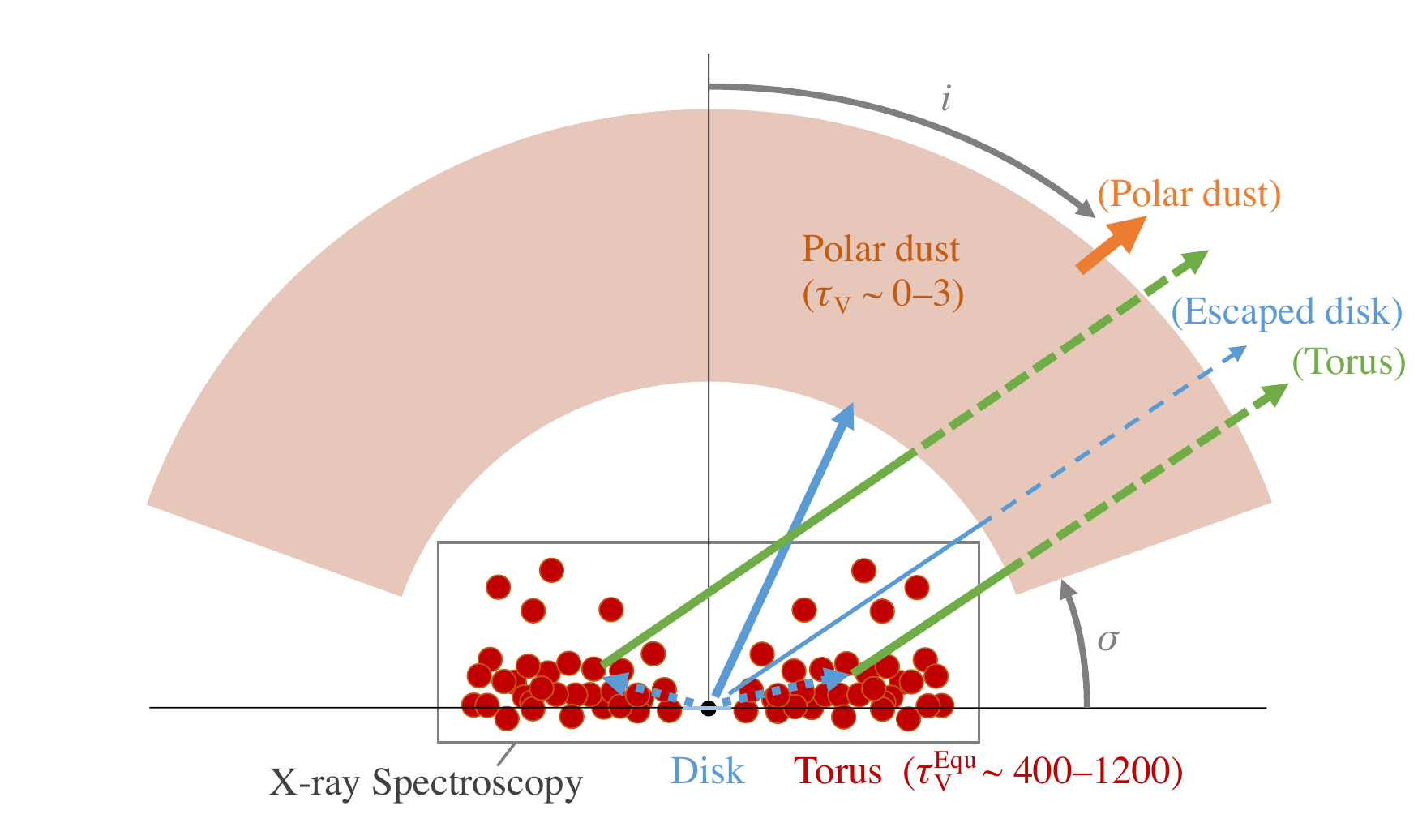}
        \caption{A schematic picture of the AGN structure assumed in the updated CLUMPY model.
        Central black circle and blue bar represent the SMBH and accretion disk, respectively.
        Brown circles show the clumps of torus, whose structure is constrained by the X-ray spectroscopy with XCLUMPY.
        Light-brown annular sector illustrates the polar dust.
        Green arrays mark the torus reflection emission.
        Blue thick (solid and dotted) arrays are the AGN disk emission, while a blue thin array is the escaped disk emission via the clump gaps.
        These emissions are attenuated by the polar dust, marked as dashed lines.
        Orange arrow shows the dust re-radiation from the polar dust.
        The symbols ``$\sigma$'' and ``$i$'' mean the torus angular width and inclination angle, respectively.
        \\}
\label{F2-geometry}
\end{figure*}

\subsubsection{Differences between CLUMPY and SKIRTOR} \label{subsub4-1-2-AGN-models}
The CLUMPY model assumes an isotropic emission of the central source commonly adopted in the literature \citep[e.g.,][]{Schartmann2005,Honig2006}.
Strong correlations between X-ray and nuclear 12~$\mu$m fluxes (i.e., least biased by extra-nuclear emission) are reported for both Seyfert 1 and 2 galaxies \citep[e.g.,][]{Gandhi2009,Asmus2015}.
When taking the X-ray emission as an isotropic tracer for the AGN luminosity, the low scatter of the mid-IR/X-ray luminosity correlation suggests that the mid-IR emission is also close to isotropic.
The reason is that the hollow cone is visible to the observer at any inclination in the inflowing disk and outflowing polar dust \citep{Honig2017,Honig2019}.
The intrinsic SED of the central AGN disk emission in CLUMPY follows the piecewise power-law distribution in \citet{Rowan-Robinson1995}, where the CLUMPY model adopts $\lambda_{\rm RJ}$ = 1~$\mu$m (wavelength marking the onset of the Rayleigh-Jeans tail).
CLUMPY model results in a Gaussian angular distribution.
A density gradient along the polar angle is proportional to exp($-\theta^2/\sigma^2$), where $\theta$ is the elevation angle and $\sigma$ is torus angular width (typically $\sim$20\degr; e.g., \citealt{Ogawa2021,Yamada2021}).
Whereas, SKIRTOR considers the anisotropy of the central source \citep{Netzer1987} and employs larger $\lambda_{\rm RJ}$ = 5~$\mu$m than that in CLUMPY. 
SKIRTOR assumes a sharp-edged angular distribution of high-density clumps embedded in a smooth component with low density.
A density gradient along the polar angle is proportional to exp($-x|$cos($90\degr - \theta$)$|$) in SKIRTOR, where $x$ is a torus density angular parameter typically fixed to 1.0 \citep[e.g.,][]{Yang2020}.

\citet{Stalevski2016} demonstrate that the difference in the input SED shape of the central engine (e.g., $\lambda_{\rm RJ}$) has little effect on the resulting dust emission: what matter is the total amount of emission from the central engine.\footnote{\url{https://sites.google.com/site/skirtorus/sed-library}}
By comparing the results obtained with the clumpy torus model of isotropic disk radiation, \citet{Stalevski2012} report that the anisotropic assumption only reduces dust fluxes by a factor of at most 2, keeping roughly the similar shape of IR SEDs.
On the other hand, the smooth component in SKIRTOR can produce a more attenuated silicate feature and have a stronger near-IR emission than without the smoothly distributed element (\citealt{Stalevski2012,Stalevski2016}; see also \citealt{Efstathiou2022}).
Moreover, the dependence of the dust distribution on $\theta$ in SKIRTOR tends to be less and may cause a stronger near-IR emission than in CLUMPY (Gaussian angular distribution).
Although we find such a trend by examining how much of the differences in the results derived from the fits with CLUMPY and SKIRTOR (see Appendix~\ref{AppendixD-SKIRTOR}), both models consequently present similar intrinsic AGN luminosities for almost all of the U/LIRGs in our sample (Figure~\ref{FD1-comp-skirtor}).
Hence, our multiwavelength SED analysis is less affected by the model assumption between CLUMPY and SKIRTOR.

\subsubsection{Implementation of CLUMPY Model} \label{subsub4-1-3-CLUMPY}
Figure~\ref{F2-geometry} shows a schematic picture of our updated CLUMPY model containing the polar dust component.
The model assumes the situation of an AGN structure consisting of (1) an accretion disk at $\ll$1~pc scale, (2) a clumpy torus at $\lesssim$10~pc scale, and (3) polar dust at $\gtrsim$10~pc scale (see Section~\ref{subsub4-2-3-polar}).
In the torus model of CLUMPY, the radial distribution of clumps is a power law with an index of $q$ (i.e., $r^{-q}$), and the angular distribution is a Gaussian distribution.
The number of clumps along a line-of-sight path ($N_{\rm clump}^{\rm LOS}$) is described by:
\begin{align}
& N_{\rm clump}^{\rm LOS}(\theta) = N_{\rm clump}^{\rm Equ}\ {\rm exp}(-\theta^2/\sigma^2),
\end{align}
where $N_{\rm clump}^{\rm Equ}$ is the number of clumps along the equatorial plane.
The updated CLUMPY considers dust absorption, re-emission, and scattering \citep{Nenkova2008a,Nenkova2008b}, except for the scattering effect in the smooth polar-dust region.
Since the total mass of polar dust is much smaller than that contained in the torus, the X-ray spectrum is almost unaffected by the polar dust component at energies above a few keV \citep[e.g.][]{Liu2019}.
Thus, the clump distribution derived from the X-ray spectroscopy with XCLUMPY \citep[e.g.,][]{Yamada2021} traces the structure of the clumpy torus without the polar dust.
A part of the disk emission is escaped via the clump gap, keeping the shape of its intrinsic AGN SED.
Considering the optical depth of a single clump of the torus with $\tau_{\rm V}$ = 40--120 in this work (corresponds to $\tau_{\rm \lambda} > 1$ in the UV-to-optical bands, the escape probability, $P_{\rm esc}$, is calculated by $P_{\rm esc}$ $\simeq$ exp($-N_{\rm clump}^{\rm LOS}$) (see also Equation A4 in \citealt{Nenkova2008a} and Section 2.2 in \citealt{Nenkova2008b}).

The distribution of the polar dust is determined by the region with $\theta \geq \sigma$ (or the region with the angle between the polar axis and edge of the torus in SKIRTOR).
By limiting the range of the polar dust temperature between 100--250~K, the polar dust model in this work reflects the extended dust at $\gtrsim$10~pc (see Section~\ref{subsub4-2-3-polar} and~\ref{sub6-5-polar-structure}).
We allow the line-of-sight optical depth ($\tau_{\rm V}$) of polar dust as a typical range in previous studies, $\tau_{\rm V} \lesssim 3$ (e.g., \citealt{Buat2021,Toba2021b}; see Section~\ref{subsub4-2-3-polar} and~\ref{sub6-2-NH-AV}).
The value is much smaller than the equatorial optical depth of the torus ($\tau_{\rm V}^{\rm Equ} \sim 400$--1200), where the number of clumps is 10 (Section~\ref{subsub4-2-2-torus}).
Here, the distributions of polar dust and the galaxy's interstellar medium (ISM) are not distinguished.
Finally, to keep energy conservation, the dust extinction and re-radiation of polar dust in the UV-to-IR bands are included in the same manner as the original X-CIGALE code.

% Table 3
\begin{deluxetable*}{lllc}
\label{T3_para}
\tabletypesize{\small}
\tablecaption{Summary of Models and Input Parameters in the UV-to-IR SED Fitting}
\tablehead{
\colhead{Model} &
\colhead{Module} &
\colhead{Parameter} &
\colhead{Value}
}
\startdata
SFH model & \texttt{shfdelayedbq} 
            & \texttt{tau\_main} (Myr) & 1000, 3000, 5000 \\
          & & \texttt{age\_main} (Myr) & 4500, 7000, 9500, 12000 \\
          & & \texttt{age\_bq} (Myr)   & 10, 20, 100 \\
          & & \texttt{r\_sfr}          & 1, 3.16, 10, 31.6, 100, 1000 \\
\hline
Stellar emission & \texttt{bc03}    
                   & \texttt{imf}         & 1 (Chabrier) \\
                 & & \texttt{metallicity} & 0.02 \\
\hline
Nebular emission & \texttt{nebular} 
                   & \texttt{logU} & $-$3.0\\
                 & & \texttt{f\_esc} & 0.0 \\
                 & & \texttt{f\_dust} & 0.0 \\
                 & & \texttt{lines\_width} (km s$^{-1}$) & 300 \\
\hline
Attenuation law &\texttt{dustatt\_modified\_starburst} 
                  & \texttt{E\_BV\_lines} & 0.1, 0.25, 0.5, 0.75, 1.0, 1.5, 2.0\\
                & & \texttt{E\_BV\_factor} & 0.44 \\
                & & \texttt{uv\_bump\_wavelength} (nm) & 217.5 \\
                & & \texttt{uv\_bump\_width} (nm) & 35.0 \\
                & & \texttt{uv\_bump\_amplitude} & 0.0 \\
                & & \texttt{powerlaw\_slope} & $-$0.8,  $-$0.4, 0.0 \\
                & & \texttt{Ext\_law\_emission\_lines} & 1 (Milky Way) \\
                & & \texttt{Rv} & 3.1 \\
\hline
Dust emission   & \texttt{dl2014} 
                  & \texttt{qpah} (= $q_{\rm pah}$) & 0.47, 1.12, 2.50 \\
                & & \texttt{umin} & 1, 5, 10, 25, 50 \\
                & & \texttt{alpha} & 2.0, 2.5\\
                & & \texttt{gamma} & 0.02 \\
\hline
AGN (torus/disk) & \texttt{CLUMPY} (Model I)  
                   & \texttt{Y\_ratio} (= $Y$) & 20 \\
                 & & \texttt{tau\_V} (= $\tau_{\rm V}$) & 40, 80, 120 \\
                 & & \texttt{N\_0} (= $N_{\rm clump}^{\rm Equ}$) & 10 \\
                 & & \texttt{q} (= $q$) & 0.5 \\
                 & & \texttt{sigma} (= $\sigma$; degree) & 15 to 70 (per bin 5; fixed) \\
                 & & \texttt{inclination} (= $i$; degree) & 30, 60, 80 (fixed) \\
                 & & \texttt{fracAGN} (= $f_{\rm AGN}$) 
                     & 0.0 (for starburst-dominant sources), or \\
                 & & & 0.05, 0.1, 0.2, 0.3, 0.4, \\
                 & & & 0.5, 0.6, 0.7, 0.8 (for AGNs)\\
\cline{2-4}
                & \texttt{SKIRTOR} (Model II; Appendix~\ref{AppendixD-SKIRTOR})   
                   & \texttt{R} (= $Y$) & 20 \\
                 & & \texttt{t} (= $\tau_{\rm {9.7}}$) & 5, 9, 11 \\
                 & & \texttt{oa} (= $\Delta$; degree) & 30, 40, 60 (fixed) \\
                 & & \texttt{i} (= $i$; degree) & 30, 60, 80 (fixed) \\
                 & & \texttt{fracAGN} (= $f_{\rm AGN}$) 
                     & 0.0 (for starburst-dominant sources), or \\
                 & & & 0.05, 0.1, 0.2, 0.3, 0.4, \\
                 & & & 0.5, 0.6, 0.7, 0.8 (for AGNs)\\
\hline
AGN (polar) & \texttt{CLUMPY/SKIRTOR} 
              & \texttt{law} & 0 (SMC) \\
            & & \texttt{EBV} (= $E$($B-V$); mag) & 0.0 (fixed), or 0.05, 0.3, 0.8\\
            & & \texttt{temperature} (= $T_{\rm polar}$; K) & 100, 150, 200, 250 \\
            & & \texttt{emissivity} (= $\beta$) & 1.6 \\
\enddata
\end{deluxetable*}

\subsection{UV-to-IR SED Fitting with CLUMPY} \label{sub4-2-SED-model}
By using the X-CIGALE code with the CLUMPY implementation, we conduct the UV-to-IR SED decomposition, whose best-fit parameters are adopted for the radio fitting (Section~\ref{sub4-3-radio-fit}).
The details of the UV-to-IR SED model are described for the host galaxy (stellar emission, nebular emission, and dust emission; Section~\ref{subsub4-2-1-galaxy}), torus (torus and escaped disk emission; Section~\ref{subsub4-2-2-torus}), and polar dust emission (Section~\ref{subsub4-2-3-polar} and~\ref{subsub4-2-4-BIC}).
The models and parameters we adopted are summarized in Table~\ref{T3_para}.

% tau_main: 1000-5000 -> Buat18(1000-5000)
% age_main: 4500--12000 -> Paspaliaris (4500-12000)
% age_bq: 10--100  -> Paspaliaris (10-100) or Buat18(10-70)+Buat19 (5-100)
% r_sfr: 1--1000 -> Paspaliaris (1-1200)

\subsubsection{Host Galaxy Model}\label{subsub4-2-1-galaxy}
We utilize a delayed star formation history (SFH) model to allow for an instantaneous burst of the starburst activities (e.g., \citealt{Ciesla2015,Nersesian2019}).
The parameter ranges of the SFH model are mainly referred from the works with CIGALE code for local Herschel-selected galaxies \citep{Buat2018} and U/LIRGs \citep{Paspaliaris2021}.
The stellar emission is modeled as a 5 Gyr simple stellar population (BC03; \citealt{Bruzual2003}).
Here, the \citet{Chabrier2003} IMF, solar metallicity, and the standard nebular emission model (see \citealt{Inoue2011}) are adopted.
We consider the attenuation law for the stellar continuum of \citet{Calzetti2000}, where the nebular emission is reddened with a Milky Way extinction curve (\citealt{Cardelli1989}).
The reddening of the nebular emission, $E$($B-V$)$_{\rm line}$ is a free parameter, while the ratio of the reddening of the emission lines and whole stellar continuum, $E$($B-V$)$_{\rm star}$/$E$($B-V$)$_{\rm line}$, was found to be 0.44 for local starbursts \citep{Calzetti2001}.
Since the contribution of the UV bump for local galaxies is small ($\sim$1/3 of that of the MW bump) and can be ignored as it affects the near-UV magnitude by only 0.1 mag \citep{Salim2018}, we fix the amplitude at zero.
Recent works report that the extinction curves show a great diversity in power-law slope (e.g., \citealt{Salim2018,Salim2020}), whose range of the slopes is covered from $-$0.8 to 0.0.
We fit the cold dust emission from the host galaxy with the physical dust model from \citet{Draine2014}, the updated model of the previous one (\citealt{Draine2007}).
The parameter ranges of \texttt{qpah} ($q_{\rm pah}$; 0.47--2.50), \texttt{umin} (1--50) and \texttt{alpha} (2.0 or 2.5) are referred from recent works (e.g., \citealt{Buat2018,Buat2021}). The dust fraction in photo-dissociation regions (PDRs) is fixed to 0.02 (e.g., \citealt{Draine2007,Dale2012,Magdis2012,Buat2018}).

\subsubsection{Torus Model (CLUMPY and SKIRTOR)} \label{subsub4-2-2-torus}
For the torus component, we adopt the CLUMPY model implemented in the X-CIGALE code (Model I). The number of clumps along the equator of $N_{\rm clump}^{\rm Equ}$ = 10, the ratio of the outer to inner radii of $Y$ = 20, and the radial clumpy distribution index of $q$ = 0.5 are assumed to make the analysis consistent with XCLUMPY \citep{Tanimoto2019}.
The torus optical depth in the $V$ band is a free parameter ($\tau_{\rm V}$; 40, 80, or 120), while the torus angular width ($\sigma$; 15\degr\ to 70\degr\ in steps of 5\degr) and inclination angle ($i$; 30\degr, 60\degr, or 80\degr) are fixed as be the closest values to the parameters of the best-fitting model in the X-ray works (\citealt{Yamada2020,Yamada2021,Ogawa2021}).\footnote{Since the reduced $\chi^2$ is larger than 10 for IRAS~F08572+3915, its inclination of 80\degr\ is chosen that present the smallest $\chi^2$.
For NGC~1068, NGC~1275, NGC~1365, Mrk~266B/Mrk~266A, and NGC~6240S/NGC~6240N, we assume $\sigma$ to be a typical value of 20\degr\ (e.g., \citealt{Yamada2021,Ogawa2021}), and select the inclination angle showing the smallest $\chi^2$ from 30\degr, 60\degr, and 80\degr.}
For the systems whose individual galaxies have different inclinations (NGC~6921/MCG+04-48-002 and NGC~7679/NGC~7682), we select $\sigma = 20^\circ$ and $i = 60^\circ$ as typical values in the X-ray results.
The UV-to-IR SEDs of starburst-dominant or hard X-ray--undetected sources are modeled without AGN component ($f_{\rm AGN}$ = 0), while allowing the AGN fraction in the IR band to free between 0.05--0.8 for AGNs.

We also perform the SED decomposition with the SKIRTOR model (Model II; see Appendix~\ref{AppendixD-SKIRTOR}).
A torus density radial parameter ($q$) and torus density angular parameter ($x$) are fixed to be 1.0 \citep[e.g.,][]{Stalevski2016,Yang2020}.\footnote{We confirm that the results and discussion are unchanged even if $q$ = 0.5 to keep consistency with CLUMPY model.}
The equatorial optical depth at 9.7~$\mu$m ($\tau_{9.7}$) is a free parameter (5, 9, or 11).
The angular distribution of clumps has a sharp edge corresponding to the angle between the equatorial plane and edge of the torus ($\Delta$).
Here, the unobscured and obscured AGNs are determined by the inclination angle above $i < 90^\circ - \Delta$ or not, respectively.
To keep fully consistency between the SKIRTOR-based AGN classification and X-ray classification in our sample (Table~\ref{T1_targets}), we convert the torus angular width ($\sigma$) of [10\degr--15\degr, 15\degr--25\degr, 25\degr--70\degr] to the $\Delta$ of [30\degr, 40\degr, 60\degr], respectively.

% Figure 3
\begin{figure*}
    \centering
    \includegraphics[keepaspectratio, scale=0.52]{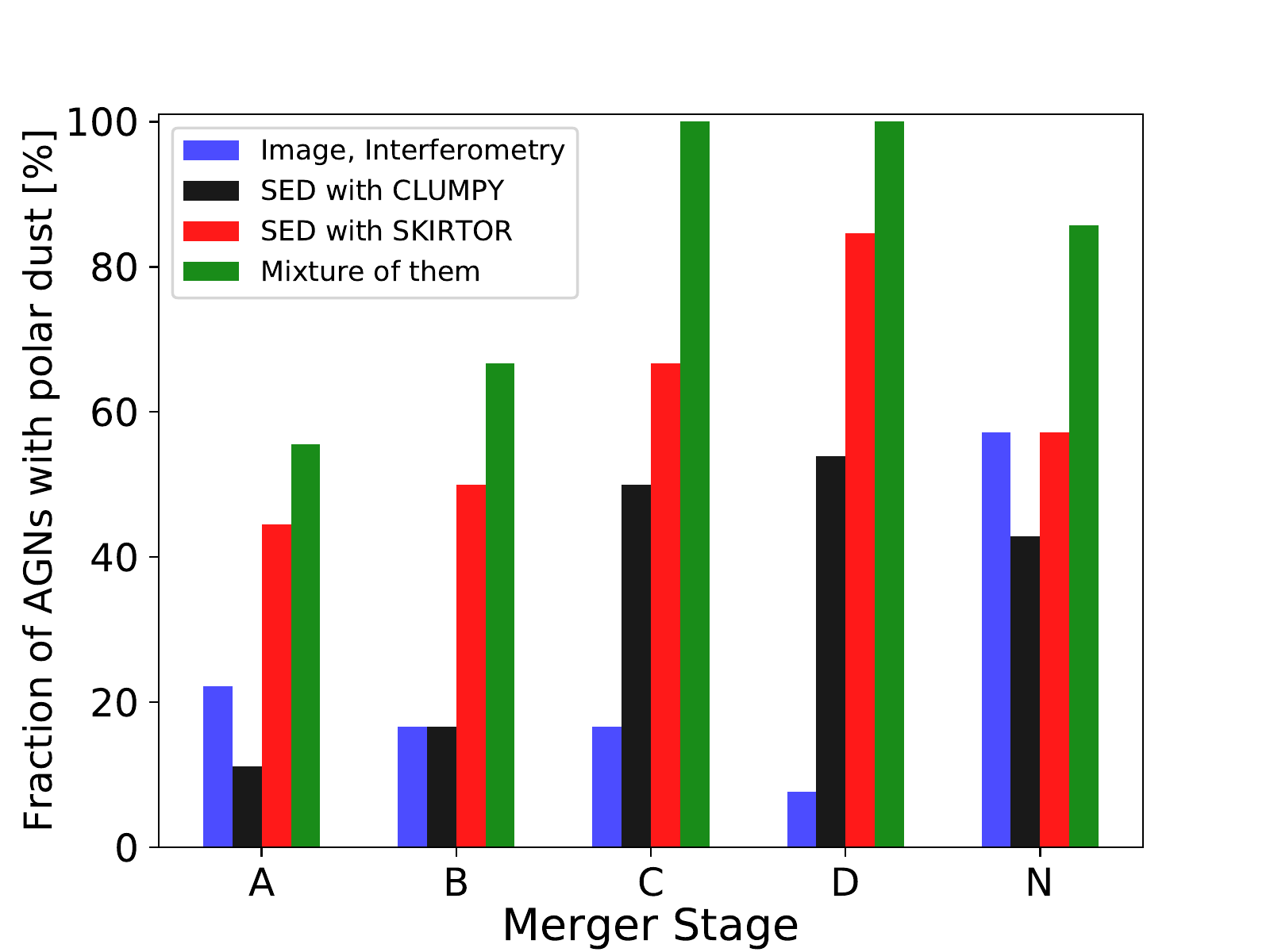}
    \includegraphics[keepaspectratio, scale=0.38]{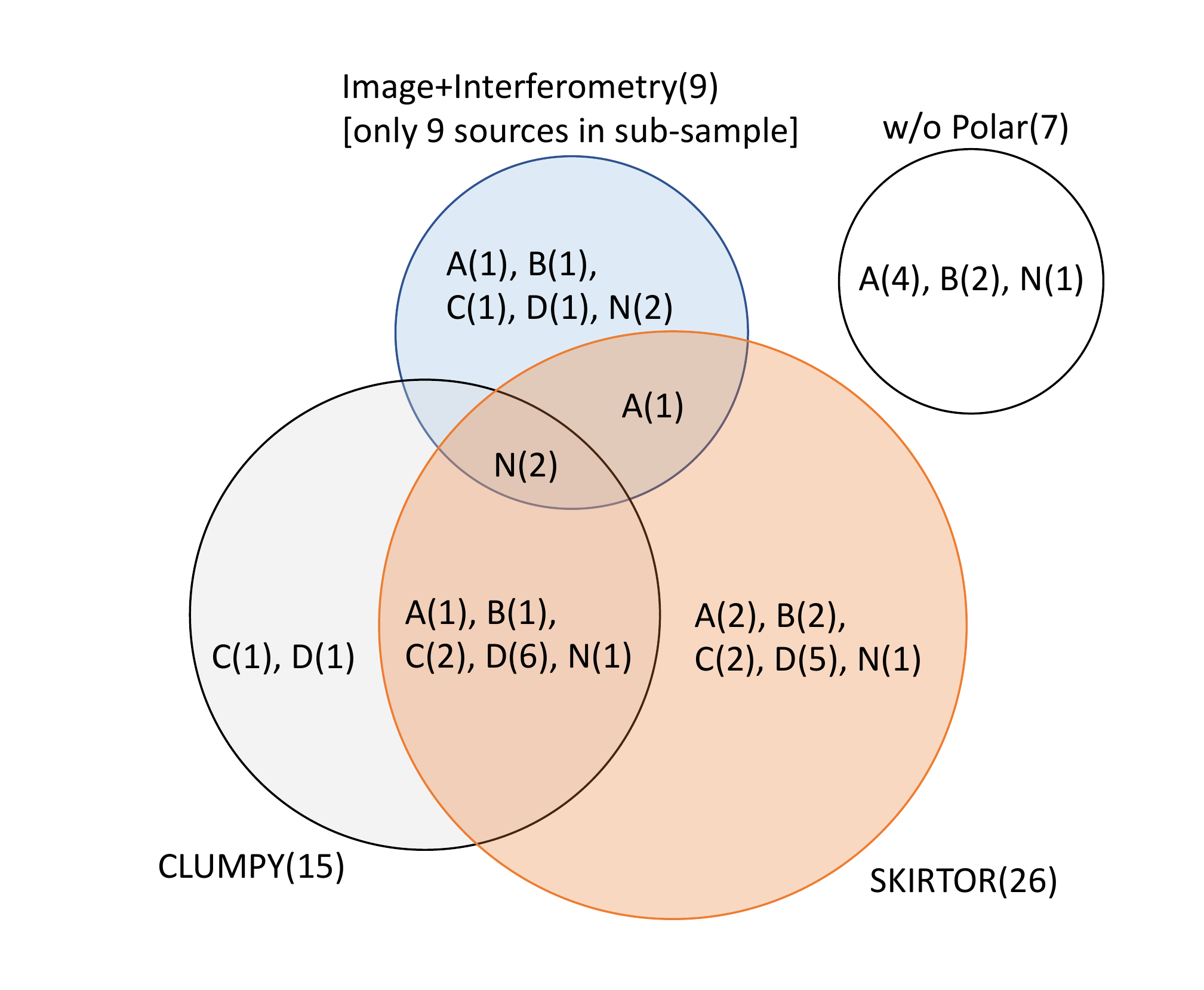}
    \caption{Left panel: Histogram of the fraction of the sources that shows some signatures of the presence of polar dust emission for the AGNs with merger stage.
    Each signature derived from the high-resolution image \citep[e.g.,][]{Asmus2019} and interferometry \citep[e.g.,][]{Lopez-Gonzaga2016a,Lopez-Gonzaga2016b} in the mid-IR band (blue), 
    the UV-to-IR SED decomposition with CLUMPY (black) and with SKIRTOR (red) are color-coded.
    Green histogram illustrates the fraction for the AGNs with one or more signs of polar dust emission (green).
    Right panel: Venn diagram of the sources with/without the signs of polar dust emission. The numbers of the sources for individual categories and merger stages are illustrated in parentheses.
    \\}
\label{F3-BIC-polar}
\end{figure*}

\subsubsection{Polar Dust Model} \label{subsub4-2-3-polar}
In the polar dust model, we adopt the Small Magellanic Cloud (SMC; \citealt{Prevot1984}) extinction curve, which is preferred from AGN observations \citep[e.g.,][]{Hopkins2004,Salvato2009,Bongiorno2012,Buat2021}.
The reddening $E$($B-V$) is free (0.05, 0.3, and 0.8; e.g., \citealt{Buat2021,Paspaliaris2021,Toba2021b,Yang2022}), or setting $E$($B-V$) = 0 if no significant AGN contribution ($f_{\rm AGN}$ = 0).
The emissivity ($\beta$) is fixed at 1.6 as a typical value \citep[e.g.,][]{Draine1984,Casey2012,Yang2020}.
The previous IR works indicate the temperature of polar dust ($T_{\rm polar}$) is $\sim$100--200~K (e.g., \citealt{Lopez-Rodriguez2018,Lyu2018,Honig2019}), while the temperature of the torus is $\gtrsim$300~K (e.g., \citealt{Tristram2009,Tristram2011}).
Recent high-spatial-resolution mid-IR observation of the nearby Seyfert 2 galaxy, the Circinus galaxy, with Very Large Telescope Interferometer also supports the similar temperatures ($\sim$370~K within the subparsec-scale central region and $\sim$200~K in outer polar dust regions; \citealt{Isbell2022}).
These temperatures are much larger than those expected for a galaxy's ISM heated only by star formation ($\sim$20--60~K; e.g., \citealt{Casey2012,Clements2018,da-Cunha2021}).
Although the higher temperature dust above 300~K should radiate the near-IR emission from a ring-like dust structure on the subparsec central region \citep[e.g.,][]{Krolik2007,Schartmann2014,Baba2018,Honig2019,Lyu2021,Gamez-Rosas2022,Matsumoto2022}, 
the material of the inner edge of a torus and accompanying polar dust would be difficult to distinguish by both near-IR and X-ray observations.
Therefore, we select the range of the polar dust temperature within 100--250~K, and clarify that (1) the estimates of the physical properties of extended ($\gtrsim$10~pc; see also Section~\ref{sub6-5-polar-structure}) polar dust that are not taken into account for the hot ($>$300~K) dust component on the subparsec region and (2) the central hot dust emission are predominantly considered as the torus emission in this study.

\subsubsection{Significance Test for Torus and Polar-dust Emission} \label{subsub4-2-4-BIC}
We examine the Bayesian information criterion (BIC; \citealt{Schwarz1978}) for two kinds of fits with and without the AGN (torus and polar dust) component.
Using the $\chi^2$ statistics, the BIC is provided by the equation as BIC = $\chi^2 + k \times {\rm ln(}N{\rm )}$, where $k$ is the number of free parameters and $N$ is the number of photometric data (\citealt{Ciesla2018,Buat2019}). 
Among the sources yielding a poor fit with reduced $\chi^2$ above 6, the three galaxies, ESO 350$-$38, ESO 374$-$IG032, and NGC 4418, show the $\Delta$BIC = BIC$_{\rm woAGN}$ $-$ BIC$_{\rm wAGN} > 6$ (significantly improvement with a posterior probability above 95\%; \citealt{Raftery1995}).
Thus, we re-classify these three galaxies as AGN candidates and select the inclination angle providing the smallest $\chi^2$.

Furthermore, we conduct the significance test for the polar dust component in the UV-to-IR SED analysis.
The left panel of Figure~\ref{F3-BIC-polar} presents the fraction of sources showing BIC $>$ 6 for a polar dust component with two kinds of torus models.
We find that the analysis with SKIRTOR provides larger fractions of the AGNs showing significant polar dust contribution than with CLUMPY.
This stronger significance of mid-IR polar dust emission can be explained by the flatter SED slope of the torus component in the SKIRTOR model due to the strong near-IR emission from smooth low-density dust (Section~\ref{subsub4-1-2-AGN-models} and Appendix~\ref{AppendixD-SKIRTOR}).
Although the sample is quite limited to the nine AGNs, the signatures of polar-dust radiation are reported by high-resolution mid-IR images and interferometry (NGC~1068, NGC~1365, NGC~3690 West, Mrk~231, MCG$-$03-34-064, NGC~5135, IC~4518W, NGC~7469, and NGC~7674; \citealt{Lopez-Gonzaga2014,Lopez-Gonzaga2016a,Lopez-Gonzaga2016b,Asmus2016,Asmus2019,Lopez-Rodriguez2017,Lopez-Rodriguez2018,Mattila2018,Gamez-Rosas2022}).\footnote{The presence of polar dust in Mrk 231 is implied by near-to-mid-IR polarimetry \citep{Lopez-Rodriguez2017}, and in NGC 3690 West by detecting the time variability of nucleus IR SED due to the tidal disruption event radiating the polar dust emission \citep{Mattila2018}.}
Six of these AGNs hosting polar dust emissions do not show any significant BIC values with these torus models (right panel of Figure~\ref{F3-BIC-polar}).
This is mainly because the available photometries of them in the near-IR (2MASS) and/or mid-IR (WISE and AKARI) bands are insufficient to give a diagnosis, which may be derived by the artifacts related to their IR emission from extended or comprehensive morphology.
So, we adopt the polar dust model for the AGNs with one or more signs from mid-IR images, interferometry, and BIC tests by SED analysis with CLUMPY and SKIRTOR.
For the seven AGNs without any signs of polar dust contribution, most of whose IR photometries are available, we treat them as AGNs without polar dust emission by applying $E$($B-V$) = 0, i.e., returning to the original torus model.
Impressively, we find that these AGNs are in early mergers and nonmergers, while all of the AGNs in late mergers show the signatures of polar dust emission, as discussed in Section~\ref{subsub6-5-2-polar-outflow}.
Finally, we list the best fitting parameters in the tables of Appendix~\ref{AppendixA-best-values} (e.g., Table~\ref{TA1_results1}--\ref{TA5_results5}).

\subsection{Radio Fitting} \label{sub4-3-radio-fit}
By fixing the best-fitting parameters obtained by the UV-to-IR SED analysis, we finally conduct the radio SED decomposition.
The radio model is composed of the continuum from the nebular emission (free-free, free-bound, and two-photon continua)\footnote{The model of the nebular component covers the UV to radio 1~m emission.
To expand the available wavelength, we assume the slope between the fluxes in the 0.8 and 1~m bands to the 1--10~m radio emission for the nebular component, where it is less dominant than the synchrotron component.}, which also radiates UV-to-IR emission (Section~\ref{subsub4-2-1-galaxy}), and independent synchrotron emission from AGN and/or starburst \citep{Boquien2019}.
We parameterized a correlation coefficient between total-IR luminosity and monochromatic radio luminosity of the synchrotron emission at 1.4~GHz ($q_{\rm IR}$ = log$L_{\rm IR}$/$L_{\rm 1.4GHz}$), and the power-law slope of synchrotron emission ($\alpha_{\rm radio}$) derived from the low-frequency radio emission in the 0.1--4~m or 0.07--3.0~GHz bands.
We choose the parameter ranges of $q_{\rm IR}$ between $-$1.50 and +3.50 with a step of 0.01 \citep[e.g.,][]{Yun2001,U2012,Vardoulaki2015}, and $\alpha_{\rm radio}$ between 0.00 and 1.50 with a step of 0.01 \citep[e.g.,][]{Murphy2013b,Toba2019b}.
If only a single radio photometry or upper limit values are available, we fix $\alpha_{\rm radio}$ to be 0.5, a median value in the $<$5~GHz bands among local U/LIRGs \citep{Murphy2013b}, consistent with the averaged value of the estimates in our study ($\alpha_{\rm radio} = $0.48$\pm$0.16).
The best-fitting parameters and radio luminosities are listed in Table~\ref{TA3_results3}, and
the combined hard X-ray to radio multiwavelength SEDs and their best-fitting models are shown in Figures~\ref{F1-SED00-F} and Appendix~\ref{AppendixE-SEDs}.

\subsection{Best-fit Parameters and Their Validity} \label{sub4-4-results}
\begin{figure}
    \centering
    \includegraphics[keepaspectratio, scale=0.42]{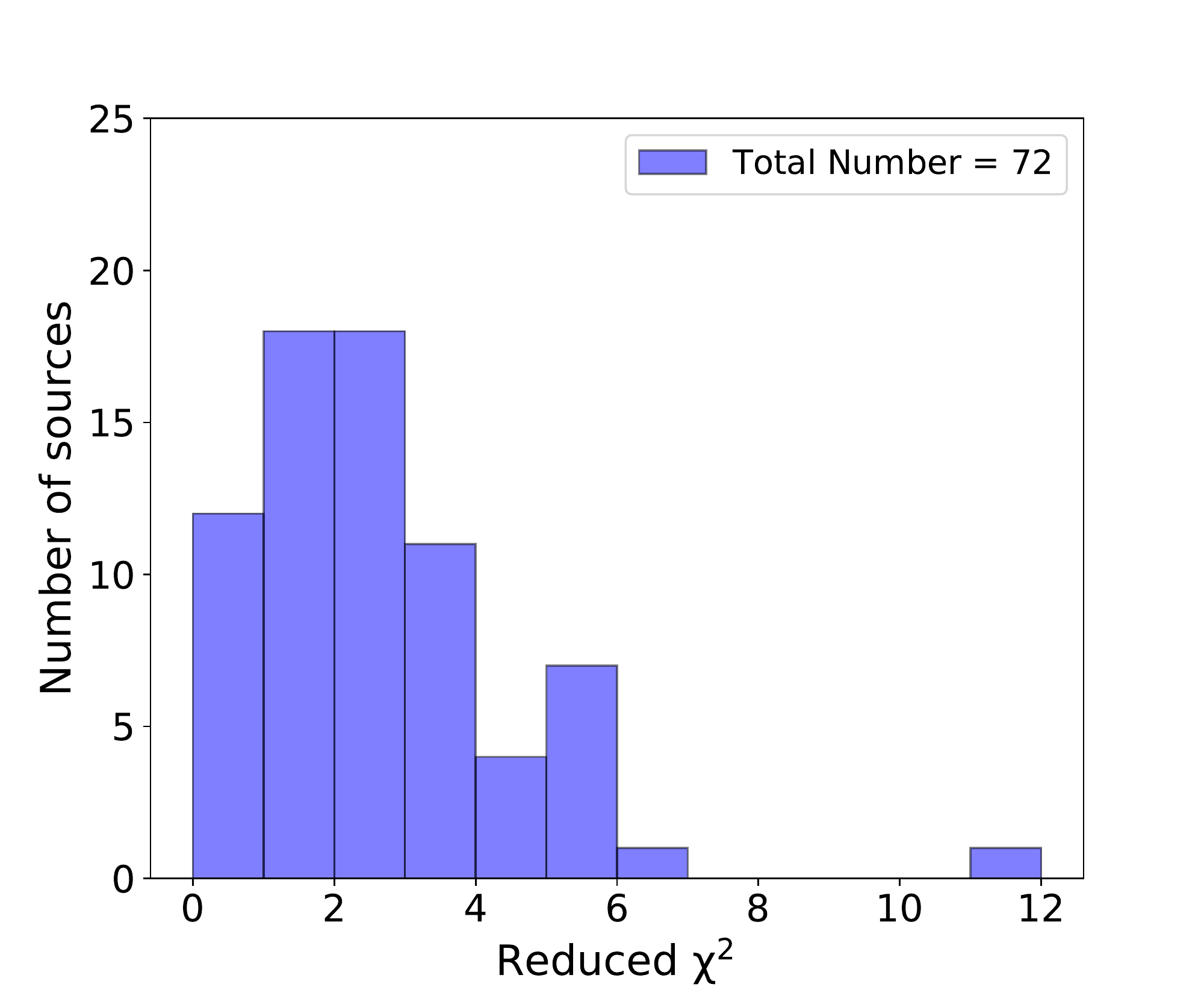}
    \caption{Histogram of the reduced $\chi^2$ derived from the UV-to-IR SED decomposition for the 72 individual targets in our sample.
    \\}
\label{F4-chi}
\end{figure}

We have performed the multiwavelength SED analysis for the local U/LIRGs, covering the hard X-ray to radio wavelength bands.
The best-fit parameters of individual sources are summarized in Tables~\ref{TA1_results1}--\ref{TA5_results5}, and the averaged values of these results are shown in Table~\ref{T4_average}.
The histogram of the reduced $\chi^2$ of the UV-to-IR SED decomposition for the 72 individual targets in our sample is displayed in Figure~\ref{F4-chi}.
The SED decomposition of some targets shows a relatively large $\chi^2$ above 5, which is mainly caused by the noisy photometric data in the UV or optical band (Figures~\ref{SED01-F}--\ref{SED18-F} in Appendix~\ref{AppendixE-SEDs}).
Conducting a mock analysis, we confirm that the best-fit parameters derived from our multiwavelength SED decomposition are less affected by photometric uncertainties (see Appendix~\ref{AppendixB-mock}).
However, the SED fitting of IRAS F08572+3915 shows a large reduced $\chi^2$ over 10 because the model presents weaker near-IR fluxes than the photometric data.
Its SED is moderately reproduced with SKIRTOR (with reduced $\chi^2$ of 6.7).
Although its AGN luminosity may be underestimated (at worse $\sim$0.4~dex), the AGN luminosities for the other sources derived with CLUMPY are well consistent with the values with SKIRTOR (Appendix~\ref{AppendixD-SKIRTOR}).
Thus, we utilize the results from the multiwavelength analysis with CLUMPY for all targets to discuss their characteristics of the host galaxy and AGN.

% Figure 4

\begin{figure}
    \centering
    \includegraphics[keepaspectratio, scale=0.42]{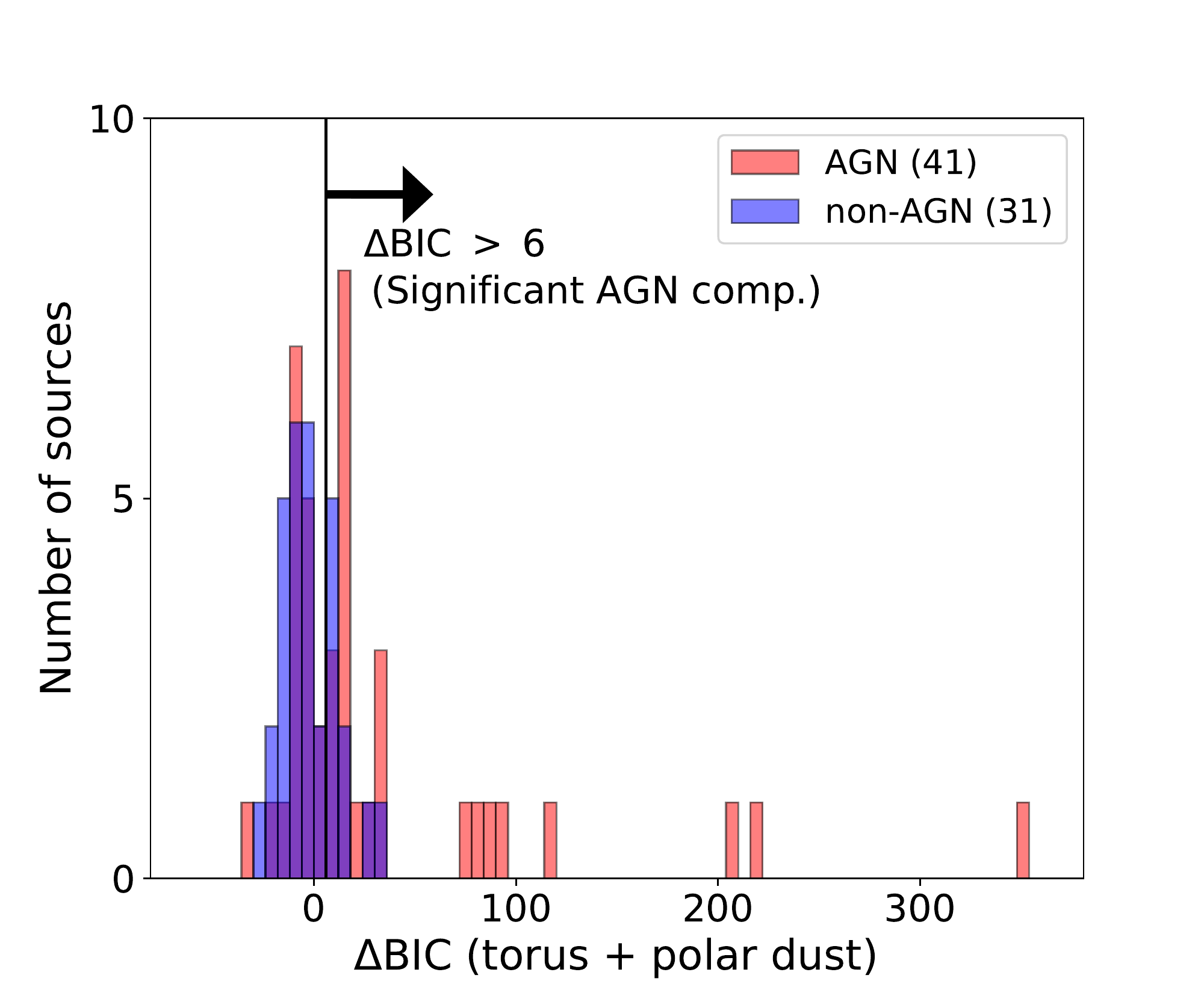}
    \caption{Histogram of $\Delta$BIC (= BIC$_{\rm woAGN} - $BIC$_{\rm wAGN}$) for the 72 individual targets in our sample.
    The black solid line shows the threshold of $\Delta$BIC = 6 as the significant differences (posterior probability above 95\%) in two fits with and without adding the AGN (torus and polar dust) component.
    The red histogram shows the number of the hard X-ray--detected AGNs and three newly identified AGN candidates. 
    The blue histogram is the number of the other sources, i.e., starburst-dominant sources or hard X-ray--undetected sources.
    \\}
\label{F5-BIC-agn}
\end{figure}

Next, we investigate the significance of the AGN feature in the multiwavelength SED decomposition.
Figure~\ref{F5-BIC-agn} shows the histogram of the $\Delta$BIC on the different fits with and without the AGN component for the 72 individual targets in our sample (see also Section~\ref{subsub4-2-4-BIC}).
Remarkably, most of the sources are distributed around the threshold of $\Delta$BIC = 6, regardless of whether the sources are the AGNs (hard X-ray--detected AGNs and three newly identified AGN candidates by the diagnostics of $\Delta$BIC $>$ 6) or the other sources (starburst-dominant or hard X-ray--undetected sources).
This indicates that the AGNs in U/LIRGs are difficult to detect only with the SED decomposition.
Therefore, the strong constraints on the AGN structure derived from the X-ray spectral analysis will effectively solve the complex SEDs in the IR band.

%Figure 5

We evaluate the effects on the results if unresolved photometries are utilized.
For the stellar mass ($M_*$) and SFR, the comparison of our results and previous works of SED analysis by using the photometry of total systems are presented in Appendix~\ref{AppendixE-SEDs}.
Figure~\ref{F6-pair} presents the differences in the results derived from the combined photometry of a total system and the separated photometries of individual galaxies for the 13 resolved pairs.
The stellar masses and SFRs are not affected by using the summed fluxes of individual galaxies.
On the other hand, for the 10 pairs hosting AGNs, the UV-to-IR torus ($L_{\rm torus}$) and polar dust luminosities ($L_{\rm polar}$), and intrinsic (bolometric) AGN disk luminosities ($L_{\rm AGN,int}$) derived from the combined photometries become larger than those from separated ones.
According to the SED models in Figures~\ref{SED01-F}--\ref{SED18-F}, the slope of the IR SEDs of the starburst component is similar to the one of the AGN (see also Section~\ref{sub5-4-averaged-SED}).
This makes it difficult to accurately extract the AGN contribution from the IR SED when the photometries of a starburst-dominant source and a galaxy hosting an AGN are mixed.
The overestimation of AGN luminosities for unresolved sources is found to be $\sim$0.2~dex, which is a small effect on the following discussion.
\\

%%%%%%%%%%%%%%%%%%%%%%%%%%%
%%% Section 5 : Results %%%
%%%%%%%%%%%%%%%%%%%%%%%%%%%
\section{Results: Multiwavelength Emission from U/LIRGs} \label{S5-results} 
Before the discussion on the properties of polar dust in U/LIRGs (Section~\ref{S6-discussion1}), 
we first study the characteristics of the multiwavelength
radiation, which are helpful to understand the detailed
environments from the nucleus to galactic scales.
We also aim to investigate unique features for future works to explore buried AGNs in U/LIRGs at higher redshift.
For these purposes, we here examine the SFR--$M_*$ relation as an indicator of the host galaxy's properties (Section~\ref{sub5-1-MS}).
Next, we evaluate the activities of AGNs and starbursts by focusing on the radio emission (Section~\ref{sub5-2-radio}), WISE IR color (Section~\ref{sub5-3-WISE-color}), and averaged multiwavelength SEDs (Section~\ref{sub5-4-averaged-SED}).
Finally, we discuss the properties of AGNs in U/LIRGs by comparing the X-ray and other wavelength luminosities corrected for the torus absorption (Section~\ref{sub5-5-X-weak}).
The main findings are summarized in Section~\ref{sub5-6-summary-SED}.

% Figure 6
\begin{figure}
    \centering
    \includegraphics[keepaspectratio, scale=0.5]{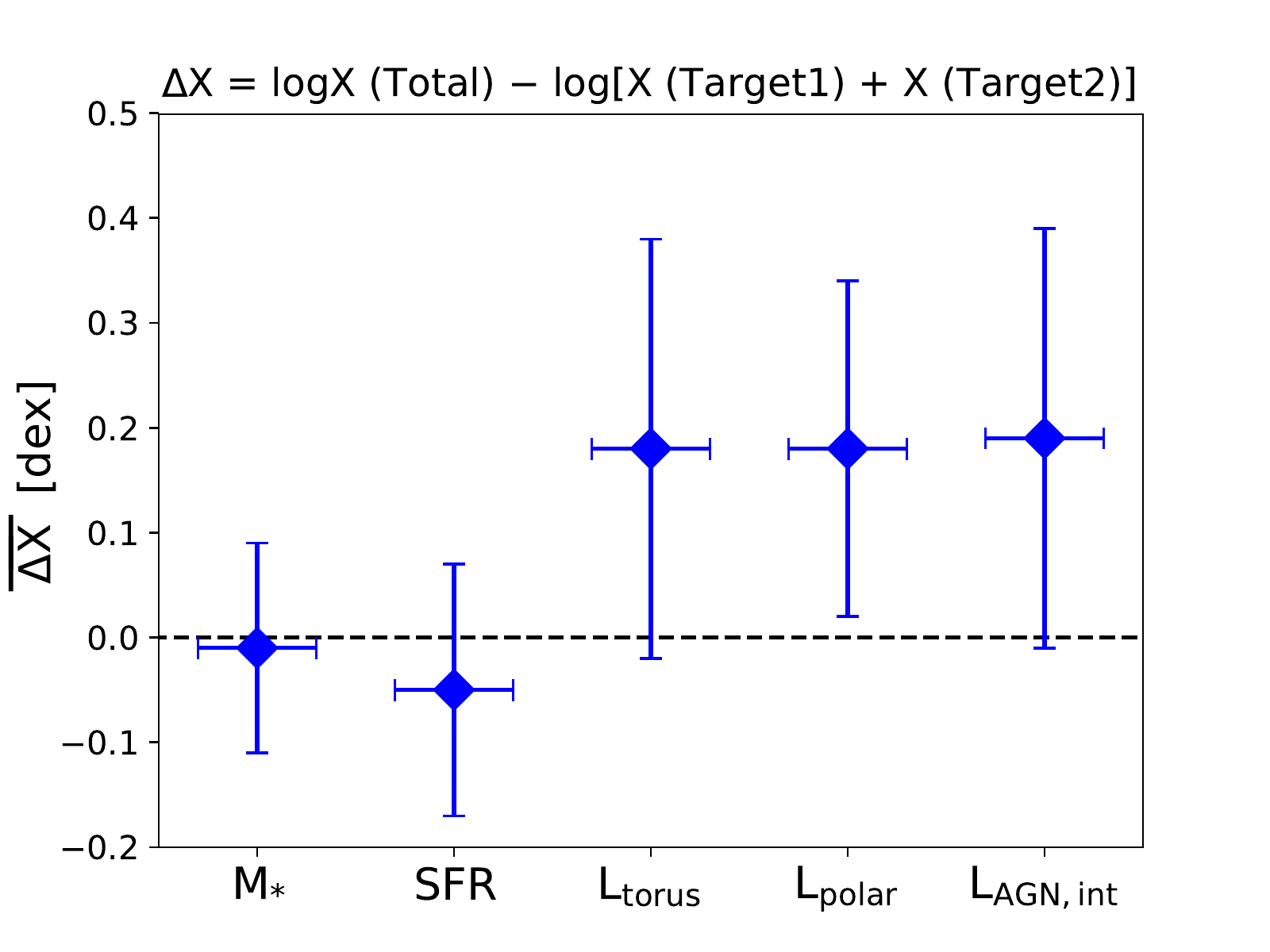}
    \caption{The differences in $M_*$, SFR, $L_{\rm torus}$, $L_{\rm polar}$, and $L_{\rm AGN,int}$ derived from the UV-to-IR SED analysis by using the combined photometry of a total system and separated photometries of the individual galaxies for 13 resolved pairs.
    Diamonds and these uncertainties denote the averaged values and 1$\sigma$ dispersions, respectively.
    \\}
\label{F6-pair}
\end{figure}

% Table 4
\begin{deluxetable*}{llccccccc}
\label{T4_average}
\tabletypesize{\scriptsize}
\tablecaption{Averaged Values of Best-fit Parameters and Accompanying Results.}
\tablehead{
\colhead{No.} &
\colhead{Parameter} &
\multicolumn{7}{c}{Merger Stage} \\
\cline{3-9}
 & & A & B & C & D & A--B & C--D & N
}
\startdata
\multicolumn{9}{c}{Free parameters (UV-to-IR SED analysis; Section~\ref{sub4-2-SED-model})} \\
\hline
1 & log(\texttt{tau\_main}) (Myr) & $3.38\pm0.23$ & $3.42\pm0.22$ & $3.50\pm0.12$ & $3.53\pm0.09$ & $3.40\pm0.23$ & $3.52\pm0.10$ & $3.38\pm0.19$ \\
2 & log(\texttt{age\_main}) (Myr) & $3.89\pm0.12$ & $3.85\pm0.10$ & $3.91\pm0.07$ & $3.89\pm0.06$ & $3.87\pm0.11$ & $3.89\pm0.07$ & $3.87\pm0.08$ \\
3 & log(\texttt{age\_bq}) (Myr)   & $1.65\pm0.22$ & $1.68\pm0.21$ & $1.55\pm0.32$ & $1.39\pm0.31$ & $1.67\pm0.22$ & $1.44\pm0.32$ & $1.61\pm0.16$ \\
4 & log(\texttt{r\_sfr})          & $1.12\pm0.79$ & $1.39\pm0.88$ & $1.57\pm0.77$ & $2.19\pm0.64$ & $1.24\pm0.84$ & $2.01\pm0.74$ & $1.26\pm0.85$ \\
5 & \texttt{E\_BV\_lines}         & $0.72\pm0.51$ & $0.99\pm0.43$ & $0.83\pm0.38$ & $1.37\pm0.29$ & $0.84\pm0.50$ & $1.22\pm0.41$ & $0.98\pm0.56$ \\
6 & \texttt{powerlaw\_slope}      & $-0.47\pm0.24$ & $-0.41\pm0.31$ & $-0.39\pm0.27$ & $-0.35\pm0.29$ & $-0.45\pm0.27$ & $-0.36\pm0.29$ & $-0.40\pm0.29$ \\
7 & $q_{\rm pah}$                 & $1.19\pm0.58$ & $1.16\pm0.68$ & $0.85\pm0.38$ & $0.80\pm0.54$ & $1.17\pm0.62$ & $0.82\pm0.50$ & $1.23\pm0.59$\\
8 & \texttt{umin}                 & $10.18\pm11.07$ & $15.49\pm15.10$ & $25.15\pm14.95$ & $29.93\pm15.15$ & $12.50\pm13.25$ & $28.57\pm15.25$ & $7.91\pm6.52$ \\
9 & \texttt{alpha}                & $2.19\pm0.20$ & $2.20\pm0.23$ & $2.18\pm0.22$ & $2.31\pm0.22$ & $2.19\pm0.22$ & $2.27\pm0.23$ & $2.17\pm0.20$ \\
10 & $\tau_{\rm V}$                & $81.51\pm8.95$ & $85.26\pm6.15$ & $74.31\pm22.58$ & $73.37\pm23.63$ & $83.01\pm8.16$ & $73.67\pm23.31$ & $67.56\pm15.52$ \\
11 & $f_{\rm AGN}$                 & $0.16\pm0.13$ & $0.11\pm0.04$ & $0.26\pm0.18$ & $0.18\pm0.19$ & $0.14\pm0.11$ & $0.20\pm0.19$ & $0.36\pm0.26$\\
12 & $E$($B-V$) (mag)              & $0.40\pm0.26$ & $0.25\pm0.17$ & $0.39\pm0.19$ & $0.33\pm0.27$ & $0.33\pm0.23$ & $0.35\pm0.25$ & $0.59\pm0.21$\\
13 & $T_{\rm polar}$ (K)           & $164.03\pm37.97$ & $173.29\pm8.66$ & $134.10\pm27.44$ & $132.57\pm33.06$ & $168.15\pm29.25$ & $133.06\pm31.40$ & $172.83\pm59.96$ \\
\hline
\multicolumn{9}{c}{Fixed Parameters (UV-to-IR SED analysis; Section~\ref{sub4-2-SED-model})} \\
\hline
14 & $\sigma$ (degree)             & $24.4\pm14.6$ & $19.2\pm1.9$ & $24.2\pm9.3$ & $19.2\pm2.7$ & $22.3\pm11.7$ & $20.8\pm6.1$ & $19.3\pm3.2$ \\
15 & $i$ (degree)                  & $54.4\pm18.9$ & $63.3\pm7.5$ & $61.7\pm16.7$ & $60.0\pm15.2$ & $58.0\pm16.0$ & $60.5\pm15.7$ & $62.9\pm21.9$ \\
\hline
\multicolumn{9}{c}{Free Parameters (radio fitting; Section~\ref{sub4-3-radio-fit})} \\
\hline
16 & $\alpha_{\rm radio}^\dagger$          & $0.54\pm0.16$ & $0.52\pm0.07$ & $0.52\pm0.14$ & $0.38\pm0.15$ & $0.54\pm0.13$ & $0.43\pm0.16$ & $0.52\pm0.16$ \\
17 & $q_{\rm ir}^\dagger$                  & $2.39\pm0.41$ & $2.26\pm0.29$ & $2.50\pm0.19$ & $2.57\pm0.27$ & $2.34\pm0.37$ & $2.55\pm0.25$ & $2.10\pm0.94$ \\
\hline
\multicolumn{9}{c}{Accompanying results} \\
\hline
18 & log($M_*$) ($M_{\odot}$)      & $10.23\pm0.91$ & $10.42\pm0.39$ & $10.43\pm0.35$ & $10.35\pm0.52$ & $10.31\pm0.74$ & $10.37\pm0.48$ & $10.54\pm0.45$ \\
19 & log(SFR) ($M_{\odot}$ yr$^{-1}$)     & $0.49\pm0.82$ & $1.06\pm0.55$ & $1.53\pm0.38$ & $1.98\pm0.32$ & $0.74\pm0.77$ & $1.86\pm0.40$ & $1.04\pm0.34$ \\
20 & log(sSFR) (yr$^{-1}$)         & $-9.74\pm1.08$ & $-9.36\pm0.80$ & $-8.90\pm0.49$ & $-8.36\pm0.54$ & $-9.58\pm0.98$ & $-8.51\pm0.58$ & $-9.50\pm0.51$ \\
21 & log($L_{\rm 1.4 GHz}$)$^\dagger$ (erg s$^{-1}$) & $38.19\pm0.88$ & $38.75\pm0.52$ & $39.09\pm0.30$ & $39.47\pm0.44$ & $38.44\pm0.80$ & $39.36\pm0.44$ & $38.97\pm0.90$ \\
22 & $q_{\rm excess}^\dagger$              & $0.12\pm0.08$ & $0.11\pm0.11$ & $0.08\pm0.02$ & $0.08\pm0.03$ & $0.11\pm0.10$ & $0.08\pm0.03$ & $0.08\pm0.02$ \\
23 & log($L_{\rm 6,AGN}$) (erg s$^{-1}$) & $42.39\pm0.95$ & $42.77\pm0.60$ & $43.30\pm0.48$ & $43.52\pm0.50$ & $42.54\pm0.85$ & $43.45\pm0.50$ & $43.18\pm0.54$ \\
24 & log($L_{\rm 12,t}$) (erg s$^{-1}$) & $43.03\pm0.83$ & $43.31\pm0.49$ & $43.93\pm0.50$ & $44.09\pm0.52$ & $43.14\pm0.73$ & $44.04\pm0.52$ & $43.74\pm0.51$ \\
25 & log($L_{\rm 12,p}$) (erg s$^{-1}$) & $43.24\pm0.29$ & $43.07\pm0.75$ & $43.10\pm0.24$ & $43.46\pm0.75$ & $43.16\pm0.55$ & $43.35\pm0.65$ & $43.34\pm0.84$ \\
26 & log($L_{\rm 12,AGN}$) (erg s$^{-1}$) & $43.21\pm0.85$ & $43.52\pm0.53$ & $44.04\pm0.40$ & $44.23\pm0.53$ & $43.34\pm0.75$ & $44.17\pm0.50$ & $43.98\pm0.56$ \\
27 & log($L_{\rm torus}$) (erg s$^{-1}$) & $43.18\pm0.80$ & $43.43\pm0.44$ & $43.98\pm0.45$ & $44.14\pm0.50$ & $43.28\pm0.69$ & $44.09\pm0.49$ & $43.80\pm0.51$ \\
28 & log($L_{\rm polar}$) (erg s$^{-1}$) & $43.62\pm0.54$ & $43.40\pm0.18$ & $44.07\pm0.39$ & $44.29\pm0.59$ & $43.53\pm0.43$ & $44.22\pm0.55$ & $43.99\pm0.46$ \\
29 & log($L_{\rm AGN,int}$) (erg s$^{-1}$) & $43.49\pm0.71$ & $43.80\pm0.36$ & $44.35\pm0.45$ & $44.53\pm0.57$ & $43.61\pm0.62$ & $44.47\pm0.54$ & $44.31\pm0.53$ \\
30 & log($M_{\rm BH}$) ($M_{\odot}$) & $7.80\pm0.41$ & $7.83\pm0.24$ & $7.34\pm0.28$ & $7.83\pm0.46$ & $7.81\pm0.35$ & $7.67\pm0.47$ & $7.79\pm0.25$ \\
31 & log($\lambda_{\rm Edd}$) & $-2.43\pm1.01$ & $-2.17\pm0.47$ & $-1.15\pm0.45$ & $-1.29\pm0.67$ & $-2.33\pm0.85$ & $-1.24\pm0.61$ & $-1.63\pm0.75$ \\
32 & log($R_{\rm polar}$) (pc) & $1.77\pm0.59$ & $1.56\pm0.05$ & $2.24\pm0.45$ & $2.35\pm0.40$ & $1.68\pm0.45$ & $2.32\pm0.42$ & $1.90\pm0.55$ \\
\enddata
\tablecomments{Comments: The table summarizes the averaged values of each parameter for individual merger stages.
(1--13) Free parameters as listed in Table~\ref{T3_para} (see Section~\ref{sub4-2-SED-model}); 
(14--15) Two fixed parameters of torus angular width and inclination angle as listed in Table~\ref{T3_para} (Section~\ref{sub4-2-SED-model});
(16--17) free parameters of the radio fitting (Section~\ref{sub4-3-radio-fit});
(18--20) stellar mass ($M_*$), SFR, and specific SFR (sSFR = SFR/$M_*$) as listed in Table~\ref{TA2_results2};
(21--22) rest-frame radio 1.4~GHz luminosity and radio-excess parameter ($q_{\rm excess}$) as listed in Table~\ref{TA3_results3} (see Section~\ref{sub5-2-radio});
(23--26) rest-frame 6~$\mu$m luminosity of AGN (torus and polar dust) component, rest-frame 12~$\mu$m luminosities of the torus, polar dust, and total AGN component (see Table~\ref{TA4_results4});
(27--32) UV-to-IR (mainly IR) luminosities (torus, polar dust, and AGN components), SMBH mass, Eddington ratio, and physical sizes of the polar dust structure (Section~\ref{sub6-5-polar-structure}) as listed in Table~\ref{TA5_results5}.
}
\tablenotetext{\dagger}{The values of the sources whose $\alpha_{\rm radio}$ is fixed at 0.5 are excluded.}
\end{deluxetable*}

% Figure 7
\begin{figure*}
    \centering
    \includegraphics[keepaspectratio, scale=0.5]{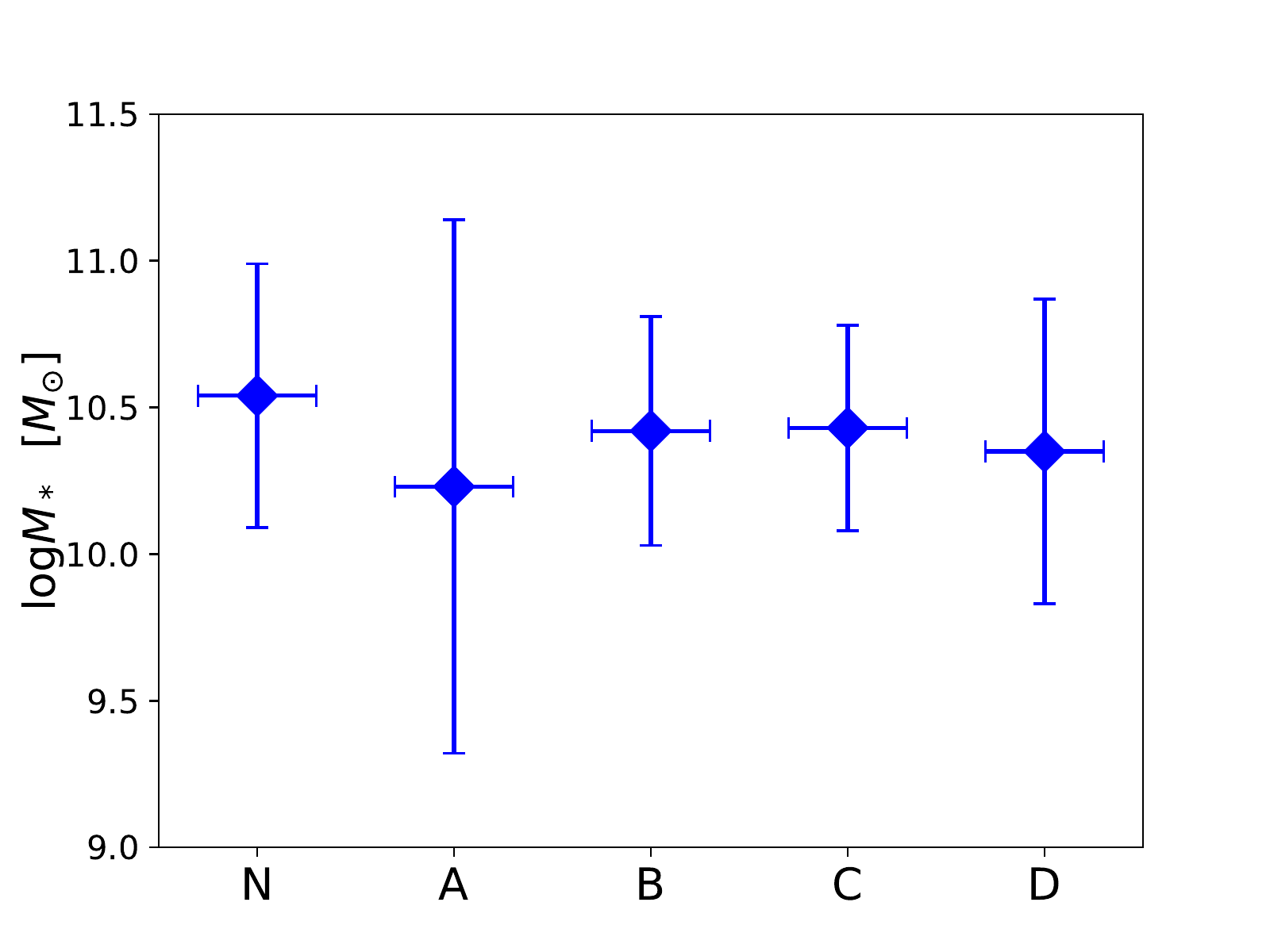}
    \includegraphics[keepaspectratio, scale=0.5]{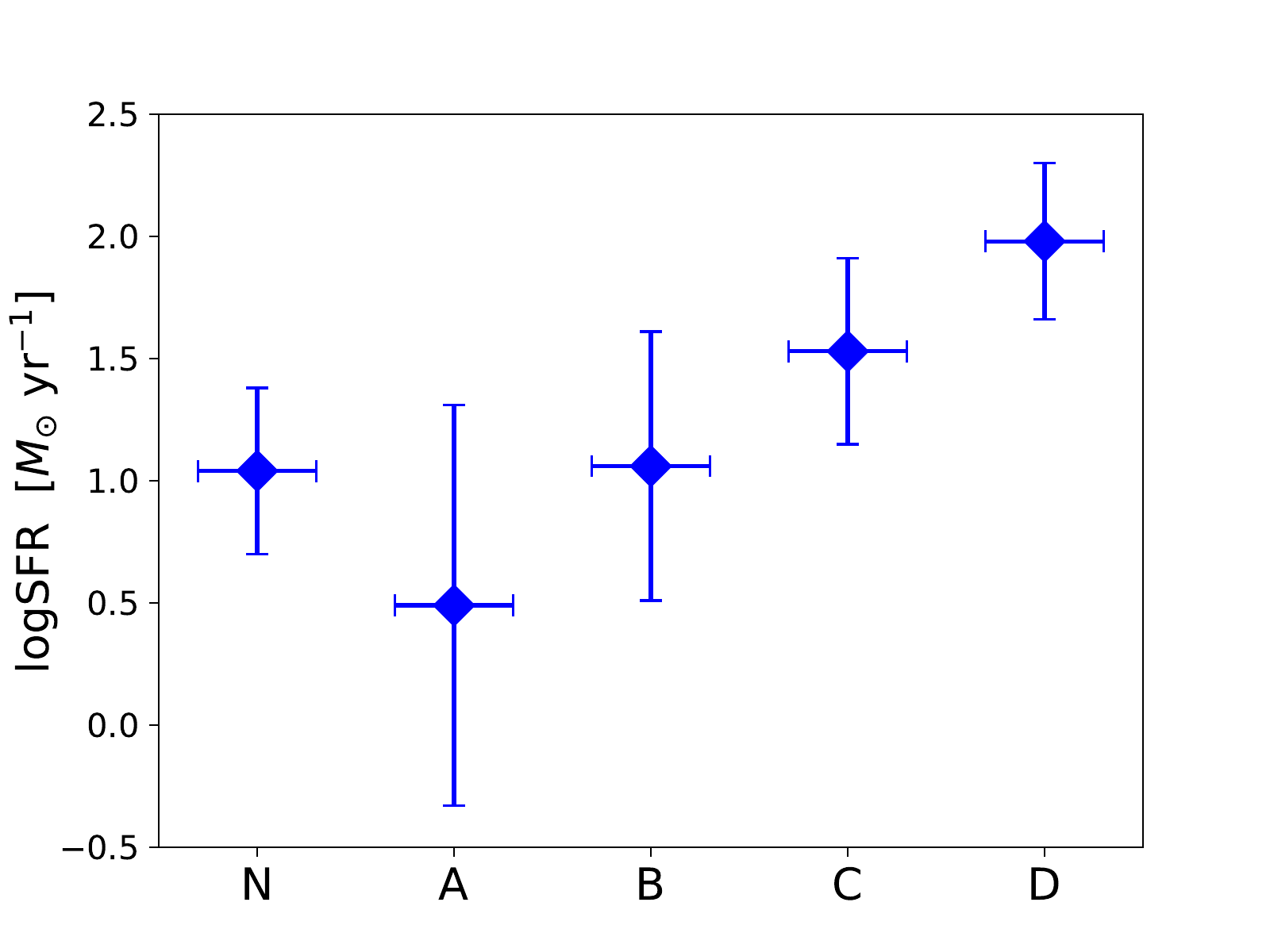}
    \caption{Left panel: averaged values of logarithmic $M_*$ compared with merger stages.
    Right panel: averaged values of logarithmic SFR compared with merger stages.
    \\}
\label{F7-graph}
\end{figure*}

% Figure 8
\begin{figure*}
    \centering
    \includegraphics[keepaspectratio, scale=0.5]{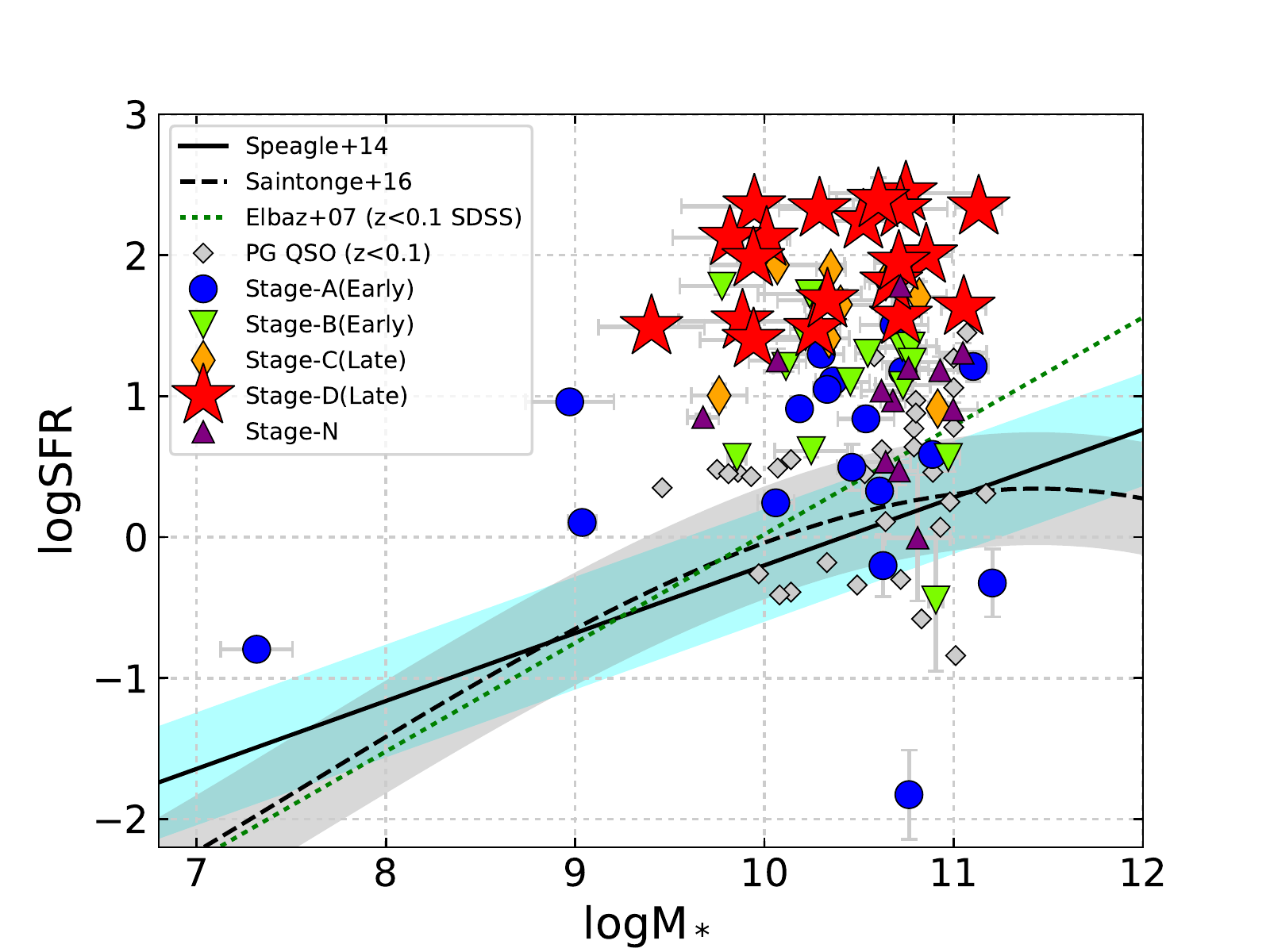}
    \includegraphics[keepaspectratio, scale=0.5]{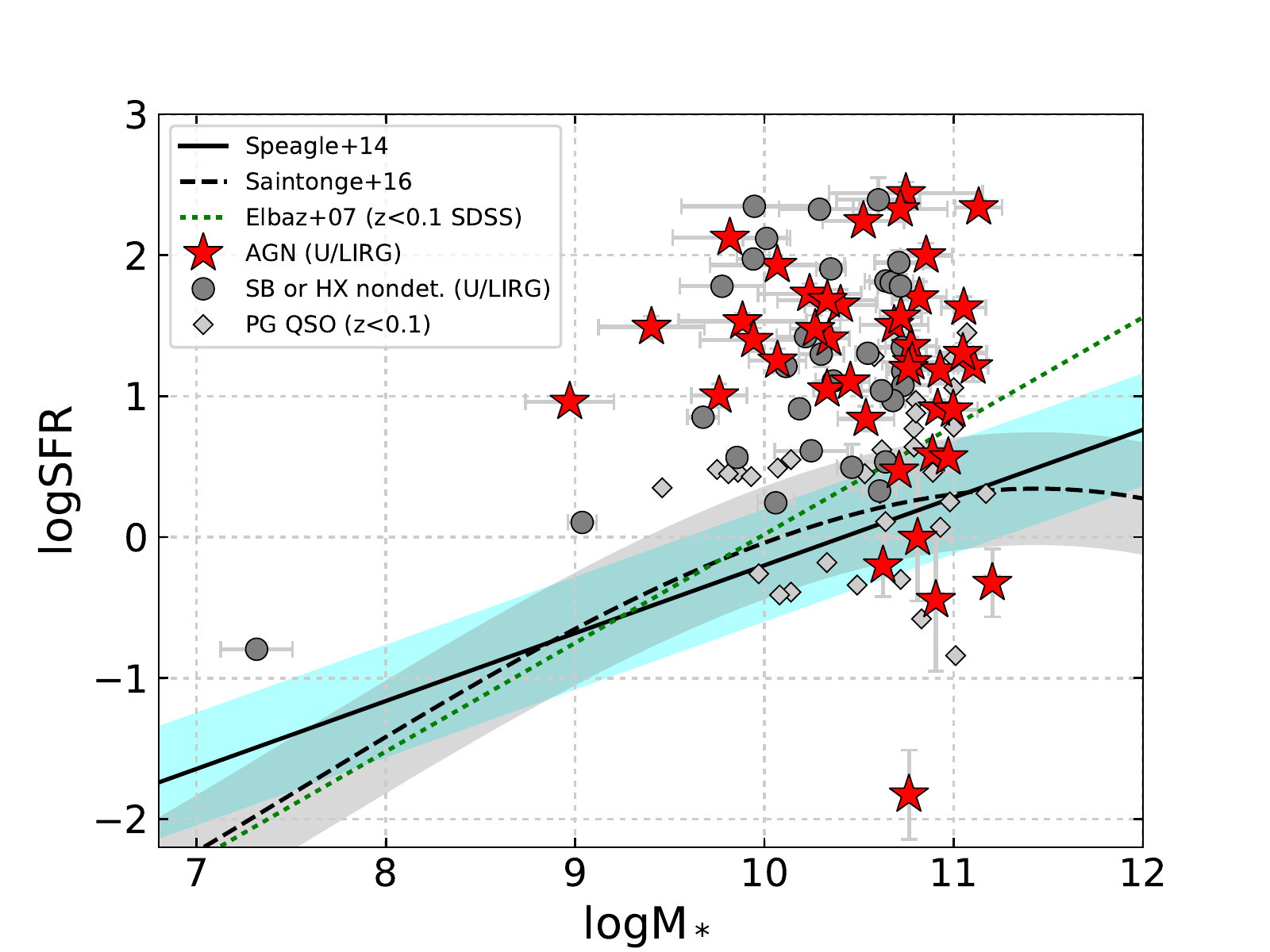}
    \caption{Left panel: logarithmic stellar mass in $M_{\odot}$ vs. logarithmic SFR in $M_{\odot}$ yr$^{-1}$. 
    The blue circles, green triangles, orange diamonds, red stars, and purple triangles represent the sources in stages A, B (early mergers), C, D (late mergers), and N (nonmergers) U/LIRGs, respectively.
    The gray diamonds illustrate the PG QSOs at $z < 0.1$, whose SFRs are estimated by the UV-to-IR SED decomposition \citep{Lyu2017}.
    The black dotted curve and cyan area denote the MS and 1$\sigma$ uncertain reported by \citet{Speagle2014}, while the dashed solid curve and shading represent those reported by \citet{Saintonge2016}.
    The green dotted line is the MS among $z < 0.1$ SDSS galaxies \citet{Elbaz2007}.
    Right panel: the same but the sources are classified by AGNs in U/LIRGs  (red star),  starburst-dominant or hard X-ray--nondetected sources in U/LIRGs (gray circle), and PG QSOs (gray diamond).
    \\}
\label{F8-MS}
\end{figure*}

\subsection{Characteristics of Host Galaxies} \label{sub5-1-MS}
We investigate the properties of host galaxies in U/LIRGs by using the measurements of stellar masses ($M_*$) and SFRs.
A strong correlation between these quantities for the majority of star-forming galaxies has been well reported \citep[e.g.,][]{Brinchmann2004,Noeske2007,Elbaz2007,Speagle2014,Saintonge2016,Tomczak2016,Pearson2018}, called the main sequence (MS).
Generally, galaxies above the MS are qualified as ``starburst'', while galaxies below the relation are passive galaxies in terms of their star-formation activity.
Thus, the comparison of these physical quantities enables us to understand the galaxy's evolution in the populations.

Figure~\ref{F7-graph} shows the averaged values of $M_*$ and SFR compared with merger stages.
The $M_*$ is almost similar for all merger stages, while the SFR increases with merger stage.
The left panel of Figure~\ref{F8-MS} presents the SFR--$M_*$ relation obtained by the multiwavelength SED analysis for our U/LIRGs color-coded by the merger stages.
We overplot the values of Palmer--Green (PG) quasars at $z < 0.1$ as gray diamonds.
For these quasars, the SFRs are estimated by the UV-to-IR SED decomposition \citep{Lyu2017} and stellar masses are referred from \citet{Xie2021} based on the high-resolution optical and near-IR images \citep{Zhang2016} or the $M_*$--$M_{\rm BH}$ relation of \citet{Greene2020}.\footnote{The values of $M_{\rm BH}$ are referred from \citet{Veilleux2009c} by averaging the results from different measurements, including spheroid luminosity, spheroid velocity dispersion, reverberation mapping, and virial relation \citep[see][]{Zhang2016}.}
For comparison, we also plot some MS relations for SDSS star-forming galaxies at $0.015 \leq z \leq 0.1$ \citep{Elbaz2007}, a large homogeneous collection of star-forming galaxies out to $z \sim 6$ (the relation at $z \approx 0$ is adopted; \citealt{Speagle2014}), and SDSS DR7 galaxies with $0.01 < z < 0.05$ and $M_{*} > 10^8 M_{\odot}$ on the basis of SFRs derived from UV and mid-IR (12 or 22~$\mu$m) luminosities \citep{Saintonge2016}.

The previous IR SED analysis of local U/LIRGs using the photometry of unresolved total systems by \citet{Shangguan2019} decouples the contributions of AGN and starburst emission and suggests the whole U/LIRGs are above the MS.
On the other hand, using the separated photometry of individual sources as possible in the SED fitting, we find that some component sources in early mergers sit on or lie below the MS.
The early mergers and PG quasars have wide ranges of SFRs above and below the MS, while the late mergers have the highest SFRs above the MS.
Except for some early mergers, the distributions of the stellar masses are comparable in both U/LIRGs and PG quasars.
As a result, the average specific SFR, sSFR (= SFR/$M_*$), is much larger in late mergers (log(sSFR/yr$^{-1}$) = $-$8.51$\pm$0.58) than those in early mergers ($-$9.58$\pm$0.98), nonmergers ($-$9.50$\pm$0.51), and PG quasars ($-$10.16$\pm$0.64).
Interestingly, this is well consistent with the recent high-resolution simulations of gas-rich major mergers \citep{Blecha2018}, which predicts the starbursts with the highest sSFR ($-$8.4 in a logarithmic scale) in the final phase of mergers.
Although the progenitor of quasars are not necessarily triggered by mergers \citep[e.g.,][]{Xie2021}, these results support that the starburst activities in U/LIRGs are triggered in early mergers, and suppressed after late mergers with high sSFR \citep[e.g.,][]{Di-Matteo2005,Hopkins2008,Ellison2013,Barrera-Ballesteros2015,Shangguan2019,U2019,Yamada2021}.

The right panel of Figure~\ref{F8-MS} denotes the SFR--$M_*$ relation for U/LIRGs with or without AGNs.
Red stars mark the hard X-ray--detected AGNs and three AGNs that are re-classified by the significance test with BIC values (Section~\ref{subsub4-2-4-BIC}), while dark gray circles mark the starburst-dominant or hard X-ray--undetected sources.
The U/LIRGs but several early mergers below the MS show similar distributions regardless of the presence or absence of AGNs.
About 50\% of the AGNs in late mergers are CT AGNs and, among the hard X-ray--undetected sources, there are few CT AGN candidates based on their 3$\sigma$ upper limits in the 8--24~keV band \citep{Yamada2021}.
These results support that the obscuration of the AGNs is not related to the difference in the SFR--$M_*$ distribution with and without AGNs.
A similar result in the IR band for local U/LIRGs of GOALS sample is reported by \citet{Shangguan2019}, who evaluate the presence of AGNs when the IR SED fit is significantly improved by adding the torus component.
Their AGN classification is based on the signature of strong mid-IR emission from the AGNs.
Thus, no matter which classifications of hard X-ray or mid-IR observations we select, the SFR--$M_*$ relation implies the small influence of the AGN activities on the starburst activities in the phase of U/LIRGs.

\subsection{Origin of Radio Emission from U/LIRGs} \label{sub5-2-radio}
Radio observation is a potential approach to reveal a mixture of starburst and AGN activities in U/LIRGs.
About 10\% of AGNs radiate strong synchrotron emission primarily from the powerful relativistic jets in the radio band, observable as radio-loud AGNs \citep[e.g.,][]{Begelman1984}.
The other AGNs are classified as radio-quiet AGNs, whose radio emissions are derived from a wide range of possible mechanisms: star formation, AGN-driven wind, free-free emission from photoionized gas, low power jets, and the innermost accretion disc coronal activities \citep[e.g.,][]{Panessa2019,Kawamuro2022}.
Thanks to the long wavelengths, radio observations can overcome the effects of dust obscuration in U/LIRGs and investigate their central engines, particularly starbursts and/or AGNs \citep[e.g.,][]{Condon1991,Lonsdale2003,Murphy2013b,Vardoulaki2015}.

\subsubsection{Radio Slope}
For star-forming galaxies, the radio emission typically appears as a power law spectrum ($S_{\nu} \propto \nu^{-\alpha}$) from thermal and non-thermal emission associated with the formation of massive stars.
The thermal emission is produced by the young massive O/B stars dominating the ionization in HII regions, and the non-thermal emission is produced by supernova remnants of the more massive stars than $\gtrsim$8~$M_{\odot}$, accelerating cosmic-ray electrons \citep[e.g.,][]{Helou1985,Condon1992}.
Under such conditions, the thermal bremsstrahlung free-free emission from the HII regions has a flat slope ($\alpha \sim 0.1$), while the non-thermal free-free emission shows a steep slope ($\alpha \sim 0.8$; e.g., \citealt{Niklas1997}).
The contribution of non-thermal emission is much larger at frequencies below 10~GHz than the thermal emission, as confirmed by the best-fit models of synchrotron (pink) and nebular free-free (gray) emission in the radio band (see Figure~\ref{F1-SED00-F} and Figures~\ref{SED01-F}--\ref{SED18-F}).
However, since the AGNs also radiate the synchrotron emission with a similar slope to the starburst emission ($\alpha \sim 0.8$; e.g., \citealt{Krolik1999,Niklas1997}; but see the study of radio-slope map by \citealt{Vardoulaki2015} and \citealt{Linden2019}), this makes difficult to separate the starburst and AGN radio emission.
For U/LIRGs, the nuclear radio emission should have flat spectra due to optically-thick free-free absorption \citep{Condon1991,Clemens2008,Leroy2011,Murphy2013a,Murphy2013b}, and thus the radio slope could not be a tracer of AGNs but an indicator of the surrounding environment of the nuclear starbursts and AGNs.

The top panels of Figure~\ref{F9-SFR-radio} display the radio slope within the 0.07--3.0~GHz band, $\alpha_{\rm radio}$, (Section~\ref{sub4-3-radio-fit}) as a function of the SFR and the fraction of the AGN luminosity in the IR band.
The previous radio study of U/LIRGs performed by \citet{Murphy2013b} reports that the median value of the radio slope at $\lesssim$5~GHz for U/LIRGs is 0.5, and the slopes decrease with merger stage.
The values in our work are well consistent with their median value and decreasing trend.
The flat slope in late mergers with high SFRs ($\alpha_{\rm radio} \sim 0.0$--0.5) can be explained by the optically-thick free-free absorption due to the rich environment of the nuclear region.\footnote{The AGNs with high-Eddington ratios also show the flat radio slope \citep{Laor2019,YangX2020}, whose contribution of the radio emission may be smaller than those of starbursts as discussed on the correlation coefficient ($q_{\rm IR}$).}
In the right panel, the hard X-ray--detected AGNs and three newly-identified AGN candidates are characterized as the positive values of the fraction of AGN luminosity.
As mentioned above, the AGN fraction (i.e., the dominance of AGN activity) is not correlated with the radio slope for U/LIRGs.

\subsubsection{Radio--IR Correlation Coefficient}
\begin{figure*}
    \centering
    \includegraphics[keepaspectratio, scale=0.45]{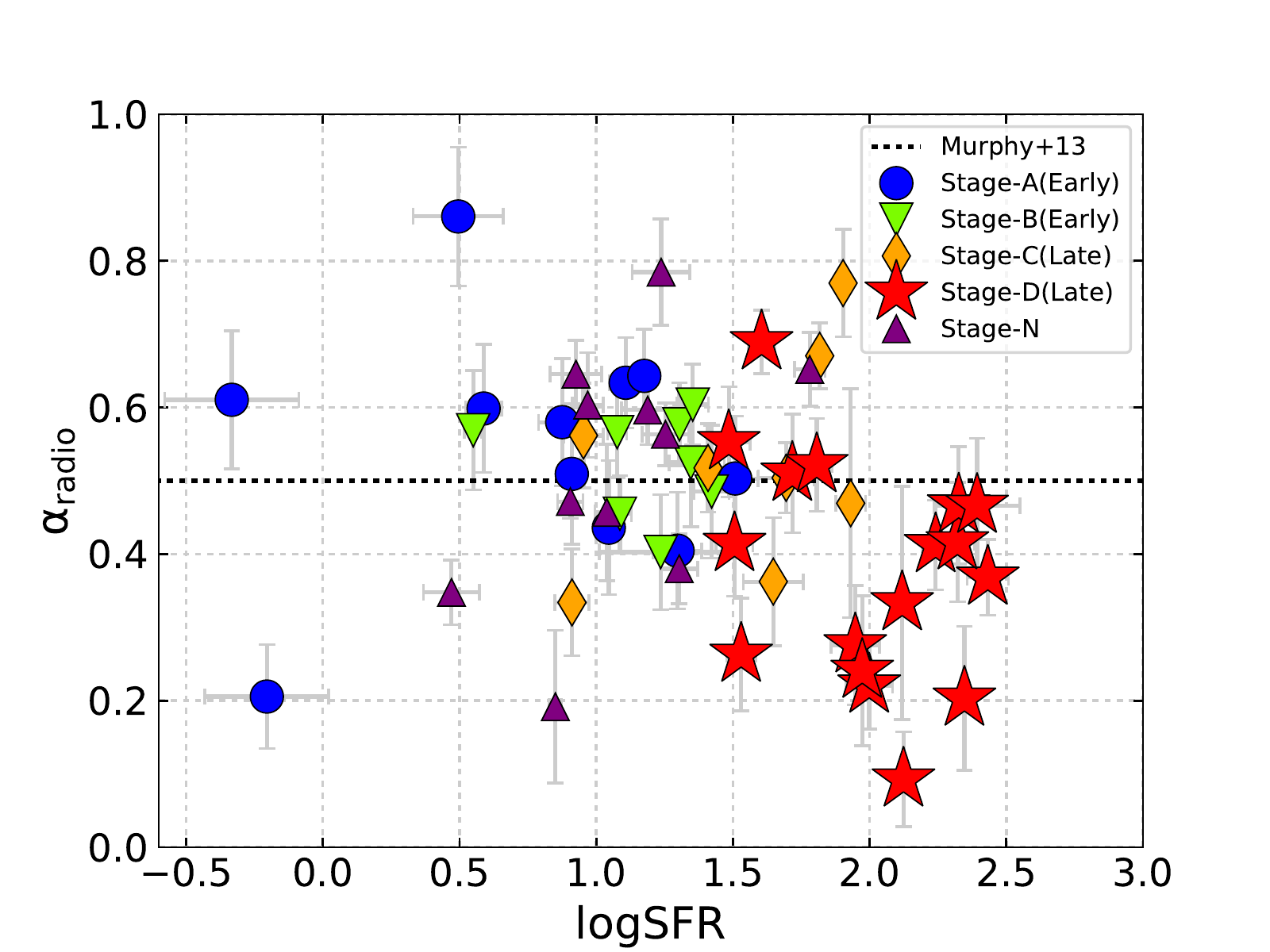}
    \includegraphics[keepaspectratio, scale=0.45]{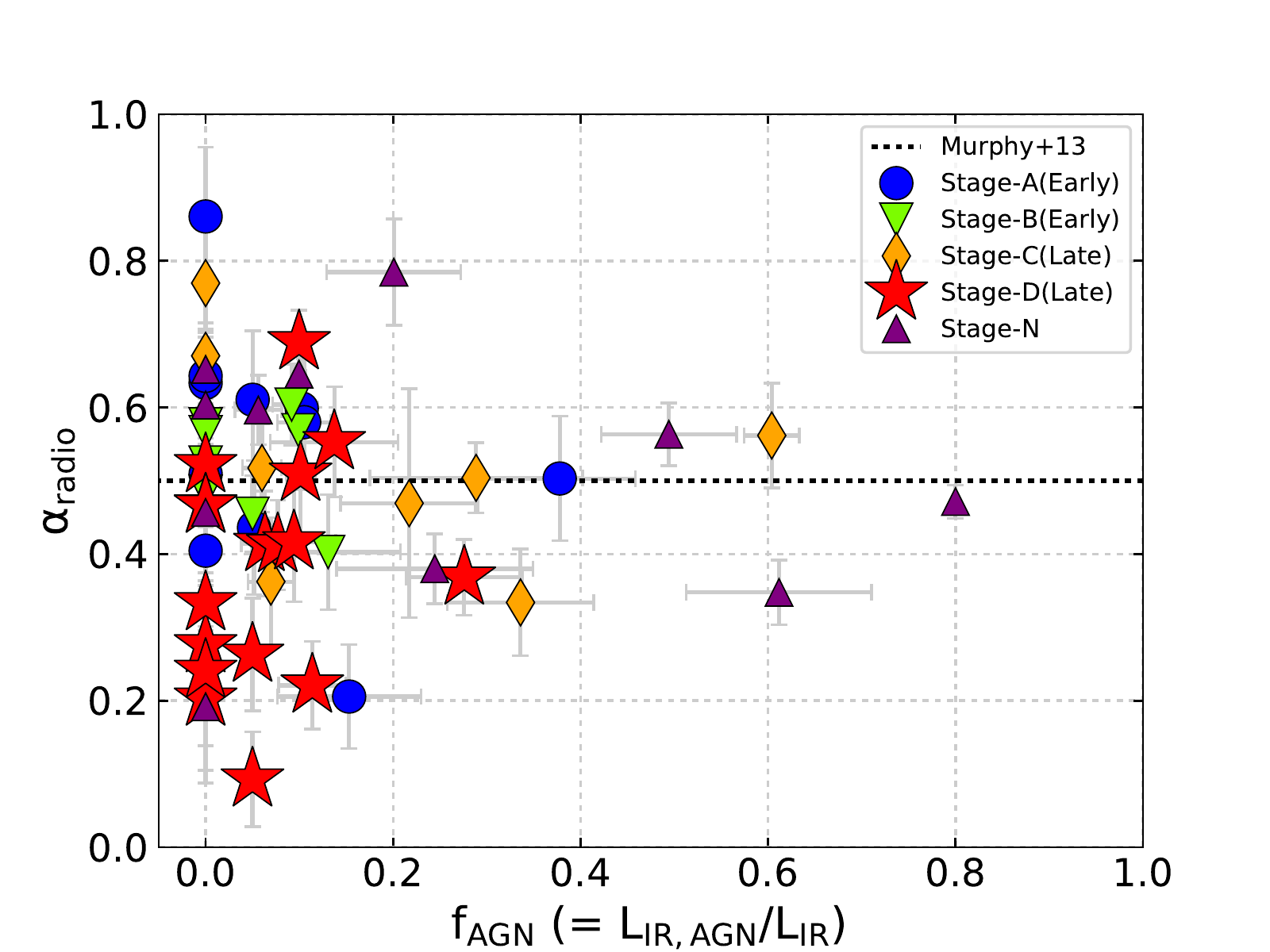}
    \includegraphics[keepaspectratio, scale=0.45]{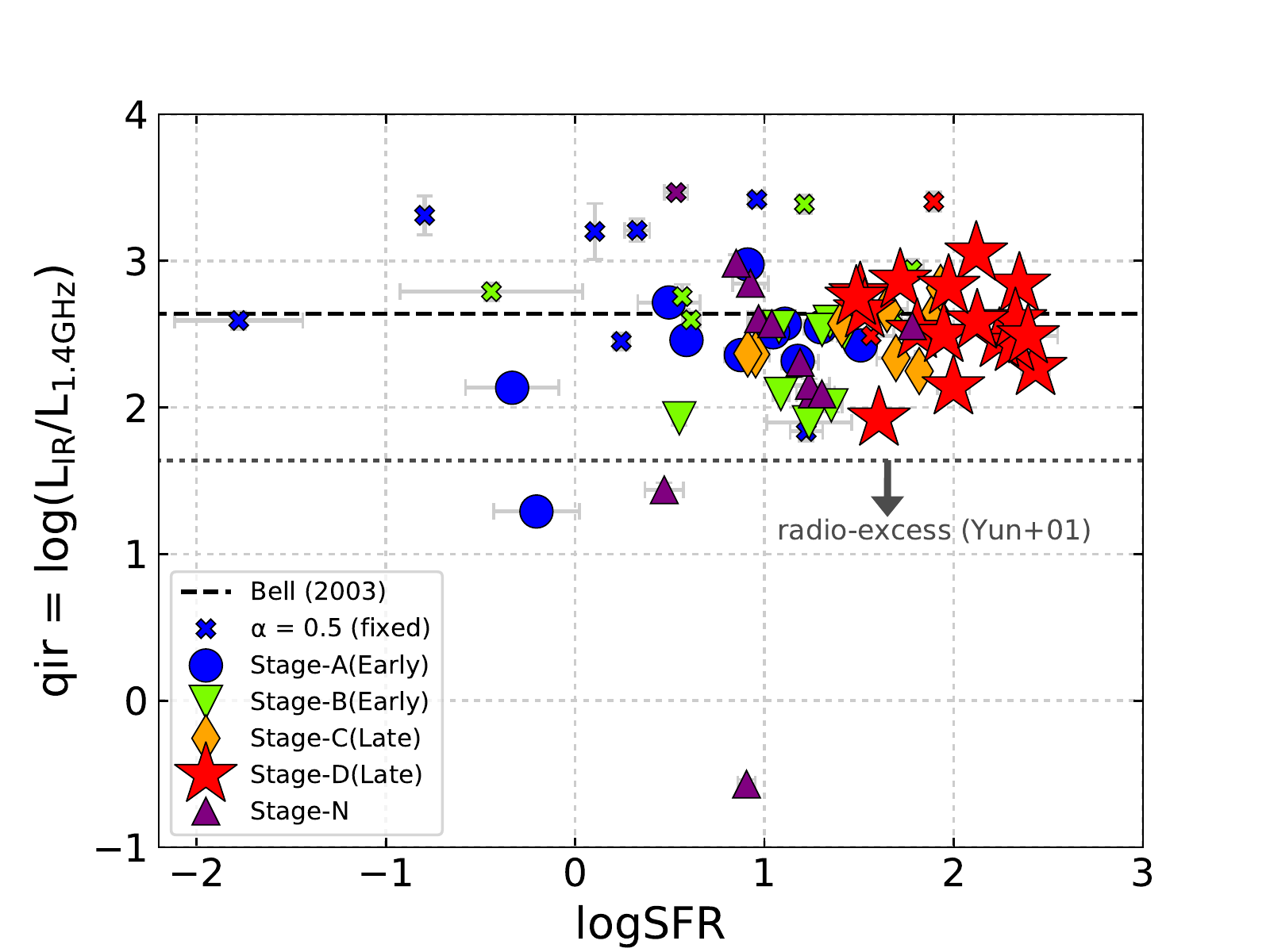}
    \includegraphics[keepaspectratio, scale=0.45]{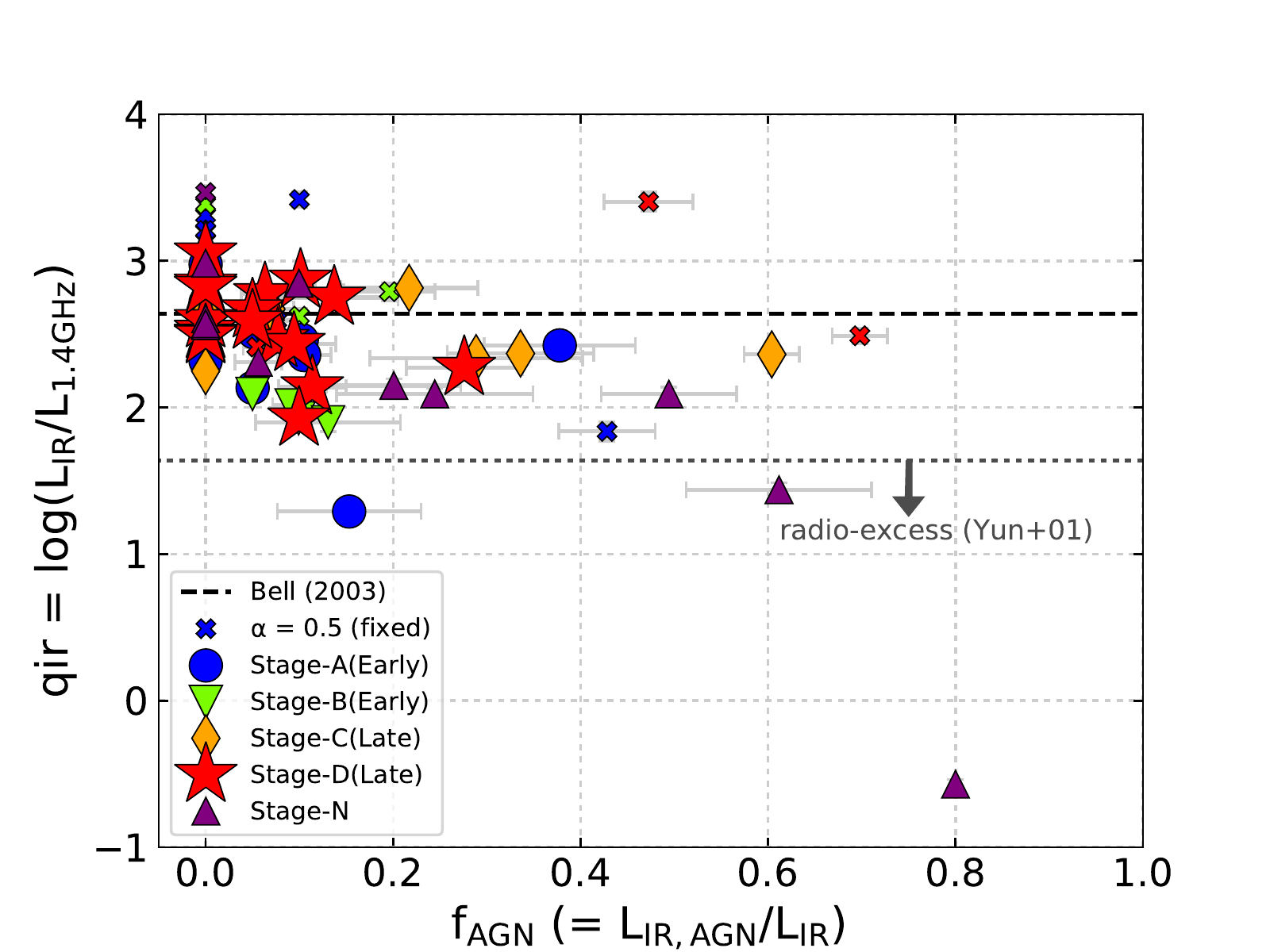}
    \caption{Top panels: logarithmic SFR vs. $\alpha_{\rm radio}$ (left) and fraction of AGN luminosity in the IR band vs. $\alpha_{\rm radio}$ (right), excluding the sources whose $\alpha_{\rm radio}$ is fixed at 0.5.
    The sources with $f_{\rm AGN} = 0.0$ mean the U/LIRGs which do not have AGNs.
    The dashed line is the median of low-frequency ($<$5~GHz) $\alpha_{\rm radio}$ among U/LIRGs reported by \citet{Murphy2013b}.
    Bottom panels: logarithmic SFR vs. $q_{\rm IR}$ (left) and fraction of AGN luminosity in the IR band vs. $q_{\rm IR}$ (right).
    The dashed line is the averaged values reported in \citet{Bell2003}, and the dotted gray line illustrates the threshold for radio-excess ($q_{\rm IR} < 1.64$) as a tracer of an AGN \citep{Yun2001,U2012}.
    Large symbols are the same in Figure~\ref{F8-MS}, while small crosses denote the sources where $\alpha_{\rm radio} = 0.5$ is assumed in the radio fitting.
    \\}
\label{F9-SFR-radio}
\end{figure*}

Alternatively, the correlation coefficient, $q_{\rm IR}$ (= log$L_{\rm IR}$/$L_{\rm 1.4 GHz}$; see Section~\ref{sub4-3-radio-fit}), can be a complementary parameter to unveil the presence of AGNs in U/LIRGs.
For star-forming galaxies, a tight correlation between radio and IR luminosities is reported \citep[e.g.,][]{Kennicutt1998,Bonzini2015}, whose luminosity ratio is expressed by $q_{\rm IR}$.
For the AGN-dominant sources, the value should be small due to the powerful synchrotron emission from the AGN \citep[e.g.,][]{Yun2001,U2012}. 
We investigate the distribution of $q_{\rm IR}$ as a function of the SFR and the fraction of the AGN luminosity (bottom panels of Figure~\ref{F9-SFR-radio}).
Except for three outliers below the criteria specified as radio-excess ($q_{\rm IR} < 1.64$, marked by gray dotted line; \citealt{Yun2001}), these U/LIRGs are distributed around the typical value of star-forming galaxies, $\sim$2.64 (black dashed line; \citealt{Bell2003}), without regard for merger stage and SFR.
Remarkably, the sources having larger fractions of AGN luminosity tend to show smaller $q_{\rm IR}$, indicating the strong radio emission due to the synchrotron radiation from the AGNs.
Not only the sources with $q_{\rm IR} < 1.64$ \citep{Yun2001}, but those with $q_{\rm IR} < 2$ have AGNs for the U/LIRGs in our sample.

\subsubsection{Radio Excess Parameter}
\begin{figure*}
    \centering
    \includegraphics[keepaspectratio, scale=0.45]{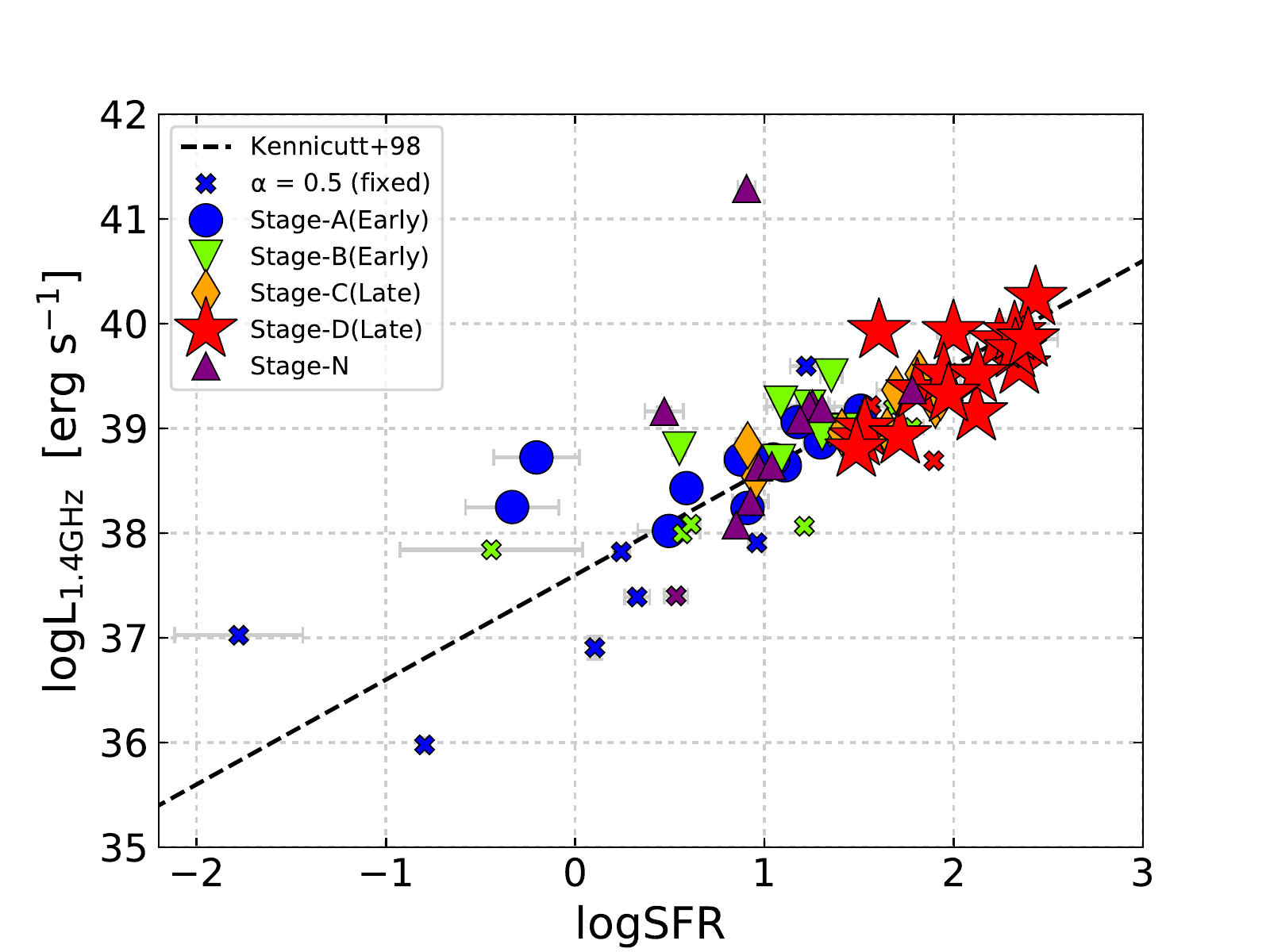}
    \includegraphics[keepaspectratio, scale=0.45]{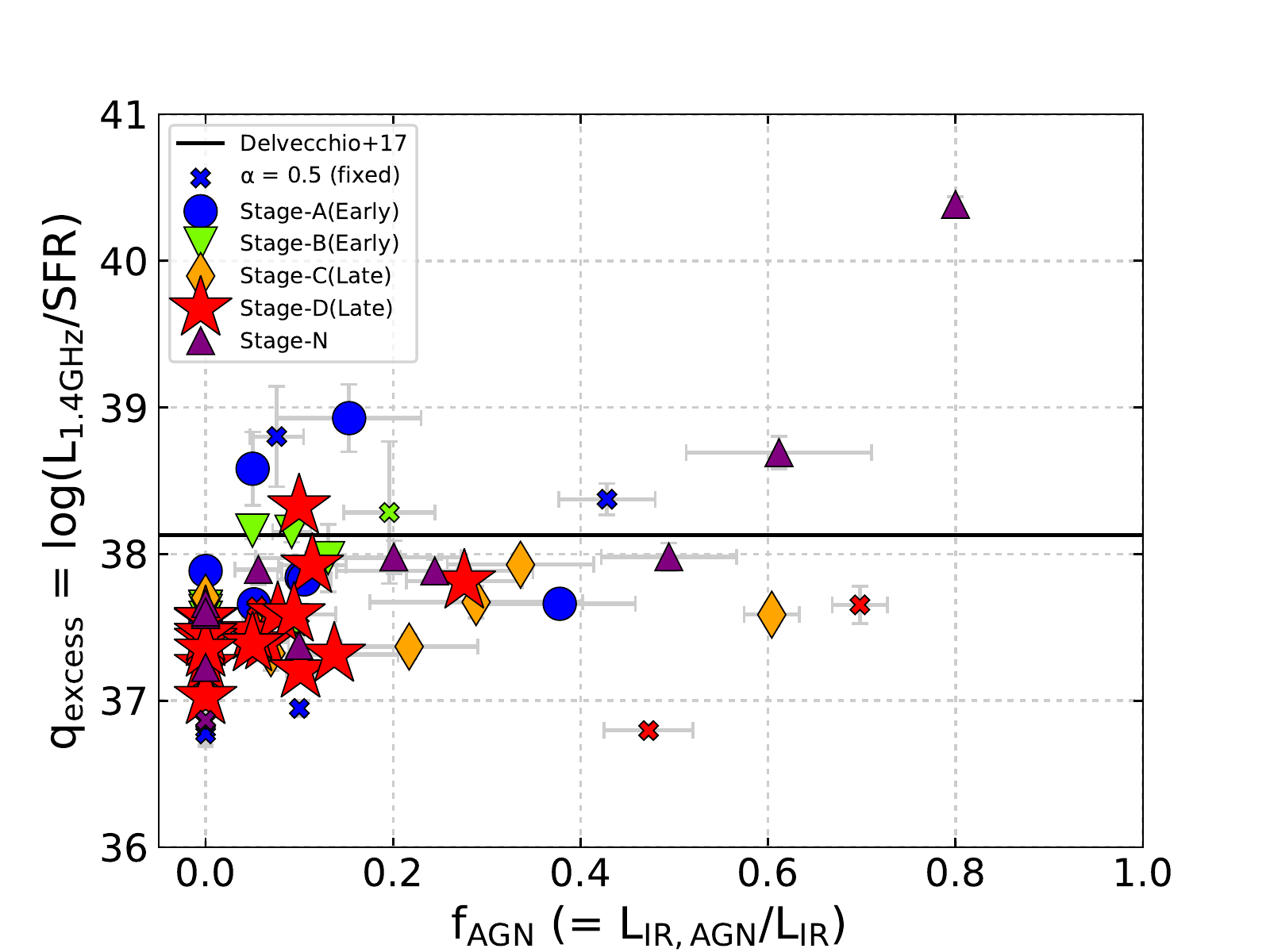}
    \caption{Left panel: logarithmic SFR vs. logarithmic radio luminosity at 1.4~GHz.
    The black dashed line is obtained by the empirical radio-far-IR relation in star-forming galaxies in \citet{Kennicutt1998}, assuming Chabrier IMF \citep[see][]{Bonzini2015}.
    Right panel: fraction of AGN luminosity in the IR band ($f_{\rm AGN}$) vs. radio excess parameter ($q_{\rm excess}$).
    The sources with $f_{\rm AGN} = 0.0$ mean the U/LIRGs which do not have AGNs.
    The black solid curve shows the threshold of radio-excess sources at $z \approx 0.0$ \citep{Delvecchio2017}.
    The symbols are the same in Figure~\ref{F9-SFR-radio}.
    \\}
\label{F10-qexcess}
\end{figure*}

In the left panel of Figure~\ref{F10-qexcess}, we also compare the SFR and radio 1.4~GHz luminosities ($L_{\rm 1.4 GHz}$) for U/LIRGs.
The black dashed line is the empirical correlation from the radio--far-IR luminosity relation and the conversion factor between SFR and far-IR luminosity for star-forming galaxies in \citet{Kennicutt1998}, assuming Chabrier IMF \citep[see also][]{Bonzini2015}:
\begin{align}
& {\rm log}(L_{\rm 1.4 GHz}/{\rm erg~s^{-1}}) = {\rm log(SFR)} + 37.60.
\end{align}
Our targets roughly follow the empirical relation.
\citet{Delvecchio2017} introduce a new diagnostics for AGNs by using the radio excess parameter, $q_{\rm excess}$ = log($L_{\rm 1.4GHz}$/SFR).
They suggest that the sources having $q_{\rm excess} > 38.130 \times (1 + z)^{0.013}$ in (erg s$^{-1}$)/($M_{\odot}$ yr$^{-1}$) are likely attributable to AGN activities.
As shown in the right panel, the radio-excess sources in our samples have $f_{\rm AGN} > 0$ (i.e., hosting AGNs) in the IR band.

% Figure 9

% Figure 10

As a whole, the radio-excess sources in U/LIRGs host AGNs detected in the IR and X-ray bands.
Whereas, the other major population of radio-quiet sources in U/LIRGs are indistinguishable from AGNs or not.
The same results are reported for AKARI-selected star-forming galaxies with and without AGNs \citep{Moric2010}, indicating that the AGNs in U/LIRGs are typical radio-quiet AGNs in star-forming galaxies (i.e., the large contribution of the radio emission from the intense starbursts).
Therefore, although sources with high $q_{\rm excess}$ values likely contain AGNs, the radio emission reflects the intensity of the starburst activities (e.g., SFR) for the majority of U/LIRGs.

\subsection{WISE Color: New Wedge of Buried AGNs} \label{sub5-3-WISE-color}
Near-IR to mid-IR observations have been thought of as a promising tool to detect the AGN emission in U/LIRGs \citep[e.g.,][]{Veilleux2009a,Petric2011,Imanishi2014,Imanishi2020,Diaz-Santos2017}.
A large part of the emission from central AGNs in U/LIRGs is absorbed by the surrounding materials such as the torus with a large covering fraction \citep[e.g.,][]{Imanishi2006,Imanishi2008,Ricci2017bMNRAS,Yamada2019} and re-radiated chiefly in the mid-IR wavelength.
The $\sim$3--30~$\mu$m SEDs of the reprocessed dust emission heated by the AGN is characterized as a steep power-law slope \citep[e.g.,][]{Laurent2000,Stern2005,Assef2010}.
Indeed, the IR color selection with WISE has been exploited to search for the optically obscured AGNs \citep[e.g.,][]{Jarrett2011,Stern2012,Mateos2012,Mateos2013,Assef2013,Assef2018,Secrest2015,Toba2015,Ellison2017,Satyapal2017,Weston2017,Goulding2018a,Pfeifle2022}.

% Figure 11
\begin{figure*}
    \centering
    \includegraphics[keepaspectratio, scale=0.45]{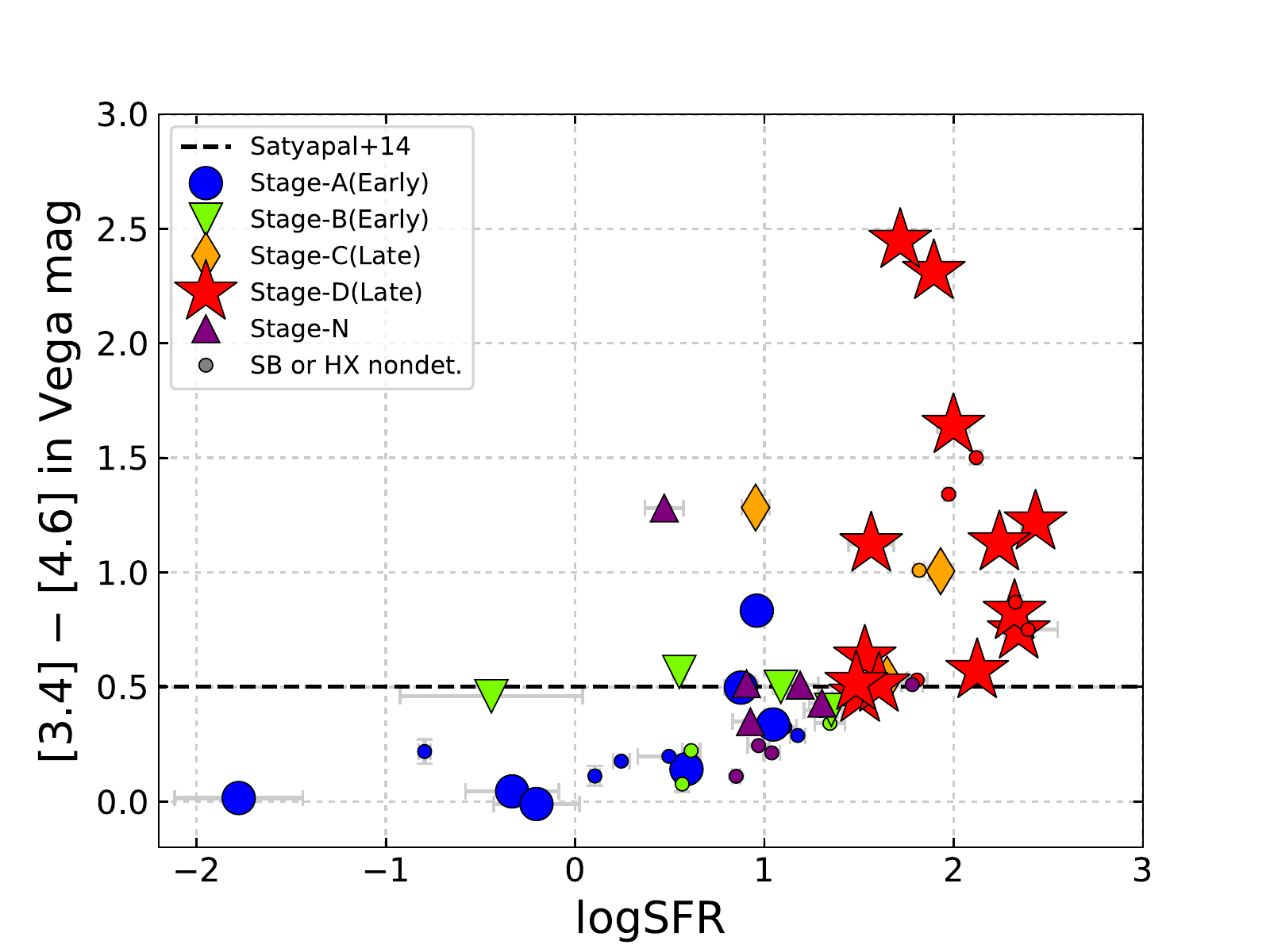}
    \includegraphics[keepaspectratio, scale=0.45]{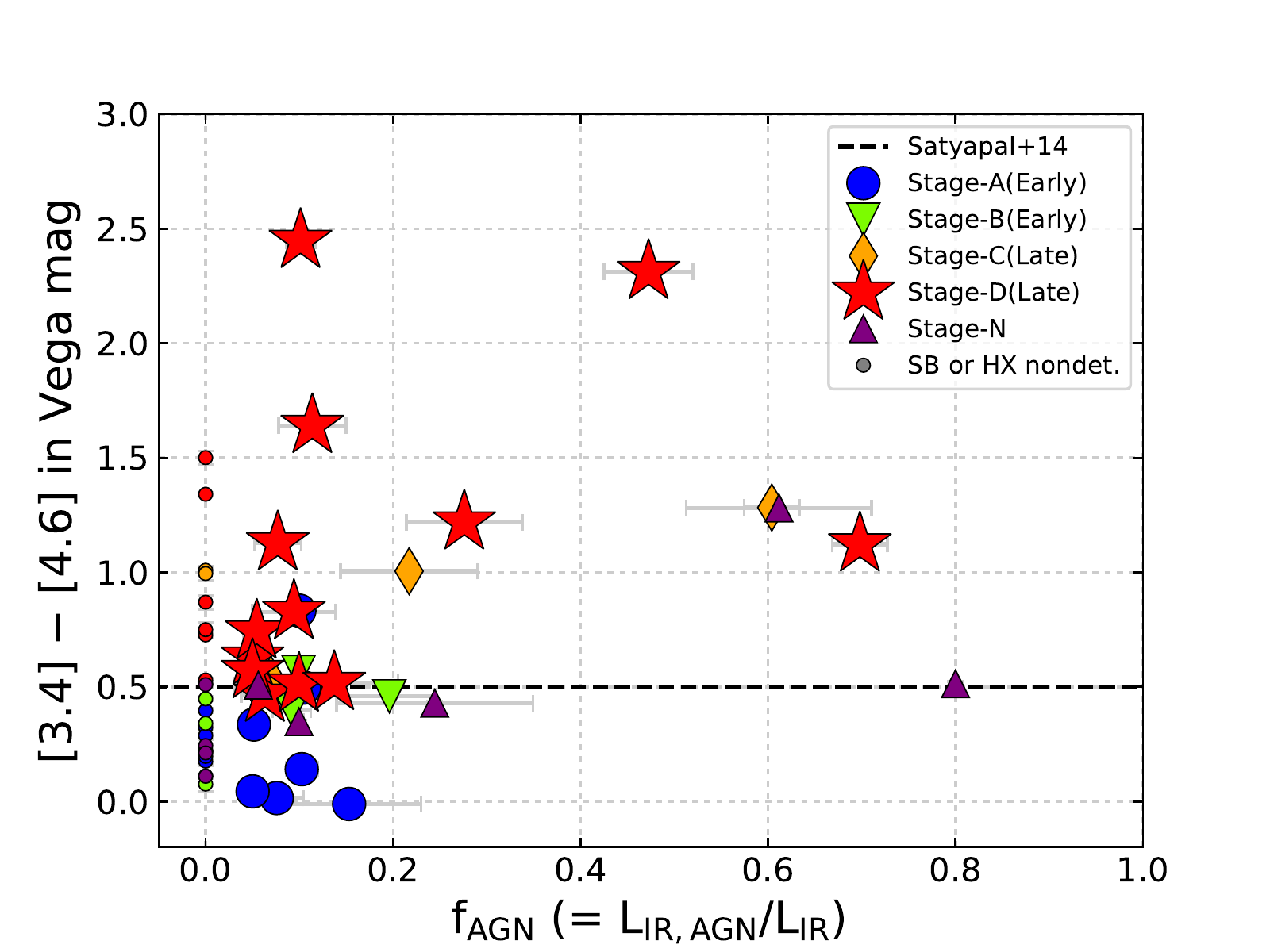}
    \caption{WISE W1--W2 color as a function of SFR (left) and the fraction of AGN luminosity in the IR band (right).
    Dashed line illustrates the ${\rm W1} - {\rm W2} > 0.5$ threshold \citep{Satyapal2014}.
    \\}
\label{F11-WISE12}
\end{figure*}

% Figure 12
\begin{figure*}
    \centering
    \includegraphics[keepaspectratio, scale=0.7]{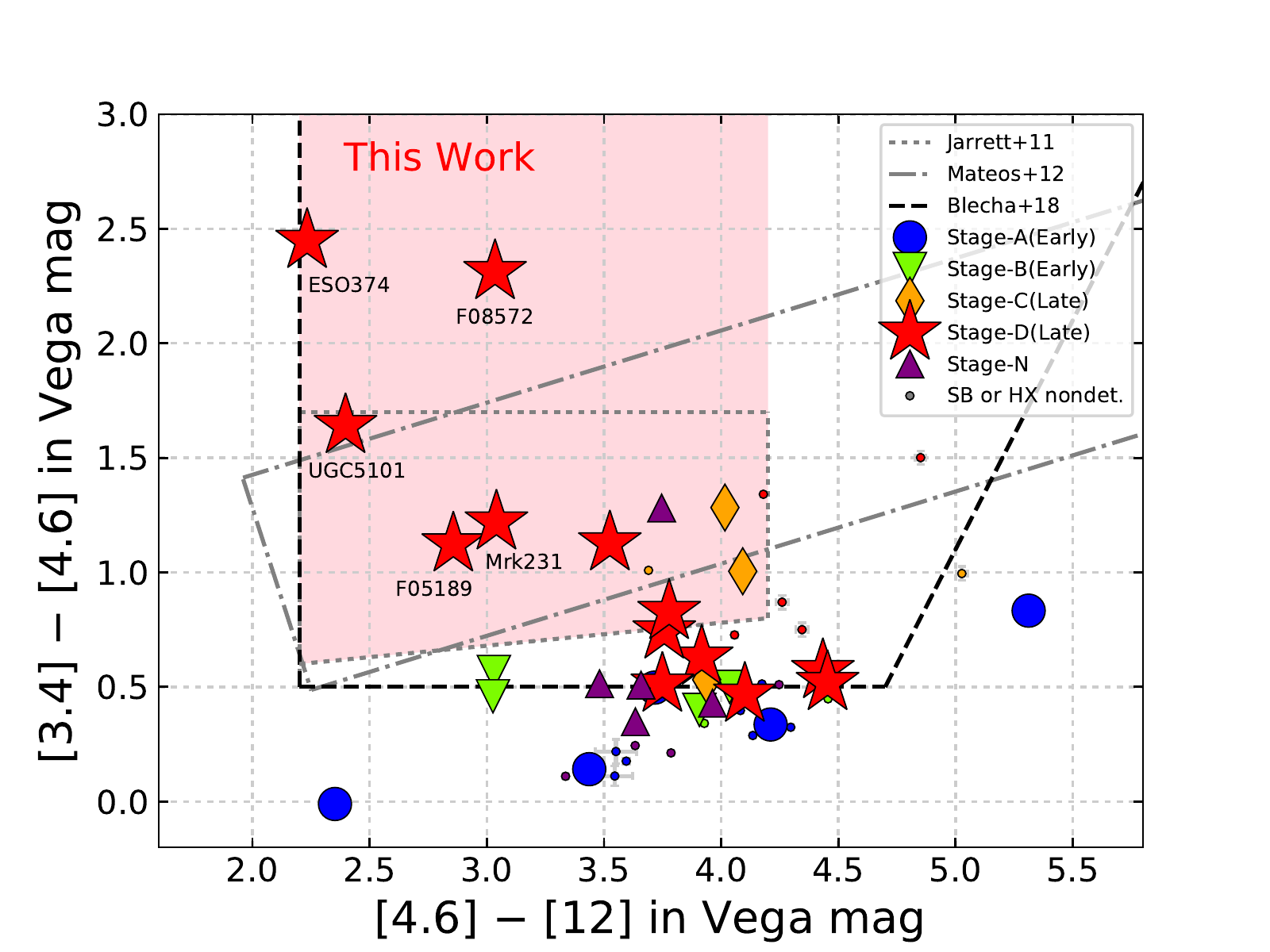}
    \caption{WISE color--color diagram.
    The AGN selection wedges are illustrated as the areas within the dotted gray \citep{Jarrett2011}, dash-dotted gray \citep{Mateos2012}, and black dashed \citep{Blecha2018} boxes, and the pink shaded area is presented in This work, respectively.
    Large symbols are the same in Figure~\ref{F8-MS}, while small circles illustrate the sources where $f_{\rm AGN} = 0.0$ (starburst-dominant or hard X-ray--nondetected sources).
    \\}
\label{F12-WISE-color}
\end{figure*}

\subsubsection{W1--W2 Color}
Figure~\ref{F11-WISE12} presents the WISE W1--W2 (or [3.4]--[4.6] in Vega magnitude) color as a function of SFR (left) and AGN fraction in the IR band (right).\footnote{The WISE photometry of Mrk 231, a stage D merger hosting the AGN with high $\lambda_{\rm Edd}$, is not utilized in the SED analysis due to the flag of w1-3sat $\approx 0.06$ (W1, W2, W3 = 7.80$\pm$0.01, 6.58$\pm$0.01, 3.54$\pm$0.01, respectively), but is only provided in Figures~\ref{F11-WISE12}--\ref{F12-WISE-color} since it will be a small effect on the WISE colors.}
Here, the WISE magnitudes are corrected for Galactic extinction (Section~\ref{sub3-8-GalExt}).
\citet{Stern2012} introduce a simple mid-IR color criterion of ${\rm [3.4]-[4.6]} > 0.8$ by using the WISE-selected sources in the Cosmic Evolution Survey (COSMOS; \citealt{Scoville2007}).
Similarly, \citet{Satyapal2014} report that a more inclusive ${\rm [3.4]-[4.6]} > 0.5$ cut identifies AGNs in low-$z$ ($z < 0.2$) mergers selected from the SDSS survey \citep{Ellison2013}.
For our U/LIRGs, we find that all late mergers have ${\rm logSFR} \gtrsim 1.5$ and ${\rm [3.4]-[4.6]} \gtrsim 0.5$, while most early mergers and nonmergers have smaller SFRs and ${\rm [3.4]-[4.6]} \lesssim 0.5$, making it easy to select the IR-luminous late mergers.
However, the comparison between WISE color and AGN fraction suggests that the hard X-ray--detected and newly-identified AGN candidates (i.e., $f_{\rm AGN} > 0$) are not distinguished from the other starburst-dominant or hard X-ray--undetected sources by the [3.4]--[4.6] color.
These results indicate that for U/LIRGs, the [3.4]--[4.6] criteria could not necessarily discriminate between the intense starbursts and the AGNs discovered with the hard X-ray observations \citep[e.g.,][]{Ricci2016,Ricci2017bMNRAS,Yamada2021}. 

\subsubsection{WISE Color--color Diagram}
Figure~\ref{F12-WISE-color} shows the WISE color--color diagram ([3.4]--[4.6] versus [4.6]--[12]) for our sample.
The AGN wedges that have been well utilized for the low-$z$ galaxies are overplotted \citep{Jarrett2011,Mateos2012,Blecha2018}.
Almost all late mergers are located within the AGN wedge that is based on the hydrodynamics and radiative transfer simulation of gas-rich mergers (dashed line; see Equation~(1) in \citealt{Blecha2018}).
In late mergers, it is notable that the values of [4.6]--[12] magnitudes decrease with the values of [3.4]--[4.6] magnitudes for the AGNs, while they increase for the other sources.
Particularly, the AGNs with [4.6]--[12] $\lesssim 3$ in stage D mergers (IRAS F05189$-$2524, IRAS F08572+3915, UGC 5101, ESO~374$-$IG032, and Mrk~231) show the excess in the $\sim$3--10~$\mu$m wavelength relative to the best-fit SED models with updated CLUMPY model (Figures~\ref{SED01-F}--\ref{SED18-F}), corresponding to the dust emission at the temperature of $\sim$300--1000~K \citep[e.g.,][]{Lyu2021}.
The contribution of the near-IR emission from the old stellar population \citep[e.g.,][]{Polletta2007} is extracted by the multiwavelength SED analysis, the origin of the $\sim$3--10~$\mu$m excess can be explainable by the additional nuclear hot dust heated by the AGNs, such as inner part of the dusty disk and/or hot polar dust within the inner parsecs \citep[e.g.,][]{Honig2013,Garcia-Bernete2017,Garcia-Bernete2022,Lyu2017,Lyu2021,Mattila2018,Mizukoshi2022}.
The near-IR excess in a sample of type 1 AGN is well reported \citep{Netzer2007,Mor2009,Mor2012} and this hot dust component can also be part of the known outflows in Mrk~231 \citep[e.g.,][]{Feruglio2015,Morganti2016,Veilleux2016}.
The strong near-IR emission from the dusty disk, hot polar dust, or both is a unique feature of buried AGNs in late mergers.

% Figure 13
\begin{figure*}
    \centering
    \includegraphics[keepaspectratio, scale=0.55]{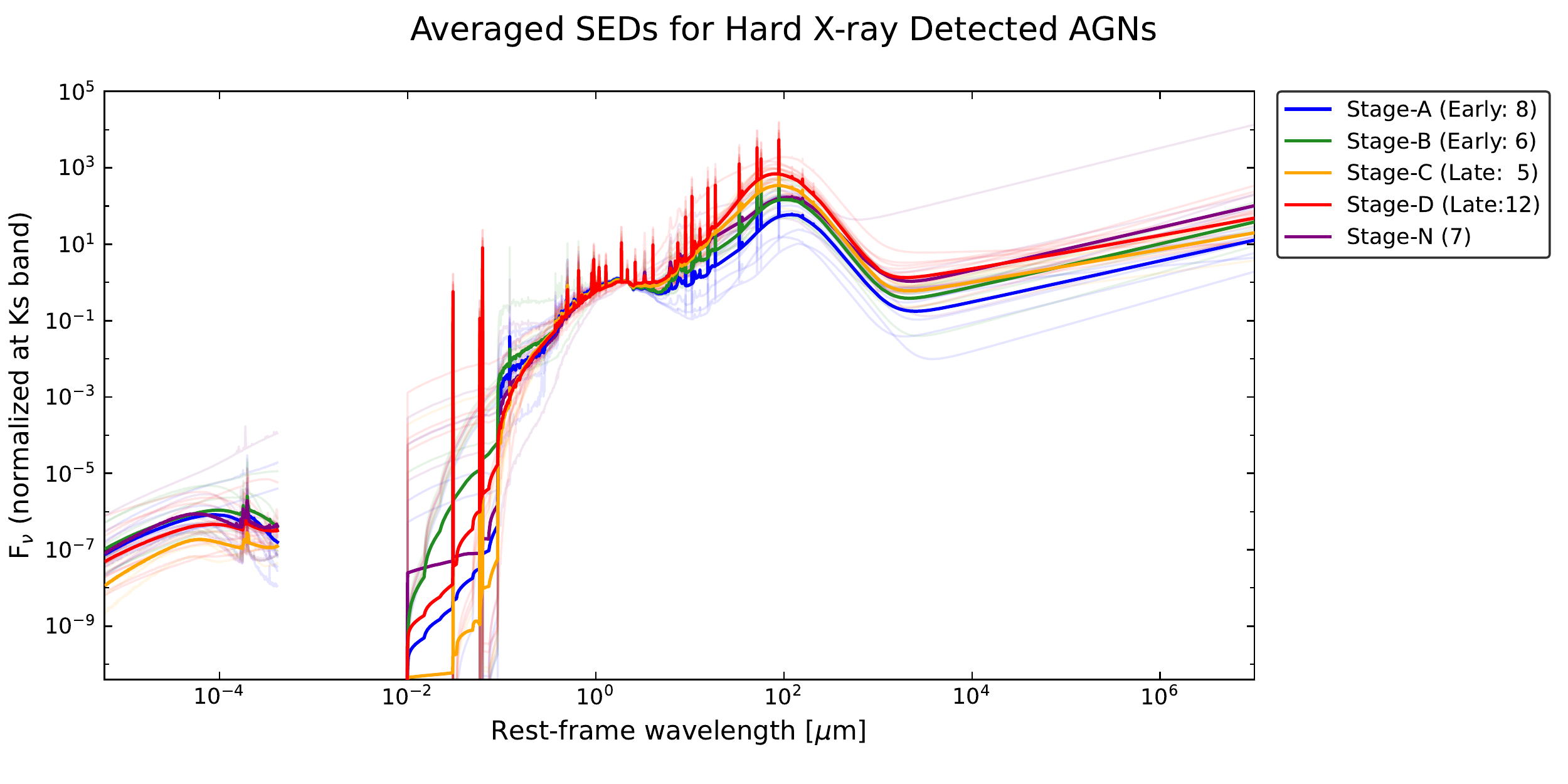}
    \includegraphics[keepaspectratio, scale=0.55]{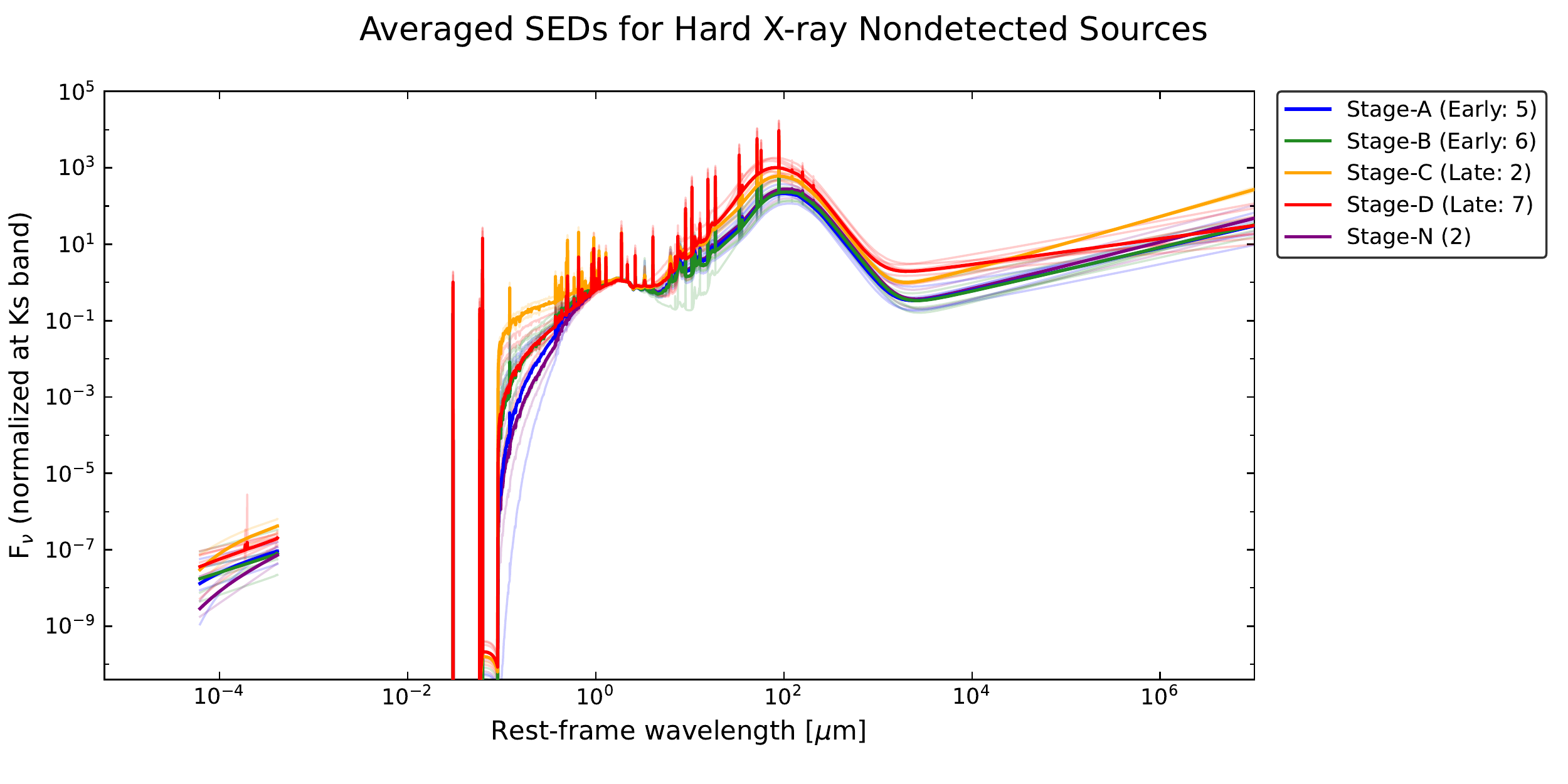}
    \caption{Top panel: rest-frame wavelength vs. flux density normalized at $K_{\rm s}$ band for the hard X-ray--detected AGNs.
    Bottom panel: the same but for the soft X-ray--detected galaxies among starburst-dominant or hard X-ray--nondetected sources.
    The solid curves represent the averaged SEDs for the sources 
    in stage A (blue), stage B (green), stage C (orange),
    stage D (red) and stage N (purple) U/LIRGs.
    Values in parentheses are the number of sources.
    The light-colored curves show the SEDs for individual sources.
    \\}
\label{F13-averaged-SED}
\end{figure*}

Although the complete selection of hard X-ray--detected AGNs by the WISE color seems to be difficult, some buried AGNs in local U/LIRGs are characterized by the near-IR excess appearing in the improved AGN color selection (pink shade);
\begin{align}
& 2.2 < {\rm [4.6]} - {\rm [12]} < 4.2, {\rm \ and} \notag\\
& {\rm [3.4]} - {\rm [4.6]} > 0.1 \times ({\rm [4.6]} - {\rm [12]}) + 0.38,
\end{align}
which are the combination of the previous AGN wedge for QSOs, Seyfert galaxies, and obscured AGNs (dotted line; see Equation~(1) in \citealt{Jarrett2011}) and the higher [3.4]--[4.6] region.
Among our local U/LIRGs within the new wedge, the selection purity of the AGNs in stage C--D mergers is 72$^{+11}_{-12}$\% (9/12).
Here the fraction and uncertainty are the 50th and 16th--84th quantiles of a binomial distribution, respectively, calculated with the beta function \citep{Cameron2011}.
By contrast, the completeness of selecting AGNs with the new wedge is low ($40\% \pm 10\%$; 9/23).
Considering that only the most luminous AGNs (e.g., IRAS F05189$-$2524, IRAS F08572+3915, and Mrk~231) tend to be selected, the low completeness suggests the difficulty of the identification of low-luminous AGNs in late mergers.
We also need to keep in mind that this criterion should be adopted for local U/LIRGs since it includes the typical criteria for QSOs and normal AGNs \citep{Jarrett2011}.
Thus, this new WISE color--color diagram for local U/LIRGs provides valuable information on the presence of the luminous AGNs hosting the dusty disk, hot polar dust, or both within the inner parsecs.

% Figure 14
\begin{figure*}
    \centering
    \includegraphics[keepaspectratio, scale=0.55]{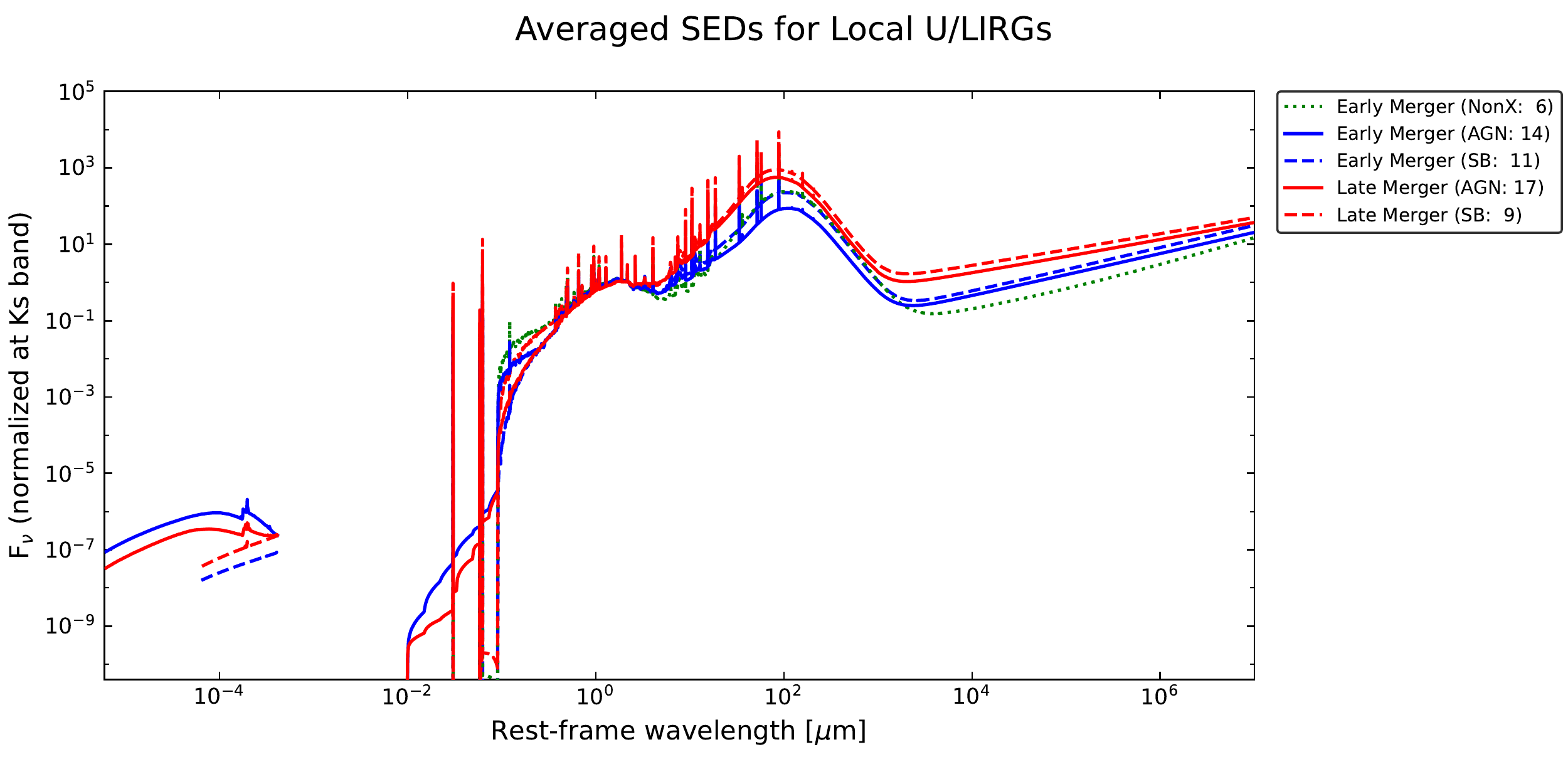}
    \includegraphics[keepaspectratio, scale=0.55]{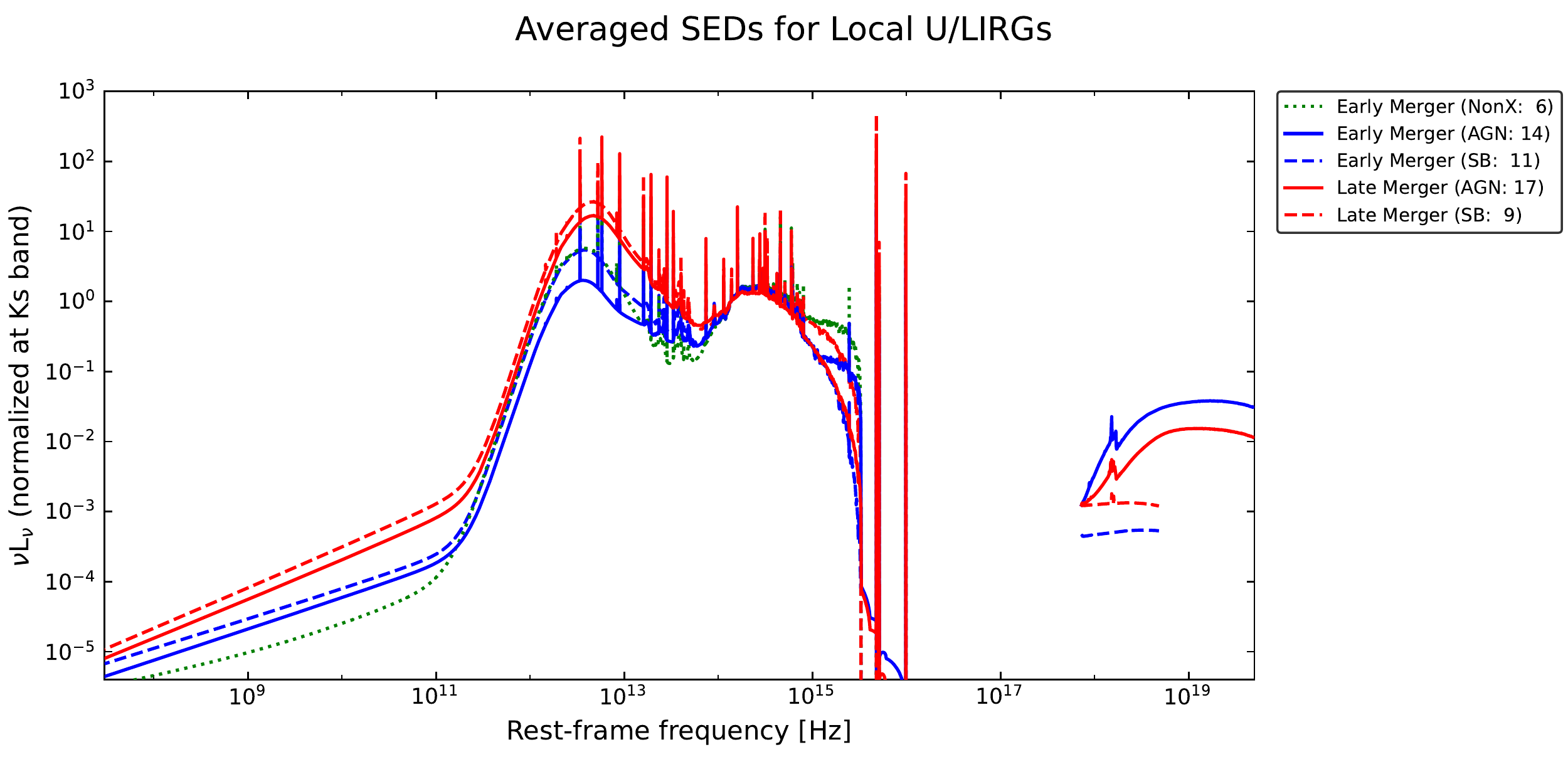}
    \caption{Top panel: rest-frame wavelength vs. flux density normalized at $K_{\rm s}$ band.
    Bottom panel: rest-frame frequency vs. luminosity normalized at $K_{\rm s}$ band.
    The dotted green curve shows the averaged SEDs for the sources whose X-ray spectra above $\gtrsim$1~keV are not obtained in stage A--B (early) mergers.
    The solid curves represent the averaged SEDs for the hard X-ray--detected AGNs 
    in stage A--B (early mergers; blue) and stage C--D (late mergers; red).
    The dashed curves are the same ones for the soft X-ray--detected galaxies among starburst-dominant or hard X-ray--nondetected sources.
    Values in parentheses are the number of sources.
    These SED templates are available in Table~\ref{T5_template}.
    \\}
\label{F14-total-SED}
\end{figure*}

% Figure 15
\begin{figure}
    \centering
    \includegraphics[keepaspectratio, scale=0.55]{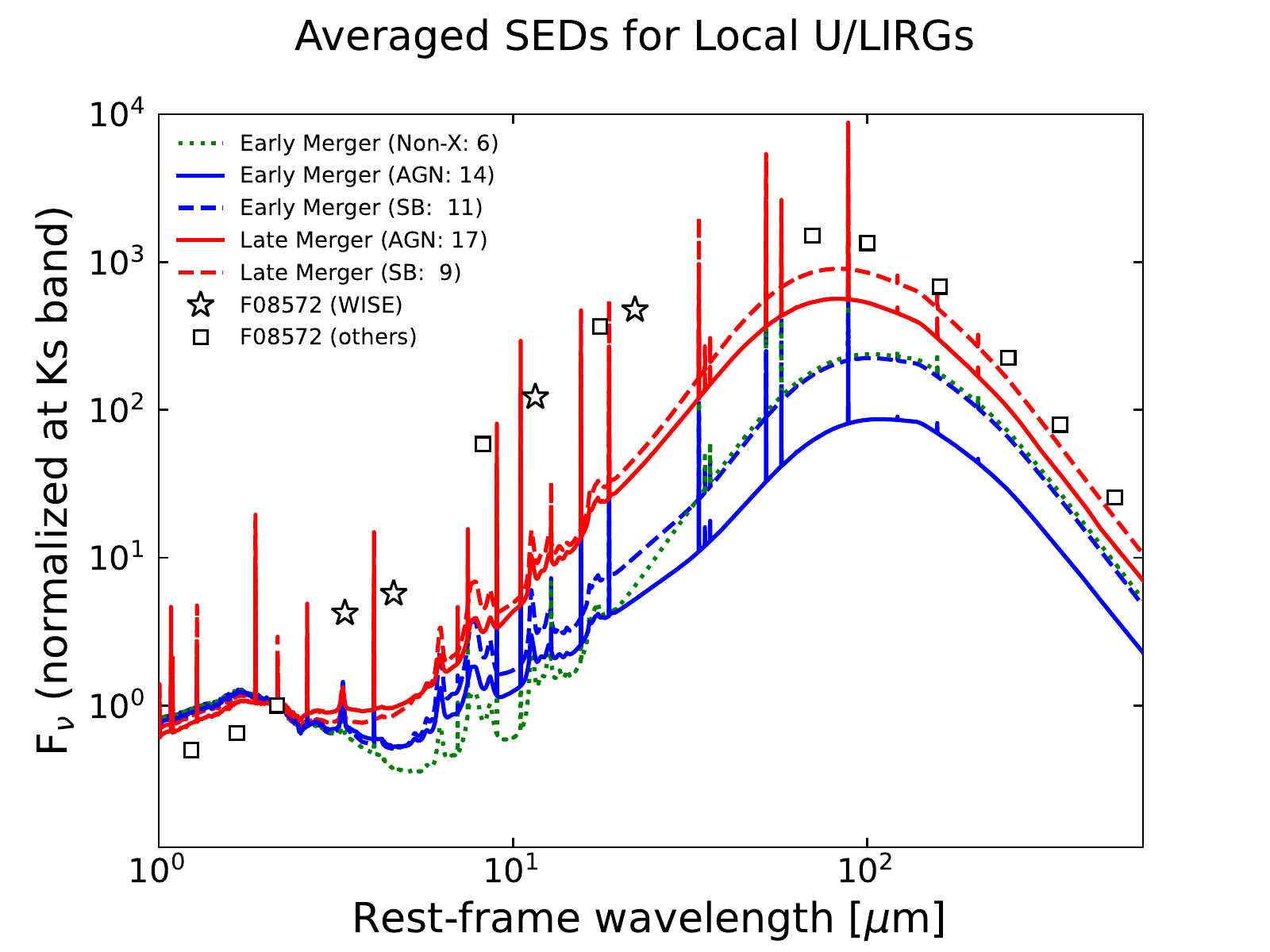}
    \caption{Enlarged picture of the IR (1--600~$\mu$m) wavelength of the left panel in Figure~\ref{F14-total-SED} (rest-frame wavelength vs. flux density normalized at $K_{\rm s}$ band).
    Stars and squares mark the photometry of WISE and other IR instruments (2MASS, AKARI, and Herschel) for the AGN in IRAS F08572+3915.
    \\}
\label{F15-IR-SED}
\end{figure}

\subsection{Averaged SEDs in U/LIRGs} \label{sub5-4-averaged-SED}
To further understand the characteristics of the hard X-ray to radio emission in local U/LIRGs, in Figure~\ref{F13-averaged-SED} we illustrate the SEDs for all sources whose X-ray spectra above $\gtrsim$1~keV are analyzed by \citet{Yamada2021}.
The $K_s$ (2.159~$\mu$m) band (near the peak of the stellar emission) is selected for the normalization since it is less affected by dust extinction and contamination from AGN-heated hot dust \citep[e.g.,][]{Kim2002,Marconi2003,Skrutskie2006,Vika2012}.\footnote{The results are almost unchanged even if the normalization is chosen in the $H$ band, which is also a reliable tracer of the stellar luminosity \citep[e.g.,][]{Marconi2003,Hainline2011}.}
The left panel shows the best-fit models of 36 single and two unresolved dual hard X-ray--detected AGNs (Mrk 266B/Mrk 266A and NGC~6240S/NGC~6240N), but excluding three newly identified AGN candidates.
Here, the complex X-ray absorption lines in NGC 1365 are ignored as it is a dramatically variable feature, mainly caused by highly ionized species of Fe in a high-velocity outflow \citep{Rivers2015}.
On the other hand, the right panel shows those of two starburst-dominant sources detected in the hard X-ray band (IC~1623A and NGC 3256) and 20 hard X-ray--undetected sources.
Since the $\lesssim$3~keV spectra are complex due to the Galactic extinction and the optically thin thermal emission in the host galaxy, we plot the best-fit models of the X-ray spectra in the 3--200~keV band for AGNs and 3--20~keV for the other sources.

These SEDs exhibit several common features in the UV-to-radio bands regardless of the presence or absence of AGNs.
First, they have a dip in the UV-to-optical wavelength, even though the SEDs are corrected for Galactic extinction.
The UV emission from the stellar populations and/or AGN disk is absorbed by the gas and dust of the host galaxy, torus, and polar dust.
Second, the IR-to-radio fluxes relative to the $K_s$~band increase with merger stage due to the re-radiation from the dust of AGNs and primarily intense starbursts (Section~\ref{sub5-1-MS} and Section~\ref{sub5-2-radio}).
Third, the increase of the IR emission causes the steep slope in the near-to-mid-IR band to appear as the features in WISE [3.4]--[4.6] color (Section~\ref{sub5-3-WISE-color}).

Figure~\ref{F14-total-SED} presents the averaged SEDs for AGNs and the other sources in both early and late mergers, whose X-ray spectra above $\gtrsim$1~keV are analyzed.
We also plot the averaged UV-to-radio SEDs for the sources whose X-ray spectra above $\gtrsim$1~keV are not obtained, all of which are in early mergers.
These averaged SED templates in Figure~\ref{F14-total-SED} are available in Table~\ref{T5_template}.
By comparing the AGNs (solid curves) and the others (dashed and dotted curves), we find that the averaged SEDs of the latter population show stronger far-IR fluxes than those of the former.
This support that the starburst-dominant sources have cooler dust than AGNs.
Whereas, the UV-to-radio SEDs of AGNs and non-AGN sources holistically resemble each other, regardless of the merger stage.
The difference in the far-IR fluxes is smaller than the scattering among the individual sources as shown in Figure~\ref{F13-averaged-SED}.
This resemblance makes it difficult to identify the AGNs in U/LIRGs.

In Figure~\ref{F15-IR-SED}, we compare the averaged SEDs and the photometry of the buried AGNs in IRAS F08572+3915, which is one of the most luminous AGNs in our sample and is well selected by the WISE color--color diagram (Section~\ref{sub5-3-WISE-color}).
We find that the source shows the flux peak at shorter IR wavelengths than in normal U/LIRGs.
This suggests the presence of a large amount of hot dust, which can be a unique feature of the luminous AGNs deeply hidden by gas and dust.
Regardless of the presence of the IR-excess features, the X-ray spectra of AGNs and others are completely different (Figure~\ref{F14-total-SED}).
The AGNs have much larger X-ray fluxes than those in the starburst-dominant or hard X-ray--undetected sources.
According to the X-ray spectral analysis in \citet{Yamada2021}, the observed X-rays of AGNs become dimmed with merger stage due to the increase of the hydrogen column density, while for the starburst-dominant sources, the soft X-rays dominated by the X-ray binary emission in the host galaxy \citep[e.g.,][]{Mineo2012a} become brighter in late mergers with high SFRs than in early mergers.
Overall, the AGNs show a significant X-ray excess relative to the other wavelength emission, as represented by the comparison between observed X-ray luminosity and SFR in Section~6.3 of \citet{Yamada2021}.
Therefore, despite the effects of the X-ray weakness (Section~\ref{sub5-5-X-weak}), the excess of X-ray fluxes relative to the UV-to-radio SEDs should be an incomparable characteristic enabling us to reveal the true energy sources (intense starbursts and/or buried AGNs) in U/LIRGs.

% Table 5
\begin{deluxetable}{llcl}
\label{T5_template}
%\tabletypesize{\small}
\tablecaption{Averaged SED Templates in Our Targets}
\tablehead{
\colhead{Column Name} &
\colhead{Format} &
\colhead{Unit} &
\colhead{Description}
}
\startdata
Class       & LONG   &        & Classification \\
Wavelength  & DOUBLE & $\mu$m & Wavelength \\
Frequency   & DOUBLE & Hz     & Frequency \\
Normed\_FNU & DOUBLE &        & Flux density at each wavelength \\
Normed\_LNU & DOUBLE &        & Luminosity at each wavelength \\
\enddata
\tablenotetext{\ }{(This table is available in its entirety in machine-readable form.)}
\end{deluxetable}

% Figure 16
\begin{figure*}
    \centering
    \includegraphics[keepaspectratio, scale=0.95]{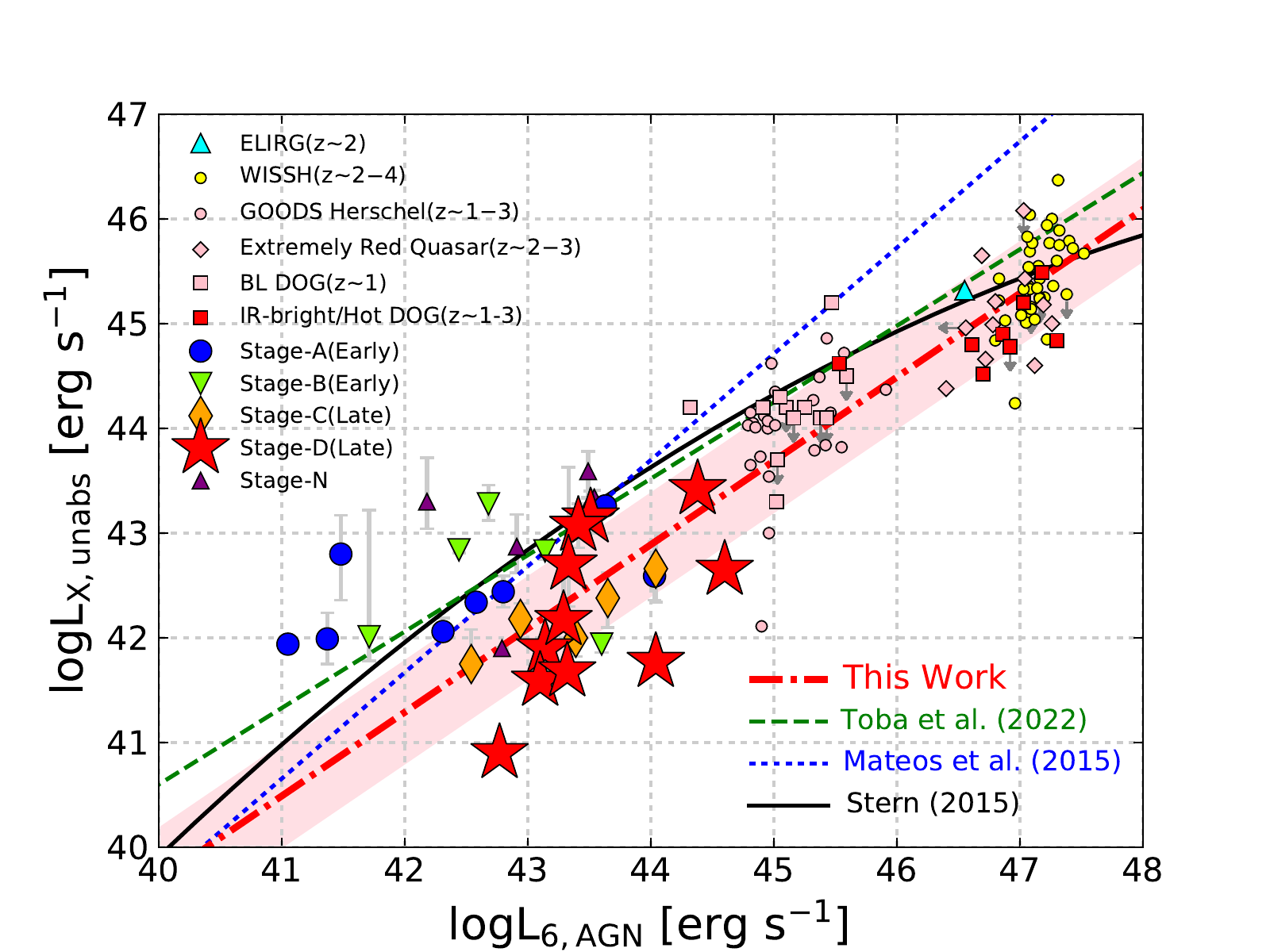}
    \caption{Rest-frame 6~$\mu$m luminosity contributed from AGNs vs. unabsorbed (absorption-corrected) rest-frame 2--10~keV luminosity for AGNs in various IR-luminous galaxies.
    The cyan triangle indicates an ELIRG at $z \sim 2$ \citep{Toba2021a}.
    The yellow circles represent WISSH quasars at $z \sim 2$--4 \citep{Martocchia2017,Zappacosta2020}.
    The pink circles indicate mid-IR luminous quasars at $z \sim 1$--3 in the GOODS--Herschel fields \citep{Del-Moro2016} while pink diamonds show extremely red quasars at $z \sim 2$--3 \citep{Goulding2018b}.
    The pink and red squares mark broad-line DOGs at $z \sim 1$ \citep{Zou2020} and IR-bright/hot DOGs at $z \sim 1$--3 \citep{Ricci2017aApJ,Vito2018,Zappacosta2018,Toba2020}, respectively.
    Their colors (cyan, yellow, pink. and red) are largely categorized by the typical $N_{\rm H}$ of the populations (see the text).
    The other symbols are the same in Figure~\ref{F8-MS}.
    The green dashed and blue dotted lines illustrate the linear relation using a sample drawn from the eROSITA Final Equatorial Depth Survey \citep{Toba2022a} and the Bright Ultra-hard XMM-Newton Survey \citep{Mateos2015}, respectively.
    The black solid curve denotes the 2D polynomial relations from \citet{Stern2015}.
    Red dash-dotted line and the pink shaded area are the best-fitting relation and its 1$\sigma$ dispersion among stage C--D mergers and the high-$z$ IR-luminous sources, respectively.
    \\}
\label{F16-L6-Lx}
\end{figure*}

\subsection{X-ray Weak AGNs in IR-luminous Galaxies} \label{sub5-5-X-weak}
We finally investigate the multiwavelength AGN luminosities corrected for the absorption of the host galaxy, torus, and polar dust.
\citet{Yamada2021} estimate the de-absorbed 2--10~keV AGN luminosities and find that the ratio of the bolometric AGN luminosity to the X-ray luminosity is quite large for the AGNs in late mergers.
This X-ray weakness has been thought of as a particular property of the AGNs in local U/LIRGs \citep[e.g.,][]{Teng2014,Teng2015}, and its origin (e.g., the optically thick failed winds launched from an inner region of the disk) is discussed in \citet{Yamada2021}.
For understanding the AGN activities in U/LIRGs through cosmic history, it is at least necessary to probe whether the X-ray weakness is a common feature for both low-$z$ and high-$z$ U/LIRGs.

\subsubsection{6 $\mu$m Luminosity vs. 2--10 keV Luminosity}
Figure~\ref{F16-L6-Lx} illustrates the relation between monochromatic luminosity of the AGN emission at rest-frame 6~$\mu$m derived from the multiwavelength SED decomposition ($L_{\rm 6,AGN}$) and unabsorbed (i.e., de-absorbed) X-ray luminosity in the rest-frame 2--10~keV band ($L_{\rm X,unabs}$).
Many studies report that the mid-IR and X-ray luminosities of the AGN component are strongly correlated over a wide luminosity ranges \citep[e.g.,][]{Lutz2004,Ramos-Almeida2007,Gandhi2009,Ichikawa2012,Ichikawa2017,Asmus2015,Mateos2015,Stern2015,Garcia-Bernete2016,Chen2017,Toba2019a,Toba2022a}.
Here, we plot the values for local U/LIRGs color-coded by merger stages, 
and overplot those for IR-luminous galaxies corresponding to high-$z$ U/LIRGs:
an extremely luminous IR galaxy (ELIRG) at $z \sim 2$ \citep{Toba2021a};
WISE/SDSS selected hyper-luminous (WISSH) quasars at $z \sim 2$--4 \citep{Martocchia2017,Zappacosta2020};
mid-IR luminous quasars at $z \sim 1$--3 from the GOODS-Herschel fields \citep{Del-Moro2016};
extremely red quasars at $z \sim 2$--3 \citep{Goulding2018b};
dust-obscured galaxies (DOGs) with broad optical/UV emission lines (BL DOGs) at $z \sim 1$ \citep{Zou2020};
and IR-bright DOGs at $z \sim 1$ \citep{Toba2020} and hot DOGs at $z \sim 1$--3 \citep{Ricci2017aApJ,Vito2018,Zappacosta2018}.\footnote{The $L_{\rm 6,AGN}$ of extremely red quasars in \citet{Goulding2018b} and hot DOGs in \citet{Ricci2017aApJ} are provided by assuming that the 6~$\mu$m emission is AGN-dominant (see e.g., \citealt{Del-Moro2016} for extremely red quasars; e.g., \citealt{Stern2014,Assef2016,Assef2020,Tsai2018,Zappacosta2018} for hot DOGs).}
The values of $L_{\rm 6,AGN}$ and $L_{\rm X,unabs}$ for high-$z$ IR-luminous galaxies are presented in their references mostly by the IR SED decomposition and X-ray spectral analysis, respectively.
The AGNs are unobscured ($\lesssim$10$^{22}$~cm$^{-2}$) for an ELIRG (cyan), 
weakly obscured ($\sim$10$^{21}$--$10^{23}$~cm$^{-2}$) for WISSH quasars (yellow),
moderately obscured ($\sim$10$^{22}$--$10^{24}$~cm$^{-2}$) for mid-IR luminous quasars in the GOODS-Herschel fields, extremely red quasars, and BL DOGs (pink),
and heavily obscured ($\gtrsim$10$^{23}$~cm$^{-2}$) for IR-bright/hot DOGs (red).
Moreover, we compare these values with the typical relation obtained from the SDSS-selected low-$z$ Seyfert galaxies \citep{Stern2015}, 
the Bright Ultra-hard XMM-Newton Survey (BUXS; \citealt{Mateos2015}), and the eROSITA Final Equatorial Depth Survey (eFEDS; \citealt{Toba2022a}).

We find that several late mergers of local U/LIRGs lie below the typical $L_{\rm 6,AGN}$--$L_{\rm X,unabs}$ relation, supporting the X-ray weakness.
For late mergers (stage C and D) and high-$z$ IR-luminous galaxies, we conduct a Bayesian maximum likelihood method of \citet{Kelly2007} between the two parameters.
The resulting linear function is
\begin{align}
& {\rm log}(L_{\rm X,unabs}) = 0.80\ {\rm log}(L_{\rm 6,AGN}) + 7.69,
\end{align}
and the 1$\sigma$ dispersion is $\pm$0.50 dex (red dashed line and pink shaded area).
The correlation coefficient is $r = 0.92 \pm 0.02$, confirming a tight correlation for low-$z$ and high-$z$ IR-luminous sources.
These AGNs in the IR-luminous sources at $z \sim 0$--4 are obscured sources and show smaller de-absorbed X-ray luminosities relative to the mid-IR luminosities than the typical relations.
In particular, most AGNs in IR-bright/hot DOGs are heavily obscured ($\gtrsim$10$^{23}$~cm$^{-2}$) and explicitly X-ray weak among them \citep[e.g.,][]{Ricci2017aApJ,Toba2020}, sharing similarity with the X-ray weak AGNs in late mergers for local U/LIRGs \citep{Yamada2021}.

% Figure 17
\begin{figure}
    \centering
    \includegraphics[keepaspectratio, scale=0.55]{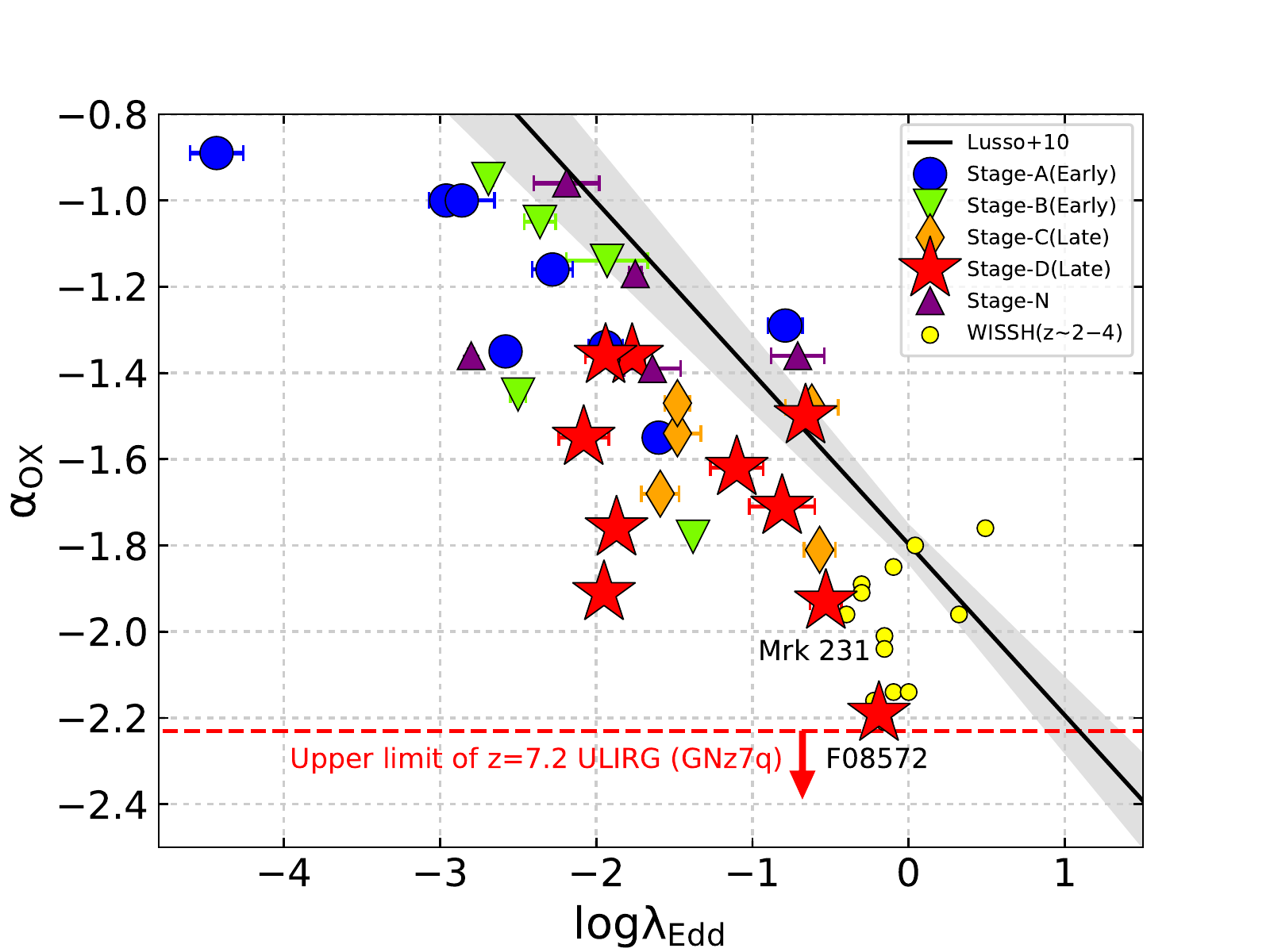}
    \caption{Logarithmic Eddington ratio vs. $\alpha_{\rm OX}$ computed using extinction-corrected 2500 \AA\ and unabsorbed 2 keV luminosities.
    The black solid curve represents the best-fit relation among local and high-$z$ AGNs, respectively \citep{Lusso2010}.
    The gray shaded area shows its 1$\sigma$ uncertainty.
    The upper limit of a ULIRG at $z = 7.2$ (GNz7q; \citealt{Fujimoto2022}) is shown as a red dashed line since its SMBH mass and Eddington ratios are not constrained.
    Yellow circles mark the WISSH quasars in $z \sim 2$--4 \citep{Zappacosta2020}.
    The other symbols are the same in Figure~\ref{F8-MS}.
    \\}
\label{F17-Edd-ox}
\end{figure}

\subsubsection{Optical to X-ray Spectral Index}
Additionally, we display the comparison between Eddington ratio ($\lambda_{\rm Edd}$) and the optical to X-ray spectral index ($\alpha_{\rm OX}$; \citealt{Tananbaum1979}) in Figure~\ref{F17-Edd-ox}.
The definition of $\alpha_{\rm OX}$ is provided by using the monochromatic luminosity at rest-frame 2500~\AA\ of the intrinsic AGN disk emission ($L_{\rm 2500,disk}$) and the de-absorbed monochromatic luminosity at rest-frame 2~keV ($L_{\rm 2 keV,unabs}$) as below:
\begin{align}
\alpha_{\rm OX} = &\ \frac{{\rm log}(L_{\rm 2 keV,unabs}/L_{\rm 2500,disk})}{{\rm log}(\nu_{\rm 2 keV}/\nu_{\rm 2500})}.
\end{align}
The $L_{\rm 2500,disk}$ is obtained by multiwavelength SED analysis in this work, and $L_{\rm 2 keV,unabs}$ is calculated from the best-fit models of the broadband X-ray spectra (Section~\ref{sub3-1-Xray}).
The $\alpha_{\rm OX}$ probes the balance between the accretion disk and hot corona activities, radiating optical and X-ray emission respectively.
Several works report a moderate correlation between the Eddington ratio and $\alpha_{\rm OX}$ \citep[e.g.,][]{Lusso2010,Chiaraluce2018} for X-ray selected AGNs and a typical relation of \citet{Lusso2010} is illustrated by a black solid line in Figure~\ref{F17-Edd-ox}.
We overplot the values of WISSH quasars (yellow circles; \citealt{Martocchia2017,Zappacosta2020}), and the upper limit of $\alpha_{\rm OX}$ ($< -$2.23) provided with deep Chandra observation for a $z = 7.2$ red quasar GHz7q, classified as U/LIRG (red dashed line; \citealt{Fujimoto2022}).

Although the Eddington ratios contain uncertainties of about $\pm$0.5 dex originating from the $M_{\rm BH}$ measurements \citep{Yamada2021}, some AGNs in stage D mergers show much smaller $\alpha_{\rm OX}$ than the typical relation.
The distribution of $\alpha_{\rm OX}$ for X-ray selected and optical-selected AGNs \citep{Just2007,Lusso2010,Dong2012,Martocchia2017,Chiaraluce2018} mostly covers from $-$1.0 to $-$2.0.
Whereas, we find that the local stage-D mergers with the largest AGN luminosities (Mrk~231 and IRAS F08572+3915), many WISSH quasars, and GHz7q lie $\alpha_{\rm OX} \lesssim -$2.
The caveat is that the X-ray luminosity of GHz7q is only corrected for Galactic absorption, but the Chandra observation of GHz7q confirms the lack of rest-frame hard X-rays, and the significant UV emission suggests the AGN is not heavily obscured.
Its bolometric AGN luminosity estimated by the optical-to-millimeter SED analysis is $(1.7\pm0.1) \times 10^{46}$~erg~s$^{-1}$, which is $\sim$7 times larger than that of IRAS F08572+3915 ($L_{\rm AGN,int} \sim 2.5 \times 10^{45}$~erg~s$^{-1}$).
Unless the AGN of GHz7q has an extreme Eddington ratio above $\lambda_{\rm Edd} \sim 10$, the upper limit of $\alpha_{\rm OX}$ suggests the possibility that the AGN is also X-ray weak.

\subsubsection{X-ray Weakness as a Common Feature}
In summary, considering the small values of $L_{\rm X,unabs}$/$L_{\rm 6,AGN}$ ratio and $\alpha_{\rm OX}$, the de-absorbed multiwavelength SEDs support that \textit{the X-ray weakness may be a common feature} among the AGNs in IR-luminous galaxies over cosmic history ($z \sim 0$--7).
Intriguingly, the strong AGN-driven outflows have been discovered from the $z \lesssim 1$ U/LIRGs \citep[e.g.,][]{Chen2020a} and the high-$z$ IR-luminous galaxies such as WISSH quasars \citep{Bischetti2017,Travascio2020,Vietri2022}, extremely red quasars \citep{Zakamska2016,Hamann2017,Goulding2018b}, and IR-bright and hot DOGs \citep[e.g.,][]{Ricci2017aApJ,Toba2017c,Toba2017d,Wu2018,Finnerty2020}.
As discussed in Section~6.1.6 of \citet{Yamada2021} for local late-stage U/LIRGs, the physical mechanism of the extreme X-ray weakness (e.g., the X-ray attenuation due to overionized optically-thick failed winds at $\sim$10$^2$--10$^3$~$r_{\rm g}$) can enhance the UV-driven winds, which may effectively trigger the massive outflows in low-$z$ and high-$z$ U/LIRGs.

% Figure 18
\begin{figure*}
    \centering
    \includegraphics[keepaspectratio, scale=0.37]{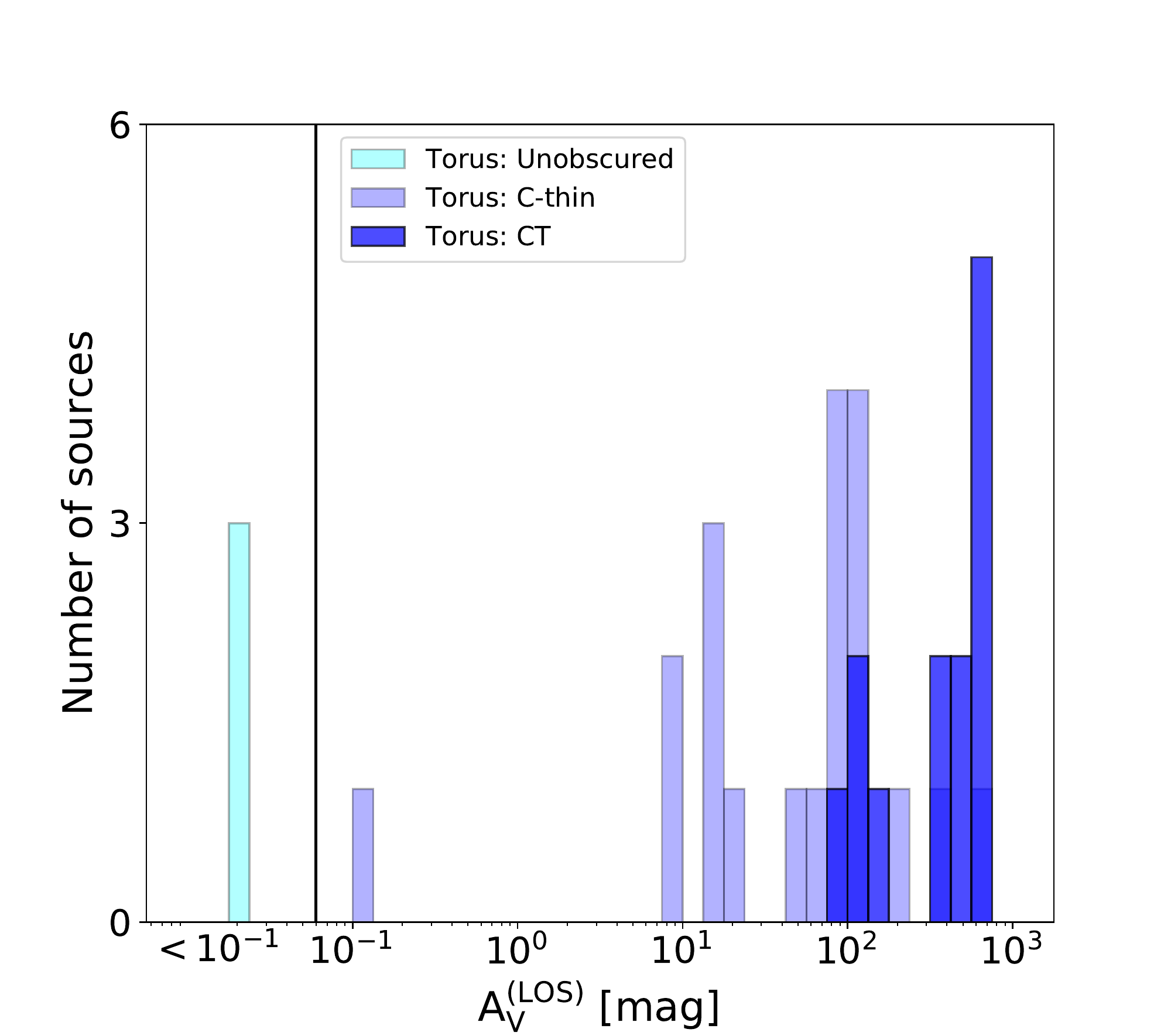}
    \includegraphics[keepaspectratio, scale=0.37]{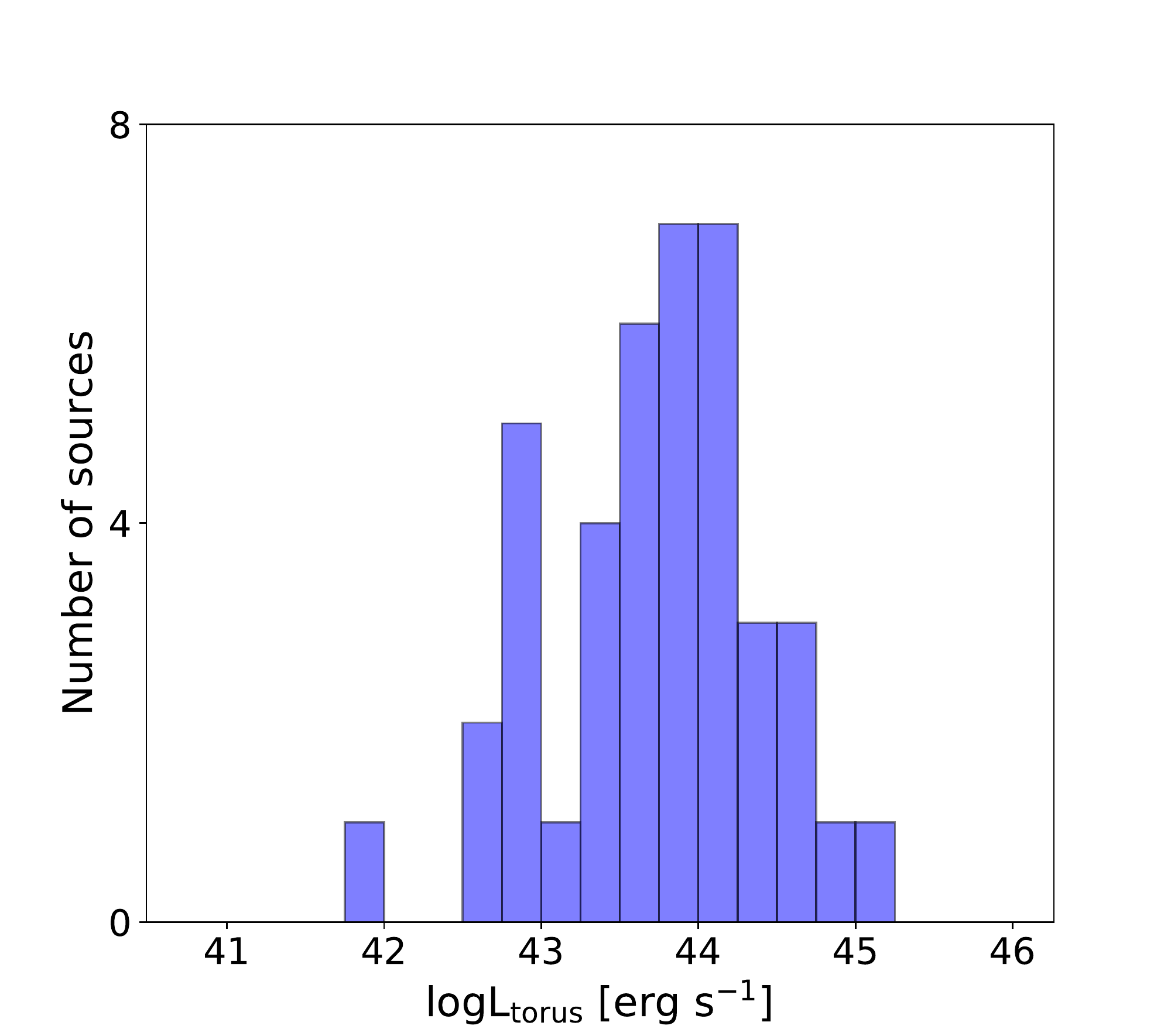}
    \includegraphics[keepaspectratio, scale=0.37]{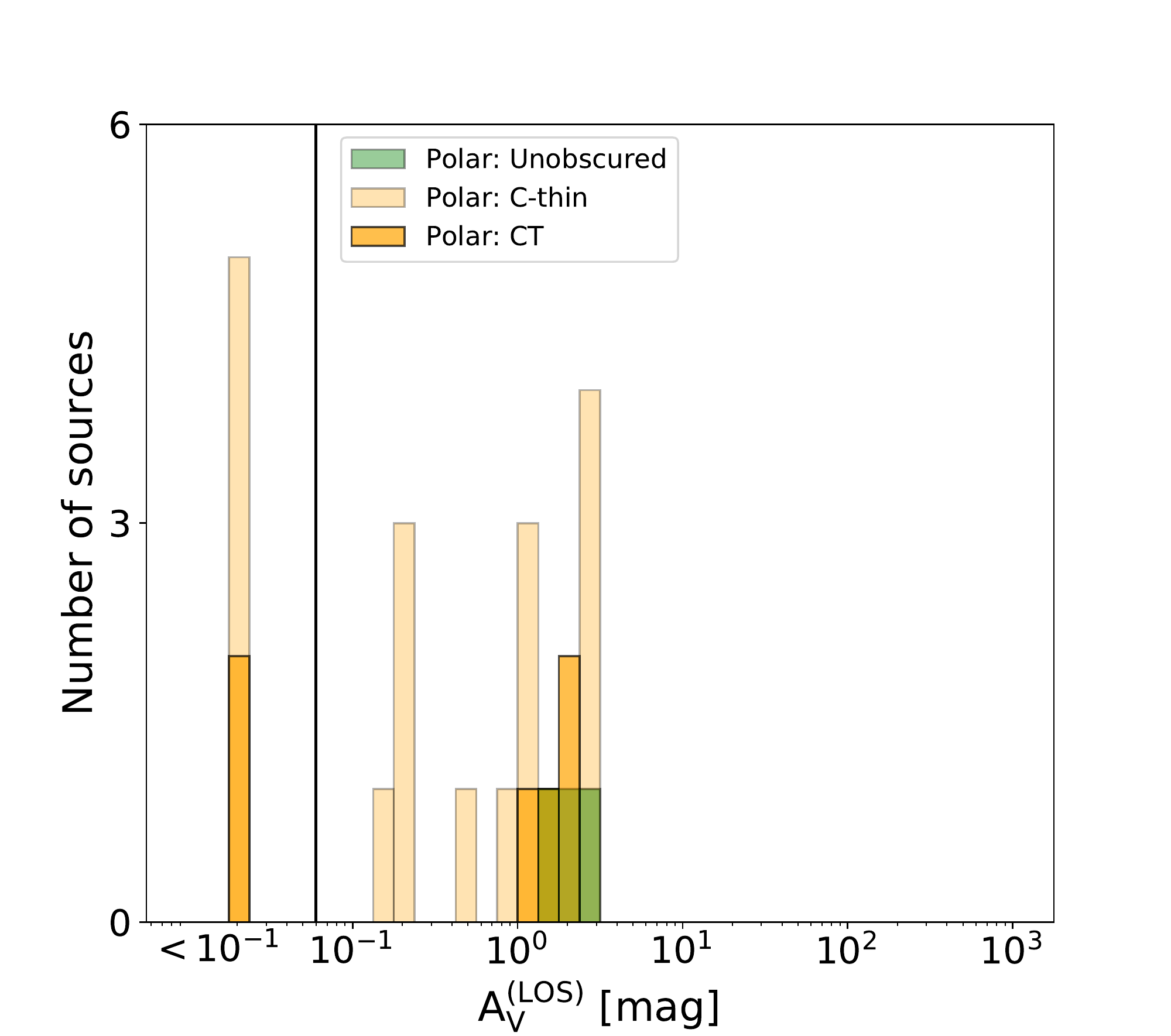}
    \includegraphics[keepaspectratio, scale=0.37]{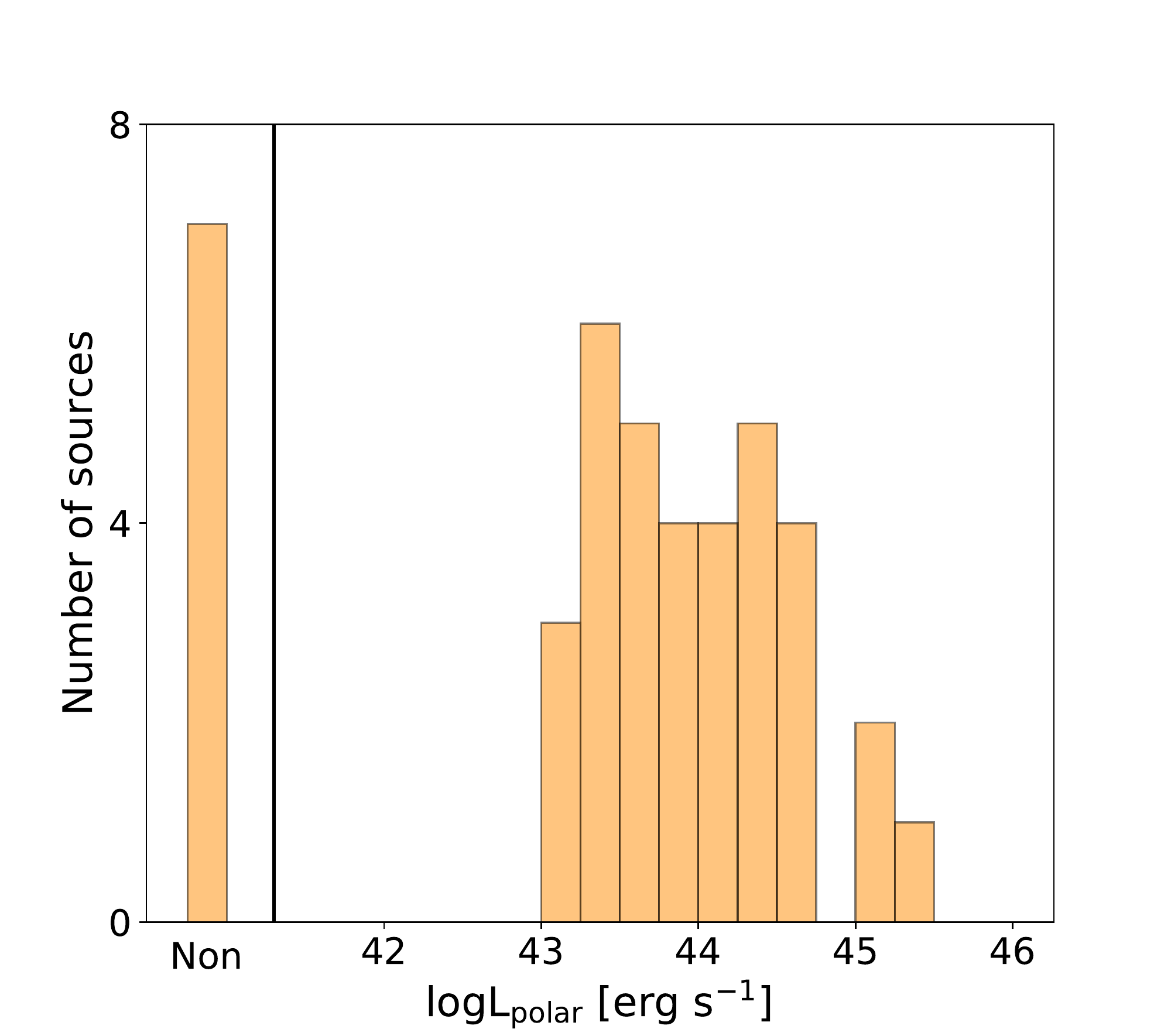}
    \caption{Top left panel: histogram of the line-of-sight extinction in the $V$ band by the torus $A_{V{\rm,torus}}^{\rm(LOS)}$ (top).
      These are divided by the X-ray AGN classification (unobscured, Compton-thin, and CT AGNs; \citealt{Yamada2021}).
      Bottom left panel: the same for the polar dust extinction, $A_{V{\rm,polar}}^{\rm(LOS)}$ (bottom).
      The seven sources showing 90$\degr - i > \sigma$ (i.e., $A_{V{\rm,polar}}^{\rm(LOS)} = 0$) are excluded.
      Top right panel: histogram of the torus (top) luminosity in the UV-to-IR band.
      Bottom right panel: the same for polar dust luminosity.
      The number of AGNs whose best-fit SED models do not include a polar dust component is shown on the left side.
    \\}
\label{F18-Av-lumin}
\end{figure*}

\subsection{Summary of Multiwavelength Features} \label{sub5-6-summary-SED}
Here, we summarize the multiwavelength features of local U/LIRGs.
The component galaxies in early mergers have a wide range of SFRs above and below the MS, while galaxies in late mergers have the highest SFRs above the MS.
Similar SFR--$M_*$ distributions of AGNs and non-AGN sources imply the small influence of the AGN activities on the star formation in the phase of U/LIRGs (Section~\ref{sub5-1-MS}).
The flat slope of radio emission in late mergers may be caused by the optically-thick free-free absorption due to the rich environment of the nuclear region.
The SFRs are correlated with radio luminosity, indicating the major origin of the radio emission is from the starburst emission except for the radio-excess AGNs (Section~\ref{sub5-2-radio}).
In the WISE color--color diagram, we propose a new wedge of the buried AGNs in late mergers. 
The strong near-IR emission of the hard X-ray--detected AGNs in late mergers suggests that they host the dusty disk and/or hot polar dust within the inner parsecs (Section~\ref{sub5-3-WISE-color}).
The averaged SEDs suggest that the excess of the X-rays is an incomparable characteristic enabling us to reveal the true energy sources (intense starbursts and/or buried AGNs) in U/LIRGs (Section~\ref{sub5-4-averaged-SED}). 
The absorption-corrected AGN SEDs show the extreme X-ray weakness relative
to the other wavelengths. This may be a common feature among the AGNs in IR-luminous galaxies over the cosmic history ($z \sim 0$--7), whose mechanism will be related to their massive outflows (Section~\ref{sub5-5-X-weak}).
In short, our results support that (1) the intense starburst and buried AGNs occur in late mergers and (2) these buried AGNs may have dusty disks and/or hot polar dust, whose X-ray weakness may be related to the massive outflows.
\\

%%%%%%%%%%%%%%%%%%%%%%%%%%%%%%%
%%% Section 6 : Discussion1 %%%
%%%%%%%%%%%%%%%%%%%%%%%%%%%%%%%
\section{Discussion I: Evolution of Polar Dust in U/LIRGs} \label{S6-discussion1}
In this Section, we investigate the structure of polar dust in U/LIRGs based on the results of the multiwavelength SED analysis.
The detailed results of the polar dust are discussed on 
the obscuration and luminosity (Section~\ref{sub6-1-polar-dust}), 
gas-to-dust ratio (Section~\ref{sub6-2-NH-AV}),
the contribution to the IR AGN luminosity (Section~\ref{sub6-3-Lpolar}),
polar dust temperature (Section~\ref{sub6-4-Tpolar}),
and the unified view of the polar dust structure in U/LIRGs (Section~\ref{sub6-5-polar-structure}).

\subsection{Obscuration and Luminosity} \label{sub6-1-polar-dust}
The top left panel of Figure~\ref{F18-Av-lumin} presents the histograms of the line-of-sight extinction in the $V$ band by the torus ($A_{V{\rm,torus}}^{\rm (LOS)}$), while the bottom left panel shows for polar dust extinction ($A_{V{\rm,polar}}^{\rm (LOS)}$).
The dust extinction of the polar dust is calculated with $R_V = 3.1$ \citep{Cardelli1989} and given when $90\degr - i > \sigma$ (Section~\ref{subsub4-1-3-CLUMPY}).
We note that the values are much smaller than those of torus due to the initial condition of E($B-V$) = 0--0.8 mag (Section~\ref{subsub4-2-2-torus} and \ref{subsub4-2-3-polar}; see e.g., \citealt{Buat2021}).
We compare the histograms of different groups of unobscured, obscured, and CT AGNs.
As expected from the dominance of the torus in the dust extinction, the $A_{V{\rm,torus}}^{\rm (LOS)}$ becomes larger with the line-of-sight AGN obscuration in the X-ray band.
Whereas, $A_{V{\rm,polar}}^{\rm (LOS)}$ does not show any positive correlation with the X-ray obscuration.
Considering that the hydrogen column densities mainly reflect the absorption by the broad line region and/or torus for obscured AGNs \citep[e.g.,][]{Tanimoto2020,Ogawa2021,Andonie2022}\footnote{Particularly for U/LIRGs, the AGNs show the dramatic variability in $N_{\rm H}$, supporting the presence of compact-scale (a few parsecs) obscurer \citep{Laha2020,Yamada2021}.}, the decorrelation between $A_{V{\rm,polar}}^{\rm (LOS)}$ and AGN types will be derived by the difference in the scales of the torus ($\sim$1~pc) and the extended polar dust emission ($\sim$10--1000~pc; see Section~\ref{sub6-5-polar-structure}) whose typical temperatures are allowed between 100--250~K (Section~\ref{subsub4-2-3-polar}).

In the top right panel, we illustrate the distributions of the integrated UV-to-IR (mainly IR) torus ($L_{\rm torus}$), and in the bottom right panel for polar dust luminosities ($L_{\rm polar}$).
They cover the ranges of log$L_{\rm torus}{\rm /(erg\ s^{-1})} \sim 42$--45 and log$L_{\rm polar}{\rm /(erg\ s^{-1})} \sim 43$--45.5, respectively.
Although seven sources show no signatures of the polar dust emission (Section~\ref{subsub4-2-4-BIC}), the distribution of the polar dust luminosities is larger than that of the torus ones.
This is in agreement with the high-spatial-resolution mid-IR imaging by \citet{Asmus2019}, who reports that a large part of the mid-IR luminosities is derived from the extended emission.

% Figure 19
\begin{figure}
    \centering
    \includegraphics[keepaspectratio, scale=0.5]{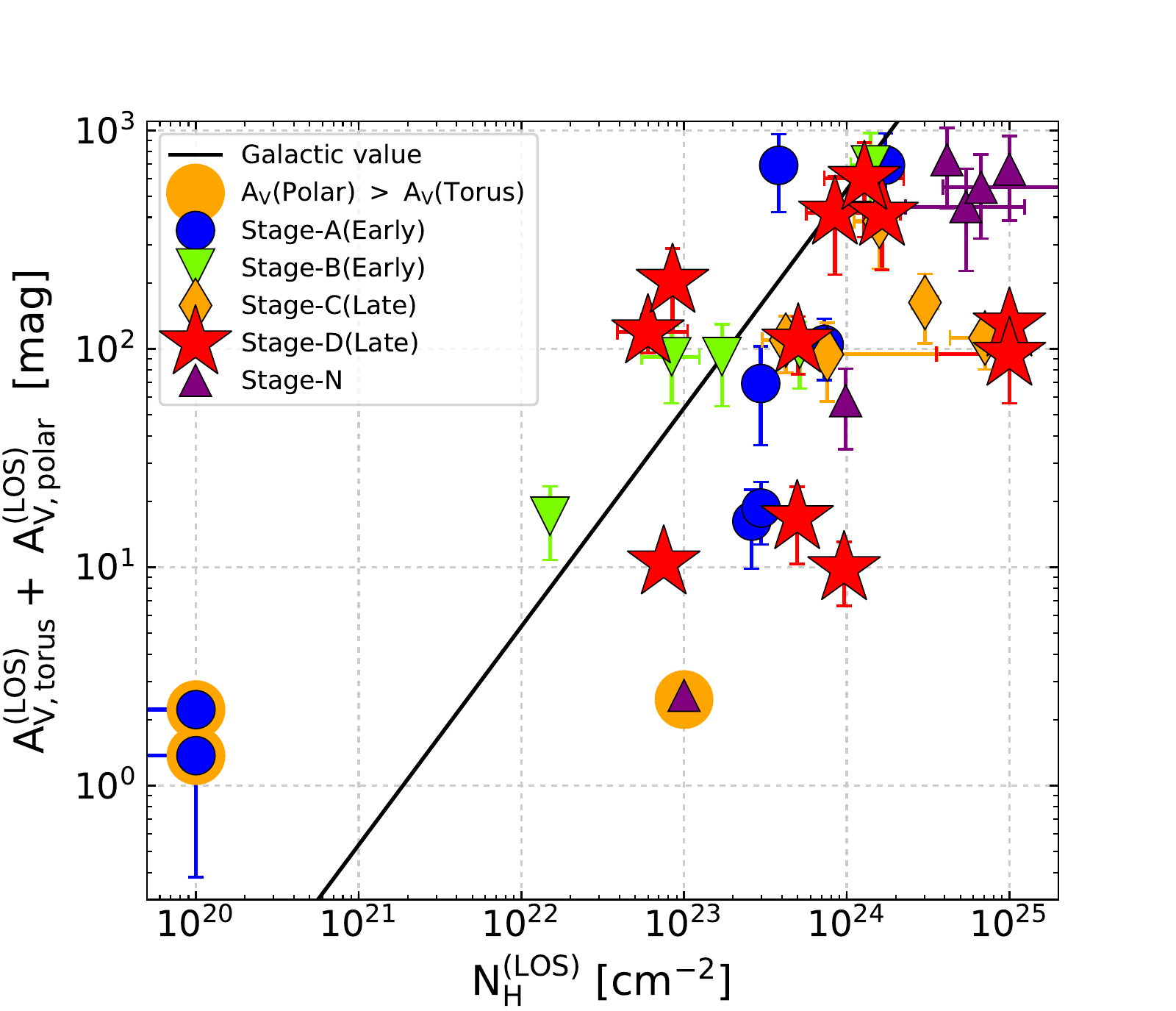}
    \caption{Line-of-sight hydrogen column density $N_{\rm H}^{\rm (LOS)}$ derived from X-ray spectra vs. $A_{V{\rm,torus}}^{\rm (LOS)}$, and polar dust, $A_{V{\rm,polar}}^{\rm (LOS)}$ (if 90$\degr - i > \sigma$, $A_{V{\rm,polar}}^{\rm (LOS)} = 0$).
    Solid line shows the Galactic gas-to-dust ratio ($N_{\rm H}$/$A_{V} = 1.87 \times 10^{21}$ cm$^{-2}$ mag$^{-1}$; \citealt{Draine2003}).
    Large orange circles mark the AGNs showing $A_{V{\rm,polar}}^{\rm (LOS)} > A_{V{\rm,torus}}^{\rm (LOS)}$.
    The other symbols are the same in Figure~\ref{F8-MS}.
    \\}
\label{F19-NH-Av}
\end{figure}

\subsection{Gas-to-dust Ratio} \label{sub6-2-NH-AV}
The comparison between X-ray absorption and dust extinction has been investigated to evaluate the gas-to-dust ratio.
The X-rays trace all material of gas and dust ($N_{\rm H}$), and the optical-to-IR emission primarily traces the dust component ($A_V$).
\citet{Maiolino2001b} report that the $N_{\rm H}$/$A_V$ ratios are larger than the Galactic value ($N_{\rm H}$/$A_V = 1.87 \times 10^{21}$~cm$^{-2}$~mag$^{-1}$; \citealt{Draine2003}), while for unobscured AGNs those are smaller \citep[see also][]{Burtscher2016}.
Recent works with the torus models of CLUMPY and XCLUMPY, but not including the polar dust component, \citep{Miyaji2019,Tanimoto2020,Ogawa2021} support that the $N_{\rm H}$/$A_V$ values are higher than or similar to the Galactic value for obscured AGNs, but smaller for unobscured AGNs.
\citet{Ogawa2021} suggest that the trend could be explained if the torus angular widths are overestimated in the IR band due to the contamination from the polar dust emission.

% Figure 20
\begin{figure*}
    \centering
    \includegraphics[keepaspectratio, scale=0.45]{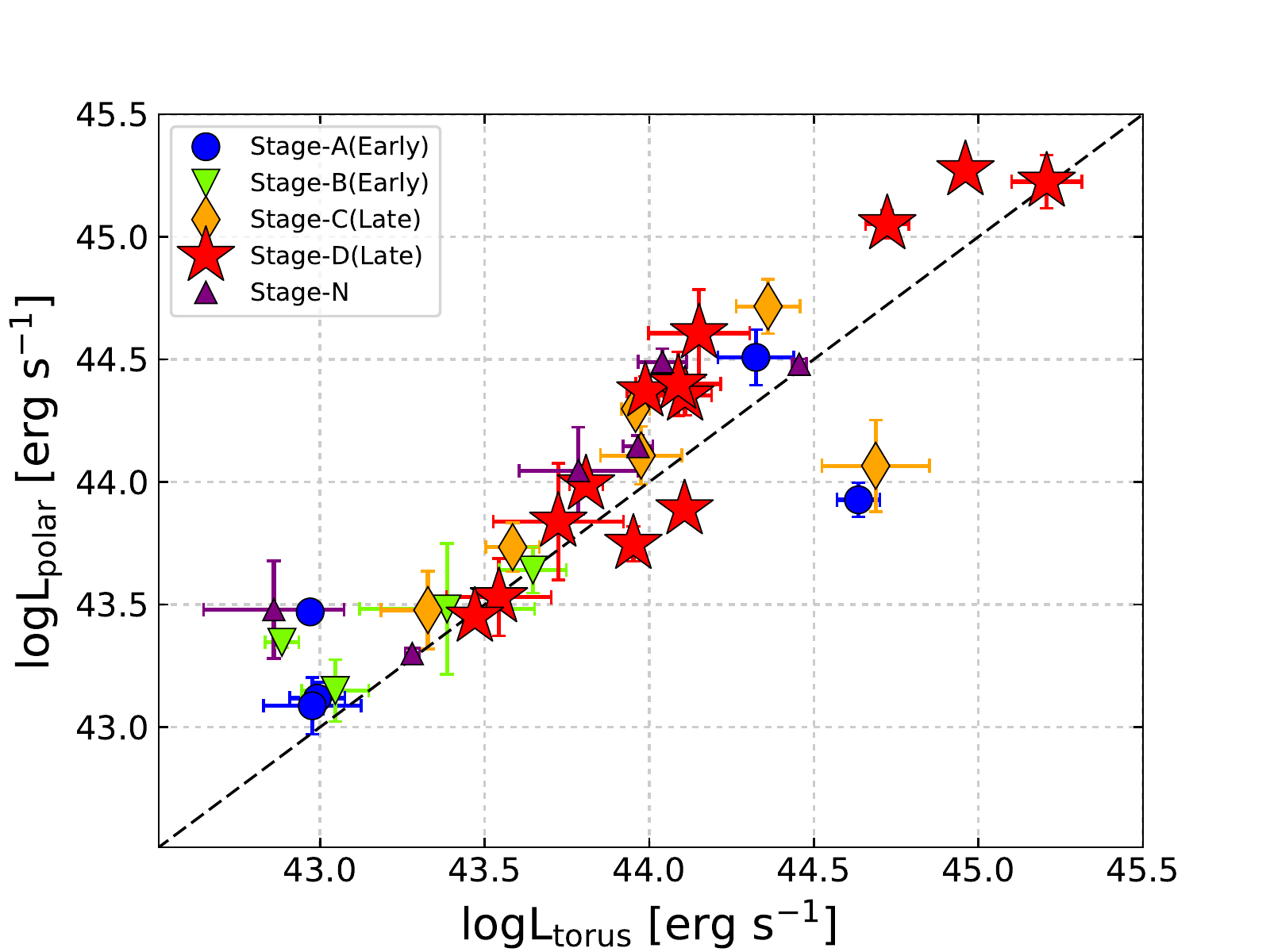}
    \includegraphics[keepaspectratio, scale=0.45]{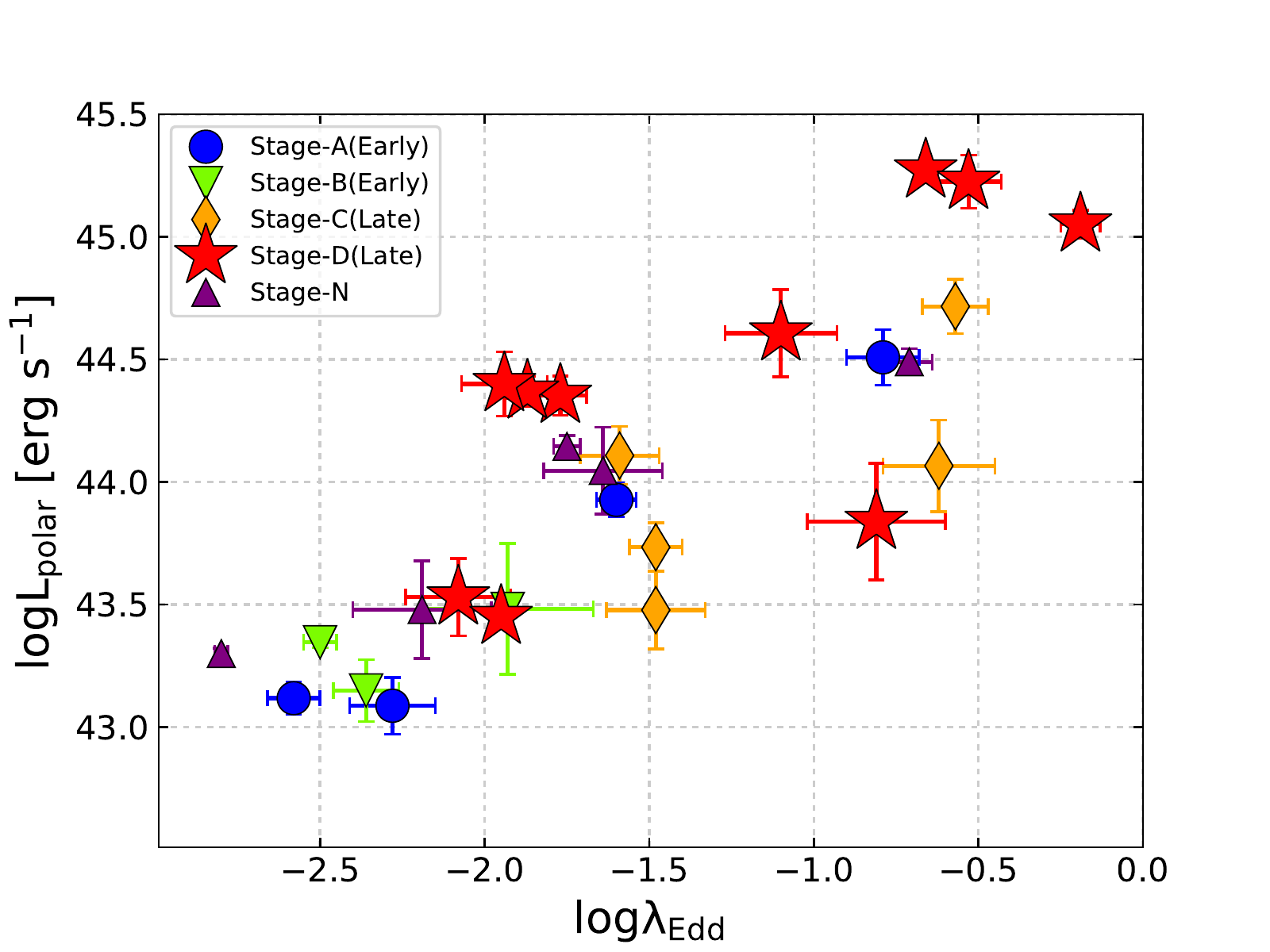}
    \caption{Left panel: logarithmic torus luminosity vs. logarithmic polar-dust luminosity in the UV-to-IR bands.
    Right panel: logarithmic Eddington ratio vs. logarithmic polar-dust luminosity.
    The symbols are the same in Figure~\ref{F8-MS}.
    \\}
    \label{F20-Lpolar}
\end{figure*}

% Figure 21
\begin{figure*}
    \centering
    \includegraphics[keepaspectratio, scale=0.45]{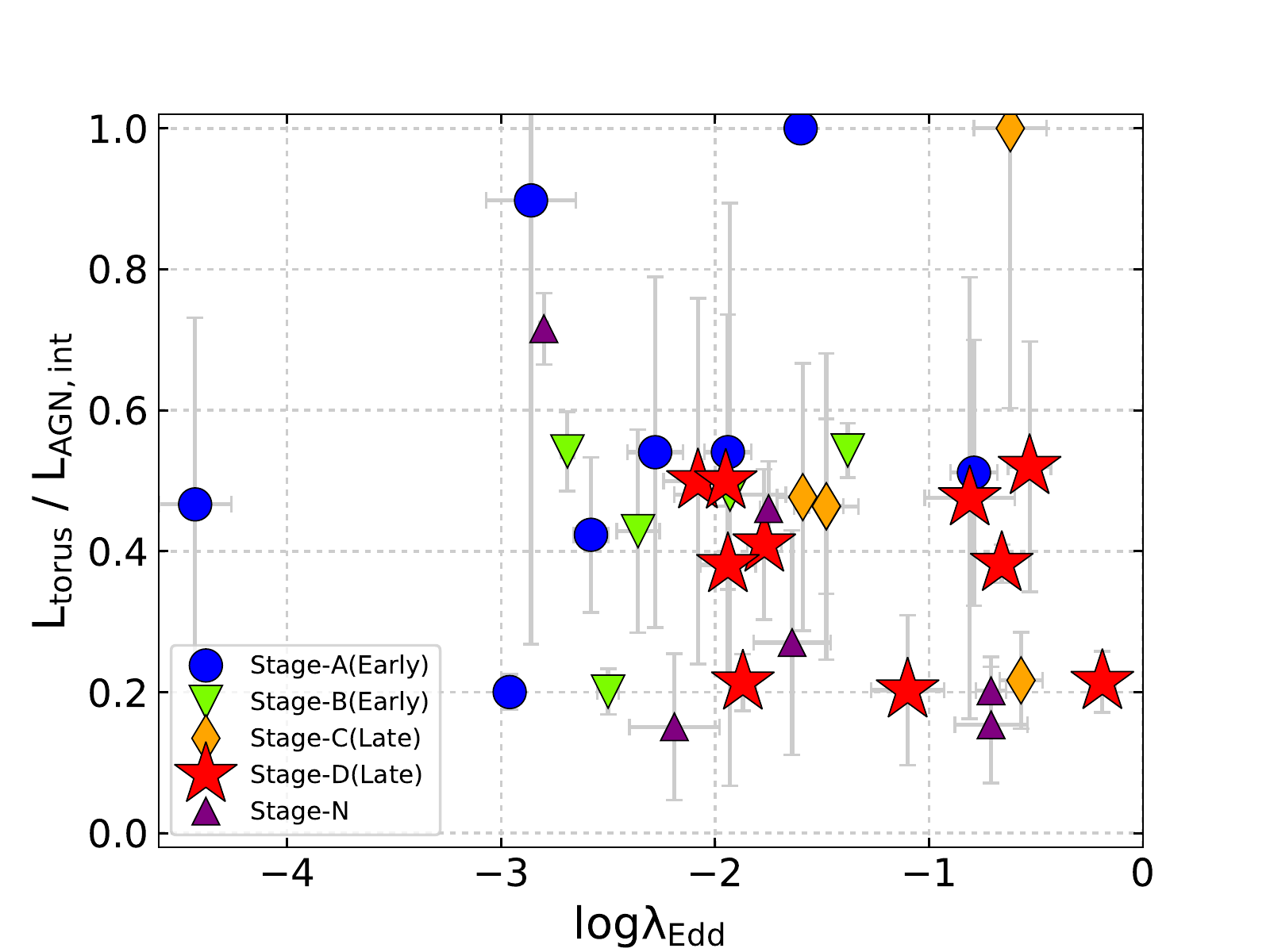}
    \includegraphics[keepaspectratio, scale=0.45]{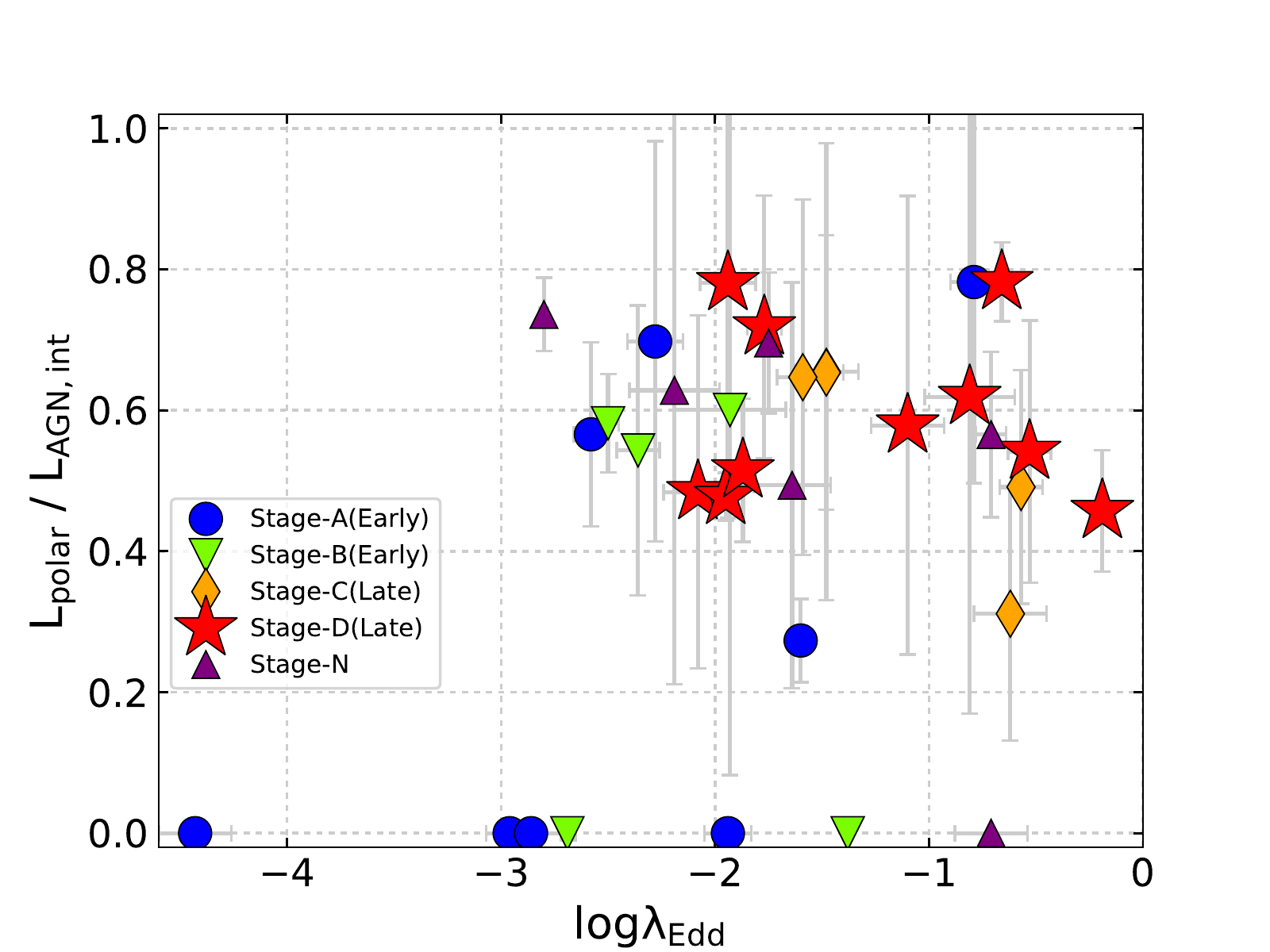}
    \caption{Left panel: logarithmic Eddington ratio vs. ratio of the torus and intrinsic AGN luminosities.
    Two AGNs whose luminosity ratio above 1 are plotted as $L_{\rm torus}/L_{\rm AGN,int} = 1$.
    Right panel: logarithmic Eddington ratio vs. ratio of polar dust and intrinsic AGN luminosities.
    The symbols are the same in Figure~\ref{F8-MS}.
    \\}
\label{F21-Edd-CF}
\end{figure*}

Our SED decomposition provides the values of $A_{V{\rm,torus}}^{\rm (LOS)}$ and $A_{V{\rm,polar}}^{\rm (LOS)}$.
In Figure~\ref{F19-NH-Av}, we compare the X-ray obscuration and dust extinction.
The three AGNs with $A_{V{\rm,torus}}^{\rm (LOS)} < A_{V{\rm,polar}}^{\rm (LOS)}$ have a wide range of gas-to-dust ratios.
As mentioned in Section~\ref{sub6-1-polar-dust}, both parameters may trace the different regions of a compact torus in X-rays and extended polar dust in the optical-to-IR bands, respectively.
Except for them, the AGNs with $N_{\rm H} > 10^{22}$~cm$^{-2}$ in our U/LIRGs show $A_{V{\rm,torus}}^{\rm (LOS)} > A_{V{\rm,polar}}^{\rm (LOS)}$, meaning the small contribution of the polar dust component to the dust extinction. 
We find that the obscured AGNs have similar or higher $N_{\rm H}$/$A_V$ ratios than the Galactic one, indicating that the materials on the compact (a few parsecs) scales in the U/LIRGs show a large gas-to-dust ratio.

Similarly, it is reported that the gas-to-dust mass ratio ($M_{\rm gas}/M_{\rm dust}$) is $\sim$200--300 for local ULIRGs \citep[e.g.,][]{Seaquist2004}, which is larger than the Galactic value.
These results may be explainable mainly by two scenarios.
The first scenario is that the $A_{V}$ decreases by the destruction of the small dust grains by supernova shocks and/or expelled by strong AGN winds \citep[e.g.,][]{Draine1979,Jones1994,Savage1996,Maiolino2001b,Gaskell2004,Liang2004,Rupke2008,Mannucci2010,Roseboom2012,Asano2013,Remy-Ruyer2014}.
The second is that the $N_{\rm H}$ increases by dust-free gas clouds covering the line-of-sight in BLR of the AGN \citep[e.g.,][]{Granato1997,Burtscher2016,Ichikawa2019a,Ogawa2021,Mizukoshi2022}.
More investigations are needed to reveal the origin of the large gas-to-dust ratios in U/LIRGs.

\subsection{Contribution to the IR AGN Luminosity} \label{sub6-3-Lpolar}
The torus and polar dust luminosities in the UV-to-IR bands are primarily determined by the total AGN luminosity times their apparent covering fractions (and/or large volumes).
Considering the averaged torus angular width is $\sigma \sim 20\degr$ \citep{Ogawa2021,Yamada2021}, the apparent covering fractions of polar dust can be expected as $\Omega/4\pi = 1 - {\rm sin}(20\degr) \sim 0.66$.
The polar dust luminosities are $\sim$2 times larger than those of the torus component (left panel of Figure~\ref{F20-Lpolar}; see also Section~\ref{sub6-1-polar-dust}).
If polar dust is likely the dust component of the AGN-driven outflows, their activities should be correlated with the Eddington ratios.
Although it is natural under the situation that the polar dust luminosities are proportional to the AGN luminosities, we confirm the positive correlation between the polar dust luminosities and Eddington ratios in our sample.

In the X-rays, it is thought that the torus covering fractions of Compton-thin matters becomes smaller with larger Eddington ratios due to the radiation pressure from the AGN \citep{Ricci2017cNature,Ricci2022}.
By analyzing the broadband X-ray spectra with XCLUMPY, \citet{Yamada2021} find that the individual torus covering fraction for AGNs in U/LIRGs follow the typical relation, except for two buried AGNs with $\lambda_{\rm Edd} \sim 1$ in stage-D mergers.
On the other hand, the torus angular width (or covering fraction) in the IR band with CLUMPY \citep{Ichikawa2015,Garcia-Bernete2019} are larger than those of X-ray results with XCLUMPY, supporting the significant polar dust emission \citep{Ogawa2021}.

Several works with other AGN models calculate the torus covering fractions in the IR band by using the conversion factor from the $L_{\rm torus}$/$L_{\rm AGN,int}$ \citep[e.g.,][]{Stalevski2016,Ichikawa2019a}.
However, considering that the polar dust is thought to be a hollow-cone structure \citep[e.g.,][]{Honig2019,Isbell2022}, it is difficult to calculate the true covering fractions of the polar dust assuming the uniform conical distribution \citep{Yang2020}.
In Figure~\ref{F21-Edd-CF}, we investigate the UV-to-IR luminosities relative to the intrinsic AGN disk luminosity ($L_{\rm AGN,int}$) for the torus and polar dust components, as a function of the Eddington ratio.
As noted above, the $L_{\rm torus}/L_{\rm AGN,int}$ ratios are smaller than those of polar dust, and both simple luminosity ratios are not well correlated with Eddington ratios.

We find that the AGNs with $L_{\rm polar}/L_{\rm AGN,int} > 0.5$ have the Eddington ratios of log$\lambda_{\rm Edd} \gtrsim -3.0$.
In the X-ray band, \citet{Fabian2008} proposed the $N_{\rm H}$--$\lambda_{\rm Edd}$ diagram to evaluate the presence of outflow, where dusty clouds are pushed away by radiation pressure against the gravitational force.
In our sample, the AGNs with higher Eddington ratios than the criteria show multi-scale outflows, such as UFO, ionized outflow, and molecular outflows \citep{Yamada2021}.
Recently, \citet{Venanzi2020} compute similar diagnostics for the IR-dominant outflows (i.e., polar dust).
When the AGN radiative acceleration and gravity are equal ($a_{\rm AGN} = a_{\rm g}$), the dusty outflows are pushed away for the AGNs with log$\lambda_{\rm Edd} \gtrsim -2.5$ assuming $N_{\rm H} \approx 10^{22}$~cm$^{-2}$ or log$\lambda_{\rm Edd} \gtrsim -3.0$ assuming $N_{\rm H} \approx 10^{21.5}$~cm$^{-2}$ \citep[see also][]{Alonso-Herrero2021}.
Thus, the decline of the polar dust luminosity for low-Eddington AGNs is consistent with the theoretical prediction of the IR-dominant dusty winds caused by the radiation pressure.

\subsection{Polar Dust Temperature} \label{sub6-4-Tpolar}
\begin{figure}
    \centering
    \includegraphics[keepaspectratio, scale=0.45]{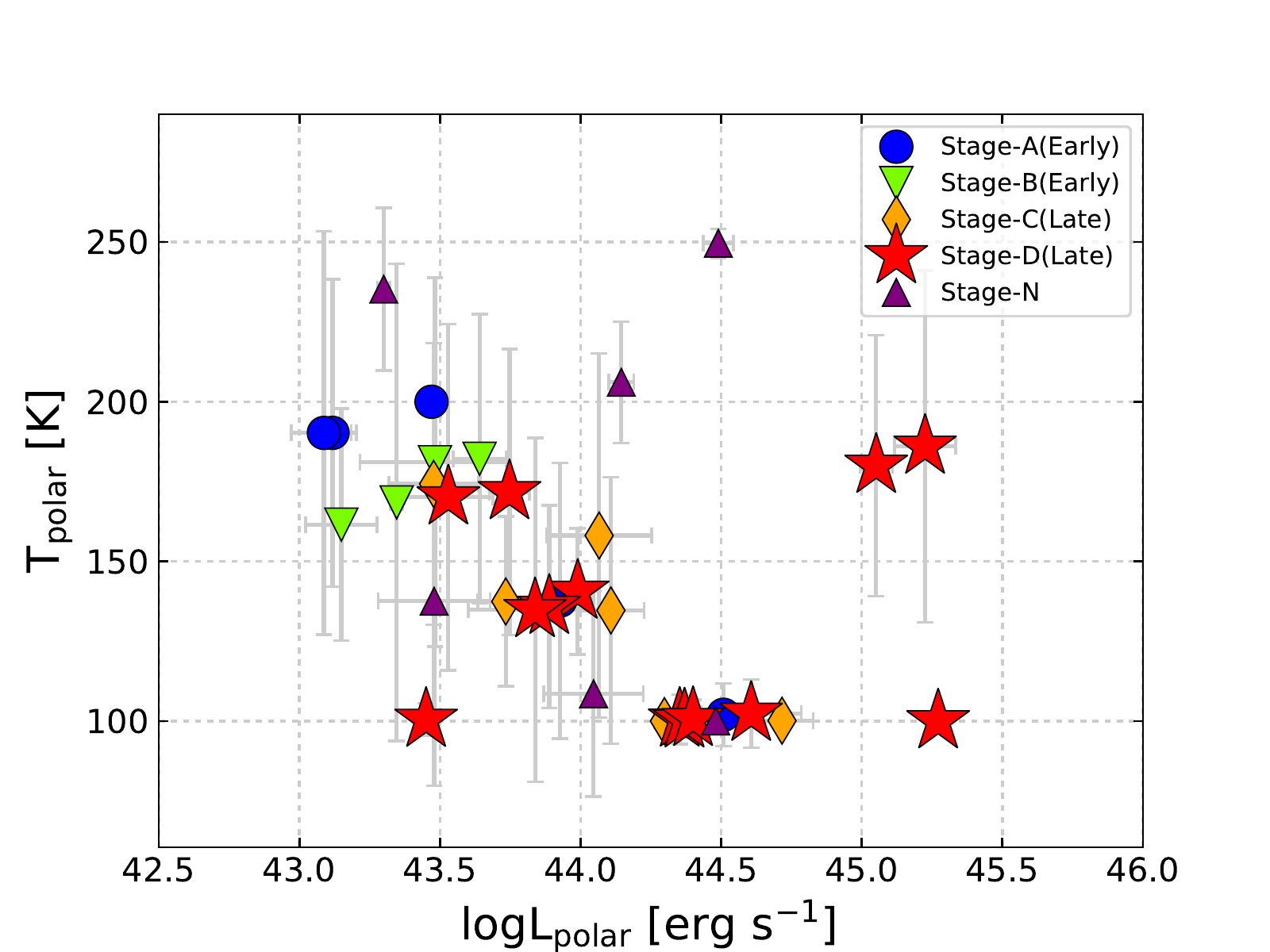}
    \caption{Logarithmic polar-dust luminosity vs. polar-dust temperature.
    The symbols are the same in Figure~\ref{F8-MS}.
    \\}
\label{F22-Tpo-Lpo}
\end{figure}

In this study, we newly constrain the polar dust temperature ($T_{\rm polar}$) for the AGNs in U/LIRGs, thanks to the combination of X-ray spectroscopy and multiwavelength SED analysis.
Figure~\ref{F22-Tpo-Lpo} provides the relation between the polar dust luminosity and its temperature.
For these AGNs with signs of the polar dust emission, we find that the averaged values of polar dust luminosity increase from early mergers (log($L_{\rm polar}/{\rm erg\ s^{-1}}) = 43.53\pm0.43$) to late mergers ($44.22\pm0.55$; see Table~\ref{T4_average}).
On the other hand, the polar dust temperature appears to decrease with merger stage.
Here, it is notable that the $T_{\rm polar}$ mostly lie in the range of $\sim$100--200~K, which are in good agreement with the researches of the IR interferometric observations and IR SED analysis as listed in Section~\ref{subsub4-2-3-polar}.
Even though the SED fitting with the range of $T_{\rm polar}$ within 100--300~K is examined, we confirm that the results are almost unchanged.

% Figure 22

% Figure 23
\begin{figure}
    \centering
    \includegraphics[keepaspectratio, scale=0.45]{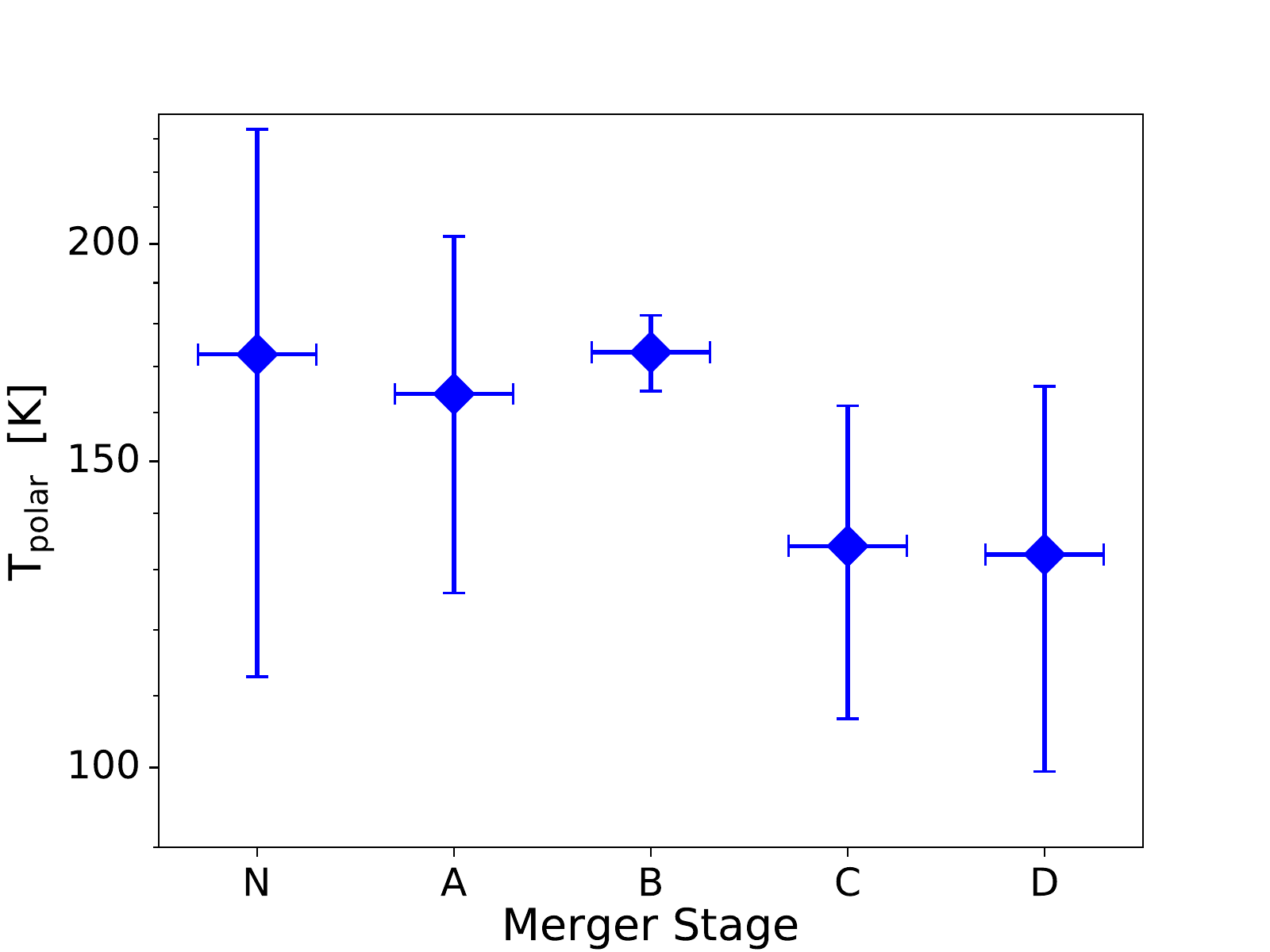}
    \caption{Differences in the polar-dust temperature with merger stage, described in the logarithmic scale.
    Diamonds and uncertainties represent the averaged value and 1$\sigma$ dispersions, respectively.
    \\}
\label{F23-Tpolar}
\end{figure}

To confirm the trend of the polar dust temperature, we also compare the averaged values with merger stages as illustrated in Figure~\ref{F23-Tpolar}.
Notably, the temperatures of late mergers ($\sim$130~K) decline from the values of early mergers or nonmergers ($\sim$170~K).
Although the allowed range is above 100~K, these averaged values are much larger than the dust temperature of the ISM in local U/LIRGs and high-$z$ submillimeter galaxies ($\sim$20--60~K; e.g., \citealt{Casey2012,Clements2018,da-Cunha2021}).
All AGNs in late mergers show significant signs of the presence of polar dust emission (Section~\ref{subsub4-2-4-BIC}).
Taking into account the high AGN luminosities (proportional to polar dust luminosity; Section~\ref{sub6-3-Lpolar}) in late mergers, the decline of the dust temperature within the inner parsecs seems to be unlikely.
The preferable reason is that the polar dust emission is radiated farther away from the center \citep[see also e.g.,][]{Lyu2018,Buat2021}, implying the development of the physical size of the polar dust structure.

\subsection{Polar Dust Structure} \label{sub6-5-polar-structure}
\subsubsection{Physical Size of the Polar Dust} \label{subsub6-5-1-polar-size}
Finally, we estimate the physical size of the polar dust for local U/LIRGs by using the polar dust temperature.
For a simple analytic model, the luminosity absorbed by dust grains at a distance ($r$) from the central radiation source, $L_{\rm abs}$, is calculated by the following equation \citep[see also e.g.,][]{Barvainis1987,Honig2011,Venanzi2020}:
\begin{align}
& L_{\rm abs} = 16 \pi r^2 Q_{\rm abs;P}(T) \sigma_{\rm SB} T^4,
\end{align}
where $Q_{\rm abs;P}(T)$ is the Planck mean absorption efficiency and $\sigma_{\rm SB}$ is Stefan–Boltzmann constant.
The polar dust model in our multiwavelength SED analysis assumes the dust emissivity of $\beta = 1.6$ \citep[e.g.,][]{Draine1984,Casey2012,Yang2020}, and the $Q_{\rm abs;P}(T)$ is proportional to $T^{1.6}$.
The typical scales of polar dust ($R_{\rm polar}$) can be estimated by solving the equation:
\begin{align}
& R_{\rm polar} = r_{\rm sub} \times (T_{\rm polar}/1500~{\rm K})^{-2.8}.
\end{align}
The $r_{\rm sub}$ is the sublimation radius at the temperature of $T_{\rm sub} = 1500$~K and can be calculated by $r_{\rm sub} = 0.4 \times (L_{\rm AGN,int}$/10$^{45}$~erg~s$^{-1}$)$^{1/2}$~pc in the CLUMPY model \citep{Nenkova2008a,Nenkova2008b}.
Table~\ref{TA5_results5} lists the estimates of $R_{\rm polar}$ for the U/LIRGs in our sample.

\begin{figure*}
    \centering
    \includegraphics[keepaspectratio, scale=0.85]{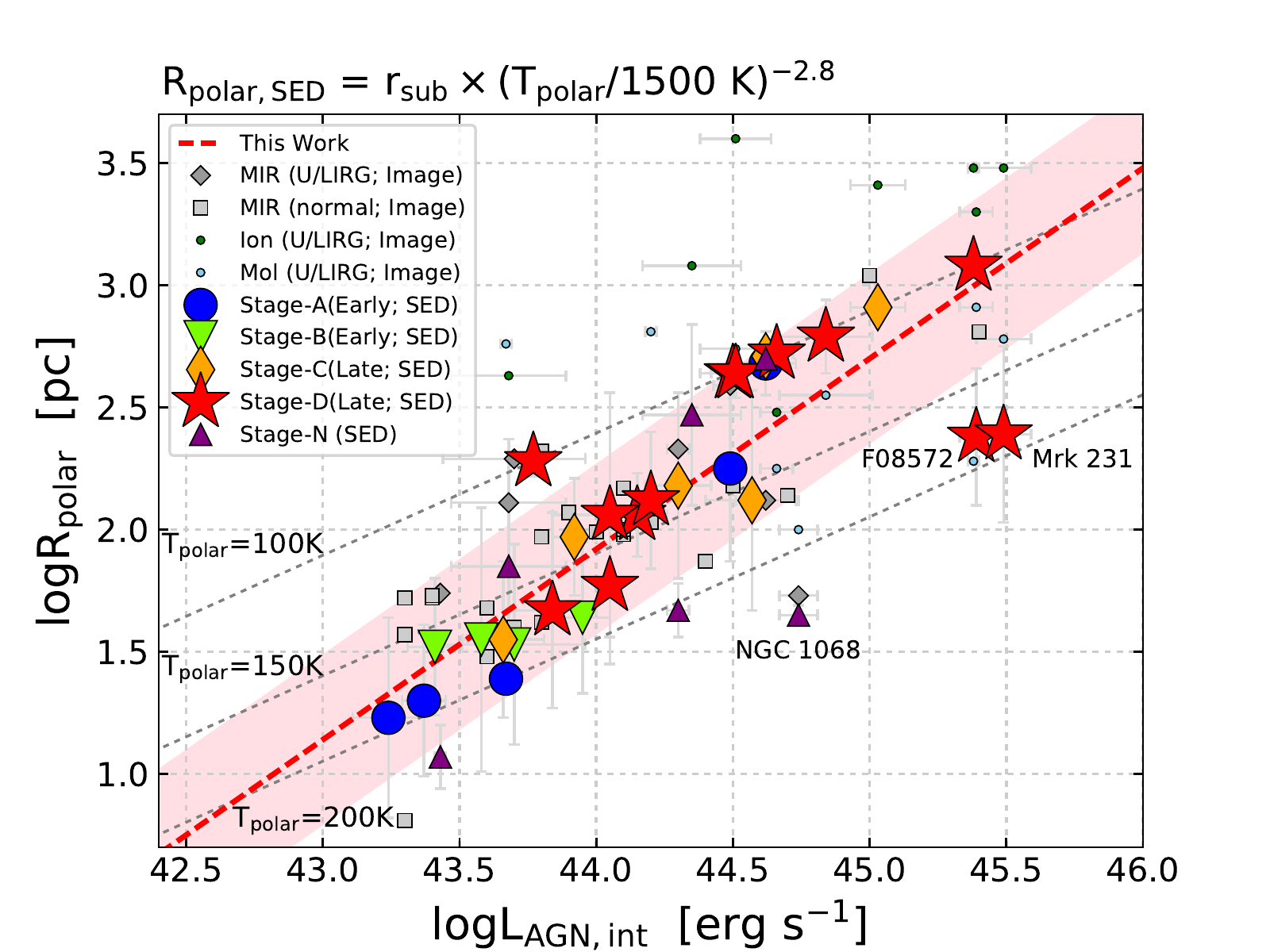}
    \includegraphics[keepaspectratio, scale=0.28]{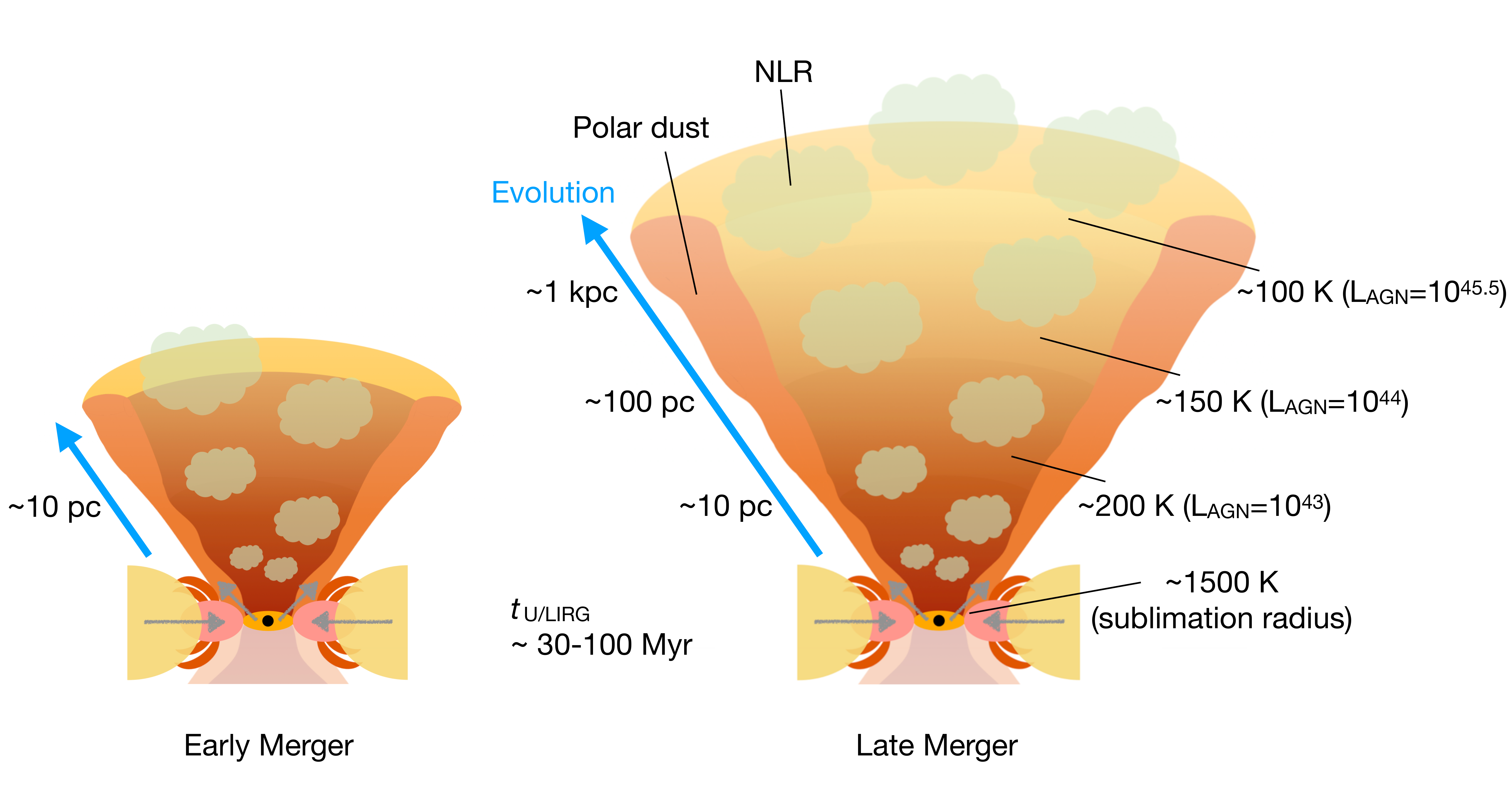}
    \caption{Top panel: logarithmic intrinsic AGN luminosity vs. logarithmic physical size of the dominant IR emission from the polar dust, estimated from the sublimation radius and polar dust temperature.
    The dashed red line and pink shade show the best-fitting linear relation and its 1$\sigma$ dispersion among the AGNs, respectively. 
    The green circles illustrate the physical size of ionized outflows measured with optical IFU observations, and light blue circles mark the size of molecular outflows measured with submillimeter observations in the subsample of our U/LIRGs \citep{Fluetsch2019,Fluetsch2021}.
    The gray diamonds and small circles denote the sizes of extended mid-IR emission for U/LIRGs and non-U/LIRG sources respectively \citep{Asmus2019}.
    For the size of molecular outflows and extended mid-IR region, the bolometric AGN luminosities of U/LIRGs are adopted to the estimates in this study, while those of non-U/LIRGs are calculated as the X-ray luminosity \citep{Asmus2019} times 20 \citep[e.g.,][]{Vasudevan2007}.
    The other symbols are the same in Figure~\ref{F8-MS}.
    Bottom panel: schematic picture of the AGN structure (polar dust, NLR, torus, failed winds, accretion disk, and SMBH) in U/LIRGs.
    \\}
\label{F24-polar-pc}
\end{figure*}

In the top panel of Figure~\ref{F24-polar-pc}, we present the relation between the intrinsic (bolometric) AGN disk luminosity and $R_{\rm polar}$.
The linear regression analysis in log–log space is performed for the targets by using the Bayesian maximum-likelihood method \citep{Kelly2007}.
The best-fitting relation (red dashed line) is
\begin{align}
& {\rm log}(R_{\rm polar}) = 0.78\ {\rm log}(L_{\rm AGN,int}/{\rm erg\ s}^{-1}) - 32.40.
\end{align}
Its 1$\sigma$ dispersion is $\pm$0.35~dex and the correlation coefficient is 0.80, implying a tight relation.

% Figure 24

It is notably that the radius of the polar dust emission measured by the high-spatial-resolution imaging in the mid-IR 12~$\mu$m band \citep{Asmus2019} for U/LIRGs (gray diamond) and normal galaxies (gray square) are well consistent with the best-fitting relation.
For NGC~1068, the compact radius of the mid-IR emission ($\sim$54~pc) relative to the AGN luminosity is in good agreement with our estimate ($R_{\rm polar} \sim 45$~pc).
The distribution of the $R_{\rm polar}$ obtained from our methods and by the mid-IR images \citep{Asmus2019} shows a steeper slope ($\sim$0.8) than the relationship with the constant temperature (0.5).
This supports that the typical polar dust temperatures decrease with AGN luminosity or physical size of the polar dust ($R_{\rm polar}$).

It should be emphasized that Equation (6) can be generally applied to an optically thin, continuous dust environment.
Whereas, the extinction of the polar dust in the $V$ band is not ($A^{\rm (LOS)}_{V{\rm ,polar}} \sim 1$--3; see Section~\ref{sub6-1-polar-dust}).
A plausible explanation of this discrepancy is that the polar dust consists of optically thin layers of otherwise optically thick dust clouds, where the temperature and emission profile is dominated by direct heating from the central source \citep{Honig2010,Honig2011}.\footnote{The polar dust temperature assuming the gray body model is different from the inverse peak-wavelength temperature ($T_{\rm peak}$) measured by Wein’s displacement law (i.e., $\lambda_{\rm peak} = 2.898 \times 10^3$~$\mu$m~K/$T_{\rm peak}$, which only applies to perfect black bodies; \citealt{Casey2012}).
Though the assumption of the emitter (e.g., gray body or smooth distribution with clumps) may cause a systematic uncertainty of the polar dust temperature, the results of the decrease in $T_{\rm polar}$ with merger stage will be almost unchanged.}
This is consistent with the radiative hydrodynamic (RHD) simulations expecting that the polar dust structure would be smoothly distributed with most substructure being filamentary and/or clumpy \citep{Wada2009,Wada2012,Schartmann2011,Dorodnitsyn2012,Chan2016,Chan2017,Dorodnitsyn2017,Williamson2019}.

If the polar dust structure is a bipolar hollow cone with the opening angle of $\theta_1$--$\theta_2$, its volume ($V_{\rm polar}$) is calculated by:
\begin{align}
V_{\rm polar} &\ = 2 \times (4/3)\pi r^3 \times 2\pi[1 - ({\rm cos}(\theta_2))]/4\pi \notag\\
 &\ \ \ \ -2 \times (4/3)\pi r^3 \times 2\pi[1 - ({\rm cos}(\theta_1))]/4\pi \notag\\
 &\ = (4/3)\pi r^3 [{\rm cos}(\theta_1) - {\rm cos}(\theta_2)].
\end{align}
The conical structure of the polar dust model in our study can be described as a simple case of $\theta_1 = 0\degr$ and $\theta_2 = 90\degr - \sigma$.
\citet{Lyu2018} introduce a power-law dust density profile, $\rho(r) \propto r^{-\alpha}$, and the value of $\alpha$ is $0 < \alpha \lesssim 2$ based on the previous observational studies \citep{Behar2009,Faucher-Giguere2012,Feruglio2015,Revalski2018}.
The total mass of the extended ($\gtrsim$10~pc) polar dust within the radius of $r$, $M_{\rm polar}(r)$, follows the mass profile of d$M_{\rm polar}$/d$r$ $\propto$ (d$V_{\rm polar}$/d$r$)$\rho(r) \propto r^{2-\alpha}$.
Assuming that the polar dust emission depends on their mass, a large part of the polar dust emission is derived from the large-scale structure at $r \sim R_{\rm polar}$.
However, the polar dust sizes for three AGNs in NGC~1068, IRAS F08572+3915, and Mrk~231 are much smaller than the best-fitting relation.
Interestingly, \citet{Feruglio2015} investigate the spatial distribution of the intense molecular CO(2$-$1) outflows in Mrk~231 with Atacama Large Millimeter/submillimeter Array (ALMA) and report that its mass profile (or the outflow filling factor) follows $\alpha \approx 2$.
This suggests that the polar dust emission in Mrk~231 is derived from the whole structure.
Although the spatial distribution for NGC~1068 and IRAS F08572+3915 is unclear, the density profile may cause the seemingly hot and compact emission from the polar dust.
Except for these three AGNs, the $R_{\rm polar}$ is thought to be the spatial scale of the polar dust detected in the mid-IR images \citep{Asmus2019}.
Thus, the best-fitting relation between AGN luminosity and $R_{\rm polar}$ suggests that the IR-luminous polar dust structure expands from a few tens parsec (early mergers) to kiloparsec scales (late mergers).

\subsubsection{Scenario of the Outflowing Polar Dust} \label{subsub6-5-2-polar-outflow}
To distinguish whether the polar dust is (1) the galactic ISM or dust near the edge of NLR being illuminated by the AGN or (2) the outflowing dusty winds launched from the inner edge of the torus, we investigate the spatial sizes of the ionized and molecular outflows for our U/LIRGs (Table~\ref{TA5_results5}).
\citet{Fluetsch2021} analyze the optical integral field unit (IFU) data of 26 local U/LIRGs with Very Large Telescope (VLT)/Multi Unit Spectroscopic Explorer (MUSE).
From the literature, they also include 31 galaxies with spatially resolved multiphase outflow information \citep[e.g.,][]{Rupke2013,Rupke2017}.
For eight U/LIRGs in our sample (small-green circles), they estimate the radius of the ionized outflow based on the extent of the broad H$\alpha$ component.
\citet{Fluetsch2019} investigate the molecular outflows primarily by using CO data from the ALMA archive in a sample of 45 local galaxies.
The radius of the molecular outflows is measured for nine U/LIRGs in our targets (light-blue circles).
We find that the sizes of ionized and molecular outflows are similar to or larger than those of the polar dust size, supporting that the materials outside the polar dust are outflowing.

Many works predict that the polar dusty outflows are launched from the surface of the inner torus \citep[e.g.,][]{Honig2010,Honig2012,Honig2013,Gallagher2015,Ishibashi2018}.
The semi-analytical disk and wind models suggest that the dusty winds can be launched by the AGN radiation pressure and the heated dust itself \citep{Venanzi2020}.
Similarly, RHD simulations also support that dusty winds are naturally driven \citep[e.g.,][]{Schartmann2014,Wada2016,Wada2018a,Williamson2019,Williamson2020}.
Some observational attempts to test this hypothesis have been performed by analyzing the IR SEDs \citep[e.g.,][]{Lyu2018} and by comparing the polar dust luminosities with the strength of the multiphase outflows \citep[e.g.,][]{Alonso-Herrero2021}. 

Since the polar dust emission does not show the emission/absorption lines, it is difficult to estimate its velocity.
For the starburst galaxy M~82, \citet{Yoshida2011} carry out the spectropolarimetry of the optical emission lines.
The outflowing dust grains are predicted to polarize the continuum and emission lines
emanating from the nuclear starburst region, acting as “moving mirrors” for nuclear light.
By comparing the velocities between the normal and polarized emission lines, they find that the velocity of the polar dust ($v_{\rm polar}$) decreases from $\sim$200~km~s$^{-1}$ at a few hundred parsecs to $\sim$20--30~km~s$^{-1}$ at 1--1.2~kpc.
For the U/LIRGs hosting nuclear starbursts and AGNs, the minimum velocity of the polar dust should be $\gtrsim30$~km~s$^{-1}$ at 1~kpc.
The typical lifetime of the U/LIRGs containing submillimeter galaxies at $z \sim 1$--2 ($t_{\rm U/LIRG}$) are $\sim$30--100~Myr \citep{Hopkins2006,Hopkins2008,Meier2010,Hickox2012,Inayoshi2018} or at most 300~Myr \citep{Swinbank2006,Meng2010,Privon2013}.
The migration length of the outflowing dusts can be roughly estimated by the velocity ($v_{\rm polar} \gtrsim 30$~km~s$^{-1}$) times the U/LIRG lifetime ($t_{\rm U/LIRG} \gtrsim 30$~Myr), i.e., $\gtrsim$1~kpc.
The smaller fraction of the AGNs with signs of polar dust emission for early mergers than that for late mergers also supports the evolution of the polar dust (Section~\ref{subsub4-2-4-BIC}).
Although the possible contribution of the polar dust emission from the non-outflowing dust in the NLR should not be rejected, the polar dust seems to be dusty winds as expected by recent simulations.
Thus, the polar dust size ($R_{\rm polar}$) increases from a few tens of parsec (early mergers) to kiloparsec scales (late mergers), indicating that the polar dust is likely the expanding (i.e., evolving) dusty outflows.

% MEMO:
% (1) ULIRG lifetime
% Hopkins2006, Hopkins2008, Inayoshi2018 : 30-100 Myr
% Meier2010: 100 Myr
% Meng2010 : 300 Myr
% Privon2013 : 100-300 Myr
%
% (2) high-z SMG lifetime (Ref: Bothwell et al. 2013, George eta l. 2014)
% Swinbank2006: 300 Myr
% Hickox2012: 100 Myr
%

% Figure 25
\begin{figure*}
    \centering
    \includegraphics[keepaspectratio, scale=0.45]{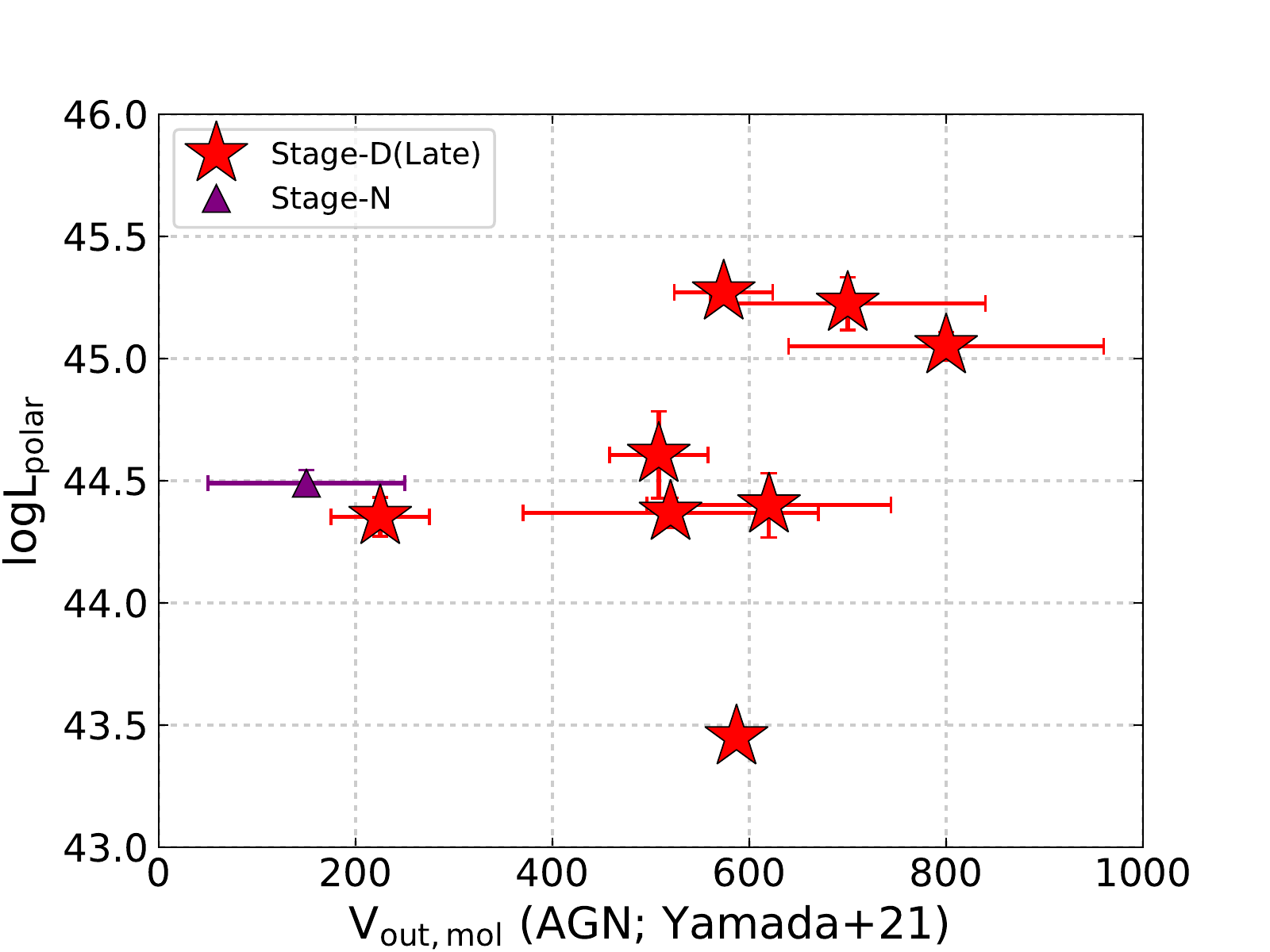}
    \includegraphics[keepaspectratio, scale=0.45]{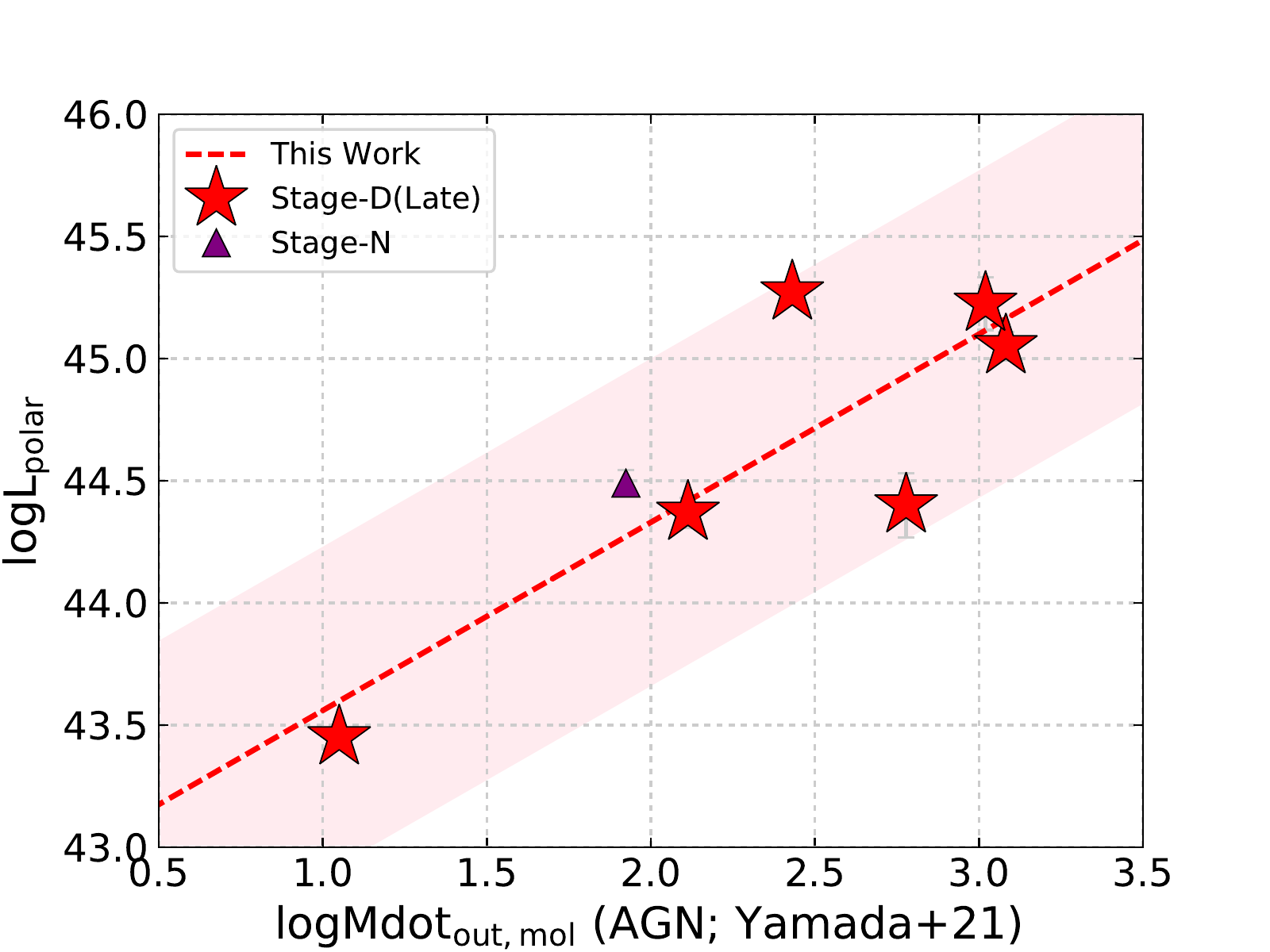}
    \caption{Polar-dust luminosity as a function of molecular outflow velocity (left) and logarithmic mass transfer rate of molecular outflow (right) referred from \citet{Yamada2021}.
    The red dashed line and pink shaded area mark the best-fitting relation and its 1$\sigma$ dispersion.
    The symbols are the same in Figure~\ref{F8-MS}.
    \\}
\label{F25-Molecular}
\end{figure*}

\subsubsection{Unified View of the Polar Dust Structure in U/LIRGs}
In the bottom panel of Figure~\ref{F24-polar-pc}, we present the schematic picture of the polar dust structure in U/LIRGs.
The polar dust sizes, estimated from the polar dust temperature and dust sublimation radius, increase with AGN luminosity (or merger stage).
The polar dust temperature decrease with the polar dust size from $\sim$200~K (a few tens of parsec) to $\sim$100~K (kiloparsec scale).
Considering the typical dust density profile of $0 < \alpha \lesssim 2$, their sizes corresponds to the outer structure of the extended polar dust (Section~\ref{subsub6-5-1-polar-size}).
The polar dust sizes are smaller than those of ionized and molecular outflows, and their expansion with merger stage can be explained by the polar dust velocity and U/LIRG lifetime (Section~\ref{subsub6-5-2-polar-outflow}).

At the parsec-scale regions, RHD simulations show that the innermost torus structure is formed by radiation-driven fountain-like outflows \citep[e.g.,][]{Wada2018a}.
A multiphase dynamic nature of torus (pink ellipses) and circumnuclear region (CND; yellow ocher) in the $r \lesssim$~tens of parsecs torus region is confirmed by recent ALMA observations \citep[e.g.,][]{Izumi2016,Izumi2018c}.
\citet{Honig2019} mentions that the polar dust is expected to be launched by the inner edge of the torus due to the AGN radiation (coming from the AGN disk or vicinity of the SMBH). 
As noted in Section~\ref{subsub6-5-1-polar-size}, the polar dust structure would be smoothly distributed with most substructure being filamentary and/or clumpy, as expected by RHD simulations \citep{Wada2012,Schartmann2011,Williamson2019}, although we simply describe it by a smooth distribution.
According to these results, the schematic picture illustrates the AGN structures, containing the polar dust, NLR, torus, failed winds around the torus, accretion disk, and SMBH in the early-to-late merging U/LIRGs.

\subsubsection{Polar Dust vs. Molecular Outflow}
This study indicates that the multiwavelength SED analysis is one of the key means to evaluate the activities of the dust components of large-scale outflows.
In Figure~\ref{F25-Molecular}, we compare the polar dust luminosities with the molecular outflow velocity ($V_{\rm out,mol}$; left panel) and mass transfer rates ($\dot{M}_{\rm out,mol}$; right panel).
\citet{Yamada2021} lists the values of $V_{\rm out,mol}$ and $\dot{M}_{\rm out,mol}$ for local U/LIRGs by referring the previous works based on the broad CO emission lines and OH absorption lines (e.g., \citealt{Gonzalez-Alfonso2017,Laha2018} and references therein).
We find that the polar dust luminosity and $V_{\rm out,mol}$ show no significant relation.
This may be due to the dispersion of the slope of the dust density profiles ($\alpha$), or the dispersion of the size of the molecular outflows relative to the polar dust size (Figure~\ref{F24-polar-pc}).
Whereas, the polar dust luminosity and mass transfer rates of the molecular outflows show a positive correlation.
This may be a reasonable result by taking into account that the polar dust luminosities depend on their mass (particularly containing the large-scale mass at $r \sim R_{\rm polar}$; Section~\ref{subsub6-5-1-polar-size}).
We calculate the best-fitting relation by using the Bayesian maximum-likelihood method as below:
\begin{align}
& {\rm log}(L_{\rm polar}) = 0.77\ {\rm log}(\dot{M}_{\rm out,mol}) + 42.79,
\end{align}
where the 1$\sigma$ dispersion is $\pm$0.67~dex and the correlation coefficient is 0.71.
Since the sample is quite limited, further studies using large samples are necessary to establish the general relation between polar dust (i.e., dusty winds) and molecular outflows.
\\

\section{Discussion II: Coevolution Process of Galaxy, SMBH, and Outflow} \label{S7-discussion2}
\subsection{Activities of Starburst and AGN} \label{sub7-1-growth-rate}
To understand the origin of the coevolution of galaxies and SMBHs, their growth rates are well studied.
Numerical simulations of galaxy mergers predict that the SFR, AGN luminosities, and obscuration of the central AGN increase with merger stage \citep[e.g.,][]{Di-Matteo2005,Hopkins2006,Hopkins2008,Narayanan2010,Angles-Alcazar2017,Blecha2018,Kawaguchi2020,Yutani2022}.
The observational studies using multiwavelength data support their increases with merger stage \citep[e.g.,][]{Imanishi2010,Lee2012,Ellison2013,Satyapal2014,Ricci2017bMNRAS,Ricci2021b,Koss2018,Pfeifle2019,Shangguan2019,Yamada2019,Yamada2021}.

\begin{figure*}
    \centering
    \includegraphics[keepaspectratio, scale=0.5]{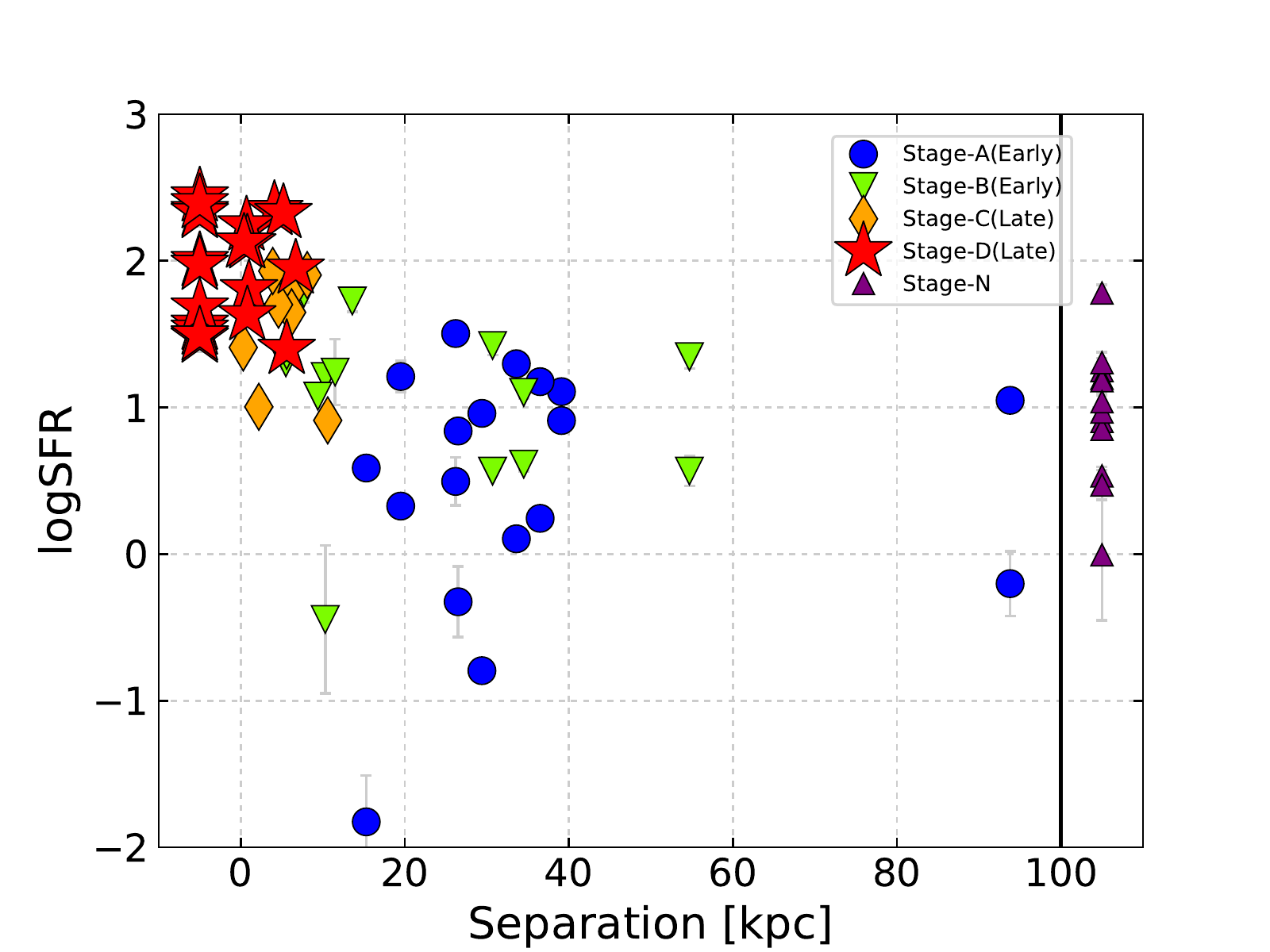}
    \includegraphics[keepaspectratio, scale=0.5]{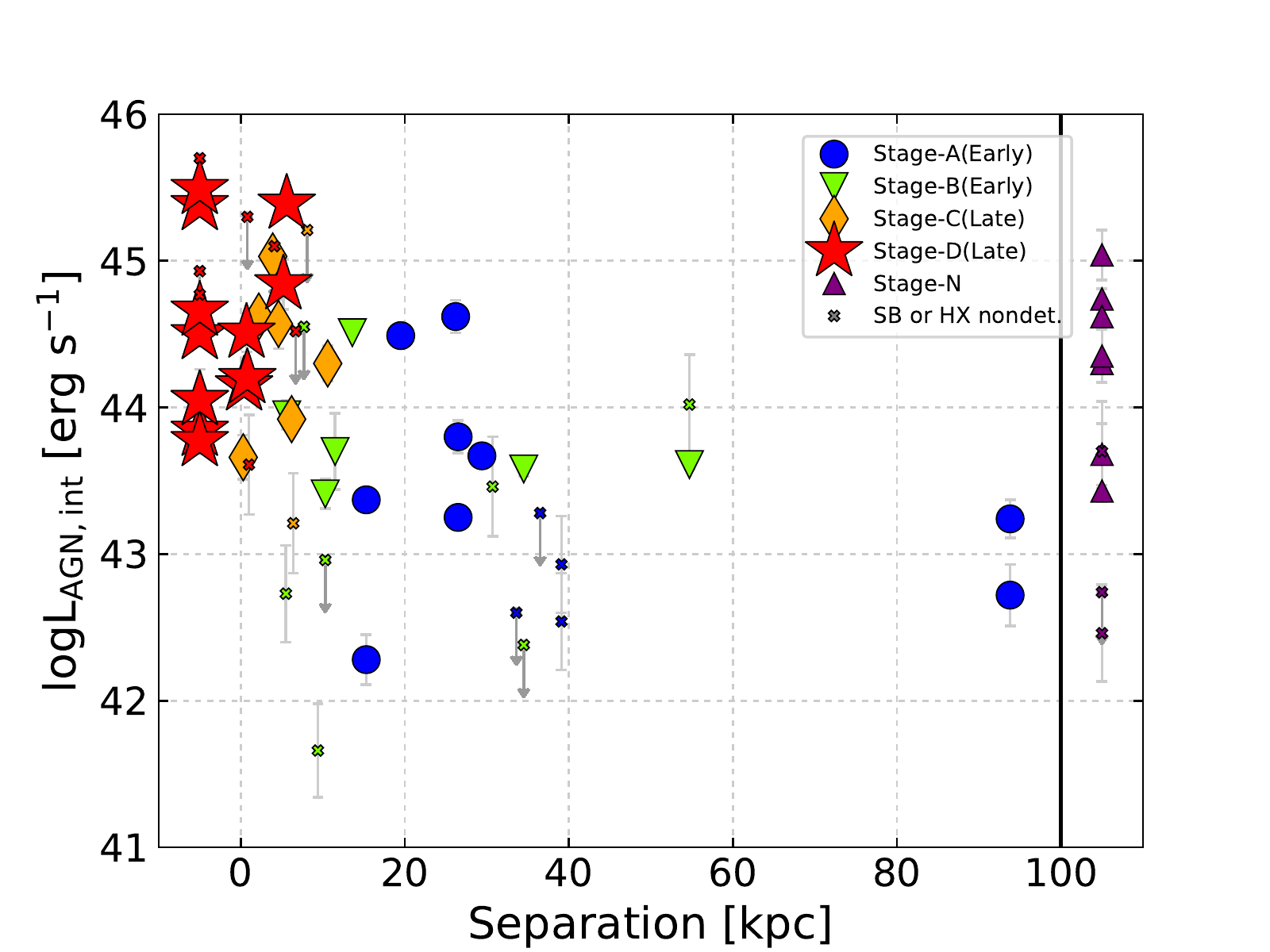}
    \caption{SFR and intrinsic AGN luminosity as a function of projected separation between two galaxies in units of kiloparsecs.
    Single nuclei in merging (stage D) and nonmerging (stage N) sources are plotted on the left (negative value) and right sides, respectively.
    Large symbols are the same in Figure~\ref{F8-MS}, while small crosses illustrate the starburst-dominant or hard X-ray--nondetected sources.
    \\}
\label{F26-separation}
\end{figure*}

By combining the broadband X-ray spectroscopy and multiwavelength SED decomposition, this study estimates the SFRs and AGN luminosities for a large sample of 72 resolved sources in U/LIRGs, which are treated separately divided into 41 hard X-ray--detected AGNs (containing 36 single AGNs, 2 unresolved dual AGN systems, and 3 newly identified AGNs) and 31 other sources (starburst-dominant or hard X-ray--undetected sources).
In Figure~\ref{F26-separation}, we investigate the distribution of SFR (left) and intrinsic (bolometric) AGN luminosity (right) as a function of projected separation between two galaxies.
For the starburst-dominant or hard X-ray--undetected sources, we refer to the AGN luminosities estimated from the [\ion{O}{4}] 25.89~$\mu$m luminosity ($L_{\rm [O\ IV]}$; small crosses) in \citet{Yamada2021}.
They confirm that these values are consistent with the 3$\sigma$ upper limits of the predicted 2--10~keV AGN luminosity from the X-ray counts by assuming $N_{\rm H} \leq 10^{25}$~cm$^{-2}$ and $L_{\rm X,unabs}$--$L_{\rm [O\ IV]}$ relation \citep{LiuTeng2014}.
We note that the dispersion of the low values of SFRs and AGN luminosities for the early mergers with the $\sim$10--40~kpc are caused by the faint companion galaxies that are not IR-luminous galaxies.
Focusing on the largest values of individual merger stages, our results support that the SFRs and AGN luminosities increase with projected separation and merger stage, consistent with the results in Section~\ref{sub5-6-summary-SED}.

% Figure 26

% Figure 27
\begin{figure*}
    \centering
    \includegraphics[keepaspectratio, scale=0.8]{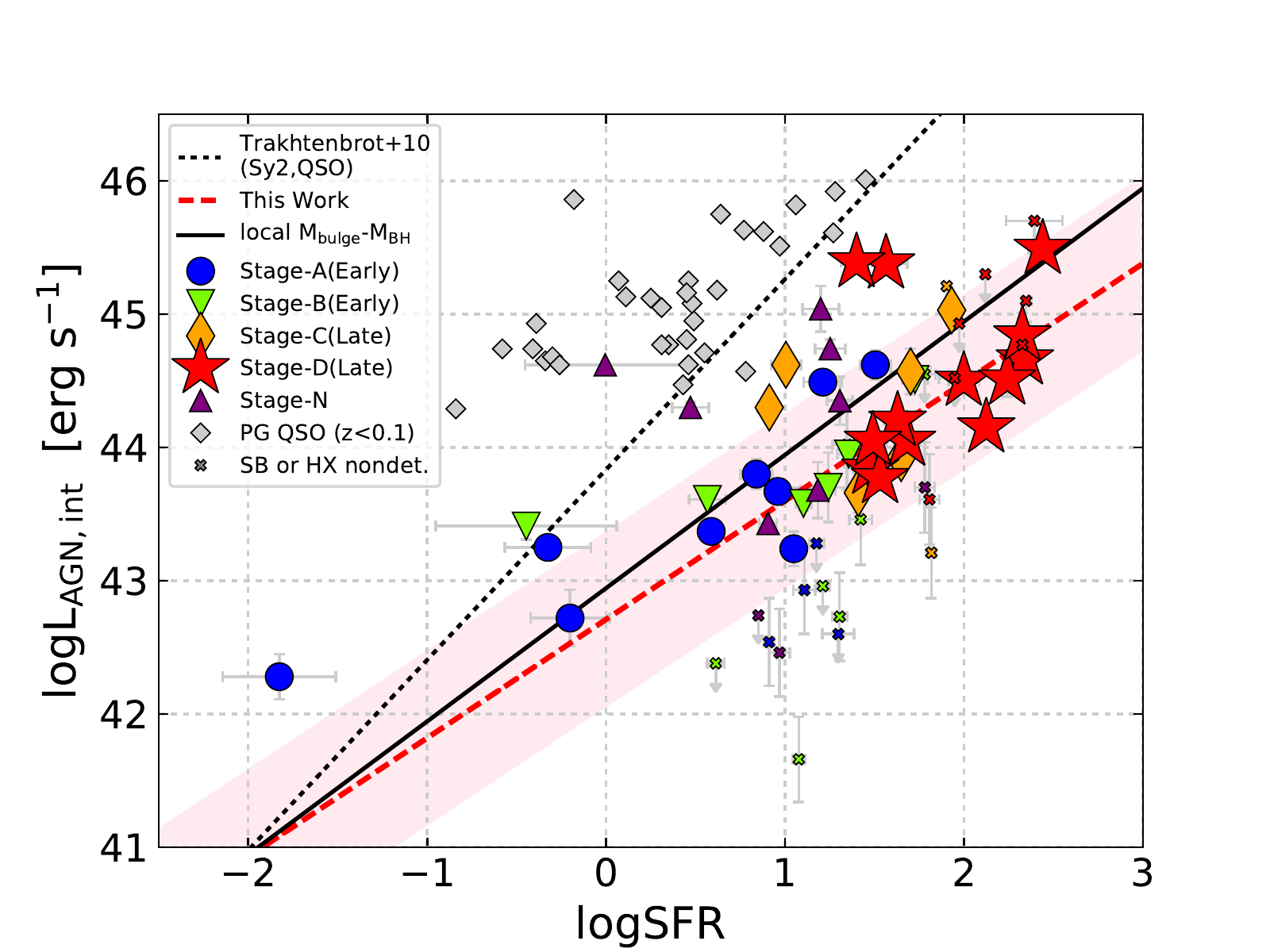}
    \caption{Logarithmic SFR vs. intrinsic AGN luminosity.
    Black dotted line denotes the typical relation among low-z Seyfert 2s, QUEST QSOs, and high-z QSOs \citep{Trakhtenbrot2010}.
    The red dashed line and pink shaded area mark the best-fitting relation and its 1$\sigma$ dispersion for the AGNs in U/LIRGs, excluding the nonmergers and AGNs with logSFR $<$ 0.
    The black solid line is the galaxy--SMBH ``simultaneous evolution'' line for $A = 200$ \citep[see also][]{Ueda2018}.
    The large symbols are the same in Figure~\ref{F8-MS}.
    The small crosses mark the starburst-dominant or hard X-ray--nondetected sources, whose $L_{\rm AGN,int}$ values are derived from the [\ion{O}{4}] luminosities \citep{Yamada2021}.
    \\}
\label{F27-SFR-AGN}
\end{figure*}

In Figure~\ref{F27-SFR-AGN}, we compare the SFRs and bolometric AGN luminosities in our sample.
By the same method in \citet{Yamada2021}, we draw the galaxy--SMBH ``simultaneous evolution'' relation (see also e.g., \citealt{Ueda2018}), where the growing systems keeping the SFR--$L_{\rm AGN,int}$ relation are expected to establish the local $M_{\rm bulge}$--$M_{\rm BH}$ relation.
Here, the relation assumes the fraction of stellar mass that are taken back to the ISM (return fraction) as $R = 0.41$ \citep{Chabrier2003}, the ratio of stellar masses to SMBH masses in the local universe as $A \sim 200$ \citep{Kormendy2013}, and a radiative efficiency as $\eta = 0.05$ \citep{Ueda2014}.
The detailed explanations are presented in \citet{Yamada2021}.
The relation is described as
\begin{align}
& {\rm log}(L_{\rm AGN,int}) = {\rm log}({\rm SFR}/M_{\odot}\ {\rm yr}^{-1}) + 42.94.
\end{align}

The hard X-ray--detected and newly identified AGNs in U/LIRGs follows the simultaneous coevolution relation, consistent with the results by \citet{Yamada2021}.
For the AGNs in our sample excluding nonmergers and outliers with ${\rm logSFR} < 0$, we calculate the best-fitting relation based on the Bayesian maximum-likelihood method:
\begin{align}
& {\rm log}(L_{\rm AGN,int}) = 0.89\ {\rm log}({\rm SFR}) + 42.71.
\end{align}
The 1$\sigma$ uncertainty is $\pm$0.66~dex and the correlation coefficient is 0.54.
As mentioned in Appendix~\ref{AppendixC-previous-works} (see Figure~\ref{FC3-comp-LAGN-L12}), the AGN luminosities in \citet{Yamada2021} are larger than the values in this work, and then their typical relation is $\sim$0.4~dex brighter in the AGN luminosity.
Their AGN luminosities are derived from the averages of four different measurements; the [\ion{O}{4}] luminosity \citep[e.g.,][]{Gruppioni2016}, bolometric AGN fraction \citep[e.g.,][]{Diaz-Santos2017}, and IR SED decomposition and Spitzer/IRS spectral
fitting \citep{Alonso-Herrero2012,Shangguan2019}.
Its systematic scatter related to the application of the averaged values is reported as about $\pm$0.27 dex.
Our best-fitting relation with the 1$\sigma$ uncertainty is overlapped on not only the relation of \citet{Yamada2021} but the simultaneous coevolution relation (black solid line). 
This supports that the AGNs in U/LIRGs are exactly in the coevolution phase of galaxies and SMBHs.

% Figure 28
\begin{figure*}
    \centering
    \includegraphics[keepaspectratio, scale=0.82]{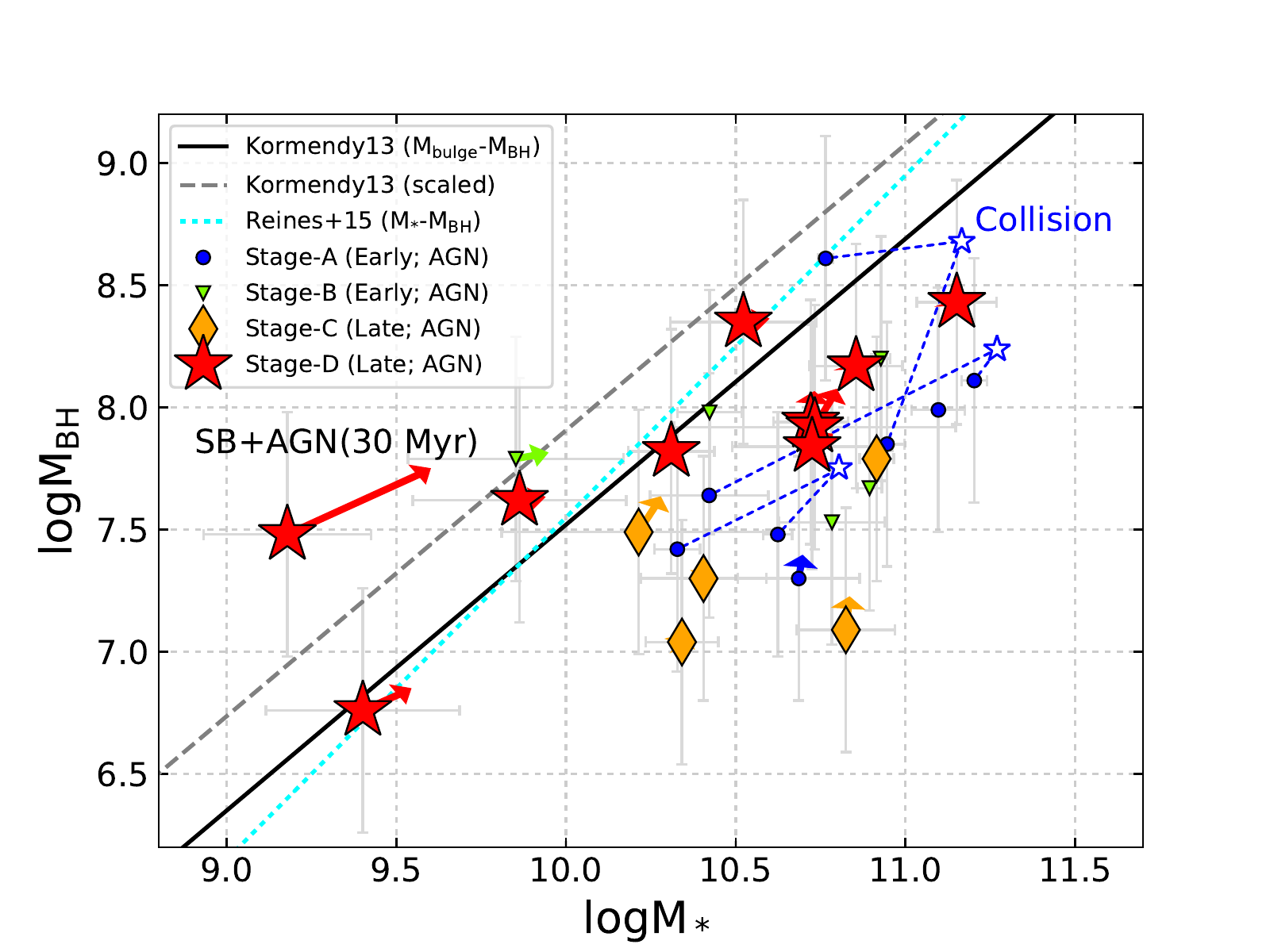}
    \caption{Logarithmic stellar mass vs. logarithmic SMBH mass referred from \citet{Yamada2021} for the AGNs in stage C (orange diamond) and stage D (red star) U/LIRGs.
    Solid arrows represent the expected masses of the galaxy and SMBH if they evolve with the constant growth rates (SFR and $\dot{M}_{\rm BH}$) for 30 Myr.
    For three dual-AGN systems in stage A mergers, the merged masses of the two sources including the mass increase for 30~Myr are marked as empty blue stars with dotted lines.
    The black solid line denotes the local bulge mass ($M_{\rm bulge}$) and SMBH mass ($M_{\rm BH}$) relation for classical bulges and elliptical galaxies \citep{Kormendy2013}.
    The dashed line shows the scaled relation that has bulge masses scaled down by 0.33 dex to consider the difference in the adopted mass-to-light ratios, 
    and blue dotted line presents the typical $M_*$--$M_{\rm BH}$ relation for classical bulges and ellipticals \citep{Reines2015}.
    \\}
\label{F28-Ms-MBH}
\end{figure*}

\citet{Yamada2021} report that the AGNs in stage D mergers show the multiphase massive outflows at subparsec to kiloparsec scales, that is, 
UFOs \citep[e.g.,][]{Feruglio2015,Tombesi2015,Mizumoto2019,Smith2019}, 
ionized outflows \citep[e.g.,][]{Fischer2013,Rich2015,Kakkad2018,Fluetsch2021,toba2022b}, and molecular outflows with velocities above 500~km~s$^{-1}$ \citep[e.g.,][]{Gonzalez-Alfonso2017,Laha2018,Fluetsch2019}.
As discussed in Section~\ref{sub6-3-Lpolar} and \ref{sub6-5-polar-structure}, the polar dust luminosities are $\sim2/3 \times L_{\rm AGN,int}$, and the physical sizes increase with AGN luminosities from a few tens of parsec to kiloparsec scales.
In short, the polar dust luminosities and sizes are the largest in the stage D mergers with high AGN luminosities.
Moreover, we plot the quantities of PG quasars at $z \lesssim 0.1$ \citep{Lyu2017} and empirical relation for local AGNs and quasars \citep{Trakhtenbrot2010}.
They lie in the AGN-dominant region above the coevolution relation, which should have strong gas outflows \citep[e.g.,][]{Woo2020}.
The coexistence of intense starbursts, luminous AGNs, and massive outflows (UFOs, ionized outflows, molecular outflows, and dusty winds) particularly in the phase of U/LIRGs support a standard AGN feedback scenario that the AGN-driven outflows suppress star-forming activities and ULIRGs eventually transit to the unobscured quasars (e.g., \citealt{Di-Matteo2005,Hopkins2008,Alexander2012,Ananna2022}; see also \citealt{Yamada2021}).

\subsection{Mass Growth of Galaxy and SMBH} \label{sub7-2-Mass}
According to the obtained SFRs and AGN luminosities, we evaluate the total growth masses of galaxies and SMBHs.
The increase in the stellar mass ($\Delta M_*$) can be calculated by the SFR times the growth timescale ($t_{\rm growth}$).
The mass accretion rates ($\dot{M}_{\rm BH}$) are estimated by using a radiative efficiency ($\eta = 0.05$; \citealt{Ueda2014}) and the light speed ($c$) as
\begin{align}
& \dot{M}_{\rm BH} = L_{\rm AGN,int} \times (1-\eta)/(\eta c^2).
\end{align}
The increase in the SMBH mass ($\Delta M_{\rm BH}$) corresponds to the $\dot{M}_{\rm BH}$ times $t_{\rm growth}$.
Here, the multiwavelength SED analysis presents the SFRs and AGN luminosities for the U/LIRGs in the various merger stages.
Since the history of these quantities during the U/LIRG lifetime ($t_{\rm U/LIRG} \sim 30$--100~Myr; Section~\ref{subsub6-5-2-polar-outflow}) is unclear, we compute the increased masses assuming $t_{\rm growth}$ is 30~Myr as a time scale of early mergers or late mergers.
For the three resolved dual AGN systems in stage A mergers (NGC 833/NGC 835, NGC 6921/MCG+04-48-002, and NGC 7679/NGC 7682), we also calculate the total mass of the two galaxies and SMBHs after the galaxy collision, containing the increased masses from their starburst and AGN activities for 30~Myr.

Figure~\ref{F28-Ms-MBH} describes the relation between the stellar masses and SMBH masses for the hard X-ray--detected AGNs in our sample.
We plot the local $M_{\rm bulge}$--$M_{\rm BH}$ relation \citep{Kormendy2013}, scaled relation assuming Chabrier IMF, and a typical $M_*$--$M_{\rm BH}$ relation for classical bulges and ellipticals \citep{Reines2015}.
The stellar masses are obtained by our SED fitting, while the SMBH masses are referred from \citet{Yamada2021}.
The distributions of these masses in our U/LIRGs at $z < 0.1$ are well consistent with the study of $z < 0.3$ U/LIRGs by \citet{Farrah2022}.
The caveat is that the SMBH masses we refer are derived from the averaged values of the four kinds of different measurements: stellar mass to SMBH mass relation \citep[e.g.,][]{Reines2015}, $M$--$\sigma_*$ relation \citep[e.g.,][]{Gultekin2009}, photometric bulge luminosity \citep[e.g.,][]{Veilleux2009c,Winter2009,Haan2011a}, and various other methods such as the flux density of old stellar emission at 2~$\mu$m \citep{Caramete2010}, velocity dispersion of [\ion{O}{3}] emission \citep{Alonso-Herrero2012}, water masers \citep{Lodato2003,Klockner2004,Izumi2016}, and so on.
Although the systematic differences appear to depend on neither the intrinsic AGN
luminosity nor Eddington ratio \citep{Yamada2021}, systematic uncertainties of the averaged SMBH masses are not negligible (at most $\pm$0.5~dex; gray error bars).
Thus, the figure does not necessarily indicate that the SMBH masses are a bit small relative to the local relations.

%Figure 29
\begin{figure*}[t]
    \centering
    \includegraphics[keepaspectratio, scale=0.9]{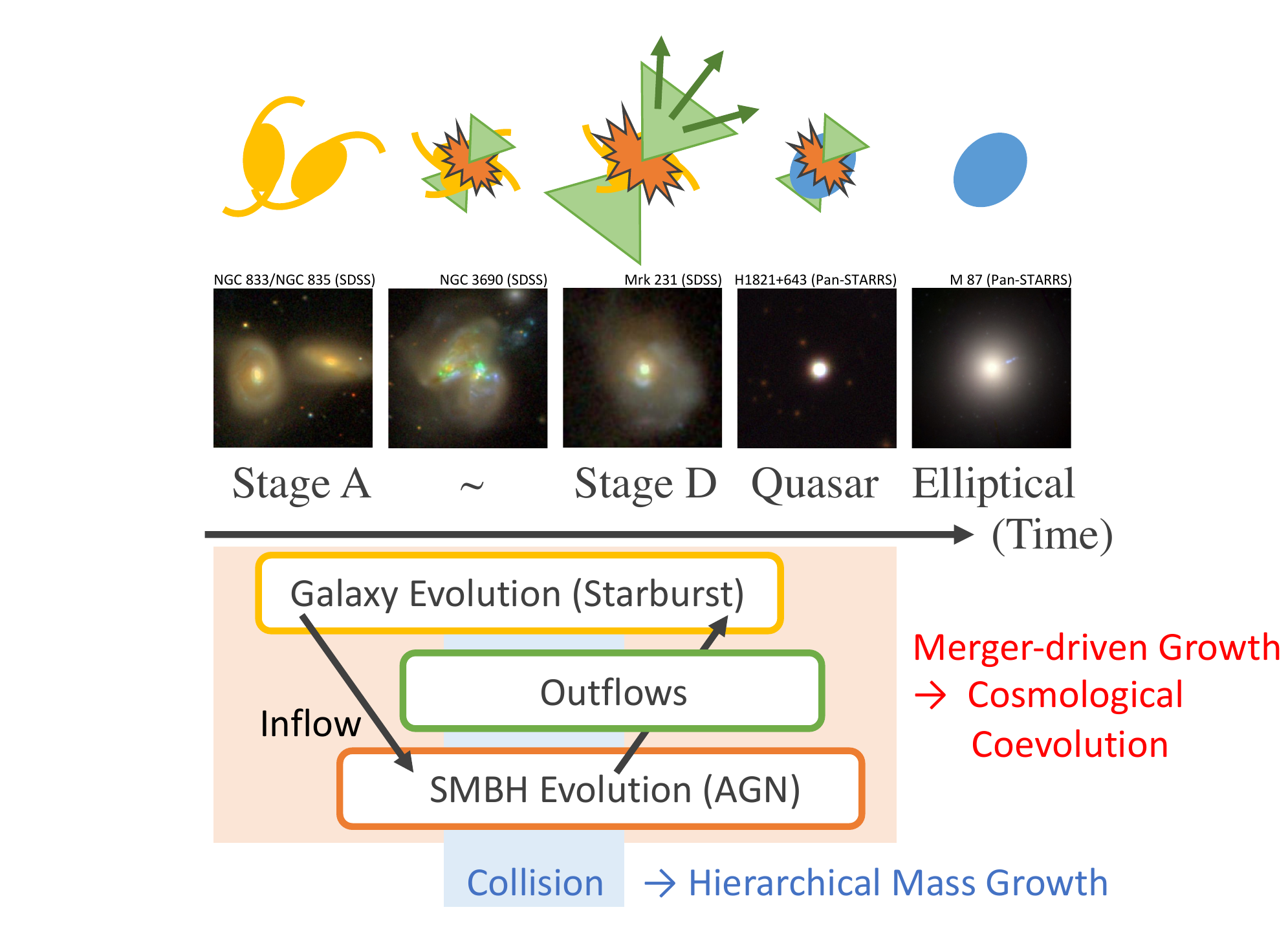}
    \caption{Schematic picture of merger-driven coevolution of galaxies and SMBHs.
    Middle panels are compiled from Pan-STARRS DR2 and SDSS DR16 images (Section~\ref{sub3-3-optical}), whose sources are merging U/LIRGs (NGC~833/NGC~835, NGC 3690, Mrk 231) and post-merger candidates (a quasar H1821+643, e.g., \citealt{Jadhav2021}; and an elliptical galaxy M87, e.g.,  \citealt{Longobardi2015}).
    \\}
\label{F29-schematic-image}
\end{figure*}

We overplot the expected masses due to the starbursts and AGN activities for 30~Myr (arrows) and the additional collision of two galaxies (empty stars).
For the two stage-D mergers with low stellar masses of log$(M_*/M_{\odot}) \sim 9.0$--9.5 (IRAS F08572+3915 and IRAS F17138$-$1017), their galaxies and SMBHs are rapidly growing thanks to the starburst and AGN activities.
We find that, however, the increased masses for the other U/LIRGs are small by these activities even though some of the stage-D U/LIRGs have the highest SFRs ($\sim$100--300~$M_{\odot}$~yr$^{-1}$) and AGN luminosities (log$L_{\rm AGN,int} \sim 45.0$--45.5).
Notably, the galaxy collision causes rapid growth for these high-mass systems.

Soltan’s argument \citep{Soltan1982}, one of the well-known
discussions on the cosmic growth of SMBHs, suggests that the SMBHs are thought to have grown primarily by the gas accretion, not SMBH--SMBH coalescence.
This requirement is imposed to explain the SMBH mass density at redshift $z$ and the accreting gas mass density from $z = \infty$ to 0 \citep{Yu2002,Shankar2004}.
Because the SMBH--SMBH coalescence conserves the SMBH mass density, the argument does not reject the presence of ubiquitous merger events \citep[e.g.,][]{Enoki2004,Jahnke2011}.
In this context, the growth events keeping the balanced SFR--$L_{\rm AGN,int}$ relation triggered by galaxy mergers (see Figure~\ref{F27-SFR-AGN}; see also e.g., \citealt{Di-Matteo2005}) should be helpful to construct the local $M_{\rm bulge}$--$M_{\rm BH}$ relation.
Whereas we need to keep in mind that, focusing on individual galaxies, the galaxy collision generates such high-mass galaxies with log$(M_*/M_{\odot} \gtrsim 10$).
Therefore, our results indicate that galaxy merger is a vital process to establish the local $M_{\rm bulge}$--$M_{\rm BH}$ relation and build up the high-mass galaxies such as local elliptical galaxies.

\subsection{Perspective of the Galaxy--SMBH Coevolution in Mergers} \label{sub7-3-Perspective}
Finally, we summarize the perspective of the merger-driven coevolution of galaxies and SMBHs as described in the schematic picture of Figure~\ref{F29-schematic-image}.
The SFRs and intrinsic (bolometric) AGN luminosities increase with merger stage, where the galaxy growth occurs first and rapid mass accretion delivered from the host galaxy to the SMBH is triggered later (Section~\ref{sub7-1-growth-rate}).
The merger process in the SFR--$L_{\rm AGN,int}$ diagram should be a key mechanism to drive the balanced growth of galaxies and SMBHs in the cosmic history \citep[e.g.,][]{Madau2014,Ueda2014,Aird2015} and then to establish the local $M_{\rm bulge}$--$M_{\rm BH}$ relation \citep[e.g.,][]{Kormendy2013}.
Moreover, galaxy collision is another important process to induce hierarchical mass growth and build up the high mass system such as local elliptical galaxies (Section~\ref{sub7-2-Mass}).

\citet{Yamada2021} find that the AGNs in stage D mergers show the multiphase outflows at subparsec to kiloparsec scales, e.g., UFOs, ionized outflows, and molecular outflows.
They also unveil that the AGNs with high Eddington ratios as $\lambda_{\rm Edd} \sim 1$  have moderate $N_{\rm H}$ ($\sim$10$^{23}$~cm$^{-2}$), where dusty clouds at parsec scales are pushed
away by radiation pressure against the gravitational force as predicted with the $N_{\rm H}$--$\lambda_{\rm Edd}$ diagram \citep[e.g.,][]{Fabian2008,Fabian2009,Ricci2017cNature,Ricci2022}.
The recent high-spatial-resolution ALMA and Chandra observations support the feedback effects on the torus by evaporating a portion of the gas \citep{Kawamuro2021}, while the AGNs in U/LIRGs are deeply buried by massive inflows and/or outflows \citep{Yamada2021}.
The X-ray weakness in U/LIRGs should make it easy to launch the massive outflows efficiently (Section~\ref{sub5-5-X-weak}).
By combining the X-ray spectroscopy and multiwavelength SED fitting, we constrain the physical parameters of polar dust structure (e.g., IR luminosity, temperature, and spatial scales) that is thought to be launched by the AGN radiation pressure and the heated dust itself (Section~\ref{sub6-5-polar-structure}; see also \citealt{Venanzi2020,Alonso-Herrero2021}).
The polar dust emission may be negligible for some low-Eddington AGNs ($\lambda_{\rm Edd} \lesssim 10^{-3}$) in early mergers (Section~\ref{subsub4-2-4-BIC} and~\ref{sub6-3-Lpolar}).
For the AGNs showing significant polar dust emission, the physical size of polar dust increases with merger stage from a few tens of parsec to kiloparsec scales.
These results indicate that the massive outflows connecting to galaxy scales are nurtured by the rapid accretion SMBHs and rich environments in U/LIRGs.

The coexistence of intense starburst, AGNs, and large-scale massive outflows in local U/LIRGs support a standard AGN feedback scenario \citep[see also][]{Chen2020a}. According to the scenario, the U/LIRGs will transit to the unobscured AGNs (Section~\ref{sub7-1-growth-rate}).
It is still unclear whether the star-forming activities are quenched due to the high specific SFRs (see e.g., Section~\ref{sub5-1-MS}) or the AGN feedback by the massive outflows.
Considering the balanced starburst and AGN activities in the merger sequence, the massive outflows would present some kind of physical connection between galaxies and SMBHs.
To completely understand the role of massive outflows in late mergers, it is needed to more researches of the mass transport mechanism and the effects on star formation in the host galaxy.
In the near future, the high-spatial-resolution IR images will be presented by the Mid-InfraRed Instrument (MIRI; \citealt{Rieke2015,Wright2015}) onboard the James Webb Space Telescope (5--28~$\mu$m), and The Mid-Infrared Multi-field Imager for gaZing at the UnKnown Universe (MIMIZUKU; \citealt{Kamizuka2020}) on The University of Tokyo Atacama Observatory (TAO; \citealt{Yoshii2010}) 6.5m ground-based telescope (2--38~$\mu$m).
Combining the multiwavelength images with such as these telescopes and ALMA will allow us to study the distribution of the warm and cold outflowing matters, and better constrain the relationship between the outflows and star-forming activities.
\\

%%%%%%%%%%%%%%%%%%%%%%%%%%%
%%% Section 8 : Summary %%%
%%%%%%%%%%%%%%%%%%%%%%%%%%%

\section{Conclusion} \label{S8-summary}
The purpose of this work is (1) to understand the features of multiwavelength emissions in U/LIRGs, (2) to reveal the structure of the outflowing polar dust, and (3) to investigate the activities of host galaxies, SMBHs, and outflows for testing whether the merger process is a vital factor to establish the local $M_{\rm bulge}$--$M_{\rm BH}$ relation.
To achieve these goals, we have performed a hard X-ray to radio multiwavelength SED decomposition for the 57 local U/LIRGs (containing 84 individual galaxies) observed with NuSTAR and/or Swift/BAT \citep{Yamada2021}.
We compile the broadband X-ray spectra and multiwavelength wide-survey catalogs in the UV to radio bands. After the cross-matching with them, we finally obtain the multiwavelength photometries of the sample consisting of 72 resolved sources and 13 duplicated systems of resolved pairs, which are spatially divided in the Herschel PACS 70~$\mu$m bands. 
We modify the latest SED-fitting code X-CIGALE by implementing the infrared (IR) CLUMPY model, allowing the multiwavelength study with the consistent X-ray torus model (XCLUMPY).
Adopting the torus parameters obtained by the X-ray fitting \citep{Yamada2021}, we constrain the properties of host galaxies, AGN tori, and polar dust.
Here, we described the main results as follows.

\begin{enumerate}

\item The sample has 36 single and two unresolved dual AGNs detected in the hard X-ray band \citep{Yamada2021}. For all targets, we perform the BIC test for two kinds of UV-to-IR SED fitting with and without AGN (torus and polar dust) components. Although it is difficult to reliably identify AGNs solely by the SED fitting, we newly identify three AGN candidates by the diagnostics of $\Delta{\rm BIC} > 6$.
We also examine the significance of the polar dust component for these AGNs.
The AGNs in late mergers support the presence of polar dust emission, while some AGNs in early mergers or nonmergers show no signatures (Section~\ref{subsub4-2-4-BIC}).

\item The component galaxies in early mergers have a wide range of SFRs above and below the main sequence (MS), while galaxies in late mergers have the highest SFRs above the MS. Similar SFR--$M_*$ distributions of AGNs and the others imply that the small influence of the AGN activities on the star formation in the phase of U/LIRGs (Section~\ref{sub5-1-MS}).

\item The flat slope of radio emission in late mergers may be caused by the optically-thick free-free absorption due to the rich environment of the nuclear region. Most AGNs in U/LIRGs are radio-quiet. The SFRs are correlated with radio luminosity, indicating that starburst emission is dominant (Section~\ref{sub5-2-radio}).

\item In the WISE color--color diagram, we propose a new wedge of the buried AGNs in late mergers, which show the $\sim$3--10~$\mu$m excess. This can be explainable by the inner part of the dusty disk and/or hot polar dust within the inner parsecs (Section~\ref{sub5-3-WISE-color}).

\item The averaged SEDs suggest that the strength of the X-rays is the best means to reveal the true energy sources (intense starbursts and/or buried AGNs) in U/LIRGs (Section~\ref{sub5-4-averaged-SED}).
The absorption-corrected AGN SEDs show X-ray weakness relative to the other wavelengths. This may be a common feature among the AGNs in IR-luminous galaxies over the cosmic history ($z \sim 0$--7), whose mechanism will be related to their massive outflows (Section~\ref{sub5-5-X-weak}).

\item Although polar-dust extinction is much smaller than torus extinction, the UV-to-IR (mainly IR) polar dust luminosities are $\sim$2 times larger than the torus ones because of the seemingly large covering fractions (and/or large volumes) of polar dust. The AGNs with the signs of polar dust emission (Section~\ref{subsub4-2-4-BIC}) have log$\lambda_{\rm Edd} \gtrsim -$3.0 (Section~\ref{sub6-3-Lpolar}).

\item The polar-dust temperature decreases with merger stages. We estimate the physical size of the polar dust by the temperature and dust sublimation radius, consistent with the high-spatial-resolution mid-IR images \citep{Asmus2019}.
Their sizes increase with AGN luminosity from a few tens of parsec (early mergers) to kiloparsec scales (late mergers), indicating that the polar dust is likely the expanding (i.e., evolving) dusty outflows (Section~\ref{sub6-5-polar-structure}).

\item The SFRs and intrinsic (bolometric) AGN luminosities increase with merger stage. The comparison between these quantities suggests that the starbursts occur first and AGNs arise later, and overall their growth rates follow the simultaneous coevolution relation that can establish the local $M_{\rm bulge}$--$M_{\rm BH}$ mass relation (Section~\ref{sub7-1-growth-rate}).

\item Considering the Soltan argument, the balanced SFR--$L_{\rm AGN,int}$ relation helps the construction of local $M_{\rm bulge}$--$M_{\rm BH}$ relation. Whereas, galaxy collision is also a key process to build up high-mass galaxies such as local elliptical galaxies (Section~\ref{sub7-2-Mass}).

\item \citet{Yamada2021} and this work suggest that the AGNs in late mergers show the multiphase outflows at subparsec to kiloparsec scales, that is, UFOs, ionized outflows, molecular outflows, and expanding dusty winds. The coexistence of intense starburst, AGNs, and large-scale massive outflows support a standard AGN feedback scenario (Section~\ref{sub7-3-Perspective}).
\\

\end{enumerate}

%\begin{acknowledgments}

We thank the anonymous referee for constructive comments and suggestions that helped improve the quality of the paper.
S.Y. acknowledges Dr. Taiki Kawamuro and Dr. Kohei Ichikawa for their helpful discussion.
This work is financially supported by JSPS KAKENHI grant numbers 
19J22216 and 22K20391 (S.Y.); 
17K05384 and 20H01946 (Y.U.); 
18J01050, 19K14759, and 22H01266 (Y.T.); 
21J13894 (S.O.); 
22J22795 (R.U.);
and 21K03632 (M.I.).
S.Y. is grateful for support from RIKEN Special Postdoctoral Researcher Program.
M.H.E. and T.M. acknowledge support by UNAM-DGAPA PAPIIT IN111319 
and CONACyT Investigaci\'on Científica B\'asica 252531.
H.M.E. also thanks support from a postdoctoral fellowship from UNAM-DGAPA.
C.R. acknowledges support from the Fondecyt Iniciacion grant 11190831 and ANID BASAL project
FB210003.

% HEASARC (Official text) & X-ray data [NuSTAR, Chandra, XMM-Newton, Suzaku, Swift]
This research has made use of data and/or software provided by the High Energy Astrophysics Science Archive Research Center (HEASARC), which is a service of the Astrophysics Science Division at NASA/GSFC.
This work makes use of data obtained from the NuSTAR Data Analysis Software (NuSTARDAS) jointly developed by the ASI Science Data Center (ASDC, Italy) and the California Institute of Technology (USA).
This publication makes use of data obtained with Chandra, supported by the Chandra X-ray Observatory Center at the Smithsonian Astrophysical Observatory, and with XMM-Newton, an ESA science mission with instruments and contributions directly funded by ESA Member States and NASA.
This study makes use of data obtained from the Suzaku satellite, a collaborative mission between the space agencies of Japan (JAXA) and the USA (NASA).
The scientific results reported in this article are based on
observations made by UK Swift Science Data Centre at the University of Leicester.

% SIMBAD, Aladin, NED, IRSA
This research makes use of the SIMBAD database, operated at CDS, Strasbourg, France \citep{Wenger2000}, and the Aladin sky atlas, developed at CDS, Strasbourg Observatory, France \citep{Bonnarel2000,Boch2014}.
Also, this research makes use of data from the NASA/IPAC Extragalactic Database (NED) and NASA/IPAC Infrared Science Archive (IRSA), operated by JPL/California Institute of Technology under contract with the National Aeronautics and Space Administration \citep[see also IPAC DOIs:][]{2MASS,WISE,AKARI}.

% GALEX
This research is based on observations with GALEX, operated for NASA by the California Institute of Technology under NASA contract NAS5-98034.
The GALEX data presented in this paper were obtained from the Mikulski Archive for Space Telescopes (MAST) at the Space Telescope Science Institute \citep[see also MAST DOI:][]{GALEX}.

% Pan-STARRS (Official text)
The Pan-STARRS1 Surveys (PS1) and the PS1 public science archive have been made possible through contributions by the Institute for Astronomy, the University of Hawaii, the Pan-STARRS Project Office, the Max-Planck Society and its participating institutes, the Max Planck Institute for Astronomy, Heidelberg and the Max Planck Institute for Extraterrestrial Physics, Garching, The Johns Hopkins University, Durham University, the University of Edinburgh, the Queen's University Belfast, the Harvard-Smithsonian Center for Astrophysics, the Las Cumbres Observatory Global Telescope Network Incorporated, the National Central University of Taiwan, the Space Telescope Science Institute, the National Aeronautics and Space Administration under Grant No. NNX08AR22G issued through the Planetary Science Division of the NASA Science Mission Directorate, the National Science Foundation Grant No. AST-1238877, the University of Maryland, Eotvos Lorand University (ELTE), the Los Alamos National Laboratory, and the Gordon and Betty Moore Foundation.

% SkyMapper (Official text)
The national facility capability for SkyMapper has been funded through ARC LIEF grant LE130100104 from the Australian Research Council, awarded to the University of Sydney, the Australian National University, Swinburne University of Technology, the University of Queensland, the University of Western Australia, the University of Melbourne, Curtin University of Technology, Monash University and the Australian Astronomical Observatory. SkyMapper is owned and operated by The Australian National University's Research School of Astronomy and Astrophysics. The survey data were processed and provided by the SkyMapper Team at ANU. The SkyMapper node of the All-Sky Virtual Observatory (ASVO) is hosted at the National Computational Infrastructure (NCI). Development and support of the SkyMapper node of the ASVO have been funded in part by Astronomy Australia Limited (AAL) and the Australian Government through the Commonwealth's Education Investment Fund (EIF) and National Collaborative Research Infrastructure Strategy (NCRIS), particularly the National eResearch Collaboration Tools and Resources (NeCTAR) and the Australian National Data Service Projects (ANDS).

% SDSS (Official text)
Funding for the SDSS IV has been provided by the Alfred P. Sloan Foundation, the U.S. Department of Energy Office of Science, and the Participating Institutions. SDSS acknowledges support and resources from the Center for High-Performance Computing at the University of Utah. The SDSS web site is www.sdss.org. SDSS is managed by the Astrophysical Research Consortium for the Participating Institutions of the SDSS Collaboration including the Brazilian Participation Group, the Carnegie Institution for Science, Carnegie Mellon University, Center for Astrophysics | Harvard \& Smithsonian (CfA), the Chilean Participation Group, the French Participation Group, Instituto de Astrofísica de Canarias, The Johns Hopkins University, Kavli Institute for the Physics and Mathematics of the Universe (IPMU) / University of Tokyo, the Korean Participation Group, Lawrence Berkeley National Laboratory, Leibniz Institut f\"{u}r Astrophysik Potsdam (AIP), Max-Planck-Institut f\"{u}r Astronomie (MPIA Heidelberg), Max-Planck-Institut für Astrophysik (MPA Garching), Max-Planck-Institut f\"{u}r Extraterrestrische Physik (MPE), National Astronomical Observatories of China, New Mexico State University, New York University, University of Notre Dame, Observat\'orio Nacional / MCTI, The Ohio State University, Pennsylvania State University, Shanghai Astronomical Observatory, United Kingdom Participation Group, Universidad Nacional Autónoma de M\'exico, University of Arizona, University of Colorado Boulder, University of Oxford, University of Portsmouth, University of Utah, University of Virginia, University of Washington, University of Wisconsin, Vanderbilt University, and Yale University.

% 2MASS (Official text)
This publication makes use of data products from the 2MASS, which is a joint project of the University of Massachusetts and the Infrared Processing and Analysis Center/California Institute of Technology, funded by the National Aeronautics and Space Administration and the National Science Foundation.
% WISE (Official text)
This publication makes use of data products from the WISE, which is a joint project of the University of California, Los Angeles, and the Jet Propulsion Laboratory/California Institute of Technology, funded by the National Aeronautics and Space Administration.
% AKARI (Official text)
This research is based on observations with AKARI, a JAXA project with the participation of ESA.
% Herschel (Official text)
Herschel is an ESA space observatory with science instruments provided by European-led Principal Investigator consortia and with important participation from NASA.

% VLA & NRAO  (Official text)
This work makes use of the data from the VLA, operated by the National Radio Astronomy Observatory (NRAO).
The NRAO is a facility of the National Science Foundation operated under a cooperative agreement by Associated Universities, Inc.
% MOST
This study makes use of data obtained from the MOST, operated with the support of the Australian Research Council and the Science Foundation for Physics within the University of Sydney.
% MWA (Official text)
This scientific work makes use of the Murchison Radio-astronomy Observatory, operated by CSIRO. We acknowledge the Wajarri Yamatji people as the traditional owners of the Observatory site. Support for the operation of the MWA is provided by the Australian Government (NCRIS), under a contract to Curtin University administered by Astronomy Australia Limited. We acknowledge the Pawsey Supercomputing Centre which is supported by the Western Australian and Australian Governments.
% GMRT (Official text)
We thank the staff of the GMRT that made these observations possible. GMRT is run by the National Centre for Radio Astrophysics of the Tata Institute of Fundamental Research.

\facilities{NuSTAR, Swift, XMM-Newton, Chandra, Suzaku,
GALEX, Pan-STARRS, SkyMapper, SDSS,
2MASS, WISE, AKARI, Herschel,
VLA, MOST, MWA, GMRT.}

\software{XCLUMPY \citep{Tanimoto2019}, HEAsoft (v6.25), XSPEC(v12.10.1; \citealt{Arnaud1996}), NuSTARDAS(v1.8.0), CIAO(v4.11), SAS (v17.0.0; \citealt{Gabriel2004}),
TOPCAT \citep{Taylor2006}, X-CIGALE \citep{Boquien2019,Yang2020}.}

%\end{acknowledgments}

\appendix
\restartappendixnumbering

\section{Best-fitting Results} \label{AppendixA-best-values}
Here, we present the best-fit parameters derived from the UV-to-IR SED decomposition, WISE W1--W4 magnitude, (Table~\ref{TA1_results1} and~\ref{TA2_results2}) and the results of the radio fitting (Table~\ref{TA3_results3}).
In Table~\ref{TA4_results4} we list the logarithmic AGN luminosities in the X-ray, optical, and IR bands.
The obscuration and UV-to-IR luminosities for the torus and polar dust component, Eddington ratios, and the spatial scales of the multiphase outflows (Section~\ref{sub6-5-polar-structure}) are summarized in Table~\ref{TA5_results5}.
The multiwavelength SEDs and the best-fit SED models are presented in Appendix~\ref{AppendixE-SEDs}.

% Table A1
%\clearpage
\startlongtable
\begin{deluxetable*}{llccccccr}
\tabletypesize{\footnotesize}
\tablecaption{Best-fitting Parameters of the UV-to-IR SED Decomposition \label{TA1_results1}}
\tablehead{
\colhead{ID} &
\colhead{Name} &
%\colhead{$M$} &
%\colhead{$D_{12}$} &
\colhead{$f_{\rm AGN}$} &
\colhead{$\tau_{\rm V}$} &
\colhead{$\sigma$} &
\colhead{$i$} &
\colhead{$E$($B-V$)$_{\rm polar}$} &
\colhead{$T_{\rm polar}$} &
\colhead{$\chi^2_{\rm red}$} \\
 & & & & (\degr) & (\degr) & (mag) & (K) & 
}
\decimalcolnumbers
\startdata
ID01 & NGC~34 & 0.06$\pm$0.03 & 95$\pm$28 & 20 & 60 & 0.07$\pm$0.12 & 170$\pm$54 & 2.1\\
ID02 & MCG$-$02-01-052/MCG$-$02-01-051 & 0.00$^{\rm (f)}$ & $\cdots$ & $\cdots$ & $\cdots$ & $\cdots$ & $\cdots$ & 1.5\\
ID03 & MCG$-$02-01-052 & 0.00$^{\rm (f)}$ & $\cdots$ & $\cdots$ & $\cdots$ & $\cdots$ & $\cdots$ & 4.5\\
ID04 & MCG$-$02-01-051 & 0.00$^{\rm (f)}$ & $\cdots$ & $\cdots$ & $\cdots$ & $\cdots$ & $\cdots$ & 0.6\\
ID05 & ESO~350$-$38 & 0.60$\pm$0.03 & 41$\pm$7 & 20 & 80 & 0.11$\pm$0.17 & 100$\pm$1 & 2.5\\
ID06 & NGC~232/NGC~235 & 0.05$\pm$0.01 & 93$\pm$28 & 20 & 60 & $\cdots$ & $\cdots$ & 1.7\\
ID07 & NGC~232 & 0.00$^{\rm (f)}$ & $\cdots$ & $\cdots$ & $\cdots$ & $\cdots$ & $\cdots$ & 2.0\\
ID08 & NGC~235 & 0.10$\pm$0.01 & 87$\pm$29 & 20 & 60 & $\cdots$ & $\cdots$ & 2.4\\
ID09 & MCG+12-02-001 & 0.06$\pm$0.02 & 96$\pm$28 & 20 & 60 & 0.55$\pm$0.29 & 174$\pm$44 & 1.8\\
ID10 & IC~1623A/IC~1623B & 0.00$^{\rm (f)}$ & $\cdots$ & $\cdots$ & $\cdots$ & $\cdots$ & $\cdots$ & 3.3\\
ID11 & NGC~833/NGC~835 & 0.10$\pm$0.01 & 87$\pm$31 & 15 & 60 & 0.55$\pm$0.29 & 201$\pm$52 & 1.7\\
ID12 & NGC~833 & 0.08$\pm$0.03 & 82$\pm$32 & 15 & 60 & $\cdots$ & $\cdots$ & 3.7\\
ID13 & NGC~835 & 0.10$\pm$0.02 & 91$\pm$30 & 15 & 60 & 0.16$\pm$0.26 & 190$\pm$48 & 1.8\\
ID14 & NGC~838 & 0.00$^{\rm (f)}$ & $\cdots$ & $\cdots$ & $\cdots$ & $\cdots$ & $\cdots$ & 1.3\\
ID15 & NGC~839 & 0.00$^{\rm (f)}$ & $\cdots$ & $\cdots$ & $\cdots$ & $\cdots$ & $\cdots$ & 3.6\\
ID16 & NGC~1068 & 0.49$\pm$0.07 & 79$\pm$33 & 20 & 80 & 0.39$\pm$0.31 & 250$\pm$5 & 5.3\\
ID17 & UGC~2608/UGC~2612 & 0.26$\pm$0.05 & 63$\pm$30 & 15 & 80 & $\cdots$ & $\cdots$ & 0.8\\
ID18 & UGC~2608 & 0.20$\pm$0.07 & 64$\pm$32 & 15 & 80 & $\cdots$ & $\cdots$ & 1.0\\
ID19 & UGC~2612 & 0.00$^{\rm (f)}$ & $\cdots$ & $\cdots$ & $\cdots$ & $\cdots$ & $\cdots$ & 4.1\\
ID20 & NGC~1275 & 0.80$\pm$0.01 & 41$\pm$8 & 20 & 30 & 0.80$\pm$0.01 & 100$\pm$1 & 5.6\\
ID21 & NGC~1365 & 0.10$\pm$0.01 & 83$\pm$32 & 20 & 30 & 0.80$\pm$0.02 & 235$\pm$25 & 1.6\\
ID22 & ESO~203$-$1 & 0.00$^{\rm (f)}$ & $\cdots$ & $\cdots$ & $\cdots$ & $\cdots$ & $\cdots$ & 1.4\\
ID23 & CGCG~468-002W/CGCG~468-002E & 0.17$\pm$0.04 & 81$\pm$32 & 15 & 60 & 0.05$\pm$0.02 & 126$\pm$35 & 2.1\\
ID24 & CGCG~468-002W & 0.20$\pm$0.05 & 85$\pm$32 & 15 & 60 & 0.06$\pm$0.07 & 162$\pm$36 & 4.0\\
ID25 & CGCG~468-002E & 0.00$^{\rm (f)}$ & $\cdots$ & $\cdots$ & $\cdots$ & $\cdots$ & $\cdots$ & 2.1\\
ID26 & IRAS~F05189$-$2524 & 0.70$\pm$0.03 & 40$\pm$3 & 15 & 60 & 0.80$\pm$0.04 & 100$\pm$1 & 3.8\\
ID27 & IRAS~F06076$-$2139/ & 0.07$\pm$0.02 & 94$\pm$28 & 20 & 60 & 0.55$\pm$0.29 & 137$\pm$27 & 0.8\\
& 2MASS~06094601$-$2140312 \\
ID28 & NGC~2623 & 0.05$\pm$0.01 & 104$\pm$21 & 20 & 60 & 0.05$\pm$0.02 & 100$\pm$5 & 3.5\\
ID29 & ESO~060$-$IG016~West/East & 0.10$\pm$0.01 & 80$\pm$31 & 20 & 60 & $\cdots$ & $\cdots$ & 1.0\\
ID30 & IRAS~F08572+3915 & 0.47$\pm$0.05 & 49$\pm$24 & 20 & 80 & 0.05$\pm$0.02 & 180$\pm$41 & 11.8\\
ID31 & UGC~5101 & 0.11$\pm$0.04 & 45$\pm$16 & 15 & 60 & 0.30$\pm$0.05 & 101$\pm$8 & 3.7\\
ID32 & MCG+08-18-012/MCG+08-18-013 & 0.00$^{\rm (f)}$ & $\cdots$ & $\cdots$ & $\cdots$ & $\cdots$ & $\cdots$ & 3.0\\
ID33 & MCG+08-18-012 & 0.00$^{\rm (f)}$ & $\cdots$ & $\cdots$ & $\cdots$ & $\cdots$ & $\cdots$ & 4.5\\
ID34 & MCG+08-18-013 & 0.00$^{\rm (f)}$ & $\cdots$ & $\cdots$ & $\cdots$ & $\cdots$ & $\cdots$ & 1.4\\
ID35 & MCG$-$01-26-013/NGC~3110 & 0.00$^{\rm (f)}$ & $\cdots$ & $\cdots$ & $\cdots$ & $\cdots$ & $\cdots$ & 1.5\\
ID36 & MCG$-$01-26-013 & 0.00$^{\rm (f)}$ & $\cdots$ & $\cdots$ & $\cdots$ & $\cdots$ & $\cdots$ & 2.3\\
ID37 & NGC~3110 & 0.00$^{\rm (f)}$ & $\cdots$ & $\cdots$ & $\cdots$ & $\cdots$ & $\cdots$ & 3.9\\
ID38 & ESO~374$-$IG032 & 0.10$\pm$0.02 & 54$\pm$25 & 20 & 30 & 0.05$\pm$0.06 & 172$\pm$45 & 5.9\\
ID39 & NGC~3256 & 0.00$^{\rm (f)}$ & $\cdots$ & $\cdots$ & $\cdots$ & $\cdots$ & $\cdots$ & 1.7\\
ID40 & IRAS~F10565+2448 & 0.00$^{\rm (f)}$ & $\cdots$ & $\cdots$ & $\cdots$ & $\cdots$ & $\cdots$ & 5.2\\
ID41 & NGC~3690~West/East & 0.29$\pm$0.11 & 88$\pm$31 & 45 & 30 & 0.58$\pm$0.31 & 158$\pm$57 & 2.6\\
ID42 & ESO~440$-$58/MCG$-$05-29-017 & 0.00$^{\rm (f)}$ & $\cdots$ & $\cdots$ & $\cdots$ & $\cdots$ & $\cdots$ & 1.7\\
ID43 & IRAS~F12112+0305 & 0.00$^{\rm (f)}$ & $\cdots$ & $\cdots$ & $\cdots$ & $\cdots$ & $\cdots$ & 5.7\\
ID44 & NGC~4418/MCG+00-32-013 & 0.29$\pm$0.03 & 117$\pm$10 & 20 & 80 & 0.65$\pm$0.23 & 150$\pm$1 & 3.2\\
ID45 & NGC~4418 & 0.10$\pm$0.01 & 82$\pm$28 & 20 & 80 & 0.63$\pm$0.24 & 200$\pm$1 & 3.4\\
ID46 & MCG+00-32-013 & 0.00$^{\rm (f)}$ & $\cdots$ & $\cdots$ & $\cdots$ & $\cdots$ & $\cdots$ & 4.9\\
ID47 & Mrk~231 & 0.28$\pm$0.06 & 79$\pm$33 & 25 & 60 & 0.42$\pm$0.29 & 186$\pm$55 & 0.3\\
ID48 & NGC~4922S/NGC~4922N & 0.34$\pm$0.08 & 81$\pm$32 & 20 & 60 & 0.38$\pm$0.19 & 135$\pm$42 & 0.8\\
ID49 & IC~860 & 0.00$^{\rm (f)}$ & $\cdots$ & $\cdots$ & $\cdots$ & $\cdots$ & $\cdots$ & 3.1\\
ID50 & IRAS~13120$-$5453 & 0.05$\pm$0.01 & 49$\pm$22 & 20 & 80 & 0.23$\pm$0.28 & 100$\pm$5 & 2.0\\
ID51 & NGC~5104 & 0.00$^{\rm (f)}$ & $\cdots$ & $\cdots$ & $\cdots$ & $\cdots$ & $\cdots$ & 2.8\\
ID52 & MCG$-$03-34-064 & 0.61$\pm$0.10 & 48$\pm$20 & 20 & 60 & 0.79$\pm$0.06 & 206$\pm$19 & 1.8\\
ID53 & NGC~5135 & 0.06$\pm$0.02 & 79$\pm$33 & 15 & 80 & 0.38$\pm$0.31 & 138$\pm$58 & 1.0\\
ID54 & Mrk~266B/Mrk~266A & 0.09$\pm$0.02 & 98$\pm$26 & 20 & 60 & 0.10$\pm$0.18 & 182$\pm$45 & 1.6\\
ID55 & Mrk~273 & 0.08$\pm$0.02 & 73$\pm$33 & 15 & 60 & 0.78$\pm$0.09 & 101$\pm$6 & 2.4\\
ID56 & IRAS~F14348$-$1447 & 0.09$\pm$0.04 & 71$\pm$33 & 20 & 80 & 0.40$\pm$0.31 & 102$\pm$11 & 2.6\\
ID57 & IRAS~F14378$-$3651 & 0.00$^{\rm (f)}$ & $\cdots$ & $\cdots$ & $\cdots$ & $\cdots$ & $\cdots$ & 2.4\\
ID58 & IC~4518A/IC~4518B & 0.13$\pm$0.08 & 79$\pm$33 & 20 & 60 & 0.39$\pm$0.31 & 181$\pm$58 & 0.3\\
ID59 & IRAS~F15250+3608 & 0.00$^{\rm (f)}$ & $\cdots$ & $\cdots$ & $\cdots$ & $\cdots$ & $\cdots$ & 2.8\\
ID60 & Arp~220W/Arp~220E & 0.05$\pm$0.01 & 110$\pm$18 & 20 & 60 & 0.72$\pm$0.18 & 141$\pm$20 & 6.2\\
ID61 & NGC~6240S/NGC~6240N & 0.10$\pm$0.01 & 104$\pm$23 & 20 & 30 & 0.05$\pm$0.06 & 136$\pm$32 & 1.4\\
ID62 & NGC~6285/NGC~6286 & 0.05$\pm$0.01 & 82$\pm$33 & 20 & 80 & 0.38$\pm$0.31 & 172$\pm$74 & 4.5\\
ID63 & NGC~6285 & 0.00$^{\rm (f)}$ & $\cdots$ & $\cdots$ & $\cdots$ & $\cdots$ & $\cdots$ & 2.3\\
ID64 & NGC~6286 & 0.05$\pm$0.01 & 82$\pm$33 & 20 & 80 & 0.44$\pm$0.31 & 168$\pm$75 & 5.5\\
ID65 & IRAS~F17138$-$1017 & 0.14$\pm$0.07 & 81$\pm$33 & 20 & 60 & 0.37$\pm$0.30 & 135$\pm$54 & 5.1\\
ID66 & IRAS~F18293$-$3413 & 0.00$^{\rm (f)}$ & $\cdots$ & $\cdots$ & $\cdots$ & $\cdots$ & $\cdots$ & 0.9\\
ID67 & IRAS~F19297$-$0406 & 0.00$^{\rm (f)}$ & $\cdots$ & $\cdots$ & $\cdots$ & $\cdots$ & $\cdots$ & 2.1\\
ID68 & NGC~6907/NGC~6908 & 0.00$^{\rm (f)}$ & $\cdots$ & $\cdots$ & $\cdots$ & $\cdots$ & $\cdots$ & 2.5\\
ID69 & NGC~6921/MCG+04-48-002 & 0.10$\pm$0.01 & 86$\pm$31 & 20 & 60 & $\cdots$ & $\cdots$ & 0.4\\
ID70 & NGC~6921 & 0.05$\pm$0.01 & 82$\pm$33 & 20 & 80 & $\cdots$ & $\cdots$ & 0.8\\
ID71 & MCG+04-48-002 & 0.11$\pm$0.03 & 91$\pm$29 & 20 & 60 & $\cdots$ & $\cdots$ & 0.7\\
ID72 & II~Zw~096/IRAS~F20550+1655~SE & 0.00$^{\rm (f)}$ & $\cdots$ & $\cdots$ & $\cdots$ & $\cdots$ & $\cdots$ & 3.7\\
ID73 & ESO~286$-$19 & 0.00$^{\rm (f)}$ & $\cdots$ & $\cdots$ & $\cdots$ & $\cdots$ & $\cdots$ & 2.3\\
ID74 & NGC~7130 & 0.24$\pm$0.10 & 79$\pm$32 & 25 & 80 & 0.38$\pm$0.31 & 109$\pm$32 & 1.6\\
ID75 & NGC~7469/IC~5283 & 0.34$\pm$0.09 & 77$\pm$33 & 15 & 30 & 0.34$\pm$0.36 & 102$\pm$11 & 0.4\\
ID76 & NGC~7469 & 0.38$\pm$0.08 & 75$\pm$33 & 15 & 30 & 0.72$\pm$0.20 & 102$\pm$10 & 0.6\\
ID77 & IC~5283 & 0.00$^{\rm (f)}$ & $\cdots$ & $\cdots$ & $\cdots$ & $\cdots$ & $\cdots$ & 1.2\\
ID78 & ESO~148$-$2 & 0.22$\pm$0.07 & 45$\pm$18 & 20 & 80 & 0.16$\pm$0.24 & 100$\pm$3 & 2.6\\
ID79 & NGC~7591 & 0.00$^{\rm (f)}$ & $\cdots$ & $\cdots$ & $\cdots$ & $\cdots$ & $\cdots$ & 2.3\\
ID80 & NGC~7674/MCG+01-59-081 & 0.41$\pm$0.04 & 71$\pm$32 & 40 & 30 & 0.06$\pm$0.07 & 137$\pm$48 & 0.7\\
ID81 & NGC~7674 & 0.43$\pm$0.05 & 60$\pm$29 & 40 & 30 & 0.06$\pm$0.08 & 138$\pm$43 & 0.6\\
ID82 & MCG+01-59-081 & 0.00$^{\rm (f)}$ & $\cdots$ & $\cdots$ & $\cdots$ & $\cdots$ & $\cdots$ & 0.3\\
ID83 & NGC~7679/NGC~7682 & 0.07$\pm$0.03 & 90$\pm$30 & 20 & 60 & 0.49$\pm$0.31 & 181$\pm$57 & 1.7\\
ID84 & NGC~7679 & 0.05$\pm$0.01 & 88$\pm$31 & 15 & 30 & 0.44$\pm$0.32 & 190$\pm$63 & 1.1\\
ID85 & NGC~7682 & 0.15$\pm$0.08 & 82$\pm$32 & 60 & 60 & $\cdots$ & $\cdots$ & 1.7\\
\enddata
\tablecomments{Comments: (1--2) ID and object name; 
(3) fraction of AGN luminosity in the IR band. 
The symbol ``(f)'' means that the value is fixed;
(4) optical depth of each clump at $V$ band;
(5--6) torus angular width and inclination; 
(7) $E$($B-V$) of the polar dust; 
(8) temperature of the polar dust; 
(9) reduced $\chi^2$ for the UV-to-IR SED decomposition.}
\tablenotetext{\ }{(This table is available in its entirety in machine-readable form.)}
\end{deluxetable*}

% Table A2
\clearpage
\startlongtable
\begin{deluxetable*}{llrrrrrrr}
\tabletypesize{\scriptsize}
\tablecaption{Properties of Host Galaxies and WISE Color \label{TA2_results2}}
\tablehead{
\colhead{ID} &
\colhead{Name} &
\colhead{log$M_*$} &
\colhead{logSFR} &
\colhead{logsSFR} &
\colhead{W1} &
\colhead{W2} &
\colhead{W3} &
\colhead{W4} \\
 & & ($M_{\odot}$) \ \ \ \ & ($M_{\odot}$ yr$^{-1}$)  & (yr) \ \ \ \ \ \
 & (mag) \ \ \ \ & (mag) \ \ \ \ & (mag) \ \ \ \ & (mag) \ \ \ \ 
}
\decimalcolnumbers
\startdata
ID01 & NGC~34 & 10.31$\pm$0.13 & 1.51$\pm$0.07 & $-$8.80$\pm$0.14 & 9.82$\pm$0.01 & 9.35$\pm$0.01 & 5.25$\pm$0.01 & 2.05$\pm$0.01 \\
ID02 & MCG$-$02-01-052/MCG$-$02-01-051 & 10.37$\pm$0.13 & 1.47$\pm$0.06 & $-$8.90$\pm$0.14 & $\cdots$\ \ \ \ \ \ & $\cdots$\ \ \ \ \ \  & $\cdots$\ \ \ \ \ \  & $\cdots$\ \ \ \ \ \ \\
ID03 & MCG$-$02-01-052 & 9.85$\pm$0.05 & 0.57$\pm$0.02 & $-$9.29$\pm$0.05 & 12.25$\pm$0.02 & 12.17$\pm$0.02 & $\cdots$\ \ \ \ \ \  & $\cdots$\ \ \ \ \ \ \\
ID04 & MCG$-$02-01-051 & 10.22$\pm$0.15 & 1.42$\pm$0.06 & $-$8.80$\pm$0.17 & 10.80$\pm$0.01 & 10.35$\pm$0.01 & 5.89$\pm$0.01 & 2.95$\pm$0.01 \\
ID05 & ESO~350$-$38 & 9.74$\pm$0.16 & 0.95$\pm$0.07 & $-$8.79$\pm$0.17 & 10.79$\pm$0.01 & 9.50$\pm$0.01 & 5.49$\pm$0.01 & 2.10$\pm$0.01 \\
ID06 & NGC~232/NGC~235 & 11.07$\pm$0.06 & 1.35$\pm$0.04 & $-$9.72$\pm$0.07 & $\cdots$\ \ \ \ \ \  & $\cdots$\ \ \ \ \ \  & $\cdots$\ \ \ \ \ \  & $\cdots$\ \ \ \ \ \ \\
ID07 & NGC~232 & 10.73$\pm$0.13 & 1.35$\pm$0.08 & $-$9.38$\pm$0.15 & 9.66$\pm$0.01 & 9.32$\pm$0.01 & 5.40$\pm$0.01 & 2.71$\pm$0.01 \\
ID08 & NGC~235 & 10.93$\pm$0.03 & 0.55$\pm$0.03 & $-$10.38$\pm$0.04 & 10.03$\pm$0.01 & 9.47$\pm$0.01 & 6.44$\pm$0.01 & 3.71$\pm$0.01 \\
ID09 & MCG+12-02-001 & 10.34$\pm$0.11 & 1.41$\pm$0.06 & $-$8.93$\pm$0.12 & $\cdots$\ \ \ \ \ \  & 8.92$\pm$0.01 & 4.49$\pm$0.01 & 1.45$\pm$0.01 \\
ID10 & IC~1623A/IC~1623B & 10.64$\pm$0.11 & 1.82$\pm$0.03 & $-$8.82$\pm$0.11 & 9.22$\pm$0.01 & 8.21$\pm$0.01 & 4.52$\pm$0.01 & 1.54$\pm$0.01 \\
ID11 & NGC~833/NGC~835 & 11.16$\pm$0.07 & 0.37$\pm$0.24 & $-$10.80$\pm$0.25 & $\cdots$\ \ \ \ \ \  & $\cdots$\ \ \ \ \ \  & $\cdots$\ \ \ \ \ \  & $\cdots$\ \ \ \ \ \ \\
ID12 & NGC~833 & 10.76$\pm$0.02 & $-$1.78$\pm$0.34 & $-$12.54$\pm$0.34 & 9.50$\pm$0.01 & 9.49$\pm$0.01 & $\cdots$\ \ \ \ \ \  & 5.71$\pm$0.04 \\
ID13 & NGC~835 & 10.95$\pm$0.05 & 0.59$\pm$0.07 & $-$10.36$\pm$0.08 & 8.91$\pm$0.01 & 8.77$\pm$0.01 & 5.33$\pm$0.01 & 3.32$\pm$0.01 \\
ID14 & NGC~838 & 10.36$\pm$0.09 & 1.11$\pm$0.06 & $-$9.26$\pm$0.11 & 9.32$\pm$0.01 & 9.00$\pm$0.01 & 4.70$\pm$0.01 & 2.14$\pm$0.01 \\
ID15 & NGC~839 & 10.19$\pm$0.04 & 0.91$\pm$0.03 & $-$9.27$\pm$0.05 & 9.60$\pm$0.01 & 9.09$\pm$0.01 & 4.91$\pm$0.01 & 1.81$\pm$0.01 \\
ID16 & NGC~1068 & 10.07$\pm$0.15 & 1.25$\pm$0.08 & $-$8.82$\pm$0.17 & $\cdots$\ \ \ \ \ \  & $\cdots$\ \ \ \ \ \  & $\cdots$\ \ \ \ \ \  & $\cdots$\ \ \ \ \ \ \\
ID17 & UGC~2608/UGC~2612 & 11.05$\pm$0.10 & 1.36$\pm$0.07 & $-$9.68$\pm$0.12 & $\cdots$\ \ \ \ \ \  & $\cdots$\ \ \ \ \ \  & $\cdots$\ \ \ \ \ \  & $\cdots$\ \ \ \ \ \ \\
ID18 & UGC~2608 & 10.75$\pm$0.11 & 1.24$\pm$0.11 & $-$9.51$\pm$0.15 & $\cdots$\ \ \ \ \ \  & 8.70$\pm$0.01 & 5.09$\pm$0.01 & 2.33$\pm$0.01 \\
ID19 & UGC~2612 & 10.64$\pm$0.13 & 0.53$\pm$0.06 & $-$10.11$\pm$0.14 & $\cdots$\ \ \ \ \ \  & 10.39$\pm$0.01 & 7.55$\pm$0.01 & 5.88$\pm$0.04 \\
ID20 & NGC~1275 & 9.70$\pm$0.20 & 0.91$\pm$0.05 & $-$8.79$\pm$0.20 & 8.03$\pm$0.01 & 7.52$\pm$0.01 & 4.04$\pm$0.01 & 1.07$\pm$0.01 \\
ID21 & NGC~1365 & 10.94$\pm$0.08 & 0.93$\pm$0.09 & $-$10.01$\pm$0.13 & 6.30$\pm$0.01 & 5.95$\pm$0.01 & 2.32$\pm$0.01 & $\cdots$\ \ \ \ \ \ \\
ID22 & ESO~203$-$1 & 9.78$\pm$0.22 & 1.78$\pm$0.06 & $-$8.00$\pm$0.23 & $\cdots$\ \ \ \ \ \  & 12.15$\pm$0.01 & 7.92$\pm$0.01 & 4.36$\pm$0.01 \\
ID23 & CGCG~468-002W/CGCG~468-002E & 10.95$\pm$0.10 & 0.87$\pm$0.20 & $-$10.07$\pm$0.22 & $\cdots$\ \ \ \ \ \  & $\cdots$\ \ \ \ \ \  & $\cdots$\ \ \ \ \ \  & $\cdots$\ \ \ \ \ \ \\
ID24 & CGCG~468-002W & 10.89$\pm$0.04 & $-$0.44$\pm$0.48 & $-$11.34$\pm$0.48 & 9.93$\pm$0.01 & 9.47$\pm$0.01 & 6.44$\pm$0.01 & $\cdots$\ \ \ \ \ \ \\
ID25 & CGCG~468-002E & 10.11$\pm$0.07 & 1.21$\pm$0.04 & $-$8.90$\pm$0.08 & $\cdots$\ \ \ \ \ \  & $\cdots$\ \ \ \ \ \  & 6.67$\pm$0.01 & 3.08$\pm$0.01 \\
ID26 & IRAS~F05189$-$2524 & 10.72$\pm$0.11 & 1.56$\pm$0.12 & $-$9.16$\pm$0.16 & 8.77$\pm$0.01 & 7.65$\pm$0.01 & 4.79$\pm$0.01 & 1.77$\pm$0.01 \\
ID27 & IRAS~F06076$-$2139/ & 10.41$\pm$0.19 & 1.65$\pm$0.11 & $-$8.76$\pm$0.21 & 11.31$\pm$0.01 & 10.78$\pm$0.01 & 6.84$\pm$0.01 & 3.81$\pm$0.01 \\
 & 2MASS~06094601$-$2140312 \\
ID28 & NGC~2623 & 9.86$\pm$0.31 & 1.53$\pm$0.05 & $-$8.33$\pm$0.32 & 10.39$\pm$0.01 & 9.76$\pm$0.01 & 5.84$\pm$0.01 & 2.60$\pm$0.01 \\
ID29 & ESO~060$-$IG016~West/East & 9.85$\pm$0.32 & 1.68$\pm$0.04 & $-$8.17$\pm$0.32 & $\cdots$\ \ \ \ \ \  & $\cdots$\ \ \ \ \ \  & $\cdots$\ \ \ \ \ \  & 3.69$\pm$0.01 \\
ID30 & IRAS~F08572+3915 & 9.18$\pm$0.25 & 1.89$\pm$0.02 & $-$7.28$\pm$0.25 & 10.25$\pm$0.02 & 7.94$\pm$0.02 & 4.90$\pm$0.02 & 2.04$\pm$0.02 \\
ID31 & UGC~5101 & 10.85$\pm$0.14 & 2.00$\pm$0.09 & $-$8.86$\pm$0.16 & 10.10$\pm$0.01 & 8.46$\pm$0.01 & 6.07$\pm$0.01 & 3.20$\pm$0.01 \\
ID32 & MCG+08-18-012/MCG+08-18-013 & 10.28$\pm$0.10 & 1.30$\pm$0.06 & $-$8.98$\pm$0.11 & $\cdots$\ \ \ \ \ \  & $\cdots$\ \ \ \ \ \  & $\cdots$\ \ \ \ \ \  & $\cdots$\ \ \ \ \ \ \\
ID33 & MCG+08-18-012 & 9.04$\pm$0.08 & 0.10$\pm$0.02 & $-$8.93$\pm$0.08 & 13.78$\pm$0.03 & 13.67$\pm$0.03 & 10.13$\pm$0.07 & $\cdots$\ \ \ \ \ \ \\
ID34 & MCG+08-18-013 & 10.30$\pm$0.12 & 1.30$\pm$0.09 & $-$9.00$\pm$0.15 & 10.84$\pm$0.01 & 10.44$\pm$0.01 & 6.36$\pm$0.01 & $\cdots$\ \ \ \ \ \ \\
ID35 & MCG$-$01-26-013/NGC~3110 & 10.68$\pm$0.12 & 1.26$\pm$0.09 & $-$9.42$\pm$0.15 & $\cdots$\ \ \ \ \ \  & $\cdots$\ \ \ \ \ \  & $\cdots$\ \ \ \ \ \  & $\cdots$\ \ \ \ \ \ \\
ID36 & MCG$-$01-26-013 & 10.06$\pm$0.10 & 0.24$\pm$0.04 & $-$9.82$\pm$0.11 & 11.42$\pm$0.01 & 11.24$\pm$0.01 & 7.64$\pm$0.01 & 5.15$\pm$0.03 \\
ID37 & NGC~3110 & 10.73$\pm$0.09 & 1.18$\pm$0.04 & $-$9.56$\pm$0.09 & 9.19$\pm$0.01 & 8.90$\pm$0.01 & 4.76$\pm$0.01 & 2.57$\pm$0.01 \\
ID38 & ESO~374$-$IG032 & 10.43$\pm$0.20 & 1.72$\pm$0.05 & $-$8.71$\pm$0.21 & 10.60$\pm$0.01 & 8.16$\pm$0.01 & 5.92$\pm$0.01 & 3.19$\pm$0.01 \\
ID39 & NGC~3256 & 10.67$\pm$0.12 & 1.81$\pm$0.06 & $-$8.86$\pm$0.13 & 7.58$\pm$0.01 & 7.05$\pm$0.01 & $\cdots$\ \ \ \ \ \  & $-$0.44$\pm$0.01 \\
ID40 & IRAS~F10565+2448 & 10.71$\pm$0.13 & 1.95$\pm$0.09 & $-$8.76$\pm$0.16 & 11.06$\pm$0.01 & $\cdots$\ \ \ \ \ \  & 6.24$\pm$0.01 & 3.17$\pm$0.01 \\
ID41 & NGC~3690~West/East & 10.82$\pm$0.14 & 1.69$\pm$0.10 & $-$9.13$\pm$0.18 & $\cdots$\ \ \ \ \ \  & $\cdots$\ \ \ \ \ \  & $\cdots$\ \ \ \ \ \  & $\cdots$\ \ \ \ \ \ \\
ID42 & ESO~440$-$58/MCG$-$05-29-017 & 10.55$\pm$0.06 & 1.30$\pm$0.04 & $-$9.24$\pm$0.08 & $\cdots$\ \ \ \ \ \  & $\cdots$\ \ \ \ \ \  & $\cdots$\ \ \ \ \ \  & $\cdots$\ \ \ \ \ \ \\
ID43 & IRAS~F12112+0305 & 9.95$\pm$0.39 & 2.35$\pm$0.03 & $-$7.60$\pm$0.39 & 12.36$\pm$0.01 & 11.63$\pm$0.01 & 7.57$\pm$0.01 & 4.28$\pm$0.01 \\
ID44 & NGC~4418/MCG+00-32-013 & 9.28$\pm$0.13 & 0.82$\pm$0.08 & $-$8.47$\pm$0.15 & $\cdots$\ \ \ \ \ \  & $\cdots$\ \ \ \ \ \  & $\cdots$\ \ \ \ \ \  & $\cdots$\ \ \ \ \ \ \\
ID45 & NGC~4418 & 8.97$\pm$0.23 & 0.96$\pm$0.05 & $-$8.01$\pm$0.24 & 10.15$\pm$0.01 & 9.32$\pm$0.01 & 4.01$\pm$0.01 & 0.40$\pm$0.01 \\
ID46 & MCG+00-32-013 & 7.32$\pm$0.19 & $-$0.80$\pm$0.03 & $-$8.11$\pm$0.19 & 14.11$\pm$0.03 & 13.89$\pm$0.04 & 10.34$\pm$0.08 & 7.77$\pm$0.16 \\
ID47 & Mrk~231 & 10.73$\pm$0.42 & 2.43$\pm$0.08 & $-$8.30$\pm$0.42 & 7.80$\pm$0.01 & 6.58$\pm$0.01 & 3.54$\pm$0.01 & 0.57$\pm$0.01 \\
ID48 & NGC~4922S/NGC~4922N & 10.92$\pm$0.05 & 0.91$\pm$0.06 & $-$10.00$\pm$0.08 & $\cdots$\ \ \ \ \ \  & $\cdots$\ \ \ \ \ \  & $\cdots$\ \ \ \ \ \  & $\cdots$\ \ \ \ \ \ \\
ID49 & IC~860 & 9.68$\pm$0.08 & 0.85$\pm$0.02 & $-$8.82$\pm$0.09 & 10.62$\pm$0.01 & 10.51$\pm$0.01 & 7.17$\pm$0.01 & 2.93$\pm$0.01 \\
ID50 & IRAS~13120$-$5453 & 11.15$\pm$0.12 & 2.34$\pm$0.07 & $-$8.81$\pm$0.14 & 9.58$\pm$0.01 & 8.83$\pm$0.01 & 5.08$\pm$0.01 & 1.91$\pm$0.01 \\
ID51 & NGC~5104 & 10.68$\pm$0.06 & 0.97$\pm$0.06 & $-$9.71$\pm$0.09 & 9.62$\pm$0.01 & 9.38$\pm$0.01 & 5.74$\pm$0.01 & 3.18$\pm$0.01 \\
ID52 & MCG$-$03-34-064 & 10.71$\pm$0.10 & 0.47$\pm$0.10 & $-$10.24$\pm$0.14 & 9.20$\pm$0.01 & 7.92$\pm$0.01 & 4.17$\pm$0.01 & 1.53$\pm$0.01 \\
ID53 & NGC~5135 & 10.93$\pm$0.17 & 1.19$\pm$0.10 & $-$9.74$\pm$0.20 & 8.61$\pm$0.01 & 8.10$\pm$0.01 & 4.44$\pm$0.01 & 1.62$\pm$0.01 \\
ID54 & Mrk~266B/Mrk~266A & 10.76$\pm$0.14 & 1.35$\pm$0.06 & $-$9.41$\pm$0.15 & 9.88$\pm$0.01 & 9.48$\pm$0.01 & 5.57$\pm$0.01 & 2.56$\pm$0.01 \\
ID55 & Mrk~273 & 10.52$\pm$0.22 & 2.24$\pm$0.05 & $-$8.28$\pm$0.22 & 10.50$\pm$0.01 & 9.37$\pm$0.01 & 5.85$\pm$0.01 & 2.16$\pm$0.01 \\
ID56 & IRAS~F14348$-$1447 & 10.73$\pm$0.24 & 2.32$\pm$0.06 & $-$8.40$\pm$0.24 & 12.12$\pm$0.01 & 11.29$\pm$0.01 & 7.52$\pm$0.01 & 3.95$\pm$0.01 \\
ID57 & IRAS~F14378$-$3651 & 10.29$\pm$0.21 & 2.33$\pm$0.03 & $-$7.96$\pm$0.21 & 12.24$\pm$0.02 & 11.37$\pm$0.02 & 7.11$\pm$0.02 & 3.42$\pm$0.02 \\
ID58 & IC~4518A/IC~4518B & 10.78$\pm$0.16 & 1.24$\pm$0.22 & $-$9.55$\pm$0.27 & $\cdots$\ \ \ \ \ \  & $\cdots$\ \ \ \ \ \  & $\cdots$\ \ \ \ \ \  & $\cdots$\ \ \ \ \ \ \\
ID59 & IRAS~F15250+3608 & 10.01$\pm$0.12 & 2.12$\pm$0.02 & $-$7.89$\pm$0.13 & 12.34$\pm$0.02 & 10.84$\pm$0.02 & 5.99$\pm$0.01 & 2.40$\pm$0.01 \\
ID60 & Arp~220W/Arp~220E & 9.82$\pm$0.30 & 2.12$\pm$0.04 & $-$7.69$\pm$0.31 & 9.71$\pm$0.01 & 9.14$\pm$0.01 & 4.70$\pm$0.01 & 0.76$\pm$0.01 \\
ID61 & NGC~6240S/NGC~6240N & 11.02$\pm$0.11 & 1.60$\pm$0.05 & $-$9.42$\pm$0.12 & 8.85$\pm$0.01 & 8.34$\pm$0.01 & 4.59$\pm$0.01 & 1.35$\pm$0.01 \\
ID62 & NGC~6285/NGC~6286 & 10.61$\pm$0.07 & 1.35$\pm$0.03 & $-$9.26$\pm$0.08 & $\cdots$\ \ \ \ \ \  & $\cdots$\ \ \ \ \ \  & $\cdots$\ \ \ \ \ \  & $\cdots$\ \ \ \ \ \ \\
ID63 & NGC~6285 & 10.25$\pm$0.19 & 0.61$\pm$0.05 & $-$9.63$\pm$0.20 & 10.89$\pm$0.01 & 10.66$\pm$0.01 & $\cdots$\ \ \ \ \ \  & 4.62$\pm$0.01 \\
ID64 & NGC~6286 & 10.42$\pm$0.09 & 1.09$\pm$0.04 & $-$9.34$\pm$0.10 & 9.90$\pm$0.02 & 9.40$\pm$0.02 & 5.37$\pm$0.01 & 3.51$\pm$0.02 \\
ID65 & IRAS~F17138$-$1017 & 9.40$\pm$0.29 & 1.48$\pm$0.08 & $-$7.92$\pm$0.30 & 9.63$\pm$0.02 & 9.11$\pm$0.02 & 4.66$\pm$0.01 & 1.84$\pm$0.01 \\
ID66 & IRAS~F18293$-$3413 & 10.72$\pm$0.12 & 1.78$\pm$0.06 & $-$8.94$\pm$0.13 & 8.88$\pm$0.01 & 8.37$\pm$0.01 & 4.12$\pm$0.01 & 1.42$\pm$0.01 \\
ID67 & IRAS~F19297$-$0406 & 10.60$\pm$0.22 & 2.39$\pm$0.16 & $-$8.21$\pm$0.27 & 11.94$\pm$0.02 & 11.19$\pm$0.02 & 6.85$\pm$0.02 & 3.46$\pm$0.02 \\
ID68 & NGC~6907/NGC~6908 & 10.73$\pm$0.14 & 1.08$\pm$0.03 & $-$9.66$\pm$0.15 & $\cdots$\ \ \ \ \ \  & $\cdots$\ \ \ \ \ \  & $\cdots$\ \ \ \ \ \  & $\cdots$\ \ \ \ \ \ \\
ID69 & NGC~6921/MCG+04-48-002 & 11.15$\pm$0.09 & 0.94$\pm$0.13 & $-$10.22$\pm$0.16 & $\cdots$\ \ \ \ \ \  & $\cdots$\ \ \ \ \ \  & $\cdots$\ \ \ \ \ \  & $\cdots$\ \ \ \ \ \ \\
ID70 & NGC~6921 & 11.20$\pm$0.04 & $-$0.33$\pm$0.25 & $-$11.54$\pm$0.25 & 8.67$\pm$0.01 & 8.62$\pm$0.01 & $\cdots$\ \ \ \ \ \  & 5.01$\pm$0.02 \\
ID71 & MCG+04-48-002 & 10.42$\pm$0.17 & 0.88$\pm$0.09 & $-$9.55$\pm$0.19 & 9.30$\pm$0.01 & 8.80$\pm$0.01 & 5.08$\pm$0.01 & 2.90$\pm$0.01 \\
ID72 & II~Zw~096/IRAS~F20550+1655~SE & 10.35$\pm$0.08 & 1.90$\pm$0.02 & $-$8.45$\pm$0.08 & 11.37$\pm$0.02 & 10.38$\pm$0.02 & 5.35$\pm$0.01 & $\cdots$\ \ \ \ \ \ \\
ID73 & ESO~286$-$19 & 9.94$\pm$0.09 & 1.97$\pm$0.02 & $-$7.97$\pm$0.09 & 11.30$\pm$0.01 & 9.96$\pm$0.01 & 5.78$\pm$0.01 & $\cdots$\ \ \ \ \ \ \\
ID74 & NGC~7130 & 11.06$\pm$0.11 & 1.30$\pm$0.07 & $-$9.76$\pm$0.13 & 8.98$\pm$0.01 & 8.55$\pm$0.01 & 4.59$\pm$0.01 & 1.74$\pm$0.01 \\
ID75 & NGC~7469/IC~5283 & 10.83$\pm$0.17 & 1.49$\pm$0.08 & $-$9.34$\pm$0.19 & $\cdots$\ \ \ \ \ \  & $\cdots$\ \ \ \ \ \  & $\cdots$\ \ \ \ \ \  & $\cdots$\ \ \ \ \ \ \\
ID76 & NGC~7469 & 10.69$\pm$0.18 & 1.51$\pm$0.08 & $-$9.18$\pm$0.20 & 8.29$\pm$0.01 & $\cdots$\ \ \ \ \ \  & 3.83$\pm$0.01 & 0.77$\pm$0.01 \\
ID77 & IC~5283 & 10.46$\pm$0.08 & 0.50$\pm$0.16 & $-$9.97$\pm$0.18 & 10.30$\pm$0.01 & 10.10$\pm$0.01 & $\cdots$\ \ \ \ \ \  & $\cdots$\ \ \ \ \ \ \\
ID78 & ESO~148$-$2 & 10.21$\pm$0.40 & 1.93$\pm$0.05 & $-$8.28$\pm$0.41 & 11.02$\pm$0.01 & 10.01$\pm$0.01 & 5.92$\pm$0.01 & 2.63$\pm$0.01 \\
ID79 & NGC~7591 & 10.62$\pm$0.11 & 1.04$\pm$0.04 & $-$9.58$\pm$0.11 & 9.54$\pm$0.01 & 9.33$\pm$0.01 & 5.54$\pm$0.01 & 2.91$\pm$0.01 \\
ID80 & NGC~7674/MCG+01-59-081 & 11.21$\pm$0.06 & 1.25$\pm$0.10 & $-$9.96$\pm$0.11 & $\cdots$\ \ \ \ \ \  & $\cdots$\ \ \ \ \ \  & $\cdots$\ \ \ \ \ \  & $\cdots$\ \ \ \ \ \ \\
ID81 & NGC~7674 & 11.10$\pm$0.08 & 1.22$\pm$0.09 & $-$9.88$\pm$0.12 & $\cdots$\ \ \ \ \ \  & 8.05$\pm$0.01 & 4.66$\pm$0.01 & 2.04$\pm$0.01 \\
ID82 & MCG+01-59-081 & 10.61$\pm$0.09 & 0.33$\pm$0.07 & $-$10.28$\pm$0.11 & $\cdots$\ \ \ \ \ \  & $\cdots$\ \ \ \ \ \  & $\cdots$\ \ \ \ \ \  & $\cdots$\ \ \ \ \ \ \\
ID83 & NGC~7679/NGC~7682 & 10.77$\pm$0.03 & 1.06$\pm$0.02 & $-$9.72$\pm$0.04 & $\cdots$\ \ \ \ \ \  & $\cdots$\ \ \ \ \ \  & $\cdots$\ \ \ \ \ \  & $\cdots$\ \ \ \ \ \ \\
ID84 & NGC~7679 & 10.33$\pm$0.07 & 1.05$\pm$0.02 & $-$9.28$\pm$0.07 & 9.63$\pm$0.01 & 9.30$\pm$0.01 & 5.09$\pm$0.01 & 2.75$\pm$0.01 \\
ID85 & NGC~7682 & 10.62$\pm$0.04 & $-$0.20$\pm$0.23 & $-$10.83$\pm$0.23 & 10.20$\pm$0.01 & 10.21$\pm$0.01 & 7.86$\pm$0.02 & 5.21$\pm$0.03 \\
\enddata
\tablecomments{Comments: (1--2) ID and object name; 
(3) logarithmic stellar mass; 
(4) logarithmic SFR; 
(5) logarithmic specific SFR (sSFR);
(6)--(9) W1, W2, W3, and W4 magnitudes (Vega) from ALLWISE catalog (Section~\ref{sub3-5-MIR} and~\ref{sub5-3-WISE-color}).
The W1 and W2 magnitudes are corrected for Galactic extinction (Section~\ref{sub3-8-GalExt}).}
\tablenotetext{\ }{(This table is available in its entirety in machine-readable form.)}
\end{deluxetable*}

% Table A3
\clearpage
\startlongtable
\begin{deluxetable*}{llccccc}
\tabletypesize{\scriptsize}
\tablecaption{Results of the Radio Fitting \label{TA3_results3}}
\tablehead{
\colhead{ID} &
\colhead{Name} &
\colhead{$\alpha_{\rm radio}$} &
\colhead{$q_{\rm ir}$} &
\colhead{log$L_{\rm 1.4GHz}$} &
\colhead{$q_{\rm excess}$} &
\colhead{$\chi^2_{\rm red}$} \\
 & & & & (erg s$^{-1}$) & &  
}
\decimalcolnumbers
\startdata
ID01 & NGC~34 & 0.41$\pm$0.07 & 2.77$\pm$0.06 & 38.91$\pm$0.04 & 21.25$\pm$0.08 & 2.1 \\
ID02 & MCG$-$02-01-052/MCG$-$02-01-051 & 0.44$\pm$0.09 & 2.54$\pm$0.05 & 39.03$\pm$0.05 & 21.42$\pm$0.08 & 1.4 \\
ID03 & MCG$-$02-01-052 & 0.5$^{\rm (f)}$ & 2.76$\pm$0.08 & 37.99$\pm$0.07 & 21.28$\pm$0.07 & 4.2 \\
ID04 & MCG$-$02-01-051 & 0.49$\pm$0.09 & 2.50$\pm$0.05 & 38.99$\pm$0.05 & 21.42$\pm$0.08 & 0.5 \\
ID05 & ESO~350$-$38 & 0.56$\pm$0.07 & 2.36$\pm$0.04 & 38.54$\pm$0.04 & 21.44$\pm$0.08 & 2.1 \\
ID06 & NGC~232/NGC~235 & 0.68$\pm$0.06 & 2.43$\pm$0.05 & 39.20$\pm$0.04 & 21.70$\pm$0.06 & 1.8 \\
ID07 & NGC~232 & 0.52$\pm$0.09 & 2.58$\pm$0.05 & 38.98$\pm$0.04 & 21.49$\pm$0.09 & 1.8 \\
ID08 & NGC~235 & 0.57$\pm$0.08 & 1.93$\pm$0.05 & 38.81$\pm$0.04 & 22.11$\pm$0.05 & 2.2 \\
ID09 & MCG+12-02-001 & 0.52$\pm$0.06 & 2.57$\pm$0.05 & 38.96$\pm$0.04 & 21.41$\pm$0.07 & 1.7 \\
ID10 & IC~1623A/IC~1623B & 0.67$\pm$0.05 & 2.25$\pm$0.05 & 39.52$\pm$0.04 & 21.56$\pm$0.05 & 2.7 \\
ID11 & NGC~833/NGC~835 & 0.61$\pm$0.07 & 2.49$\pm$0.05 & 38.41$\pm$0.05 & 21.90$\pm$0.25 & 1.6 \\
ID12 & NGC~833 & 0.5$^{\rm (f)}$ & 2.59$\pm$0.06 & 37.02$\pm$0.06 & 22.66$\pm$0.34 & 3.1 \\
ID13 & NGC~835 & 0.60$\pm$0.09 & 2.46$\pm$0.05 & 38.43$\pm$0.05 & 21.70$\pm$0.08 & 1.6 \\
ID14 & NGC~838 & 0.63$\pm$0.06 & 2.57$\pm$0.05 & 38.65$\pm$0.04 & 21.39$\pm$0.08 & 1.2 \\
ID15 & NGC~839 & 0.51$\pm$0.10 & 2.98$\pm$0.06 & 38.24$\pm$0.05 & 21.18$\pm$0.05 & 3.3 \\
ID16 & NGC~1068 & 0.56$\pm$0.04 & 2.09$\pm$0.05 & 39.23$\pm$0.04 & 21.83$\pm$0.09 & 3.6 \\
ID17 & UGC~2608/UGC~2612 & 0.79$\pm$0.07 & 2.25$\pm$0.05 & 39.37$\pm$0.04 & 21.86$\pm$0.08 & 0.8 \\
ID18 & UGC~2608 & 0.78$\pm$0.07 & 2.15$\pm$0.05 & 39.22$\pm$0.04 & 21.83$\pm$0.11 & 1.0 \\
ID19 & UGC~2612 & 0.5$^{\rm (f)}$ & 3.47$\pm$0.03 & 37.40$\pm$0.02 & 20.72$\pm$0.07 & 4.0 \\
ID20 & NGC~1275 & 0.47$\pm$0.02 & $-$0.57$\pm$0.03 & 41.29$\pm$0.03 & 24.24$\pm$0.05 & 3.6 \\
ID21 & NGC~1365 & 0.65$\pm$0.05 & 2.85$\pm$0.04 & 38.30$\pm$0.03 & 21.23$\pm$0.10 & 2.3 \\
ID22 & ESO~203$-$1 & 0.5$^{\rm (f)}$ & 2.94$\pm$0.07 & 39.01$\pm$0.06 & 21.08$\pm$0.08 & 1.3 \\
ID23 & CGCG~468-002W/CGCG~468-002E & 0.5$^{\rm (f)}$ & 2.76$\pm$0.06 & 38.47$\pm$0.05 & 21.45$\pm$0.20 & 2.0 \\
ID24 & CGCG~468-002W & 0.5$^{\rm (f)}$ & 2.79$\pm$0.06 & 37.84$\pm$0.05 & 22.14$\pm$0.48 & 3.4 \\
ID25 & CGCG~468-002E & 0.5$^{\rm (f)}$ & 3.39$\pm$0.06 & 38.07$\pm$0.03 & 20.71$\pm$0.05 & 2.3 \\
ID26 & IRAS~F05189$-$2524 & 0.5$^{\rm (f)}$ & 2.49$\pm$0.05 & 39.22$\pm$0.05 & 21.51$\pm$0.13 & 3.7 \\
ID27 & IRAS~F06076$-$2139/ & 0.36$\pm$0.09 & 2.67$\pm$0.05 & 38.97$\pm$0.05 & 21.18$\pm$0.12 & 0.7 \\
 & 2MASS~06094601$-$2140312 \\
ID28 & NGC~2623 & 0.26$\pm$0.08 & 2.66$\pm$0.05 & 39.00$\pm$0.04 & 21.32$\pm$0.07 & 3.1 \\
ID29 & ESO~060$-$IG016~West/East & 0.5$^{\rm (f)}$ & 2.62$\pm$0.05 & 39.19$\pm$0.05 & 21.36$\pm$0.06 & 1.0 \\
ID30 & IRAS~F08572+3915 & 0.5$^{\rm (f)}$ & 3.40$\pm$0.07 & 38.69$\pm$0.04 & 20.65$\pm$0.04 & 11.3 \\
ID31 & UGC~5101 & 0.22$\pm$0.06 & 2.14$\pm$0.05 & 39.92$\pm$0.04 & 21.78$\pm$0.09 & 3.3 \\
ID32 & MCG+08-18-012/MCG+08-18-013 & 0.41$\pm$0.08 & 2.56$\pm$0.05 & 38.85$\pm$0.04 & 21.41$\pm$0.07 & 2.7 \\
ID33 & MCG+08-18-012 & 0.5$^{\rm (f)}$ & 3.20$\pm$0.19 & 36.91$\pm$0.11 & 20.66$\pm$0.11 & 3.9 \\
ID34 & MCG+08-18-013 & 0.40$\pm$0.08 & 2.55$\pm$0.05 & 38.87$\pm$0.04 & 21.42$\pm$0.10 & 1.3 \\
ID35 & MCG$-$01-26-013/NGC~3110 & 0.64$\pm$0.05 & 2.31$\pm$0.05 & 39.06$\pm$0.04 & 21.66$\pm$0.10 & 1.4 \\
ID36 & MCG$-$01-26-013 & 0.5$^{\rm (f)}$ & 2.45$\pm$0.06 & 37.82$\pm$0.05 & 21.43$\pm$0.07 & 2.2 \\
ID37 & NGC~3110 & 0.64$\pm$0.06 & 2.32$\pm$0.05 & 39.06$\pm$0.04 & 21.74$\pm$0.06 & 3.4 \\
ID38 & ESO~374$-$IG032 & 0.51$\pm$0.08 & 2.87$\pm$0.05 & 38.92$\pm$0.04 & 21.06$\pm$0.06 & 5.3 \\
ID39 & NGC~3256 & 0.52$\pm$0.06 & 2.52$\pm$0.06 & 39.36$\pm$0.05 & 21.41$\pm$0.08 & 1.3 \\
ID40 & IRAS~F10565+2448 & 0.28$\pm$0.08 & 2.50$\pm$0.05 & 39.51$\pm$0.04 & 21.42$\pm$0.10 & 4.8 \\
ID41 & NGC~3690~West/East & 0.50$\pm$0.05 & 2.34$\pm$0.05 & 39.37$\pm$0.04 & 21.53$\pm$0.11 & 2.2 \\
ID42 & ESO~440$-$58/MCG$-$05-29-017 & 0.58$\pm$0.06 & 2.53$\pm$0.04 & 38.95$\pm$0.03 & 21.49$\pm$0.05 & 1.5 \\
ID43 & IRAS~F12112+0305 & 0.20$\pm$0.10 & 2.84$\pm$0.06 & 39.59$\pm$0.04 & 21.10$\pm$0.05 & 5.2 \\
ID44 & NGC~4418/MCG+00-32-013 & 0.5$^{\rm (f)}$ & 3.21$\pm$0.07 & 37.89$\pm$0.05 & 20.93$\pm$0.10 & 3.1 \\
ID45 & NGC~4418 & 0.5$^{\rm (f)}$ & 3.42$\pm$0.05 & 37.91$\pm$0.03 & 20.80$\pm$0.06 & 3.2 \\
ID46 & MCG+00-32-013 & 0.5$^{\rm (f)}$ & 3.31$\pm$0.13 & 35.98$\pm$0.07 & 20.63$\pm$0.07 & 4.2 \\
ID47 & Mrk~231 & 0.37$\pm$0.05 & 2.27$\pm$0.05 & 40.25$\pm$0.04 & 21.67$\pm$0.09 & 0.3 \\
ID48 & NGC~4922S/NGC~4922N & 0.33$\pm$0.07 & 2.37$\pm$0.05 & 38.84$\pm$0.04 & 21.78$\pm$0.07 & 0.9 \\
ID49 & IC~860 & 0.19$\pm$0.10 & 2.98$\pm$0.06 & 38.08$\pm$0.04 & 21.08$\pm$0.05 & 2.8 \\
ID50 & IRAS~13120$-$5453 & 0.5$^{\rm (f)}$ & 2.41$\pm$0.05 & 39.98$\pm$0.04 & 21.49$\pm$0.08 & 1.9 \\
ID51 & NGC~5104 & 0.60$\pm$0.07 & 2.60$\pm$0.05 & 38.63$\pm$0.04 & 21.52$\pm$0.07 & 2.6 \\
ID52 & MCG$-$03-34-064 & 0.35$\pm$0.04 & 1.44$\pm$0.05 & 39.16$\pm$0.04 & 22.55$\pm$0.11 & 2.5 \\
ID53 & NGC~5135 & 0.60$\pm$0.05 & 2.31$\pm$0.05 & 39.08$\pm$0.04 & 21.75$\pm$0.10 & 0.9 \\
ID54 & Mrk~266B/Mrk~266A & 0.60$\pm$0.06 & 2.01$\pm$0.04 & 39.50$\pm$0.04 & 22.01$\pm$0.07 & 1.5 \\
ID55 & Mrk~273 & 0.41$\pm$0.06 & 2.46$\pm$0.05 & 39.83$\pm$0.04 & 21.44$\pm$0.06 & 2.2 \\
ID56 & IRAS~F14348$-$1447 & 0.42$\pm$0.08 & 2.43$\pm$0.05 & 39.91$\pm$0.04 & 21.44$\pm$0.08 & 2.3 \\
ID57 & IRAS~F14378$-$3651 & 0.47$\pm$0.08 & 2.60$\pm$0.05 & 39.75$\pm$0.04 & 21.28$\pm$0.04 & 2.1 \\
ID58 & IC~4518A/IC~4518B & 0.40$\pm$0.08 & 1.90$\pm$0.06 & 39.21$\pm$0.06 & 21.83$\pm$0.23 & 0.7 \\
ID59 & IRAS~F15250+3608 & 0.33$\pm$0.16 & 3.05$\pm$0.07 & 39.15$\pm$0.05 & 20.88$\pm$0.05 & 2.6 \\
ID60 & Arp~220W/Arp~220E & 0.09$\pm$0.06 & 2.59$\pm$0.05 & 39.51$\pm$0.04 & 21.24$\pm$0.06 & 5.3 \\
ID61 & NGC~6240S/NGC~6240N & 0.69$\pm$0.04 & 1.93$\pm$0.04 & 39.93$\pm$0.04 & 22.18$\pm$0.06 & 1.5 \\
ID62 & NGC~6285/NGC~6286 & 0.44$\pm$0.05 & 2.23$\pm$0.04 & 39.29$\pm$0.04 & 21.79$\pm$0.05 & 4.2 \\
ID63 & NGC~6285 & 0.5$^{\rm (f)}$ & 2.60$\pm$0.06 & 38.09$\pm$0.05 & 21.33$\pm$0.07 & 2.1 \\
ID64 & NGC~6286 & 0.46$\pm$0.05 & 2.09$\pm$0.04 & 39.25$\pm$0.04 & 22.01$\pm$0.06 & 5.0 \\
ID65 & IRAS~F17138$-$1017 & 0.55$\pm$0.08 & 2.75$\pm$0.06 & 38.80$\pm$0.04 & 21.17$\pm$0.09 & 4.6 \\
ID66 & IRAS~F18293$-$3413 & 0.65$\pm$0.05 & 2.56$\pm$0.04 & 39.37$\pm$0.03 & 21.44$\pm$0.07 & 0.8 \\
ID67 & IRAS~F19297$-$0406 & 0.47$\pm$0.09 & 2.49$\pm$0.05 & 39.85$\pm$0.04 & 21.31$\pm$0.16 & 1.9 \\
ID68 & NGC~6907/NGC~6908 & 0.57$\pm$0.06 & 2.55$\pm$0.05 & 38.69$\pm$0.04 & 21.47$\pm$0.06 & 2.8 \\
ID69 & NGC~6921/MCG+04-48-002 & 0.59$\pm$0.08 & 2.32$\pm$0.05 & 38.84$\pm$0.04 & 21.76$\pm$0.14 & 0.4 \\
ID70 & NGC~6921 & 0.61$\pm$0.09 & 2.13$\pm$0.05 & 38.25$\pm$0.05 & 22.44$\pm$0.25 & 0.7 \\
ID71 & MCG+04-48-002 & 0.58$\pm$0.09 & 2.36$\pm$0.05 & 38.70$\pm$0.05 & 21.68$\pm$0.10 & 0.6 \\
ID72 & II~Zw~096/IRAS~F20550+1655~SE & 0.77$\pm$0.07 & 2.65$\pm$0.06 & 39.23$\pm$0.05 & 21.18$\pm$0.05 & 3.1 \\
ID73 & ESO~286$-$19 & 0.24$\pm$0.10 & 2.82$\pm$0.07 & 39.32$\pm$0.06 & 21.20$\pm$0.06 & 2.2 \\
ID74 & NGC~7130 & 0.38$\pm$0.05 & 2.09$\pm$0.04 & 39.19$\pm$0.03 & 21.74$\pm$0.08 & 1.3 \\
ID75 & NGC~7469/IC~5283 & 0.53$\pm$0.08 & 2.37$\pm$0.05 & 39.19$\pm$0.04 & 21.56$\pm$0.09 & 0.4 \\
ID76 & NGC~7469 & 0.50$\pm$0.08 & 2.42$\pm$0.05 & 39.17$\pm$0.04 & 21.51$\pm$0.09 & 0.5 \\
ID77 & IC~5283 & 0.86$\pm$0.09 & 2.72$\pm$0.06 & 38.02$\pm$0.05 & 21.38$\pm$0.17 & 1.1 \\
ID78 & ESO~148$-$2 & 0.47$\pm$0.16 & 2.81$\pm$0.08 & 39.30$\pm$0.06 & 21.22$\pm$0.08 & 2.4 \\
ID79 & NGC~7591 & 0.46$\pm$0.09 & 2.57$\pm$0.05 & 38.64$\pm$0.05 & 21.46$\pm$0.06 & 2.1 \\
ID80 & NGC~7674/MCG+01-59-081 & 0.43$\pm$0.06 & 1.73$\pm$0.05 & 39.75$\pm$0.05 & 22.36$\pm$0.11 & 0.8 \\
ID81 & NGC~7674 & 0.5$^{\rm (f)}$ & 1.84$\pm$0.07 & 39.60$\pm$0.06 & 22.23$\pm$0.11 & 0.6 \\
ID82 & MCG+01-59-081 & 0.5$^{\rm (f)}$ & 3.21$\pm$0.08 & 37.39$\pm$0.06 & 20.92$\pm$0.09 & 0.6 \\
ID83 & NGC~7679/NGC~7682 & 0.32$\pm$0.08 & 2.21$\pm$0.05 & 39.02$\pm$0.04 & 21.82$\pm$0.05 & 1.5 \\
ID84 & NGC~7679 & 0.44$\pm$0.09 & 2.51$\pm$0.05 & 38.70$\pm$0.04 & 21.51$\pm$0.05 & 1.0 \\
ID85 & NGC~7682 & 0.21$\pm$0.07 & 1.29$\pm$0.05 & 38.72$\pm$0.04 & 22.78$\pm$0.23 & 1.5 \\
\enddata
\tablecomments{Comments: (1--2) ID and object name; 
(3) radio spectral index of the power-law synchrotron emission. 
The symbol ``(f)'' means that the value is fixed; 
(4) the far-IR/radio correlation coefficient;
(5) rest-frame radio 1.4~GHz luminosity;
(6) radio-excess parameter $q_{\rm excess}$ = log($L_{\rm 1.4GHz}$/SFR); 
(7) reduced $\chi^2$ for the combination of the radio fitting and UV-to-IR SED decomposition.}
\tablenotetext{\ }{(This table is available in its entirety in machine-readable form.)}
\end{deluxetable*}

% Table A4
\clearpage
\startlongtable
\begin{deluxetable*}{llccccccccc}
\tabletypesize{\scriptsize}
\tablecaption{Summary of the logarithmic AGN luminosities \label{TA4_results4}}
\tablehead{
\colhead{ID} &
\colhead{Name} &
\colhead{$L_{\rm 6,AGN}$} &
\colhead{$L_{\rm 12,t}$} &
\colhead{$L_{\rm 12,p}$} &
\colhead{$L_{\rm 12,AGN}$} &
\colhead{$L_{\rm 12,nuc}$} &
\colhead{$L_{\rm X,unabs}$} &
\colhead{$L_{\rm 2keV,unabs}$} &
\colhead{$L_{\rm 2500,disk}$} &
\colhead{$\alpha_{\rm OX}$}
}
\decimalcolnumbers
\startdata
ID01 & NGC~34 & 43.14 & 43.46 & 43.70 & 43.89 & 43.10 & 41.90$_{-0.10}^{+0.10}$ & 41.63 & 43.07 & $-$1.55\\
ID05 & ESO~350$-$38 & 43.23 & 43.87 & 42.93 & 43.92 & $\cdots$ & $\cdots$ & $\cdots$ & 43.84 & $\cdots$\\
ID06 & NGC~232/NGC~235 & 43.01 & 43.59 & $\cdots$ & 43.59 & $\cdots$ & 43.28$_{-0.16}^{+0.18}$ & 42.98 & 43.24 & $-$1.10\\
ID08 & NGC~235 & 42.68 & 43.25 & $\cdots$ & 43.25 & 43.29 & 43.28$_{-0.16}^{+0.18}$ & 42.97 & 42.84 & $-$0.95\\
ID09 & MCG+12-02-001 & 42.54 & 43.15 & 43.09 & 43.42 & $\cdots$ & 41.75$_{-0.08}^{+0.33}$ & 41.48 & 42.88 & $-$1.54\\
ID11 & NGC~833/NGC~835 & 42.49 & 42.77 & 43.32 & 43.43 & $\cdots$ & 42.33$_{-0.24}^{+0.28}$ & 42.04 & 42.54 & $-$1.19\\
ID12 & NGC~833 & 41.37 & 41.95 & $\cdots$ & 41.95 & $\cdots$ & 41.99$_{-0.24}^{+0.25}$ & 41.78 & 41.51 & $-$0.89\\
ID13 & NGC~835 & 42.31 & 42.89 & 42.75 & 43.13 & $\cdots$ & 42.06$_{-0.03}^{+0.13}$ & 41.67 & 42.59 & $-$1.35\\
ID16 & NGC~1068 & 43.54 & 43.93 & 44.52 & 44.62 & 43.80 & 43.34$_{-0.08}^{+0.07}$ & $\cdots$ & 43.96 & $\cdots$\\
ID17 & UGC~2608/UGC~2612 & 43.84 & 44.48 & $\cdots$ & 44.48 & $\cdots$ & 43.59$_{-0.25}^{+0.19}$ & 43.48 & 44.55 & $-$1.41\\
ID18 & UGC~2608 & 43.49 & 44.12 & $\cdots$ & 44.12 & $\cdots$ & 43.59$_{-0.25}^{+0.19}$ & 43.33 & 44.26 & $-$1.36\\
ID20 & NGC~1275$^{\dagger}$ & 43.88 & 44.43 & 43.10 & 44.45 & 44.21 & $\cdots$ & 43.19 & 43.84 & $-$1.25\\
ID21 & NGC~1365 & 42.79 & 43.23 & 43.44 & 43.65 & 42.51 & 41.90$_{-0.01}^{+0.01}$ & 41.71 & 42.65 & $-$1.36\\
ID23 & CGCG~468-002W/CGCG~468-002E & 42.98 & 43.55 & 42.39 & 43.58 & $\cdots$ & 42.84$_{-0.03}^{+0.04}$ & 42.49 & 43.14 & $-$1.25\\
ID24 & CGCG~468-002W & 42.44 & 42.97 & 42.83 & 43.21 & $\cdots$ & 42.84$_{-0.03}^{+0.04}$ & 42.51 & 42.64 & $-$1.05\\
ID26 & IRAS~F05189$-$2524 & 44.38 & 44.92 & 43.89 & 44.95 & 44.87 & 43.42$_{-0.02}^{+0.01}$ & 43.30 & 44.60 & $-$1.50\\
ID27 & IRAS~F06076$-$2139/ & 42.94 & 43.55 & 43.50 & 43.82 & $\cdots$ & 42.18$_{-0.14}^{+0.15}$ & 41.92 & 43.14 & $-$1.47\\
 & 2MASS~06094601$-$2140312\\
ID28 & NGC~2623 & 42.77 & 43.37 & 42.11 & 43.39 & 43.61 & 40.90$_{-0.11}^{+0.11}$ & 40.64 & 43.00 & $-$1.91\\
ID29 & ESO~060$-$IG016~West/East & 43.60 & 44.14 & $\cdots$ & 44.14 & $\cdots$ & 41.94$_{-0.08}^{+0.07}$ & 41.69 & 43.73 & $-$1.78\\
ID30 & IRAS~F08572+3915 & 44.04 & 44.70 & 44.79 & 45.05 & 45.13 & 41.77$_{-0.07}^{+0.06}$ & 41.52 & 44.61 & $-$2.19\\
ID31 & UGC~5101 & 43.51 & 44.05 & 42.97 & 44.08 & 44.35 & 43.15$_{-0.15}^{+0.25}$ & 42.77 & 43.72 & $-$1.36\\
ID38 & ESO~374$-$IG032 & 43.39 & 43.93 & 43.43 & 44.05 & $\cdots$ & $\cdots$ & $\cdots$ & 43.28 & $\cdots$\\
ID41 & NGC~3690~West/East & 44.04 & 44.70 & 42.89 & 44.70 & $\cdots$ & 42.66$_{-0.32}^{+0.34}$ & 42.55 & 43.80 & $-$1.48\\
ID44 & NGC~4418/MCG+00-32-013 & 42.33 & 43.11 & 43.55 & 43.68 & $\cdots$ & $\cdots$ & $\cdots$ & 43.32 & $\cdots$\\
ID45 & NGC~4418 & 42.24 & 42.75 & 43.49 & 43.56 & 43.70 & $\cdots$ & $\cdots$ & 42.89 & $\cdots$\\
ID47 & Mrk~231 & 44.60 & 45.20 & 44.92 & 45.38 & $\cdots$ & 42.65$_{-0.02}^{+0.02}$ & 42.28 & 44.71 & $-$1.93\\
ID48 & NGC~4922S/NGC~4922N & 43.39 & 44.02 & 42.86 & 44.05 & $\cdots$ & 42.00$_{-0.18}^{+0.20}$ & 41.74 & 43.52 & $-$1.68\\
ID50 & IRAS~13120$-$5453 & 43.29 & 43.94 & 43.00 & 43.99 & $\cdots$ & 42.17$_{-0.11}^{+0.59}$ & 41.91 & 43.88 & $-$1.76\\
ID52 & MCG$-$03-34-064 & 43.49 & 43.95 & 44.18 & 44.38 & 44.00 & 43.27$_{-0.07}^{+0.06}$ & 43.08 & 43.53 & $-$1.17\\
ID53 & NGC~5135 & 42.18 & 42.82 & 42.04 & 42.89 & 43.23 & 43.30$_{-0.26}^{+0.42}$ & 43.01 & 42.90 & $-$0.96\\
ID54 & Mrk~266B/Mrk~266A & 43.07 & 43.56 & 43.69 & 43.93 & $\cdots$ & $\cdots$ & 42.71 & 43.17 & $-$1.18\\
ID55 & Mrk~273 & 43.41 & 43.94 & 42.91 & 43.98 & $\cdots$ & 43.07$_{-0.21}^{+0.21}$ & 42.78 & 43.73 & $-$1.36\\
ID56 & IRAS~F14348$-$1447 & 43.33 & 44.05 & 43.27 & 44.12 & $\cdots$ & 42.70$_{-0.40}^{+0.93}$ & 42.46 & 44.07 & $-$1.62\\
ID58 & IC~4518A/IC~4518B & 43.14 & 43.39 & 43.81 & 43.95 & 43.54 & 42.83$_{-0.08}^{+0.06}$ & 42.55 & 42.93 & $-$1.14\\
ID60 & Arp~220W/Arp~220E & 43.10 & 43.70 & 43.64 & 43.97 & $\cdots$ & 41.59$_{-0.02}^{+0.02}$ & 41.33 & 43.38 & $-$1.79\\
ID61 & NGC~6240S/NGC~6240N & 43.45 & 44.01 & 43.52 & 44.14 & $\cdots$ & $\cdots$ & 43.57 & 43.42 & $-$0.94\\
ID62 & NGC~6285/NGC~6286 & 42.10 & 42.94 & 42.20 & 43.01 & $\cdots$ & 42.01$_{-0.23}^{+1.21}$ & 41.64 & 42.98 & $-$1.51\\
ID64 & NGC~6286 & 41.71 & 42.54 & 41.96 & 42.64 & $\cdots$ & 42.01$_{-0.23}^{+1.21}$ & 41.63 & 42.80 & $-$1.45\\
ID65 & IRAS~F17138$-$1017 & 43.32 & 43.92 & 42.80 & 43.95 & $\cdots$ & 41.68$_{-0.06}^{+0.09}$ & 41.41 & 43.27 & $-$1.71\\
ID69 & NGC~6921/MCG+04-48-002 & 42.96 & 43.52 & $\cdots$ & 43.52 & $\cdots$ & 42.96$_{-0.46}^{+0.40}$ & 42.64 & 43.12 & $-$1.19\\
ID70 & NGC~6921 & 41.48 & 42.47 & $\cdots$ & 42.47 & $\cdots$ & 42.80$_{-0.44}^{+0.37}$ & 42.49 & 42.48 & $-$1.00\\
ID71 & MCG+04-48-002 & 42.80 & 43.37 & $\cdots$ & 43.37 & $\cdots$ & 42.44$_{-0.15}^{+0.15}$ & 42.12 & 43.02 & $-$1.34\\
ID74 & NGC~7130 & 42.91 & 43.71 & 42.75 & 43.75 & 43.18 & 42.87$_{-0.25}^{+0.31}$ & 42.55 & 43.58 & $-$1.39\\
ID75 & NGC~7469/IC~5283 & 43.76 & 44.35 & 43.20 & 44.38 & $\cdots$ & 43.26$_{-0.02}^{+0.03}$ & 43.06 & 43.85 & $-$1.30\\
ID76 & NGC~7469 & 43.63 & 44.21 & 43.07 & 44.24 & 43.83 & 43.26$_{-0.02}^{+0.03}$ & 43.07 & 43.84 & $-$1.29\\
ID78 & ESO~148$-$2 & 43.65 & 44.29 & 43.35 & 44.34 & $\cdots$ & 42.38$_{-0.28}^{+0.24}$ & 42.13 & 44.25 & $-$1.81\\
ID80 & NGC~7674/MCG+01-59-081 & 44.04 & 44.58 & 42.55 & 44.59 & $\cdots$ & 42.59$_{-0.23}^{+0.33}$ & 42.29 & 43.73 & $-$1.55\\
ID81 & NGC~7674 & 44.03 & 44.54 & 43.53 & 44.58 & 44.26 & 42.59$_{-0.23}^{+0.33}$ & 42.28 & 43.71 & $-$1.55\\
ID83 & NGC~7679/NGC~7682 & 42.67 & 43.16 & 43.47 & 43.65 & $\cdots$ & 42.49$_{-0.07}^{+0.08}$ & 42.21 & 42.70 & $-$1.19\\
ID84 & NGC~7679 & 42.58 & 42.96 & 43.34 & 43.49 & 42.74 & 42.34$_{-0.04}^{+0.06}$ & 42.06 & 42.47 & $-$1.16\\
ID85 & NGC~7682 & 41.05 & 42.13 & $\cdots$ & 42.13 & $\cdots$ & 41.94$_{-0.06}^{+0.06}$ & 41.68 & 41.94 & $-$1.00\\
\enddata
\tablecomments{Comments: (1--2) ID and object name; 
(3) logarithmic rest-frame 6~$\mu$m luminosity of AGN (torus and polar dust) component ($L_{\rm 6,AGN}$);
(4)--(6) logarithmic rest-frame 12~$\mu$m luminosities of torus component ($L_{\rm 12,t}$), polar component ($L_{\rm 12,p}$), and both components ($L_{\rm 12,AGN}$);
(7) logarithmic nuclear 12~$\mu$m luminosity ($L_{\rm 12,nuc}$) referred from \citet{Asmus2014,Asmus2015};
(8)--(9) logarithmic unabsorbed (absorption-corrected) AGN luminosities in the rest-frame 2--10~keV ($L_{\rm X,unabs}$) and 2~keV bands ($L_{\rm 2keV,unabs}$);
(10) logarithmic extinction-corrected intrinsic AGN disk luminosity in the rest-frame 2500~\AA\ band ($L_{\rm 2500,disk}$);
(11) X-ray to optical spectral index computed by $\alpha_{\rm OX} \equiv$ log($L''_{\rm 2keV}$/$L''_{\rm 2500}$)/log($\nu_{\rm 2keV}$/$\nu_{\rm 2500}$), where $L''_{\rm 2keV}$ = $L_{\rm 2keV,unabs}$/$\nu_{\rm 2keV}$ and $L''_{\rm 2500}$ = $L_{\rm 2500,disk}$/$\nu_{\rm 2500}$.}
\tablenotetext{\dagger}{Large fractions of the mid-IR and X-ray luminosities in NGC~1275 could be explained by the jet emission \citep[e.g.,][]{Hitomi-Collaboration2018,Rani2018}.}
\tablenotetext{\ }{(This table is available in its entirety in machine-readable form.)}
\end{deluxetable*}

% Table A5
%\clearpage
%\startlongtable
\begin{longrotatetable}
\begin{deluxetable*}{llcccccccccc}
\tabletypesize{\scriptsize}
\tablecaption{Summary of the Extinction, AGN Luminosities, and Polar Dust Sizes \label{TA5_results5}}
\tablehead{
\colhead{ID} &
\colhead{Name} &
\colhead{$N_{\rm H}^{\rm (LOS)}$} &
\colhead{$A_{V{\rm,torus}}^{\rm (LOS)}$} &
\colhead{$A_{V{\rm,polar}}^{\rm (LOS)}$} &
\colhead{log$L_{\rm torus}$} &
\colhead{log$L_{\rm polar}$} &
\colhead{log$L_{\rm AGN,int}$} &
\colhead{log$M_{\rm BH}$} &
\colhead{log$\lambda_{\rm Edd}$} &
\colhead{log$R_{\rm polar}$} &
\colhead{log$R_{\rm (mir/io/mo)}$} \\
&  & (10$^{22}$ cm$^{-2}$) & (mag) & (mag) & (erg s$^{-1}$) & (erg s$^{-1}$) & (erg s$^{-1}$) & ($M_{\odot}$) & & (pc) & (pc) 
}
\decimalcolnumbers
\startdata
ID01 & NGC~34 & 50.3$_{-9.9}^{+10.3}$ & 108.3$\pm$32.0 & 0.2$\pm$0.4 & 43.54$\pm$0.16 & 43.53$\pm$0.16 & 43.84$\pm$0.16 & 7.82 & $-$2.08$\pm$0.16 & 1.67$\pm$0.40 & $\cdots$/$\cdots$/$\cdots$\\
ID05 & ESO~350$-$38 & $\cdots$ & 346.6$\pm$58.5 & 0 & 43.96$\pm$0.04 & 44.30$\pm$0.04 & 44.62$\pm$0.04 & $\cdots$ & $\cdots$ & 2.71$\pm$0.02 & $\cdots$/$\cdots$/$\cdots$\\
ID06 & NGC~232/NGC~235 & 51.4$_{-6.1}^{+8.3}$ & 106.7$\pm$32.3 & $\cdots$ & 43.75$\pm$0.02 & $\cdots$ & 44.01$\pm$0.02 & 8.20 & $-$2.29$\pm$0.02 & $\cdots$ & $\cdots$/$\cdots$/$\cdots$\\
ID08 & NGC~235 & 51.4$_{-6.1}^{+8.3}$ & 99.4$\pm$33.7 & $\cdots$ & 43.35$\pm$0.03 & $\cdots$ & 43.61$\pm$0.03 & 8.20 & $-$2.69$\pm$0.03 & $\cdots$ & $\cdots$/$\cdots$/$\cdots$\\
ID09 & MCG+12-02-001 & 707.5$_{-277.1}^{+268.4}$ & 110.3$\pm$31.5 & 1.7$\pm$0.9 & 43.33$\pm$0.14 & 43.48$\pm$0.16 & 43.66$\pm$0.15 & 7.04 & $-$1.48$\pm$0.15 & 1.55$\pm$0.32 & $\cdots$/$\cdots$/$\cdots$\\
ID11 & NGC~833/NGC~835 & $\cdots$ & 17.4$\pm$6.2 & 1.7$\pm$0.9 & 42.91$\pm$0.07 & 43.17$\pm$0.04 & 43.31$\pm$0.06 & $\cdots$ & $\cdots$ & 1.21$\pm$0.31 & $\cdots$/$\cdots$/$\cdots$\\
ID12 & NGC~833 & 26.1$_{-1.1}^{+1.2}$ & 16.2$\pm$6.4 & $\cdots$ & 41.95$\pm$0.17 & $\cdots$ & 42.28$\pm$0.17 & 8.61 & $-$4.43$\pm$0.17 & $\cdots$ & $\cdots$/$\cdots$/$\cdots$\\
ID13 & NGC~835 & 29.8$_{-1.5}^{+0.8}$ & 18.2$\pm$5.9 & 0.5$\pm$0.8 & 42.99$\pm$0.08 & 43.12$\pm$0.07 & 43.37$\pm$0.08 & 7.85 & $-$2.58$\pm$0.08 & 1.30$\pm$0.31 & $\cdots$/$\cdots$/$\cdots$\\
ID16 & NGC~1068 & $\sim$1000 & 664.8$\pm$278.0 & 0 & 44.04$\pm$0.07 & 44.49$\pm$0.05 & 44.74$\pm$0.07 & 7.35 & $-$0.71$\pm$0.07 & 1.65$\pm$0.04 & 1.73/$\cdots$/2.00\\
ID17 & UGC~2608/UGC~2612 & 540$_{-310}^{+700}$ & 436.8$\pm$209.2 & $\cdots$ & 44.52$\pm$0.09 & $\cdots$ & 45.33$\pm$0.09 & 7.64 & $-$0.42$\pm$0.09 & $\cdots$ & $\cdots$/$\cdots$/$\cdots$\\
ID18 & UGC~2608 & 540$_{-310}^{+700}$ & 447.1$\pm$220.0 & $\cdots$ & 44.22$\pm$0.16 & $\cdots$ & 45.04$\pm$0.17 & 7.64 & $-$0.71$\pm$0.17 & $\cdots$ & $\cdots$/$\cdots$/$\cdots$\\
ID20 & NGC~1275$^{\dagger}$ & $\cdots$ & 0.1$\pm$0.1 & 2.5$\pm$0.1 & 44.46$\pm$0.02 & 44.48$\pm$0.02 & 44.62$\pm$0.02 & $\cdots$ & $\cdots$ & 2.70$\pm$0.01 & $\cdots$/$\cdots$/$\cdots$\\
ID21 & NGC~1365 & $\sim$10 & 0.1$\pm$0.1 & 2.5$\pm$0.1 & 43.28$\pm$0.02 & 43.30$\pm$0.02 & 43.43$\pm$0.02 & 8.13 & $-$2.80$\pm$0.02 & 1.07$\pm$0.13 & 1.74/$\cdots$/$\cdots$\\
ID23 & CGCG~468-002W/CGCG~468-002E & 1.5$_{-0.1}^{+0.1}$ & 16.2$\pm$6.4 & 0.2$\pm$0.1 & 43.55$\pm$0.12 & 43.64$\pm$0.12 & 43.91$\pm$0.12 & 7.67 & $-$1.86$\pm$0.12 & 2.07$\pm$0.34 & $\cdots$/$\cdots$/$\cdots$\\
ID24 & CGCG~468-002W & 1.5$_{-0.1}^{+0.1}$ & 16.9$\pm$6.3 & 0.2$\pm$0.2 & 43.05$\pm$0.10 & 43.15$\pm$0.13 & 43.41$\pm$0.10 & 7.67 & $-$2.36$\pm$0.10 & 1.52$\pm$0.28 & $\cdots$/$\cdots$/$\cdots$\\
ID26 & IRAS~F05189$-$2524 & 7.5$_{-0.1}^{+0.1}$ & 8.0$\pm$0.5 & 2.5$\pm$0.1 & 44.96$\pm$0.02 & 45.27$\pm$0.02 & 45.38$\pm$0.02 & 7.94 & $-$0.66$\pm$0.02 & 3.08$\pm$0.01 & $\cdots$/3.48/2.28\\
ID27 & IRAS~F06076$-$2139/ & 42.2$_{-12.0}^{+24.0}$ & 107.7$\pm$31.7 & 1.7$\pm$0.9 & 43.59$\pm$0.08 & 43.73$\pm$0.10 & 43.92$\pm$0.08 & 7.30 & $-$1.48$\pm$0.08 & 1.97$\pm$0.24 & $\cdots$/$\cdots$/$\cdots$\\
 & 2MASS~06094601$-$2140312\\
ID28 & NGC~2623 & 6.0$_{-2.1}^{+4.5}$ & 119.3$\pm$23.9 & 0.2$\pm$0.1 & 43.47$\pm$0.02 & 43.45$\pm$0.02 & 43.77$\pm$0.02 & 7.62 & $-$1.95$\pm$0.02 & 2.28$\pm$0.06 & $\cdots$/$\cdots$/$\cdots$\\
ID29 & ESO~060$-$IG016~West/East & 8.4$_{-2.9}^{+4.0}$ & 91.9$\pm$35.7 & $\cdots$ & 44.25$\pm$0.02 & $\cdots$ & 44.51$\pm$0.02 & 7.79 & $-$1.38$\pm$0.02 & $\cdots$ & $\cdots$/$\cdots$/$\cdots$\\
ID30 & IRAS~F08572+3915 & 84.6$_{-28.3}^{+129.3}$ & 417.9$\pm$199.3 & 0 & 44.72$\pm$0.07 & 45.05$\pm$0.06 & 45.39$\pm$0.06 & 7.48 & $-$0.19$\pm$0.06 & 2.38$\pm$0.28 & $\cdots$/3.30/2.91\\
ID31 & UGC~5101 & 96.3$_{-1.6}^{+4.3}$ & 8.9$\pm$3.2 & 0.9$\pm$0.2 & 44.11$\pm$0.08 & 44.35$\pm$0.08 & 44.50$\pm$0.08 & 8.17 & $-$1.77$\pm$0.08 & 2.64$\pm$0.10 & $\cdots$/$\cdots$/$\cdots$\\
ID38 & ESO~374$-$IG032 & $\cdots$ & 0.1$\pm$0.1 & 0.2$\pm$0.2 & 43.95$\pm$0.03 & 43.75$\pm$0.07 & 44.05$\pm$0.04 & $\cdots$ & $\cdots$ & 1.77$\pm$0.32 & $\cdots$/$\cdots$/$\cdots$\\
ID41 & NGC~3690~West/East & 302.6$_{-49.7}^{+60.6}$ & 161.0$\pm$56.9 & 1.8$\pm$1.0 & 44.69$\pm$0.16 & 44.07$\pm$0.19 & 44.57$\pm$0.17 & 7.09 & $-$0.62$\pm$0.17 & 2.12$\pm$0.45 & $\cdots$/$\cdots$/$\cdots$\\
ID44 & NGC~4418/MCG+00-32-013 & $\cdots$ & 991.4$\pm$86.0 & 0 & 43.38$\pm$0.03 & 43.90$\pm$0.03 & 44.10$\pm$0.03 & $\cdots$ & $\cdots$ & 1.95$\pm$0.01 & $\cdots$/$\cdots$/$\cdots$\\
ID45 & NGC~4418 & $\cdots$ & 695.0$\pm$238.8 & 0 & 42.97$\pm$0.02 & 43.47$\pm$0.02 & 43.67$\pm$0.02 & $\cdots$ & $\cdots$ & 1.39$\pm$0.01 & $\cdots$/$\cdots$/2.76\\
ID47 & Mrk~231 & 8.5$_{-0.2}^{+0.2}$ & 202.3$\pm$83.7 & 1.3$\pm$0.9 & 45.21$\pm$0.11 & 45.23$\pm$0.11 & 45.49$\pm$0.10 & 7.92 & $-$0.53$\pm$0.10 & 2.39$\pm$0.36 & $\cdots$/3.48/2.78\\
ID48 & NGC~4922S/NGC~4922N & 75.9$_{-21.6}^{+285.1}$ & 93.1$\pm$37.0 & 1.2$\pm$0.6 & 43.98$\pm$0.12 & 44.11$\pm$0.12 & 44.30$\pm$0.12 & 7.79 & $-$1.59$\pm$0.12 & 2.18$\pm$0.38 & $\cdots$/$\cdots$/$\cdots$\\
ID50 & IRAS~13120$-$5453 & 164.8$_{-20.5}^{+21.6}$ & 413.0$\pm$183.4 & 0 & 43.99$\pm$0.06 & 44.37$\pm$0.06 & 44.66$\pm$0.06 & 8.43 & $-$1.87$\pm$0.06 & 2.72$\pm$0.07 & $\cdots$/2.48/2.25\\
ID52 & MCG$-$03-34-064 & 98.4$_{-2.4}^{+2.8}$ & 55.3$\pm$23.1 & 2.5$\pm$0.2 & 43.97$\pm$0.04 & 44.14$\pm$0.04 & 44.30$\pm$0.04 & 7.95 & $-$1.75$\pm$0.04 & 1.67$\pm$0.11 & 2.33/$\cdots$/$\cdots$\\
ID53 & NGC~5135 & 670$_{-280}^{+1660}$ & 548.1$\pm$228.5 & 0 & 42.86$\pm$0.21 & 43.48$\pm$0.20 & 43.68$\pm$0.21 & 7.77 & $-$2.19$\pm$0.21 & 1.85$\pm$0.52 & 2.11/2.63/$\cdots$\\
ID54 & Mrk~266B/Mrk~266A & $700/6.8$ & 111.9$\pm$30.2 & 0.3$\pm$0.5 & 43.65$\pm$0.10 & 43.64$\pm$0.09 & 43.95$\pm$0.10 & $\cdots$ & $\cdots$ & 1.64$\pm$0.31 & $\cdots$/$\cdots$/$\cdots$\\
ID55 & Mrk~273 & 49.6$_{-2.8}^{+5.5}$ & 14.4$\pm$6.5 & 2.4$\pm$0.3 & 44.09$\pm$0.13 & 44.40$\pm$0.13 & 44.51$\pm$0.13 & 8.35 & $-$1.94$\pm$0.13 & 2.64$\pm$0.10 & $\cdots$/3.60/2.74\\
ID56 & IRAS~F14348$-$1447 & 128.3$_{-55.4}^{+95.0}$ & 602.6$\pm$277.5 & 0 & 44.15$\pm$0.15 & 44.61$\pm$0.18 & 44.84$\pm$0.17 & 7.84 & $-$1.10$\pm$0.17 & 2.79$\pm$0.15 & $\cdots$/$\cdots$/2.55\\
ID58 & IC~4518A/IC~4518B & 17.1$_{-0.9}^{+0.5}$ & 90.9$\pm$37.4 & 1.2$\pm$1.0 & 43.39$\pm$0.27 & 43.48$\pm$0.27 & 43.70$\pm$0.26 & 7.53 & $-$1.93$\pm$0.26 & 1.53$\pm$0.41 & 2.29/$\cdots$/$\cdots$\\
ID60 & Arp~220W/Arp~220E & 1000.0$_{-22.1}^{+n}$ & 126.3$\pm$20.0 & 2.2$\pm$0.6 & 43.81$\pm$0.05 & 43.99$\pm$0.03 & 44.15$\pm$0.04 & $\cdots$ & $\cdots$ & 2.06$\pm$0.17 & $\cdots$/$\cdots$/$\cdots$\\
ID61 & NGC~6240S/NGC~6240N & $147/155$ & 0.1$\pm$0.1 & 0.2$\pm$0.2 & 44.11$\pm$0.02 & 43.89$\pm$0.02 & 44.20$\pm$0.02 & $\cdots$ & $\cdots$ & 2.12$\pm$0.28 & $\cdots$/$\cdots$/2.81\\
ID62 & NGC~6285/NGC~6286 & 140.2$_{-34.6}^{+26.7}$ & 690.5$\pm$276.1 & 0 & 43.06$\pm$0.05 & 43.50$\pm$0.02 & 43.75$\pm$0.05 & 7.98 & $-$2.33$\pm$0.05 & 1.61$\pm$0.53 & $\cdots$/$\cdots$/$\cdots$\\
ID64 & NGC~6286 & 140.2$_{-34.6}^{+26.7}$ & 694.8$\pm$277.4 & 0 & 42.88$\pm$0.05 & 43.35$\pm$0.02 & 43.58$\pm$0.05 & 7.98 & $-$2.50$\pm$0.05 & 1.55$\pm$0.54 & $\cdots$/$\cdots$/$\cdots$\\
ID65 & IRAS~F17138$-$1017 & 1000.0$_{-643.5}^{+n}$ & 93.2$\pm$38.0 & 1.1$\pm$0.9 & 43.72$\pm$0.20 & 43.84$\pm$0.24 & 44.05$\pm$0.21 & 6.76 & $-$0.81$\pm$0.21 & 2.06$\pm$0.50 & $\cdots$/$\cdots$/$\cdots$\\
ID69 & NGC~6921/MCG+04-48-002 & $\cdots$ & 98.9$\pm$35.0 & $\cdots$ & 43.63$\pm$0.02 & $\cdots$ & 43.89$\pm$0.02 & $\cdots$ & $\cdots$ & $\cdots$ & $\cdots$/$\cdots$/$\cdots$\\
ID70 & NGC~6921 & 173.1$_{-31.9}^{+27.8}$ & 693.8$\pm$275.7 & $\cdots$ & 42.55$\pm$0.03 & $\cdots$ & 43.25$\pm$0.04 & 8.11 & $-$2.96$\pm$0.04 & $\cdots$ & $\cdots$/$\cdots$/$\cdots$\\
ID71 & MCG+04-48-002 & 72.7$_{-8.0}^{+14.8}$ & 104.5$\pm$32.8 & $\cdots$ & 43.53$\pm$0.11 & $\cdots$ & 43.80$\pm$0.11 & 7.64 & $-$1.94$\pm$0.11 & $\cdots$ & $\cdots$/$\cdots$/$\cdots$\\
ID74 & NGC~7130 & 413.3$_{-7.4}^{+13.1}$ & 734.0$\pm$292.7 & 0 & 43.78$\pm$0.18 & 44.05$\pm$0.18 & 44.35$\pm$0.18 & 7.89 & $-$1.64$\pm$0.18 & 2.47$\pm$0.37 & $\cdots$/3.08/$\cdots$\\
ID75 & NGC~7469/IC~5283 & $<$0.01 & 0.0$\pm$0.1 & 1.1$\pm$1.1 & 44.36$\pm$0.11 & 44.44$\pm$0.14 & 44.63$\pm$0.11 & 7.30 & $-$0.77$\pm$0.11 & 2.68$\pm$0.14 & $\cdots$/$\cdots$/$\cdots$\\
ID76 & NGC~7469 & $<$0.01 & 0.0$\pm$0.1 & 2.2$\pm$0.6 & 44.32$\pm$0.12 & 44.51$\pm$0.11 & 44.62$\pm$0.11 & 7.30 & $-$0.79$\pm$0.11 & 2.68$\pm$0.13 & 2.12/$\cdots$/$\cdots$\\
ID78 & ESO~148$-$2 & 158.5$_{-47.6}^{+45.0}$ & 383.4$\pm$151.1 & 0 & 44.36$\pm$0.10 & 44.72$\pm$0.11 & 45.03$\pm$0.10 & 7.49 & $-$0.57$\pm$0.10 & 2.91$\pm$0.06 & $\cdots$/3.41/$\cdots$\\
ID80 & NGC~7674/MCG+01-59-081 & 29.6$_{-1.9}^{+2.5}$ & 81.1$\pm$36.2 & 0.2$\pm$0.2 & 44.66$\pm$0.05 & 43.95$\pm$0.05 & 44.51$\pm$0.05 & 7.99 & $-$1.58$\pm$0.05 & 2.26$\pm$0.42 & $\cdots$/$\cdots$/$\cdots$\\
ID81 & NGC~7674 & 29.6$_{-1.9}^{+2.5}$ & 69.1$\pm$33.1 & 0.2$\pm$0.2 & 44.64$\pm$0.06 & 43.93$\pm$0.07 & 44.49$\pm$0.06 & 7.99 & $-$1.60$\pm$0.06 & 2.25$\pm$0.38 & 2.59/$\cdots$/$\cdots$\\
ID83 & NGC~7679/NGC~7682 & $\cdots$ & 103.3$\pm$34.2 & 1.5$\pm$1.0 & 43.15$\pm$0.16 & 43.28$\pm$0.18 & 43.48$\pm$0.16 & $\cdots$ & $\cdots$ & 1.42$\pm$0.39 & $\cdots$/$\cdots$/$\cdots$\\
ID84 & NGC~7679 & $<$0.01 & 0.0$\pm$0.1 & 1.4$\pm$1.0 & 42.98$\pm$0.15 & 43.09$\pm$0.12 & 43.24$\pm$0.13 & 7.42 & $-$2.28$\pm$0.13 & 1.23$\pm$0.41 & $\cdots$/$\cdots$/$\cdots$\\
ID85 & NGC~7682 & 38.2$_{-7.0}^{+6.5}$ & 691.4$\pm$268.8 & $\cdots$ & 42.67$\pm$0.22 & $\cdots$ & 42.72$\pm$0.21 & 7.48 & $-$2.86$\pm$0.21 & $\cdots$ & $\cdots$/$\cdots$/$\cdots$\\
\enddata
\tablecomments{Comments: (1--2) ID and object name; 
(3) line-of-sight hydrogen column density derived from X-ray spectral analysis (Section~\ref{sub3-1-Xray}; \citealt{Yamada2021});
(4)--(5) line-of-sight extinction in the $V$ band due to torus, $A_{V{\rm,torus}}^{\rm (LOS)}$, and polar dust, $A_{V{\rm,torus}}^{\rm (LOS)}$;
(6)--(8) logarithmic UV-to-IR luminosities of torus component ($L_{\rm torus}$), polar dust component ($L_{\rm polar}$), and intrinsic AGN disk component ($L_{\rm AGN,int}$);
(9) SMBH mass estimated from the averaged value of four independent measurements in \citet{Yamada2021};
(10) logarithmic Eddington ratio ($L_{\rm AGN,int}$/$L_{\rm Edd}$).
The Eddington luminosity is defined as $L_{\rm Edd} = 1.26 \times 10^{38} M_{\rm BH}/M_{\odot}$;
(11) physical size of polar dust, estimated by the polar dust temperature and dust sublimation radius (Section~\ref{subsub6-5-1-polar-size});
(12) physical sizes of the polar dust emission detected in the mid-IR band ($R_{\rm mir}$; \citealt{Asmus2019}), 
radius of ionized outflows ($R_{\rm io}$; \citealt{Fluetsch2021}), 
and radius of molecular outflows ($R_{\rm mo}$; \citealt{Fluetsch2019}).}
\tablenotetext{\dagger}{Large fractions of the X-ray and mid-IR luminosities in NGC~1275 could be explained by the jet emission \citep[e.g.,][]{Hitomi-Collaboration2018,Rani2018}.}
\tablenotetext{\ }{(This table is available in its entirety in machine-readable form.)}
\end{deluxetable*}
\end{longrotatetable}

\restartappendixnumbering

\section{Mock Analysis} \label{AppendixB-mock}
We evaluate the constrainability of a model parameter by the ``mock analysis'' of X-CIGALE.
This is implemented in the previous version of CIGALE \citep[see e.g.,][]{Buat2012,Buat2014,Ciesla2015,Lo-Faro2017,Boquien2019,Toba2020c,Yang2020}.
X-CIGALE uses the photometric data for each object based on the best-fit parameters. It allows the fluxes to vary within the uncertainties of the observations by adding a value taken from a Gaussian distribution with the same standard deviation as indicated by the photometric data.

\begin{figure*}
    \centering
    \includegraphics[keepaspectratio, scale=0.5]{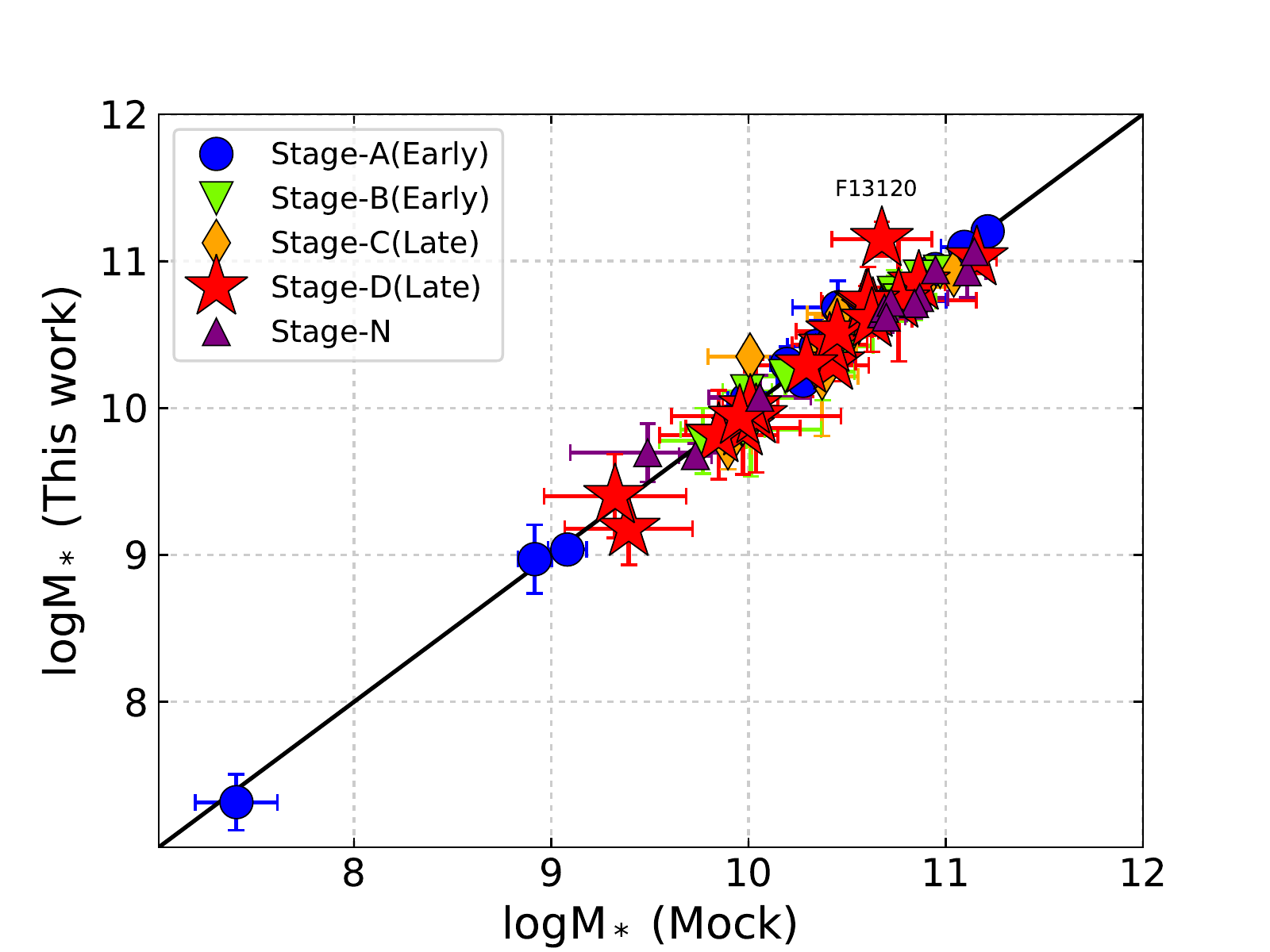}
    \includegraphics[keepaspectratio, scale=0.5]{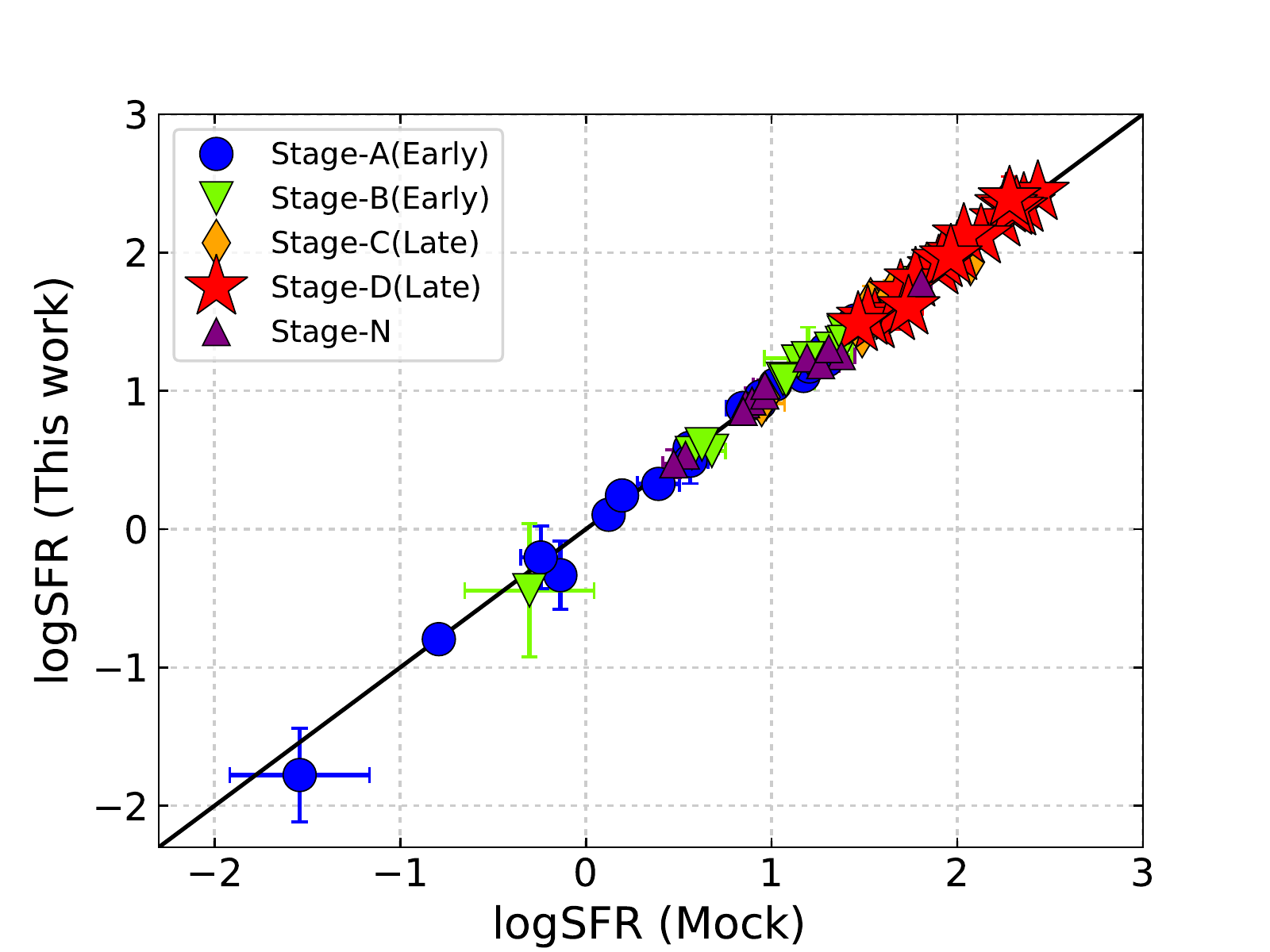}
    \includegraphics[keepaspectratio, scale=0.5]{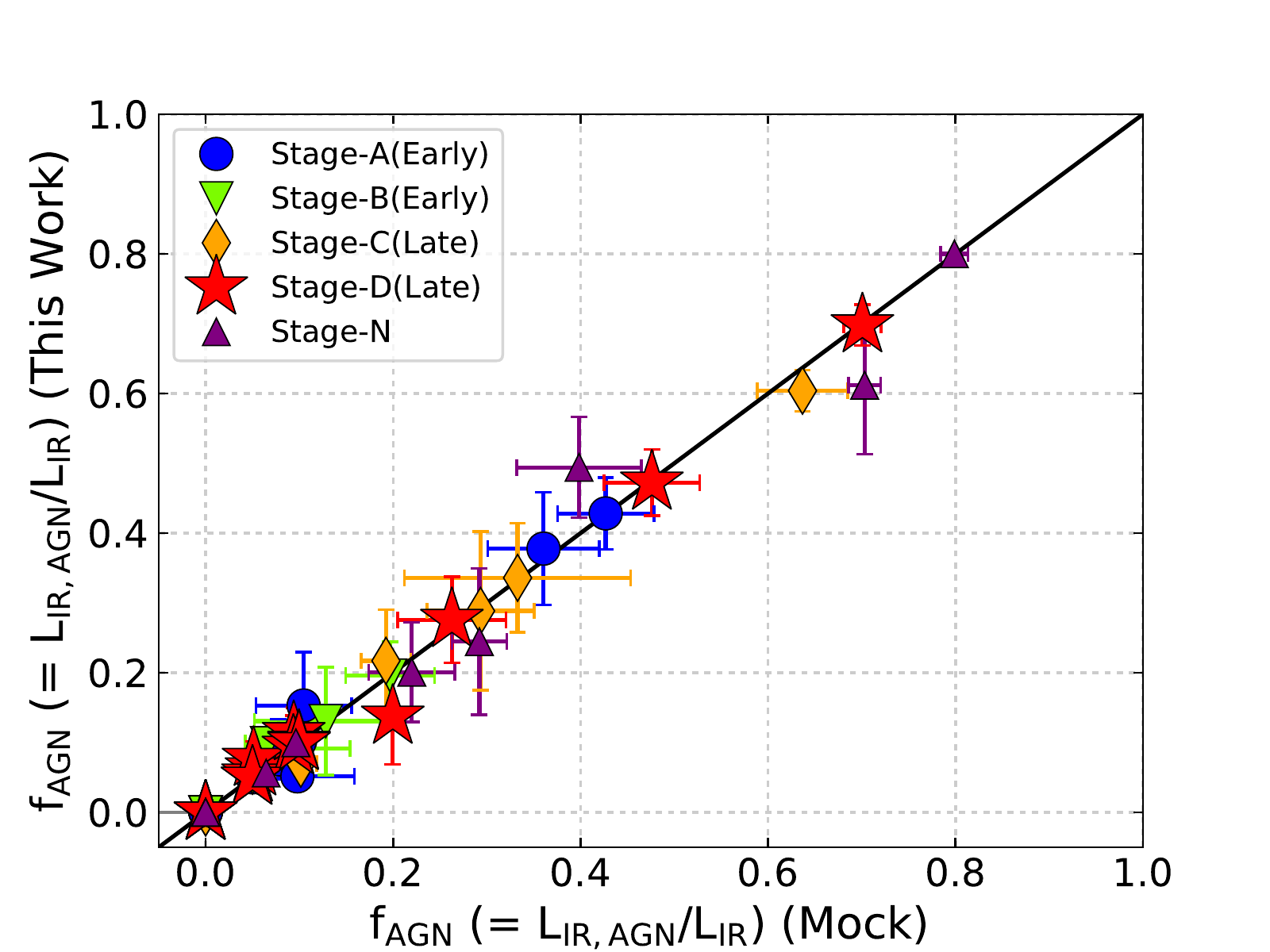}
    \includegraphics[keepaspectratio, scale=0.5]{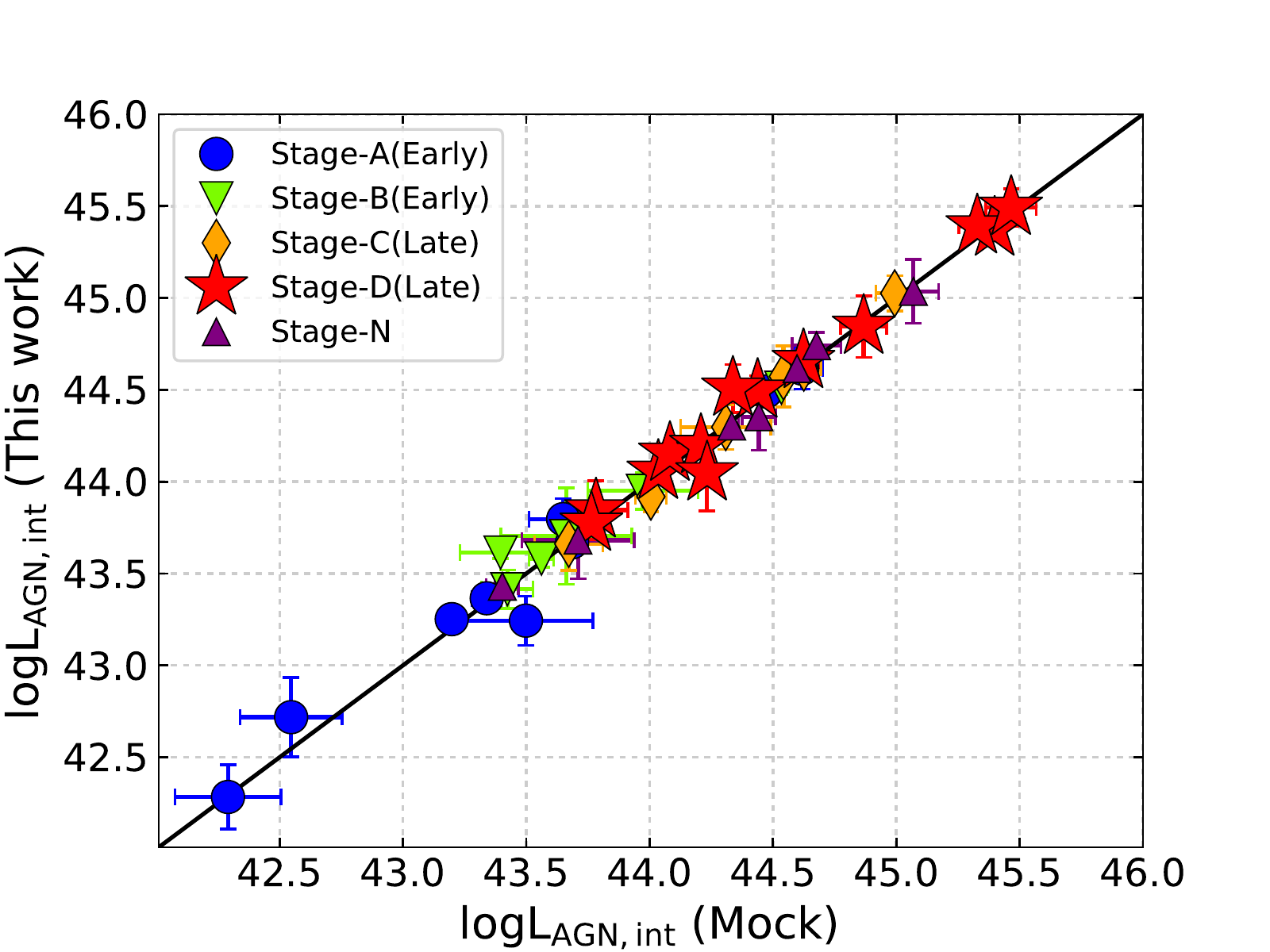}
    \includegraphics[keepaspectratio, scale=0.5]{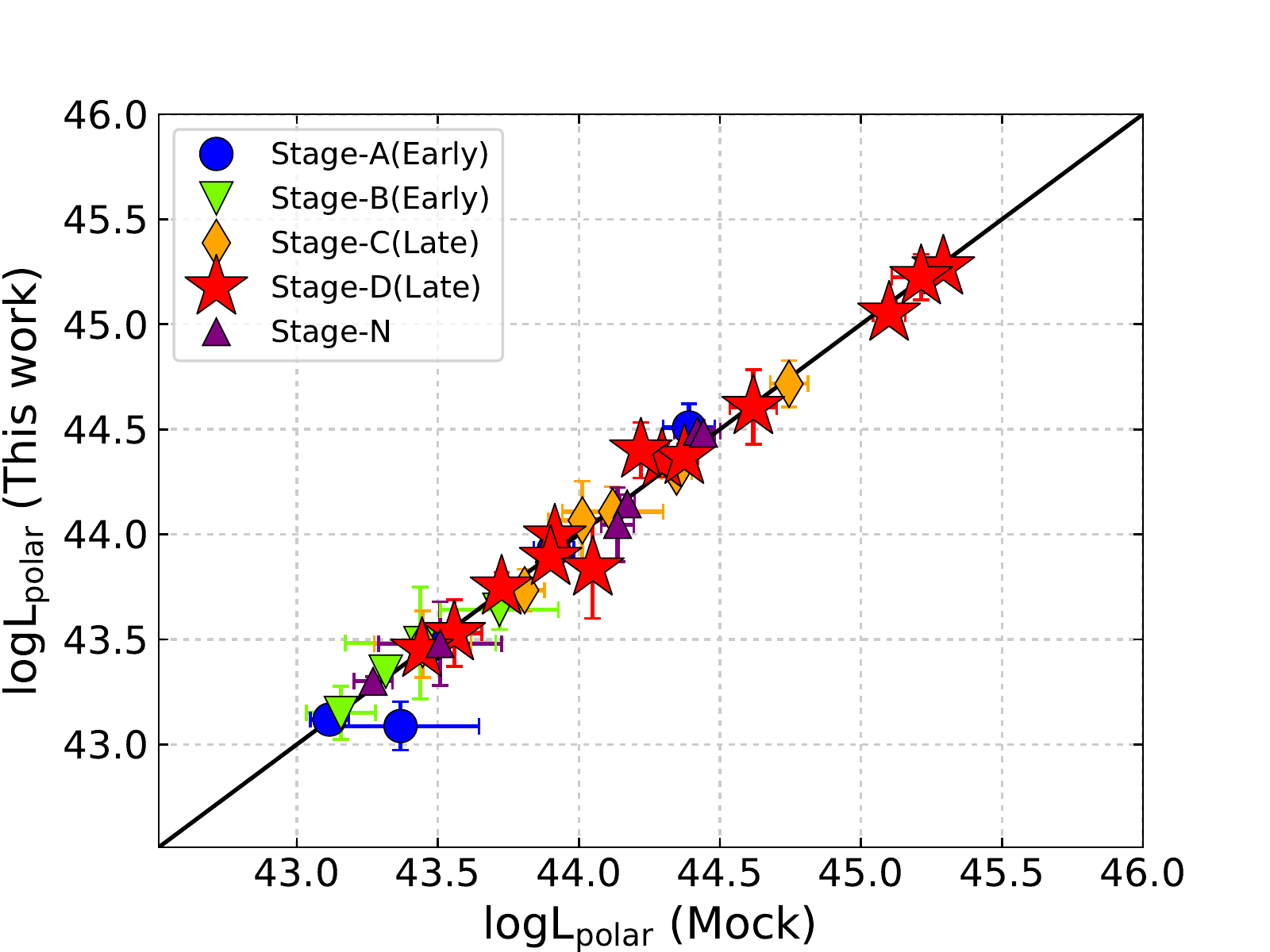}
    \includegraphics[keepaspectratio, scale=0.5]{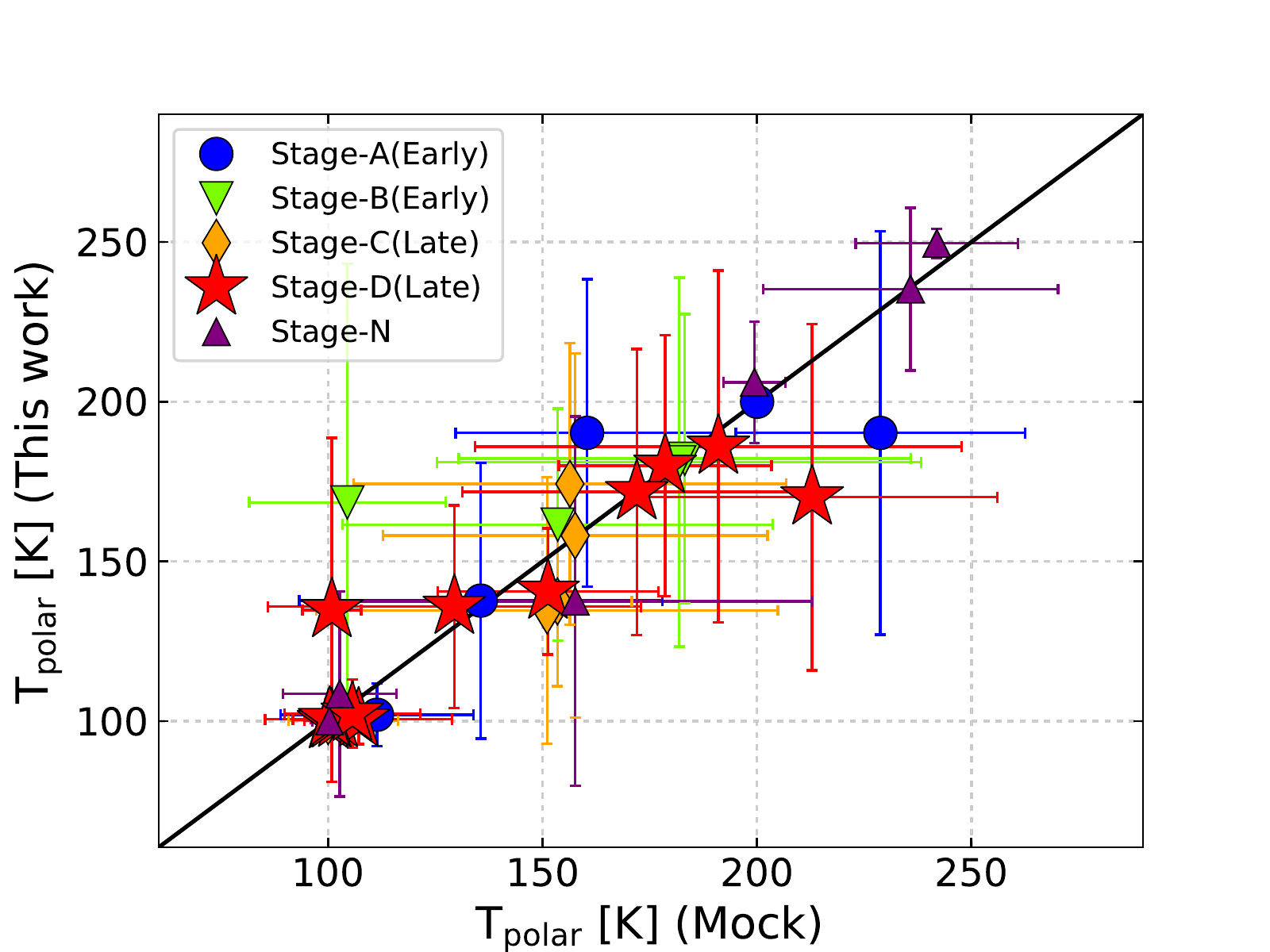}
    \caption{
    Comparison of the best-fit parameters and those derived from the mock analysis for the $M_*$ (top left), SFR (top right), AGN fraction (middle left), intrinsic AGN luminosity (middle right), polar dust luminosity (bottom left), and polar dust temperature (bottom right).
    The symbols are the same in Figure~\ref{F8-MS}.
    \\}
\label{FB1-mock}
\end{figure*}

Figure~\ref{FB1-mock} shows the results of the mock analysis with the best-fit values compared to the mock values for the main parameters in this study.
We find that the mean differences (the best-fit values $-$ mock values) are small; $\Delta$log$M_*$ = $-0.01 \pm 0.11$, $\Delta$logSFR = $-0.01 \pm 0.07$, $\Delta f_{\rm AGN}$ = $0.00 \pm 0.03$, $\Delta$log$L_{\rm AGN, int}$ = $0.01 \pm 0.08$, $\Delta$log$L_{\rm polar}$ = $-0.01 \pm 0.08$, and $\Delta T_{\rm polar}$ = $0.21  \pm 18.19$.
The deviation of the differences in $T_{\rm polar}$ is relatively large but consistent with the range of 1$\sigma$ uncertainty of each object.
Therefore, we conclude that these quantities are not sensitive to photometric uncertainties thanks to the multiwavelength SED analysis with the self-consistent AGN model.
\\

% Figure B1

\restartappendixnumbering

\section{Comparison with Previous SED Studies} \label{AppendixC-previous-works}
\begin{figure*}
    \centering
    \includegraphics[keepaspectratio, scale=0.45]{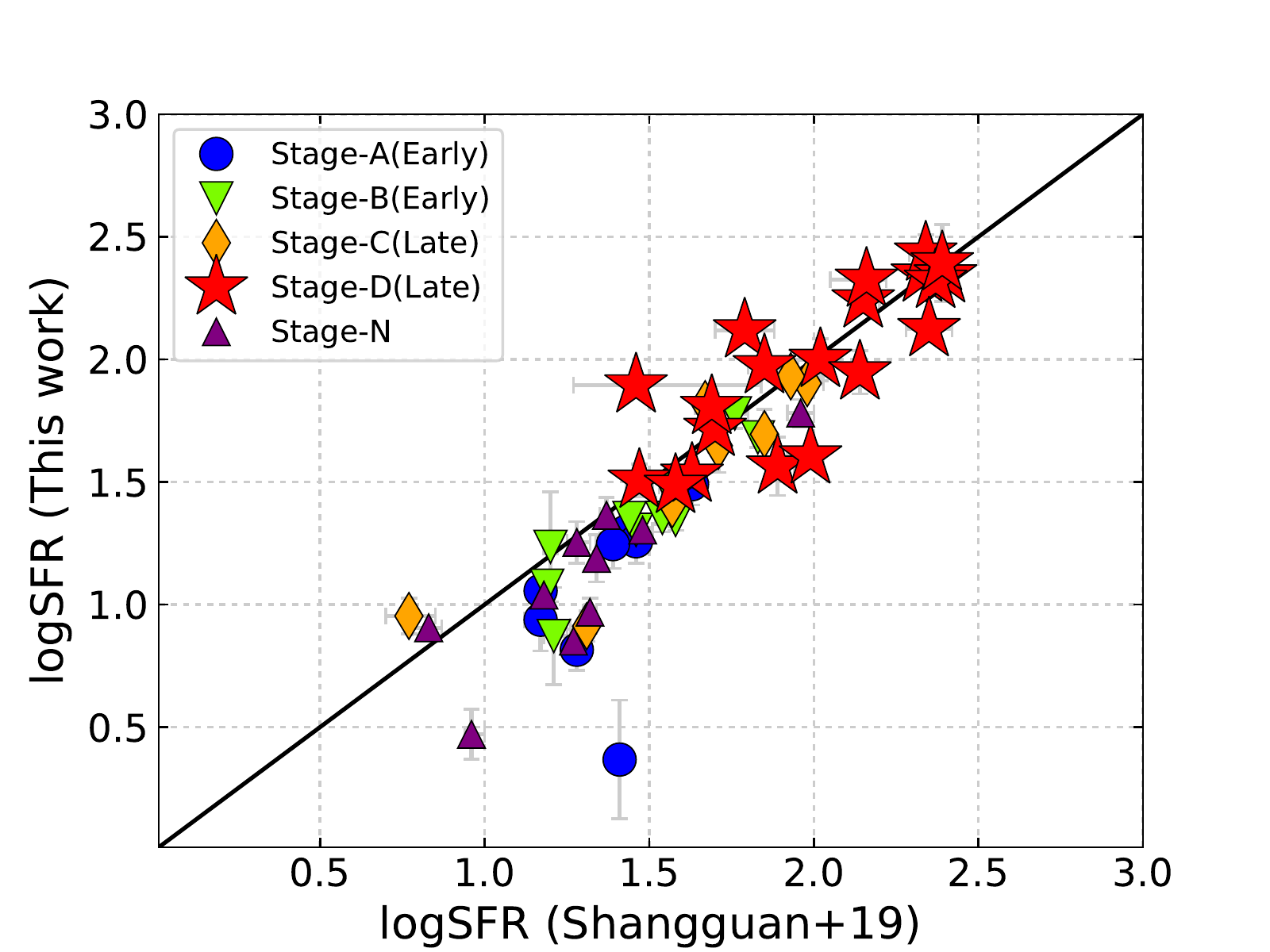}
    \includegraphics[keepaspectratio, scale=0.45]{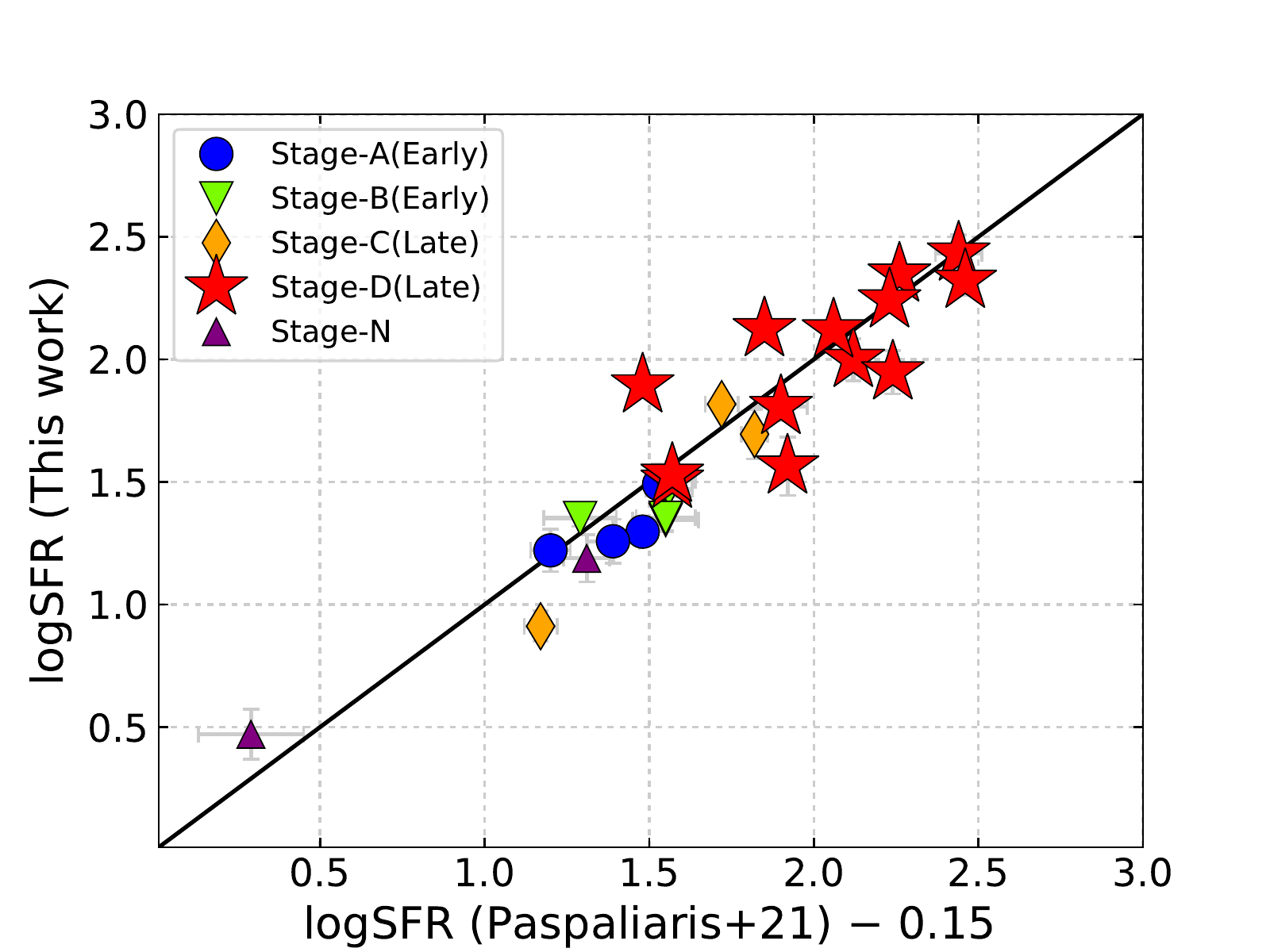}
    \caption{Left panel: comparison of SFRs estimated in \citet{Shangguan2019} and this work.
    Right panel: comparison of SFRs estimated in \citet{Paspaliaris2021} and this work.
    The former values are converted from \citet{Salpeter1955} IMF to \citet{Chabrier2003} IMF with 0.15~dex.
    The symbols are the same in Figure~\ref{F8-MS}.
    \\}
\label{FC1-comp-SFR}
\end{figure*}

\begin{figure*}
    \centering
    \includegraphics[keepaspectratio, scale=0.45]{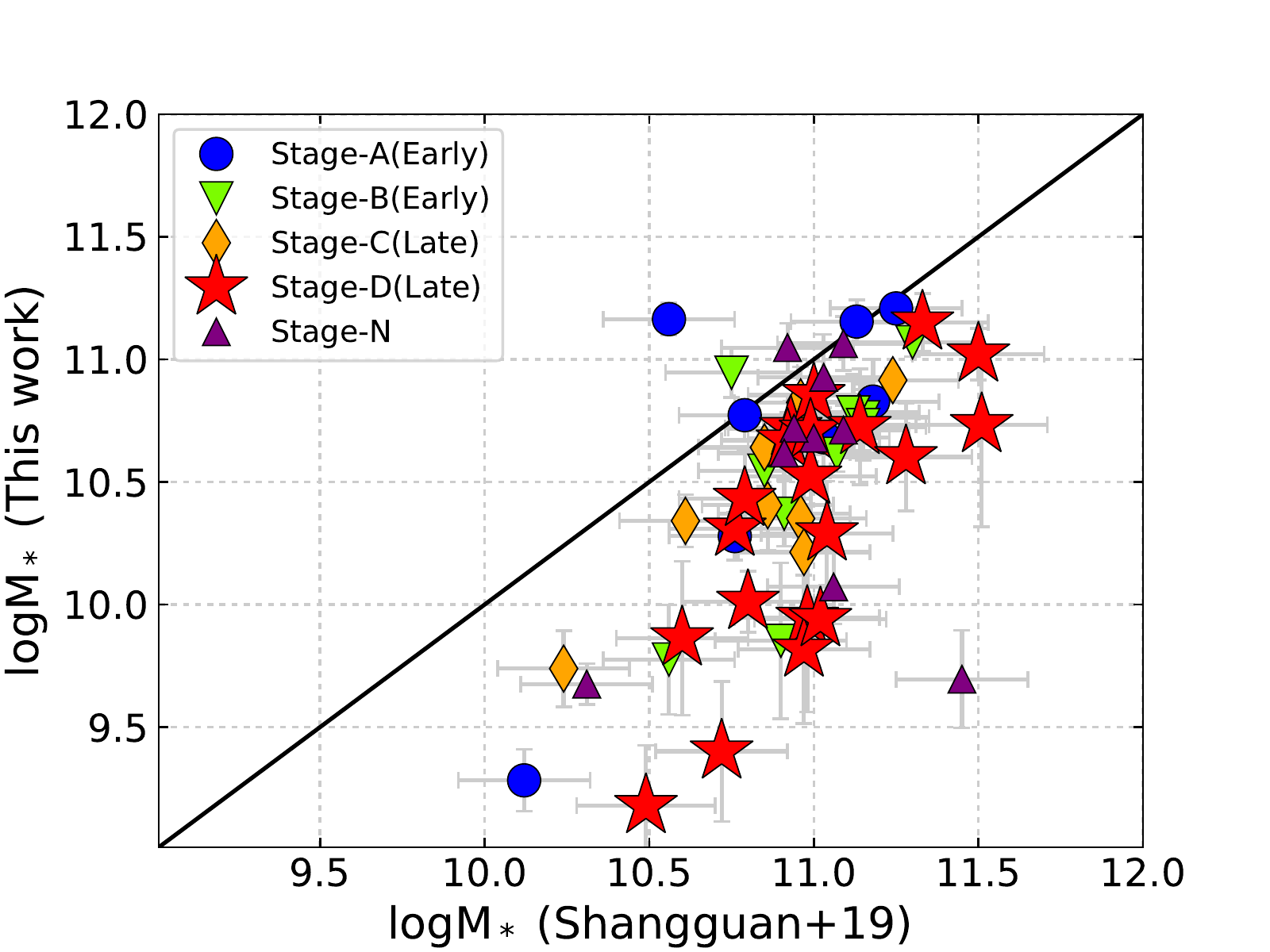}
    \includegraphics[keepaspectratio, scale=0.45]{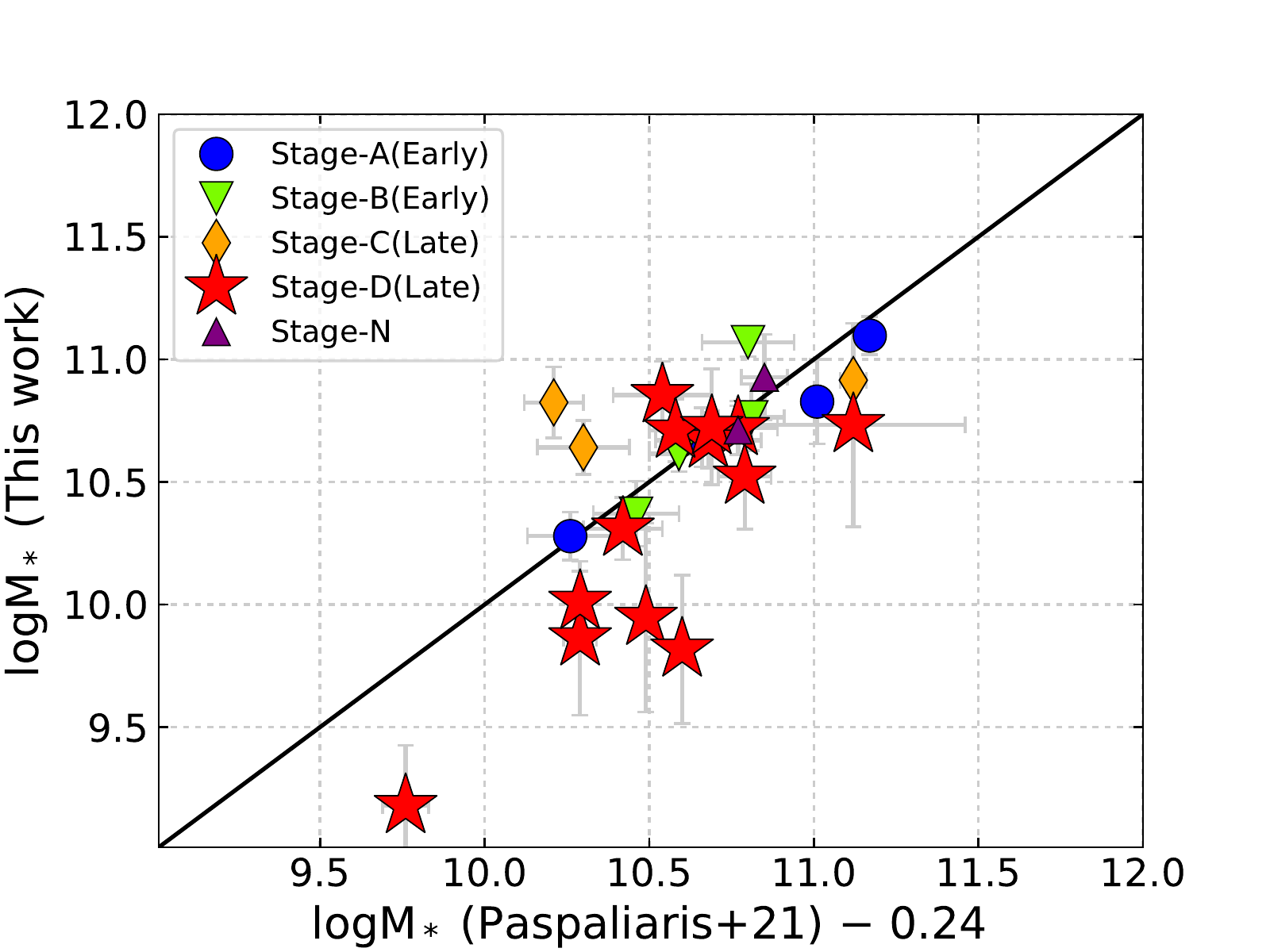}
    \caption{Left panel: comparison of $M_*$ estimated in \citet{Shangguan2019} and this work.
    Right panel: comparison of $M_*$ estimated in \citet{Paspaliaris2021} and this work.
    The former values are converted from \citet{Salpeter1955} IMF to \citet{Chabrier2003} IMF with 0.24~dex.
    The symbols are the same in Figure~\ref{F8-MS}.
    \\}
\label{FC2-comp-Ms}
\end{figure*}

Here, we examine the difference in the results of this study and previous SED works.
For the local U/LIRGs in GOALS sample, \citet{Shangguan2019} performed the IR (1--500~$\mu$m) SED fitting and \citet{Paspaliaris2021} carried out the UV-to-submillimeter SED analysis.
Our estimates of the SFRs are well consistent with their works (Figure~\ref{FC1-comp-SFR}).
Whereas, we find that their stellar mass is smaller than those of our results (Figure~\ref{FC2-comp-Ms}).
For the results of \citet{Shangguan2019}, this may be explained by that the lack of optical photometric data that causes the overestimation of the fraction of the old-stellar population.
In this case, the estimates of the luminosity from each star become small, and thus the total stellar mass will be overestimated.
Although it is unclear why are the estimates of stellar mass by \citet{Paspaliaris2021} also larger, this may be affected by the difference in the contribution of the IR polar dust emission in our analysis that applies the torus parameters obtained from the X-ray fitting.

\begin{figure*}
    \centering
    \includegraphics[keepaspectratio, scale=0.45]{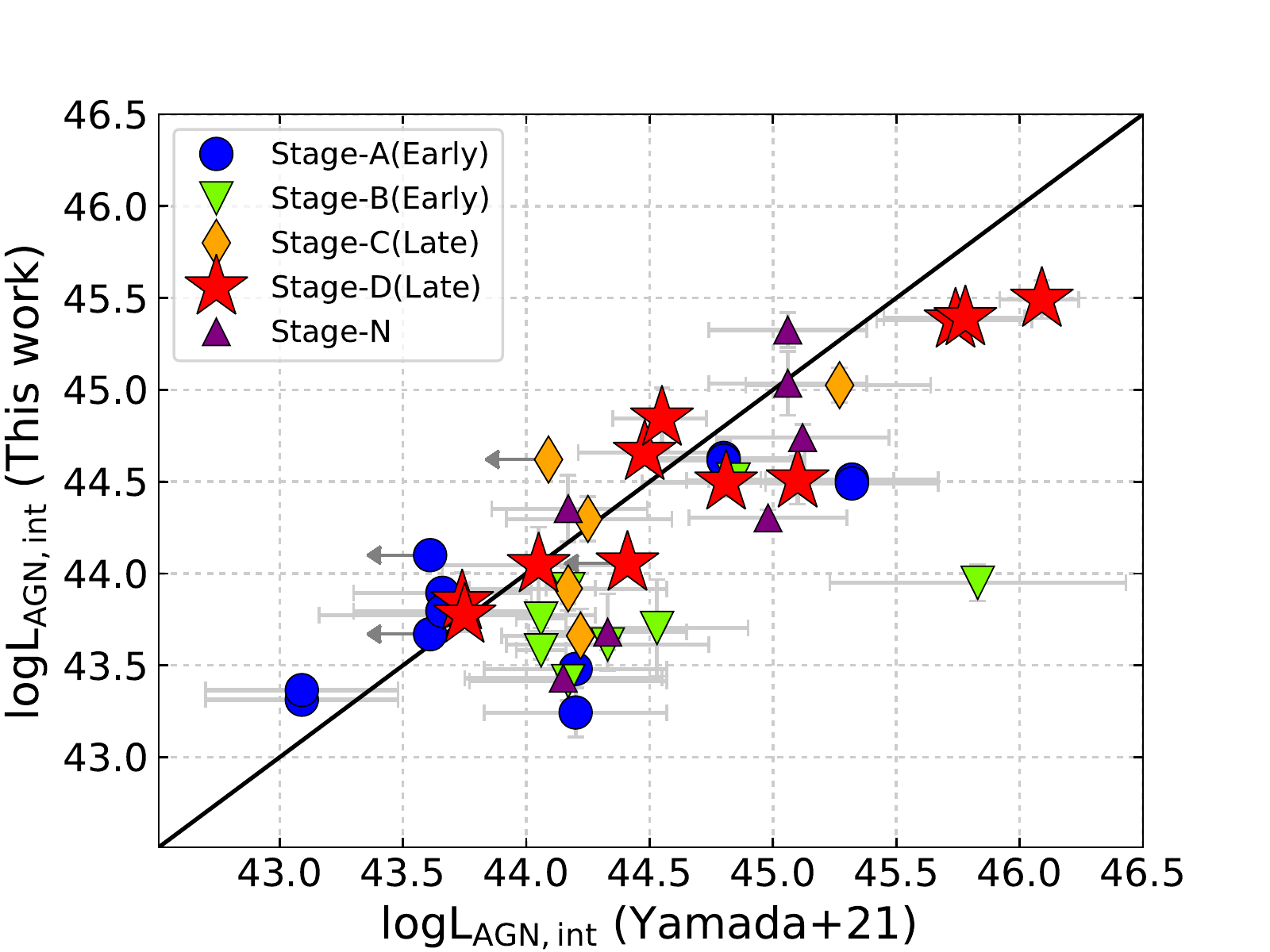}
    \includegraphics[keepaspectratio, scale=0.45]{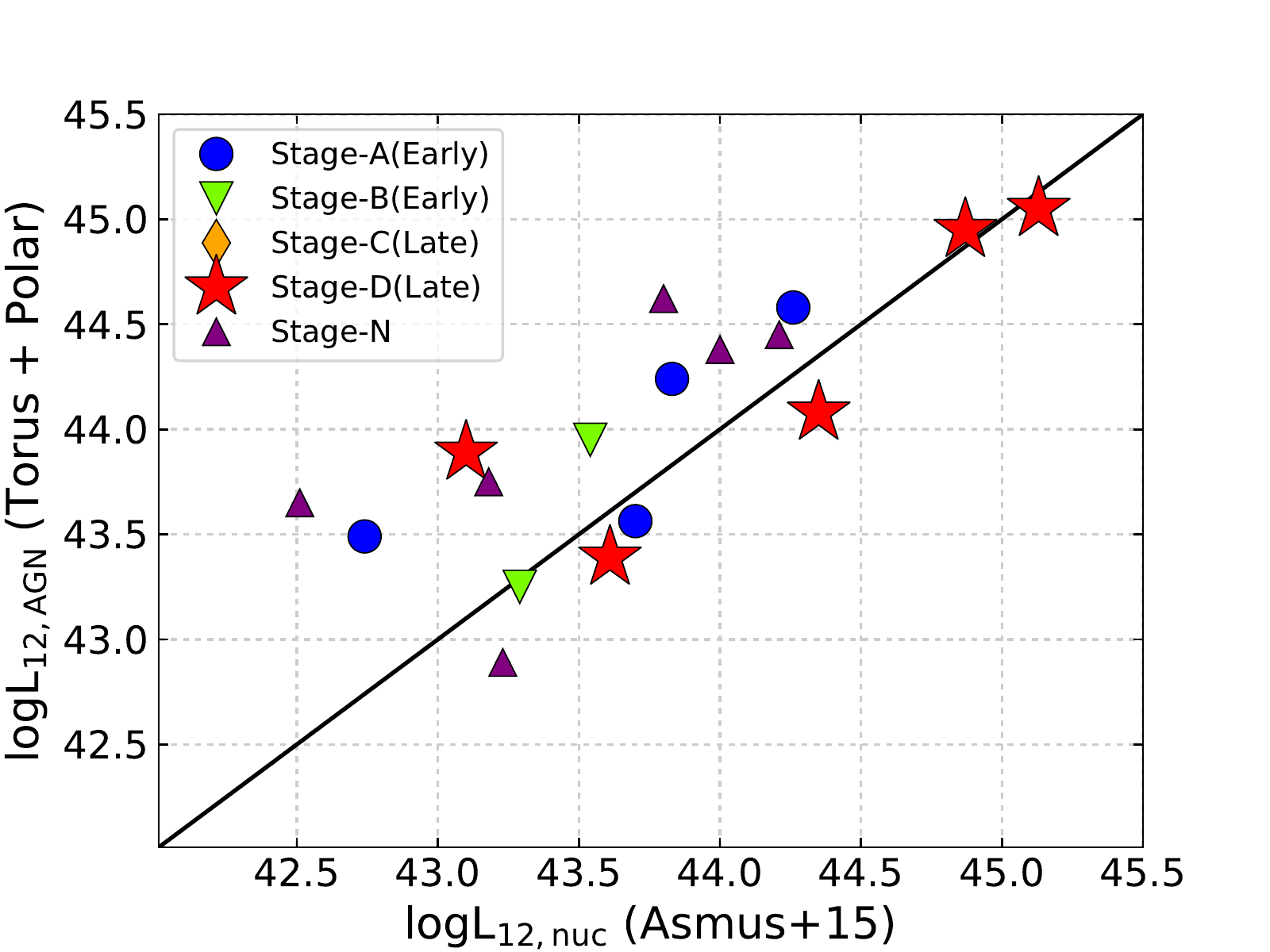}
    \caption{Left panel: comparison of $L_{\rm AGN,int}$ estimated in \citet{Yamada2021} and this work.
    Right panel: comparison with the AGN (torus and polar-dust components) 12~$\mu$m luminosity from UV-to-IR SED analysis ($L_{\rm 12 \mu m, AGN}$) and the nuclear 12~$\mu$m luminosity ($L_{\rm 12 \mu m, nuc}$; \citealt{Asmus2015}).
    \\}
\label{FC3-comp-LAGN-L12}
\end{figure*}

In the left panel of Figure~\ref{FC3-comp-LAGN-L12}, we compare the AGN luminosities in this study and those in \citet{Yamada2021}.
Their AGN luminosities are estimated by the averaged values of four different methods (e.g., the [\ion{O}{4}] luminosity, bolometric AGN fraction, and IR SED analysis).
Its typical scatter is reported as $\sim$0.27~dex when we adopt the averaged value.
Particularly, the estimates using the [\ion{O}{4}] luminosities may be larger than those from other methods, which may be due to the contamination from the intense starburst emission in U/LIRGs (but see also \citealt{Yamada2019}).
Even if the intrinsic AGN luminosities given by \citet{Yamada2021} are adopted, the main discussion on the SFR--$L_{\rm AGN,int}$ relation is unchanged (see Section~\ref{sub7-1-growth-rate}).
The right panel of Figure~\ref{FC3-comp-LAGN-L12} presents the comparison between the 12~$\mu$m luminosity of the AGN component (this study) and the nuclear 12~$\mu$m luminosities estimated with the high-spatial-resolution mid-IR images \citep{Asmus2019}.
Although there are a few differences, the multiwavelength SED analysis extracts the AGN emission that is consistent with the imaging studies.
\\

%Figure C1

% Figure C2

%Figure C3

\restartappendixnumbering

\section{Comparison with the Results of SKIRTOR Model} \label{AppendixD-SKIRTOR}
\begin{figure*}
    \centering
    \includegraphics[keepaspectratio, scale=0.45]{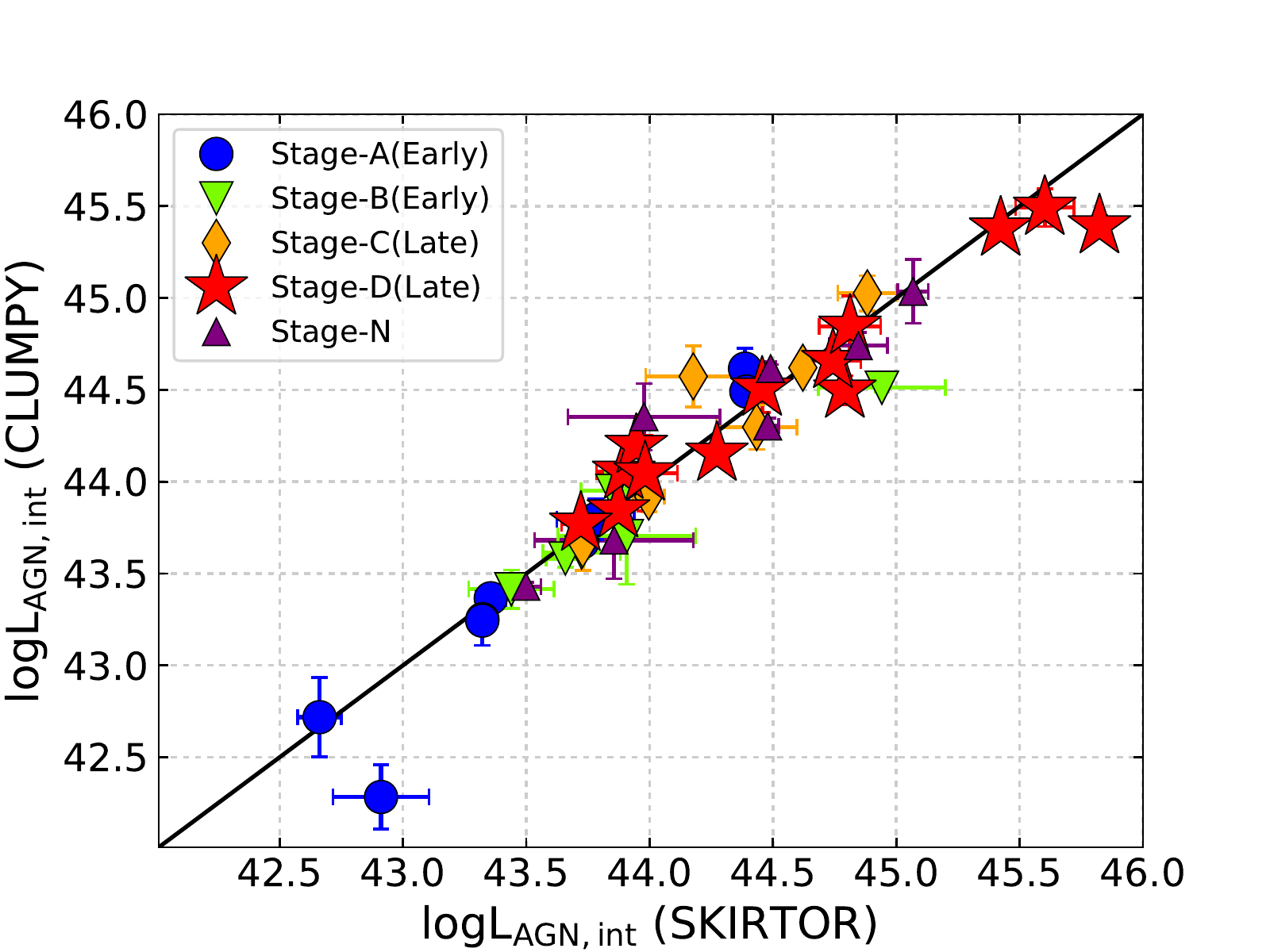}
    \includegraphics[keepaspectratio, scale=0.45]{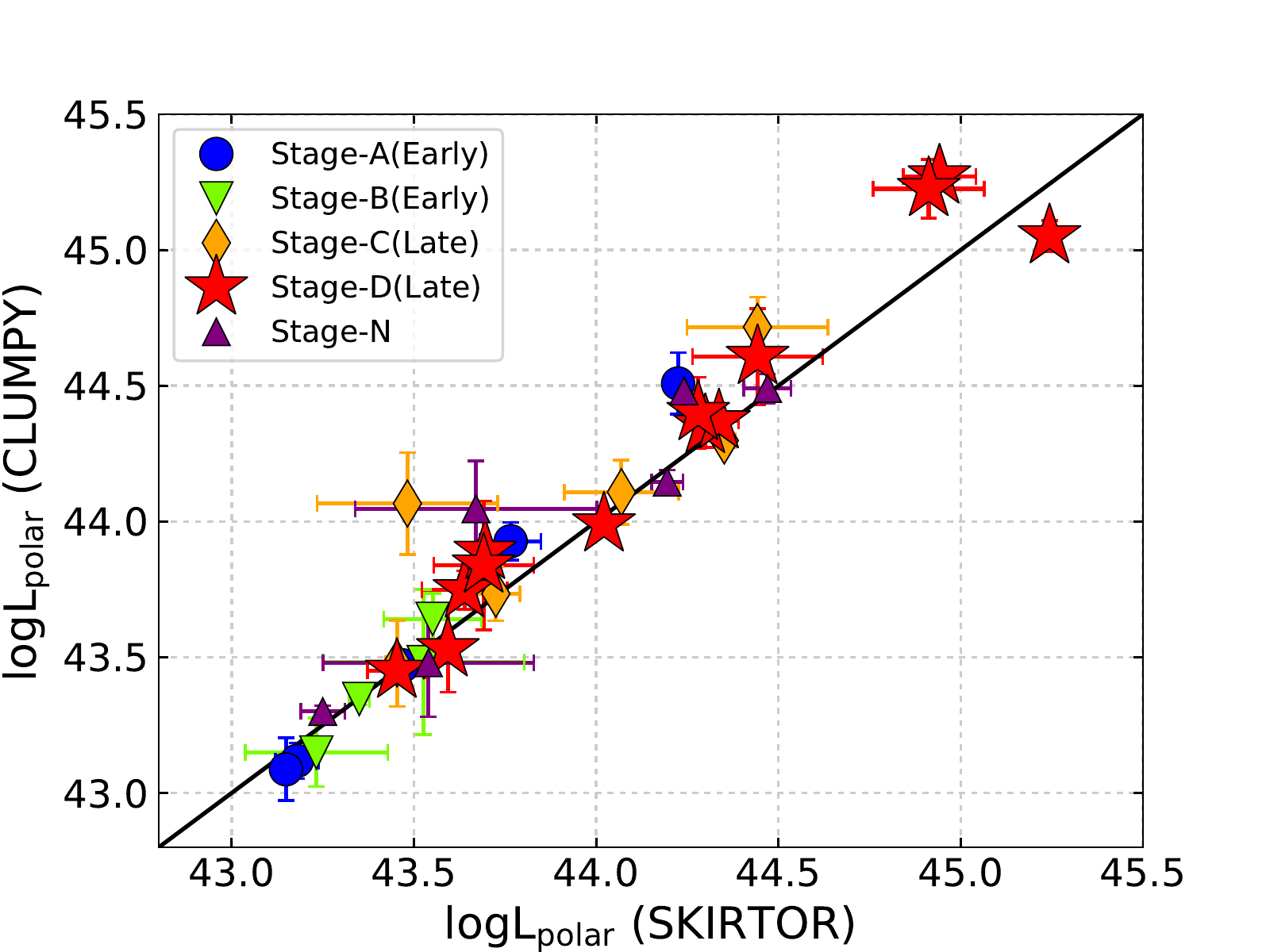}
    \caption{Left panel: comparison of intrinsic (bolometric) AGN luminosities derived with the CLUMPY model and SKIRTOR model.
    Right panel: comparison of polar-dust luminosities derived with CLUMPY model and SKIRTOR model.
    The symbols are the same in Figure~\ref{F8-MS}.
    \\}
\label{FD1-comp-skirtor}
\end{figure*}

In this study, we also performed the SED decomposition by using the AGN model as the SKIRTOR model (\citealt{Stalevski2016}) instead of the CLUMPY model.
The $\sigma$ values are converted from torus angular width ($\sigma$) of [10\degr--15\degr, 15\degr--25\degr, 25\degr--70\degr] to angle between the equatorial plane and edge of the torus ($\Delta$) of [30\degr, 40\degr, 60\degr], respectively.
The smooth component in the SKIRTOR model shields the AGN disk emission at subparsec scales and re-radiates the strong near-IR emission relative to the clumpy torus model \citep{Stalevski2012,Stalevski2016}.
In fact, the UV-to-IR SEDs of IRAS F08572+3915 that is not reproduced with CLUMPY ($\chi^2_{\rm red} = 11.8$) due to the near-IR bump (as illustrated in the WISE color--color diagram in Section~\ref{sub5-3-WISE-color}), while it is well fitted with SKIRTOR model ($\chi^2_{\rm red} = 6.7$).
The intrinsic AGN luminosities of IRAS F08572+3915 estimated with CLUMPY (log$L_{\rm AGN,int} = 45.39 \pm 0.06$) is smaller than that from SKIRTOR ($45.82 \pm 0.02$).
NGC~833, a stage A merger, is another outlier that shows the gap between the estimates with CLUMPY (log$L_{\rm AGN,int} = 42.28 \pm 0.17$) and SKIRTOR ($42.91 \pm 0.20$; left panel of Figure~\ref{FD1-comp-skirtor}), but the reason for the difference should be the poor photometry in the 5--70~$\mu$m band (see Figure~\ref{SED01-F}).
Except for the IRAS F08572+3915 and NGC~833, all sources show similar AGN luminosities with these models.
Identically, the polar dust luminosities constrained with the SKIRTOR model are well consistent with the values with CLUMPY (right panel of Figure~\ref{FD1-comp-skirtor}).

% Figure D1

% Figure D2
\begin{figure*}
    \centering
    \includegraphics[keepaspectratio, scale=0.45]{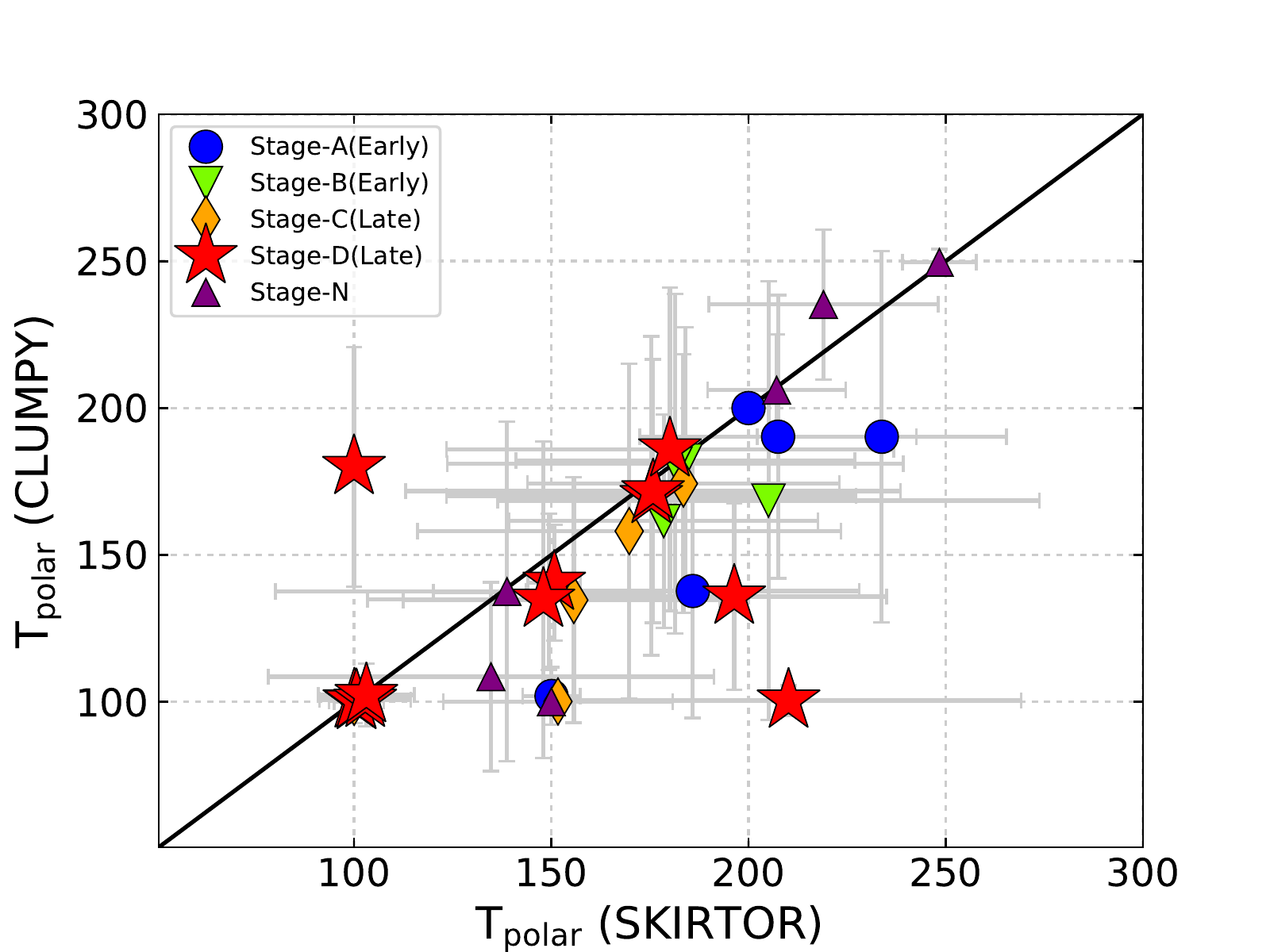}
    \includegraphics[keepaspectratio, scale=0.45]{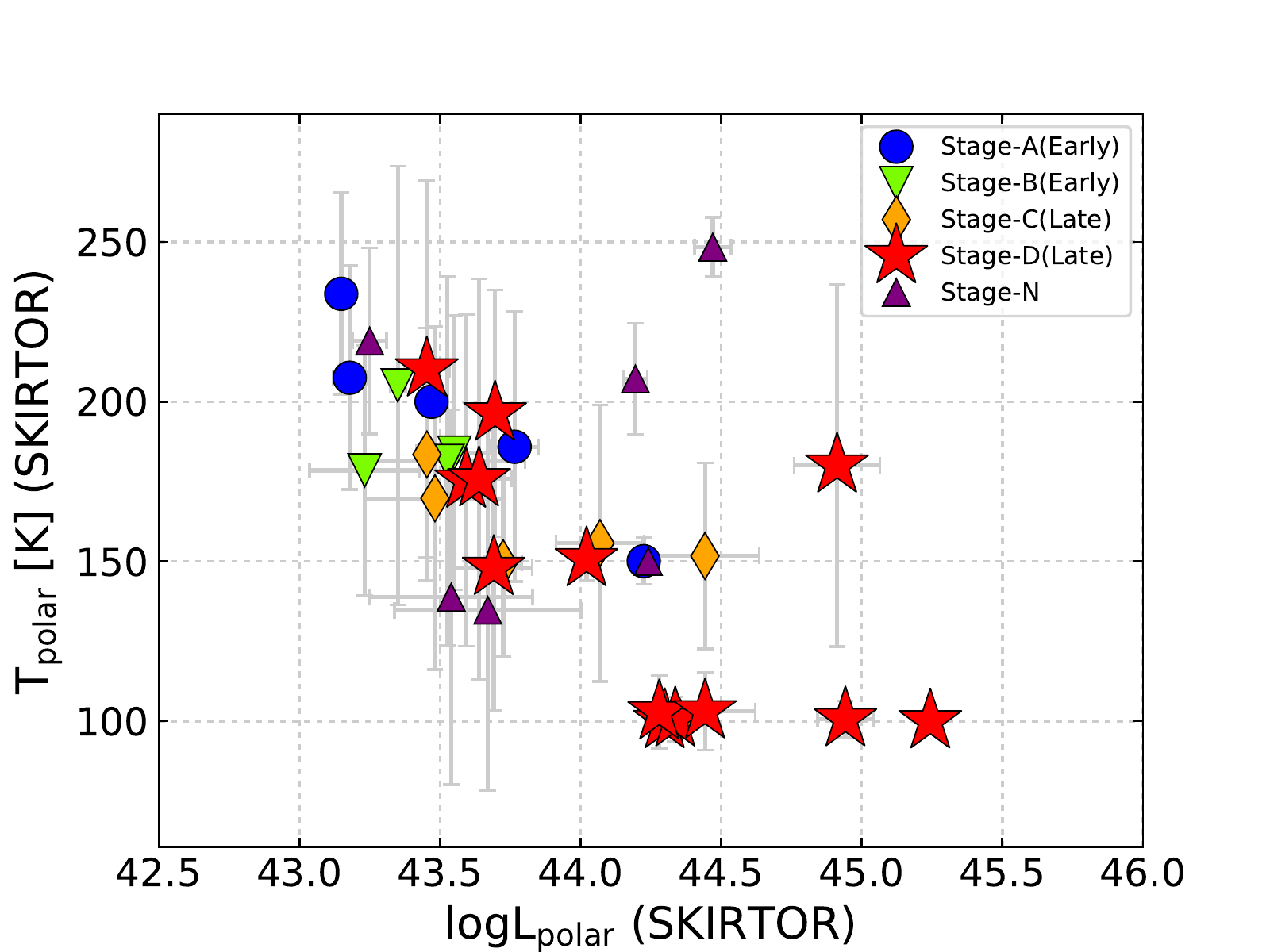}
    \caption{Left panel: comparison of polar dust temperatures derived with the CLUMPY model and SKIRTOR model.
    Right panel: logarithmic polar-dust luminosity vs. polar-dust temperature when SKIRTOR model is adopted.
    The symbols are the same in Figure~\ref{F8-MS}.
    \\}
\label{FD2-skirtor-Tpo}
\end{figure*}

Additionally, we investigate the polar dust temperature derived from the fits with SKIRTOR (left panel of Figure~\ref{FD2-skirtor-Tpo}).
Due to the strong near-IR emission, the SEDs of the torus with SKIRTOR are flatter than with CLUMPY.
In this case, a large part of the mid-IR emission will be modeled by the polar dust emission.
This causes a larger fraction of the AGNs to show the signs of polar dust emission with SKIRTOR than with CLUMPY (Section~\ref{subsub4-2-4-BIC}).
Hence, the polar temperature from the SKIRTOR model is larger than those from the CLUMPY model.
In the right panel of Figure~\ref{FD2-skirtor-Tpo}, we find that the results with SKIRTOR, that is, the trends of the large polar dust luminosities and low temperatures in late mergers, are consistent with the results with CLUMPY.
We note that the estimates of the polar dust features by fixing the torus parameters with SKIRTOR are not self-consistent since the geometry of the XCLUMPY and SKIRTOR is different.
Thus, it is preferred to adopt the results with CLUMPY particularly when we discuss the polar dust structure.
\\

\restartappendixnumbering

\section{Multiwavelength SED Decomposition} \label{AppendixE-SEDs}
\begin{figure*}
    \centering
   \includegraphics[keepaspectratio, scale=0.34]{Fnu-ID01-SED.pdf}
   \includegraphics[keepaspectratio, scale=0.34]{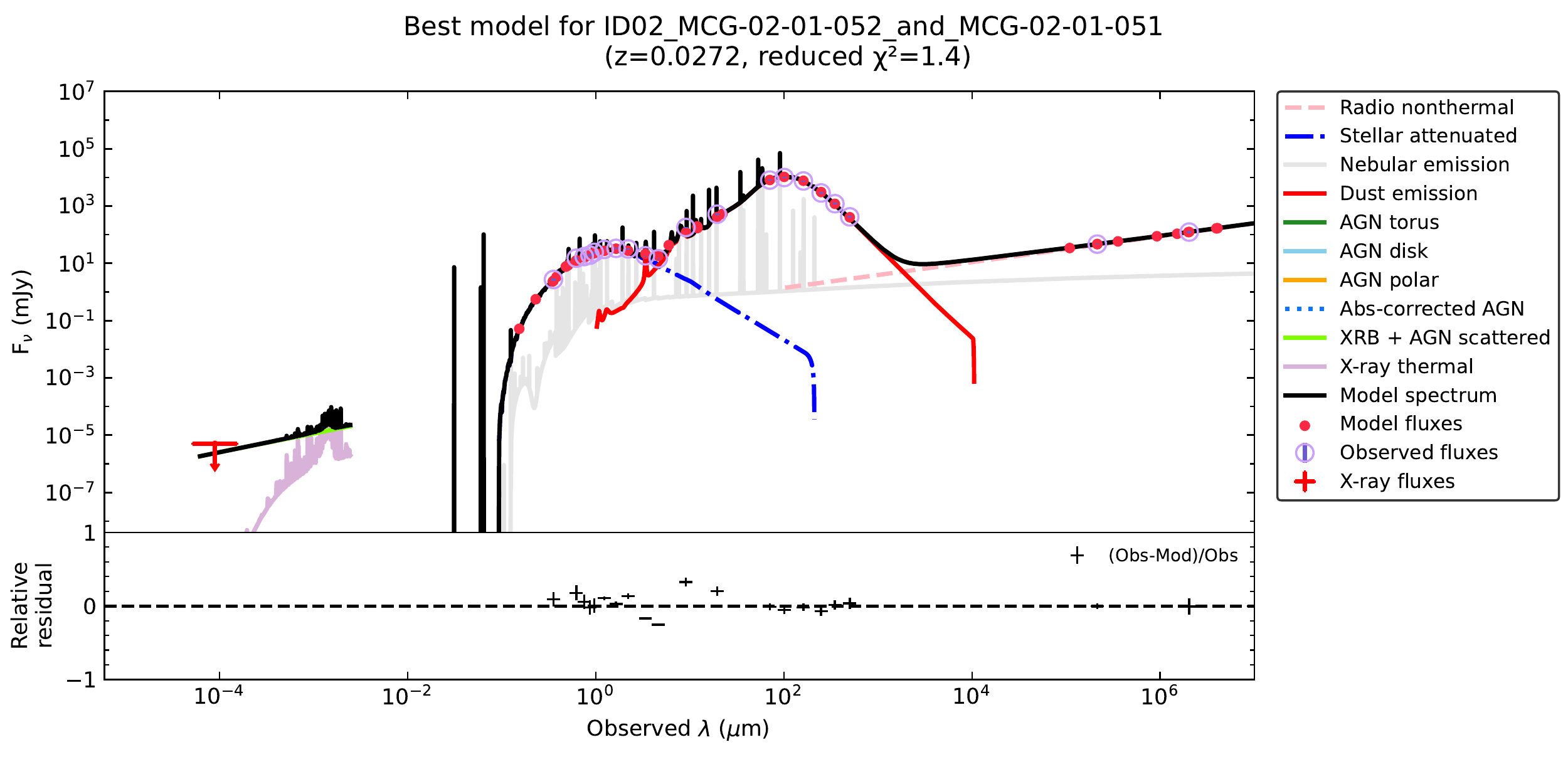}
   \includegraphics[keepaspectratio, scale=0.34]{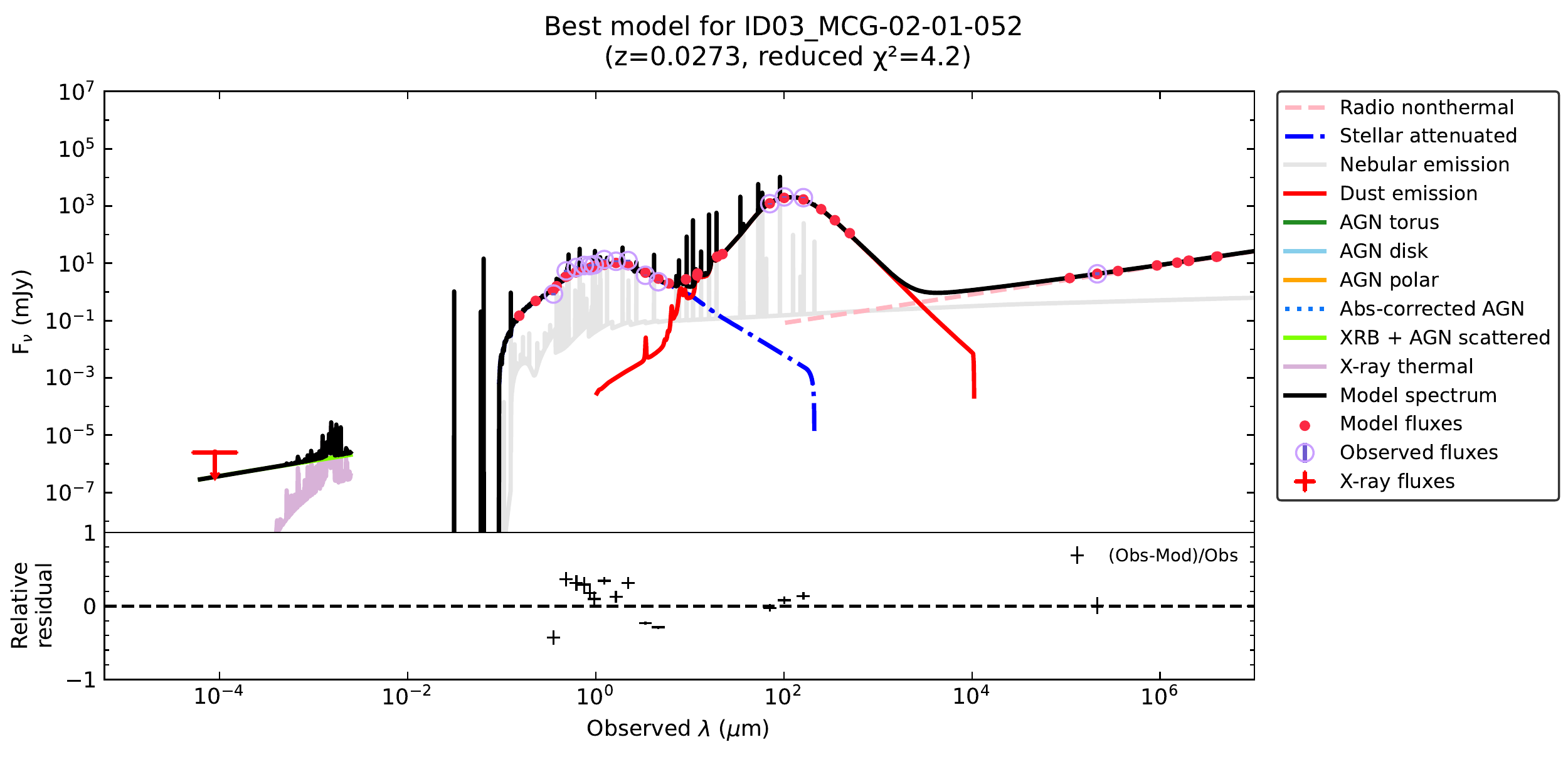}
   \includegraphics[keepaspectratio, scale=0.34]{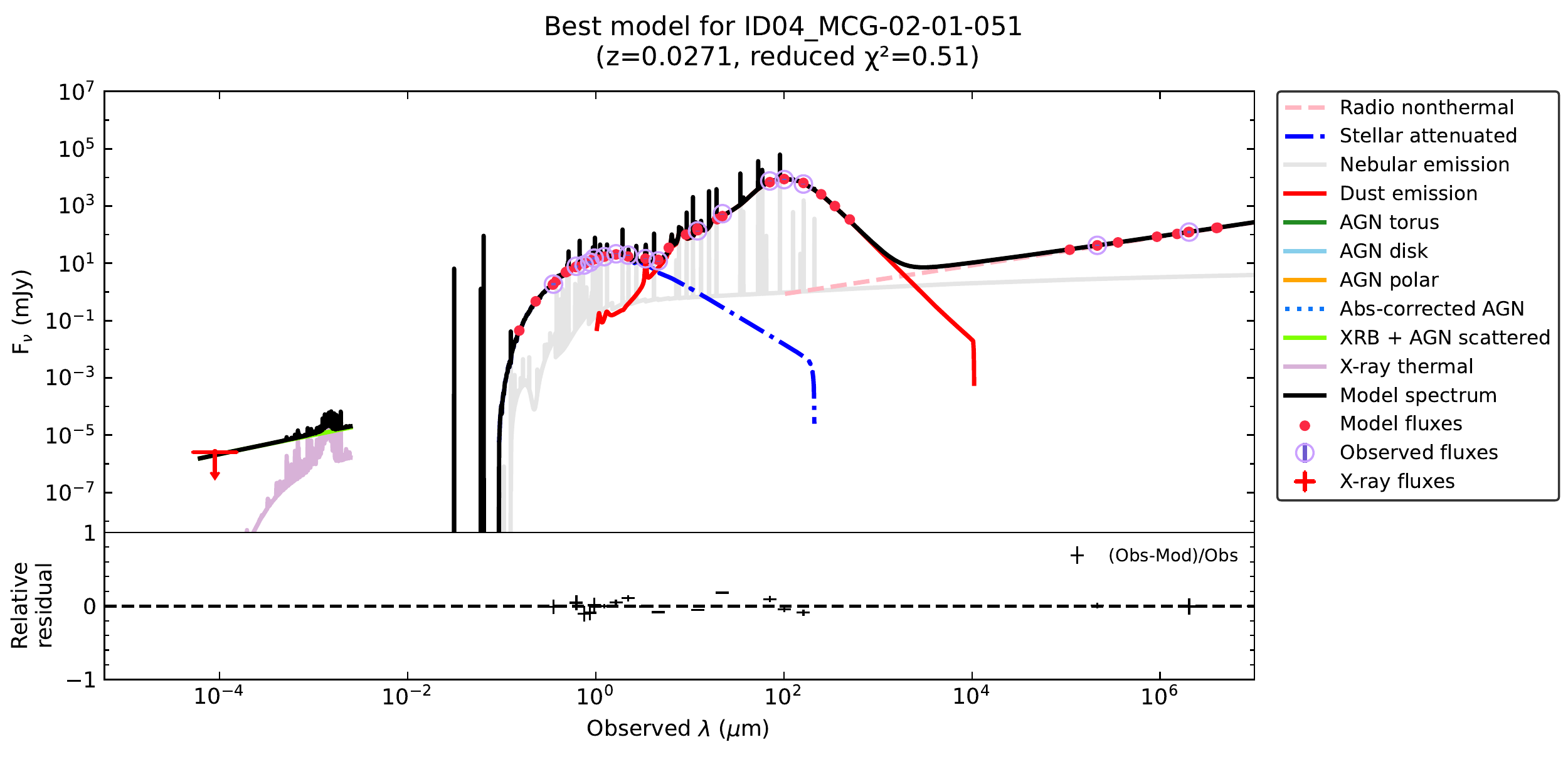}
   \caption{The hard-X-ray-to-radio SEDs and the best fitting models in units of flux density for our sample.
        The bottom panels show the residuals in the UV-to-radio bands.
        The individual curves are the same as in Figure~\ref{F1-SED00-F}.
        Purple and red circles represent the observed and model flux densities, respectively. 
        Red crosses in the X-ray band denote the NuSTAR spectra (or, Swift/BAT for NGC~235 and XMM-Newton/MOS for IC~5283 and MCG+01-59-081).
        The red arrow and green triangles mark the 5$\sigma$ upper limit.\\
        (The complete figure set of 85 images is available.)}
\label{SED01-F}
\end{figure*}

\begin{figure*}
    \centering
   \includegraphics[keepaspectratio, scale=0.34]{Lum-ID01-SED.pdf}
   \includegraphics[keepaspectratio, scale=0.34]{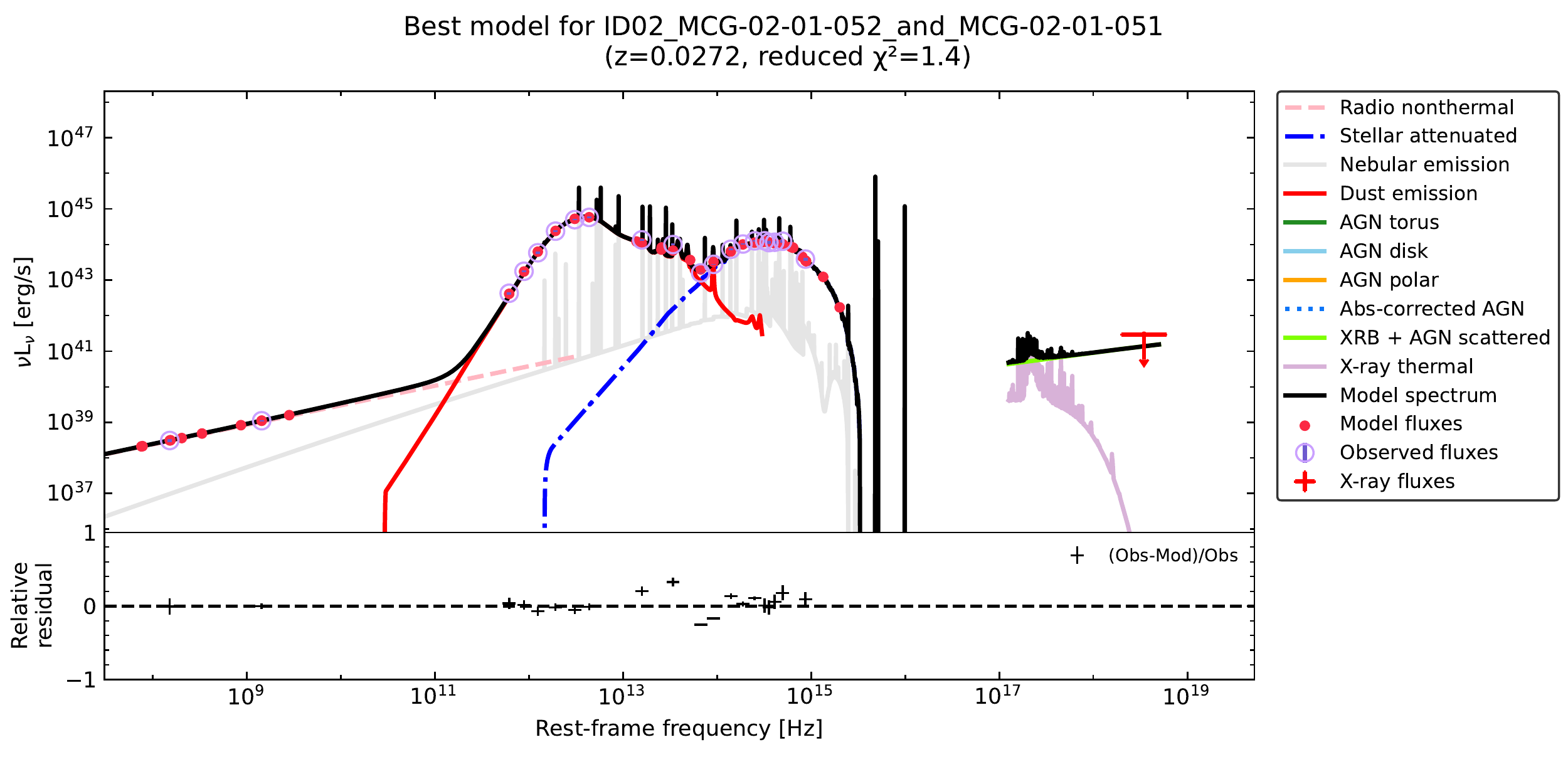}
   \includegraphics[keepaspectratio, scale=0.34]{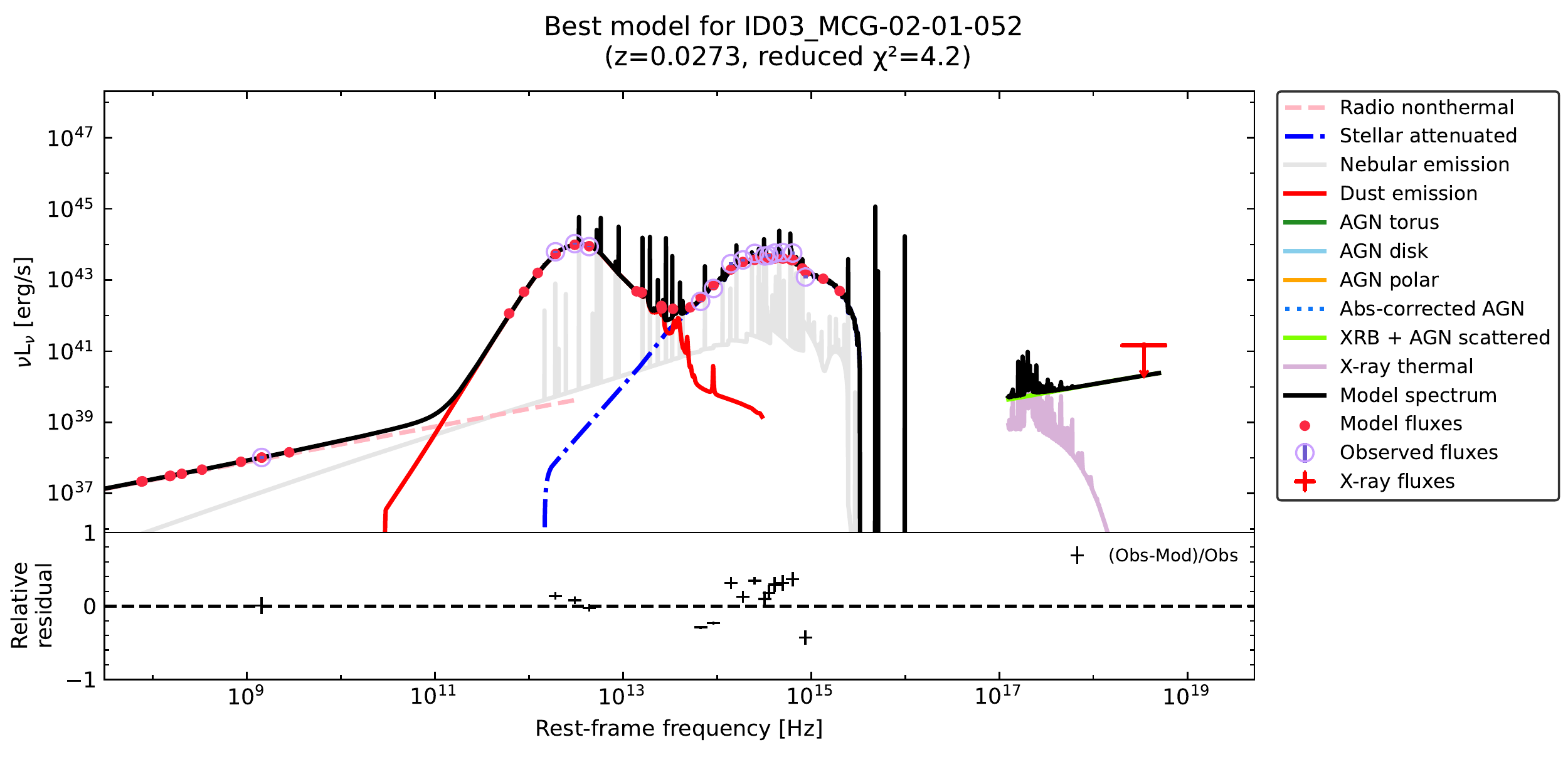}
   \includegraphics[keepaspectratio, scale=0.34]{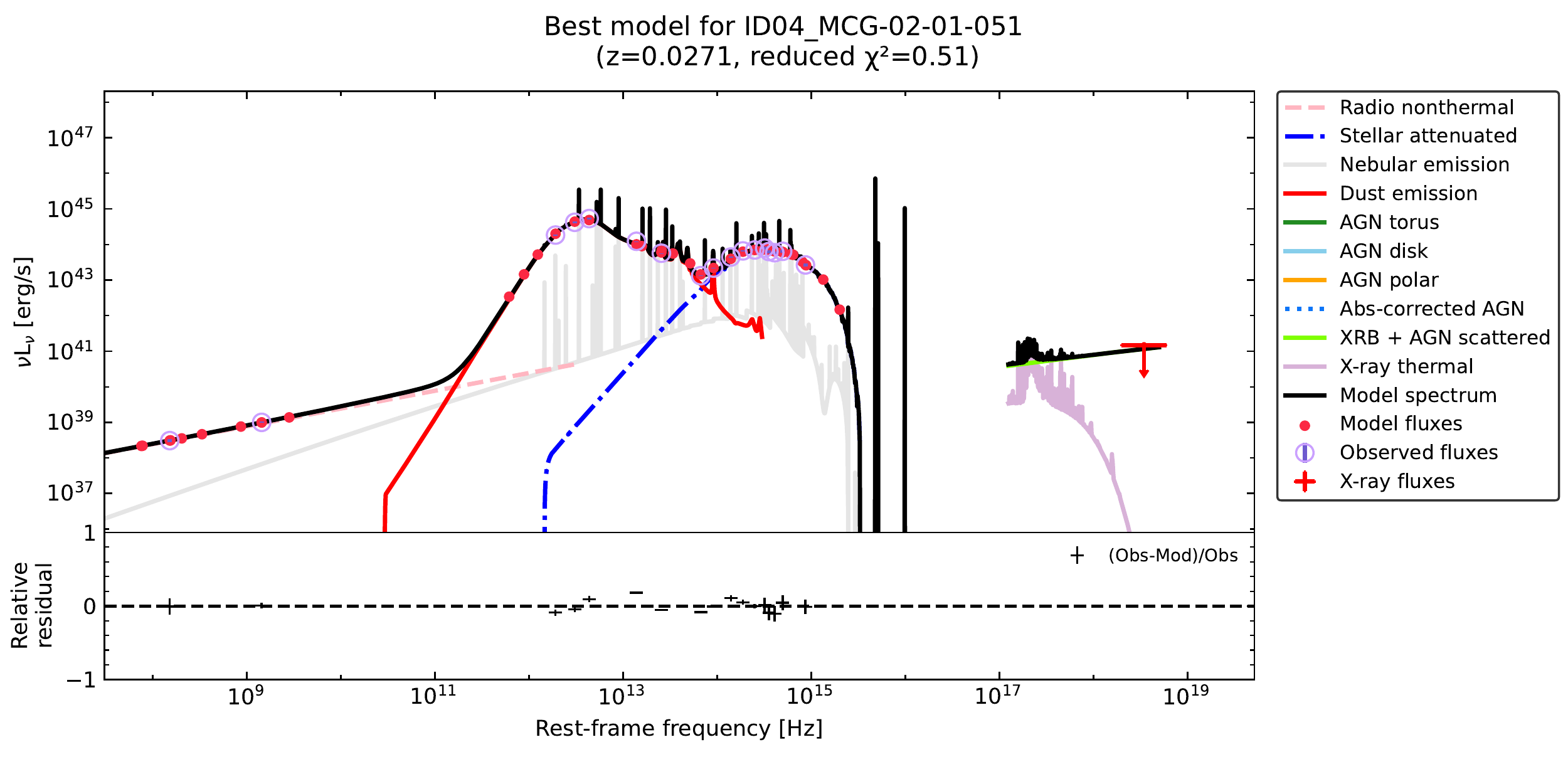}
   \includegraphics[keepaspectratio, scale=0.34]{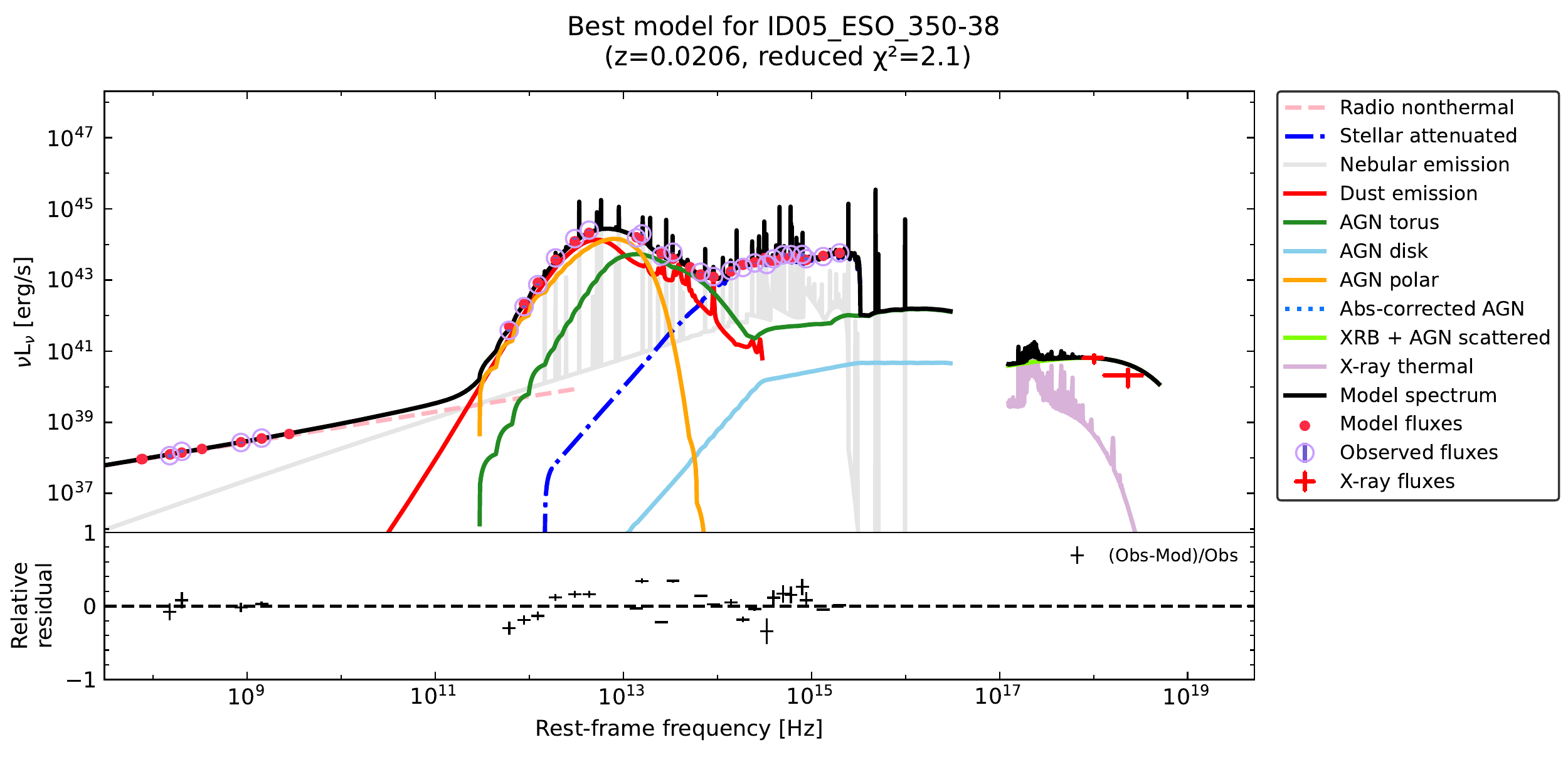}
   \includegraphics[keepaspectratio, scale=0.34]{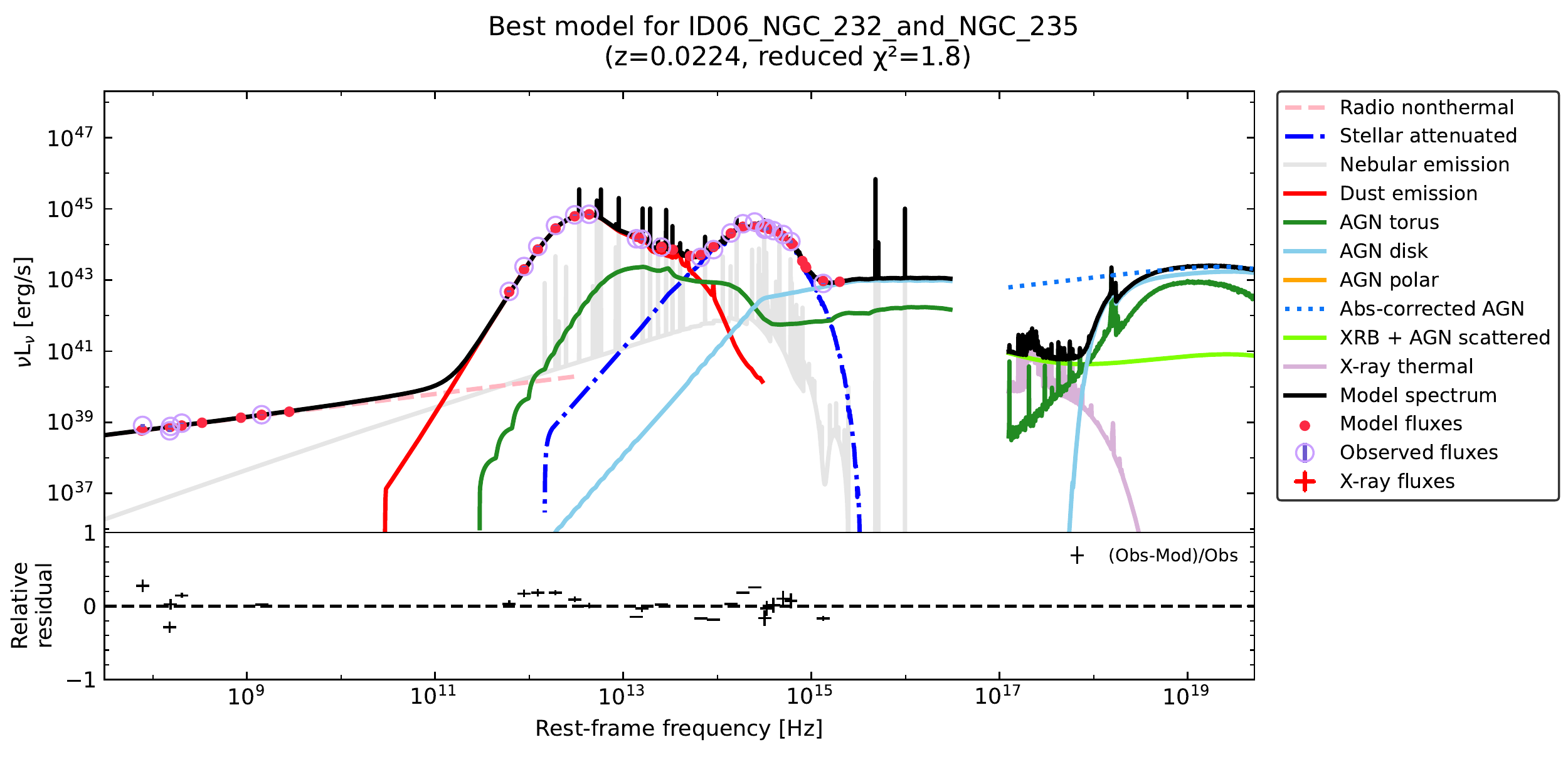}
   \includegraphics[keepaspectratio, scale=0.34]{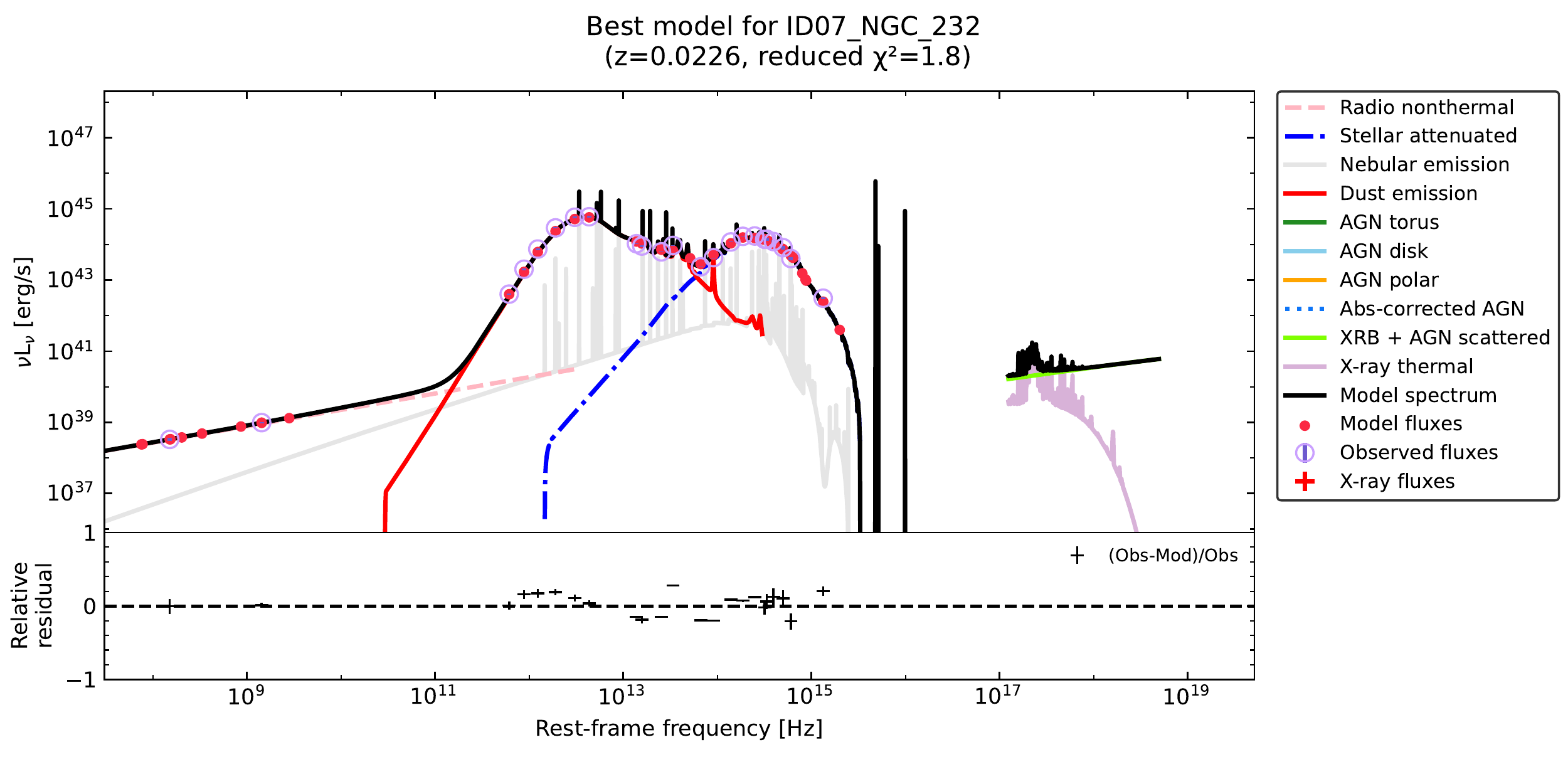}
   \includegraphics[keepaspectratio, scale=0.34]{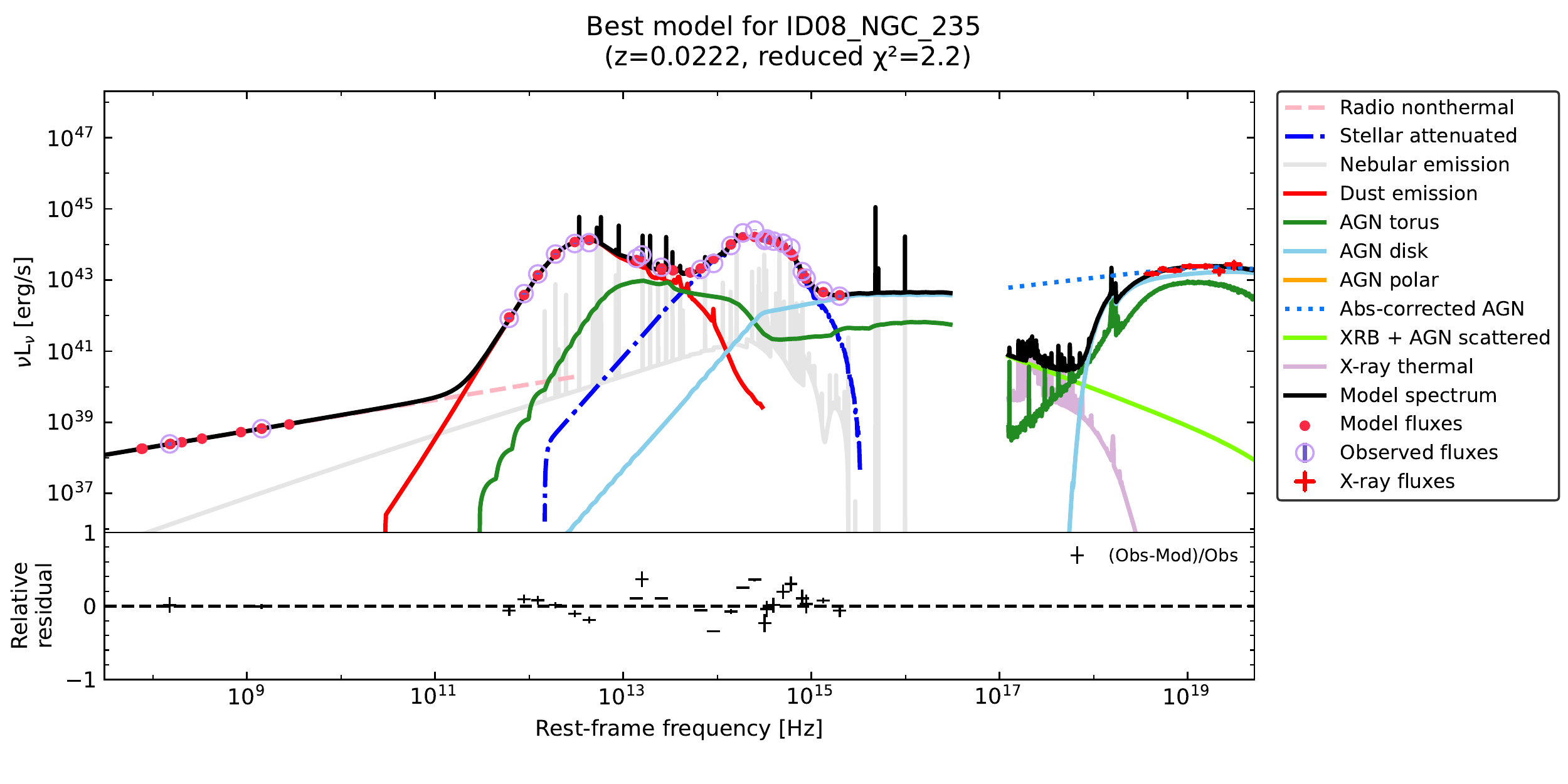}
   \includegraphics[keepaspectratio, scale=0.34]{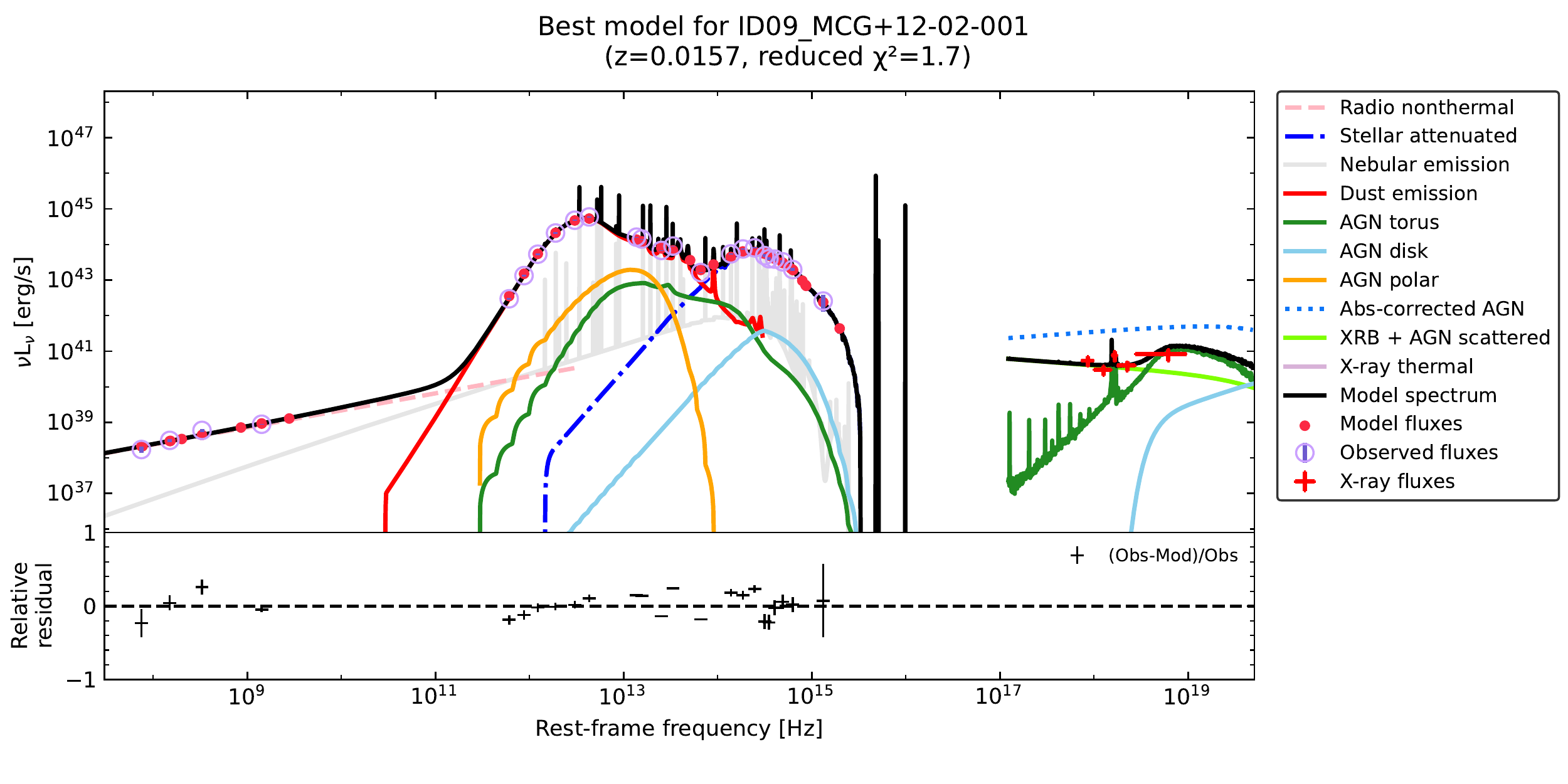}
   \includegraphics[keepaspectratio, scale=0.34]{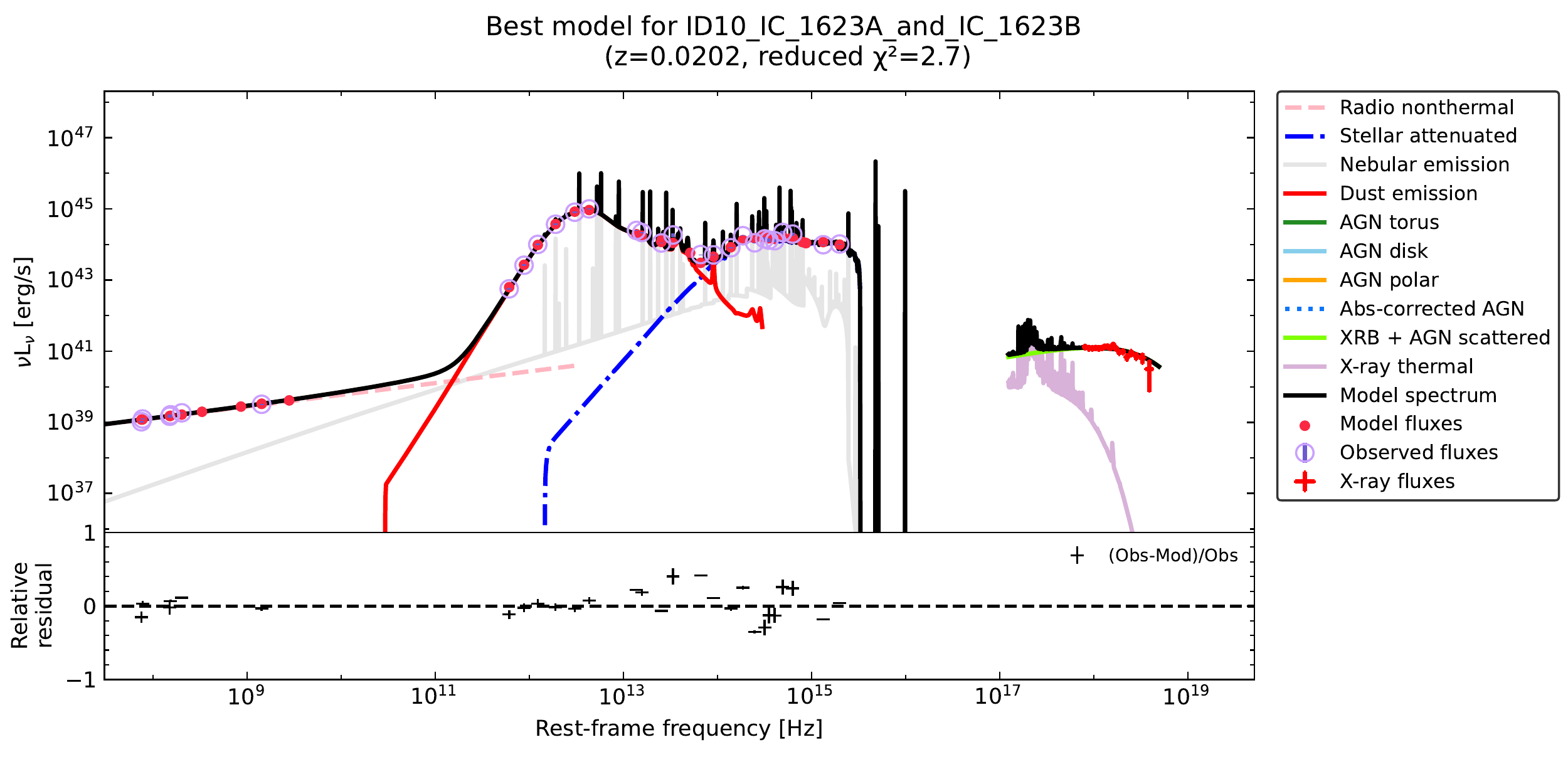}
        \caption{The hard-X-ray-to-radio SEDs and the best fitting models in units of luminosity for the sample of ID=1--10.
        The bottom panels show the residuals in the UV-to-radio bands.
        The individual curves are the same as in Figure~\ref{F1-SED00-F}.
        Purple and red circles represent the observed and model flux densities, respectively. 
        Red crosses in the X-ray band denote the NuSTAR spectra (or, Swift/BAT for NGC~235 and XMM-Newton/MOS for IC~5283 and MCG+01-59-081).
        The red arrow and green triangles mark the 5$\sigma$ upper limit.}
\label{SED10-F}
\end{figure*}

\begin{figure*}
    \centering
   \includegraphics[keepaspectratio, scale=0.34]{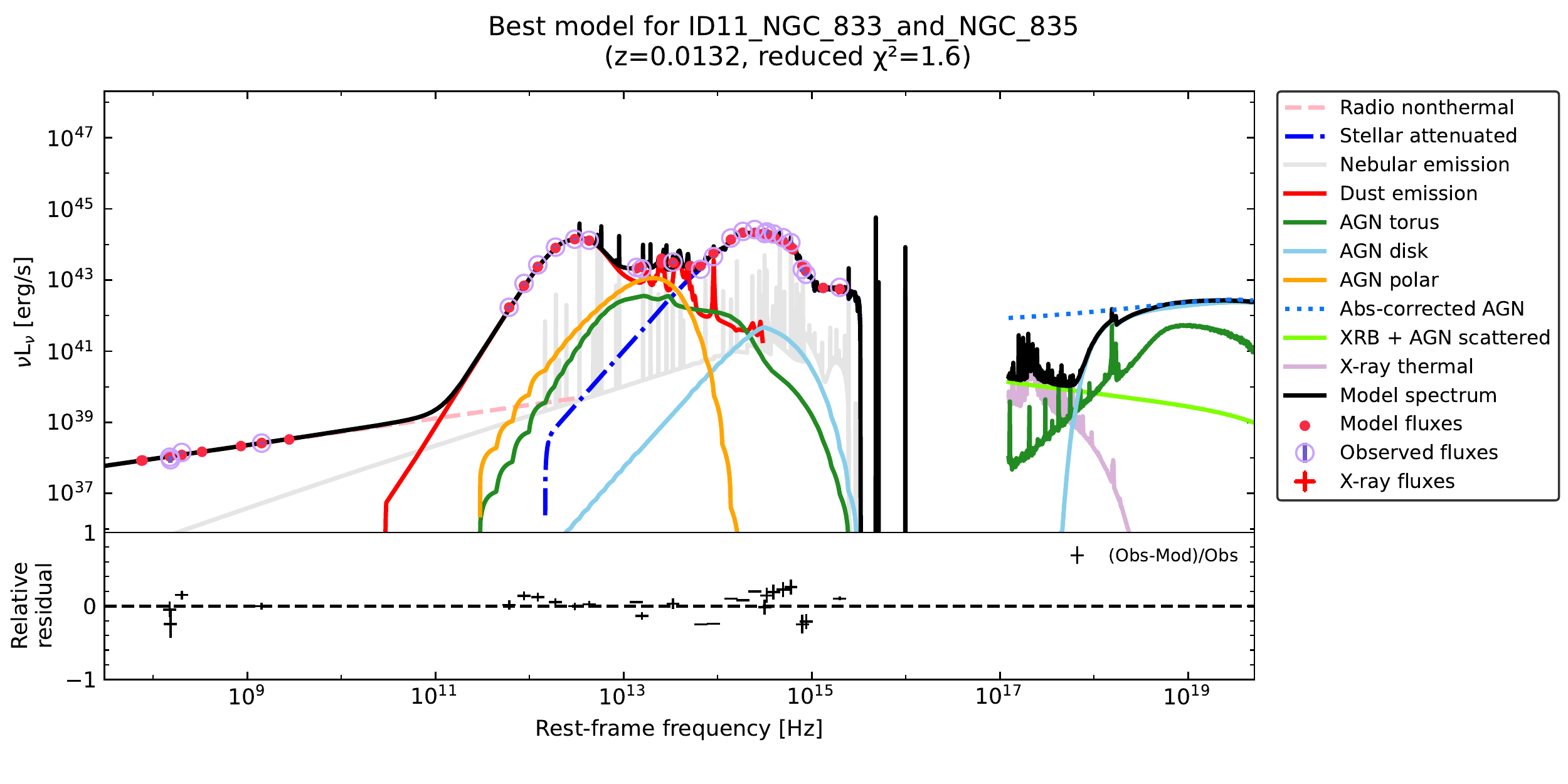}
   \includegraphics[keepaspectratio, scale=0.34]{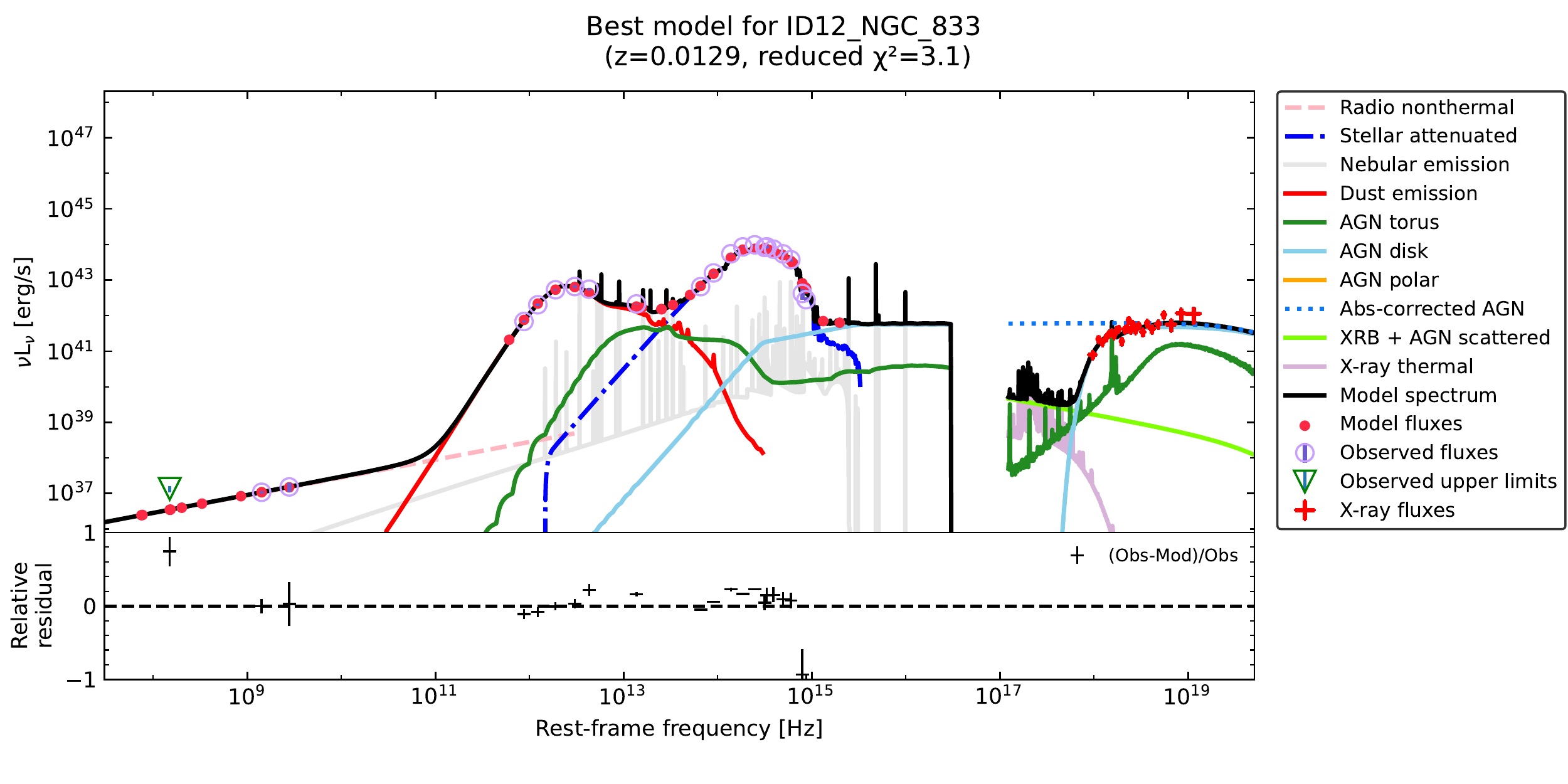}
   \includegraphics[keepaspectratio, scale=0.34]{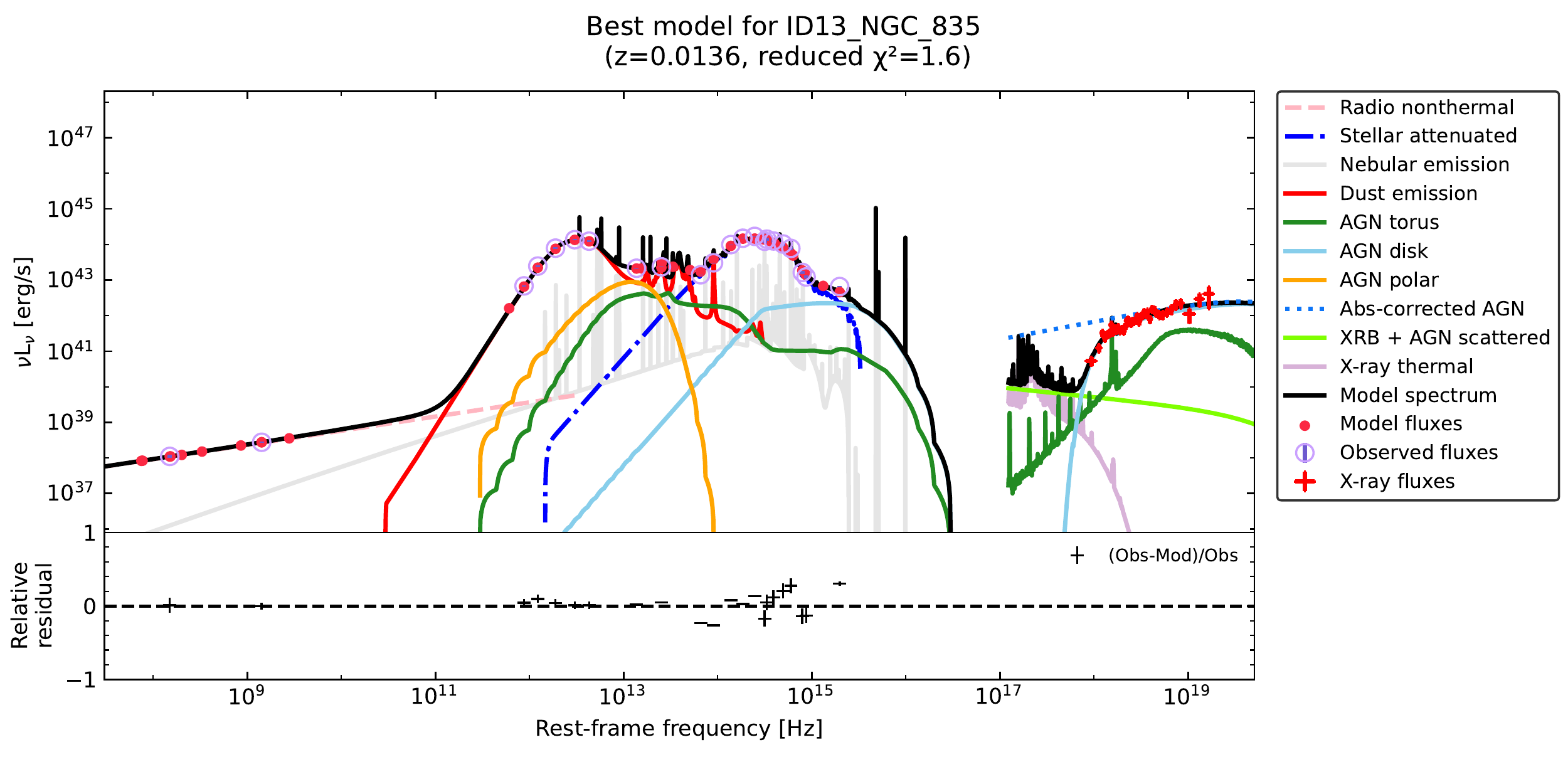}
   \includegraphics[keepaspectratio, scale=0.34]{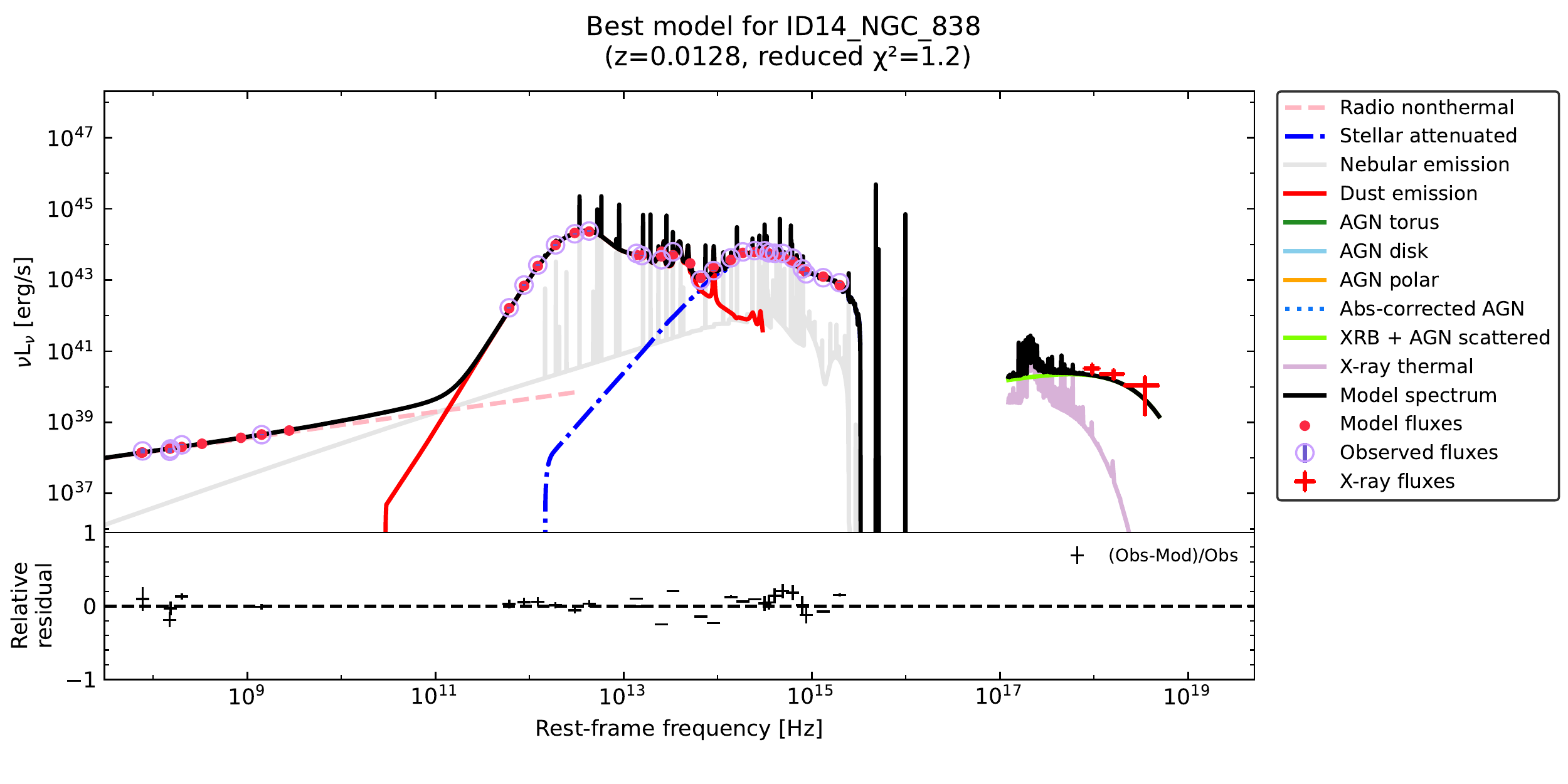}
   \includegraphics[keepaspectratio, scale=0.34]{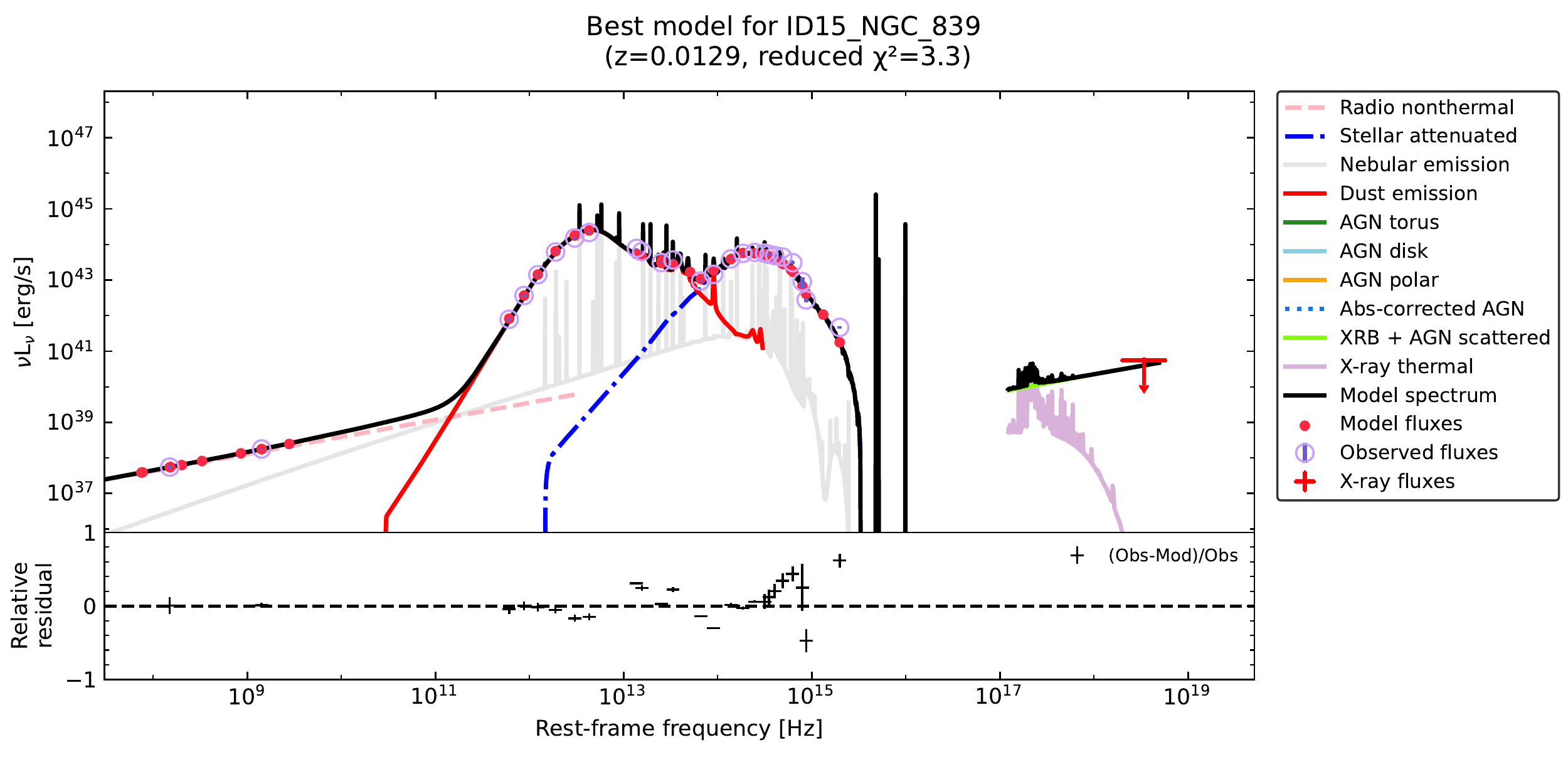}
   \includegraphics[keepaspectratio, scale=0.34]{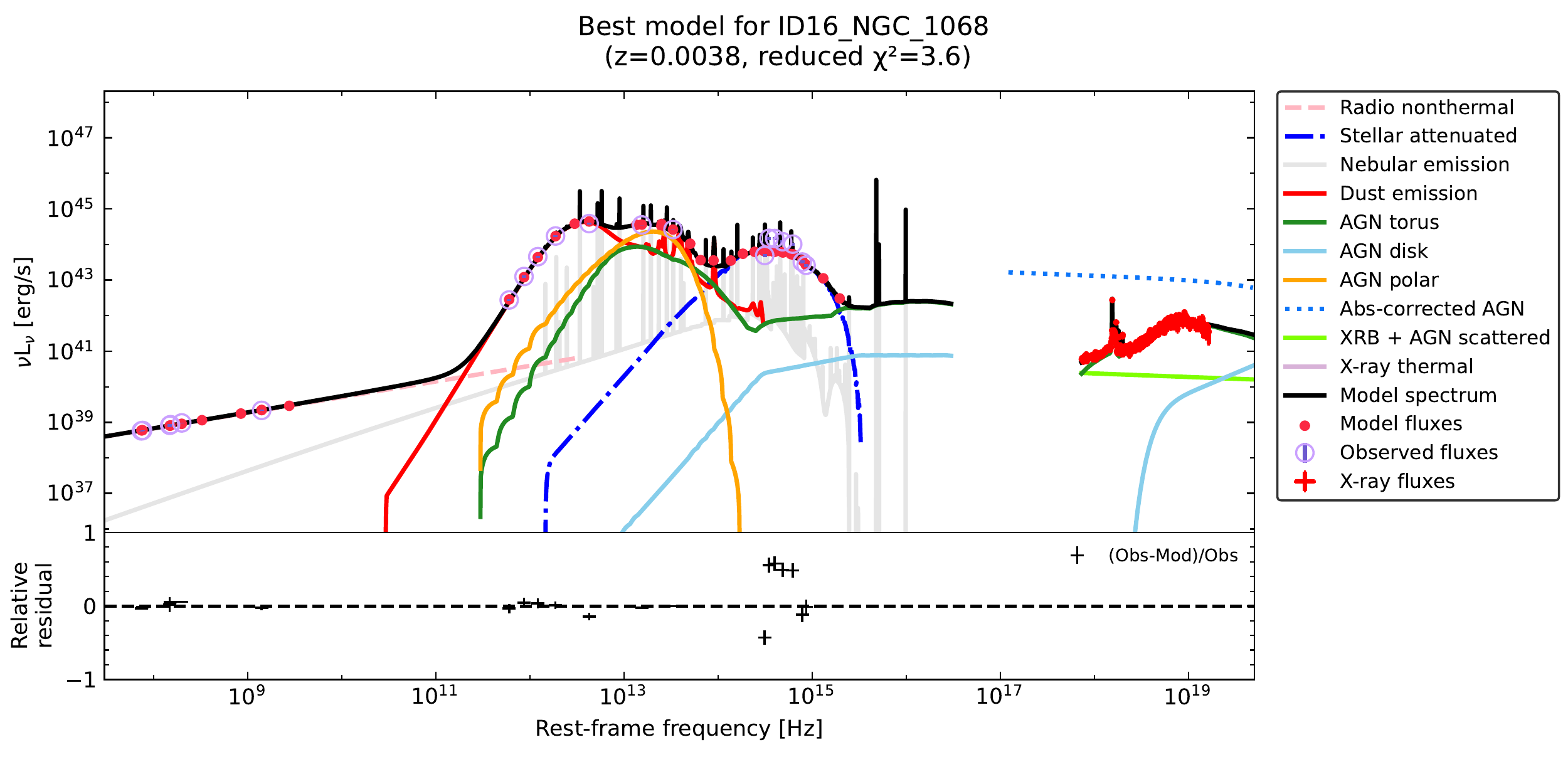}
   \includegraphics[keepaspectratio, scale=0.34]{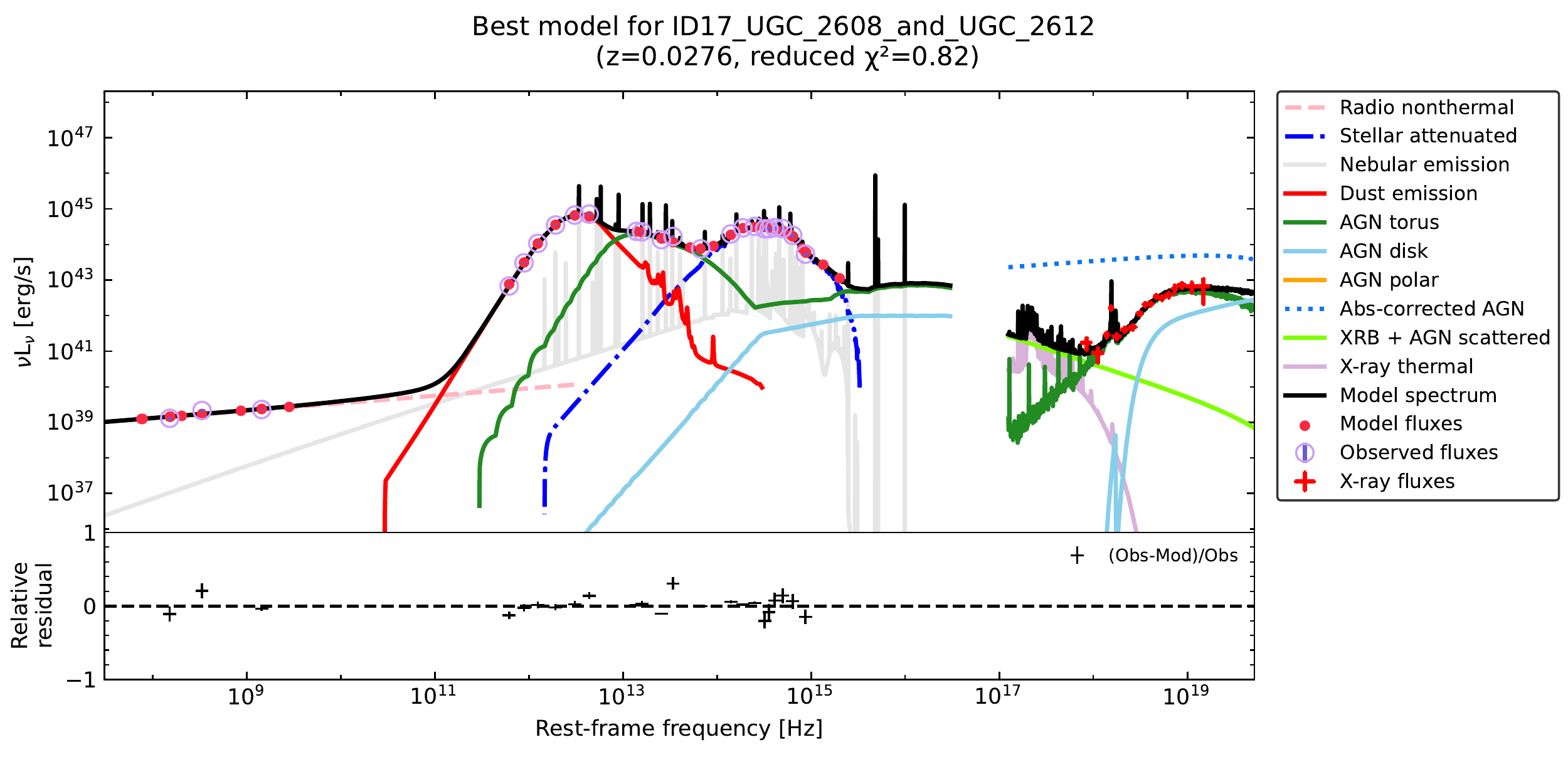}
   \includegraphics[keepaspectratio, scale=0.34]{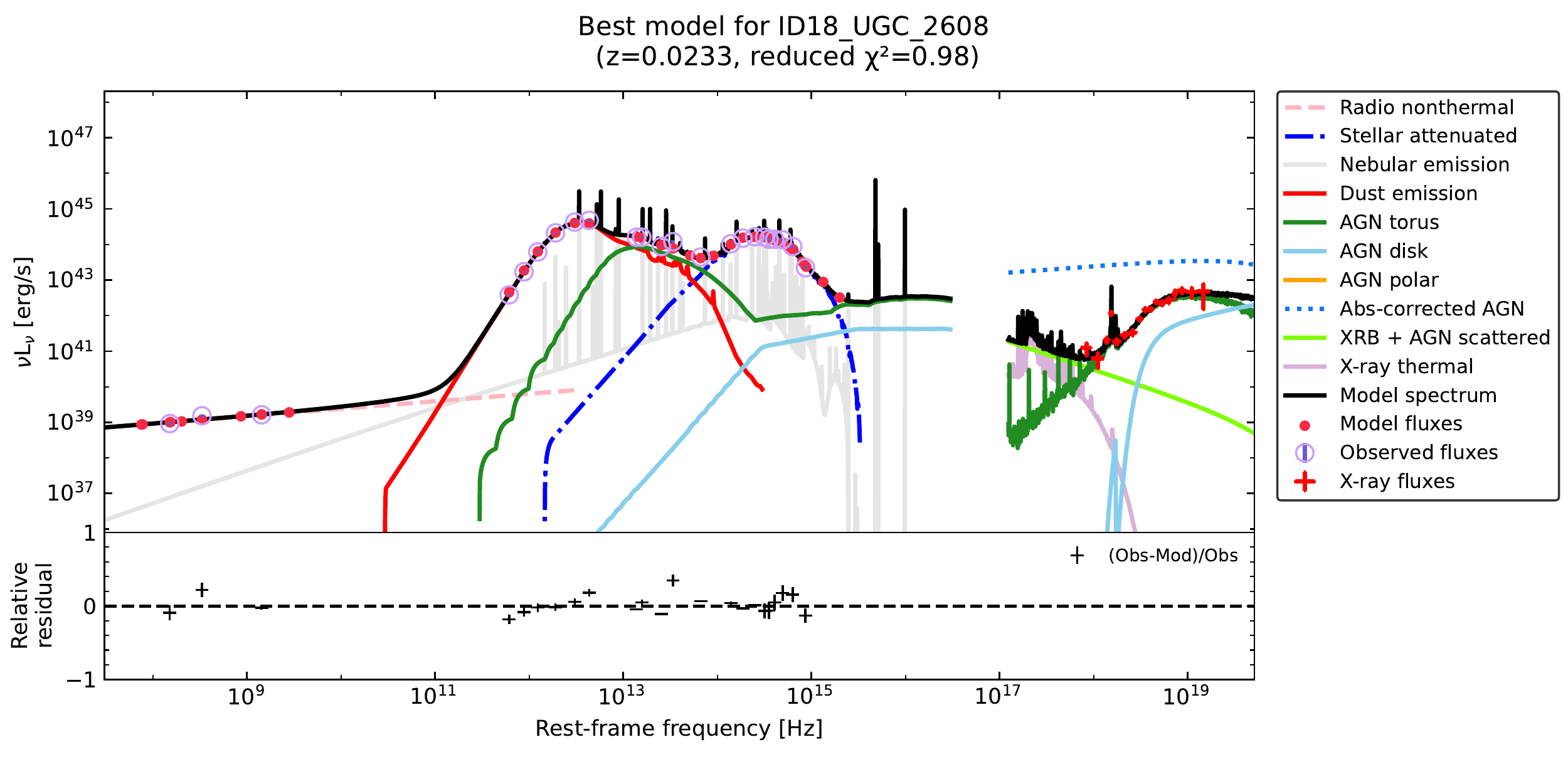}
   \includegraphics[keepaspectratio, scale=0.34]{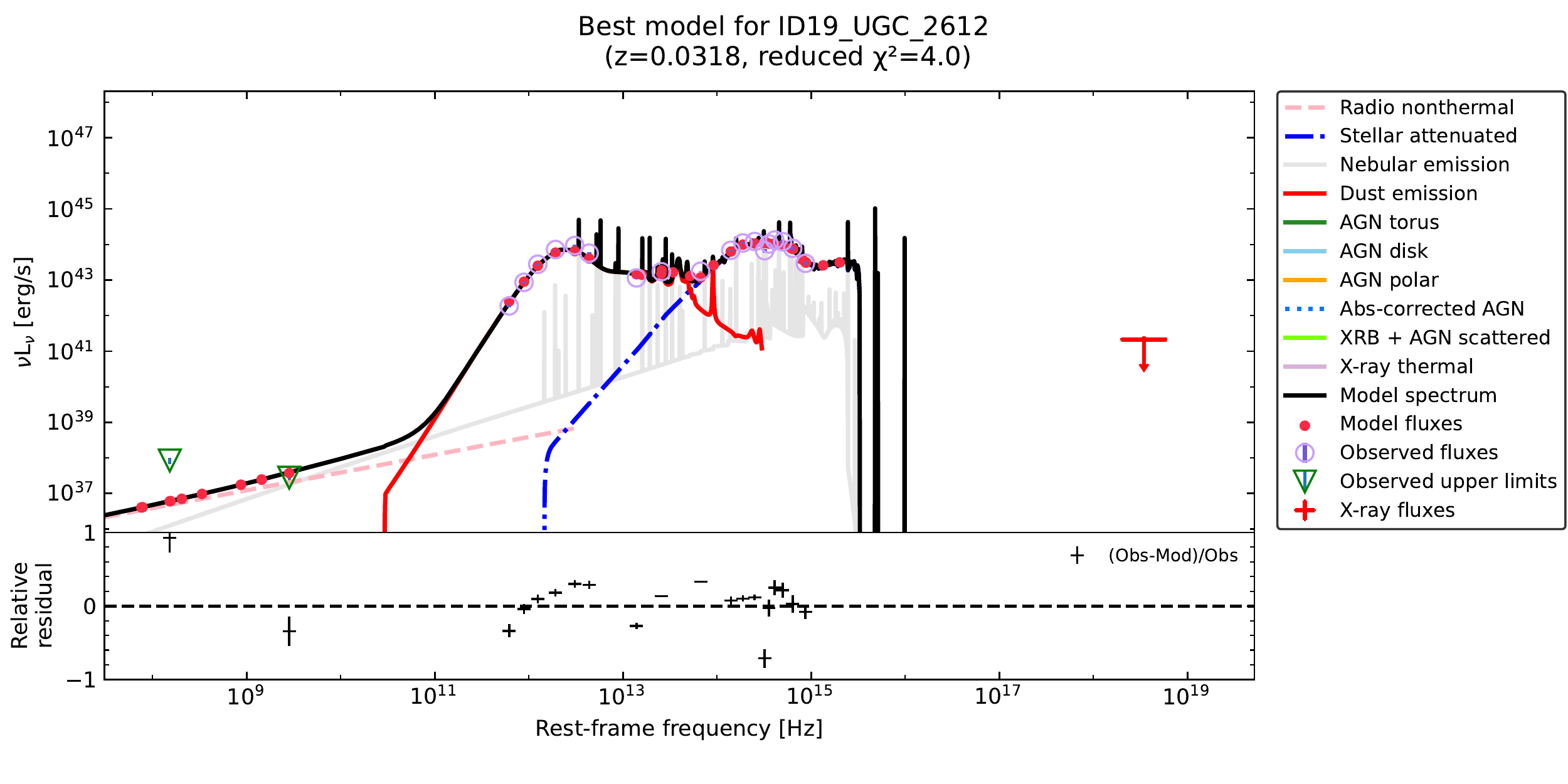}
   \includegraphics[keepaspectratio, scale=0.34]{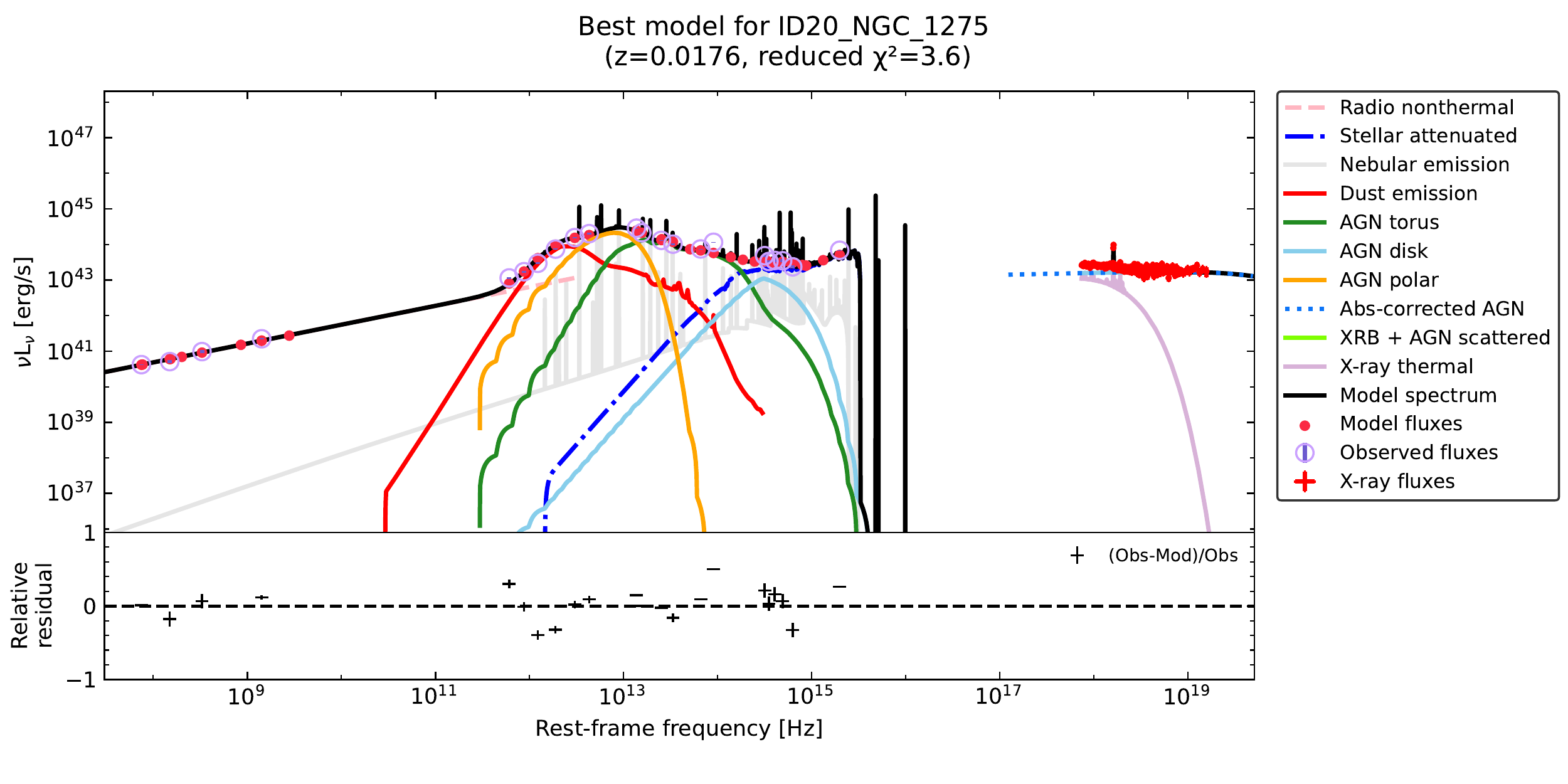}
        \caption{The same as in Figure~\ref{SED10-F} but for the sample of ID=11--20.}
\label{SED11-F}
\end{figure*}

\begin{figure*}
    \centering
   \includegraphics[keepaspectratio, scale=0.34]{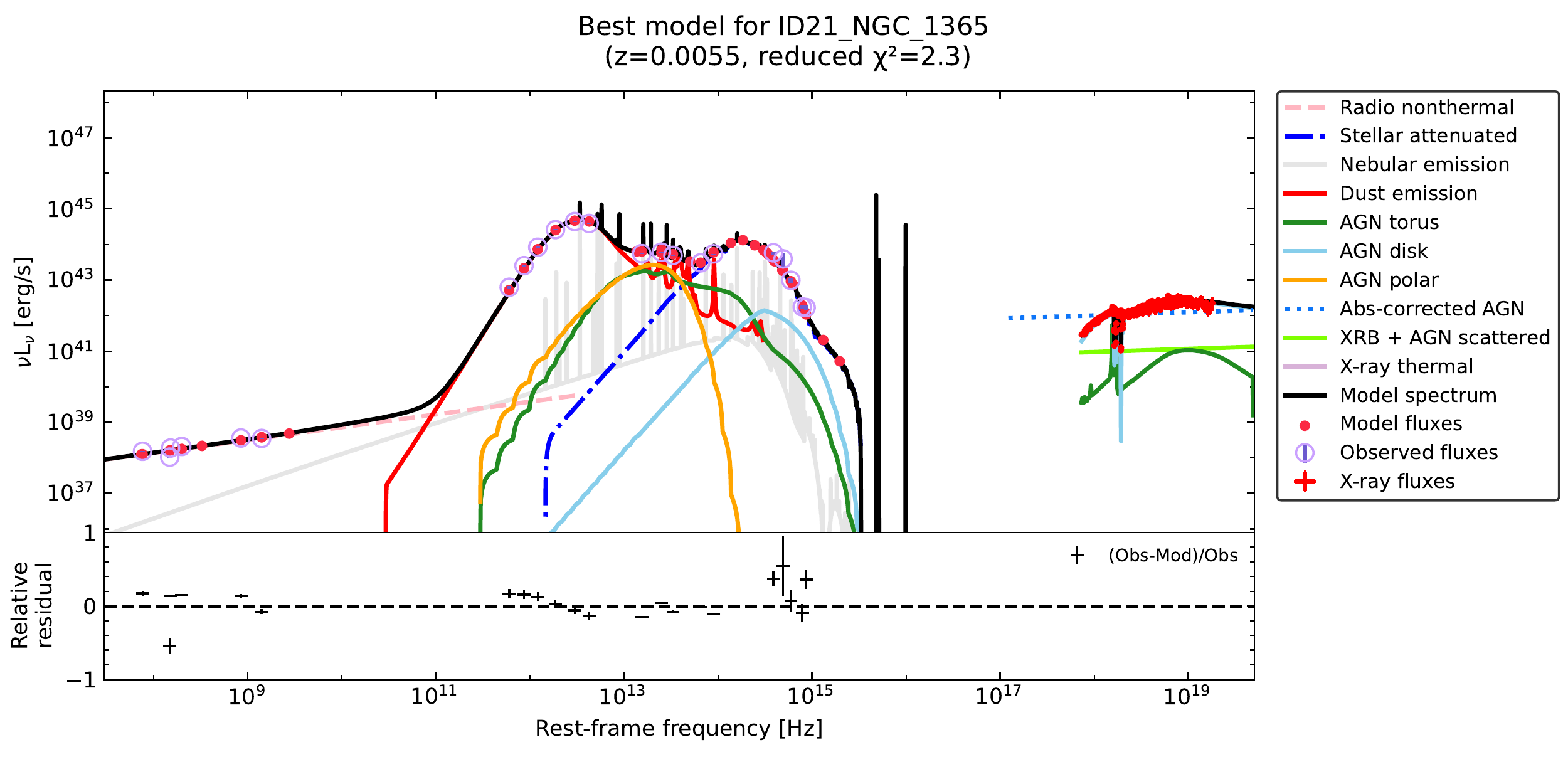}
   \includegraphics[keepaspectratio, scale=0.34]{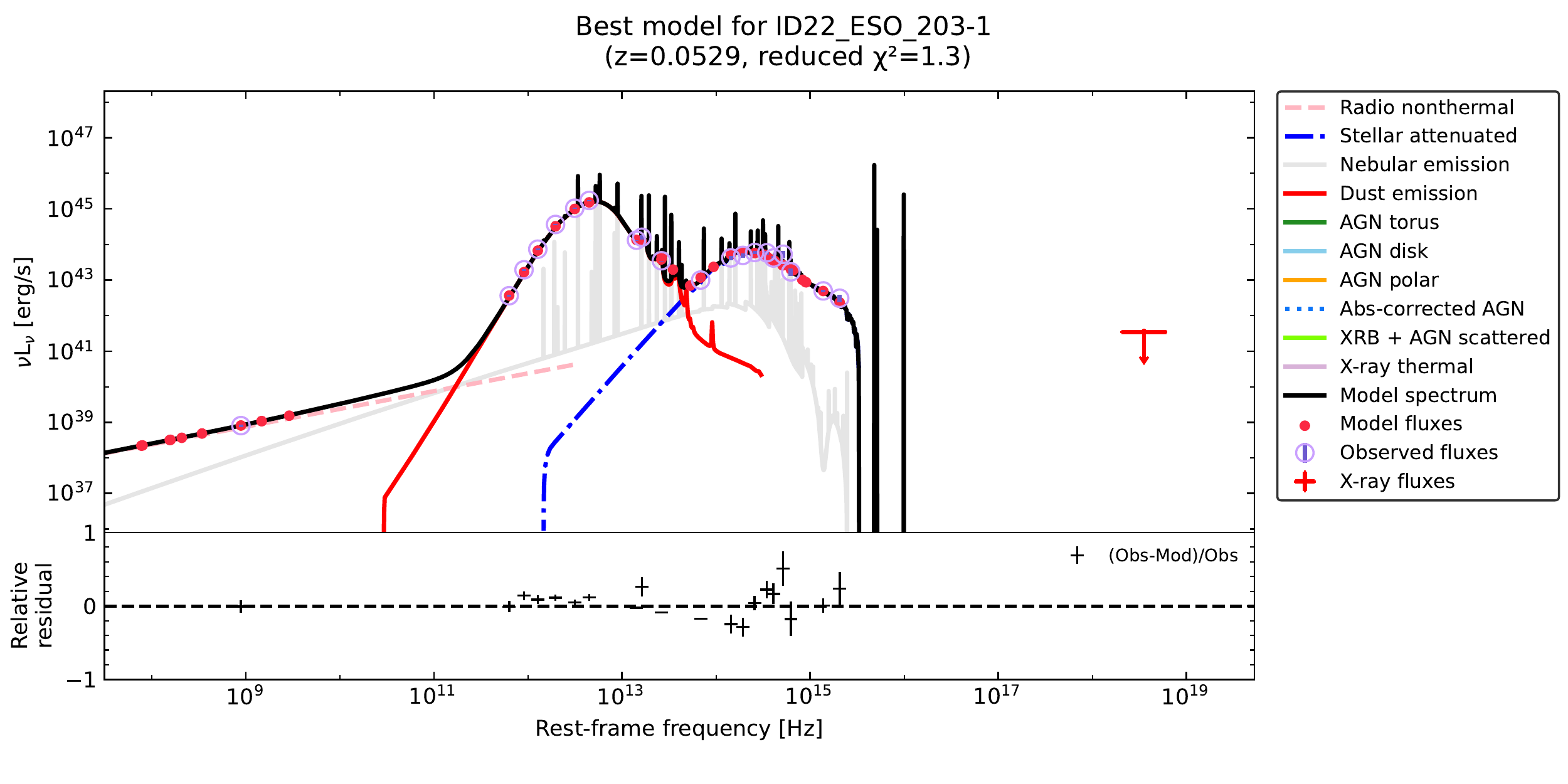}
   \includegraphics[keepaspectratio, scale=0.34]{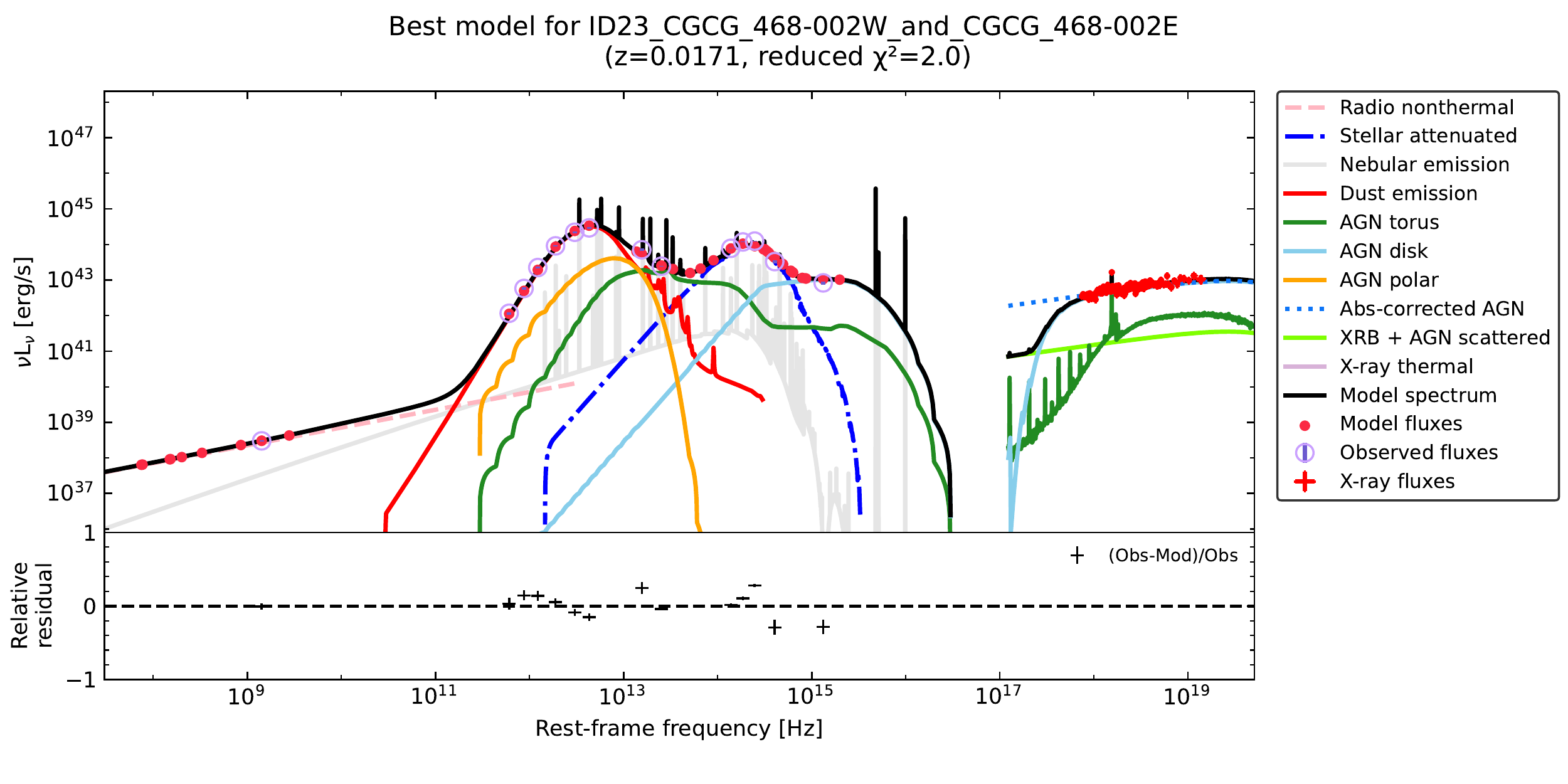}
   \includegraphics[keepaspectratio, scale=0.34]{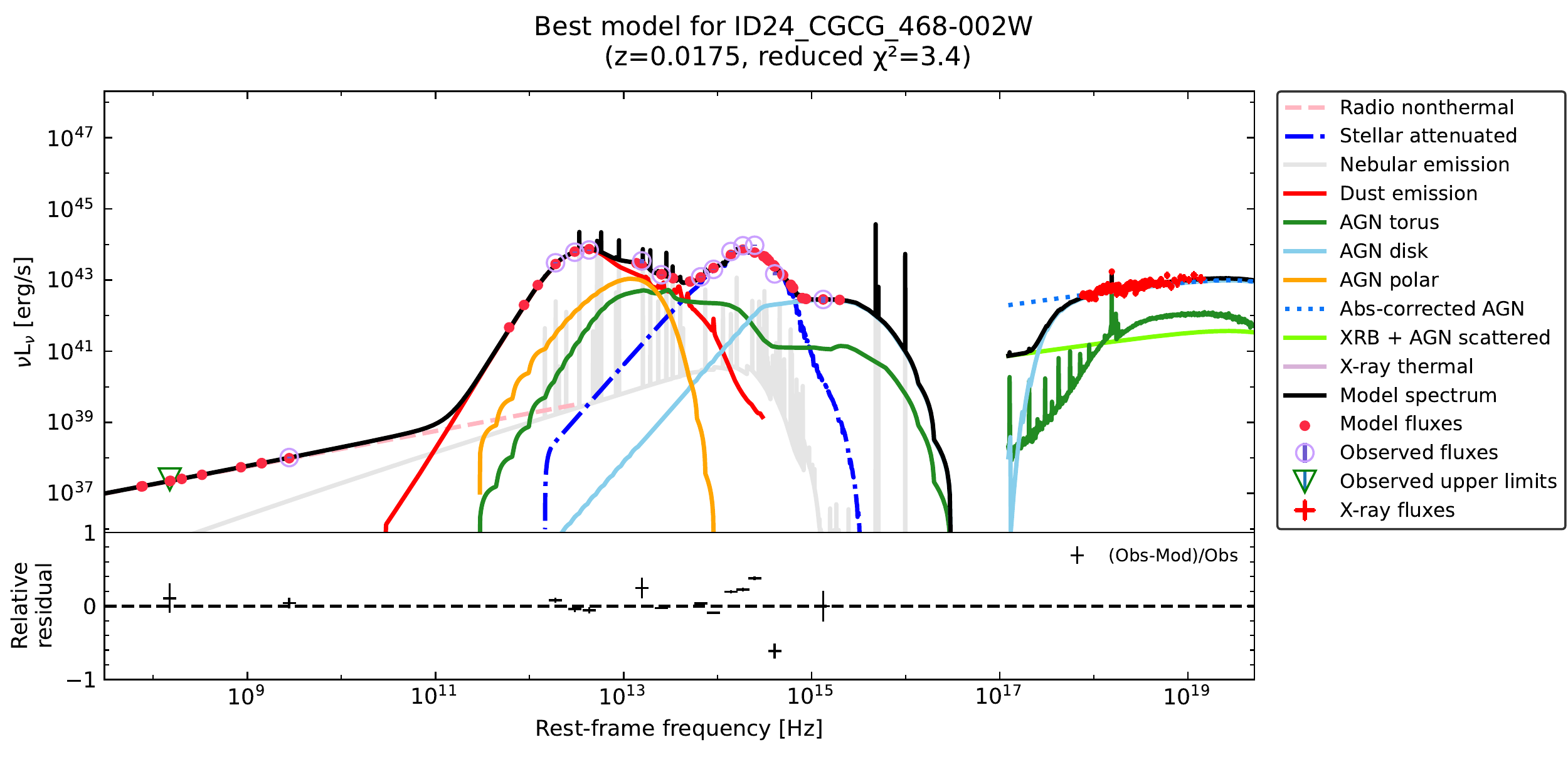}
   \includegraphics[keepaspectratio, scale=0.34]{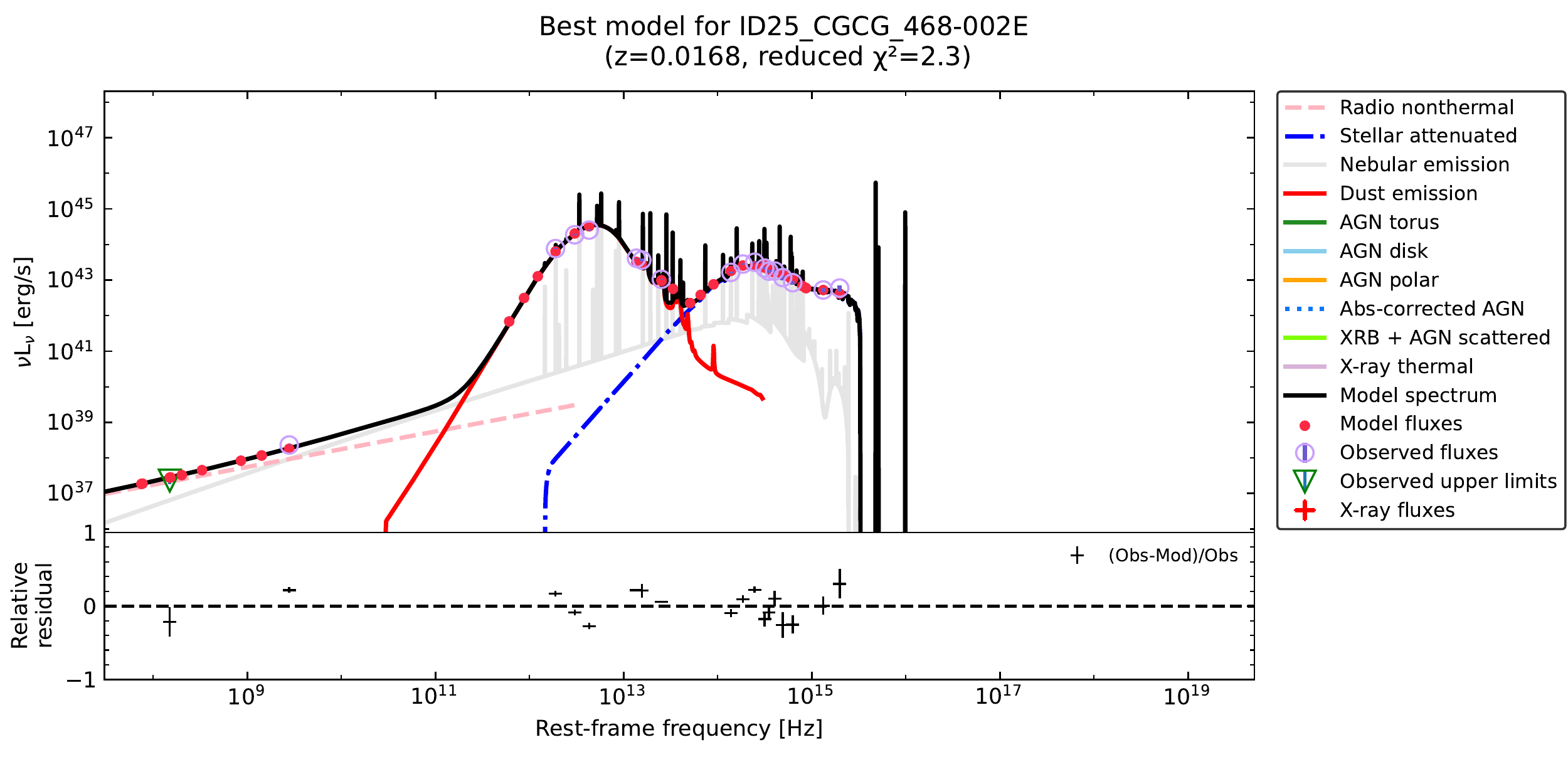}
   \includegraphics[keepaspectratio, scale=0.34]{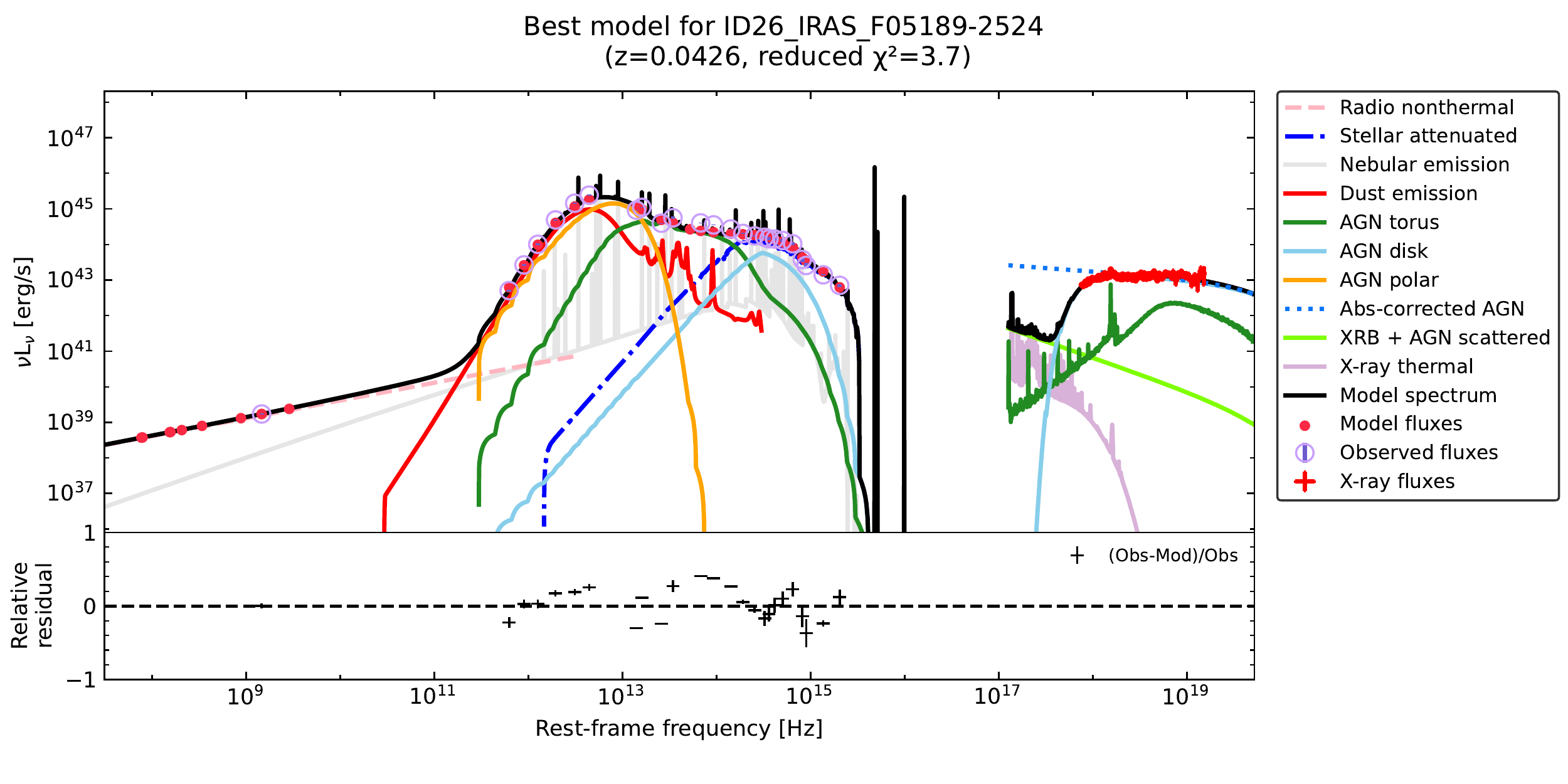}
   \includegraphics[keepaspectratio, scale=0.34]{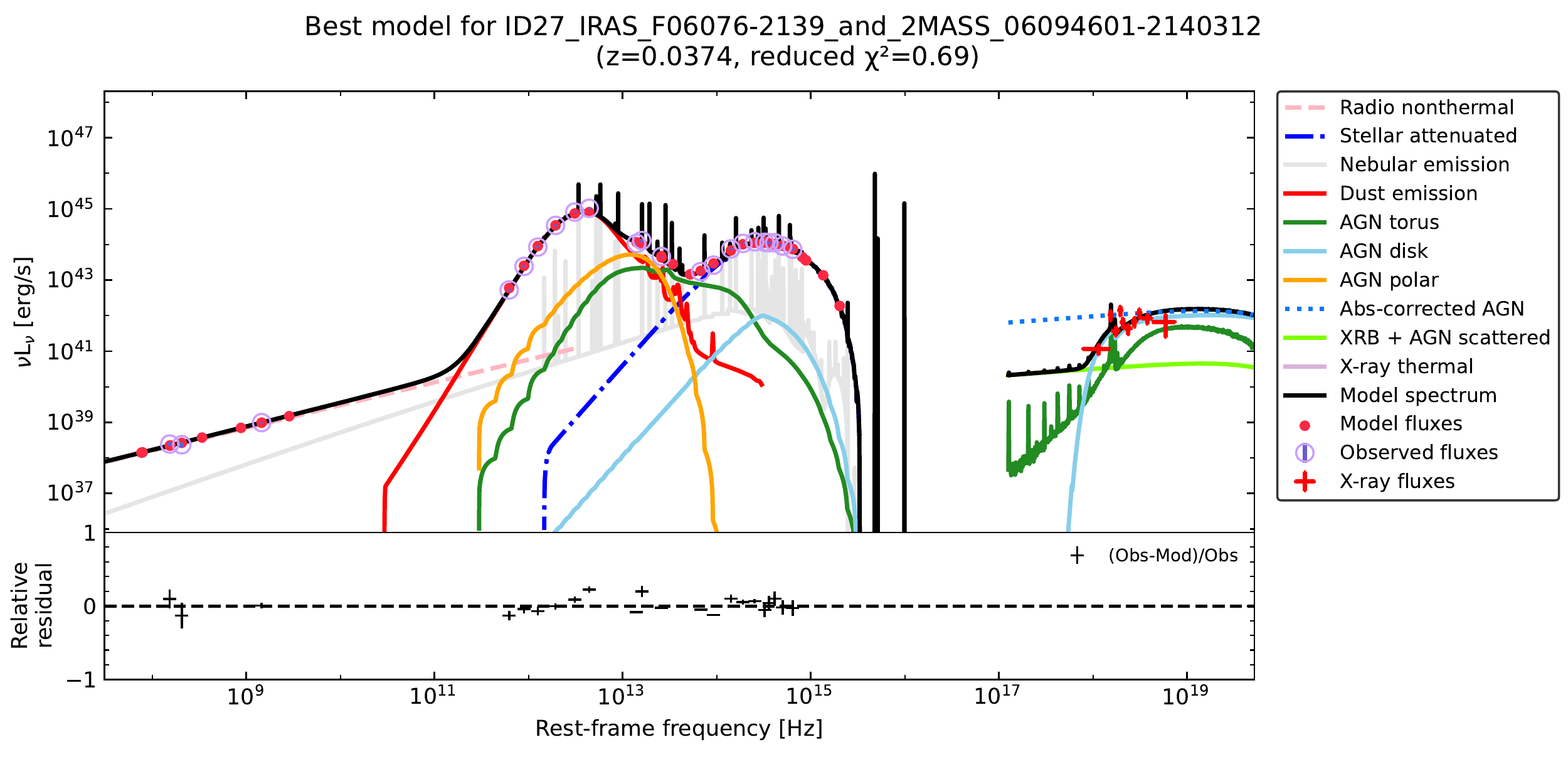}
   \includegraphics[keepaspectratio, scale=0.34]{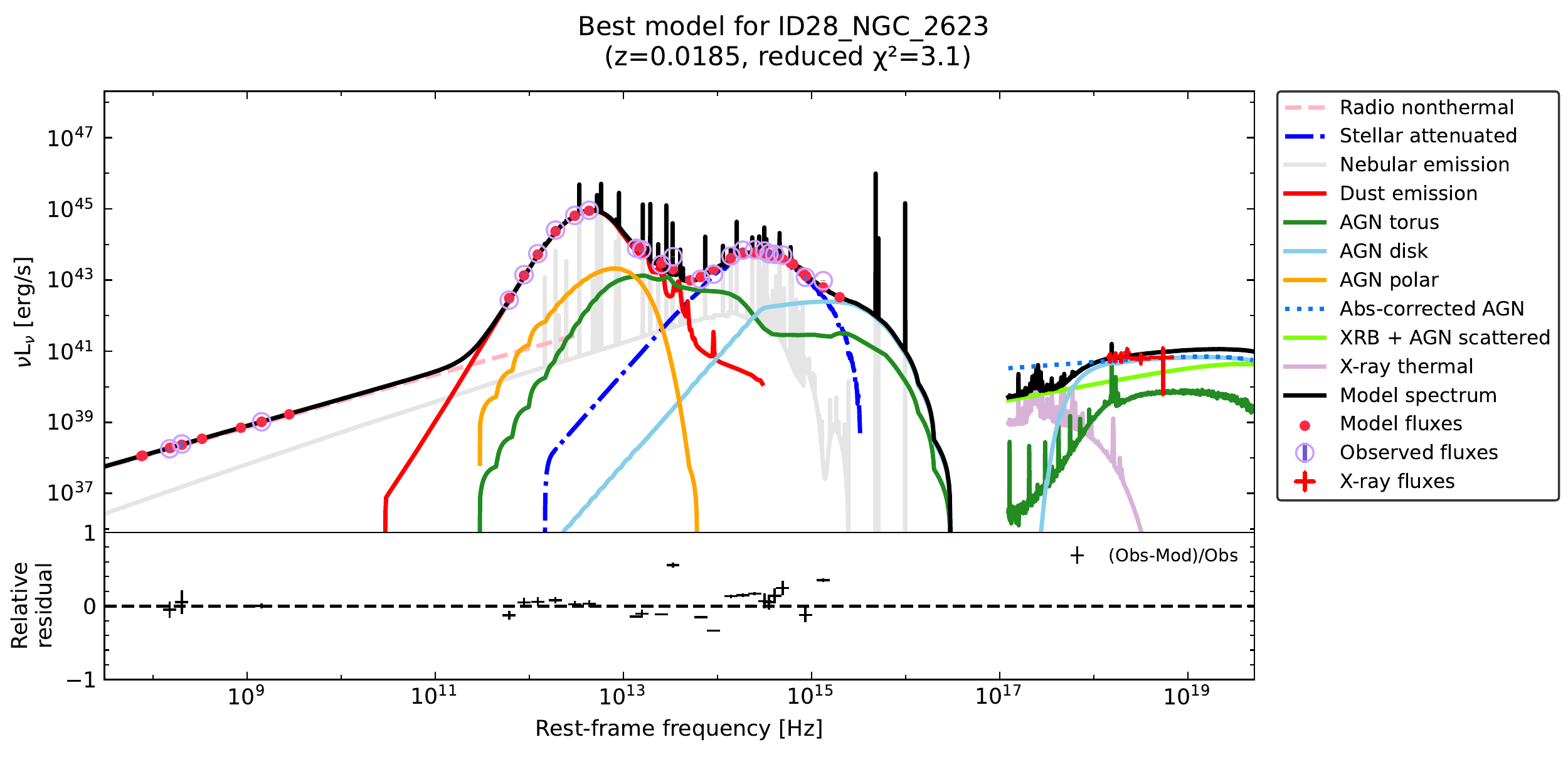}
   \includegraphics[keepaspectratio, scale=0.34]{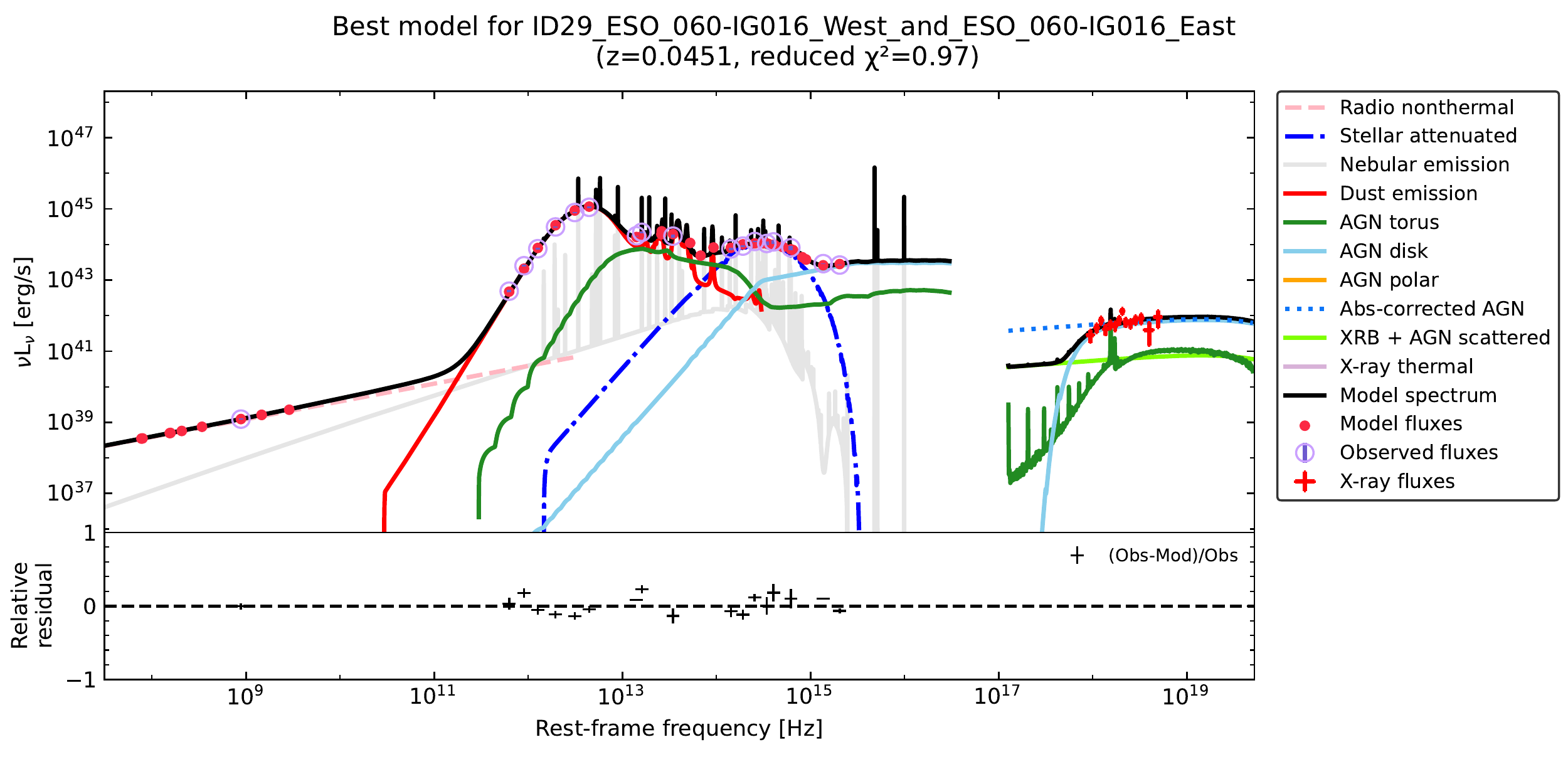}
   \includegraphics[keepaspectratio, scale=0.34]{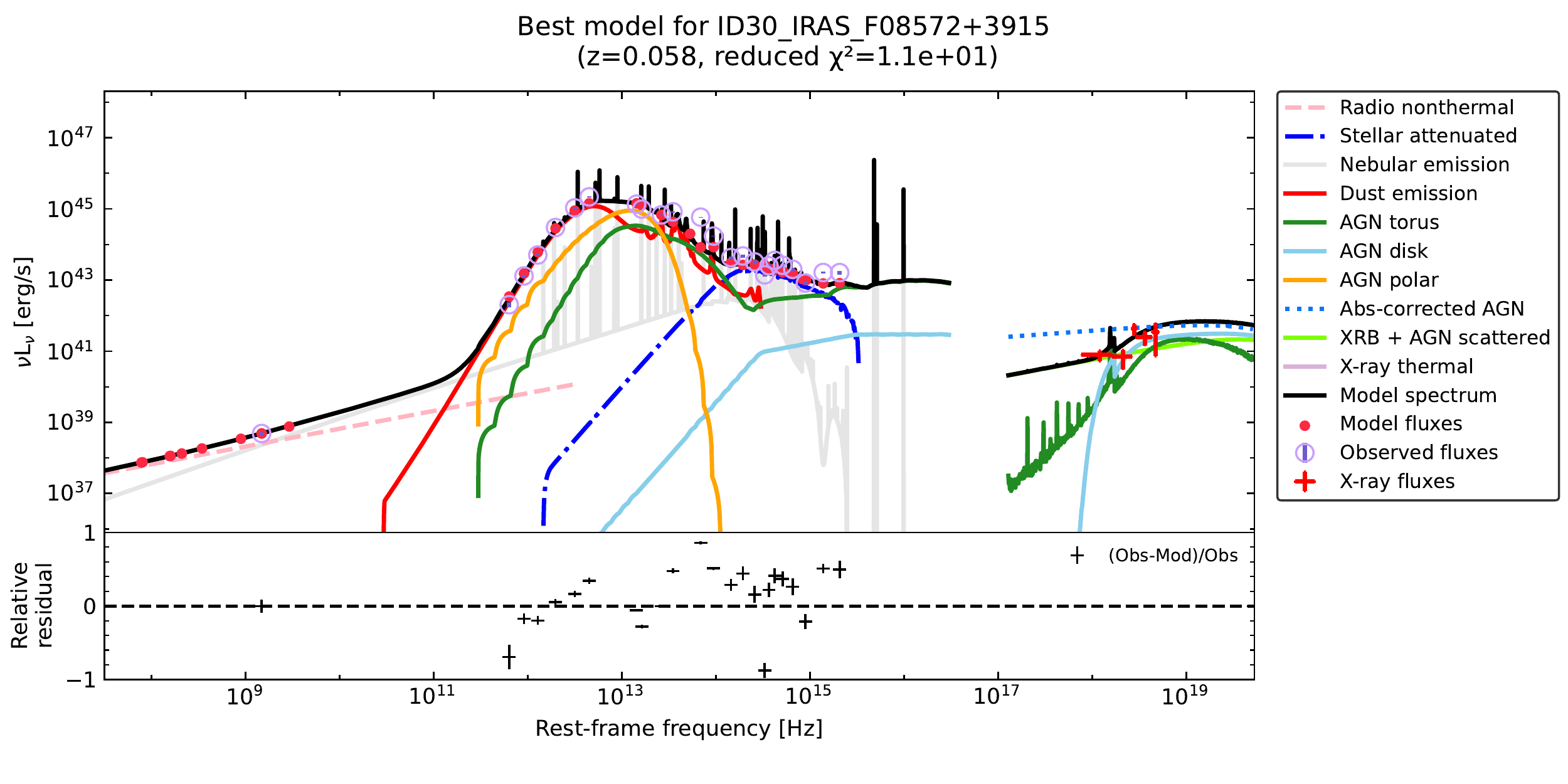}
        \caption{The same as in Figure~\ref{SED10-F} but for the sample of ID=21--30.}
\label{SED12-F}
\end{figure*}

\begin{figure*}
    \centering
   \includegraphics[keepaspectratio, scale=0.34]{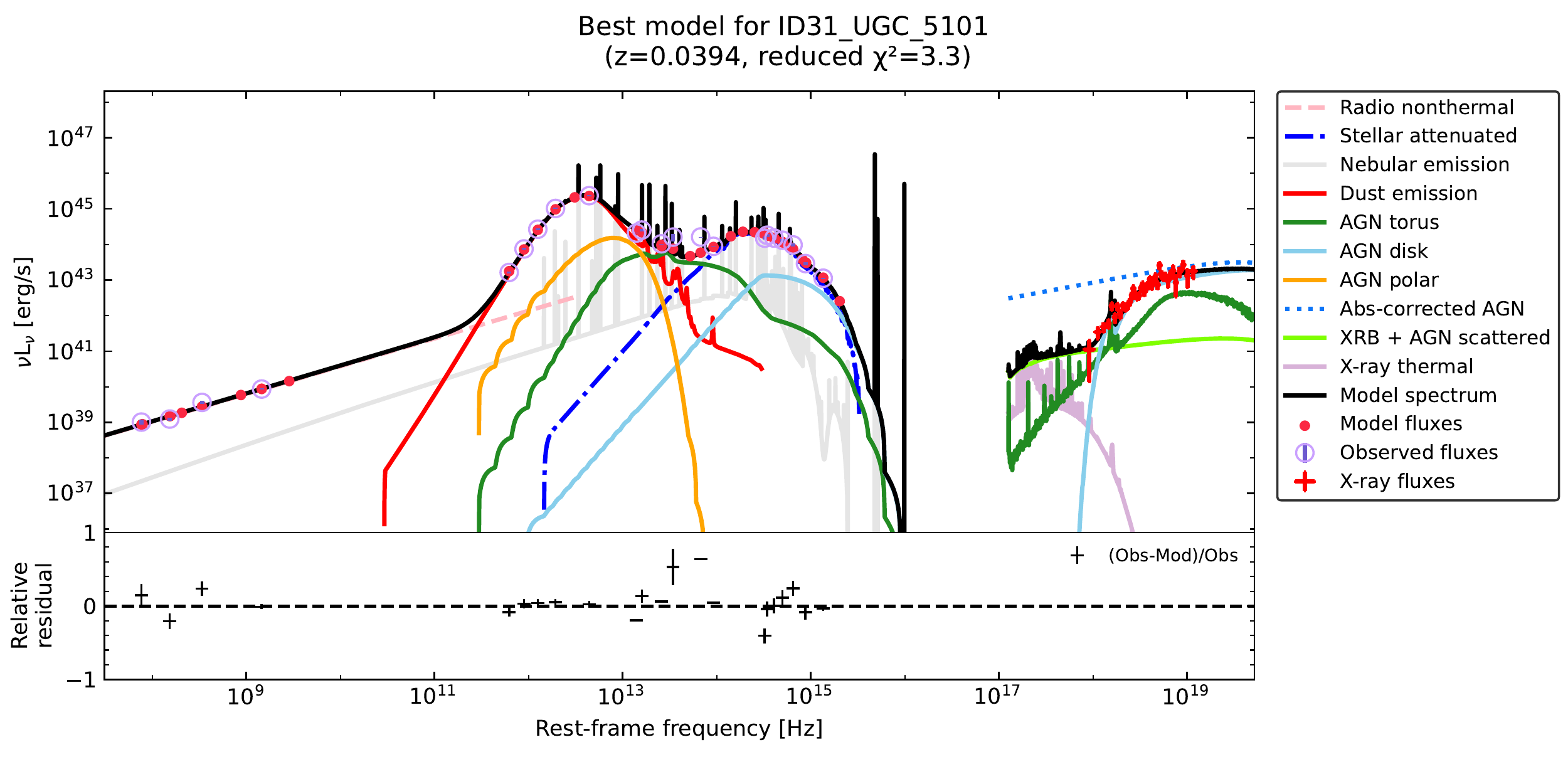}
   \includegraphics[keepaspectratio, scale=0.34]{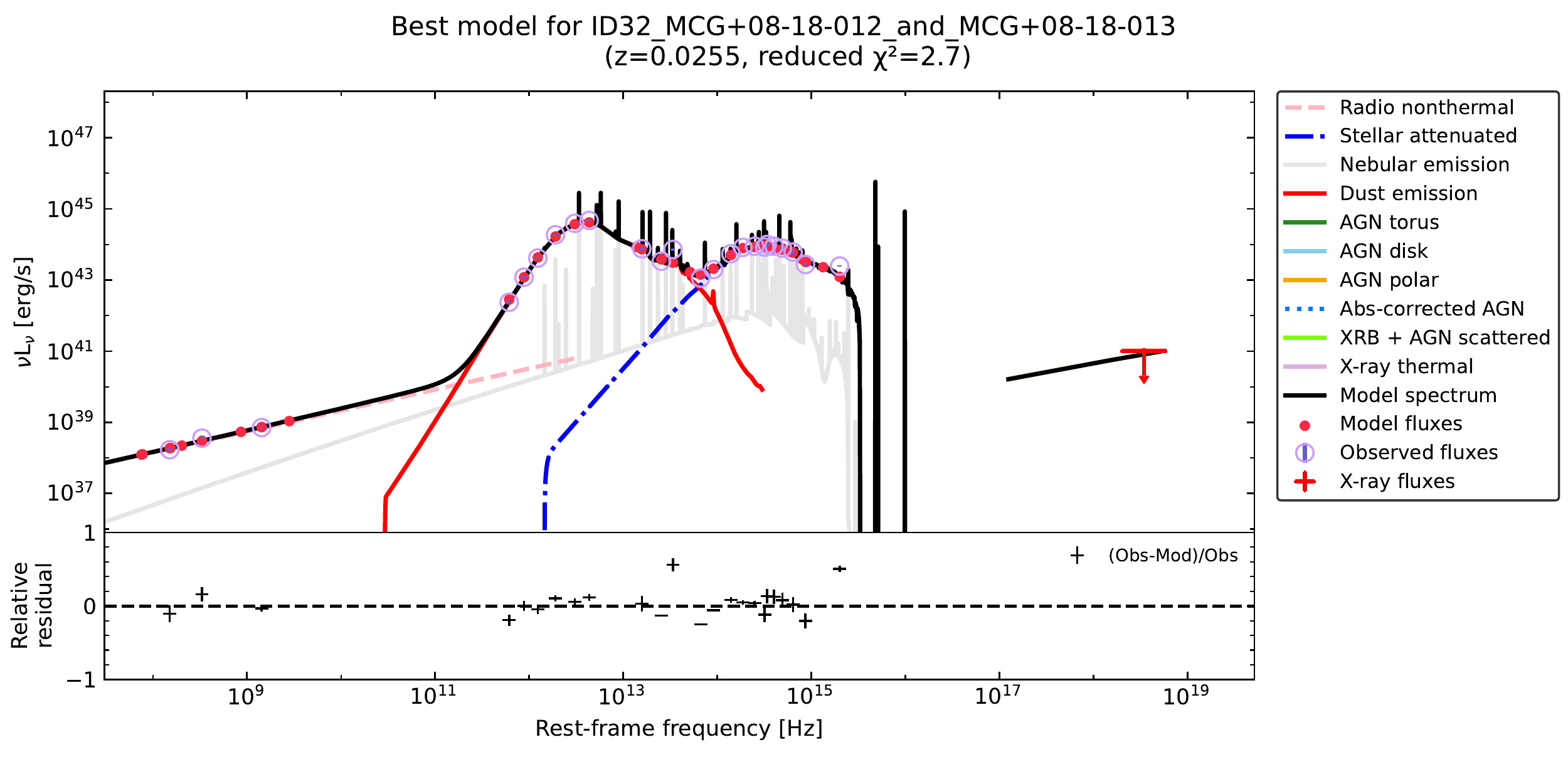}
   \includegraphics[keepaspectratio, scale=0.34]{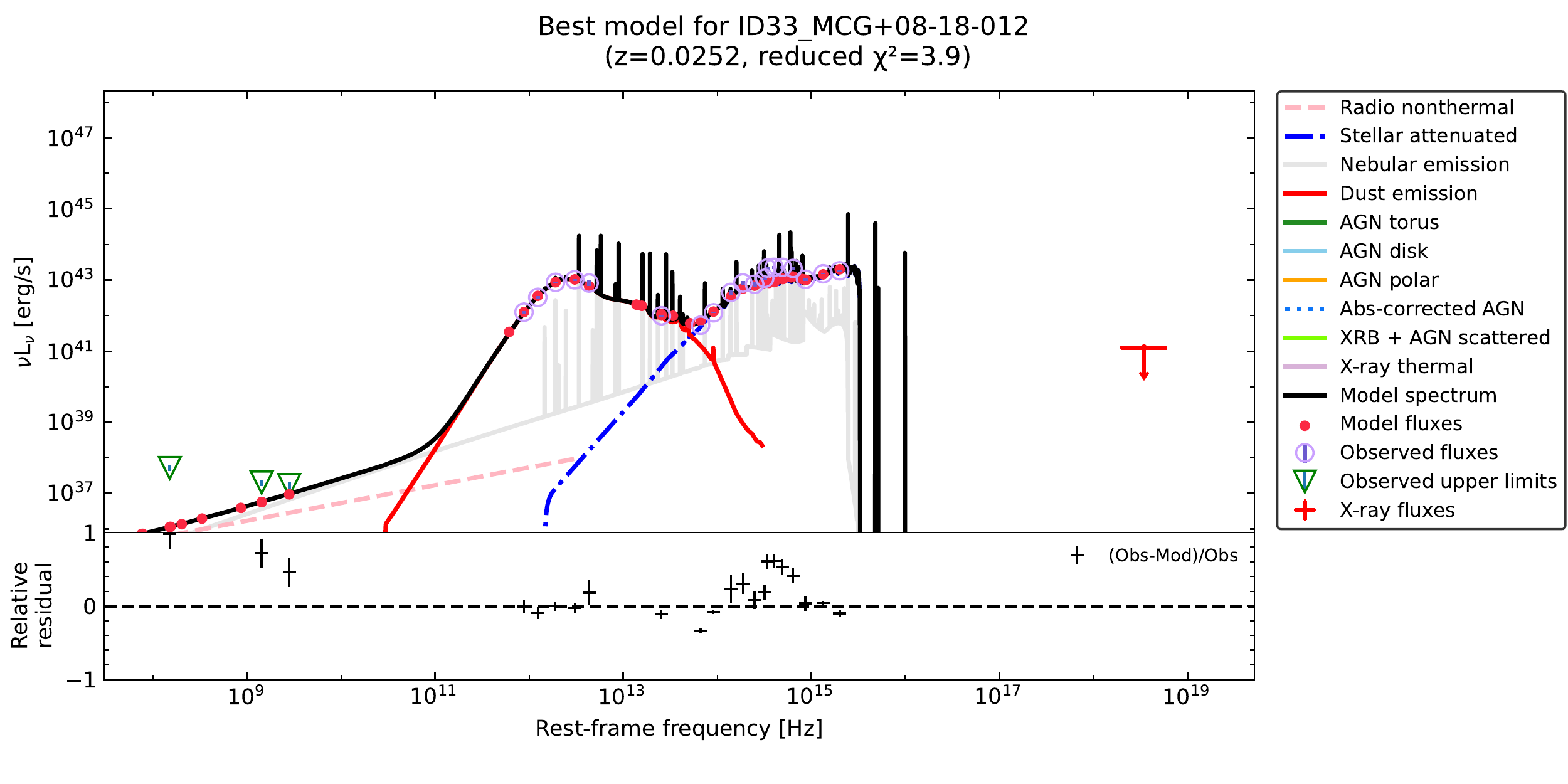}
   \includegraphics[keepaspectratio, scale=0.34]{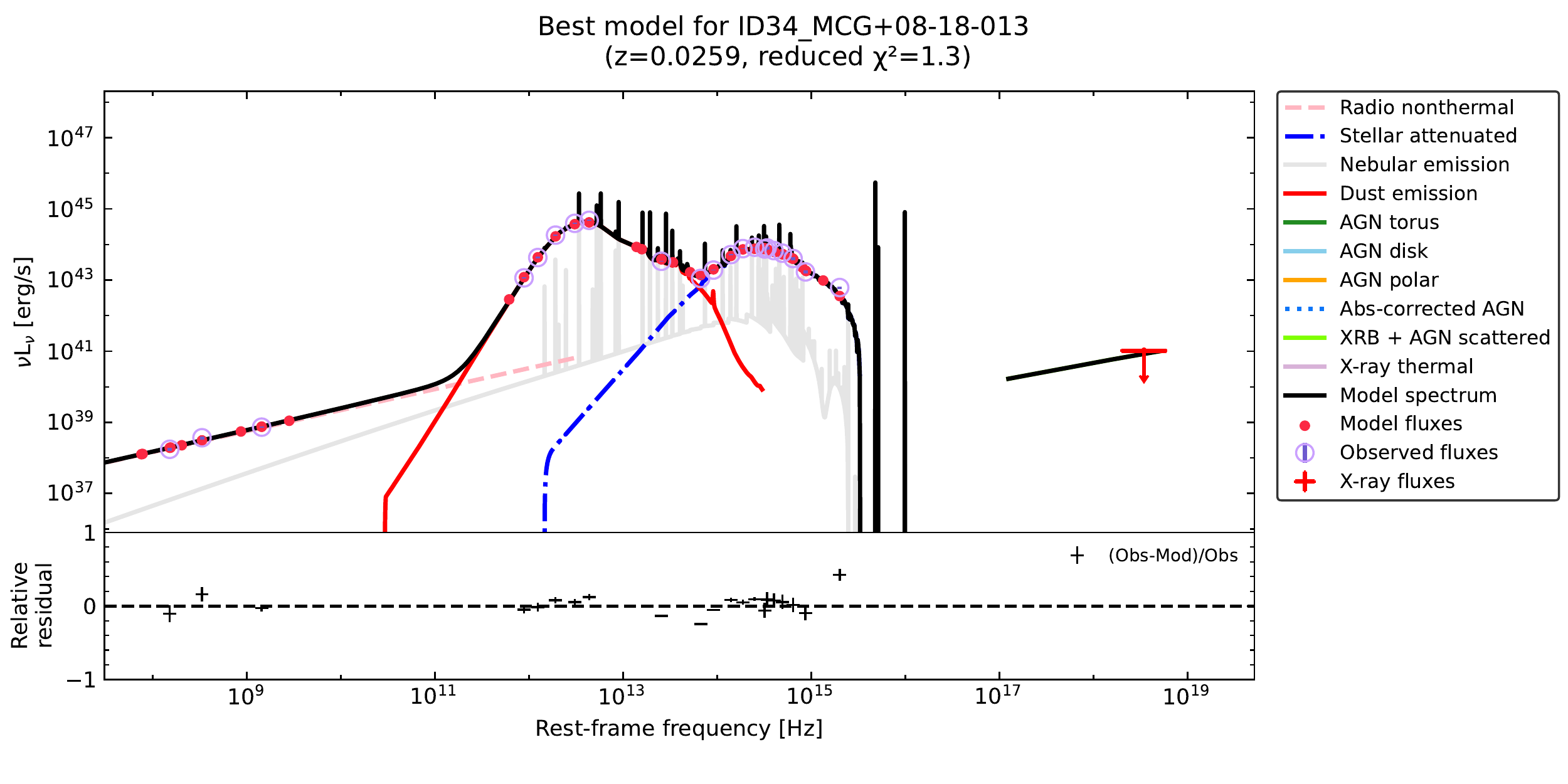}
   \includegraphics[keepaspectratio, scale=0.34]{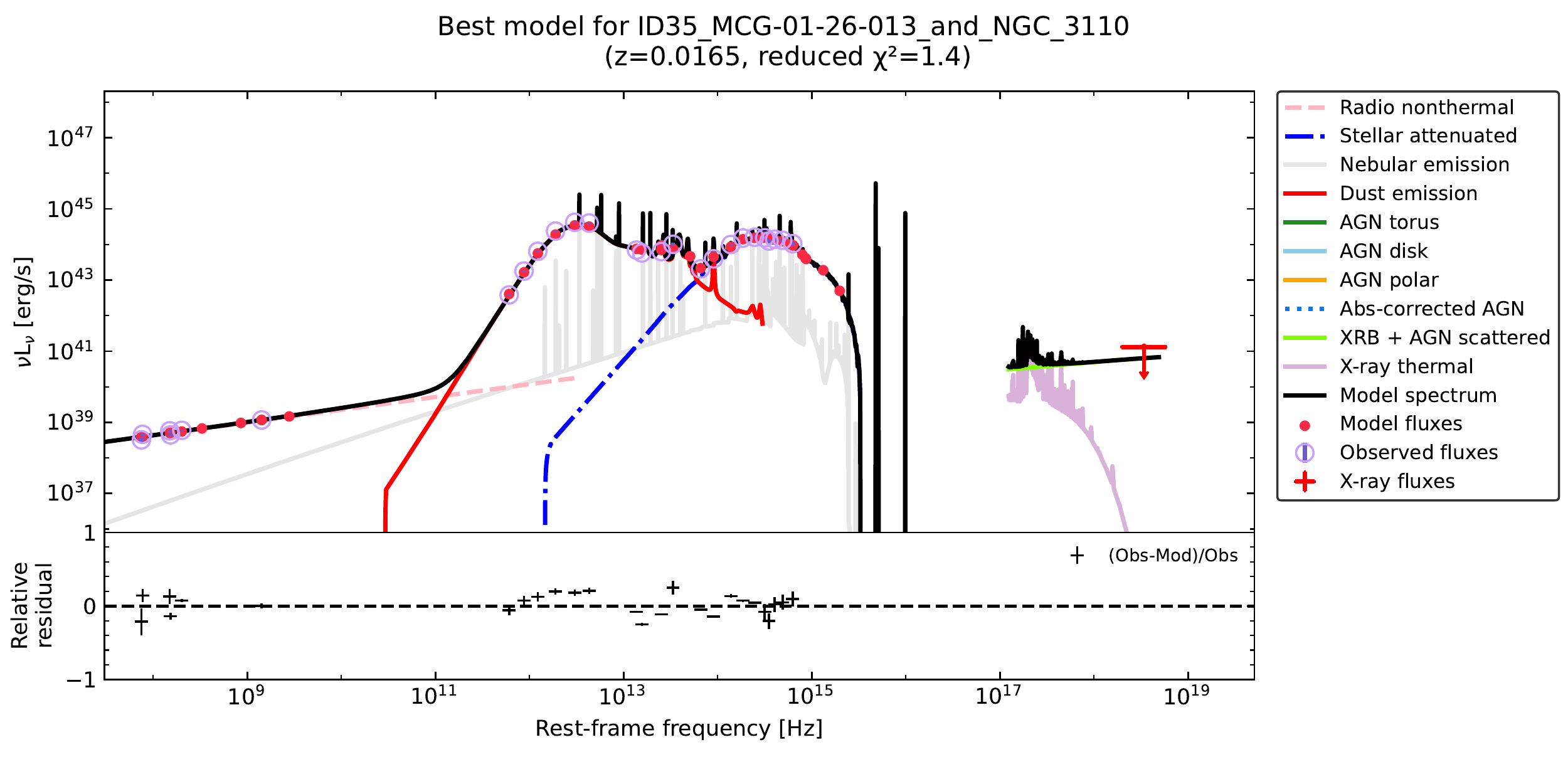}
   \includegraphics[keepaspectratio, scale=0.34]{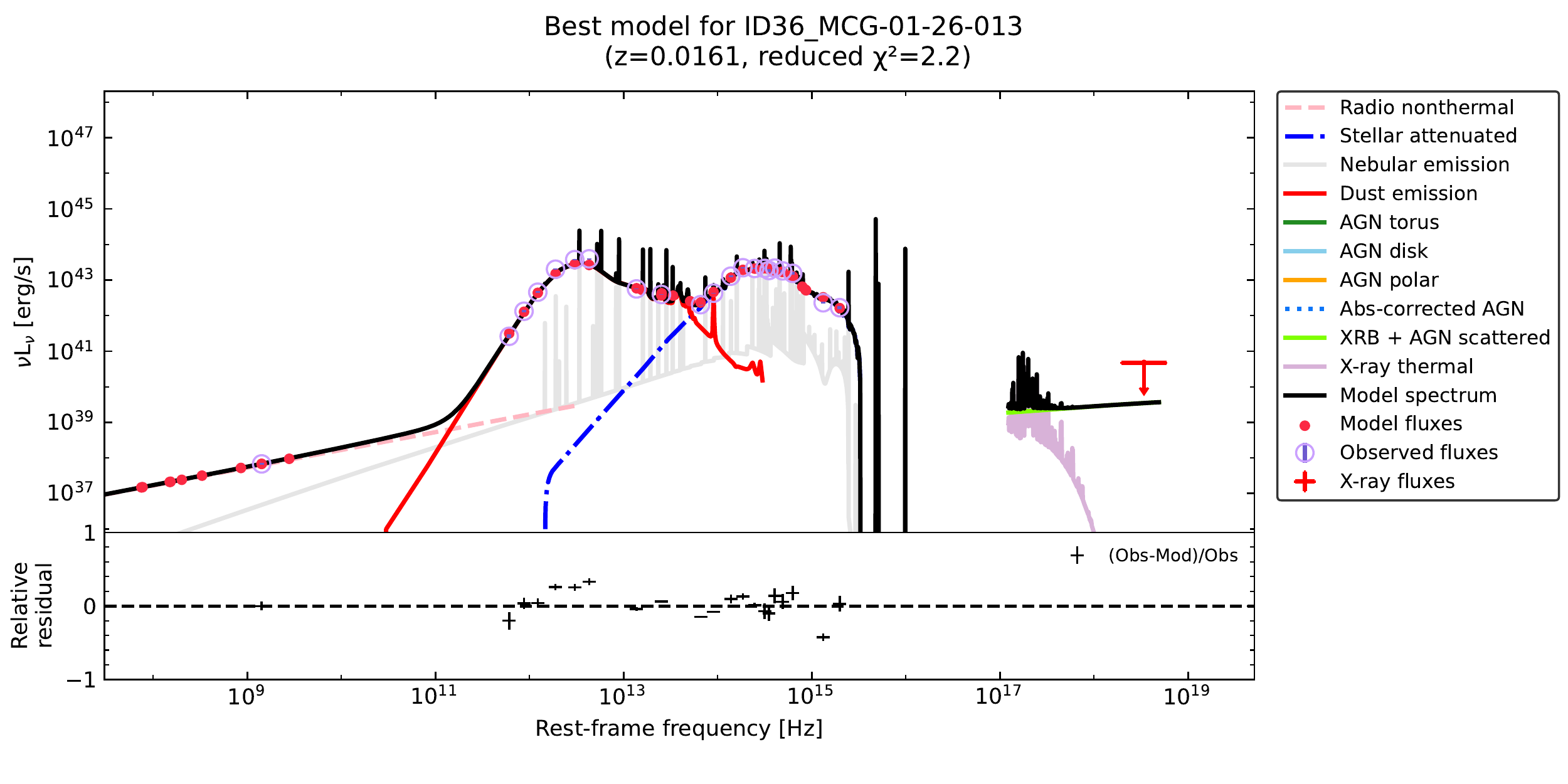}
   \includegraphics[keepaspectratio, scale=0.34]{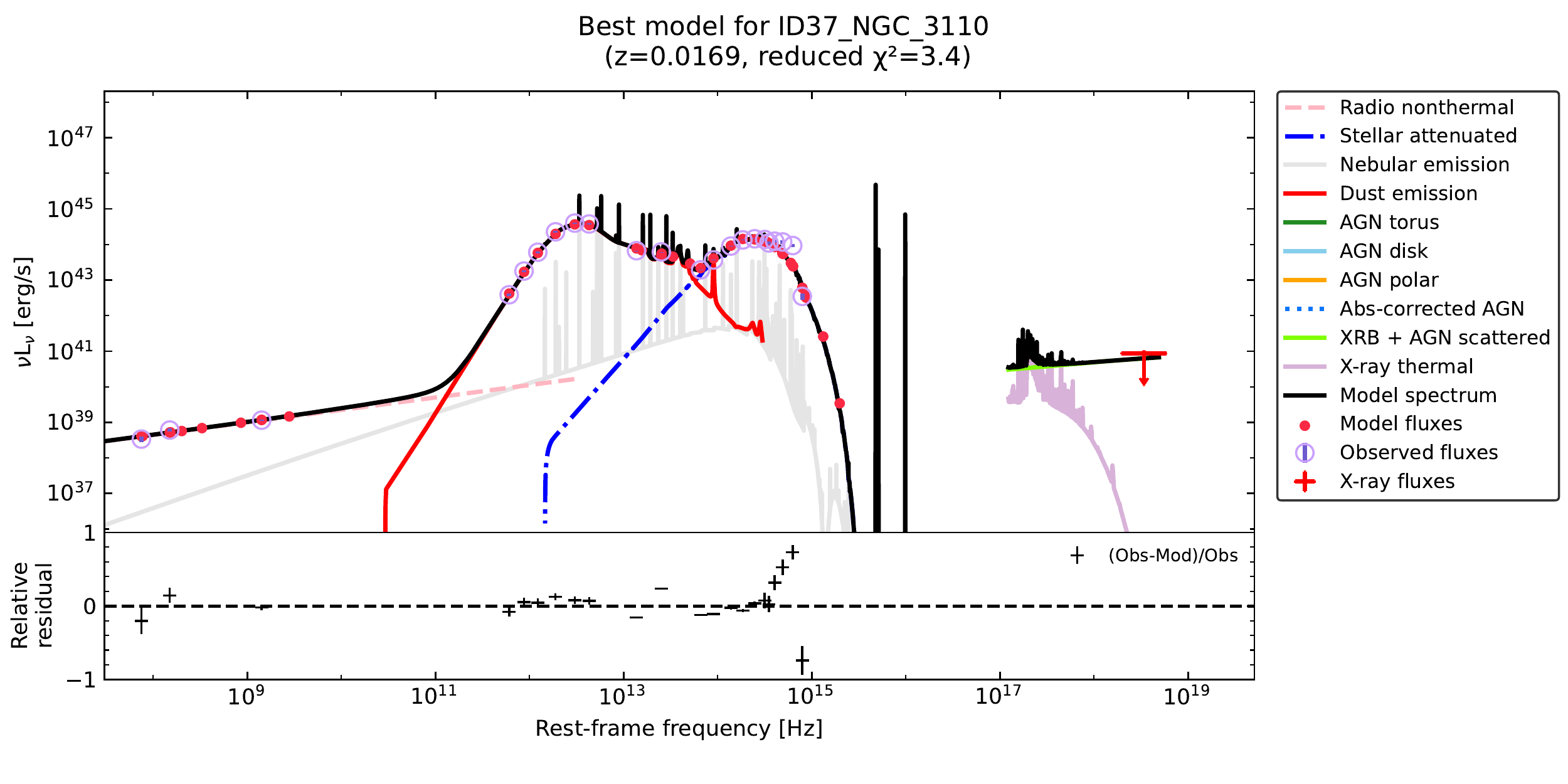}
   \includegraphics[keepaspectratio, scale=0.34]{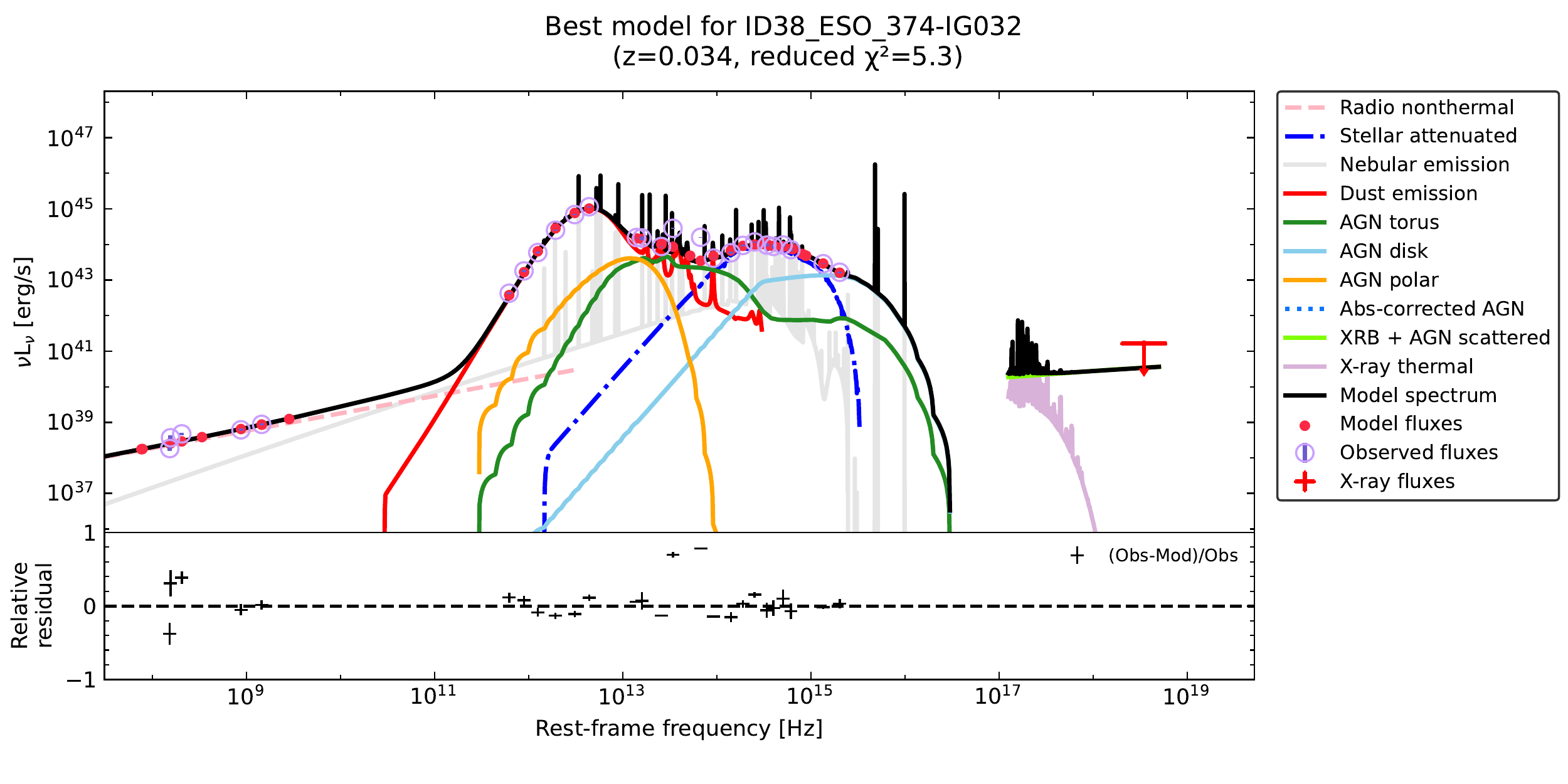}
   \includegraphics[keepaspectratio, scale=0.34]{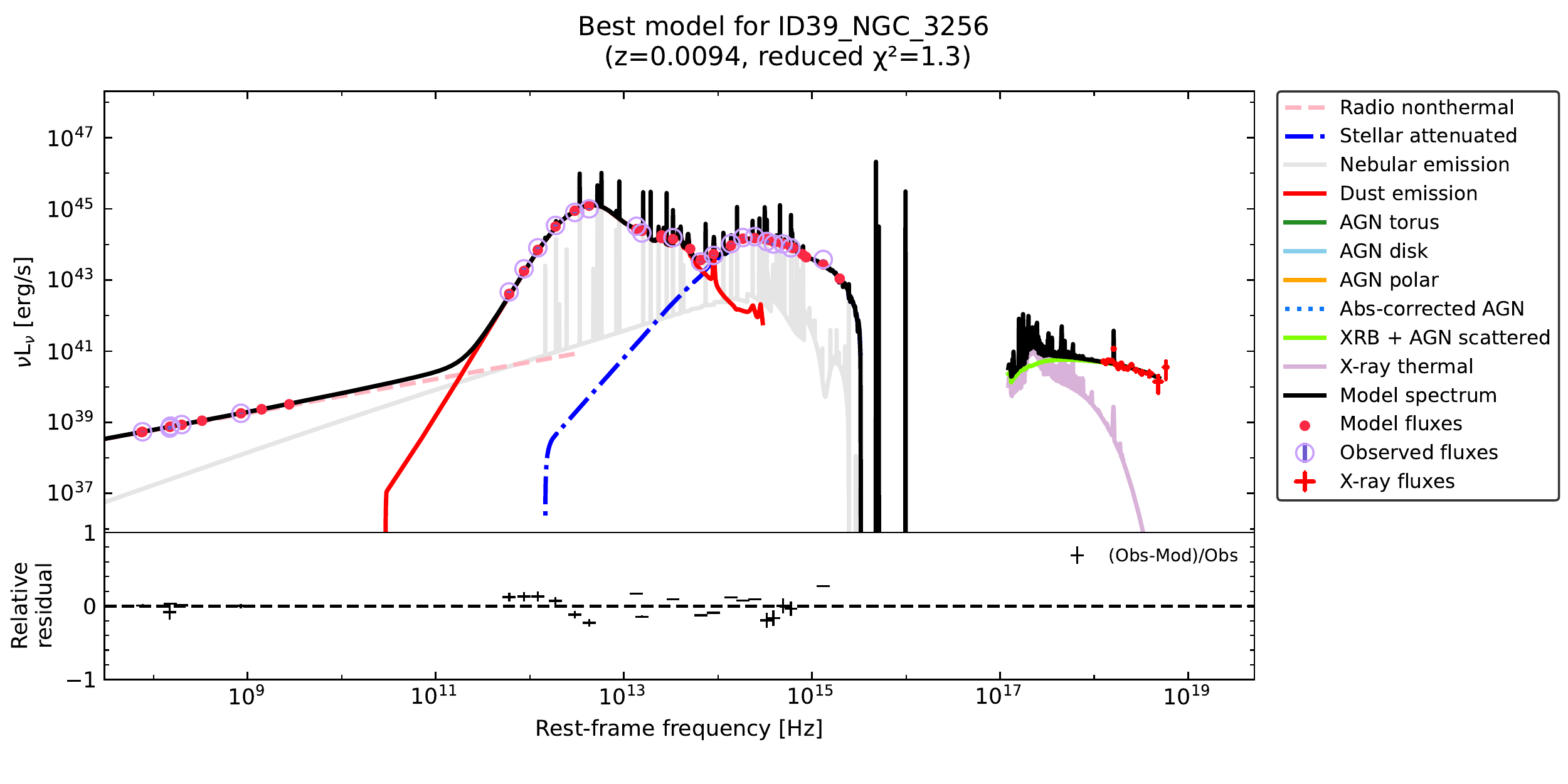}
   \includegraphics[keepaspectratio, scale=0.34]{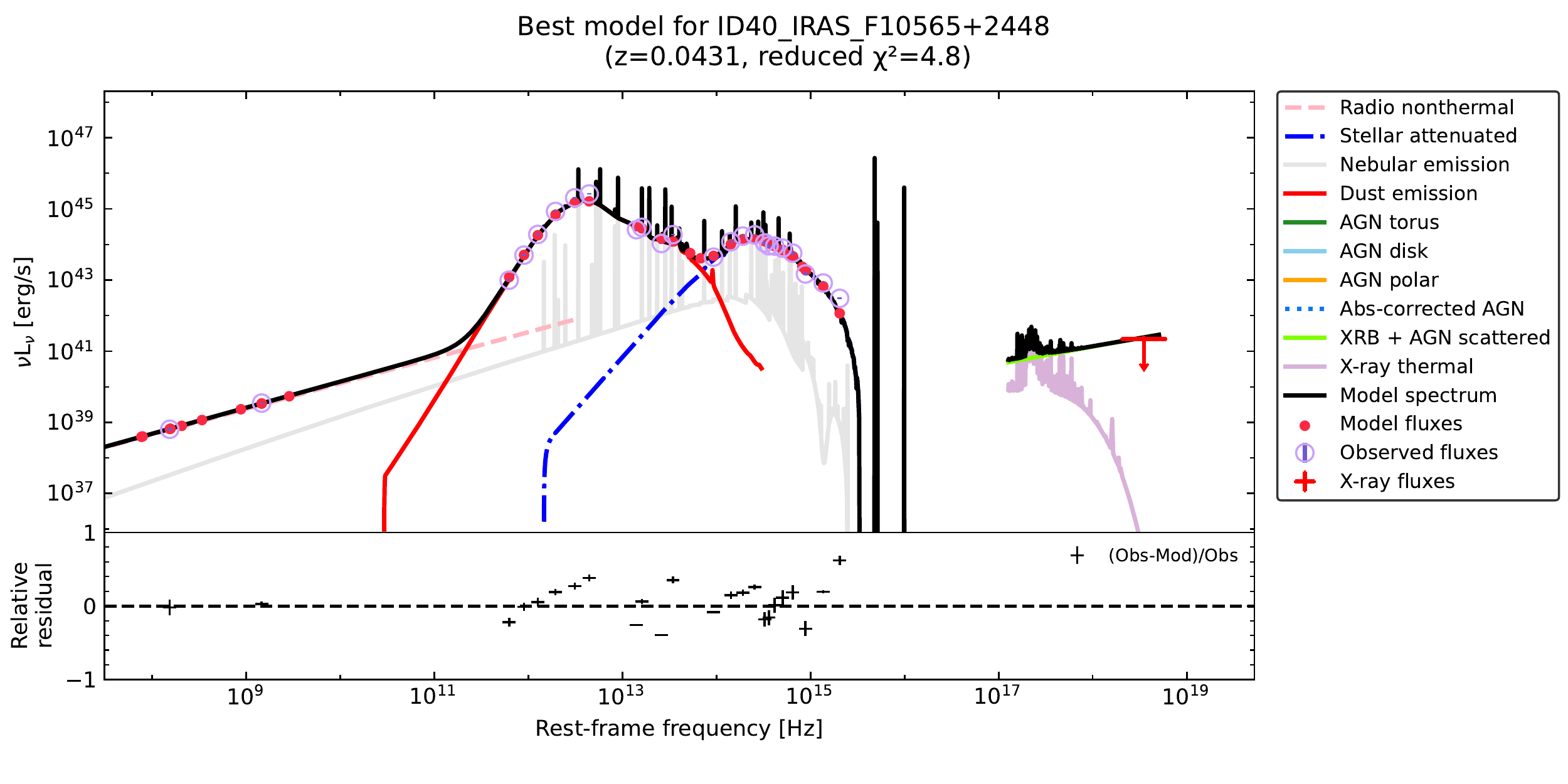}
        \caption{The same as in Figure~\ref{SED10-F} but for the sample of ID=31--40.}
\label{SED13-F}
\end{figure*}

\begin{figure*}
    \centering
   \includegraphics[keepaspectratio, scale=0.34]{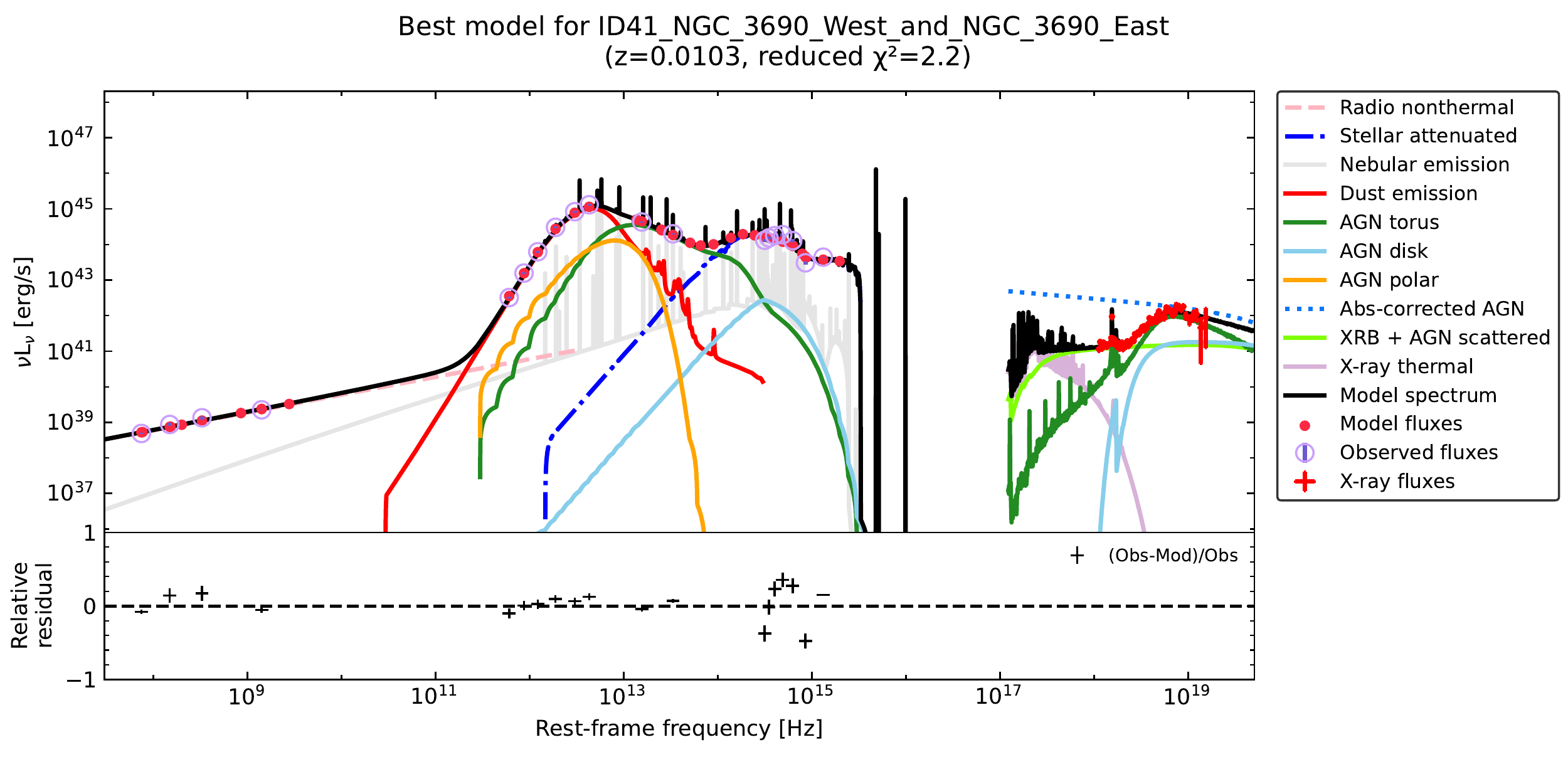}
   \includegraphics[keepaspectratio, scale=0.34]{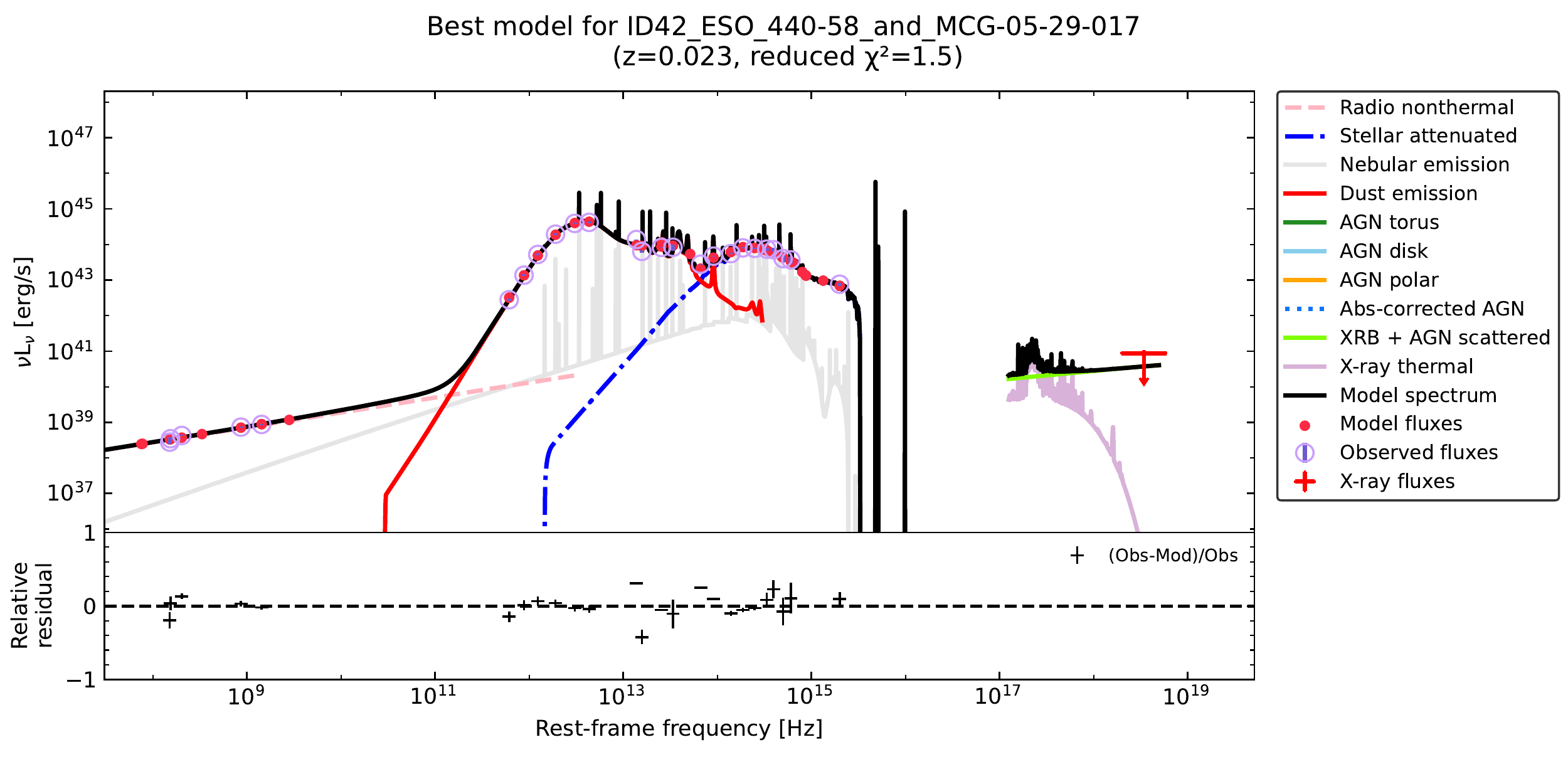}
   \includegraphics[keepaspectratio, scale=0.34]{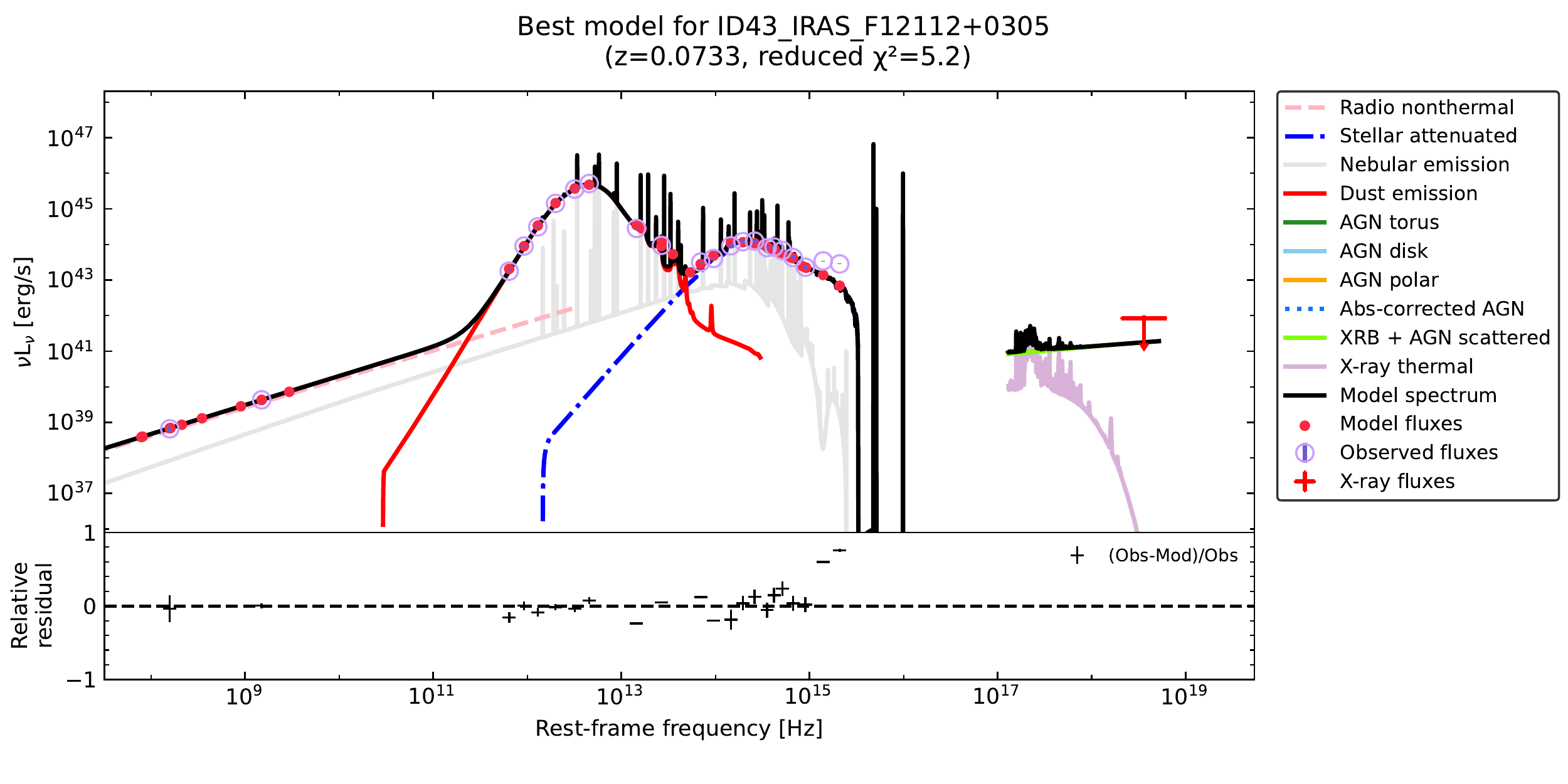}
   \includegraphics[keepaspectratio, scale=0.34]{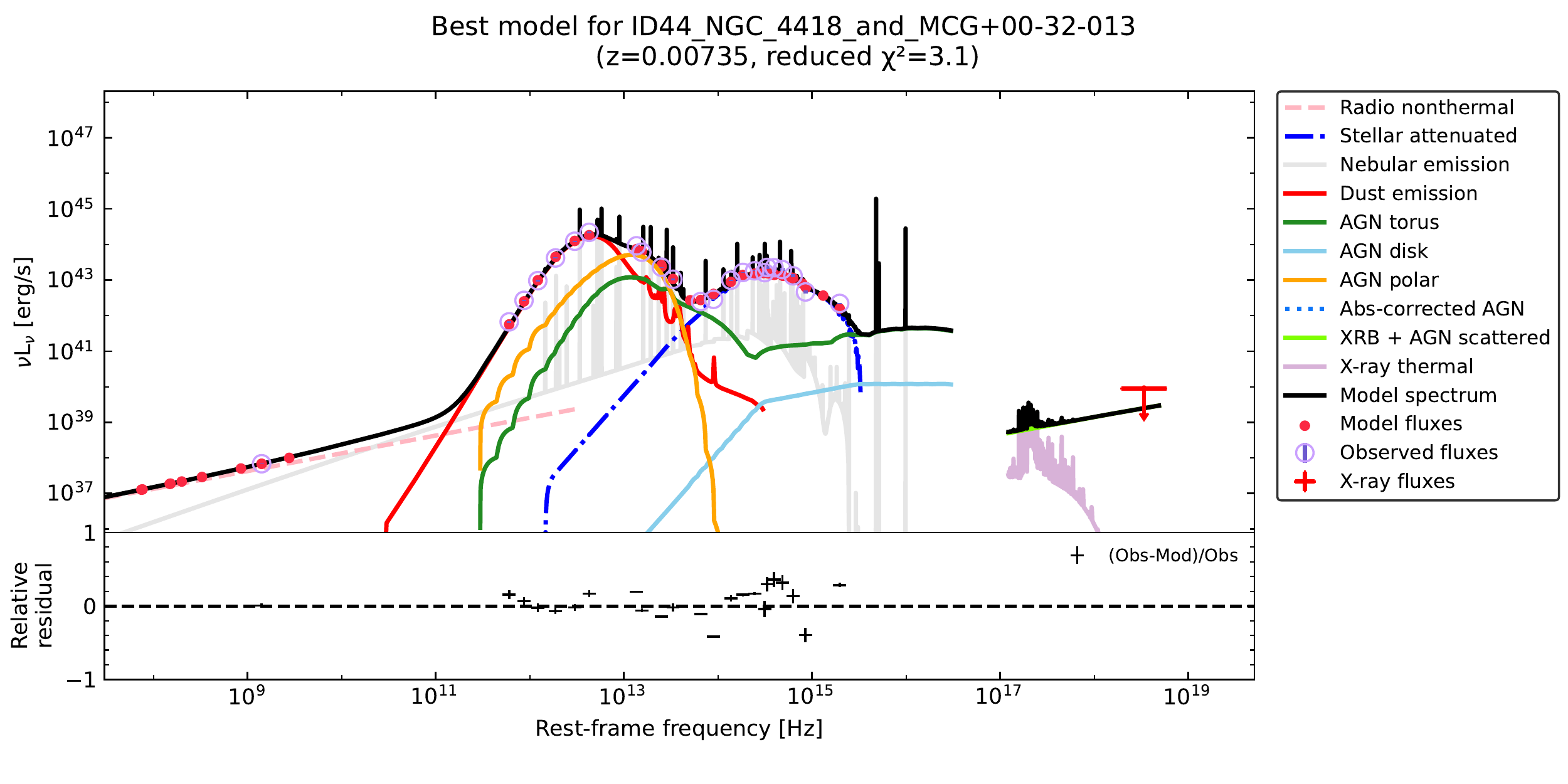}
   \includegraphics[keepaspectratio, scale=0.34]{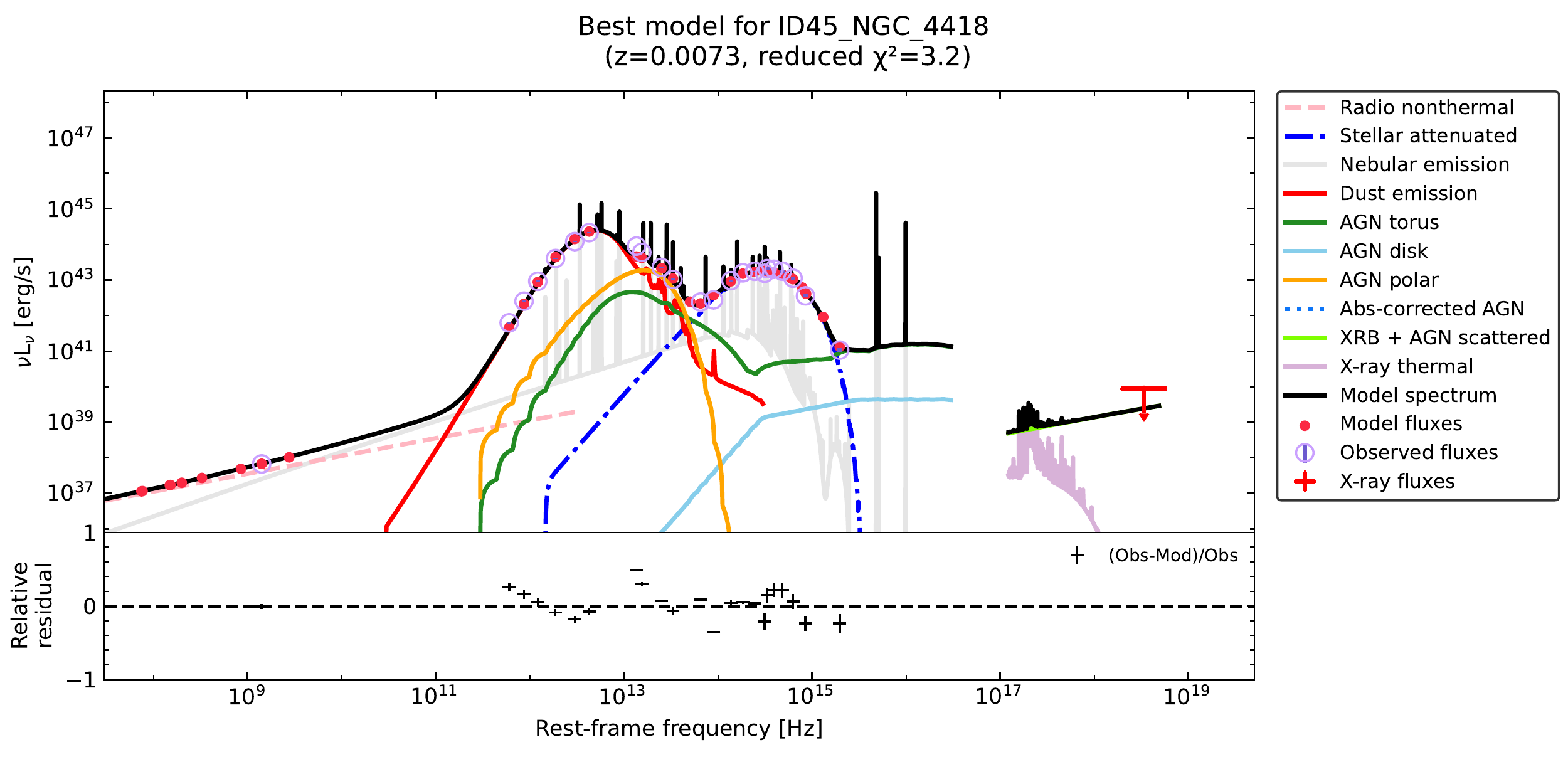}
   \includegraphics[keepaspectratio, scale=0.34]{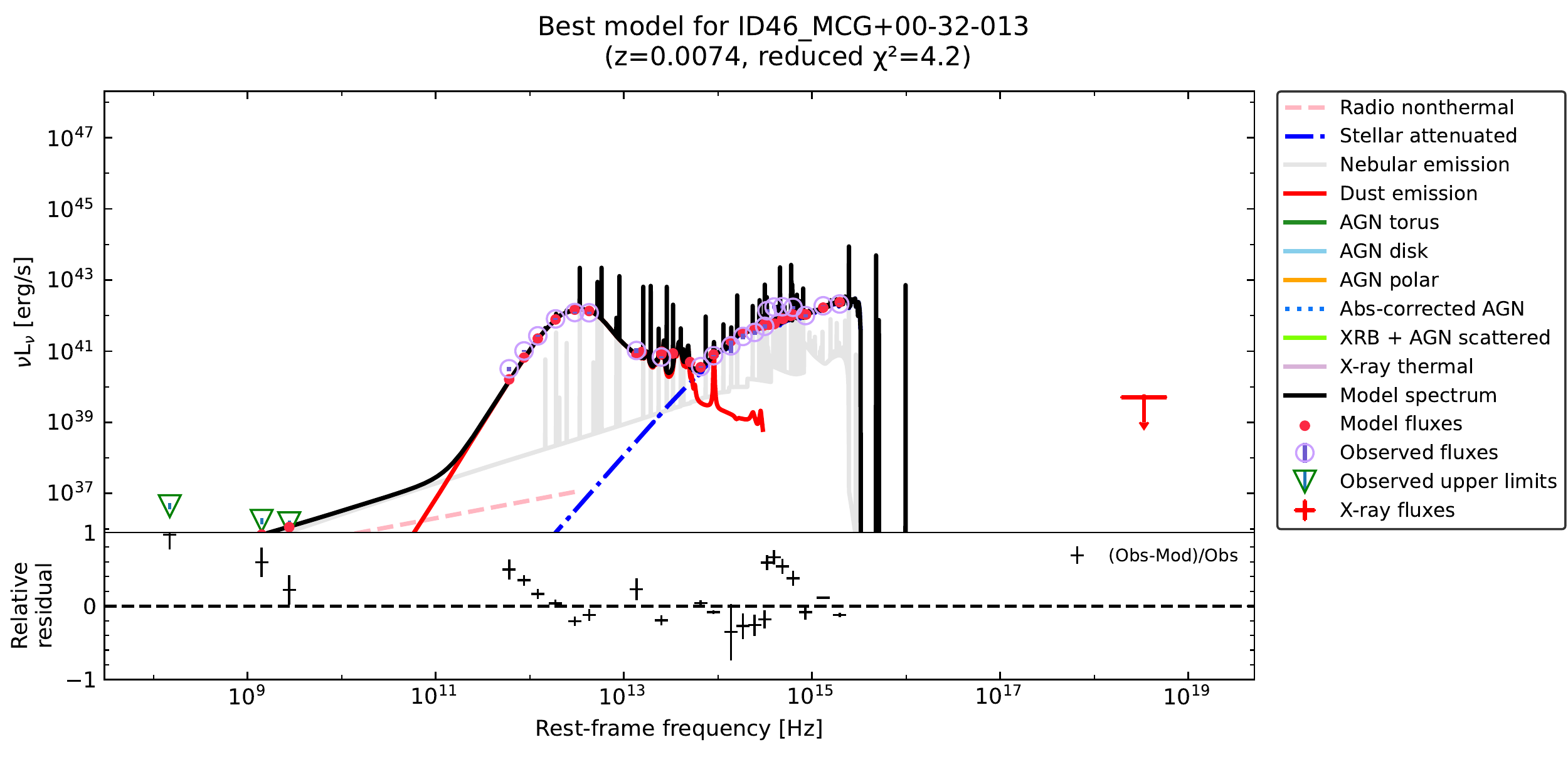}
   \includegraphics[keepaspectratio, scale=0.34]{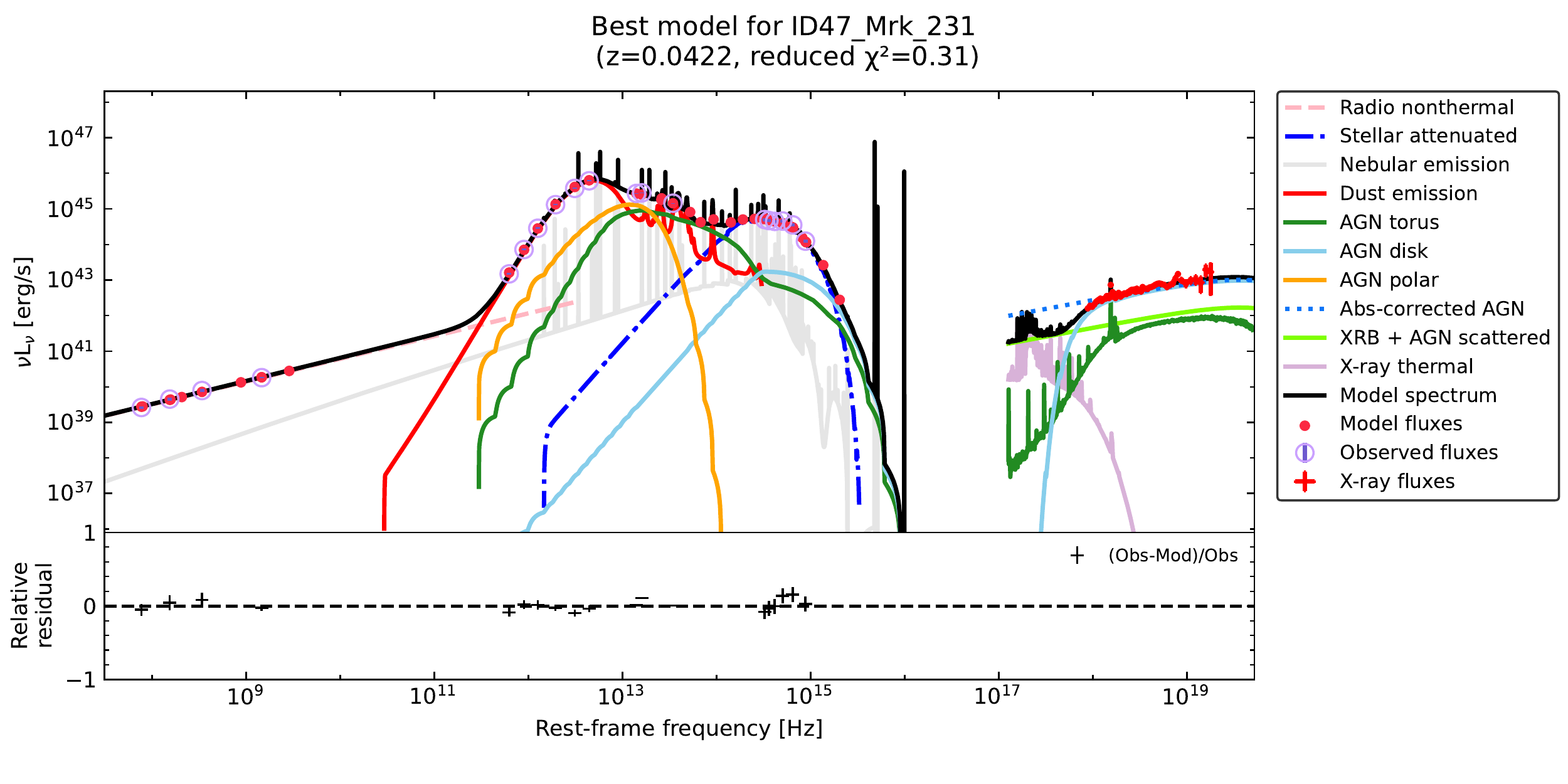}
   \includegraphics[keepaspectratio, scale=0.34]{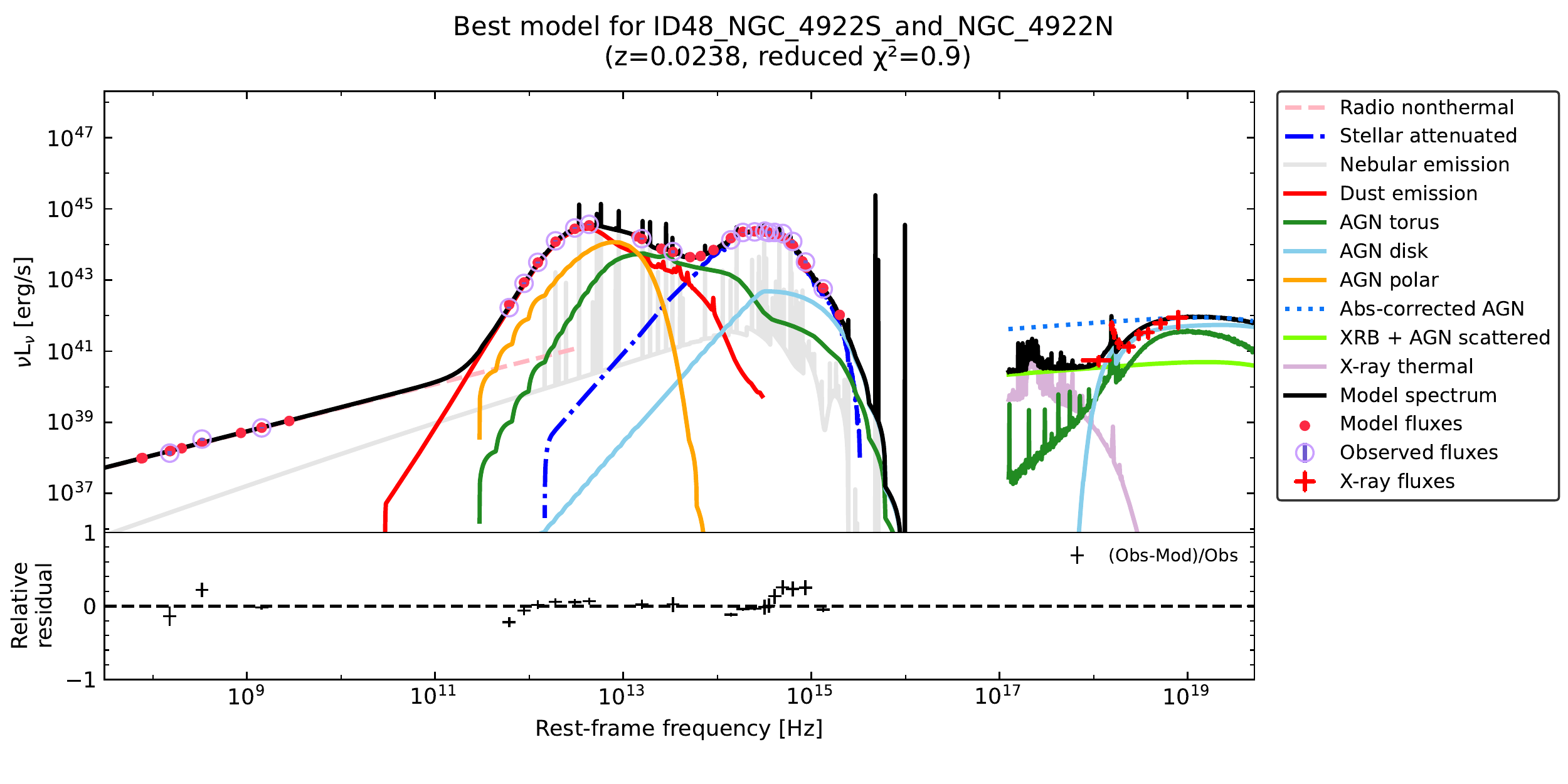}
   \includegraphics[keepaspectratio, scale=0.34]{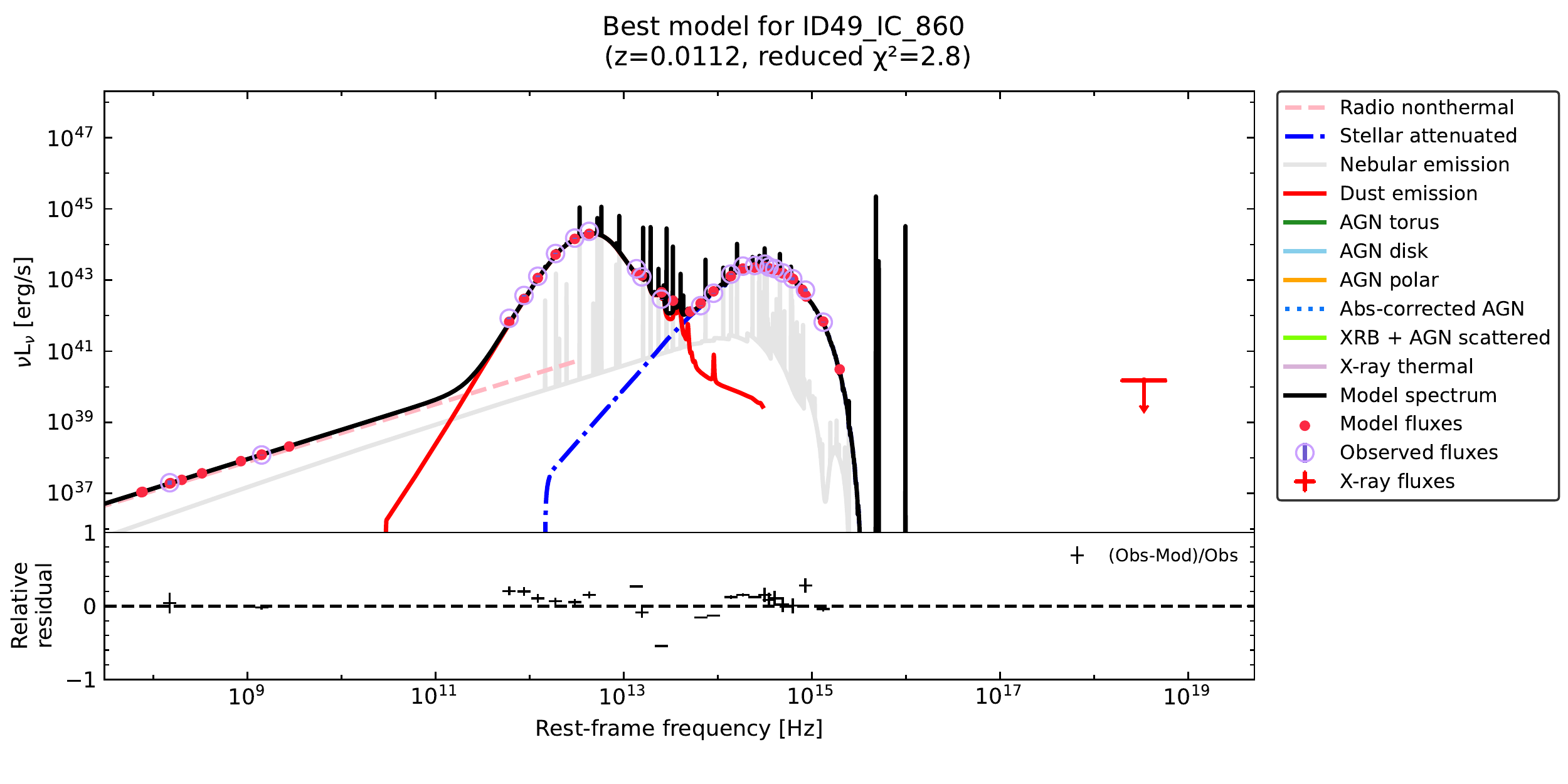}
   \includegraphics[keepaspectratio, scale=0.34]{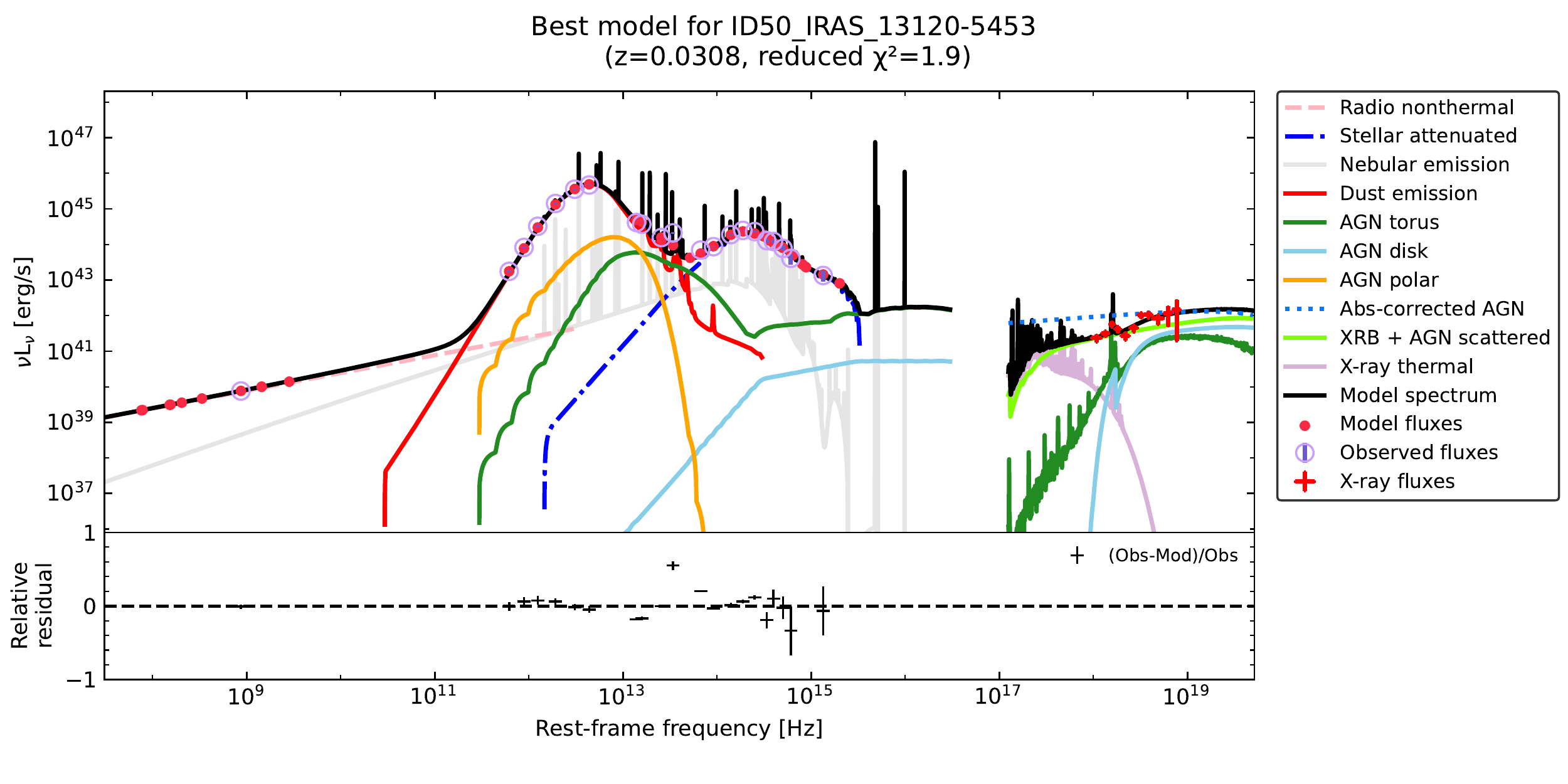}
        \caption{The same as in Figure~\ref{SED10-F} but for the sample of ID=41--50.}
\label{SED14-F}
\end{figure*}

\begin{figure*}
    \centering
   \includegraphics[keepaspectratio, scale=0.34]{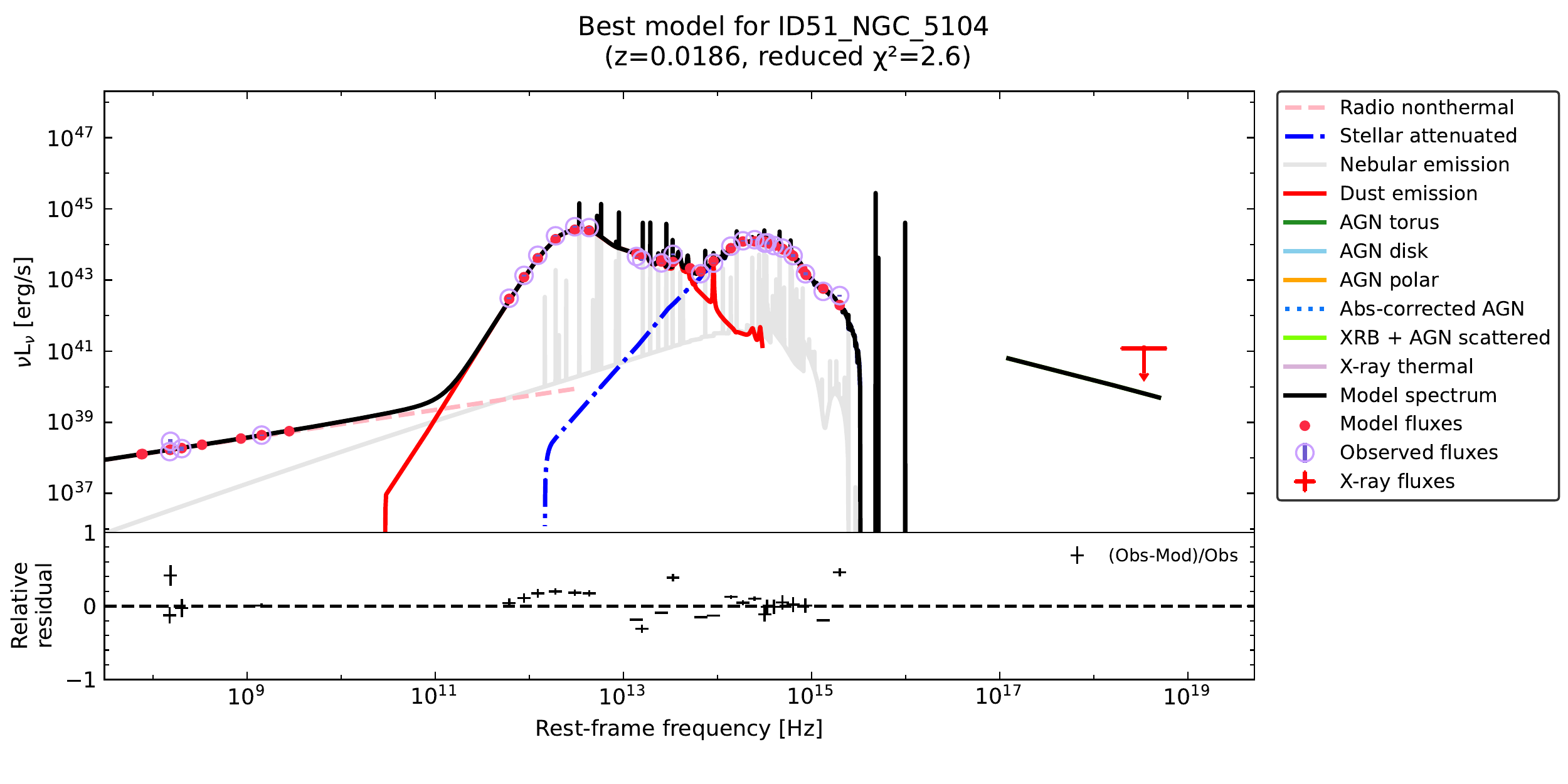}
   \includegraphics[keepaspectratio, scale=0.34]{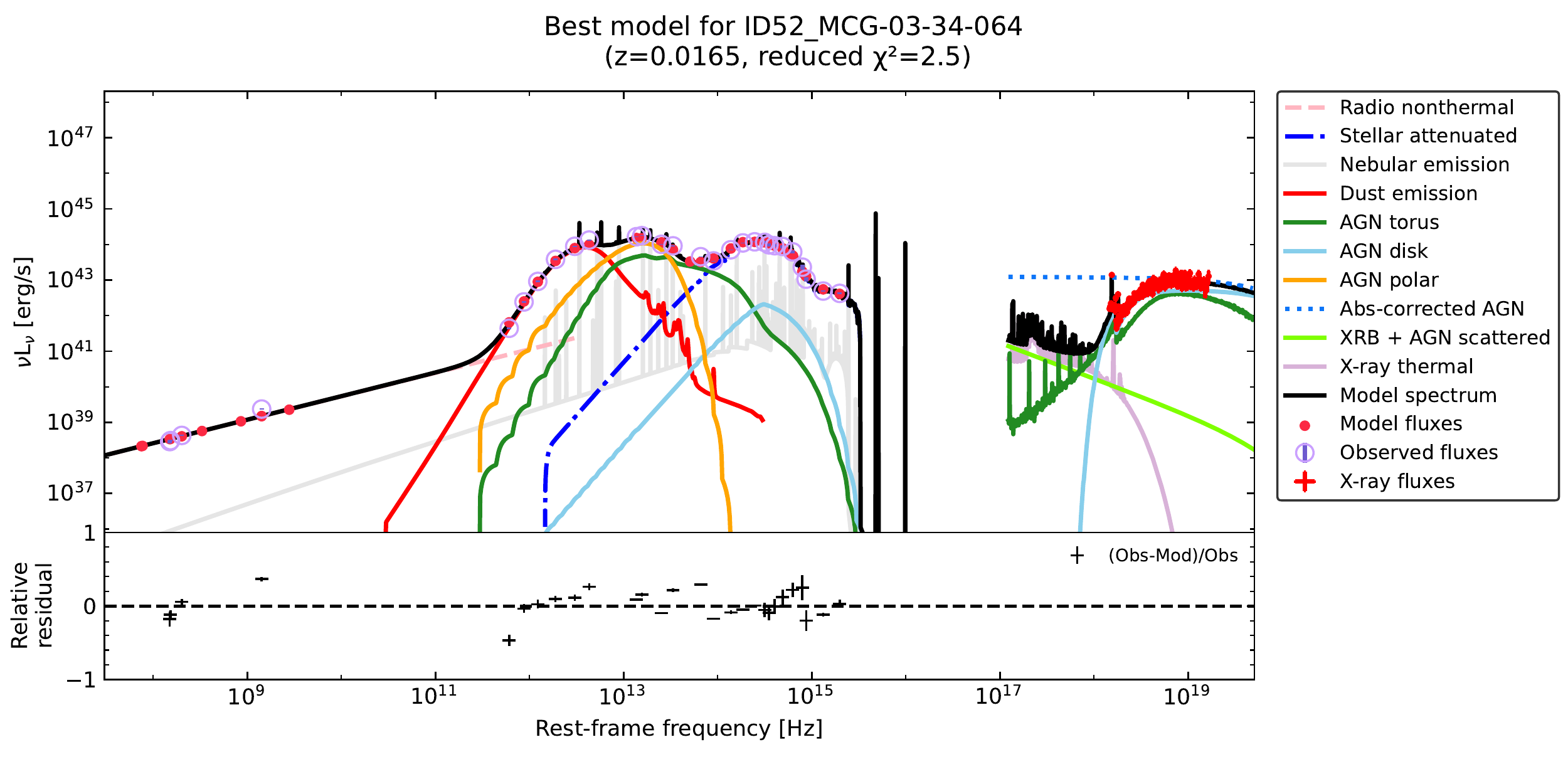}
   \includegraphics[keepaspectratio, scale=0.34]{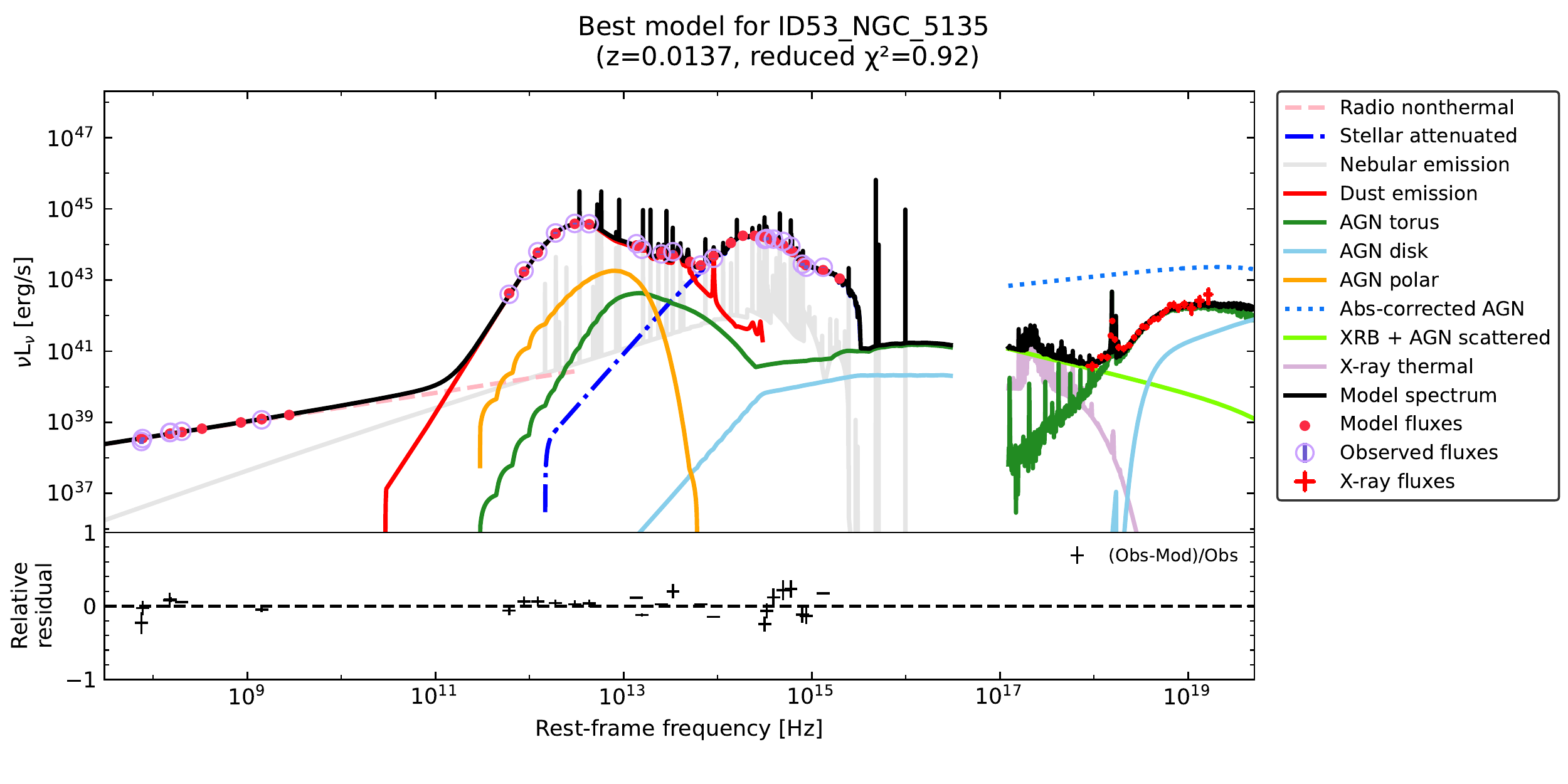}
   \includegraphics[keepaspectratio, scale=0.34]{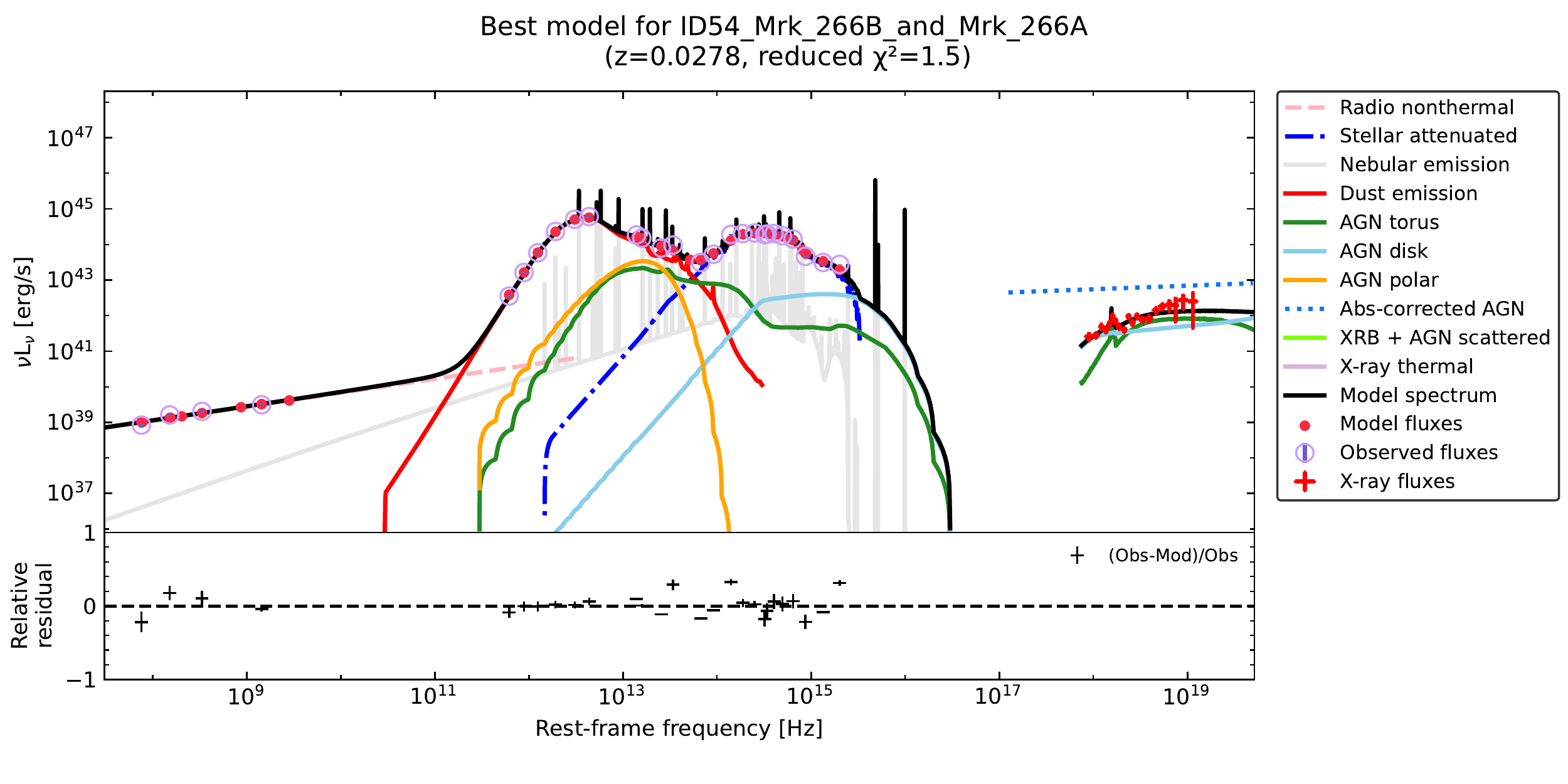}
   \includegraphics[keepaspectratio, scale=0.34]{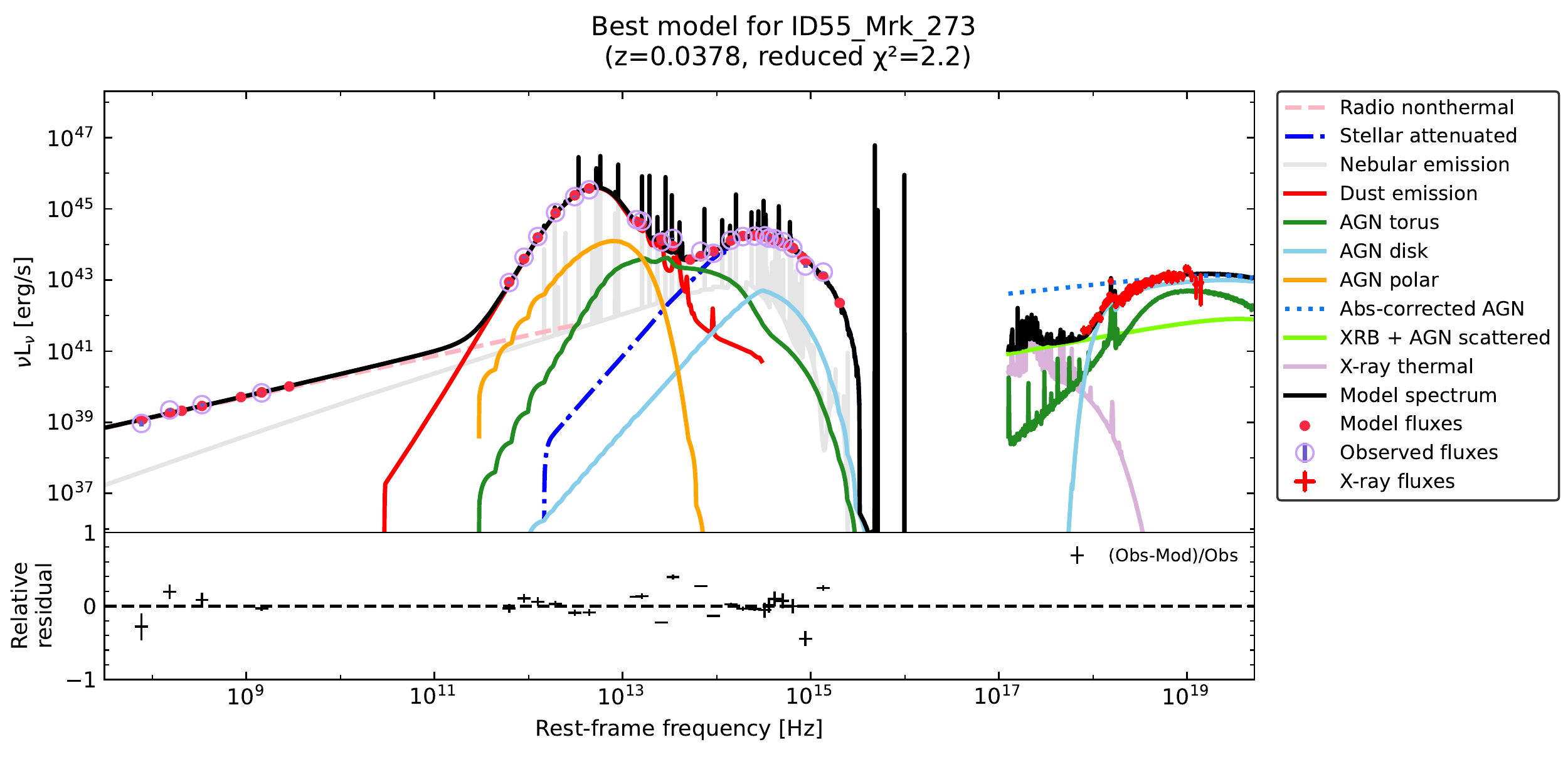}
   \includegraphics[keepaspectratio, scale=0.34]{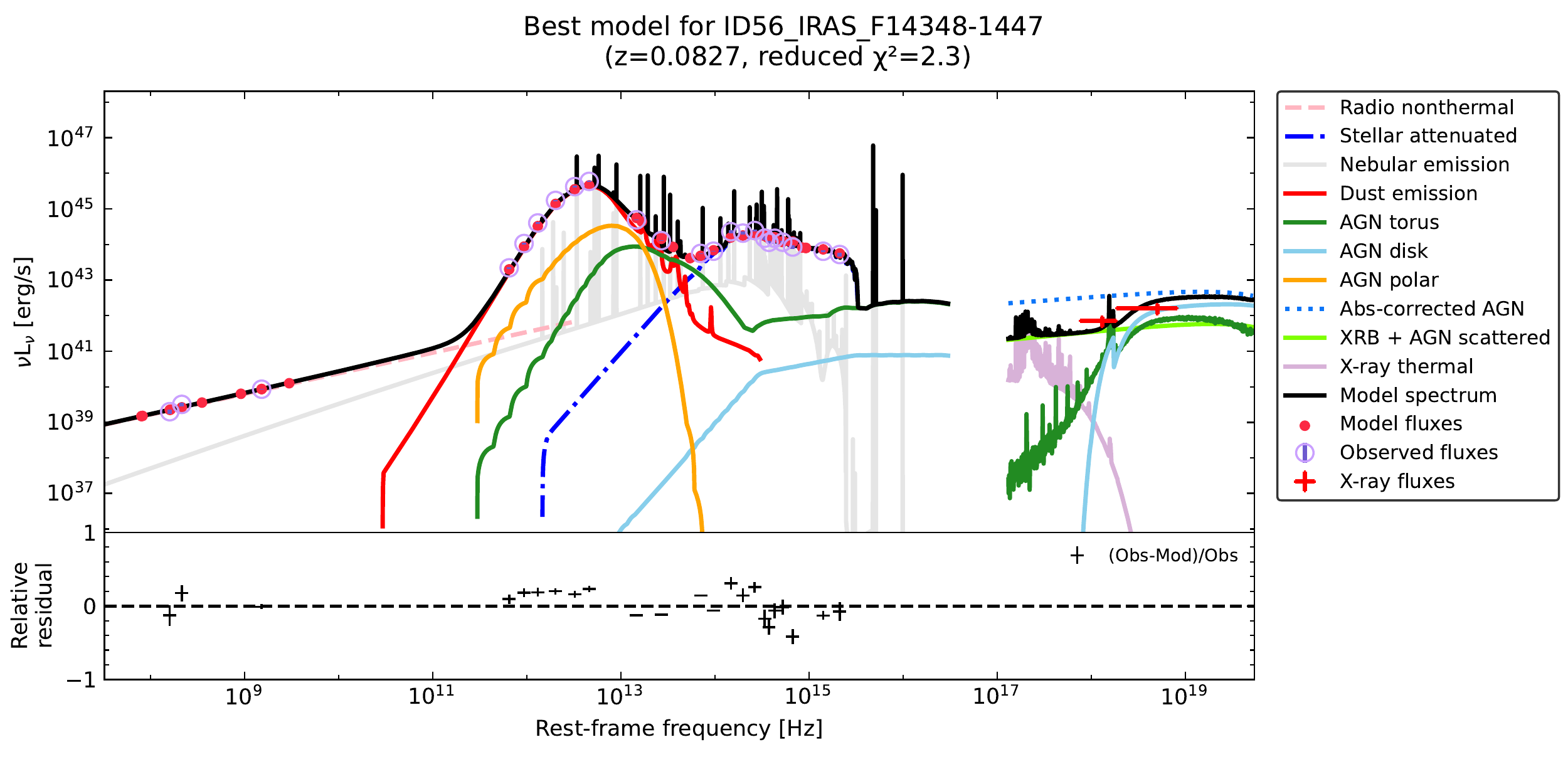}
   \includegraphics[keepaspectratio, scale=0.34]{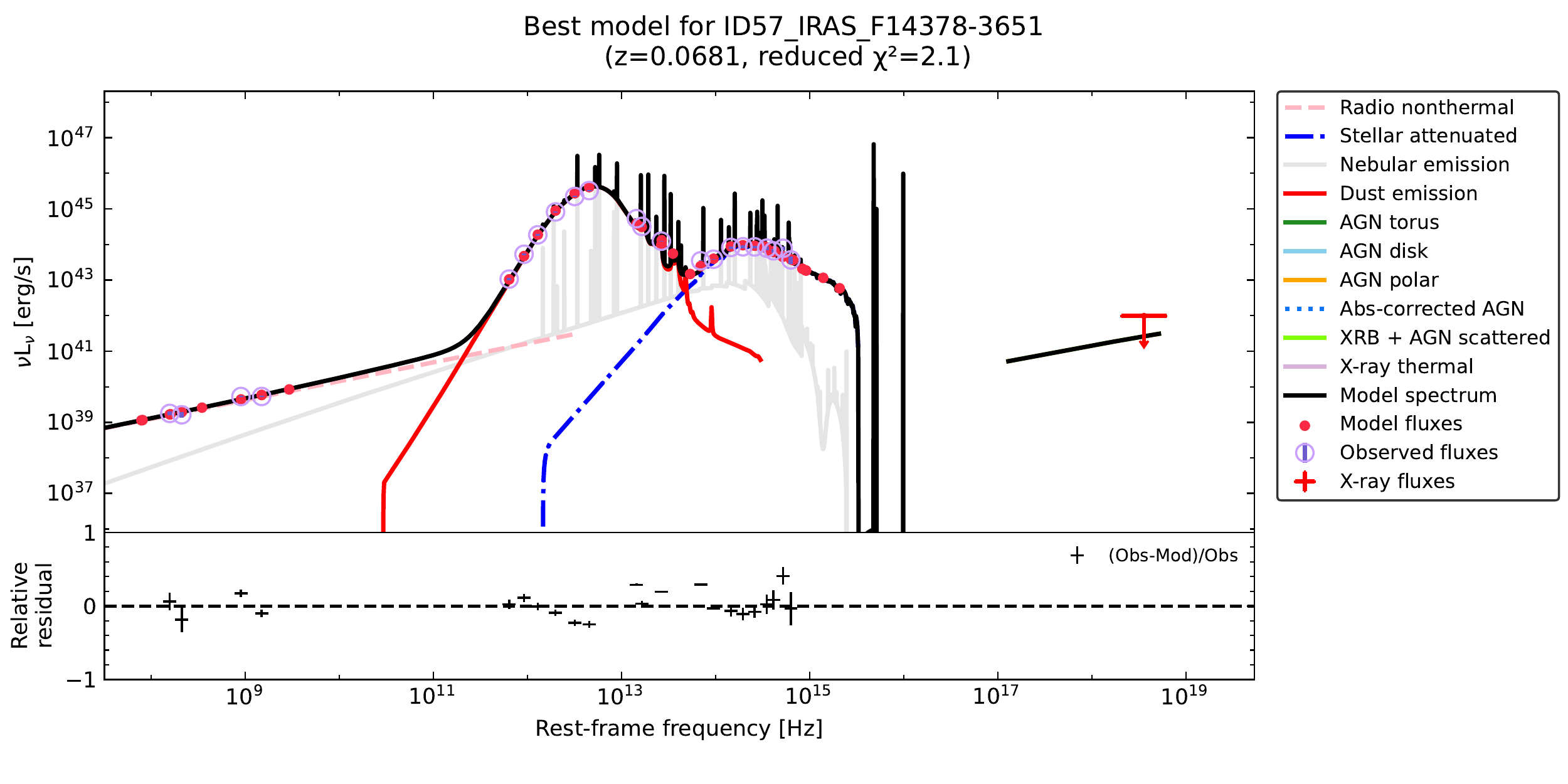}   
   \includegraphics[keepaspectratio, scale=0.34]{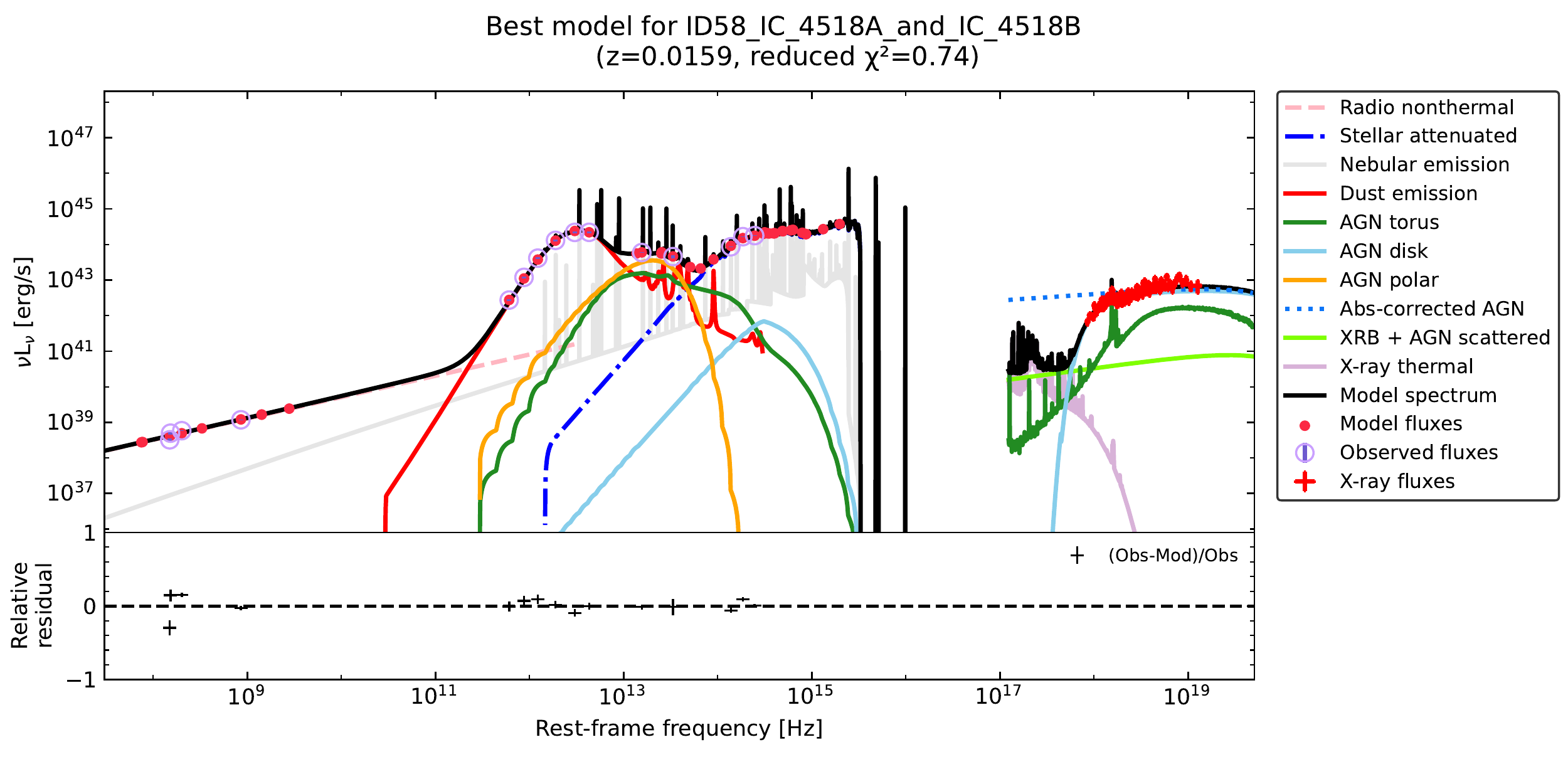}
   \includegraphics[keepaspectratio, scale=0.34]{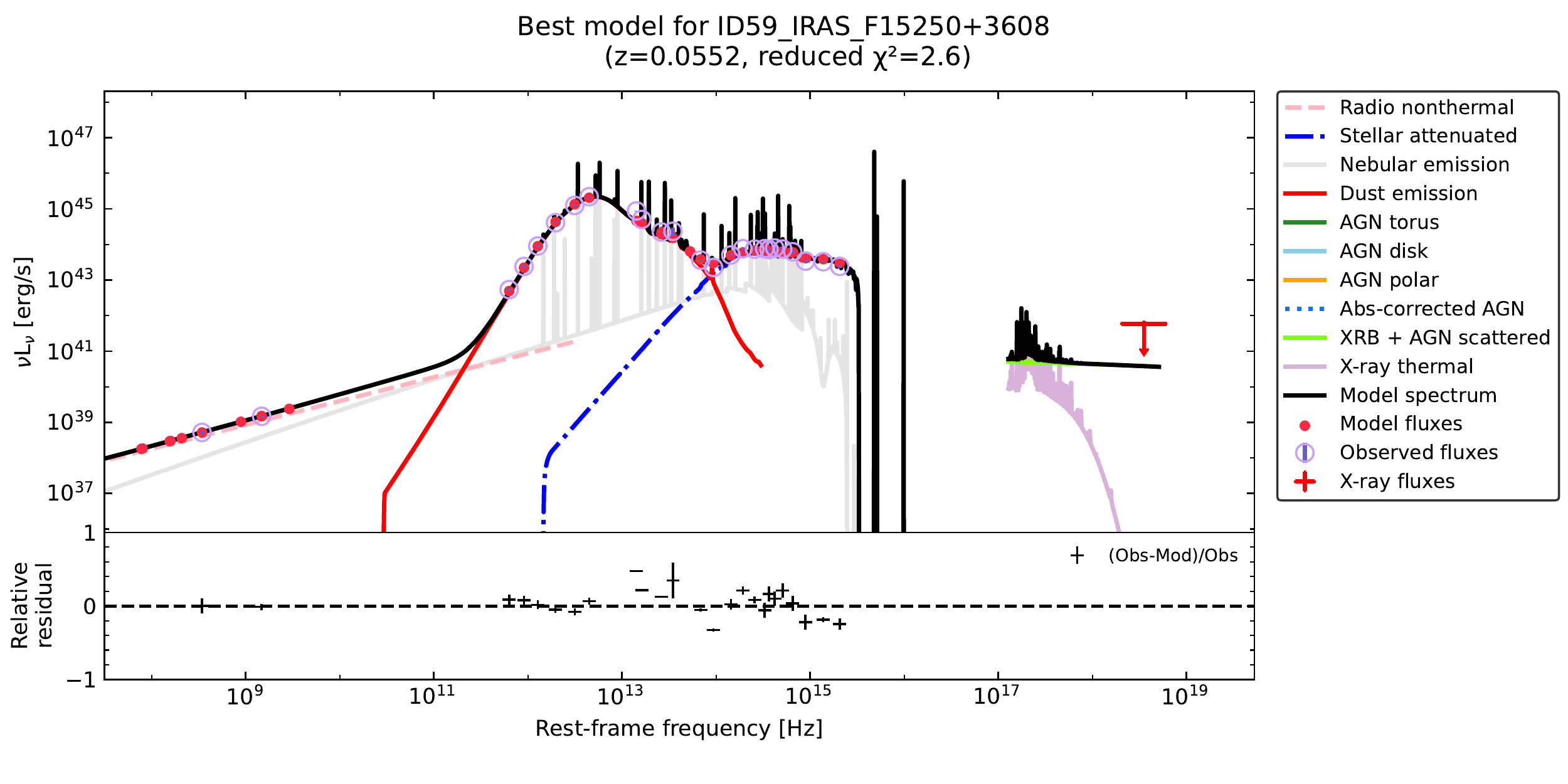}
   \includegraphics[keepaspectratio, scale=0.34]{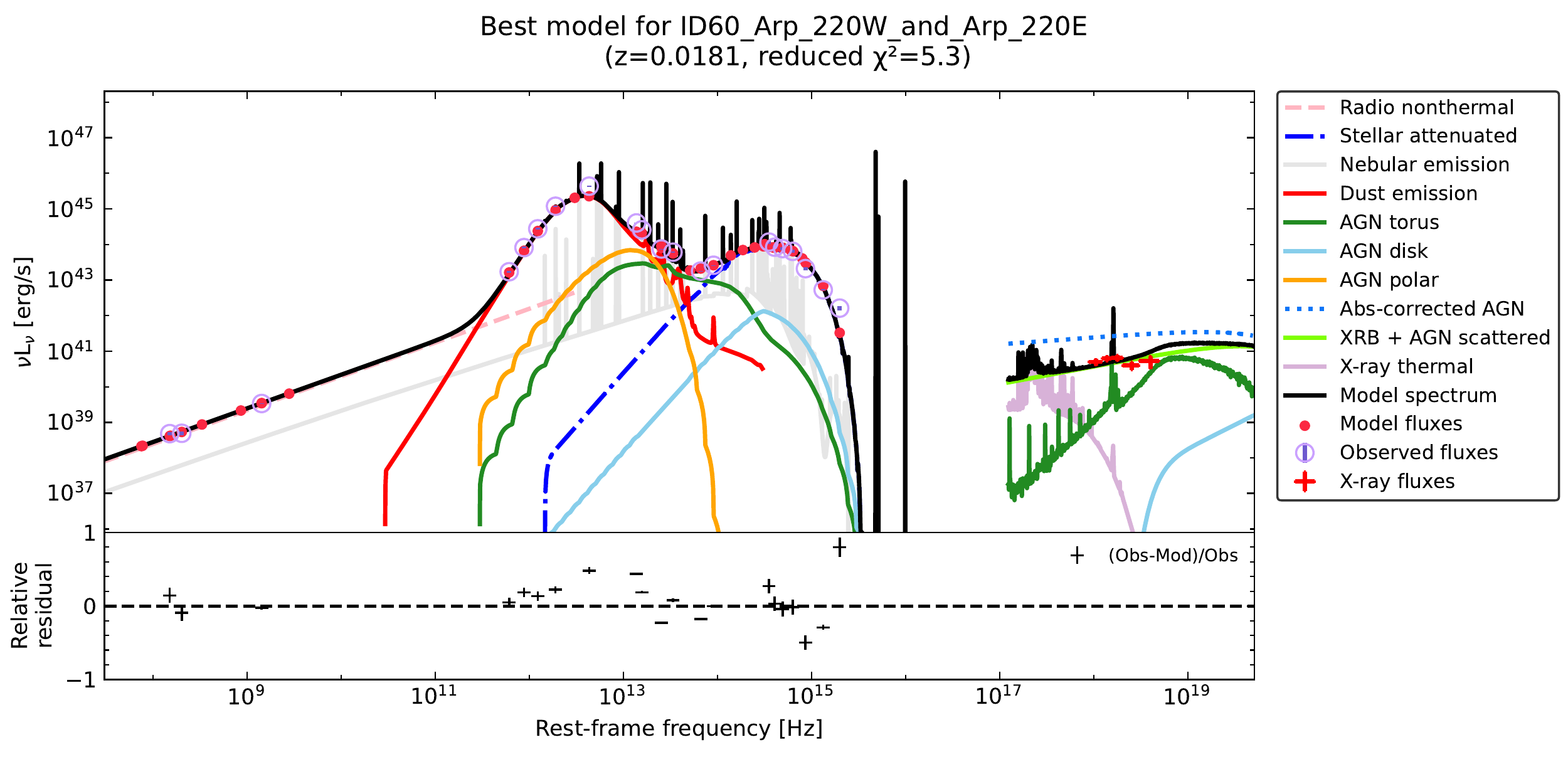}
        \caption{The same as in Figure~\ref{SED10-F} but for the sample of ID=51--60.}
\label{SED15-F}
\end{figure*}

\begin{figure*}
    \centering
   \includegraphics[keepaspectratio, scale=0.34]{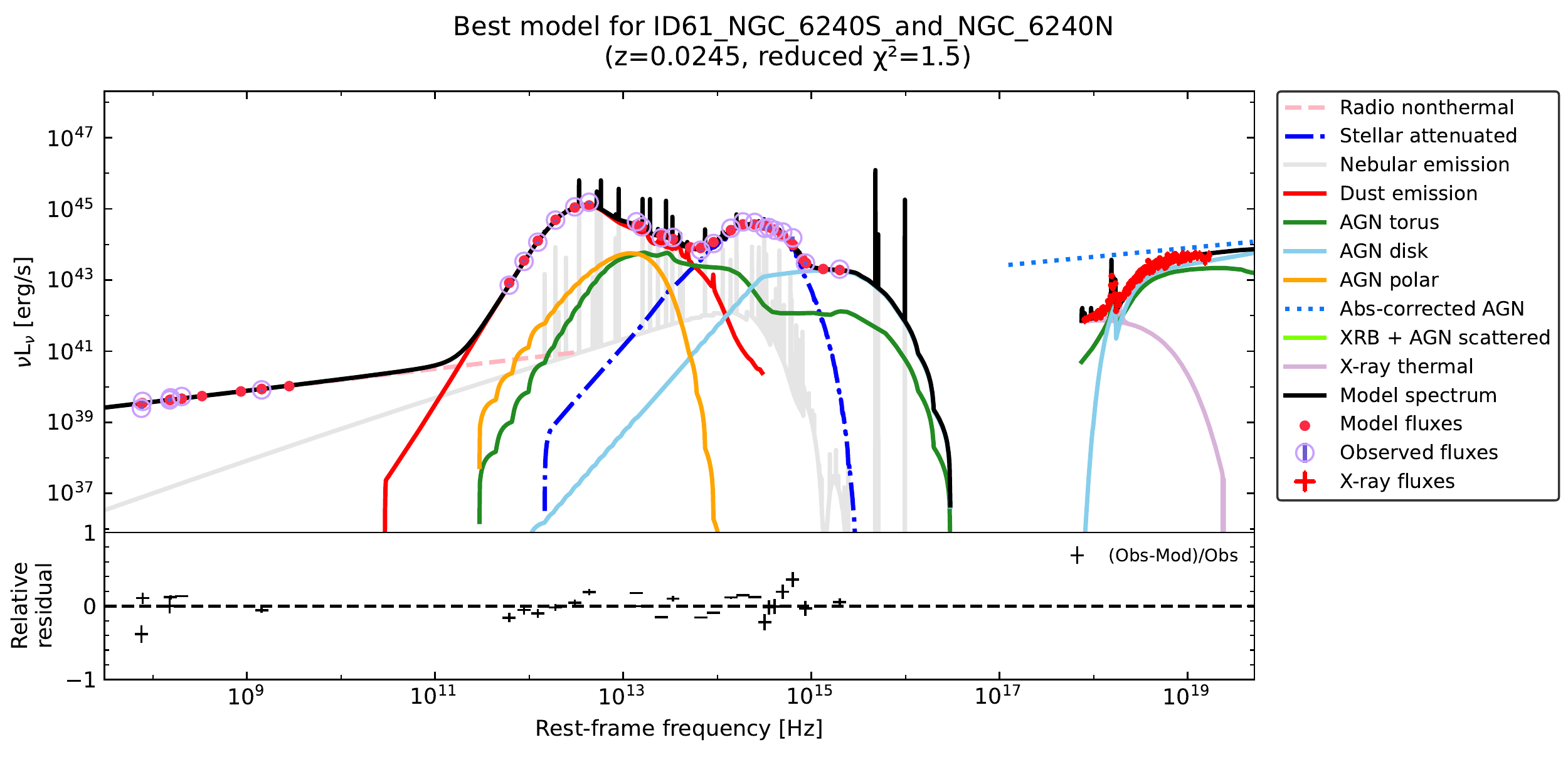}
   \includegraphics[keepaspectratio, scale=0.34]{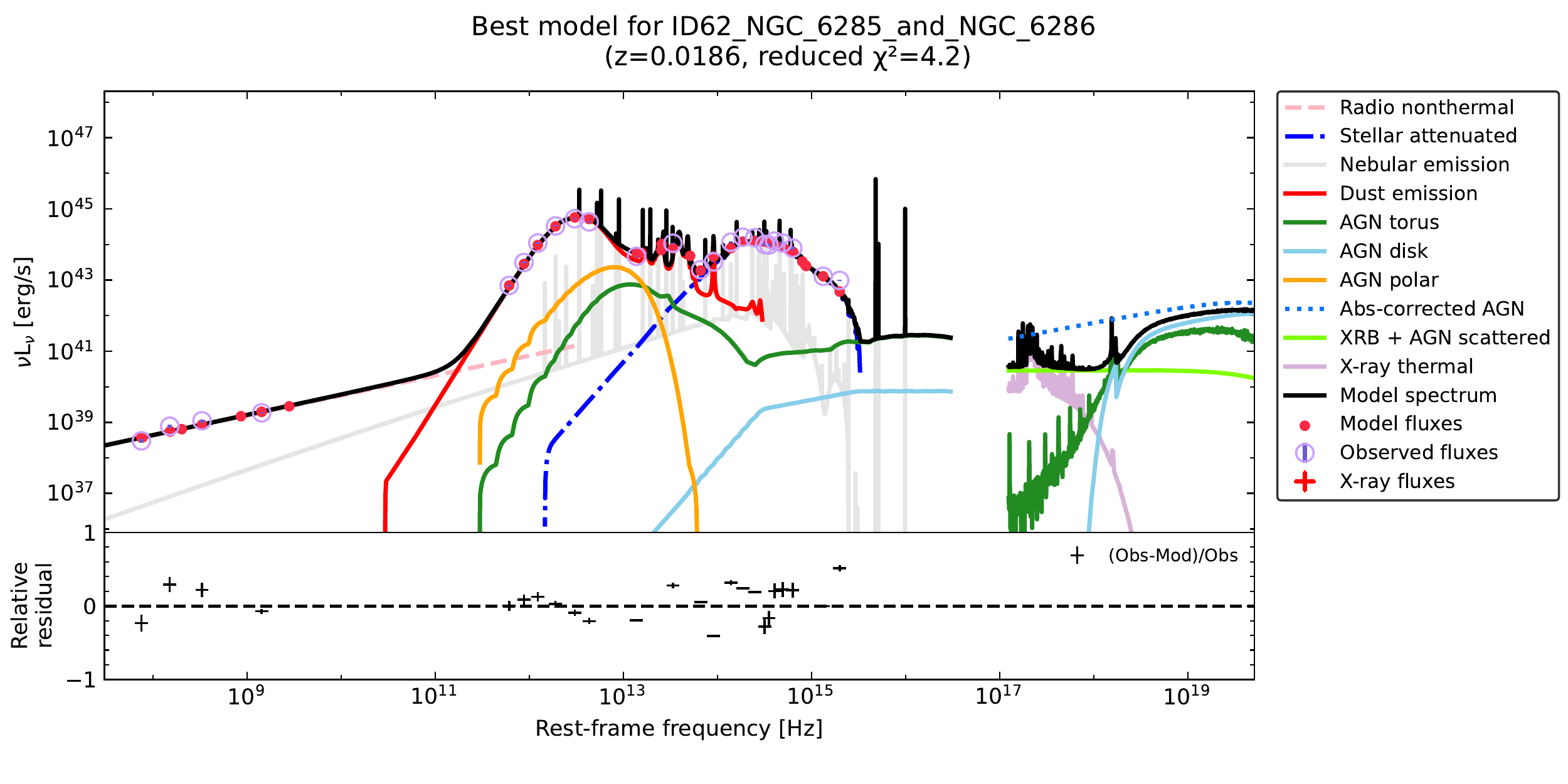}
   \includegraphics[keepaspectratio, scale=0.34]{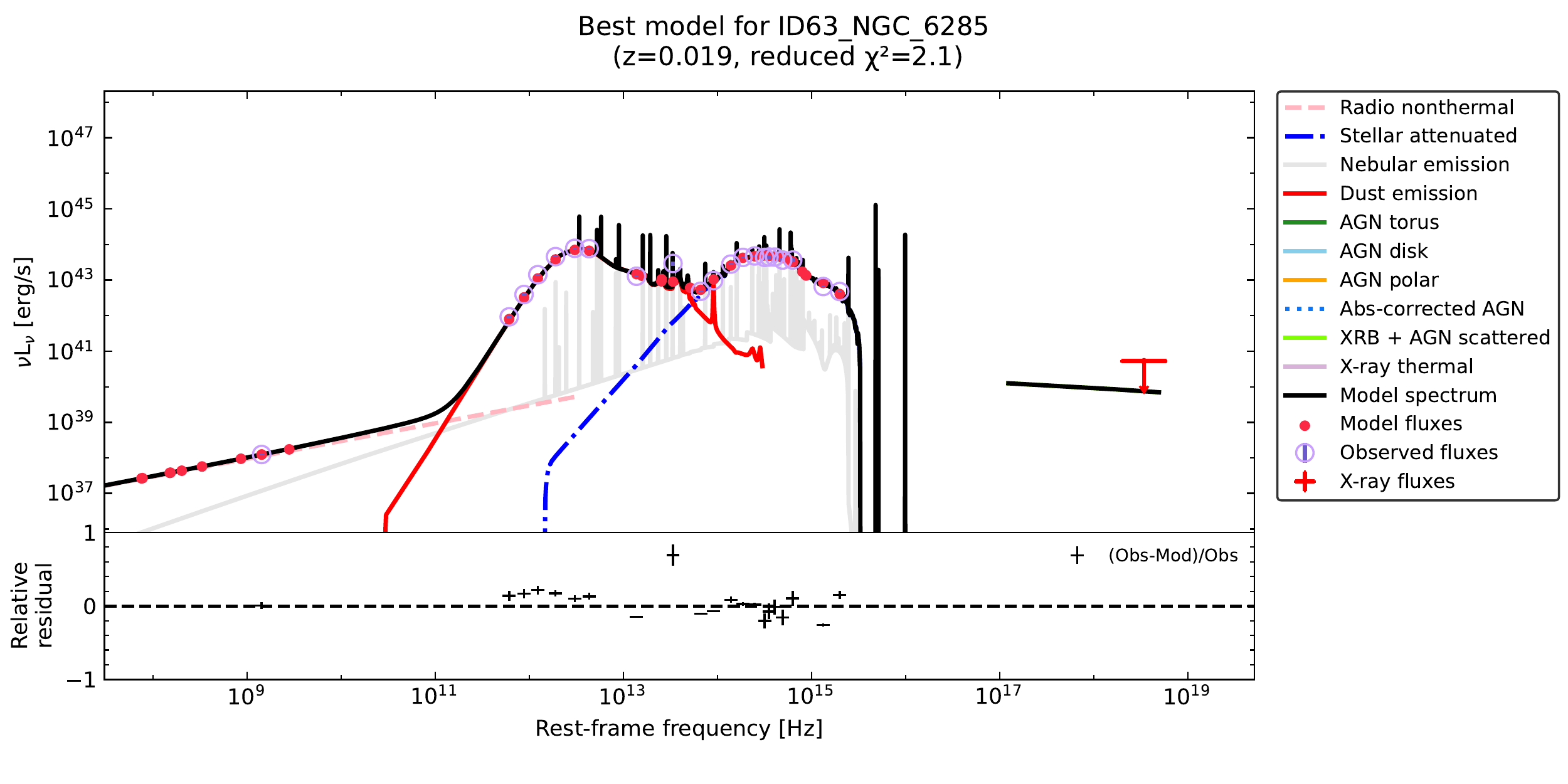}
   \includegraphics[keepaspectratio, scale=0.34]{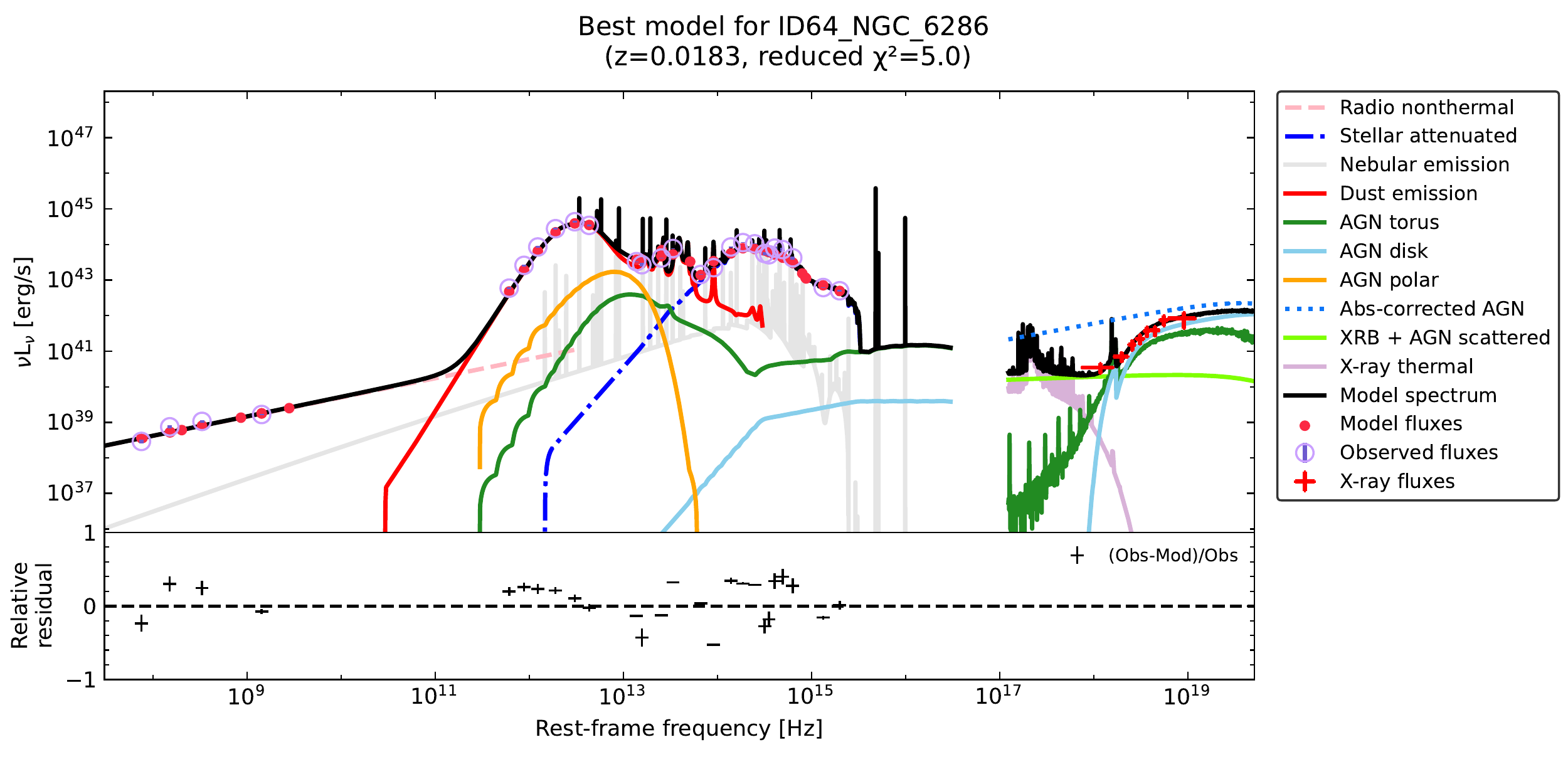}
   \includegraphics[keepaspectratio, scale=0.34]{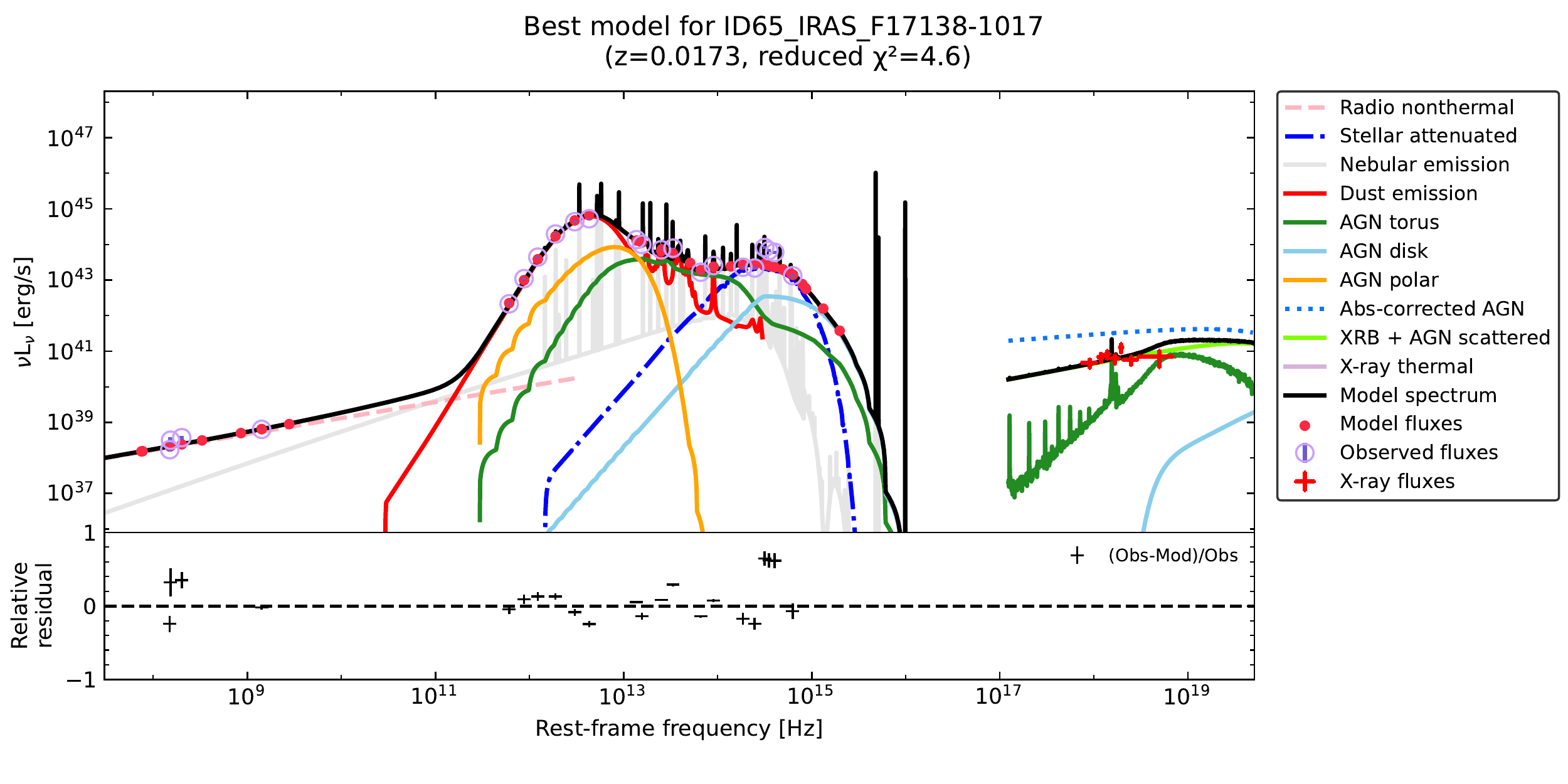}
   \includegraphics[keepaspectratio, scale=0.34]{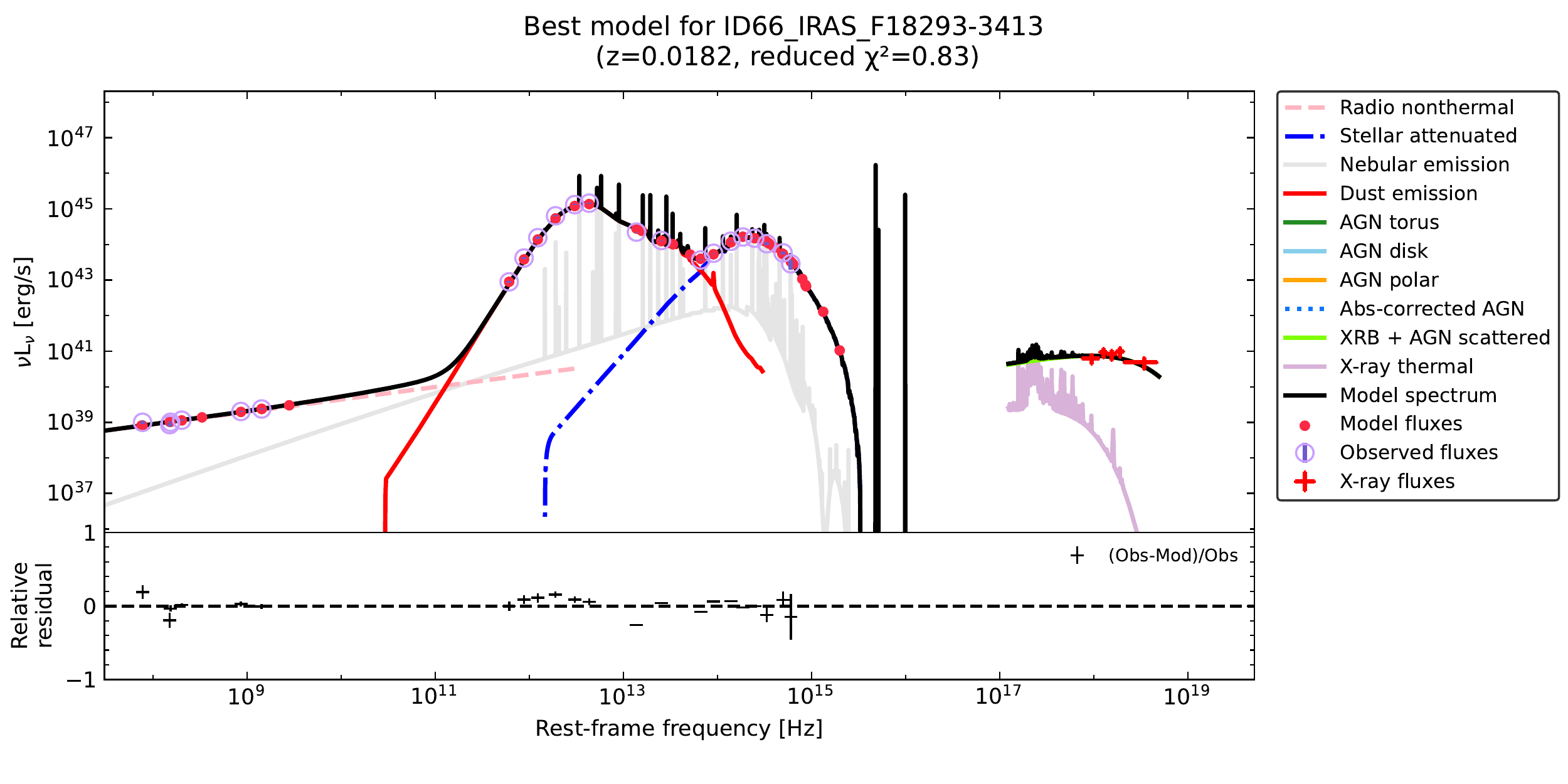}
   \includegraphics[keepaspectratio, scale=0.34]{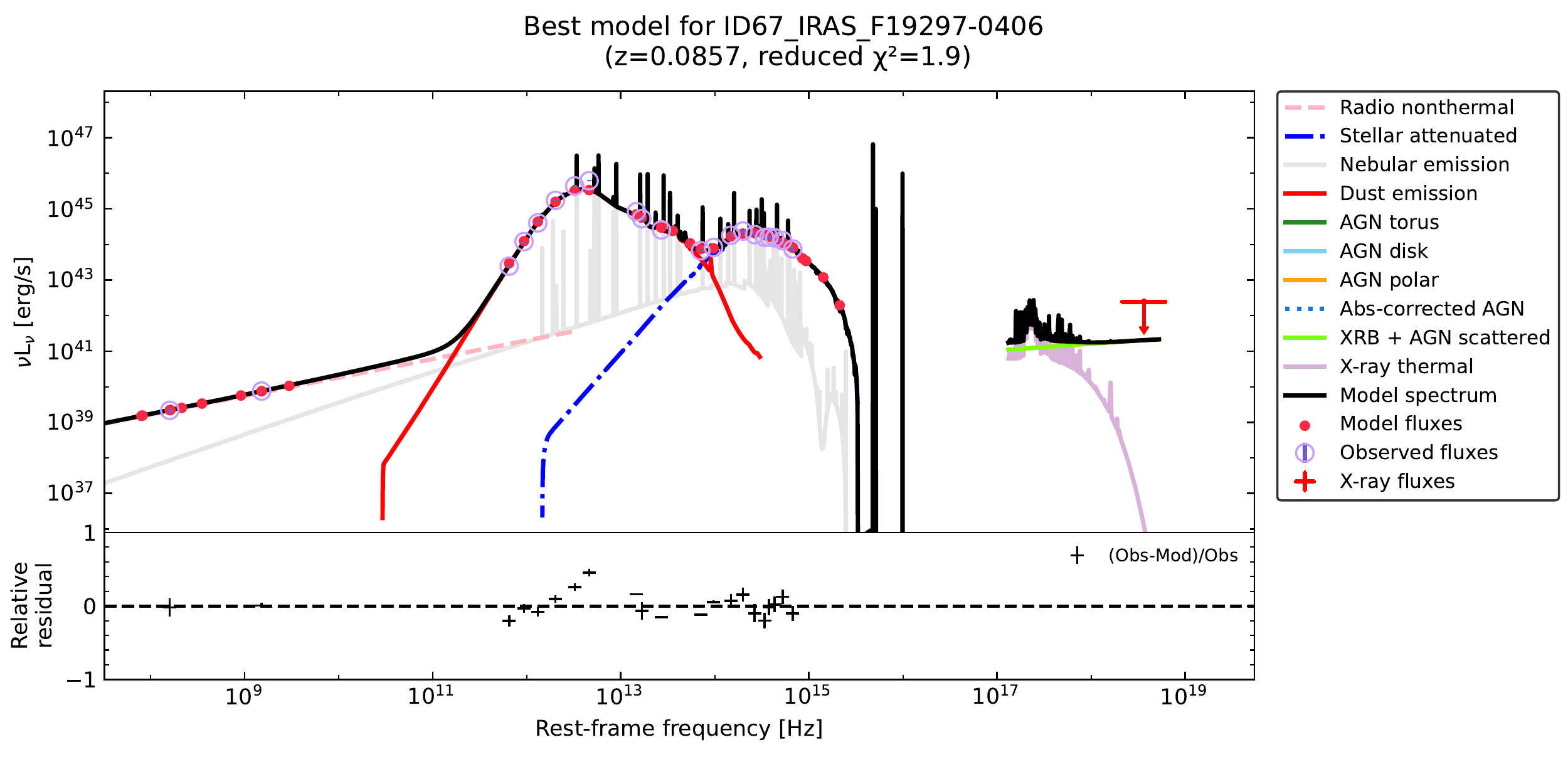}
   \includegraphics[keepaspectratio, scale=0.34]{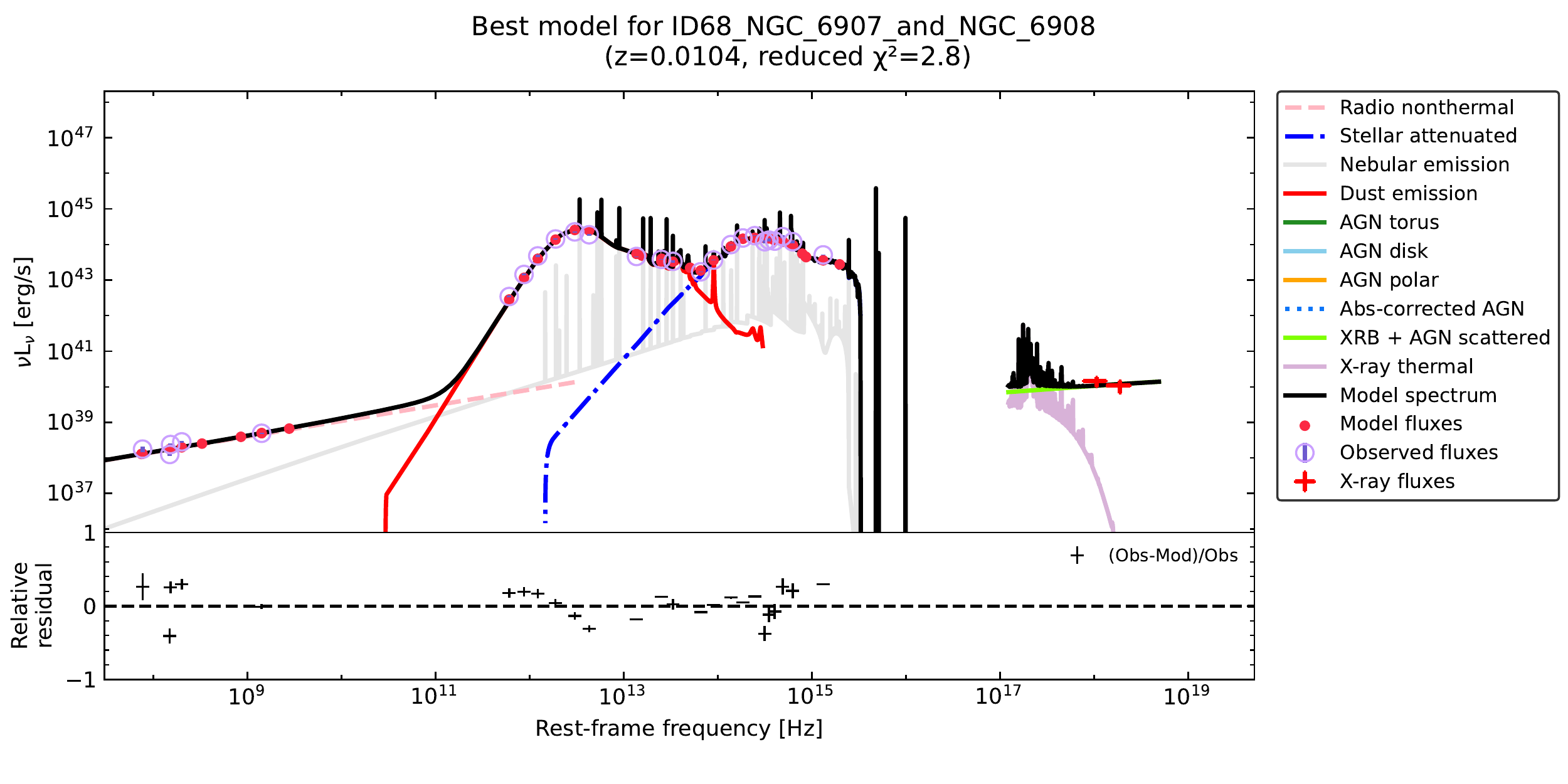}
   \includegraphics[keepaspectratio, scale=0.34]{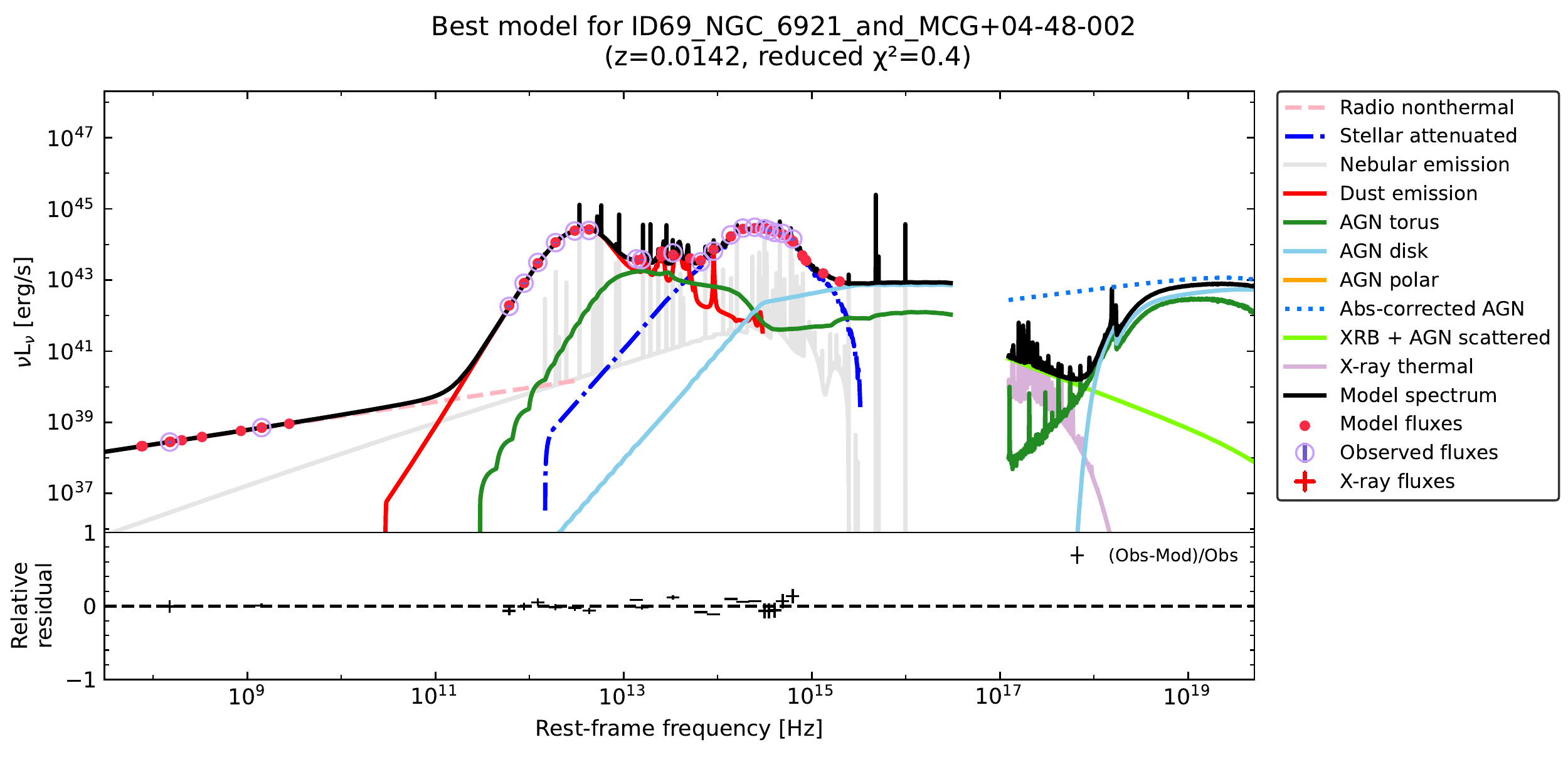}
   \includegraphics[keepaspectratio, scale=0.34]{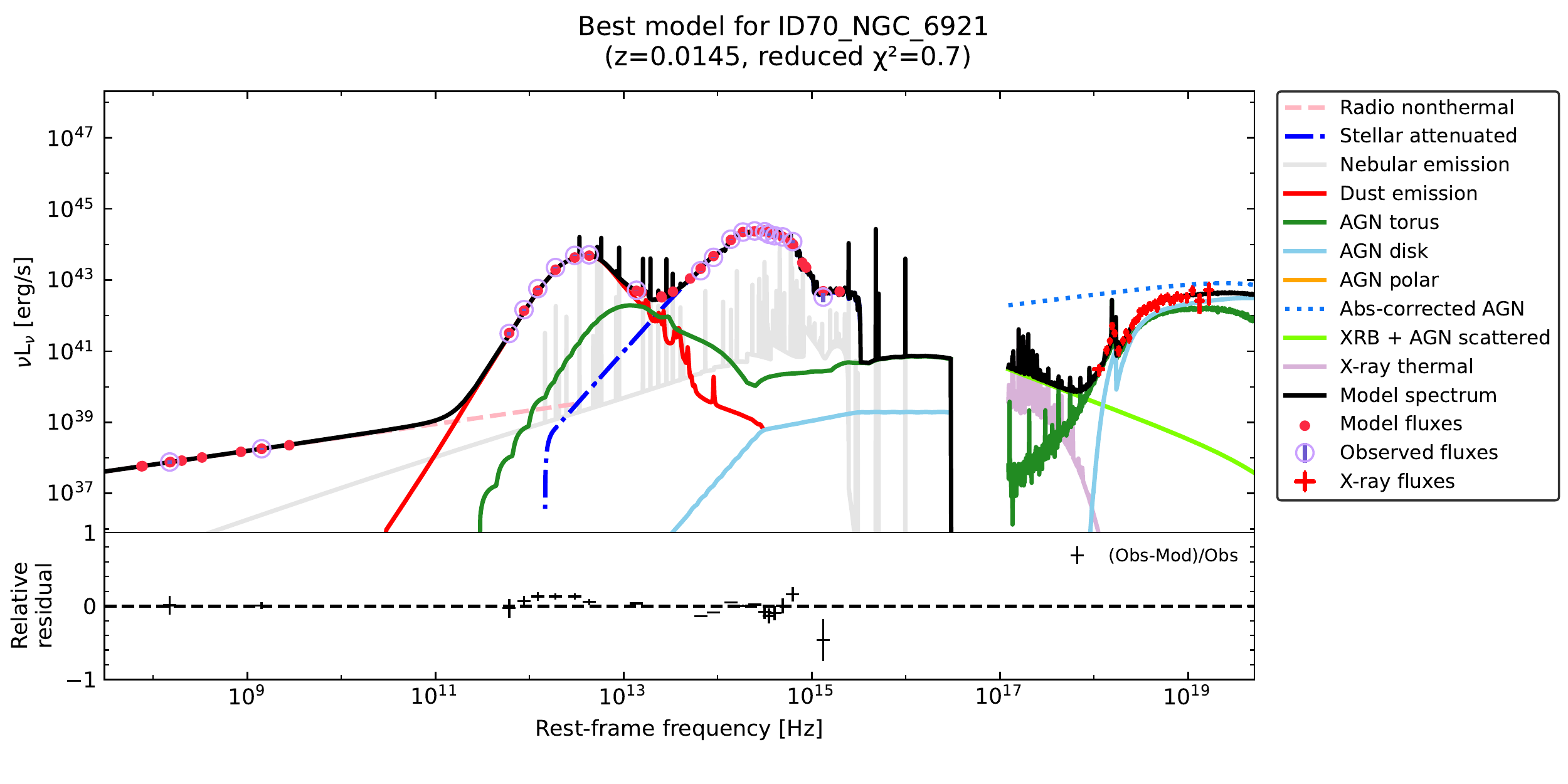}
        \caption{The same as in Figure~\ref{SED10-F} but for the sample of ID=61--70.}
\label{SED16-F}
\end{figure*}

\begin{figure*}
    \centering
   \includegraphics[keepaspectratio, scale=0.34]{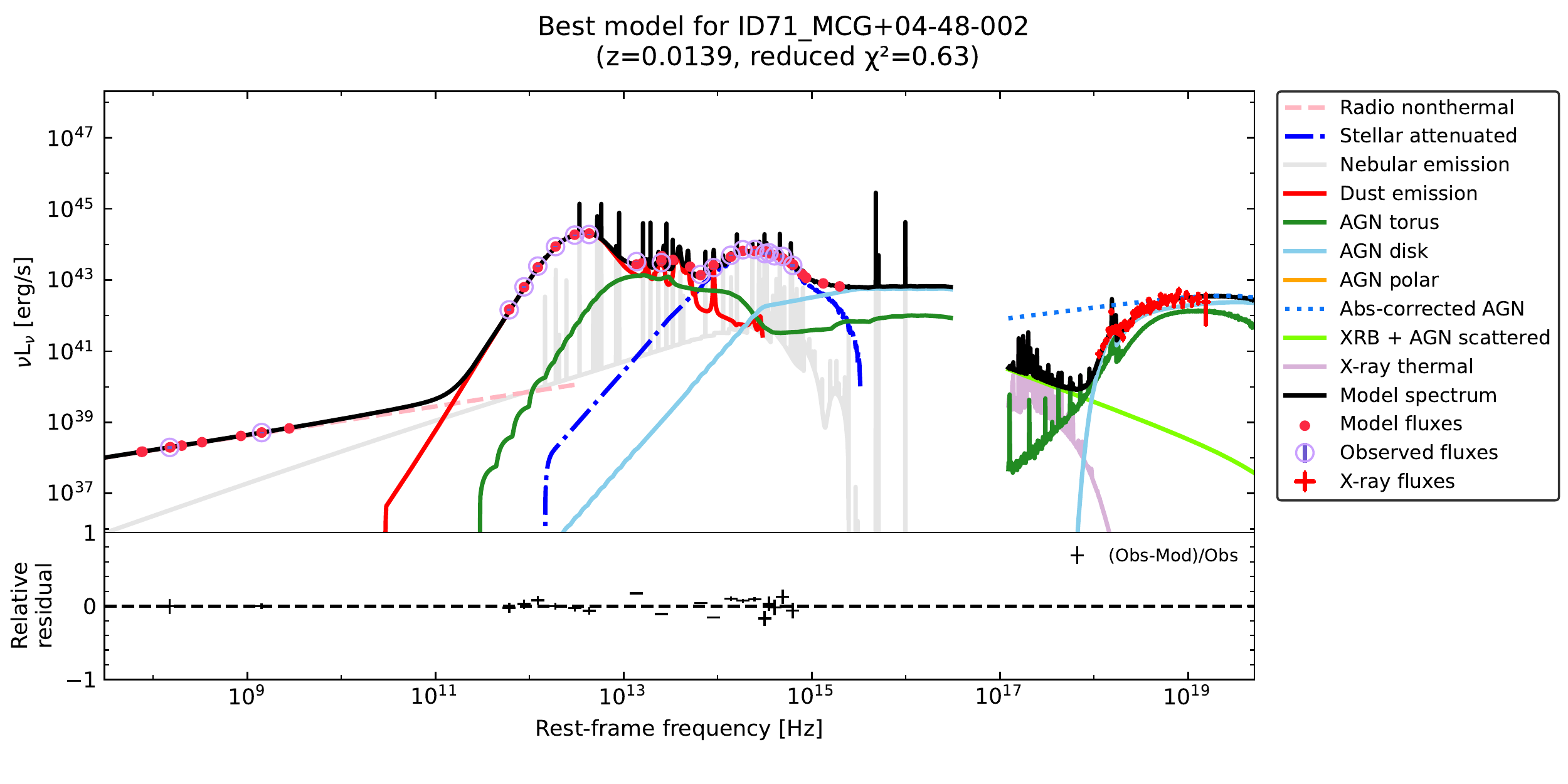}
   \includegraphics[keepaspectratio, scale=0.34]{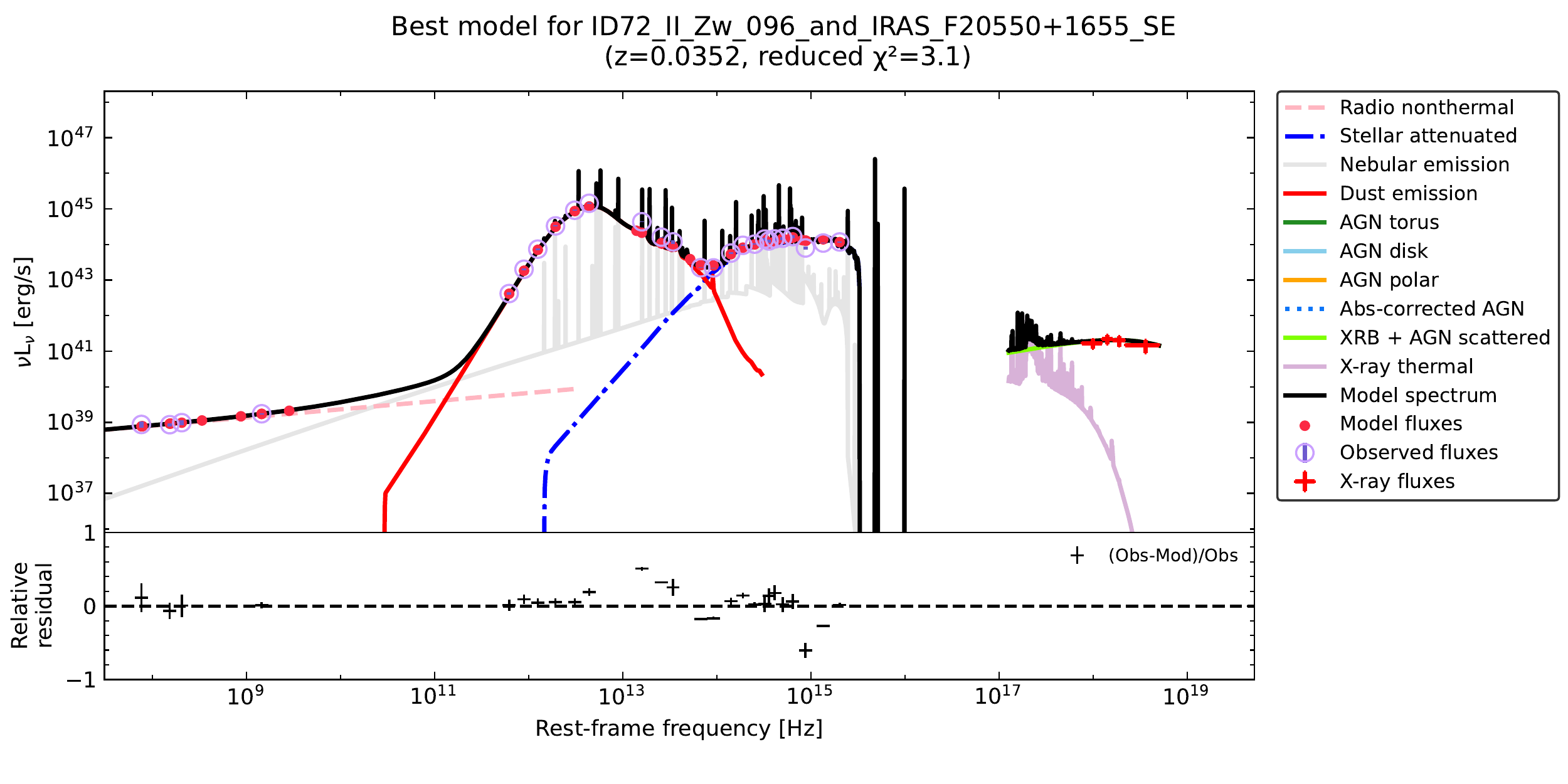}
   \includegraphics[keepaspectratio, scale=0.34]{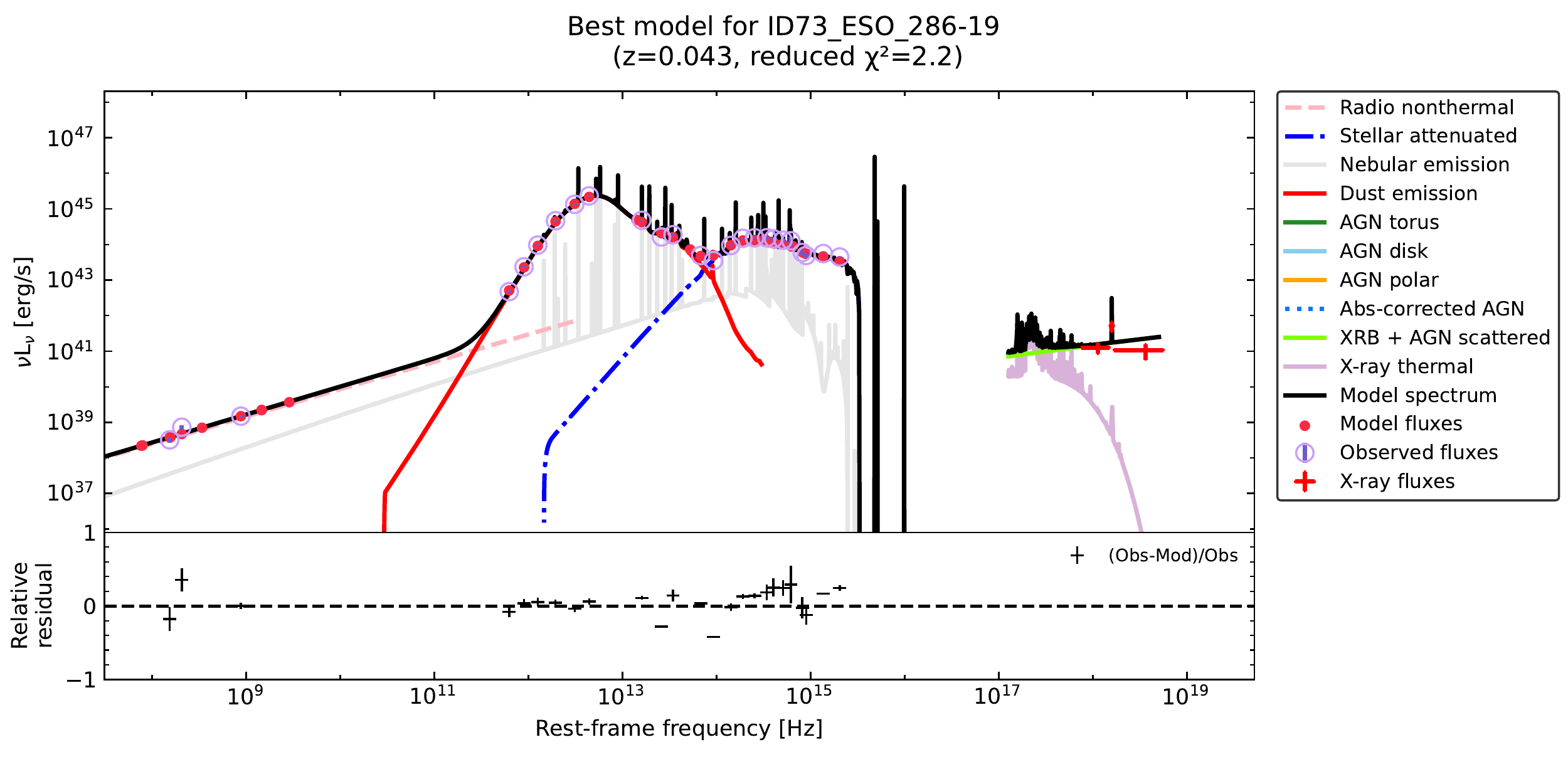}
   \includegraphics[keepaspectratio, scale=0.34]{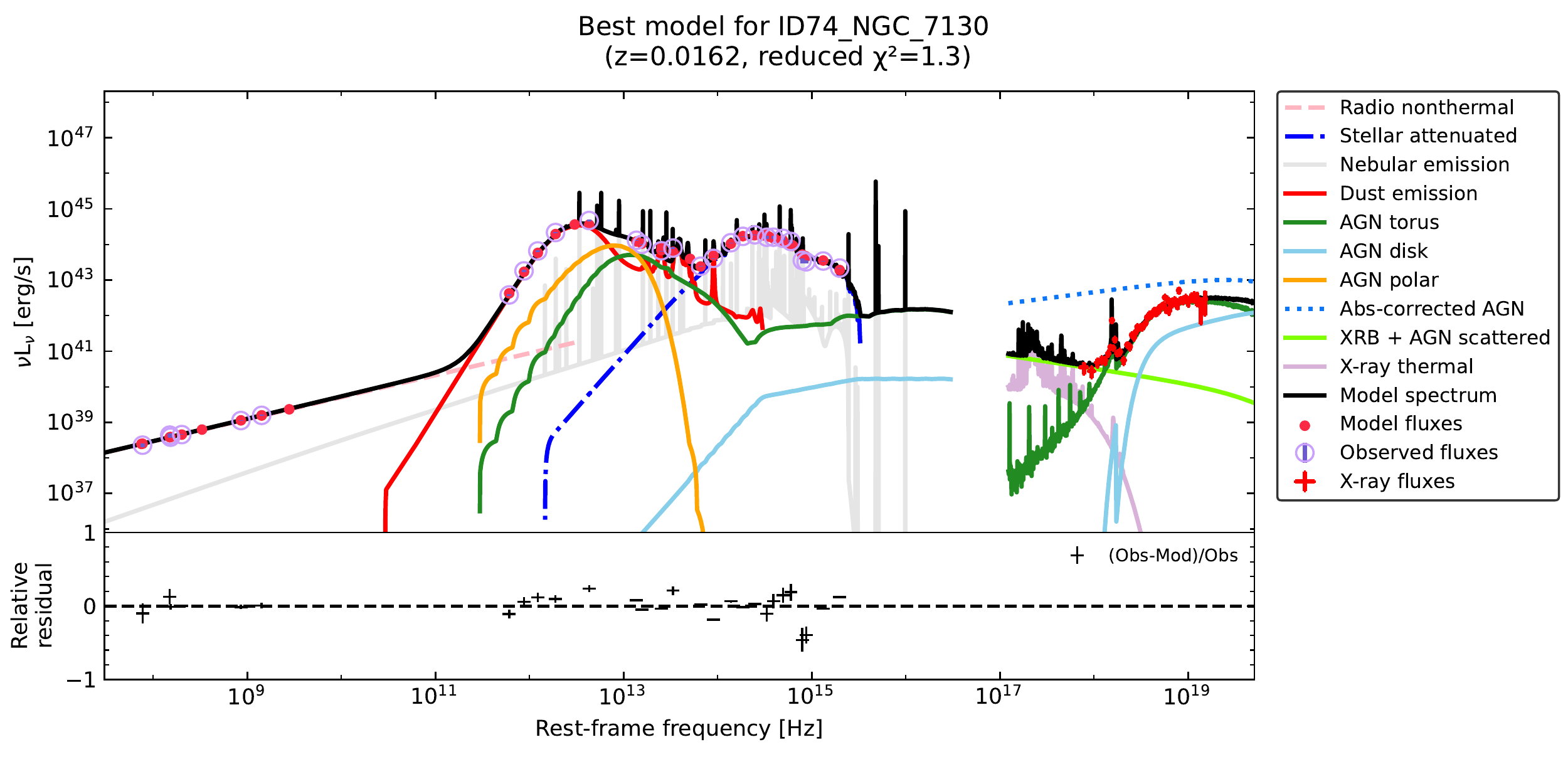}
   \includegraphics[keepaspectratio, scale=0.34]{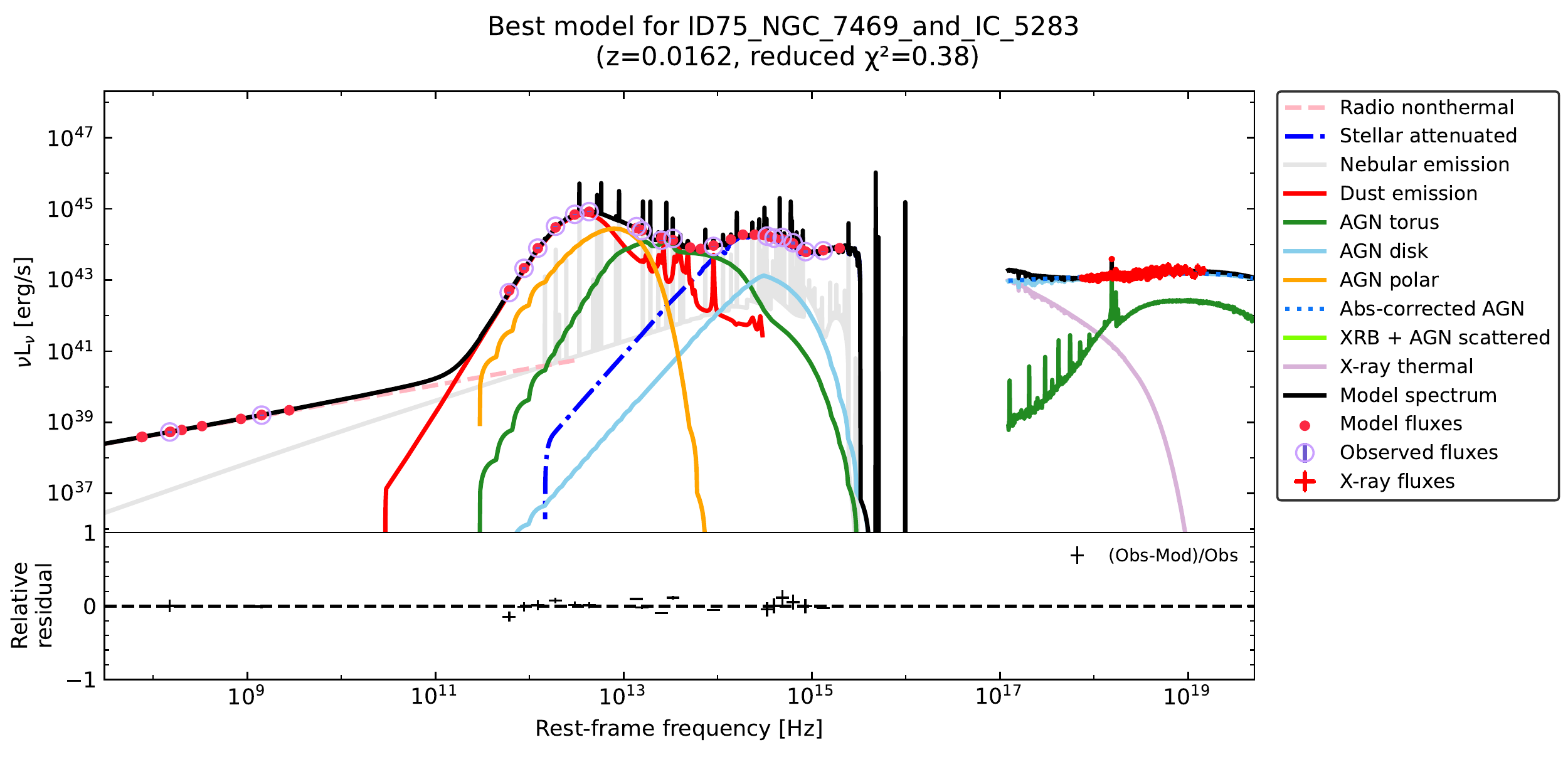}
   \includegraphics[keepaspectratio, scale=0.34]{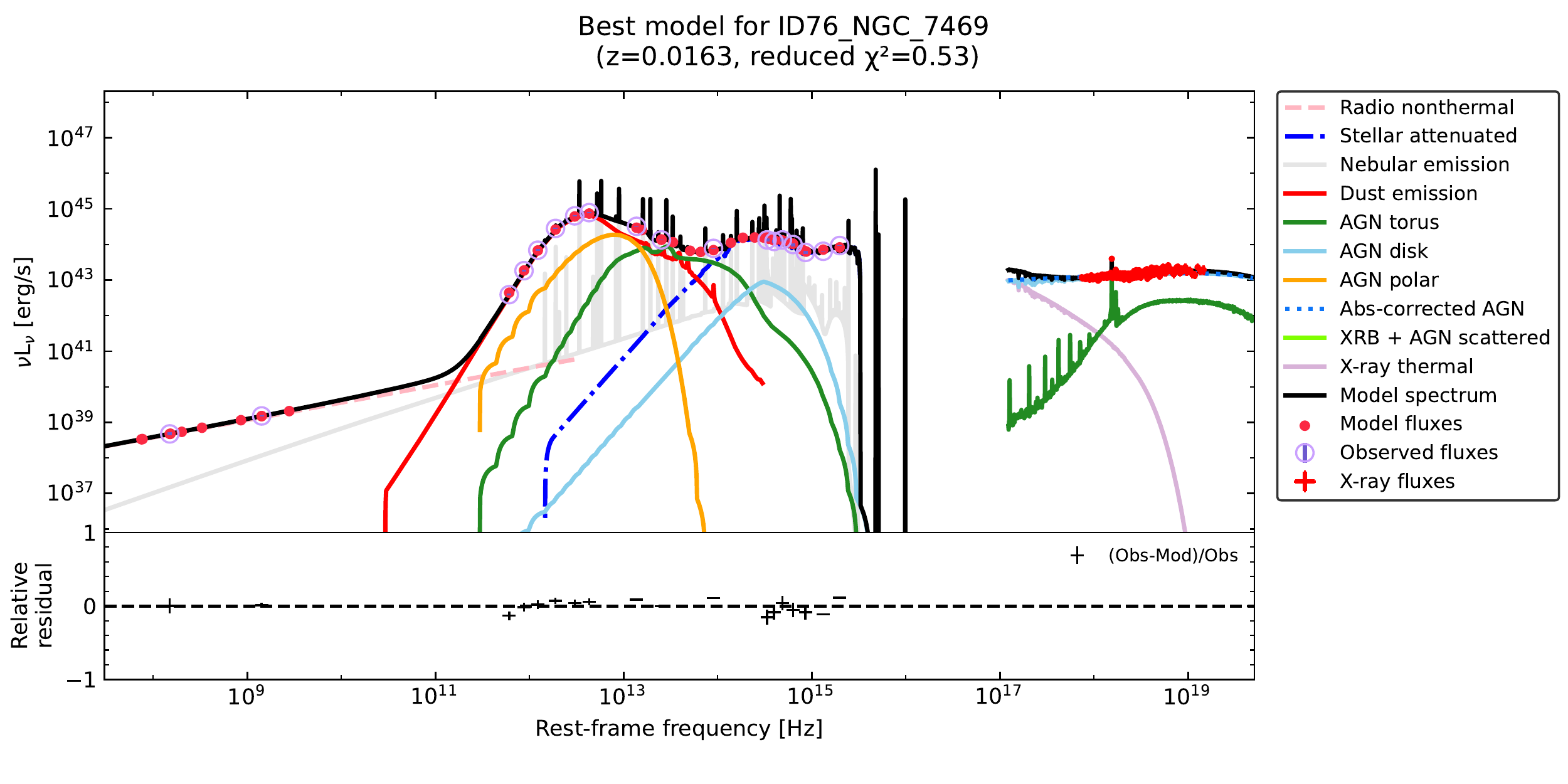}
   \includegraphics[keepaspectratio, scale=0.34]{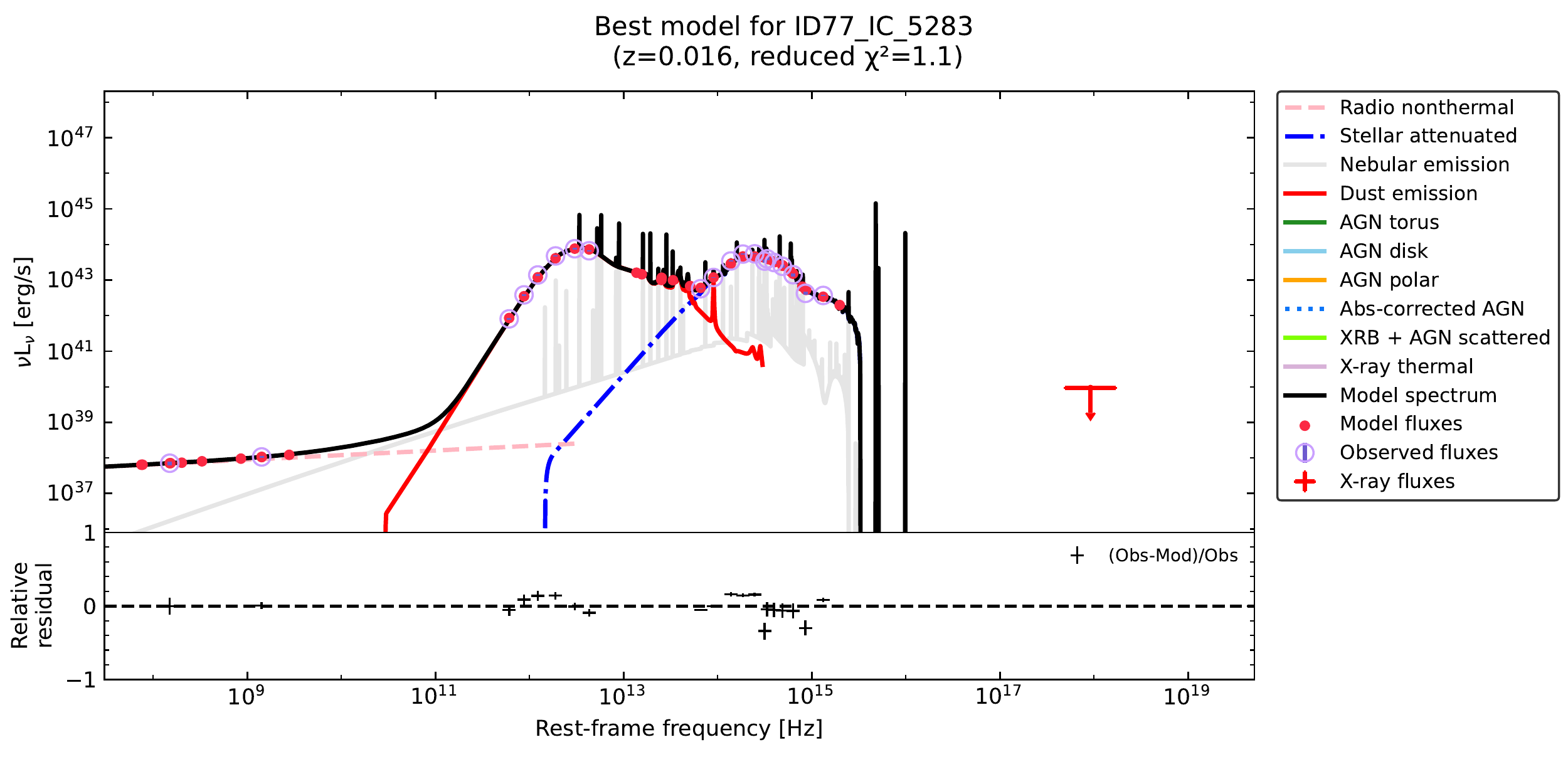}
   \includegraphics[keepaspectratio, scale=0.34]{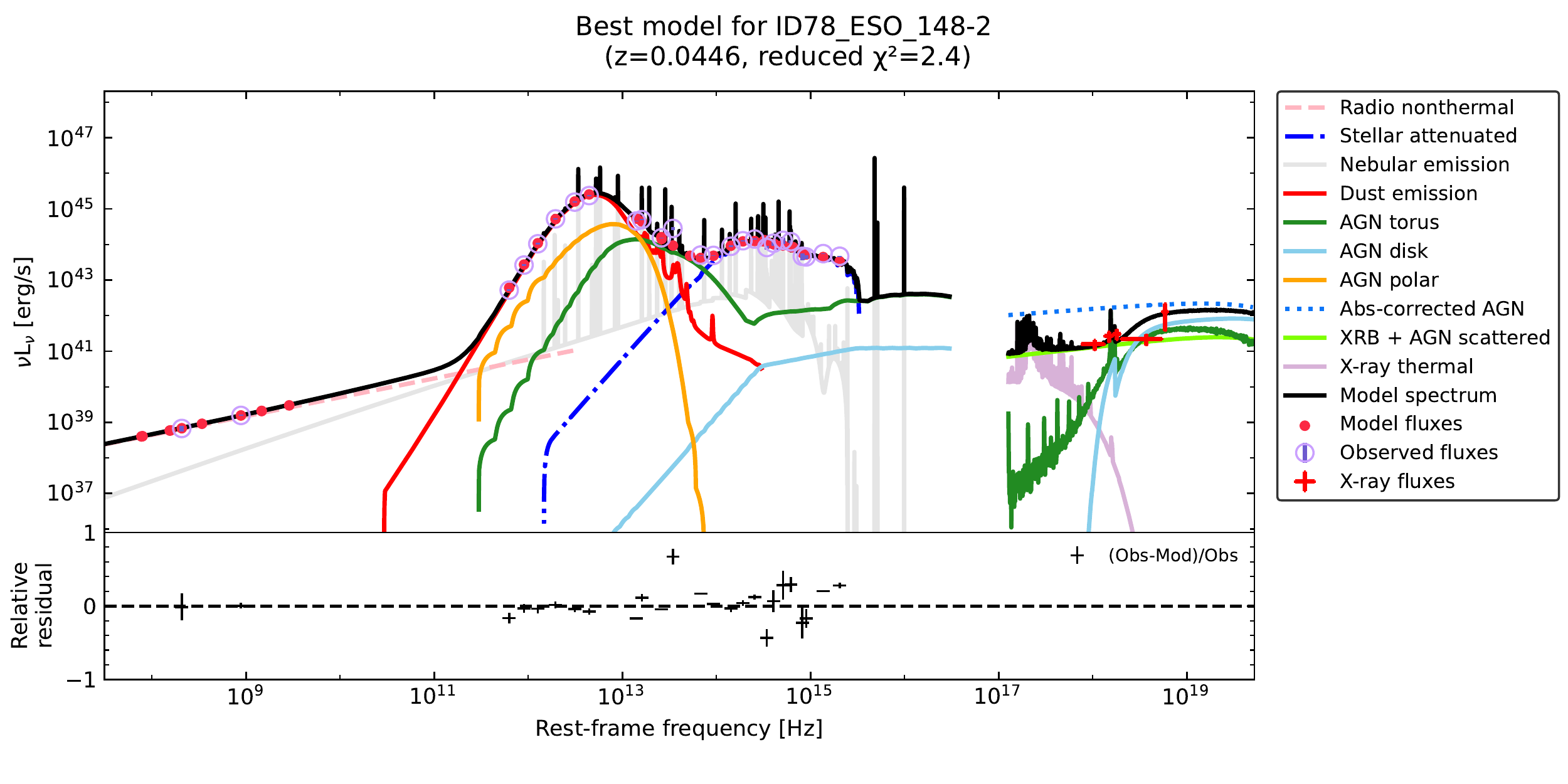}
   \includegraphics[keepaspectratio, scale=0.34]{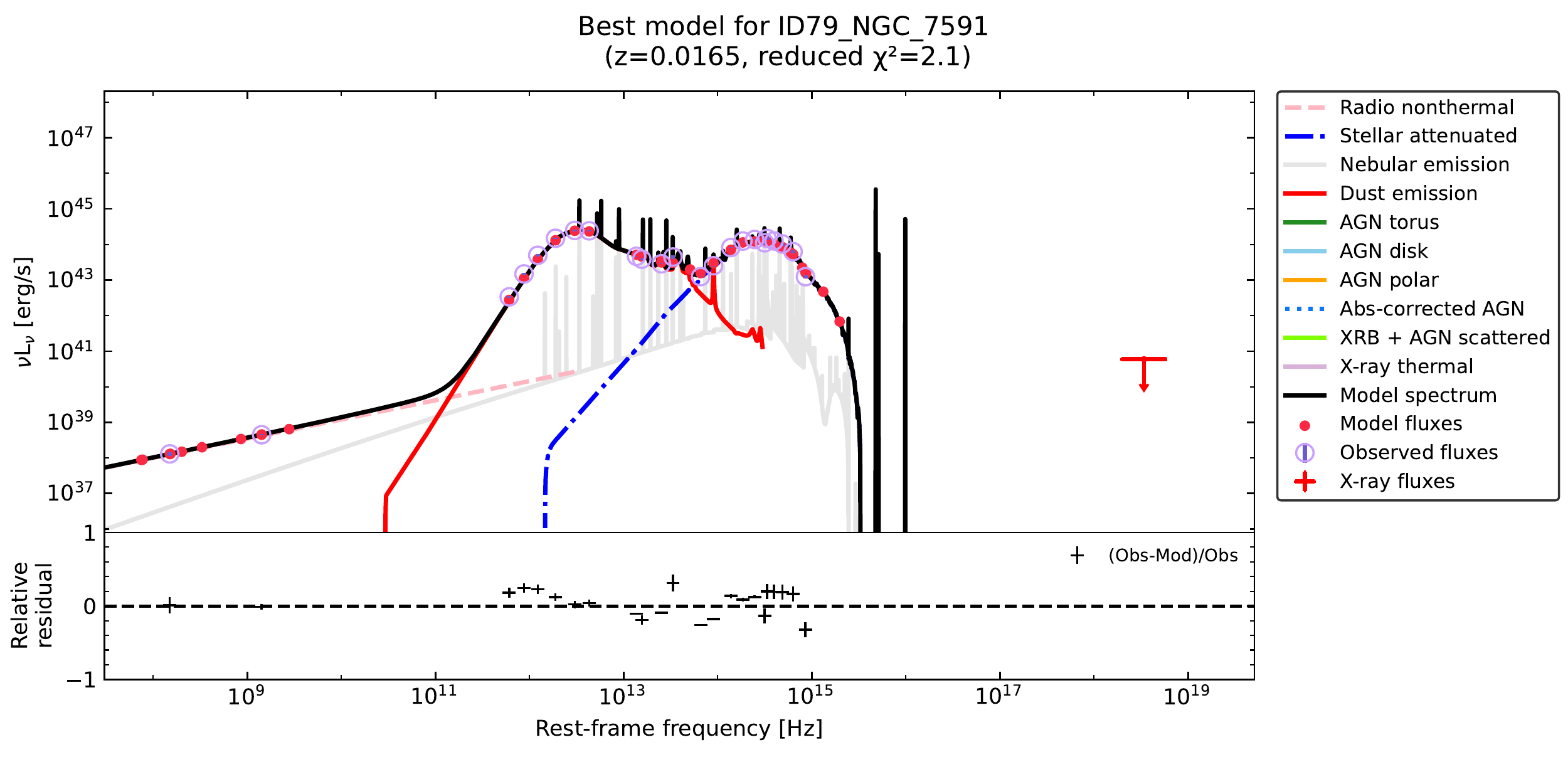}
   \includegraphics[keepaspectratio, scale=0.34]{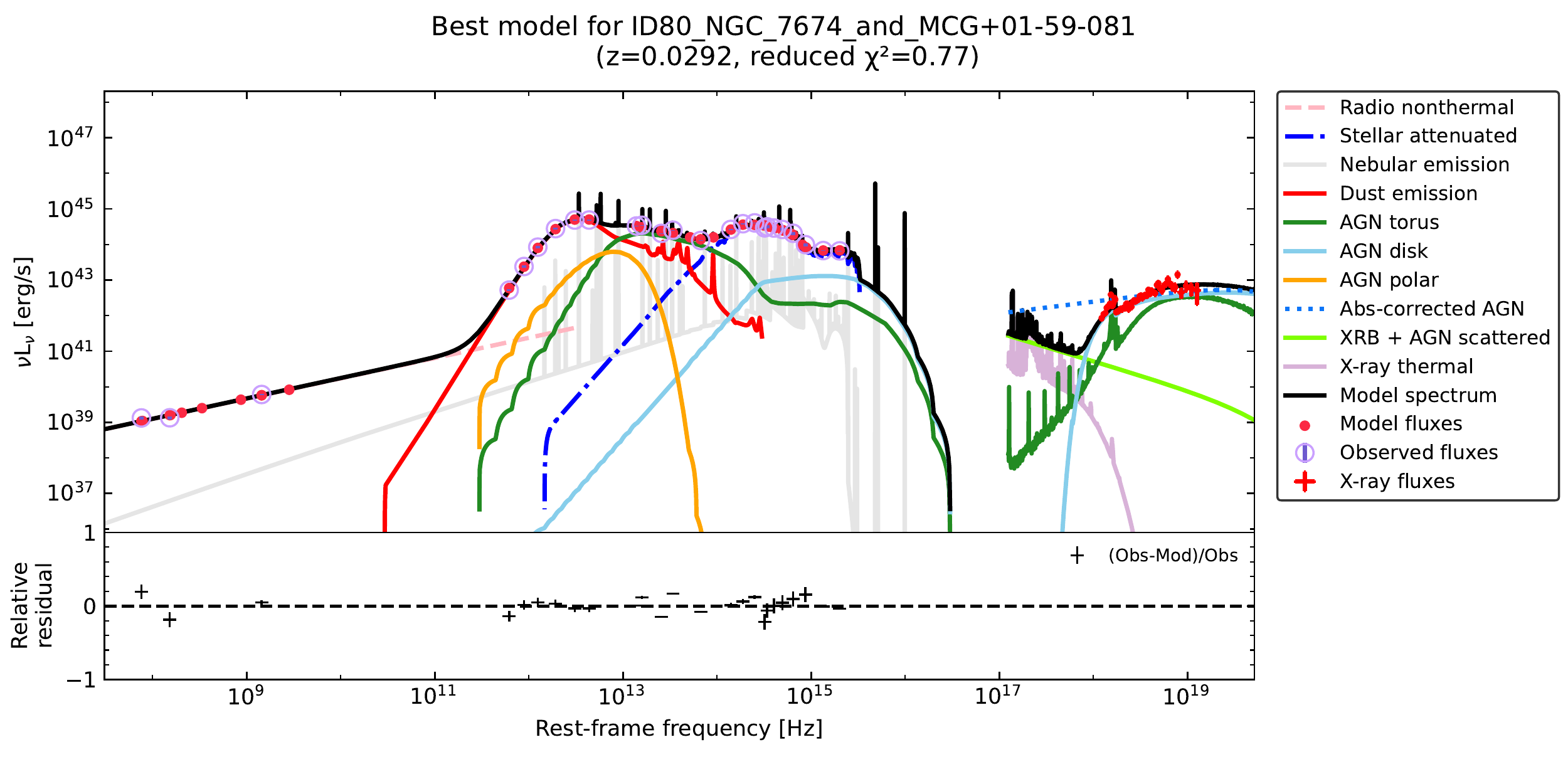}
        \caption{The same as in Figure~\ref{SED10-F} but for the sample of ID=71--80.}
\label{SED17-F}
\end{figure*}

\begin{figure*}
    \centering
   \includegraphics[keepaspectratio, scale=0.34]{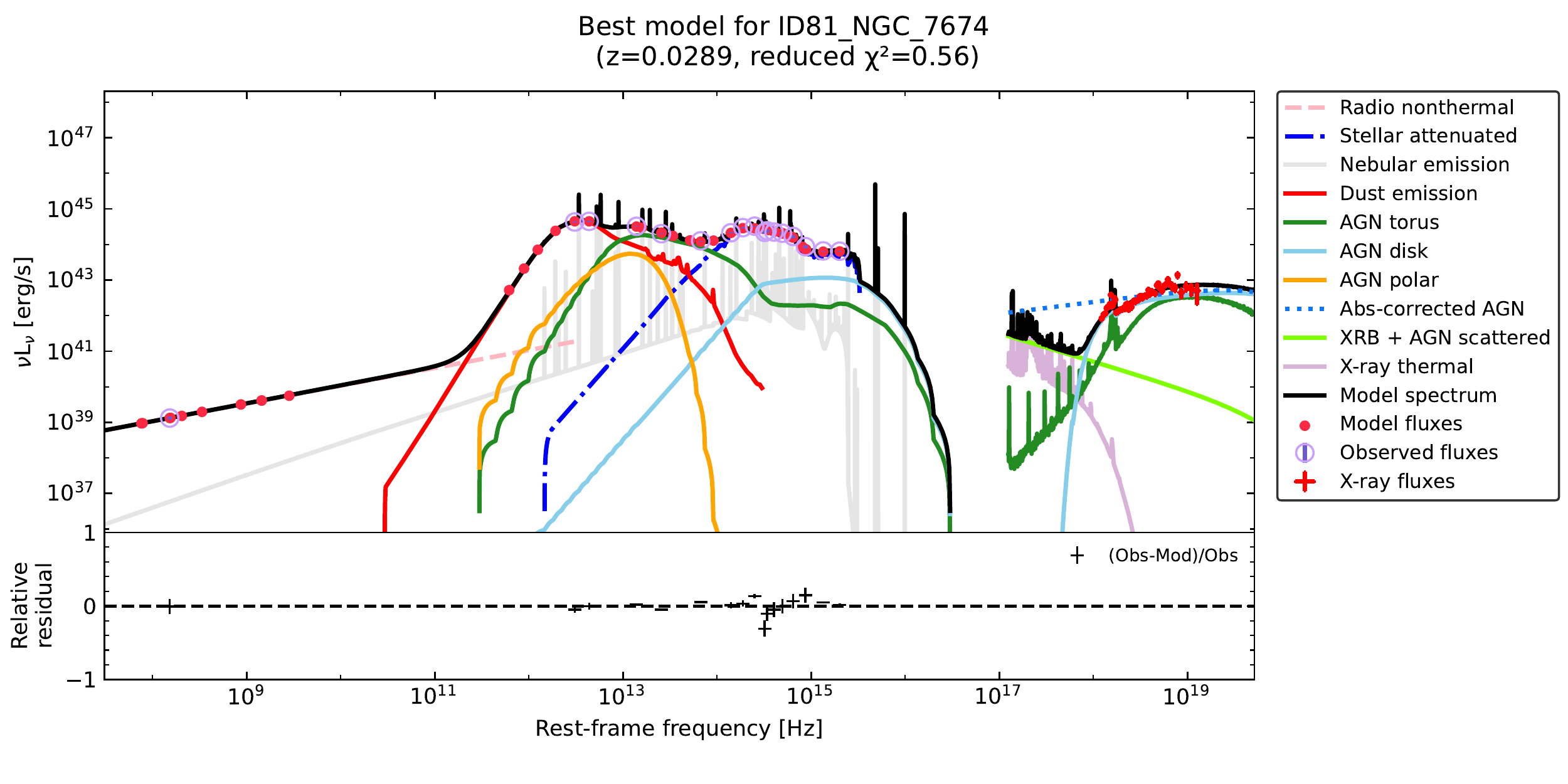}
   \includegraphics[keepaspectratio, scale=0.34]{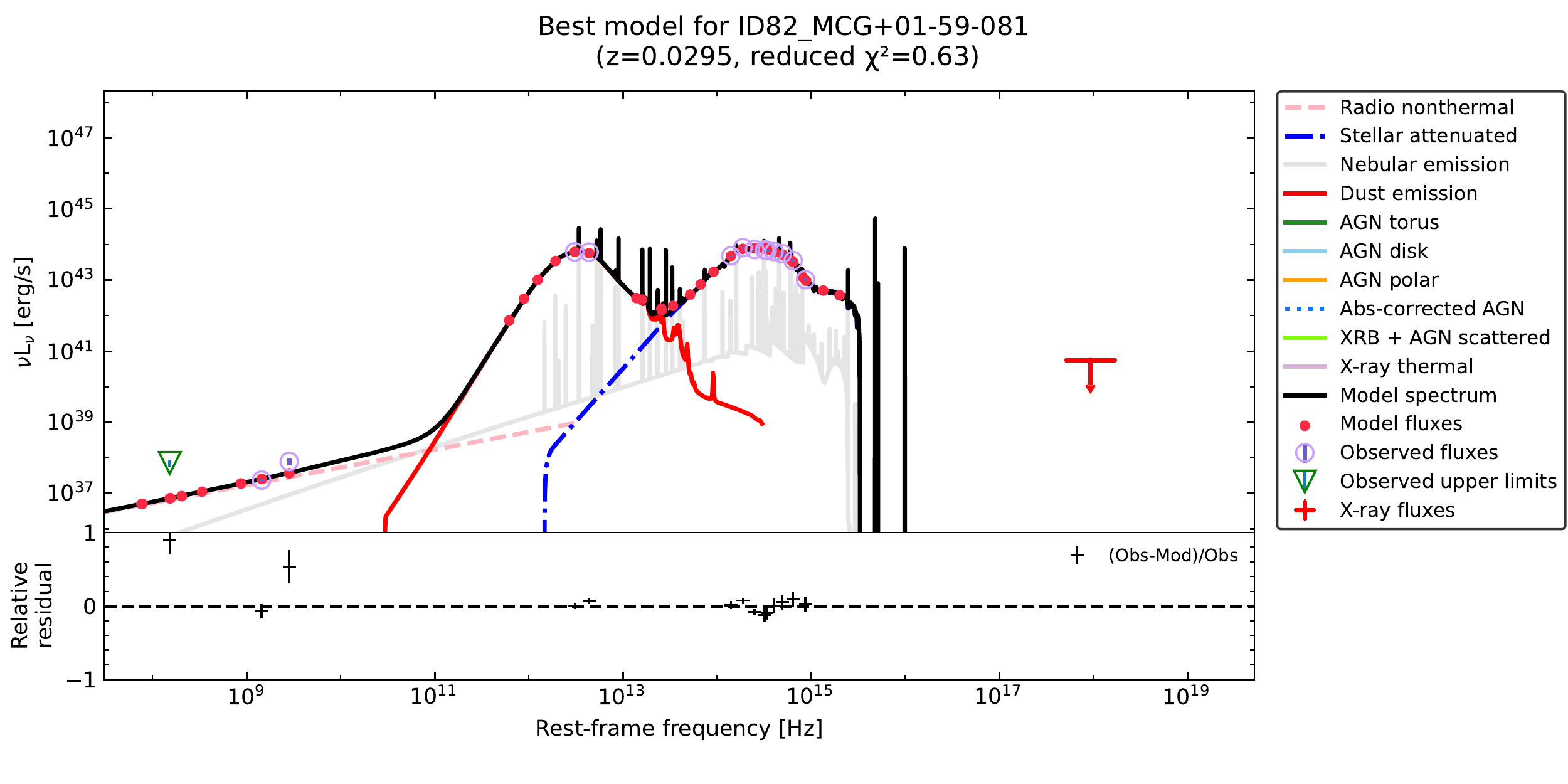}
   \includegraphics[keepaspectratio, scale=0.34]{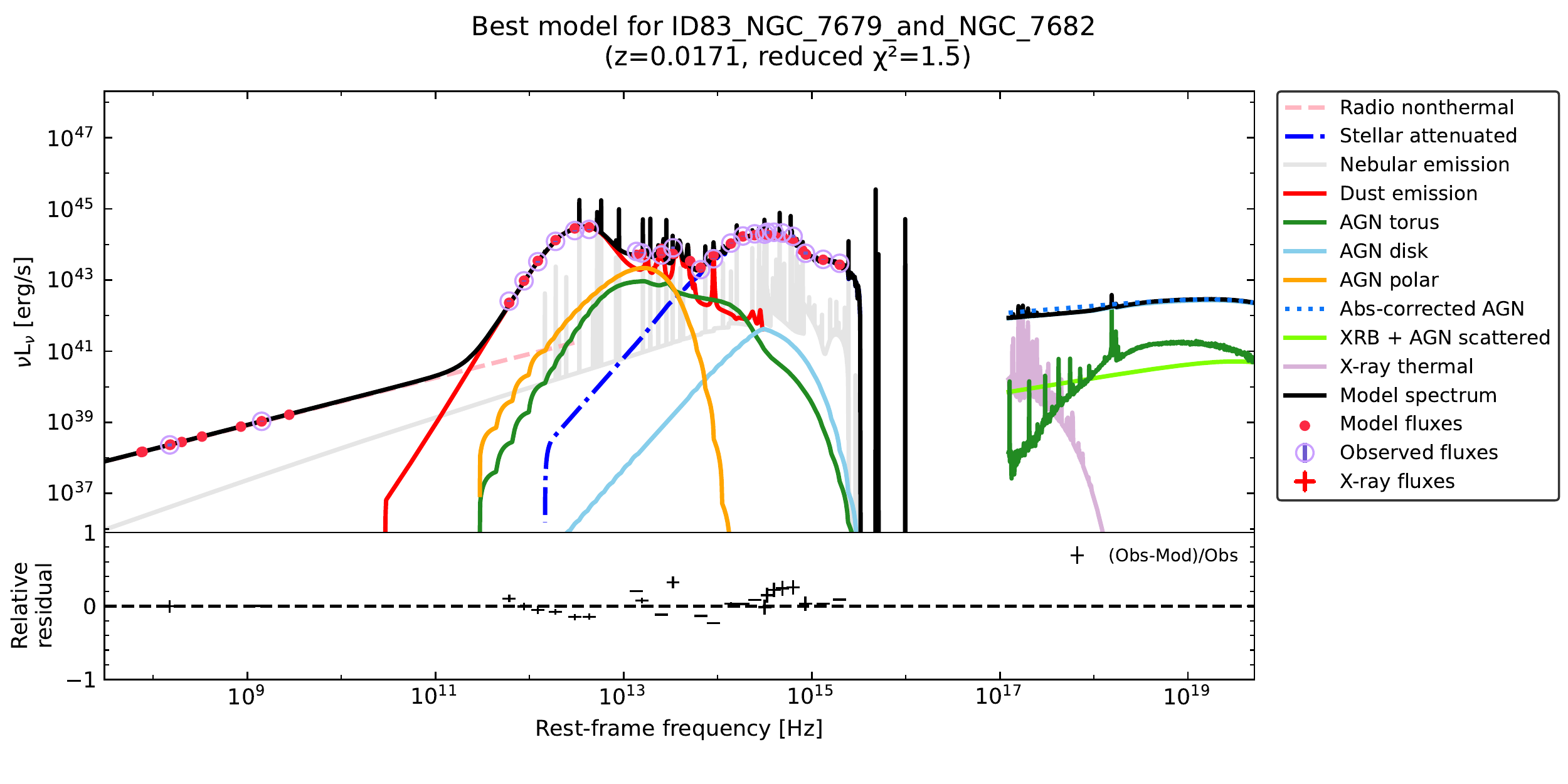}
   \includegraphics[keepaspectratio, scale=0.34]{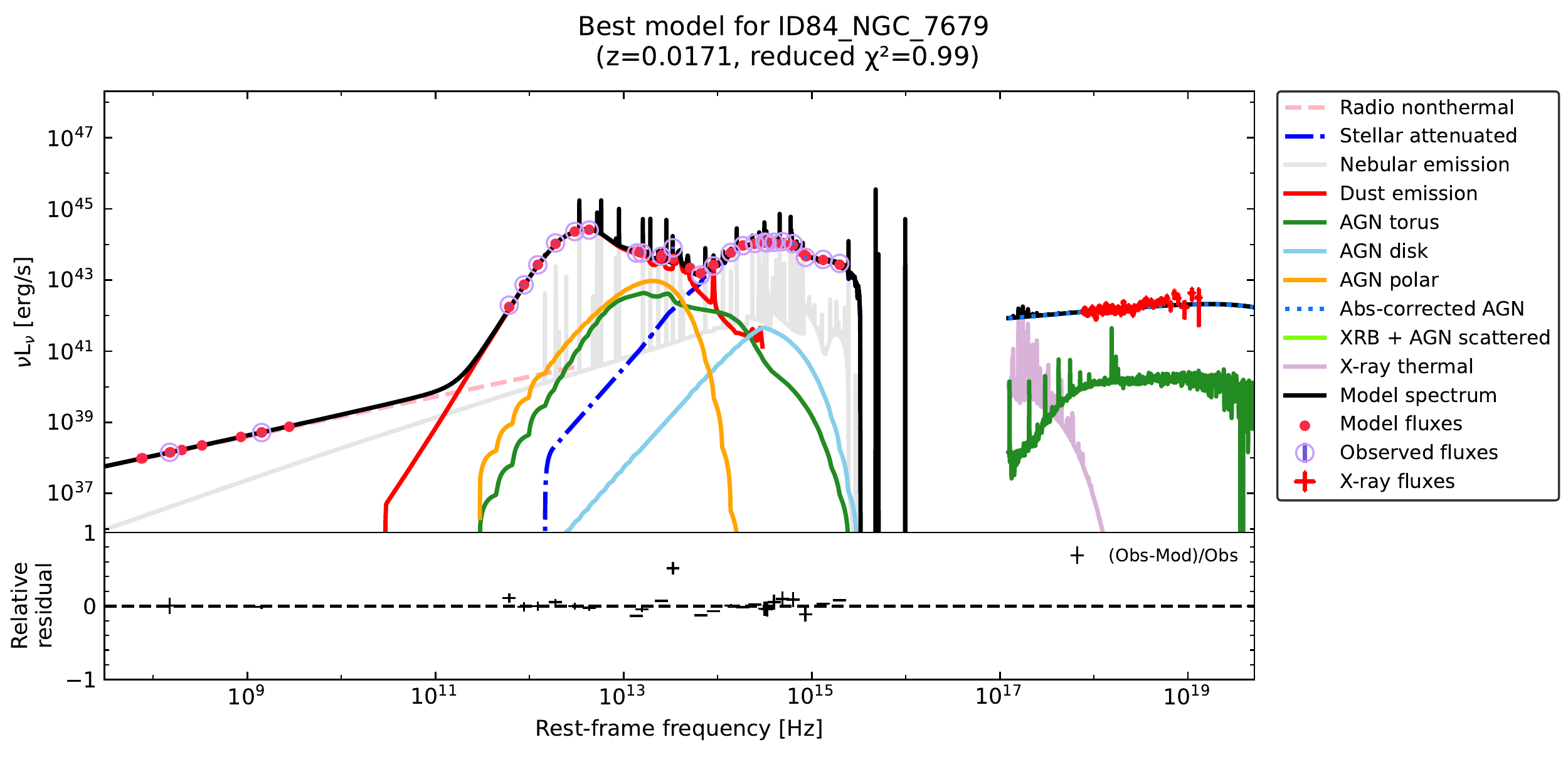}
   \includegraphics[keepaspectratio, scale=0.34]{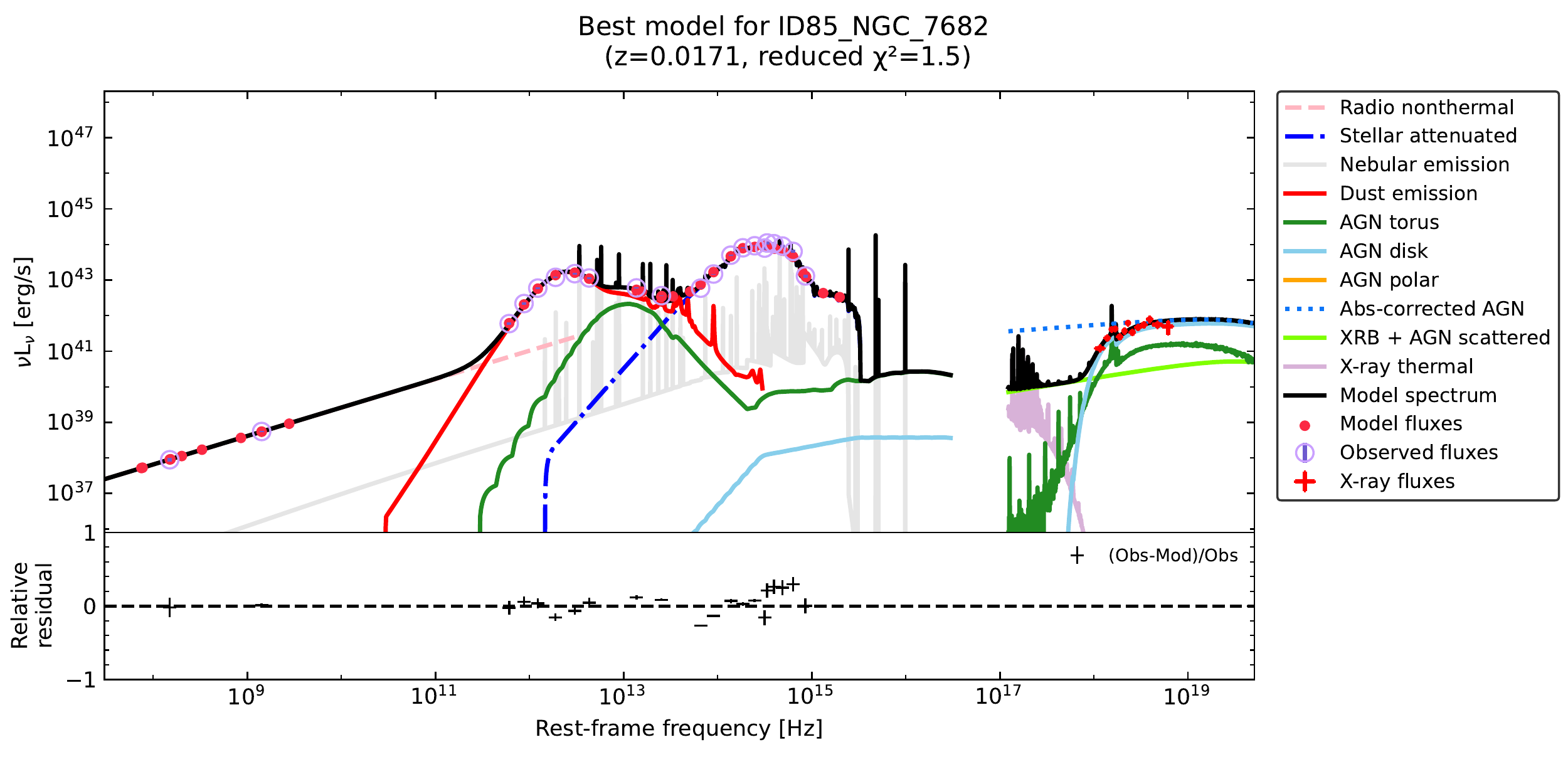}
        \caption{The same as in Figure~\ref{SED10-F} but for the sample of ID=81--85.}
\label{SED18-F}
\end{figure*}

For convenience on the multiwavelength studies in our sample, we present the hard-X-ray-to-radio SEDs and the best fitting models at the observed frame in units of flux density in Figures~\ref{SED01-F}.
We also illustrate the SEDs and the best fitting models at the rest frame in units of luminosity in Figures~\ref{SED10-F}--\ref{SED18-F}.
The photometry data are summarized in Table~\ref{TE1_photo1}--\ref{TE3_photo3}.

% Table E1
\clearpage
\begin{scriptsize}
\begin{longrotatetable}
\begin{deluxetable*}{llcccccccccccc}
\tablecaption{Photometries of our targets in mJy ($\lambda < $1~$\mu$m) \label{TE1_photo1}}
\tablehead{
\colhead{ID} & 
\colhead{Name} & 
\colhead{$E$($B-V$)} & 
\colhead{FUV} & 
\colhead{NUV} & 
\colhead{$u$} & 
\colhead{$v$} & 
\colhead{F\_$uv$} & 
\colhead{$g$} & 
\colhead{$r$} & 
\colhead{$i$} & 
\colhead{$z$} & 
\colhead{F\_$griz$} & 
\colhead{$y$} \\
 &  &  & 152.8~nm & 231.0~nm &  &  &  & 481.1~nm & 615.6~nm & 750.4~nm & 866.9~nm & P & 961.3~nm\\
 &  &  &  &  & 350.0~nm & 387.9~nm & SM & 501.6~nm & 607.7~nm & 773.3~nm & 912.0~nm & SM & \\
 &  &  &  &  & 354.3~nm &  & SD & 477.0~nm & 623.1~nm & 762.5~nm & 913.4~nm & SD & \\
(1)&(2)&(3)&(4)&(5)&(6)&(7)&(8)&(9)&(10)&(11)&(12)&(13)&(14)
}
\startdata
\hline
ID01 & NGC 34 & 0.0268 & 0.598 & 1.463 & 3.26 & 6.72 & SM & 18.22 & 23.15 & 32.05 & 36.59 & SM & 41.72\\
 &  &   & 0.017  & 0.016  & 0.77  & 1.89 &   & 1.86  & 2.33  & 3.27  & 3.70 &   & 4.18\\
ID02 & MCG-02-01-052/ & 0.0361 & $\cdots$ & $\cdots$ & 2.72 & $\cdots$ & SD & $\cdots$ & 15.05 & 17.16 & 19.26 & P & 24.05\\
 &     MCG-02-01-051 &  & $\cdots$ & $\cdots$  & 0.27 & $\cdots$ &  & $\cdots$  & 1.54  & 1.74  & 1.93 &   & 2.43\\
ID03 & MCG-02-01-052 & 0.0361 & $\cdots$ & $\cdots$ & 0.84 & $\cdots$ & SD & 5.44 & 7.15 & 8.41 & 8.37 & P & 8.95\\
 &  &  & $\cdots$ & $\cdots$  & 0.09 & $\cdots$ &   & 0.55  & 0.78  & 0.85  & 0.85 &   & 0.92\\
\enddata
\tablecomments{
Columns: (1) target ID; (2) target name; (3) Galactic extinction estimated 
by \citet{Schlegel1998}; (4--5) FUV and NUV flux densities
from the GALEX database (GR6plus7; \citealt{Bianchi2017}); (6--8) $u$-band and $v$-band flux densities, and the facility of these values,
 respectively. The adopted data of Pan-STARRS DR2 (P; e.g., \citealt{Chambers2016}), SkyMapper DR1 and DR2 (SK; e.g., \citealt{Wolf2018}), and
 SDSS DR16 (SD; \citealt{Ahumada2020}) are listed in Column (8); (9--13) $g$-band, $r$-band, $i$-band, $z$-band flux densities and the facility of these
values, respectively; (14) $y$-band flux density from Pan-STARRS. The Pan-STARRS photometries of $g$--$y$ bands are converted from the
original Kron magnitudes by multiplying 100/90 (for 10\% missing fluxes or 0.115 mag). The SkyMapper photometries of $u$--$z$
bands are converted from the original Petrosian magnitudes by multiplying 1.127 (e.g., 0.13 mag in case of typical Sersic index of 3;
\citealt{Graham2005}; \citealt{Haan2011a}). The flux densities are in units of mJy, and corrected for the Galactic extinction of Column (3), by
applying the band passes as noted in the text. Their 1$\sigma$ uncertainties are provided in the row just below. The uncertainties of $u$--$y$ bands
are corrected by adding 10\% of their flux densities to consider the dispersions among different kinds of measurements in their facilities.}
\tablenotetext{\ }{(This table is available in its entirety in machine-readable form.)}
\end{deluxetable*}
\end{longrotatetable}
\end{scriptsize}

% Table E2
\begin{tiny}
\begin{longrotatetable}
\begin{deluxetable*}{llcccccccccccc}
\tablecaption{Photometries of our targets in mJy ($\lambda = $1--200~$\mu$m) \label{TE2_photo2}}
\tablehead{
\colhead{ID} &
\colhead{Name} &
\colhead{$J$} &
\colhead{$H$} &
\colhead{$K_{\rm s}$} &
\colhead{W1} &
\colhead{W2} &
\colhead{S9W} &
\colhead{W3} &
\colhead{L18W} &
\colhead{W4} &
\colhead{PACS blue} &
\colhead{PACS green} &
\colhead{PACS red} \\
 &  & 1.235~$\mu$m & 1.662~$\mu$m & 2.159~$\mu$m & 3.353~$\mu$m & 4.603~$\mu$m & 8.228~$\mu$m & 11.56~$\mu$m & 17.61~$\mu$m & 22.09~$\mu$m & 70~$\mu$m & 100~$\mu$m & 160~$\mu$m\\
(1)&(2)&(3)&(4)&(5)&(6)&(7)&(8)&(9)&(10)&(11)&(12)&(13)&(14)
}
\startdata
\hline
ID01 & NGC 34 & $\cdots$ & $\cdots$ & $\cdots$ & 35.67 & 30.43 & 269 & 245.26 & $\cdots$ & 1240.58 & 18210 & 17680 & 10650\\
 &  & $\cdots$ & $\cdots$ & $\cdots$ & 0.20 & 0.17 & 26 & 1.36 & $\cdots$ & 6.86 & 840 & 810 & 480\\
ID02 & MCG-02-01-052/ & 30.59 & 33.25 & 31.82 & 18.32 & 14.43 & 178 & $\cdots$ & 517 & $\cdots$ & 8038 & 9876 & 7496\\
 &     MCG-02-01-051 & 0.78 & 1.20 & 1.34 & 0.11 & 0.09 & 11 & $\cdots$ & 33 & $\cdots$ & 403 & 495 & 375\\
ID03 & MCG-02-01-052 & 13.53 & 11.78 & 12.44 & 3.82 & 2.27 & $\cdots$ & $\cdots$ & $\cdots$ & $\cdots$ & 1206 & 2081 & 1960\\
 &  & 0.65 & 0.96 & 1.05 & 0.08 & 0.05 & $\cdots$ & $\cdots$ & $\cdots$ & $\cdots$ & 62 & 105 & 99\\
\enddata
\tablecomments{Columns: (1) target ID; (2) target name; (3--5) $J$-band, $H$-band, and $K_{\rm s}$-band flux densities from the 2MASS extended catalog, except for IRAS F17138--1017
from point-source catalog (e.g., \citealt{Skrutskie2006}); (6--7), (9), and (11) W1, W2, W3, and W4 flux densities from ALLWISE catalog (updated version on
February 16, 2021; \citealt{Cutri2014}) respectively; (8) and (10) S9W and L18W flux densities from AKARI/IRC mid-IR all-sky survey \citep{Ishihara2010};
(12--14) flux densities in PACS blue, green, and red bands from Herschel/PACS data (\citealt{Chu2017}); The flux densities are in units of mJy. The Galactic extinction
in $J$, $H$, $K_{\rm s}$, W1, and W2 bands are corrected by applying the band passes as noted in the text. Their 1$\sigma$ uncertainties are provided in the row just below.}
\tablenotetext{\ }{(This table is available in its entirety in machine-readable form.)}
\end{deluxetable*}
\end{longrotatetable}
\end{tiny}

% Table E3
\begin{tiny}
\begin{longrotatetable}
\begin{deluxetable*}{llccccccccccccc}
\tablecaption{Photometries of our targets in mJy ($\lambda > $200~$\mu$m) \label{TE3_photo3}}
\tablehead{
\colhead{ID} &
\colhead{Name} &
\colhead{PSW} &
\colhead{PMW} &
\colhead{PLW} &
\colhead{NVSS} &
\colhead{SUMSS} &
\colhead{WENSS} &
\colhead{GLEAM1} &
\colhead{GLEAM2} &
\colhead{TGSS} &
\colhead{GLEAM3} &
\colhead{VLSSr} &
\colhead{VLASS} &
\colhead{FIRST} \\
 &  & 250~$\mu$m & 350~$\mu$m & 500~$\mu$m & 214.1~mm & 355.6~mm & 922.1~mm & 1.50~m & 1.99~m & 2.03~m & 3.94~m & 4.06~m & 99.9~mm & 214.1~mm\\
 &  &  &  &  & 1.4~GHz & 843~MHz & 325.125~MHz & 200.5~MHz & 150.5~MHz & 147.5~MHz & 76~MHz & 73.8~MHz & 3~GHz  & 1.4~GHz\\
(1)&(2)&(3)&(4)&(5)&(6)&(7)&(8)&(9)&(10)&(11)&(12)&(13)&(14)&(15)
}
\startdata
\hline
ID01 & NGC 34 & 3573 & 1239 & 339 & 66.9 & $\cdots$ & $\cdots$ & 162.1 & 212.3 & 125.5 & $\cdots$ & $\cdots$ & [41.3] & $\cdots$\\
 &  & 214 & 75 & 21 & 2.5 & $\cdots$ & $\cdots$ & 10.6 & 26.5 & 14.3 & $\cdots$ & $\cdots$ & [1.5] & $\cdots$\\
ID02 & MCG-02-01-052/ & 2862 & 1209 & 419 & 46.7 & $\cdots$ & $\cdots$ & $\cdots$ & $\cdots$ & 123.2 & $\cdots$ & $\cdots$ & [12.7] & [33.6]\\
 &     MCG-02-01-051 & 189 & 81 & 33 & 1.8 & $\cdots$ & $\cdots$ & $\cdots$ & $\cdots$ & 13.6 & $\cdots$ & $\cdots$ & [0.4] & [3.4]\\
ID03 & MCG-02-01-052 & $\cdots$ & $\cdots$ & $\cdots$ & 4.3 & $\cdots$ & $\cdots$ & $\cdots$ & $\cdots$ & $\cdots$ & $\cdots$ & $\cdots$ & [$<$0.44] & [$<$1.0]\\
 &  & $\cdots$ & $\cdots$ & $\cdots$ & 0.5 & $\cdots$ & $\cdots$ & $\cdots$ & $\cdots$ & $\cdots$ & $\cdots$ & $\cdots$ & $\cdots$ & $\cdots$\\
\enddata
\tablecomments{
Columns: (1) target ID; (2) target name; (3--5) flux densities in PSW, PMW, and PLW bands from Herschel/SPIRE data (\citealt{Chu2017}); (6) 1.4~GHz flux
densities from NVSS catalog \citep{Condon1998}; (7) 843~MHz flux densities from SUMSS catalog \citep{Mauch2003}; (8) 325.125~MHz flux densities
from WENSS catalog \citep{Rengelink1997}; (9--10) and (12) 200.5~MHz (170--231~MHz), 150.5~MHz (147--154~MHz), and 76~MHz (72--80~MHz) flux
densities from GLEAM catalog \citep{Hurley-Walker2017}. The values of $<$5$\sigma$ of GLEAM catalog are removed; (11) 147.5~MHz flux densities from TGSS
catalog \citep{Intema2017}; (13) 73.8~MHz flux densities from VLSSr \citep{Lane2014}; (14) 3~GHz flux densities with the high angular resolution (2\farcs5)
from VLASS catalog \citep{Gordon2021}; (15) 1.4~GHz flux densities with the high angular resolution (5\arcsec) from FIRST catalog (\citealt{Becker1995};
\citealt{Helfand2015}).
Since the uncertainties of FIRST flux densities were not provided, we assume them as 10\% uncertainties.
The values in the square brackets for the VLASS and FIRST data represent the flux densities and uncertainties that are not utilized in the radio fits (see Section~\ref{sub3-7-Radio} in details);
The flux densities are in units of mJy. Their 1$\sigma$ uncertainties are provided in the row just below.}
\tablenotetext{\ }{(This table is available in its entirety in machine-readable form.)}
\end{deluxetable*}
\end{longrotatetable}
\end{tiny}

\ifnum0=1

\clearpage
\begin{scriptsize}
\begin{longrotatetable}
\begin{deluxetable*}{llcccccccccccc}
\tablecaption{Photometries of our targets in mJy ($\lambda < $1~$\mu$m) \label{TE1_photo1}}
\tablehead{
\colhead{ID} & 
\colhead{Name} & 
\colhead{$E$($B-V$)} & 
\colhead{FUV} & 
\colhead{NUV} & 
\colhead{$u$} & 
\colhead{$v$} & 
\colhead{F\_$uv$} & 
\colhead{$g$} & 
\colhead{$r$} & 
\colhead{$i$} & 
\colhead{$z$} & 
\colhead{F\_$griz$} & 
\colhead{$y$} \\
 &  &  & 152.8~nm & 231.0~nm &  &  &  & 481.1~nm & 615.6~nm & 750.4~nm & 866.9~nm & P & 961.3~nm\\
 &  &  &  &  & 350.0~nm & 387.9~nm & SM & 501.6~nm & 607.7~nm & 773.3~nm & 912.0~nm & SM & \\
 &  &  &  &  & 354.3~nm &  & SD & 477.0~nm & 623.1~nm & 762.5~nm & 913.4~nm & SD & \\
(1)&(2)&(3)&(4)&(5)&(6)&(7)&(8)&(9)&(10)&(11)&(12)&(13)&(14)
}
\startdata
\hline
ID01 & NGC 34 & 0.0268 & 0.598 & 1.463 & 3.26 & 6.72 & SM & 18.22 & 23.15 & 32.05 & 36.59 & SM & 41.72\\
 &  &   & 0.017  & 0.016  & 0.77  & 1.89 &   & 1.86  & 2.33  & 3.27  & 3.70 &   & 4.18\\
ID02 & MCG-02-01-052/ & 0.0361 & $\cdots$ & $\cdots$ & 2.72 & $\cdots$ & SD & $\cdots$ & 15.05 & 17.16 & 19.26 & P & 24.05\\
 &     MCG-02-01-051 &  & $\cdots$ & $\cdots$  & 0.27 & $\cdots$ &  & $\cdots$  & 1.54  & 1.74  & 1.93 &   & 2.43\\
ID03 & MCG-02-01-052 & 0.0361 & $\cdots$ & $\cdots$ & 0.84 & $\cdots$ & SD & 5.44 & 7.15 & 8.41 & 8.37 & P & 8.95\\
 &  &  & $\cdots$ & $\cdots$  & 0.09 & $\cdots$ &   & 0.55  & 0.78  & 0.85  & 0.85 &   & 0.92\\
ID04 & MCG-02-01-051 & 0.0360 & $\cdots$ & $\cdots$ & 1.87 & $\cdots$ & SD & $\cdots$ & 7.90 & 8.75 & 10.89 & P & 15.10\\
 &  &  & $\cdots$ & $\cdots$  & 0.19 & $\cdots$ &  & $\cdots$  & 0.80  & 0.92  & 1.09 &   & 1.53\\
\enddata
\tablecomments{
Columns: (1) target ID; (2) target name; (3) Galactic extinction estimated 
by \citet{Schlegel1998}; (4--5) FUV and NUV flux densities
from the GALEX database (GR6plus7; \citealt{Bianchi2017}); (6--8) $u$-band and $v$-band flux densities, and the facility of these values,
 respectively. The adopted data of Pan-STARRS DR2 (P; e.g., \citealt{Chambers2016}), SkyMapper DR1 and DR2 (SK; e.g., \citealt{Wolf2018}), and
 SDSS DR16 (SD; \citealt{Ahumada2020}) are listed in Column (8); (9--13) $g$-band, $r$-band, $i$-band, $z$-band flux densities and the facility of these
values, respectively; (14) $y$-band flux density from Pan-STARRS. The Pan-STARRS photometries of $g$--$y$ bands are converted from the
original Kron magnitudes by multiplying 100/90 (for 10\% missing fluxes or 0.115 mag). The SkyMapper photometries of $u$--$z$
bands are converted from the original Petrosian magnitudes by multiplying 1.127 (e.g., 0.13 mag in case of typical Sersic index of 3;
\citealt{Graham2005}; \citealt{Haan2011a}). The flux densities are in units of mJy, and corrected for the Galactic extinction of Column (3), by
applying the band passes as noted in the text. Their 1$\sigma$ uncertainties are provided in the row just below. The uncertainties of $u$--$y$ bands
are corrected by adding 10\% of their flux densities to consider the dispersions among different kinds of measurements in their facilities.}
\end{deluxetable*}
\end{longrotatetable}
\end{scriptsize}

\clearpage
\begin{tiny}
\begin{longrotatetable}
\begin{deluxetable*}{llcccccccccccc}
\tablecaption{Photometries of our targets in mJy ($\lambda = $1--200~$\mu$m) \label{TE2_photo2}}
\tablehead{
\colhead{ID} &
\colhead{Name} &
\colhead{$J$} &
\colhead{$H$} &
\colhead{$K_{\rm s}$} &
\colhead{W1} &
\colhead{W2} &
\colhead{S9W} &
\colhead{W3} &
\colhead{L18W} &
\colhead{W4} &
\colhead{PACS blue} &
\colhead{PACS green} &
\colhead{PACS red} \\
 &  & 1.235~$\mu$m & 1.662~$\mu$m & 2.159~$\mu$m & 3.353~$\mu$m & 4.603~$\mu$m & 8.228~$\mu$m & 11.56~$\mu$m & 17.61~$\mu$m & 22.09~$\mu$m & 70~$\mu$m & 100~$\mu$m & 160~$\mu$m\\
(1)&(2)&(3)&(4)&(5)&(6)&(7)&(8)&(9)&(10)&(11)&(12)&(13)&(14)
}
\startdata
\hline
ID01 & NGC 34 & $\cdots$ & $\cdots$ & $\cdots$ & 35.67 & 30.43 & 269 & 245.26 & $\cdots$ & 1240.58 & 18210 & 17680 & 10650\\
 &  & $\cdots$ & $\cdots$ & $\cdots$ & 0.20 & 0.17 & 26 & 1.36 & $\cdots$ & 6.86 & 840 & 810 & 480\\
ID02 & MCG-02-01-052/ & 30.59 & 33.25 & 31.82 & 18.32 & 14.43 & 178 & $\cdots$ & 517 & $\cdots$ & 8038 & 9876 & 7496\\
 &     MCG-02-01-051 & 0.78 & 1.20 & 1.34 & 0.11 & 0.09 & 11 & $\cdots$ & 33 & $\cdots$ & 403 & 495 & 375\\
ID03 & MCG-02-01-052 & 13.53 & 11.78 & 12.44 & 3.82 & 2.27 & $\cdots$ & $\cdots$ & $\cdots$ & $\cdots$ & 1206 & 2081 & 1960\\
 &  & 0.65 & 0.96 & 1.05 & 0.08 & 0.05 & $\cdots$ & $\cdots$ & $\cdots$ & $\cdots$ & 62 & 105 & 99\\
ID04 & MCG-02-01-051 & 17.06 & 21.47 & 19.37 & 14.50 & 12.15 & $\cdots$ & 135.90 & $\cdots$ & 540.54 & 7501 & 8382 & 5870\\
 &  & 0.44 & 0.73 & 0.82 & 0.08 & 0.08 & $\cdots$ & 0.75 & $\cdots$ & 2.99 & 329 & 364 & 244\\
\enddata
\tablecomments{Columns: (1) target ID; (2) target name; (3--5) $J$-band, $H$-band, and $K_{\rm s}$-band flux densities from the 2MASS extended catalog, except for IRAS F17138--1017
from point-source catalog (e.g., \citealt{Skrutskie2006}); (6--7), (9), and (11) W1, W2, W3, and W4 flux densities from ALLWISE catalog (updated version on
February 16, 2021; \citealt{Cutri2014}) respectively; (8) and (10) S9W and L18W flux densities from AKARI/IRC mid-IR all-sky survey \citep{Ishihara2010};
(12--14) flux densities in PACS blue, green, and red bands from Herschel/PACS data (\citealt{Chu2017}); The flux densities are in units of mJy. The Galactic extinction
in $J$, $H$, $K_{\rm s}$, W1, and W2 bands are corrected by applying the band passes as noted in the text. Their 1$\sigma$ uncertainties are provided in the row just below.}
\end{deluxetable*}
\end{longrotatetable}
\end{tiny}

\clearpage
\begin{tiny}
\begin{longrotatetable}
\begin{deluxetable*}{llccccccccccccc}
\tablecaption{Photometries of our targets in mJy ($\lambda > $200~$\mu$m) \label{TE3_photo3}}
\tablehead{
\colhead{ID} &
\colhead{Name} &
\colhead{PSW} &
\colhead{PMW} &
\colhead{PLW} &
\colhead{NVSS} &
\colhead{SUMSS} &
\colhead{WENSS} &
\colhead{GLEAM1} &
\colhead{GLEAM2} &
\colhead{TGSS} &
\colhead{GLEAM3} &
\colhead{VLSSr} &
\colhead{VLASS} &
\colhead{FIRST} \\
 &  & 250~$\mu$m & 350~$\mu$m & 500~$\mu$m & 214.1~mm & 355.6~mm & 922.1~mm & 1.50~m & 1.99~m & 2.03~m & 3.94~m & 4.06~m & 99.9~mm & 214.1~mm\\
 &  &  &  &  & 1.4~GHz & 843~MHz & 325.125~MHz & 200.5~MHz & 150.5~MHz & 147.5~MHz & 76~MHz & 73.8~MHz & 3~GHz  & 1.4~GHz\\
(1)&(2)&(3)&(4)&(5)&(6)&(7)&(8)&(9)&(10)&(11)&(12)&(13)&(14)&(15)
}
\startdata
\hline
ID01 & NGC 34 & 3573 & 1239 & 339 & 66.9 & $\cdots$ & $\cdots$ & 162.1 & 212.3 & 125.5 & $\cdots$ & $\cdots$ & 41.3 & $\cdots$\\
 &  & 214 & 75 & 21 & 2.5 & $\cdots$ & $\cdots$ & 10.6 & 26.5 & 14.3 & $\cdots$ & $\cdots$ & 1.6 & $\cdots$\\
ID02 & MCG-02-01-052/ & 2862 & 1209 & 419 & 46.7 & $\cdots$ & $\cdots$ & $\cdots$ & $\cdots$ & 123.2 & $\cdots$ & $\cdots$ & 12.7 & 33.6\\
 &     MCG-02-01-051 & 189 & 81 & 33 & 1.8 & $\cdots$ & $\cdots$ & $\cdots$ & $\cdots$ & 13.6 & $\cdots$ & $\cdots$ & 0.4 & 3.4\\
ID03 & MCG-02-01-052 & $\cdots$ & $\cdots$ & $\cdots$ & 4.3 & $\cdots$ & $\cdots$ & $\cdots$ & $\cdots$ & $\cdots$ & $\cdots$ & $\cdots$ & $<$0.44 & $<$1.0\\
 &  & $\cdots$ & $\cdots$ & $\cdots$ & 0.5 & $\cdots$ & $\cdots$ & $\cdots$ & $\cdots$ & $\cdots$ & $\cdots$ & $\cdots$ & $\cdots$ & $\cdots$\\
ID04 & MCG-02-01-051 & $\cdots$ & $\cdots$ & $\cdots$ & 42.4 & $\cdots$ & $\cdots$ & $\cdots$ & $\cdots$ & 123.2 & $\cdots$ & $\cdots$ & 12.7 & 33.6\\
 &  & $\cdots$ & $\cdots$ & $\cdots$ & 1.7 & $\cdots$ & $\cdots$ & $\cdots$ & $\cdots$ & 13.6 & $\cdots$ & $\cdots$ & 0.4 & 3.4\\
\enddata
\tablecomments{
Columns: (1) target ID; (2) target name; (3--5) flux densities in PSW, PMW, and PLW bands from Herschel/SPIRE data (\citealt{Chu2017}); (6) 1.4~GHz flux
densities from NVSS catalog \citep{Condon1998}; (7) 843~MHz flux densities from SUMSS catalog \citep{Mauch2003}; (8) 325.125~MHz flux densities
from WENSS catalog \citep{Rengelink1997}; (9--10) and (12) 200.5~MHz (170--231~MHz), 150.5~MHz (147--154~MHz), and 76~MHz (72--80~MHz) flux
densities from GLEAM catalog \citep{Hurley-Walker2017}. The values of $<$5$\sigma$ of GLEAM catalog are removed; (11) 147.5~MHz flux densities from TGSS
catalog \citep{Intema2017}; (13) 73.8~MHz flux densities from VLSSr \citep{Lane2014}; (14) 3~GHz flux densities with the high angular resolution (2\farcs5)
from VLASS catalog \citep{Gordon2021}; (15) 1.4~GHz flux densities with the high angular resolution (5\arcsec) from FIRST catalog (\citealt{Becker1995};
\citealt{Helfand2015}); The flux densities are in units of mJy. Their 1$\sigma$ uncertainties are provided in the row just below.
}
\end{deluxetable*}
\end{longrotatetable}
\end{tiny}

\fi

\bibliography{reference}{}
\bibliographystyle{aasjournal}



\end{document}